\documentclass[
    aps
    ,prb
    ,reprint
    ,amsmath
    ,amssymb
    ,superscriptaddress
    ,bibnotes
]{revtex4-2}

\pdfoutput = 1

\usepackage{hyperref}
\usepackage{amsfonts}
\usepackage{dsfont}
\usepackage{braket}
\usepackage{color}
\usepackage{colortbl}
\usepackage{graphicx}
\usepackage{xtab}
\usepackage{booktabs}
\usepackage{longtable}
\usepackage{multirow}
\usepackage{graphics}

\DeclareMathOperator{\Rep}{Rep}
\DeclareMathOperator{\sRep}{sRep}

\DeclareMathOperator{\Tr}{Tr}

\newcommand{\SL}{\mathrm{SL}_2(\mathbb{Z})}
\newcommand{\Gth}{\Gamma_\theta}

\newcommand{\ZZ}{{\mathbb Z}}

\newcommand{\cA}{{\mathcal A}}
\newcommand{\cB}{{\mathcal B}}

\newcommand{\cC}{{\mathcal C}}
\newcommand{\cF}{{\mathcal F}}
\newcommand{\cM}{{\mathcal M}}

\newcommand{\cE}{{\mathcal E}}

\newcommand{\hS}{\hat{S}}
\newcommand{\hT}{\hat{T}}

\usepackage{subfiles}

\begin{document}

\title{Modular extension of topological orders from congruence representations}

\author{Donghae Seo}
\thanks{These two authors contributed equally.}
\affiliation{Department of Physics, Pohang University of Science and Technology, Pohang, Gyeongbuk, 37673, Korea}

\author{Minyoung You}
\thanks{These two authors contributed equally.}
\affiliation{Asia-Pacific Center for Theoretical Physics, Pohang, Gyeongbuk, 37673, Korea}
\affiliation{POSTECH Basic Science Research Institute, Pohang, Gyeongbuk, 37673, Korea}

\author{Gil Young Cho}
\email{gilyoungcho@postech.ac.kr}
\affiliation{Department of Physics, Pohang University of Science and Technology, Pohang, Gyeongbuk, 37673, Korea}
\affiliation{Asia-Pacific Center for Theoretical Physics, Pohang, Gyeongbuk, 37673, Korea}
\affiliation{Center for Artificial Low Dimensional Electronic Systems, Institute for Basic Science (IBS), Pohang, Gyeongbuk, 37673, Korea}

\author{Hee-Cheol Kim}
\email{heecheol@postech.ac.kr}
\affiliation{Department of Physics, Pohang University of Science and Technology, Pohang, Gyeongbuk, 37673, Korea}
\affiliation{Asia-Pacific Center for Theoretical Physics, Pohang, Gyeongbuk, 37673, Korea}

\date{\today}

\begin{abstract}
{We present an efficient method to compute the modular extension of both fermionic topological orders and $\mathbb{Z}_2$-symmetric bosonic topological orders in two spatial dimensions, basing on congruence representations of $\SL$ and its subgroups.  To demonstrate the validity of our approach, we provide explicit calculations for topological orders with rank up to 10 for the fermionic cases and up to 6 for the bosonic cases. Along the way, we clarify the relation between fermionic rational conformal field theories, which live on the boundary of the corresponding fermionic topological orders, and modular extensions. In particular, we show that the $\SL$ representation of the R-R sector can be determined from the NS-NS sector using the modular extensions.}
\end{abstract}


\maketitle

\tableofcontents 

\section{Introduction}

{Topological orders are gapped phases of matter at zero temperature beyond the Landau paradigm, which are characterized by topology-dependent ground state degeneracy, non-Abelian geometric phases, and long-range entanglement \cite{Wen1989:degeneracy,Wen1990:rigidstates,KesWen1993:fqh,Wen2002:quantumorder,Wen2017:zoo}. Fractional quantum Hall states \cite{TsuStoGos1982:fqhe,KesWen1993:fqh,StoTsuGos1999:fqhe} and gapped quantum spin liquid states \cite{Wen1989:degeneracy,Kit2003:faulttolerant,SavBal2017:qsl} are the best known examples. The intriguing connection between topological order and both topological quantum field theory (TQFT) \cite{Wit1988:tqft} and rational conformal field theory (RCFT) \cite{MooSei1989:cft} has garnered significant attention, not only from the field of condensed matter physics but also from high energy physics and mathematics. The impetus driving the investigation of topological order, however, transcends its theoretical profundity. Interestingly, (2+1)-dimensional topological orders host exotic point-like excitations called {anyons}, which play a pivotal role in the realization of the topological quantum computation \cite{NaySimSte-etal2008:tqc,KitLau2009:tqc,Wan2010:tqc,SteLin2013:tqc}.}

{The trivial topological order is a universality class of states that can be adiabatically connected to a product state without closing the gap. A topological order is called \emph{invertible} if, after being stacked with another topological order, a state in that order can be adiabatically connected to a state in the trivial topological order. Universal properties of (2+1)-dimensional topological orders up to invertible topological orders are captured and formulated by braided fusion category (BFC) theory \cite{BerNeu2015:category,Wen2015:bosonic,LanKonWen2016:fermionic,KonZha2022:invitation}. Roughly speaking, BFCs are mathematical structures formed by the equivalence classes of anyons with their fusion and braiding data. In particular, (2+1)-dimensional bosonic and fermionic topological orders without symmetry are described by modular tensor categories (MTCs) \cite{Wen2015:bosonic,NgRowWanWen2023:reconstruction} and super-modular tensor categories (super-MTCs) \cite{LanKonWen2016:fermionic,BruGalHag-etal2017:16fold,ChoKimSeoYou2022:classification}, respectively. When taking global symmetry into account, the classification of topological orders becomes more intricate, leading to the concepts of symmetry-protected topological orders (SPTs) and symmetry-enriched topological orders (SETs) \cite{Wen2017:zoo}, whose bulk excitations are described by {BFCs with M\"uger center $\Rep(G)$} \cite{LanKonWen2017:symmetryenriched}. Here, $\Rep(G)$ denotes the category of representations of the symmetry group $G$, which is a symmetric fusion category (SFC). (The M\"uger center is a subcategory of excitations that exhibit trivial braiding with every other excitation.) We will call such BFCs ``$G$-BFCs" for brevity. }

{Each BFC have a pair of $r$-dimensional symmetric matrices $(S,T)$ associated to it. Here, the dimension $r$ is called \emph{rank}, which is equal to the number of inequivalent anyon types. The matrices $S$ and $T$ encode the mutual and self statistics of anyons, respectively. While $T$ is always unitary and diagonal, $S$ may be degenerate. For MTCs, $S$ matrices are unitary and, in this case, the pair $(S,T)$ are called \emph{modular data}. In this manuscript, we use the term ``modular data'' to describe any pair $(S,T)$, regardless of whether it corresponds to an MTC or not. Modular data are gauge-invariant and prove to be valuable tools for the analysis and classification of topological orders. Indeed, numerous previous works \cite{Wen2015:bosonic,LanKonWen2016:fermionic,LanKonWen2017:symmetryenriched,ChoGanKim2020:mtheory,ChoKimSeoYou2022:classification,NgRowWanWen2023:reconstruction,integralrank12,NgRowWen2023:rank11} that investigated topological orders heavily relied on modular data. However, it should be emphasized that in certain cases, a given set of modular data may not uniquely determine a BFC \cite{LanKonWen2017:symmetryenriched,MigSch2021:notdeterminedby}. In other words, multiple distinct BFCs can share the same $(S,T)$. For MTCs, such cases are known to occur only when the rank $e$ is sufficiently high, typically $r \geq 49$ \cite{MigSch2021:notdeterminedby}, and it is believed that in the cases of sufficiently row ranks, modular data can uniquely determine an MTC. On the other hand, there is an effort to overcome the limitations of $(S,T)$ by introducing higher-genus invariants \cite{WenWen2019:highergenus}. For generic BFCs, such ambiguity can be encountered even in low ranks \cite{LanKonWen2017:symmetryenriched}.}

{Recently, it was realized that the modular data are closely related to \emph{congruence representations} of $\SL$ or one of its subgroups \cite{BonRowWanZha2018:congruence,NgRowWanWen2023:reconstruction,ChoKimSeoYou2022:classification}. Specifically, the modular data of an MTC and a super-MTC form a \emph{projective} congruence representation of $\SL$ and $\Gth < \SL$, respectively. Moreover, all congruence representations of $\SL$ were classified recently \cite{NgWanWil2023;symmetric}, allowing a systematic classification of modular data by constructing candidate data from the representations and then checking consistency conditions. Using this approach, the modular data of MTCs and super-MTCs were classified up to rank 11 (and partially up to rank 12) \cite{ng2023classification} and 10 \cite{ChoKimSeoYou2022:classification}, respectively. A natural extension of these previous studies would be classifying modular data of $G$-BFCs, which characterize SPTs and SETs. In fact, the modular data of $G$-BFCs do not form group representations by themselves, thus it is impossible to directly apply a similar approach mentioned above. However, a given $G$-BFC can be mapped to an MTC in two ways: symmetry breaking and modular extension. Symmetry breaking is a map from a given $G$-BFC to its underlying MTC \cite{LanKonWen2017:symmetryenriched}. In this work, we make the symmetry breaking procedure explicit for the $G = \mathbb{Z}_2$ case in terms of modular data. Since the modular data of MTCs can be classified by projective congruence representations of $\SL$, a systematic study of modular data of $\mathbb{Z}_2$-BFCs is then possible. In this way, we can classify $\mathbb{Z}_2$-BFCs without imposing any upper bound on fusion coefficients or total quantum dimension.}

{Mathematically, a modular extension is an MTC that contains a super-MTC or a $G$-BFC. Physically, it corresponds to a gauged version of a given super-MTC or $G$-BFC. If a given $(d+1)$-dimensional topological order can be realized on a $d$-dimensional lattice, then the topological order is called \emph{anomaly-free}. In terms of category theory, this anomaly-free condition is translated into the condition of existence of a modular extension \cite{LanKonWen2016:fermionic,LanKonWen2016:mex,LanKonWen2017:symmetryenriched}. For super-MTCs or $\mathbb{Z}_2$-BFCs, it is known that they always admit modular extensions and thus are anomaly-free. However, it is possible that a given ``modular data" of these BFCs may turn out to be invalid, indicating that it cannot actually realize a super-MTC or a $\mathbb{Z}_2$-BFC. The existence of modular extensions serves as a necessary condition for confirming the validity of candidate modular data \cite{LanKonWen2017:symmetryenriched}. Consequently, by explicitly computing the modular extensions, we can rule out such invalid modular data. 

Moreover, modular extensions are closely related to the boundary theory of fermionic topological orders as we will explain in Sec.~\ref{sec:FRCFT}. However, a systematic approach to compute the modular data of modular extensions has remained elusive. In this work, we provide a systematic procedure for computing the modular data of fermionic topological orders and $\mathbb{Z}_2$-SETs, by using a hidden structure of modular data. Roughly speaking, the modular data of these topological orders can be decomposed via a basis change to different ``sectors,'' each of which transforms in a congruence representation of $\SL$ or one of its index-$3$ congruence subgroups. We compute the modular data of all 16 modular extensions of each super-MTC found in Ref.~\cite{ChoKimSeoYou2022:classification} and all modular extensions of each $\mathbb{Z}_2$-BFC we found. Importantly, the fact that each super-MTC admits 16 modular extensions is consistent with the previous theorem \cite{BruGalHag-etal2017:16fold} and serves as a strong evidence for the validity of the super-MTCs given in Ref.~\cite{ChoKimSeoYou2022:classification}. We also find that each one of the modular extensions of the new classes of super-MTCs indeed aligns with those previously given in Ref.~\cite{RowSolZha2023:neargroup}.}

In Sec.~\ref{sec:background}, we provide an overview of related concepts, including the mathematical description of (2+1)-dimensional topological orders, the relation between modular data and congruence representations, the algebraic consistency conditions that BFCs satisfy, the relation between gauging and modular extensions, and the relation between fermionic topological orders and both TQFT and RCFT. In Sec.~\ref{sec:method}, we introduce our method for classifying modular data of $\mathbb{Z}_2$-BFCs and compute modular extensions of super-MTCs and $\mathbb{Z}_2$-BFCs from congruence representations of $\SL$ and its congruence subgroups. We summarize the result of this work and make some comments on observations in Sec.~\ref{sec:result}. The lists of modular extensions of super-MTCs and $\mathbb{Z}_2$-BFCs are presented in Appendix~\ref{app:ftomex} and Appendix~\ref{app:Z2modular extension}, respectively.

\section{\label{sec:background} Background}

In this section we introduce some essential background. We focus mainly on the physical intuition and significance of the concepts and concrete formulae, while details are omitted. For more detailed and mathematical explanation, readers are encouraged to see, for example, Refs.~\cite{BakKir2001:lectures,Mac2013:workingmathematician,BerNeu2015:category,KonZha2022:invitation}.

\subsection{Category-theoretic description of (2+1)-dimensional topological orders}

Within the framework of category theory, the characterization of a (2+1)-dimensional topological order denoted as $\mathsf{C}$ is achieved through the association with a BFC $\mathcal{C}$. Roughly speaking, a BFC consists of a set of labels for different anyon types and their fusion/braiding rules. The fusion and braiding rules should satisfy some consistency equations known as the pentagon and hexagon equations. These BFCs are in other words BFCs with M\"uger center $\Rep(G)$, where the SFC 
is denoted by $\mathcal{E}$ \cite{LanKonWen2016:fermionic}. Here, the global symmetry of $\mathsf{C}$ is characterized by $\mathcal{E}$, which can be understood as a sub-BFC of local pointlike excitations. For example, when $\mathsf{C}$ possesses a finite onsite bosonic symmetry, denoted by a group $G$, its local pointlike excitations carry group representations of $G$, thereby giving rise to $\mathcal{E} = \Rep(G)$, i.e., the category of group representations of $G$. On the other hand, if the symmetry contains the fermion-number parity, being denoted by $G^f$, then $\mathcal{E} = \sRep(G^f)$, i.e., the category of group super-representations of $G^f$.

Being the simplest case, (2+1)-dimensional bosonic topological orders without symmetry have $\mathcal{E} = \mathcal{B}_0$, where $\mathcal{B}_0$ is the category of finite-dimensional Hilbert spaces. In this case, BFCs describing these topological orders become MTCs by themselves, i.e., all pointlike excitations are nonlocal, thus braid nontrivially among themselves. In juxtaposition, (2+1)-dimensional fermionic topological orders without symmetry have $\mathcal{E} = \mathcal{F}_0 \equiv \sRep(\mathbb{Z}_2^f)$ where the nontrivial element of $\mathbb{Z}_2^f \simeq \mathbb{Z}_2$ represents the fermion-number parity. When this is the case, the BFCs describing these topological orders are called super-MTCs \cite{BruGalHag-etal2017:16fold}. If a nontrivial (2+1)-dimensional topological order is endowed with a nontrivial bosonic (fermionic) symmetry group $G$ ($G^f$), then such $\mathsf{C}$ is called SET \cite{Wen2017:zoo} and described by a $G$-BFC ($G^f$-BFC). In contrast, if the topological order is trivial, then such $\mathsf{C}$ is called SPT \cite{Wen2017:zoo}.


\subsection{Modular data and congruence representation}

Though a BFC is in principle defined by the solutions of the pentagon and hexagon equations, i.e., a set of $F$- and $R$-tensors \cite{BakKir2001:lectures}, these tensors are gauge-dependent quantities and difficult to classify. As noted in the introduction, it is more convenient to work with the modular data $(S, T)$ which are gauge-invariant. 


Let us explain in more  detail the meaning of modular data. First, the elements of the first row of $S$-matrix are real and none of them are equal to zero. The $a$-th element in the first row divided by $S_{11}$ corresponds to the \emph{quantum dimension} $d_a$ of anyon $a$, i.e., $d_a = S_{1a}/S_{11}$. The vacuum is labeled by $1$ and has $d_1 = 1$. The normalization factor is written as $S_{11} = 1/D$ and $D \equiv \sqrt{\sum_i d_i^2}$ is called \emph{total quantum dimension}. This total quantum dimension $D$ is an important physical observable captured by the topological entanglement entropy \cite{KitPre2006:entropy,LevWen2006:entropy}. The nondiagonal entries of $S_{ab}$ capture information about the \emph{mutual braiding} of anyons $a$ and $b$. On the other hand, the $T$-matrix is diagonal. The $a$-th diagonal element of it is given by $T_{aa} = e^{2\pi i \theta_a}$, where $\theta_a$ is the \emph{topological spin} of anyon $a$. The topological spin of the vacuum is always given by $\theta_1 = 0$.

The modular data have an interesting connection to congruence representations, and this fact has been exploited recently to make progress on the classification problem \cite{NgRowWanWen2023:reconstruction,ChoKimSeoYou2022:classification, NgRowWen2023:rank11}. For an MTC, the modular data form a projective congruence representation of $\SL$ \cite{NgRowWanWen2023:reconstruction}. Since congruence representations of $\SL$ are classified \cite{NgWanWil2023;symmetric}, the modular data of an MTC can be established through a two-step process. First, we identify candidate $S$- and $T$-matrices from the classification table of congruence representations. Then, we test that the pair ($S,T$) satisfies all the necessary consistency conditions, thereby confirming it as modular data\cite{NgRowWanWen2023:reconstruction}. 

Things are similar for super-MTCs, but with some modification. The modular data of a super-MTC always can be decomposed as \cite{BonRowWanZha2018:congruence}
\begin{equation}
    S = \frac{1}{2}
    \begin{pmatrix}
        1 & 1 \\
        1 & 1 
    \end{pmatrix} 
    \otimes \hat{S}, 
    \quad T = 
    \begin{pmatrix}
        1 & 0 \\
        0 & -1
    \end{pmatrix}
    \otimes \hat{T}.
\end{equation}
Here, due to sign ambiguity, only $\hat{T}^2$ is well defined. Physically, this corresponds to extracting the local fermion part from the modular data, while leaving only the information of nonlocal anyons that braid nontrivially among themselves in $(\hat{S},\hat{T}^2)$. Importantly, $\hat{S}$ is unitary and, together with $\hat{T}^2$, generates a projective congruence representation of $\Gth$ \cite{BonRowWanZha2018:congruence}, which is an index-3 congruence subgroup of $\SL$. Since the congruence representations of $\Gth$ can be obtained from those of $\SL$, a similar approach can be applied to compute the modular data of super-MTCs \cite{ChoKimSeoYou2022:classification}.

At  first glance, it seems that the modular data of a general BFC other than an MTC or a super-MTC does not have such a connection because they do not form a representation of any group. However, by ``breaking the symmetry" of a given BFC \cite{LanKonWen2017:symmetryenriched}, we can get a modular data of an MTC. Hence, they are somehow indirectly connected to the congruence representations of $\SL$. Another connection with congruence representations comes through modular extensions \cite{LanKonWen2016:mex}. In Sec.~\ref{sec:Z2-enriched} and Sec.~\ref{sec:Z2-mex}, we introduce the procedures to obtain the  modular data of both symmetry-broken MTCs and modular extensions for bosonic $G = \mathbb{Z}_2$ case. Using the method, we classify (2+1)-dimensional bosonic $\mathbb{Z}_2$-SETs and their modular extensions up to rank 6.

\subsection{\label{subsec:consistency} Consistency conditions for modular data}

The modular data of BFCs should satisfy a set of algebraic consistency conditions \cite{Wen2015:bosonic,LanKonWen2016:fermionic}. The conditions listed below are necessary conditions;  a set of sufficient conditions in not known. However, we believe that they are stringent enough to considerably narrow down the candidates for valid topological orders.

\paragraph{Verlinde's formula.} The $S$-matrix of a BFC satisfies
\begin{equation} \label{eq:Verlinde-BFC}
    \frac{S_{il} S_{jl}}{S_{1l}} = \sum_k N^{ij}_k S_{kl}
\end{equation} 
where $N^{ij}_k$ are nonnegative integers called \emph{fusion coefficients}. The coefficients further satisfy 
\begin{equation}
\begin{aligned}
    N^{ij}_k = N^{ji}_k, \quad N^{1i}_j = \delta_{ij}, \\
    \sum_k N^{ik}_1 N^{kj}_1 = \delta_{ij}, \quad \sum_k N^{ij}_k N_k = N_i N_j,
\end{aligned}
\end{equation}
where $(N_i)_{jk} \equiv N^{ij}_k$. These coefficients define fusion rules of anyons, e.g., $a \times b = \sum_c N^{ab}_c c$ where $a$, $b$, and $c$ denote anyons. If a given BFC is an MTC, then Eq.~(\ref{eq:Verlinde-BFC}) can be written as
\begin{equation}
    N^{ij}_k = \sum_l \frac{S_{il} S_{jl} S_{kl}^*}{S_{1l}}.
\end{equation}
Hence, for an MTC, the fusion coefficients are uniquely determined by its $S$-matrix.

\paragraph{Rational condition.} Define 
\begin{multline}
    V_{ijkl}^r = N^{ij}_r N^{kl}_{\bar{r}} + N^{il}_r N^{jk}_{\bar{r}} + N^{ik}_r N^{jl}_{\bar{r}} \\
    -(\delta_{ir} + \delta_{jr} + \delta_{kr} + \delta_{lr}) \sum_m N^{ij}_m N^{kl}_{\bar{m}},
\end{multline}
then
\begin{equation}
    \sum_{r} V_{ijkl}^r \theta_r \in \mathbb{Z}.
\end{equation}

\paragraph{Balancing equation.} The elements of $S$-matrix are given by
\begin{equation}
    S_{ij} = \frac{1}{D} \sum_k N^{ij}_k e^{2\pi i (\theta_i + \theta_j - \theta_k)} d_k
\end{equation}
where $D = \sqrt{\sum_i d_i^2}$ is the total quantum dimension.

\paragraph{Frobenius-Schur indicator.} The quantity 
\begin{equation}
    \nu_k = \frac{1}{D^2} \sum_{ij} N^{ij}_k d_i d_j \cos[4\pi(\theta_i - \theta_j)],
\end{equation}
called the Frobenius-Schur indicator, satisfies $\nu_k \in \mathbb{Z}$ if $k = \bar{k}$.

\paragraph{Weak modularity.} Define
\begin{equation}
    \Theta = \frac{1}{D} \sum_i e^{2\pi i \theta_i} d_i^2,
\end{equation}
then 
\begin{equation}
    S^\dagger T S = \Theta T^\dagger S^\dagger T^\dagger.
\end{equation}
Importantly, $\Theta = |\Theta| e^{2\pi i \frac{c}{8}}$ where $c$ is the chiral central charge mod 8. From this relation, one can compute $c$ of a given topological order. For super-MTCs, however, $|\Theta| = 0$, thus one cannot rely on this method and should compute their modular extensions. Recently, it was proposed that the chiral central charge of super-MTCs can be obtained without computing modular extensions if one uses congruence representations of $\Gth$ \cite{ChoKimSeoYou2022:classification}.

\subsection{\label{sec:gauging} Gauging and modular extension}

Gauging is a promotion of a global symmetry to a gauge symmetry. In terms of category theory, the gauging process is understood as computing modular extensions of a given BFC with M\"uger center $\Rep(G)$. A modular extension $\mathcal{M}$ of a UBFC with M\"uger center $\Rep(G)$ $\mathcal{E}$, denoted as $\mathcal{C}$, is a UMTC where $\mathcal{C}$ is faithfully embedded \cite{LanKonWen2016:fermionic,LanKonWen2016:mex}. A modular extension $\mathcal{M}$ of $\mathcal{C}$ is minimal if the total quantum dimension of $\mathcal{M}$ is twice of that of $\mathcal{C}$. In this paper, ``modular extension'' will always mean a minimal modular extension. 

It was recently proven that every super-MTC admits a modular extension \cite{JohReu2023:mex}, which is in mathematics literature called spin MTC. Interestingly, it was proven in advance that if a super-MTC has a modular extension, then there should be 16 different modular extensions \cite{LanKonWen2016:mex}. These have distinct $c$ mod $8$, and are related to each other by stacking with the $p+ip$ superconductor, the generator of invertible fermionic topological orders \cite{BruGalHag-etal2017:16fold}. For a topological order with bosonic symmetry described by $\mathcal{E} = \Rep(G)$, the modular extensions all have the same $c$ mod $8$, and are related to each other by stacking with $G$-SPTs \cite{LanKonWen2016:mex}. For our purposes, since we work with fermionic or $\ZZ_2$-symmetric modular data given in terms of a degenerate $S$-matrix, modular extension in practice will simply mean adding in additional anyons to make $S$ non-degenerate.

{
For fermionic topological orders specified by $(\cC, c)$ where $\cC$ is a super-MTC and $c$ is the chiral central charge, the corresponding modular extension $\cM$ (which is fixed by $c$ mod $8$) can be considered a bosonzied description of the same phase, obtained by gauging $\ZZ_2$-fermion parity symmetry. 

Gauging a bosonic $\ZZ_2$-symmetry gives rise to an emergent $\ZZ_2$ one-form symmetry, which is generated by the Wilson line operator corresponding to the SFC $\cE = \Rep(\ZZ_2).$ We will refer to this line operator as the $\ZZ_2$-charge, or as ``$q$.'' We note that, in the formalism of Ref.~\cite{BarBonCheWan2019:prb}, a $G$-SET is described by an MTC together with a permutation action $\rho$ of $G$ on the anyons, plus the data of symmetry fractionalization, which is classified by $H^2_\rho(G, \cA)$. In our formalism, a $G$-SET is described instead by a BFC with $\Rep(G)$ as its M\"uger center. This formalism automatically takes into account the symmetry fractionalization data, i.e., different symmetry fractionaliztion classes correspond to different BFCs. While it is difficult to see the exact correspondence concretely, in principle the BFC-based classification takes into account all possible symmetry fractionalization patterns \cite{lan2018classification}.

We note, however, that these BFCs may not always be distinguishable solely based on the modular data; we may need $R$- and $F$-tensors in addition to the $S$- and $T$-matrices to tell them apart \cite{LanKonWen2017:symmetryenriched}. Sometimes, different symmetry fractionalization classes lead to the same degenerate $S$ but different fusion rules (recall that a degenerate $S$-matrix only partially fixes the fusion rules); sometimes, they may lead to the same $S$ and the same fusion rules. In such cases, it is expected that modular extension will resolve the  ambiguity: two BFCs which share the same modular data will nevertheless give rise to distinct modular data after modular extension. At the level of modular data, this manifests itself as the same degenerate modular data leading to 
more than one class of non-degenerate modular data after extension (here, modular data of different classes cannot be related to each other via stacking with $G$-SPTs). }

For a given topological order described by $\mathcal{C}$ to be anomaly-free, i.e., realizable on a lattice model in the same dimension, $\mathcal{C}$ must have a modular extension, or in more physical terms, the symmetry must be ``gaugable." Despite their significance in shedding light on the intricate interplay between topological orders and symmetry, a systematic method for computing modular extensions has not yet been developed. In Sec.~\ref{sec:method}, we introduce an algorithmic method for computing modular data of modular extensions. Using this method, we shall compute modular data of modular extensions of super-MTCs up to rank 10 and those of $\mathbb{Z}_2$-BFCs up to rank 6.

\subsection{More on (2+1)-dimensional fermionic topological orders}

\subsubsection{\label{subsubsec:structure} Structure of modular extensions}
A modular extension of a super-MTC are given by a spin MTC, which is a regular MTC containing a fermionic quasiparticle, i.e., an anyon $\psi$ such that $\psi \times \psi = 1$ and $\theta_\psi = -1$. If there are multiple anyons which have this property, we need to specify a distinguished fermion.

The presence of a fermion $\psi$ gives rise to the following structure in the spin MTC \cite{BruGalHag-etal2017:16fold, BonRowWanZha2018:congruence}:
\begin{itemize}

    \item Each anyon $\alpha$ has mutual braiding $\pm 1$ with $\psi$.  
    We divide the anyons into two sectors $\cC_{\rm NS} \oplus \cC_{\rm R}$ based on whether they braid trivially or nontrivially with $\psi$.

    \item In the trivial-braiding sector $\cC_{\rm NS}$, $\psi \alpha := \alpha \times \psi$ is always a distinct anyon from $\alpha$. We can thus (non-canonically) divide the anyons in $\cC_{\rm NS}$ into two sets $\Pi_0$ and $\psi \Pi_0$ of equal size.

    \item In the nontrivial-braiding sector $\cC_{\rm R}$, $\alpha \times \psi$ can be either disintct from $\alpha$ or equal to $\alpha$. We refer to the former case as a long orbit (with respect to fusion with $\psi$), while the latter case is referred to as a short orbit (the anyon ``absorbs'' $\psi$). Long orbits can again be divided into two sets of equal size, $\Pi_v$ and $\psi \Pi_v$; we call the set of short orbits  $\Pi_\sigma$.
    
\end{itemize}
The modular data then have the form
\begin{equation} \label{eq:spinMTC}
\begin{aligned}
    S^{\rm spin} &= 
    \begin{pmatrix}
        \frac{1}{2}\hat{S} & \frac{1}{2}\hat{S} & A & A & X \\
        \frac{1}{2}\hat{S} & \frac{1}{2}\hat{S} & -A & -A & -X \\
        A^T & -A^T & B & -B & 0 \\
        A^T & -A^T & -B & B & 0 \\
        X^T & -X^T & 0 & 0 & 0
    \end{pmatrix}, \\
    T^{\rm spin} &=
    \begin{pmatrix}
        \hat{T} & 0 & 0 & 0 & 0 \\
        0 & -\hat{T} & 0 & 0 & 0 \\
        0 & 0 & \hat{T}_v & 0 & 0 \\
        0 & 0 & 0 & \hat{T}_v & 0 \\
        0 & 0 & 0 & 0 & \hat{T}_\sigma
    \end{pmatrix}.
\end{aligned}
\end{equation}
in the basis $\Pi = \Pi_0 \cup \psi \Pi_0 \cup \Pi_v \cup \psi \Pi_v \cup \Pi_\sigma$. 
These matrices are written in a block form \cite{BonRowWanZha2018:congruence}
\begin{equation} \label{eq:spinsector}
\begin{aligned}
    \tilde{S}^{\rm spin} &= 
    \begin{pmatrix}
        \hat{S} & 0 & 0 & 0 & 0 \\
        0 & 0 & 2A & 2X & 0 \\
        0 & 2A^T & 0 & 0 & 0 \\
        0 & 2X^T & 0 & 0 & 0 \\
        0 & 0 & 0 & 0 & B
    \end{pmatrix}, \\
    \tilde{T}^{\rm spin} &=
    \begin{pmatrix}
        0 & \hat{T} & 0 & 0 & 0 \\
        \hat{T} & 0 & 0 & 0 & 0 \\
        0 & 0 & \hat{T}_v & 0 & 0 \\
        0 & 0 & 0 & \hat{T}_\sigma & 0 \\
        0 & 0 & 0 & 0 & \hat{T}_v
    \end{pmatrix}.
\end{aligned}
\end{equation}
in the basis $\tilde{\Pi} = \Pi_0^+ \cup \Pi_0^- \cup \Pi_v^+ \cup \Pi_\sigma \cup \Pi_v^-$ where $\Pi_0^\pm = \{X \pm X^\psi | X \in \Pi_0\}$ and $\Pi_v^\pm = \{Y \pm Y^\psi | Y\in\Pi_v\}$.

\subsubsection{Torus Hilbert space of (2+1)-dimensional spin topological quantum field theory}
\label{sec:Hilbert_space}

It is well known that an MTC defines a (2+1)-dimensional TQFT, which assigns a Hilbert space of states to each closed 2-manifold and linear maps between such Hilbert spaces to cobordisms between 2-manifolds. On the torus, given an MTC $\mathcal{C}$, each state can be labeled by an anyon $a \in \mathcal{C}$; we denote this as $\ket{a} \in \mathcal{H}(T^2)$. From this point of view, the $S$- and $T$-matrices then tell us how these states on the torus transform into each other under modular transformations of the torus.

In the fermionic case, there should also be a correspondence between the categorical description of anyons and the transformation properties of the torus Hilbert space of a TQFT. Because of fermions, the TQFT should depend on the spin structure on manifolds; such a TQFT is called a spin TQFT. On the torus, we have four spin structures, specifying whether the boundary conditions for fermions are antiperiodic (Neveu-Schwarz; NS) or periodic (Ramond; R) along the two cycles of the torus. We shall label the four tori as NS-NS, NS-R, R-NS, and R-R.

Given a spin MTC $\mathcal{M}$, or a modular extension of a super-MTC, we can construct the Hilbert space on the four tori by condensing the distinguished fermion $\psi$. States in each sector take the form \cite{DelGaiGom2021:anomaly}:
\begin{itemize}
    \item NS-NS: \begin{equation}
        \frac{1}{\sqrt{2}} \left( |a \rangle + |a \times \psi \rangle \right) 
    \end{equation} where $a \in \Pi_0$.

    \item NS-R: \begin{equation}
        \frac{1}{\sqrt{2}} \left( |a \rangle - |a \times \psi \rangle \right) 
    \end{equation} where $a \in \Pi_0$.

    \item R-NS: \begin{equation}
        \frac{1}{\sqrt{2}} \left( |x \rangle + |x \times \psi \rangle \right) 
    \end{equation} where $x \in \Pi_v$ and 
    \begin{equation}
        |m\rangle 
    \end{equation}
    where $m \in \Pi_\sigma$.

    \item R-R: \begin{equation}
        \frac{1}{\sqrt{2}} \left( |x \rangle - |x \times \psi \rangle \right) 
    \end{equation} where $x \in \Pi_v$  \footnote{In Ref.~\cite{DelGaiGom2021:anomaly}, there are also \emph{puncture states} in the R-R sector, which are odd under fermion parity. These states are irrelevant for our purposes --- they do not contribute to the modular extension, or give rise to fermionic RCFT characters --- so we ignore them.}.
\end{itemize}

We see that, given an MTC $\cC$, the basis change of Eq.~(\ref{eq:spinsector}) corresponds precisely to the sector basis: $\Pi_{\text{NS-NS}} = \Pi_0^+, \Pi_{\text{NS-R}} = \Pi_0^-, \Pi_{\text{R-NS}} = \Pi_v^+ \cup \Pi_\sigma, \Pi_{\text{R-R}} = \Pi_v^-$. Treating each sector as a block, we can re-write Eq.~(\ref{eq:spinsector}) as:
\begin{equation}
\begin{aligned}
    \tilde{S} &= 
    \begin{pmatrix}
        \hS & 0 & 0 \\
        0 & 0& \hS' \\
        0 & \hS'^T & 0    
    \end{pmatrix} \oplus S_{\text{R-R}}, \\
    \tilde{T} &= \begin{pmatrix} 0 & \hT  & 0  \\ \hT & 0&  0  \\ 0 & 0 & T_{\text{R-NS}}  \end{pmatrix} \oplus   T_{\text{R-R}}.
    \label{eq:spinsector2}
\end{aligned}
\end{equation}
As expected, $S$ takes the NS-R sector to R-NS sector and vice versa, while $T$ takes the NS-NS sector to NS-R sector and vice versa. The first three sectors mix under $\SL$, while the R-R sector transforms into itself under $\SL$. 

In Appendix~\ref{app:induced}, we show that the $\SL$ representation of the first three sectors is in fact the induced representation of the $\Gth$-representation of the NS-NS sector, which defines the corresponding super-MTC.

\subsubsection{Relation to two-dimensional fermionic rational conformal field theory}
\label{sec:FRCFT}

Let us begin by recalling some well-known facts about RCFT and their realtion to MTCs 
\cite{MooSei1989:cft}. In an RCFT, the torus partition function can be expressed as
\begin{equation}
    Z(\tau,\bar{\tau}) := \Tr\left[q^{L_0 - \frac{c}{24}} \bar{q}^{\bar{L}_0 - \frac{c}{24}}\right] = \sum_{i,j} M_{ij} \chi_i(\tau) \bar{\chi}_j(\bar{\tau})
\end{equation}
where $\chi_i(\tau)$ are \emph{characters}, $i = 1, \cdots, N$ where $N$ is finite, and $M_{ij}$ is some matrix. The characters $\chi_i(\tau)$ transform covariantly under $\SL$,
\begin{equation}
\begin{aligned}
    \chi_i(\tau + 1) &= \sum_j T_{ij} \chi_j(\tau), \\
    \chi_i(-1/\tau) &= \sum_j S_{ij} \chi_j(\tau),
\end{aligned}
\end{equation}
while $Z(\tau, \bar{\tau})$ is invariant.

Bulk $S$ and $T$ form a projective representations of $\SL$, whereas boundary $S$ and $T$ form a linear representation of $\SL$. They are related via
\begin{equation} \label{eq:MTC-CFT}
\begin{aligned}
    S_{\rm CFT} &= S_{\rm MTC}, \\
    T_{\rm CFT} &= e^{-2\pi i c/24} T_{\rm MTC},
\end{aligned}
\end{equation}
where $c$ is the chiral central charge. Recall that bulk $S$ and $T$ determine $c$ mod $8$ but not mod $24$.

In a fermionic conformal field theory \cite{HsiNakTac2021:fermionicminimal,Kul2021:fermionicminimal,DuaLeeLeeLi2023:classification}, the partition function depends on the spin structure. We label the four spin structures on the torus as NS-NS, NS-R, R-NS, and R-R (or, equivalently, NS, $\widetilde{\text{NS}}$, R, and $\widetilde{\text{R}}$), where NS and R denote antiperiodic and periodic boundary conditions respectively. NS-R ($\widetilde{\text{NS}}$) denotes antiperiodic along the spatial direction and periodic along the temporal direction, while R-NS (R) denotes periodic along the spatial direction and anatiperiodic along the temporal direction. The partition functions are:
\begin{equation}
\begin{aligned}
    Z^{\rm NS}(\tau,\bar{\tau}) &= \Tr_{\mathcal{H}_{\rm NS}}\left[q^{L_0 - \frac{c}{24}} \bar{q}^{\bar{L}_0 - \frac{c}{24}}\right] \\
    Z^{\widetilde{\rm{NS}}}(\tau,\bar{\tau}) &= \Tr_{\mathcal{H}_{\rm NS}}\left[(-1)^F q^{L_0 - \frac{c}{24}} \bar{q}^{\bar{L}_0 - \frac{c}{24}}\right] \\
      Z^{\rm R}(\tau,\bar{\tau}) &= \Tr_{\mathcal{H}_{\rm R}}\left[q^{L_0 - \frac{c}{24}} \bar{q}^{\bar{L}_0 - \frac{c}{24}}\right] \\
    Z^{\widetilde{\rm{R}}}(\tau,\bar{\tau}) &= \Tr_{\mathcal{H}_{\rm R}}\left[(-1)^F q^{L_0 - \frac{c}{24}} \bar{q}^{\bar{L}_0 - \frac{c}{24}}\right] 
\end{aligned}
\end{equation}

In a fermionic RCFT, the partition function of each sector can be written in terms of a finite number of characters:  $\chi_i^{\rm NS}(\tau)$ for the NS sector, $\chi_i^{\widetilde{\rm NS} }(\tau)$ for the $\widetilde{ {\rm NS}}$ sector, etc. Since we are interested in the interplay between RCFT and MTC (which captures the chiral information about RCFTs), for the rest of the paper we focus on the characters and not the full partition functions.

The simplest example of a fermionic RCFT is the free Majorana fermion CFT. It has one character per sector (except for the $\widetilde{\text{R}}$ sector which is empty), given by 
\begin{equation}
\begin{aligned}
    \chi^{\rm NS}(\tau) &= \sqrt{\frac{\theta_3(\tau)}{\eta(\tau)}} \\
    \chi^{\widetilde{\rm NS}}(\tau) &= \sqrt{\frac{\theta_4(\tau)}{\eta(\tau)}} \\
    \chi^{\rm R}(\tau) &= \sqrt{\frac{\theta_2(\tau)}{\eta(\tau)}} \\
    \chi^{\widetilde{\rm R}}(\tau) &= 0.
\end{aligned}
\end{equation}
It is well-known that these can be written in terms of the characters of a bosonic RCFT, in this case the Ising CFT. The Ising CFT has three chracters $\chi_h$ with conformal dimension $h = 0, \frac{1}{2}, \frac{1}{16}$. We can write the above characters as
\begin{equation}
\begin{aligned}
    \chi^{\rm NS} &= \frac{1}{2} \left(\chi_0 + \chi_{\frac{1}{2}}\right) \\
    \chi^{\widetilde{\rm NS}} &= \frac{1}{2} \left(\chi_0 - \chi_{\frac{1}{2}}\right) \\
    \chi^{\widetilde{\rm R}} &= \frac{\sqrt{2}}{2} \chi_{\frac{1}{16}}.
\end{aligned}
\end{equation}
Such a process is referred to as \emph{fermionization}. Similarly to how each bosonic RCFT character corresponded to a basis state on the torus (labeled by an anyon) of the bulk TQFT, in a fermionic RCFT a character corresponds to a basis state in a particular sector. The basis states, as we have seen in Sec. \ref{sec:Hilbert_space}, are written as  a linear combination of bosonic  basis states. 

On the other hand, reversing the above  gives us the Ising characters in terms of the fermionic characters. If we know the fermionic characters, we can \emph{bosonize} the theory to obtain bosonic RCFT characters.

A  less trivial example is given by the WZW $\mathrm{SU}(2)_6$ model, which has seven characters with conformal weight $h = 0, \frac{3}{32}, \frac{1}{4}, \frac{15}{32}, \frac{3}{4},\frac{35}{32}, \frac{3}{2}$, respectively. The corresponding MTC can be thought of as a modular extension of the super-MTC $\mathrm{PSU}(2)_6$, which contains  $h = \{ 0,  \frac{3}{2} , \frac{1}{4}, \frac{3}{4} \}$; $h = \{ \frac{3}{32}, \frac{35}{32}, \frac{15}{32} \}$ correspond to anyons which are added in to form the modular extension. Given the seven characters of $\mathrm{SU}(2)_6$, the linear combination
\begin{equation}
    \begin{aligned}
 \chi^{\rm NS}_0 =  \chi_0 +\chi_{\frac{3}{2}}, \ \chi^{\rm NS}_\frac{1}{4} = \chi_{\frac{1}{4}} + \chi_{\frac{3}{4}}  \\ 
\chi^{\widetilde{\rm{NS}} }_0 = \chi_0 - \chi_{\frac{3}{2}}, \ \chi^{\widetilde{\rm{NS}}}_\frac{1}{4} =  \chi_{\frac{1}{4}} - \chi_{\frac{3}{4}}\\
\chi^{\rm R-NS}_\frac{3}{32} = \chi_{\frac{3}{32}} + \chi_{\frac{35}{32}}, \ \chi^{\rm R-NS}_\frac{15}{32}  = \chi_{\frac{15}{32}}\\
\chi^{\widetilde{\rm{R}}}_\frac{3}{32} =  \chi_{\frac{3}{32}} - \chi_\frac{35}{32}
 \end{aligned}
\end{equation}
gives rise to a fermionic RCFT with two characters in each sector, except for the R-R sector which has a single character \cite{BaeLee2021:supersymmetry}. 

If we begin with a bosonic RCFT and fermionize it, we will automatically obtain the fermionic characters in every sector. Conversely, if we know the fermionic characters in every sector, we can obtain the bosonized theory. In practice, however, we may only have partial information in the fermionic side. See, for example, Refs. ~\cite{BaeDuaLee-etal2021:frcft, BaeDuaLee-etal2022:bootstrap, DuaLeeLeeLi2023:classification}, which classify characters in the NS-NS and R-NS sectors. The constraints of fermionic RCFTs actually allow us to obtain the NS-R and R-NS sector characters immediately via modular transformations from the NS-NS sector. However, the R-R sector cannot be obtained in such a manner. Thus, the question arises: given the NS-NS sector characters, what can we say about the R-R sector?

Ref.~\cite{BenLin2020:lessons} initiated the investigation of this question. Although they focused on partition functions rather than the characters, the basic idea is one which we shall follow: given a fermionic RCFT, there should exist a consistent bosonization, and this fact can be used to constrain the R-R sector. 
As we have seen earlier, on the spin-TQFT level, bosonization essentially corresponds to modular extension. Hence, if we can compute the modular extensions, we could treat this question systematically, and give an answer at least at the level of representations.

More specifically, suppose we are given the NS-NS characters of a fermionic RCFT. The NS-NS sector determines the associated $\Gth$ representation $\rho$ and the chiral central charge $c$. $\rho$ determines a super-MTC, and $c$ specifies a modular extension. This modular extension gives the $\SL$-representation for all the sectors; in particular, the representation of the R-R sector, which was otherwise not reliably available, can be determined in this way. 
    
\section{\label{sec:method} Method}

In this section, we elaborate the methods for computing modular data of modular extensions of super-MTCs and $\mathbb{Z}_2$-BFCs, and classiying modular data of $\mathbb{Z}_2$-BFCs. Our methods take advantage from congruence representations of $\SL$ which are completely classified recently \cite{NgWanWil2023;symmetric}. To make the connection to representation theory more manifest, we assume that modular data form \emph{linear} congrunece representations, rather than projective congruence representations, i.e., we assume that the $T$-matrices carry the chiral central charge factor $e^{-2\pi i \frac{c}{24}}$. In other words, we adopt the convention from CFT as shown in Eq.~(\ref{eq:MTC-CFT}). Furthermore, for brevity, all representations henceforth are assumed to be congruence representations.

\subsection{Modular extensions of (2+1)-dimensional fermionic topological orders}

A modular extension of a super-MTC is called spin MTC, which is an MTC with a distinguished fermion $\psi$. For each super-MTC, there always exists a modular extension \cite{JohReu2023:mex}, and there are always 16 different modular extensions \cite{BonRowWanZha2018:congruence}. Since a spin MTC is an MTC, it can be captured by the classification of (2+1)-dimensional bosonic topological orders without symmetry. However, they are in general of very high ranks: for a super-MTC of rank $r$, its modular extensions can have ranks between $\frac{3}{2}r$ and $2r$. Furthermore, in principle, if there are more than one anyons with fermionic self statistics, then one has to choose which should be $\psi$, and each choice leads to a different super-MTC. Therefore, it would be more efficient if we can compute modular extensions from the data of each super-MTC.

As explained in Sec.~\ref{subsubsec:structure}, the modular data of a modular extension can be written in a block form in the $\tilde{\Pi}$ basis. Interestingly, the upper blocks of the matrices in Eq.~(\ref{eq:spinsector}) are equivalent to the generators of a representation induced from $\Gth$ to $\SL$. Specifically, for given $\hat{S}$ and $\hat{T}^2$, the induced representation is formed by
\begin{equation} \label{eq:indrep}
    S^{\rm ind} = 
    \begin{pmatrix}
        \hat{S} & 0 & 0 \\
        0 & 0 & \hat{S}^2 \\
        0 & \mathds{1} & 0
    \end{pmatrix}, \quad
    T^{\rm ind} =
    \begin{pmatrix}
        0 & \hat{T}^2 & 0 \\
        \mathds{1} & 0 & 0 \\
        0 & 0 & (\hat{S}\hat{T}^2)^{-1}
    \end{pmatrix}.
\end{equation}
Since the modular data are symmetric matrices, the induced represenation shown in Eq.~(\ref{eq:indrep}) should be symmetrized. The symmetrization is done by conjugating $S^{\rm ind}$ and $T^{\rm ind}$ with a unitary matrix, yielding
\begin{equation} \label{eq:symrep}
    S^{\rm sym} = 
    \begin{pmatrix}
        \hat{S} & 0 & 0 \\
        0 & 0 & M \\
        0 & M^T & 0
    \end{pmatrix}, \quad
    T^{\rm sym} = 
    \begin{pmatrix}
        0 & \hat{T} & 0 \\
        \hat{T} & 0 & 0 \\
        0 & 0 & W
    \end{pmatrix},
\end{equation}
where $M = \hat{T}\hat{S}P^\dagger A^\dagger$ with a unitary matrix $P$ that diagonalizes $\hat{S}^3\hat{T}^{-2}$ and $A$ given by $A^TA = P^* \hat{S}\hat{T}^2\hat{S}^3 P^\dagger$, and $W = V P \hat{S}^3 \hat{T}^{-2} P^\dagger V^\dagger$ is a diagonal matrix with an orthogonal matrix $V$ that diagonalizes $A^TA$. Then, the matrices in Eq.~(\ref{eq:symrep}) are equivalent to the upper blocks of $\tilde{S}^{\rm spin}$ and $\tilde{T}^{\rm spin}$ up to orthogonal transformations that act nontrivially only on the same $T$-eigenvalue subspaces.

Next, we note that $\hat{T}_v$ is contained in $W$ though the explicit entries are not known a priori. This is consistent with the fact that the rank of a modular extension of a given rank-$r$ super-MTC is between $\frac{3r}{2}$ and $2r$. We thus check for all possible choices of $\hat{T}_v$ that are contained in $W$ whether there exists a $\SL$-representation with whose $T$-matrix is equal to the choice $\hat{T}_v$. For given candidates of $\SL$-representations, we construct candidate modular data in the sector basis. Lastly, we go back to the $\Pi$ basis and check the consistency conditions given in Sec.~\ref{subsec:consistency}.

One of 16 different modular extensions is determined by the chiral central charge factor $e^{-2\pi i \frac{c}{12}}$ of $\hat{T}^2$ that we begin with. Suppose that we are given a modular data $(\hat{S},\hat{T}^2)$. As we mentioned above, $\hat{T}^2$ is assumed to carry the chiral central charge factor $e^{-2\pi i \frac{c}{12}}$. Note that we have chosen $c$ mod $12$ rather than mod  $\frac{1}{2}$. Indeed, there are 24 different $\Gth$-representations that give rise to the same fermion-quotient modular data, whose phase factors differ by the factor $e^{2\pi i \frac{1}{24}}$. In addition, since the chiral central charge of MTCs are defined mod $8$, there are threefold choices for $c$ for a modular extension. The choice of a chiral central charge mod $8$ fixes the specific  modular extension of the 16 possibilities.

\subsection{\label{sec:Z2-enriched} Classification of (2+1)-dimensional \texorpdfstring{$\mathbb{Z}_2$}{Z\_2}-enriched bosonic topological orders}

(2+1)-dimensional $\mathbb{Z}_2$-enriched bosonic topological orders are classified by $\mathbb{Z}_2$-BFCs. For a given bosonic topological order $\mathsf{C}$ with an onsite $\mathbb{Z}_2$ symmetry, each indecomposable local excitation is labeled by an irreducible representation of $\mathbb{Z}_2$. Since there is only one nontrivial irreducible representation, there is only one nontrivial label for local excitation, denoted by $q$. The anyons of $\mathsf{C}$ are then categorzied by whether they are acted trivially or nontrivially by the $\mathbb{Z}_2$ symmetry. 

Suppose that an anyon $a$ is acted nontrivially by the $\mathbb{Z}_2$ symmetry. In other words, the anyons change the anyon label when they fuse with $q$, i.e., $a \times q = a^q \neq a$. We say that such anyons like $a$ are in the $A$-sector, while $a^q$ are in the $A^q$-sector. Since $q \times q = 1$, distinction between $A$- and $A^q$-sector is arbitrary. In contrast, the anyons that remain unchanged under fusion with $q$ are said to be in the $B$-sector, i.e., for $b$ in the $B$-sector, $b \times q = b^q = b$. The modular data $S^{\mathbb{Z}_2}$ and $T^{\mathbb{Z}_2}$ should satisfy $S^{\mathbb{Z}_2}_{ij} = S^{\mathbb{Z}_2}_{i^qj^q}$ and $T^{\mathbb{Z}_2}_{ij} = T^{\mathbb{Z}_2}_{i^qj^q}$, thus they can be written as the following block form:
\begin{equation} \label{eq:MTCoverRepZ2}
    S^{\mathbb{Z}_2} = 
    \begin{pmatrix}
        A & A & B \\
        A & A & B \\
        B^T & B^T & C
    \end{pmatrix}, \quad 
    T^{\mathbb{Z}_2} = 
    \begin{pmatrix}
        T_A & 0 & 0 \\
        0 & T_A & 0 \\
        0 & 0 & T_B
    \end{pmatrix}.
\end{equation}
Now, we break the $\mathbb{Z}_2$ symmetry to get an MTC. We note that the anyons in the $A$-sector are merely doubled by the $\mathbb{Z}_2$ symmetry action, while those in the $B$-sector can be understood as a composite of the anyons exchanged by the action. For example, let us consider the toric code MTC, with anyon labels $\{1,e,m,f\}$, and a $\mathbb{Z}_2$ symmetry that exchanges $e$ and $m$. Then, after enriching the $\mathbb{Z}_2$ symmetry, the anyons in the $A$-sector and the $A^q$-sector are $\{1,f\}$ and $\{q,f^q\}$, respectively, while the only anyon in the $B$-sector is $e \oplus m$. Breaking the symmetry is basically undoing this, which leads us to
\begin{equation}
    S^{\rm SB} = 
    \begin{pmatrix}
        2A & B & B \\
        B^T & \frac{C+K}{2} & \frac{C-K}{2} \\
        B^T & \frac{C-K}{2} & \frac{C+K}{2}
    \end{pmatrix}, \quad 
    T^{\rm SB} = 
    \begin{pmatrix}
        T_A & 0 & 0 \\
        0 & T_B & 0 \\
        0 & 0 & T_B
    \end{pmatrix}.
\end{equation}
Here, the square matrix $K$ is an extra data which cannot directly be computed from Eq.~(\ref{eq:MTCoverRepZ2}). By an orthogonal transformation, the matrices in Eq.~(\ref{eq:MTCoverRepZ2}) are simulataneously block-diagonalized:
\begin{equation} \label{eq:SBbd}
    \tilde{S}^{\rm SB} = 
    \begin{pmatrix}
        2A & \sqrt{2}B & 0 \\
        \sqrt{2}B^T & C & 0 \\
        0 & 0 & K
    \end{pmatrix}, \quad 
    \tilde{T}^{\rm SB} = 
    \begin{pmatrix}
        T_A & 0 & 0 \\
        0 & T_B & 0 \\
        0 & 0 & T_B
    \end{pmatrix}
\end{equation}
where the upper and the lower blocks form a $\SL$-representation, respectively.

From a known list of $\SL$-representations, we get candidates for the upper blocks in Eq.~(\ref{eq:SBbd}) and reconstruct the matrices in Eq.~(\ref{eq:MTCoverRepZ2}) from them. Then, we numerically solve Verlinde's formula shown in Eq.~(\ref{eq:Verlinde-BFC}) to get nonnegative integer solutions $N^{ij}_k$ that satisfy the other consistency conditions explained in Sec.~\ref{subsec:consistency} as well. It is important to emphasize that one modular data candidate can yields different sets of solutions. Furthermore, though a solution satisfies all the consistency conditions, it can be invalid by disallowing modular extensions. Thus, we explicitly compute the modular data of modular extensions of $\mathbb{Z}_2$-BFCs.

The $\SL$-representation we start out with may be reducible. We note that our method may miss certain modular data coming from unresolved $\SL$ representations \cite{NgRowWanWen2023:reconstruction}. We also note that, direct sums only involving 2d and 1d irreps lead to the symmetry-broken MTC being integral, and since integral MTCs have been classified up to rank 12 \cite{integralrank12}, we do not need to construct the representation separately in these cases; we can simply take the modular data of symmetry-broken MTCs of Ref. \cite{integralrank12} and obtain the corresponding $\ZZ_2$-BFCs.

\subsection{\label{sec:Z2-mex} Modular extension of (2+1)-dimensional \texorpdfstring{$\mathbb{Z}_2$}{Z\_2}-enriched bosonic topological orders}

As we explained above, satisfying all the consistency conditions given in Sec.~\ref{subsec:consistency} does not guarantee that a candidate modular data is valid: if a candidate modular data does not admit a modular extension, then the candidate is invalid. Thus, we explicitly compute the modular data of modular extensions of the candidate $\mathbb{Z}_2$-BFCs. Physically, a modular extension corresponds to a gauged SET. 

Similarly to the fermionic modular extension case, the starting point is to write the modular data of the modular extension in a block form. To do so, we again decompose the label set of anyons into disjoint subsets: $\Pi = \Pi_{00} \cup \Pi_{10} \cup \Pi_{01} \cup \Pi_{11}$. Here, $\Pi_{00} \cup \Pi_{10}$ and $\Pi_{01} \cup \Pi_{11}$ are the sets for the anyons that braid trivially and nontrivially with the $\mathbb{Z}_2$ charge $q$, respectively, and $\Pi_{00} \cup \Pi_{01}$ and $\Pi_{10} \cup \Pi_{11}$ are the sets for the anyons that are variant and invariant under the fusion with $q$, respectively. For example, if $a \in \Pi_{10}$, then $a \times q = a$ and $M_{aq} \equiv \theta_{a\times q} / \theta_a \theta_q = 1$. Overall, we have
\begin{equation}
\begin{aligned}
    S^{\mathbb{Z}_2\text{-gauged}} &=
    \begin{pmatrix}
        A & A & B & X & X \\
        A & A & B & -X & -X \\
        B^T & B^T & C & 0 & 0 \\
        X^T & -X^T & 0 & Y & -Y \\
        X^T & -X^T & 0 & -Y & Y
    \end{pmatrix}, \\
    T^{\mathbb{Z}_2\text{-gauged}} &=
    \begin{pmatrix}
        T_A & 0 & 0 & 0 & 0 \\
        0 & T_A & 0 & 0 & 0 \\
        0 & 0 & T_B & 0 & 0 \\
        0 & 0 & 0 & T_X & 0 \\
        0 & 0 & 0 & 0 & -T_X
    \end{pmatrix}.
\end{aligned}
\end{equation}
These matrices can be simultaneously block-diagonalized by an orthogonal transformation
\begin{equation} \label{eq:Z2sector}
\begin{aligned}
    \tilde{S}^{\mathbb{Z}_2\text{-gauged}} &= 
    \begin{pmatrix}
        2A & \sqrt{2}B & 0 & 0 & 0 \\
        \sqrt{2}B^T & C & 0 & 0 & 0 \\
        0 & 0 & 0 & 2X & 0 \\
        0 & 0 & 2X^T & 0 & 0 \\
        0 & 0 & 0 & 0 & 2Y
    \end{pmatrix}, \\
    \tilde{T}^{\mathbb{Z}_2\text{-gauged}} &= 
    \begin{pmatrix}
        T_A & 0 & 0 & 0 & 0 \\
        0 & T_B & 0 & 0 & 0 \\
        0 & 0 & T_A & 0 & 0 \\
        0 & 0 & 0 & 0 & T_X \\
        0 & 0 & 0 & T_X & 0
    \end{pmatrix},
\end{aligned}
\end{equation}
where the upper blocks of $\tilde{S}^{\mathbb{Z}_2\text{-gauged}}$ and $\tilde{T}^{\mathbb{Z}_2\text{-gauged}}$ are the same as the upper blocks of $\tilde{S}^{\rm SB}$ and $\tilde{T}^{\rm SB}$ in Eq.~(\ref{eq:SBbd}). Importantly, the lower blocks of $\tilde{S}^{\mathbb{Z}_2\text{-gauged}}$ and $\tilde{T}^{\mathbb{Z}_2\text{-gauged}}$ form an induced representation from a $\Gamma_0(2)$-representation, whose $T$-matrix is given by $T_A$.

$\Gamma_0(2)$ is an index-3 subgroup of $\SL$, which is generated by $\mathfrak{s}\mathfrak{t}^2\mathfrak{s}$ and $\mathfrak{t}$. Given a $\Gamma_0(2)$-representation $\pi$, its induced representation is generated by
\begin{equation}
\label{eq:Z2ind}
\begin{aligned}
    S^{\rm ind} &= 
    \begin{pmatrix}
        0 & \left(\pi(\mathfrak{s}\mathfrak{t}^2\mathfrak{s})\pi(\mathfrak{t})\right)^2 & 0 \\
        \pi(\mathds{1}) & 0 & 0 \\
        0 & 0 & \pi(\mathfrak{s}\mathfrak{t}^2\mathfrak{s})\pi(\mathfrak{t})
    \end{pmatrix}, \\
    T^{\rm ind} &= 
    \begin{pmatrix}
        \pi(\mathfrak{t}) & 0 & 0 \\
        0 & 0 & \pi(\mathds{1}) \\
        0 & \pi(\mathfrak{s}\mathfrak{t}^2\mathfrak{s})\left(\pi(\mathfrak{s}\mathfrak{t}^2\mathfrak{s})\pi(\mathfrak{t})\right)^2 & 0
    \end{pmatrix}.
\end{aligned}
\end{equation}
Again, these matrices are equivalent to the lower blocks of the matrices in Eq.~(\ref{eq:Z2sector}). We could do a similar procedure as the fermionic modular extension case, i.e., symmetrizing the induced representation. That is one possible way, but we proceed in a different way; we compute the eigenvalues of $T^{\rm ind}$ and search $\SL$-representations with the same set of $T$-eigenvalues. Since the $\SL$-representations are all given in a symmetric form, we do not need to concern ourselves with the symmetrization of the induced representation.

Given the $\SL$-representations with the desired $T$-eigenvalues, we construct the candidate modular data in the basis of Eq.~(\ref{eq:Z2sector}) and go back to the $\Pi$ basis. By checking the consistency conditions given in Sec.~\ref{subsec:consistency}, we get modular data of modular extensions of $\mathbb{Z}_2$-BFCs.

\section{\label{sec:result} Result and discussion}

We list our results in Appendices \ref{app:ftomex} and \ref{app:Z2modular extension}. Here, we make some observations and comments on our results. 

\subsection{Fermionic modular extensions}

It had been known that unitary super-MTCs always admit 16 unitary modular extensions \cite{LanKonWen2016:mex, JohReu2023:mex}. We find that they also admit 16 pseudo-unitary modular extensions, where the only difference is that the quantum dimensions of the anyon in $\cC_{\rm R}$ come with a minus sign. In terms of modular data, they differ by conjugating with a sign-diagonal matrix whose negative elements are on the rows corresponding to the added anyons. So each unitary super-MTC admits 16 pairs of modular extensions, where each pair contians a unitary and a pseudo-unitary MTC. 

We find that non-unitary super-MTCs also admit 16 pairs of modular extensions. Both MTCs in the pair are non-unitary, but the two are related again by a minus sign on the quantum dimensions of the anyons of the $\cC_{\rm R}$-sector. Similarly to the unitary cases, they are connected by conjugating with a sign-diagonal matrix. This leads us to conjecture that, just as for unitary super-MTCs, non-unitary super-MTCs also admit 16 pairs of modular extensions.

In particular, every entry belonging to the new classes of  modular data found in Ref. \cite{ChoKimSeoYou2022:classification}, with $D^2 = 472.379 $ and $475.151$, do have modular extensions (on the level of modular data). This is consistent with the explicit realization of these modular data via super-MTCs carried out  for specific examples from these classes in Ref.  \cite{RowSolZha2023:neargroup}. 

We find that for the vast majority of cases, a given super-MTC leads to the modular extensions of two different ranks; in such a case, those modular extensions whose $c$ differ by an integer have the same rank. For example, for the rank 8 super-MTC $\mathrm{PSU}(2)_{14}$ (rank 8 \# 66 in our table), modular extensions with $c = \frac{1}{8} + k$, ($k \in \ZZ$) have rank 13, such as the $\mathrm{SO}(7)_{-3}$ MTC which corresponds to $c = \frac{1}{8}$, but those modular extensions with $c = \frac{1}{8} + k + \frac{1}{2}$ have rank 15, such as the $\mathrm{SU}(2)_{14}$ MTC with $c = \frac{21}{8}$. There are exceptions to this, however: for $\mathrm{PSU}(2)_4$ at rank 4, every modular extension has rank 7, while for $\mathrm{PSU}(2)_6 \boxtimes_{\cF_0} \mathrm{PSU}(2)_6$ (the fermionic stacking of two $\mathrm{PSU}(2)_6$ super-MTCs) of rank 8, every modular extension is of rank 14.

We present the chiral central charge $c$, quantum dimensions $d_i$, and topological spins $\theta_i$ of modular extensions for each super-MTC in Appendix~\ref{app:ftomex}. We always choose $D > 0$ when computing $\Theta$. If we had chosen $D < 0$, which is a conventional choice for non-unitary MTCs, then the central charge would have shifted by $4$.

For the rank-8 super-MTCs \#7--\#10 \cite{ChoKimSeoYou2022:classification}, we failed to compute the modular data of modular extensions of them within our methodology since their R-NS sector spins are threefold degenerate. This degeneracy allow the freedom of 3-dimensional orthogonal transformations keeping the $T$-matrix diagonal. Fortunately, for unitary super-MTCs \#7 and \#8, each one of their modular extensions are known \cite{ChoKimSeoYou2022:classification}, and we find that each one of modular extensions of \#9 and \#10 corresponds to $5^B_{\#8} \boxtimes 4^B_{3}$ and $5^B_{\#7} \boxtimes 4^B_{-3}$, respectively \cite{NgRowWanWen2023:reconstruction,Wen2015:bosonic}. Starting from the known modular extension, we can get another with the chiral central charge shifted by $\frac{1}{2}$, by stacking it with the Ising MTC and condensing a boson. By doing the same procedure successively, we can obtain all 16 modular extensions.

\subsubsection{The Ramond-Ramond sector of fermionic rational conformal field theory}

We mentioned in Sec. \ref{sec:FRCFT} that modular extensions can be used to gain information about the R-R sector from the NS-NS sector. Armed with our data of modular extensions, we can now see how this works in practice. 

Let us illustrate this with a pair of examples: the $c = \frac{21}{8}$ entry and the $c = \frac{225}{8}$ entry from the Rank four table of Ref.~\cite{DuaLeeLeeLi2023:classification}.   It has four characters in the NS-NS sector, and we identify the corresponding super-MTC as $\mathrm{PSU}(2)_{14}$. The $c = \frac{21}{8}$ theory actually corresponds to the $\mathrm{SU}(2)_{14}$ modular extension, which is of rank 15. Performing the basis change of Eq.~\ref{eq:spinsector}, we obtain the R-R sector representation as  
\begin{equation}
\begin{aligned}
        S_{\text{R-R}} &= \frac{1}{2}\begin{pmatrix}
            1 &  - 1 & -\sqrt{2} \\ -1 &  1 & -\sqrt{2} \\ -\sqrt{2} &  -\sqrt{2} & 0 
        \end{pmatrix}, \\
        T_{\text{R-R}} &= e^{-2 \pi i /16} \begin{pmatrix}
             1& 0 & 0 \\ 0 &  -1 & 0 \\ 0& 0& e^{2 \pi i \frac{3}{16}} 
        \end{pmatrix}. 
\end{aligned}
    \end{equation}
On the other hand, the $c = \frac{225}{8}$ theory has $c = \frac{33}{8}$ mod $8$ and the corresponding modular extension is of rank 13. The corresponding R-R sector representation is a trivial 1-dimensional representation, so this has a single R-R character which is constant. 

We note the Ref.~\cite{bae2021energy} has proved a simple criterion for whether the R-R sector partition function is a constant or not in terms of the R-NS sector exponents, without making use of modular extension --- so the whole modular extension is not necessary if we are interested simply in whether the R-R sector is a constant or not. Our approach, however, allows us to obtain the exact representation of the R-R sector, not just whether it is a constant.

We may also mention that Ref.~\cite{KawYah2023:codecft} has given a construction of a class of fermionic RCFTs, which automatically constructs the R-R sector as well as the other sectors, though the MTCs corresponding to this class of theories are all Abelian. Our approach works even for RCFTs whose corresponding MTCs contain non-Abelian anyons.

\subsection{Classification of \texorpdfstring{$\mathbb{Z}_2$}{Z\_2}-symmetry enriched topological orders and modular extensions}

Because (2+1)-dimensional $\ZZ_2$-SPTs are classified by $H^3(\ZZ_2, U(1)) = \ZZ_2$ \cite{Chen_2013}, we expect that $\mathbb{Z}_2$-BFCs will have two modular extensions \cite{BarBonCheWan2019:prb, LanKonWen2016:mex}. However, in certain cases it may happen that an SET ``absorbs'' an SPT under stacking \cite{Cheng_2020}. Such  a phenomenon was already observed in Ref.~\cite{LanKonWen2017:symmetryenriched}, for some rank-5 BFCs which come from rank-4 MTCs and lead to rank-9 modular extension. In such a case, a BFC has only one modular extension. We observe the same phenomenon for some rank-6 BFCs leading to a rank 10 modular extension: for rank-6 BFC labeled ``$6_1^{\zeta^1_2}$; self-dual,'' and similar, we find only a single modular extension. In Appendix~\ref{app:absorption}, we provide a simple test in terms of the modular extension for when this occurs. 

It may also happen that two distinct SETs (and hence BFCs) are indistinguishable from their modular data --- they will have different $R$- and $F$-symbols, but the same $S$, $T$. This can happen if there are different symmetry fractionalization classes for the same symmetry action on anyons. Sometimes, two BFCs with the same $S, T$ may be distinguished based on the fusion rules, but sometimes even the fusion rules fail to distinguish them, and one needs to take the modular extension to distinguish them, as discussed for certain rank 5 examples in Ref. \cite{LanKonWen2017:symmetryenriched}.

We find that for $6_1^{\zeta^1_2}$ (and similar) from Ref. \cite{LanKonWen2017:symmetryenriched}, there are two different BFCs with the same $S$ and $T$, but that they can be distinguished by the fusion rules. The theory labeled ``$6_1^{\zeta^1_2}$; self-dual'' has self-dual fusion rules (every anyon is its own antiparticle) leads to a single modular extension (which absorbs stacking with $\ZZ_2$-SPTs); the other, non-self-dual  fusion rule leads to a different set of modular extensions.

\section{Conclusion}

In this work, we computed modular extensions of super-MTCs and $\mathbb{Z}_2$-BFCs up to rank 10 and 6 using induced representations from the congruence representations of $\Gth$ and $\Gamma_0(2)$, respectively. Furthermore, we classified $\mathbb{Z}_2$-BFCs up to rank 6 by their modular data, using congruence representations of $\SL$.

Our method for computing modular data of modular extensions is much more efficient than previous approaches based on finding plausible fusion coefficients and checking the other consistency conditions. High-rank modular extensions were inaccessible in the previous approach. We also classified the modular data of $\mathbb{Z}_2$-BFCs without setting any upper limit of fusion coefficients or total quantum dimensions. Therefore, at least up to unresolved cases, we can argue that the classification is complete up to rank 6.

The fact that the modular data form congruence representations of $\SL$ or its subgroups is advantageous for studying topological orders. First, apart from general representations of $\SL$, the congruence representations are those of finite groups and much restricted, leading to the recent complete classification \cite{NgWanWil2023;symmetric}. Furthermore, congruence representations of the congruence subgroups of $\SL$ can be obtained from those of $\SL$, using the concept of induced representations \cite{ChoKimSeoYou2022:classification}. These advantages lead to recent progress in classification of modular data \cite{NgRowWanWen2023:reconstruction,ChoKimSeoYou2022:classification}.

While we have worked out the systematic relationship between the R-R sector of fermionic RCFTs and modular extensions of super-MTCs, this still does not determine the  R-R sector characters themselves (i.e., the concrete functions of the modular parameter $\tau$) from the NS-NS sector characters. It would be interesting to find out a way to use our representation information about the R-R sector to compute the R-R sector characters explicitly.

Our methods may be generalized to symmetry groups other than $\ZZ_2$, though the challenge is to find the relevant congruence subgroups of $\SL$ and their representations. For more complicated groups, such as $\ZZ_2 \times \ZZ_2$, the `t Hooft anomaly given by $H^4(G, U(1))$ can be nontrivial and hence there may be an obstruction to finding a modular extension even when the modular data define a genuine $G$-BFC. Since such an anomaly defines a (3+1)D topological order, this may lead to a computationally efficient approach to studying (3+1)D topological orders through BFCs/anomalous (2+1)D SETs, an example of which was Ref. \cite{Wang_2017}.

\acknowledgments
We thank Xiao-Gang Wen for motivating us to investigate modular extensions. We acknowledge the financial support by Samsung Science and Technology Foundation under Project Number SSTF-BA2002-05, the NRF of Korea (Grant No. RS-2023-00208291, No. 2023M3K5A1094810, No. 2023M3K5A1094813, and No. 2023R1A2C1006542) funded by the Korean Government (MSIT), the Air Force Office of Scientific Research under Award No. FA2386-22-1-4061, and Institute of Basic Science under project code IBS-R014-D1. M.Y. is also supported by the Basic Science Research Institute Fund, whose NRF grant number is 2021R1A6A1A10042944.

\appendix

\section{Induced representation inside the modular extension}
\label{app:induced}

Given the NS-NS sector $\Gth$-representation, we can determine the representations of the NS-R and R-NS sectors, since these sectors are connected via modular transformations. This has been shown in Appendix C of Ref. \cite{you2023gapped}. Here, we give an alternative, more physical, argument in terms of RCFT characters. 

Given $\chi^{\rm NS}_i (\tau)$, we can define a  basis of characters in the NS-R and R-NS sectors simply by transforming the NS-NS sector basis characters: 
\begin{equation}
    \begin{aligned}
\chi^{\widetilde{\rm NS}}_i (\tau) &:= \chi^{\rm NS}(t \cdot \tau),
\\ \chi^{\rm R}_i(\tau) &:= \chi^{\rm NS}(st \cdot \tau).
    \end{aligned}
\end{equation}
Note that this basis may not be the basis in which the $\SL$-representation is symmetric; however, once we obtain the $\SL$-representation, we can simply do a basis change via a unitary matrix appropriately. 

Now, the action of $S$ is
\begin{equation}
    \begin{aligned}
        \chi^{\rm NS}_i (s \cdot \tau) &= \hS_{ij} \chi^{\rm NS}_j (\tau), \\
        \chi^{\widetilde{\rm NS}}_i (s \cdot \tau) &= \chi^{\rm NS}(st \cdot \tau)  = \chi^{\rm R}_i(\tau), \\
       \chi^{\rm R}_i(s \cdot \tau) &= \chi^{\rm NS}_i (s^2 t \cdot \tau) =\chi^{\rm NS}_i(t s^2 \cdot \tau) \\ &= (\hS^2)_{ij} \chi^{\rm NS}_i(t \cdot \tau)  = (\hS^2)_{ij} \chi^{\widetilde{\rm NS}}_j (\tau).
    \end{aligned}
\end{equation}
In other words, acting on the basis $\begin{pmatrix}
    \chi^{\rm NS}_i(\tau) \\ 
    \chi^{\widetilde{\rm NS}}_i (\tau)\\
 \chi^{\rm R}_i(\tau)
\end{pmatrix}$, we have 
\begin{equation}
    S = 
    \begin{pmatrix}
        \hat{S} & 0 & 0 \\
        0 & 0 & \hat{S}^2 \\
        0 & \mathds{1} & 0
    \end{pmatrix}.
\end{equation}
The action of $T$ is
\begin{equation}
    \begin{aligned}
        \chi^{\rm NS}_i (t \cdot \tau) &=  \chi^{\widetilde{\rm NS}}_i, \\
       \chi^{\widetilde{\rm NS}}_i (t \cdot \tau) &=  
         \chi^{\rm NS}(t^2 \cdot \tau)  = \hT_{ij} \chi^{\rm NS}_j(\tau), \\
        \chi_i^{\rm R} (t \cdot \tau) &= \chi_i^{\rm NS} (t st \cdot \tau)    = \chi_i^{\rm NS} (st (st^2)^{-1} \cdot \tau) \\&= (\hS \hT^2)^{-1}_{ij} \chi_i^{\rm NS} (st \cdot \tau) =   (\hS \hT^2)^{-1}_{ij} \chi_i^{\rm R} (\tau).
    \end{aligned}
\end{equation}
This leads to 
\begin{equation} 
    T =
    \begin{pmatrix}
        0 & \hat{T}^2 & 0 \\
        \mathds{1} & 0 & 0 \\
        0 & 0 & (\hat{S}\hat{T}^2)^{-1}
    \end{pmatrix}.
\end{equation}

This is nothing but the $\SL$-representation induced from the $\Gth$-representation $(\hS, \hT^2)$. Since this induced representation is the $\SL$-representation acting on the NS-NS, NS-R, and R-NS sectors, it is nothing but the first block of Eq.~(\ref{eq:spinsector2}), i.e., a part of the $\SL$-representation of the modular extension, simply in a different basis. 

In fact, $T_{\text{R-NS}} = \begin{pmatrix}
    T_v & 0 \\ 0& T_\sigma
\end{pmatrix}$ can be easily computed from the above: it is simply the eigenvalues of $(\hS \hT^2)^{-1}.$ The form of Eq.~(\ref{eq:spinsector2}) corresponds to a basis of R-NS characters which are diagonal under $T$.

For $\ZZ_2$-symmetry-enriched phases, the same principle applies. If we are given an MTC which contains a $\ZZ_2$-charge (i.e., an anyon $g$ with $\theta_g = 1$ and $g \times g = 1$), we can condense this and obtain states on the different sectors of the torus \cite{DelGaiGom2021:anomaly}. The torus sectors are given by the $\ZZ_2$-holonomies along the two cycles. In this case, the $(0,0)$-sector of the torus transforms into itself under $\SL$, while the $(0,1)$-, $(1,0)$-, and $(1,1)$-sectors transform into each other under $\SL$. Hence, if we take the $(0,1)$-sector,  which carries a representation of $\Gamma_0(2)$, and form the induced representation Eq.~(\ref{eq:Z2ind}), this will be the lower block of the $\SL$-representation Eq.~(\ref{eq:Z2sector}) of the modular extension.

\section{Test for the absorption of \texorpdfstring{$\ZZ_2$}{Z2}-symmetry protected topological order}
\label{app:absorption}

Gauged $\ZZ_2$-SPT corresponds to the Double Semion (DS) topological order, which  has anyons $\{1, s \bar{s}, s, \bar{s} \}$ with spins $\{0 , 0, \frac{1}{4}, -\frac{1}{4} \}$. This is a modular extension of $\Rep(\ZZ_2)$, with the first two spins $\{0 ,0 \}$ belonging to $\Rep(\ZZ_2).$ Note that $s \bar{s}$ plays the role of the $\ZZ_2$-charge.

Hence, once we have a modular extension $\cC$ (which has the $\ZZ_2$-charge $q$) of a $\ZZ_2$-BFC $\cB$, in order to obtain another modular extension $\cC'$, we stack with DS and condense the composite boson  $(q, s\bar{s})$ in $\cC \boxtimes {\rm DS}.$ \cite{LanKonWen2016:mex, LanKonWen2017:symmetryenriched}.

An anyon $(\alpha, x)$ of the stacked theory $\cC \boxtimes {\rm DS}$ survives the condensation (i.e. is not confined) if it has trivial braiding with $(q, s\bar{s})$. This can only be that case if  one of the following hold: 
\begin{itemize}
    \item $\alpha$ has trivial braiding with $q$ and $x$ has trivial braiding with $s\bar{s}$ (i.e., $x = 1$ or $s\bar{s}$) 
    
    \item $\alpha$ has $-1$ braiding with $q$ and $x$ is $s$ or $\bar{s}$
\end{itemize}
The first type of anyons are  exactly those which come from the $\ZZ_2$-BFC $\cB$. Both types of anyons comes in pairs $\alpha$ and $\alpha^q := q \times \alpha$, which have spins $\theta_{\alpha^q} = \theta_{\alpha}$ for the first type and $\theta_{\alpha^q} = - \theta_\alpha$ for the second type. In the stacked theory, 
\begin{align}
    (\alpha, 1) \times (\alpha, s \bar{s}) = (\alpha, 1) \\
    (\alpha, s) \times (q, s\bar{s}) = (\alpha^q, \bar{s}).
\end{align} 
Thus, under the condensation of $(q, s\bar{s})$, $(\alpha, 0)$ and $(\alpha, s\bar{s})$ are identified, leading to exactly the anyons $\alpha$ of the new modular extension $\cC'$. The anyons $(\alpha, s)$ and $(\alpha^q, \bar{s})$ are also identified. Note that these indeed have the same topological spin, as required, since $\theta_s = \theta_\alpha + \frac{1}{4}  = \theta_{\alpha^q} - \frac{1}{4}$. 

Hence, given a modular extension $\cC$ of $\cB$, the the other modular extension $\cC'$ will have anyons  given by the following procedure: the anyons of $\cC$ in the original $\ZZ_2$-BFC $\cB$ will remain the same, while the anyons of $\cC$ in the sector with nontrivial braiding with $q$ will have their spins shifted by $\frac{1}{4}$. If this shift leads to the same set of spins, then the modular extension absorbs the stacking with the $\ZZ_2$-SPT.

\begin{widetext}

\section{\label{app:ftomex} List of fermionic modular extensions}

Below, we present lists of modular extensions of primitive super-MTCs and super-MTCs that are obtained by stacking super-MTCs only (i.e., with no bosonic MTC factor). The numbers for super-MTCs are retrieved from Ref.~\cite{ChoKimSeoYou2022:classification}. For each set of modular extensions of a rank-$r$ super-MTC, the first $r$ objects are those from the rank-$r$ super-MTC with the $(\frac{r}{2}+1)$-th object being the distinguished fermion $\psi$, while the other objects are added to form a modular extension. Besides the presented modular extensions, there are 16 more modular extensions with opposite sign on the quantum dimensions of the added objects. Here, we show only one of each pair.

\paragraph*{Rank 4; \#7}

\begin{enumerate}

\item $c = \frac{1}{4}$, $(d_i,\theta_i)$ = \{$(1., 0)$, $(2.41421, -\frac{1}{4})$, $(1., \frac{1}{2})$, $(2.41421, \frac{1}{4})$, $(2.61313, \frac{7}{32})$, $(1.84776, -\frac{5}{32})$, $(1.84776, -\frac{5}{32})$\}

\item $c = \frac{3}{4}$, $(d_i,\theta_i)$ = \{$(1., 0)$, $(2.41421, -\frac{1}{4})$, $(1., \frac{1}{2})$, $(2.41421, \frac{1}{4})$, $(2.61313, -\frac{3}{32})$, $(1.84776, \frac{9}{32})$, $(1.84776, \frac{9}{32})$\}

\item $c = \frac{5}{4}$, $(d_i,\theta_i)$ = \{$(1., 0)$, $(2.41421, -\frac{1}{4})$, $(1., \frac{1}{2})$, $(2.41421, \frac{1}{4})$, $(2.61313, \frac{11}{32})$, $(1.84776, -\frac{1}{32})$, $(1.84776, -\frac{1}{32})$\}

\item $c = \frac{7}{4}$, $(d_i,\theta_i)$ = \{$(1., 0)$, $(2.41421, \frac{1}{4})$, $(1., \frac{1}{2})$, $(2.41421, -\frac{1}{4})$, $(2.61313, \frac{1}{32})$, $(1.84776, \frac{13}{32})$, $(1.84776, \frac{13}{32})$\}

\item $c = \frac{9}{4}$, $(d_i,\theta_i)$ = \{$(1., 0)$, $(2.41421, \frac{1}{4})$, $(1., \frac{1}{2})$, $(2.41421, -\frac{1}{4})$, $(2.61313, \frac{15}{32})$, $(1.84776, \frac{3}{32})$, $(1.84776, \frac{3}{32})$\}

\item $c = \frac{11}{4}$, $(d_i,\theta_i)$ = \{$(1., 0)$, $(2.41421, \frac{1}{4})$, $(1., \frac{1}{2})$, $(2.41421, -\frac{1}{4})$, $(2.61313, \frac{5}{32})$, $(1.84776, -\frac{15}{32})$, $(1.84776, -\frac{15}{32})$\}

\item $c = \frac{13}{4}$, $(d_i,\theta_i)$ = \{$(1., 0)$, $(2.41421, -\frac{1}{4})$, $(1., \frac{1}{2})$, $(2.41421, \frac{1}{4})$, $(2.61313, -\frac{13}{32})$, $(1.84776, \frac{7}{32})$, $(1.84776, \frac{7}{32})$\}

\item $c = \frac{15}{4}$, $(d_i,\theta_i)$ = \{$(1., 0)$, $(2.41421, \frac{1}{4})$, $(1., \frac{1}{2})$, $(2.41421, -\frac{1}{4})$, $(2.61313, \frac{9}{32})$, $(1.84776, -\frac{11}{32})$, $(1.84776, -\frac{11}{32})$\}

\item $c = \frac{17}{4}$, $(d_i,\theta_i)$ = \{$(1., 0)$, $(2.41421, \frac{1}{4})$, $(1., \frac{1}{2})$, $(2.41421, -\frac{1}{4})$, $(2.61313, -\frac{9}{32})$, $(1.84776, \frac{11}{32})$, $(1.84776, \frac{11}{32})$\}

\item $c = \frac{19}{4}$, $(d_i,\theta_i)$ = \{$(1., 0)$, $(2.41421, \frac{1}{4})$, $(1., \frac{1}{2})$, $(2.41421, -\frac{1}{4})$, $(2.61313, \frac{13}{32})$, $(1.84776, -\frac{7}{32})$, $(1.84776, -\frac{7}{32})$\}

\item $c = \frac{21}{4}$, $(d_i,\theta_i)$ = \{$(1., 0)$, $(2.41421, \frac{1}{4})$, $(1., \frac{1}{2})$, $(2.41421, -\frac{1}{4})$, $(2.61313, -\frac{5}{32})$, $(1.84776, \frac{15}{32})$, $(1.84776, \frac{15}{32})$\}

\item $c = \frac{23}{4}$, $(d_i,\theta_i)$ = \{$(1., 0)$, $(2.41421, -\frac{1}{4})$, $(1., \frac{1}{2})$, $(2.41421, \frac{1}{4})$, $(2.61313, -\frac{15}{32})$, $(1.84776, -\frac{3}{32})$, $(1.84776, -\frac{3}{32})$\}

\item $c = \frac{25}{4}$, $(d_i,\theta_i)$ = \{$(1., 0)$, $(2.41421, -\frac{1}{4})$, $(1., \frac{1}{2})$, $(2.41421, \frac{1}{4})$, $(2.61313, -\frac{1}{32})$, $(1.84776, -\frac{13}{32})$, $(1.84776, -\frac{13}{32})$\}

\item $c = \frac{27}{4}$, $(d_i,\theta_i)$ = \{$(1., 0)$, $(2.41421, -\frac{1}{4})$, $(1., \frac{1}{2})$, $(2.41421, \frac{1}{4})$, $(2.61313, -\frac{11}{32})$, $(1.84776, \frac{1}{32})$, $(1.84776, \frac{1}{32})$\}

\item $c = \frac{29}{4}$, $(d_i,\theta_i)$ = \{$(1., 0)$, $(2.41421, \frac{1}{4})$, $(1., \frac{1}{2})$, $(2.41421, -\frac{1}{4})$, $(2.61313, \frac{3}{32})$, $(1.84776, -\frac{9}{32})$, $(1.84776, -\frac{9}{32})$\}

\item $c = \frac{31}{4}$, $(d_i,\theta_i)$ = \{$(1., 0)$, $(2.41421, -\frac{1}{4})$, $(1., \frac{1}{2})$, $(2.41421, \frac{1}{4})$, $(2.61313, -\frac{7}{32})$, $(1.84776, \frac{5}{32})$, $(1.84776, \frac{5}{32})$\}

\end{enumerate}

\paragraph*{Rank 4; \#8}

\begin{enumerate}

\item $c = \frac{1}{4}$, $(d_i,\theta_i)$ = \{$(1., 0)$, $(-0.414214, \frac{1}{4})$, $(1., \frac{1}{2})$, $(-0.414214, -\frac{1}{4})$, $(-1.08239, -\frac{1}{32})$, $(0.765367, \frac{3}{32})$, $(0.765367, \frac{3}{32})$\}

\item $c = \frac{3}{4}$, $(d_i,\theta_i)$ = \{$(1., 0)$, $(-0.414214, -\frac{1}{4})$, $(1., \frac{1}{2})$, $(-0.414214, \frac{1}{4})$, $(-1.08239, \frac{5}{32})$, $(0.765367, \frac{1}{32})$, $(0.765367, \frac{1}{32})$\}

\item $c = \frac{5}{4}$, $(d_i,\theta_i)$ = \{$(1., 0)$, $(-0.414214, -\frac{1}{4})$, $(1., \frac{1}{2})$, $(-0.414214, \frac{1}{4})$, $(-1.08239, \frac{3}{32})$, $(0.765367, \frac{7}{32})$, $(0.765367, \frac{7}{32})$\}

\item $c = \frac{7}{4}$, $(d_i,\theta_i)$ = \{$(1., 0)$, $(-0.414214, \frac{1}{4})$, $(1., \frac{1}{2})$, $(-0.414214, -\frac{1}{4})$, $(-1.08239, \frac{9}{32})$, $(0.765367, \frac{5}{32})$, $(0.765367, \frac{5}{32})$\}

\item $c = \frac{9}{4}$, $(d_i,\theta_i)$ = \{$(1., 0)$, $(-0.414214, \frac{1}{4})$, $(1., \frac{1}{2})$, $(-0.414214, -\frac{1}{4})$, $(-1.08239, \frac{7}{32})$, $(0.765367, \frac{11}{32})$, $(0.765367, \frac{11}{32})$\}

\item $c = \frac{11}{4}$, $(d_i,\theta_i)$ = \{$(1., 0)$, $(-0.414214, -\frac{1}{4})$, $(1., \frac{1}{2})$, $(-0.414214, \frac{1}{4})$, $(-1.08239, \frac{13}{32})$, $(0.765367, \frac{9}{32})$, $(0.765367, \frac{9}{32})$\}

\item $c = \frac{13}{4}$, $(d_i,\theta_i)$ = \{$(1., 0)$, $(-0.414214, -\frac{1}{4})$, $(1., \frac{1}{2})$, $(-0.414214, \frac{1}{4})$, $(-1.08239, \frac{11}{32})$, $(0.765367, \frac{15}{32})$, $(0.765367, \frac{15}{32})$\}

\item $c = \frac{15}{4}$, $(d_i,\theta_i)$ = \{$(1., 0)$, $(-0.414214, \frac{1}{4})$, $(1., \frac{1}{2})$, $(-0.414214, -\frac{1}{4})$, $(-1.08239, -\frac{15}{32})$, $(0.765367, \frac{13}{32})$, $(0.765367, \frac{13}{32})$\}

\item $c = \frac{17}{4}$, $(d_i,\theta_i)$ = \{$(1., 0)$, $(-0.414214, -\frac{1}{4})$, $(1., \frac{1}{2})$, $(-0.414214, \frac{1}{4})$, $(-1.08239, \frac{15}{32})$, $(0.765367, -\frac{13}{32})$, $(0.765367, -\frac{13}{32})$\}

\item $c = \frac{19}{4}$, $(d_i,\theta_i)$ = \{$(1., 0)$, $(-0.414214, \frac{1}{4})$, $(1., \frac{1}{2})$, $(-0.414214, -\frac{1}{4})$, $(-1.08239, -\frac{11}{32})$, $(0.765367, -\frac{15}{32})$, $(0.765367, -\frac{15}{32})$\}

\item $c = \frac{21}{4}$, $(d_i,\theta_i)$ = \{$(1., 0)$, $(-0.414214, \frac{1}{4})$, $(1., \frac{1}{2})$, $(-0.414214, -\frac{1}{4})$, $(-1.08239, -\frac{13}{32})$, $(0.765367, -\frac{9}{32})$, $(0.765367, -\frac{9}{32})$\}

\item $c = \frac{23}{4}$, $(d_i,\theta_i)$ = \{$(1., 0)$, $(-0.414214, -\frac{1}{4})$, $(1., \frac{1}{2})$, $(-0.414214, \frac{1}{4})$, $(-1.08239, -\frac{7}{32})$, $(0.765367, -\frac{11}{32})$, $(0.765367, -\frac{11}{32})$\}

\item $c = \frac{25}{4}$, $(d_i,\theta_i)$ = \{$(1., 0)$, $(-0.414214, -\frac{1}{4})$, $(1., \frac{1}{2})$, $(-0.414214, \frac{1}{4})$, $(-1.08239, -\frac{9}{32})$, $(0.765367, -\frac{5}{32})$, $(0.765367, -\frac{5}{32})$\}

\item $c = \frac{27}{4}$, $(d_i,\theta_i)$ = \{$(1., 0)$, $(-0.414214, \frac{1}{4})$, $(1., \frac{1}{2})$, $(-0.414214, -\frac{1}{4})$, $(-1.08239, -\frac{3}{32})$, $(0.765367, -\frac{7}{32})$, $(0.765367, -\frac{7}{32})$\}

\item $c = \frac{29}{4}$, $(d_i,\theta_i)$ = \{$(1., 0)$, $(-0.414214, \frac{1}{4})$, $(1., \frac{1}{2})$, $(-0.414214, -\frac{1}{4})$, $(-1.08239, -\frac{5}{32})$, $(0.765367, -\frac{1}{32})$, $(0.765367, -\frac{1}{32})$\}

\item $c = \frac{31}{4}$, $(d_i,\theta_i)$ = \{$(1., 0)$, $(-0.414214, -\frac{1}{4})$, $(1., \frac{1}{2})$, $(-0.414214, \frac{1}{4})$, $(-1.08239, \frac{1}{32})$, $(0.765367, -\frac{3}{32})$, $(0.765367, -\frac{3}{32})$\}

\end{enumerate}

\paragraph*{Rank 6; \#17}

\begin{enumerate}

\item $c = 0$, $(d_i,\theta_i)$ = \{$(1., 0)$, $(3.73205, 0)$, $(2.73205, -\frac{1}{6})$, $(1., \frac{1}{2})$, $(3.73205, \frac{1}{2})$, $(2.73205, \frac{1}{3})$, $(2.73205, \frac{1}{4})$, $(4.73205, 0)$, $(2.73205, -\frac{5}{12})$, $(2.73205, -\frac{5}{12})$\}

\item $c = \frac{1}{2}$, $(d_i,\theta_i)$ = \{$(1., 0)$, $(3.73205, 0)$, $(2.73205, -\frac{1}{6})$, $(1., \frac{1}{2})$, $(3.73205, \frac{1}{2})$, $(2.73205, \frac{1}{3})$, $(3.8637, -\frac{17}{48})$, $(3.34607, \frac{1}{16})$, $(1.93185, \frac{5}{16})$, $(3.34607, \frac{1}{16})$, $(1.93185, \frac{5}{16})$\}

\item $c = 1$, $(d_i,\theta_i)$ = \{$(1., 0)$, $(3.73205, 0)$, $(2.73205, -\frac{1}{6})$, $(1., \frac{1}{2})$, $(3.73205, \frac{1}{2})$, $(2.73205, \frac{1}{3})$, $(2.73205, \frac{3}{8})$, $(4.73205, \frac{1}{8})$, $(2.73205, -\frac{7}{24})$, $(2.73205, -\frac{7}{24})$\}

\item $c = \frac{3}{2}$, $(d_i,\theta_i)$ = \{$(1., 0)$, $(3.73205, 0)$, $(2.73205, -\frac{1}{6})$, $(1., \frac{1}{2})$, $(3.73205, \frac{1}{2})$, $(2.73205, \frac{1}{3})$, $(3.8637, -\frac{11}{48})$, $(3.34607, \frac{3}{16})$, $(1.93185, \frac{7}{16})$, $(3.34607, \frac{3}{16})$, $(1.93185, \frac{7}{16})$\}

\item $c = 2$, $(d_i,\theta_i)$ = \{$(1., 0)$, $(3.73205, 0)$, $(2.73205, \frac{1}{3})$, $(1., \frac{1}{2})$, $(3.73205, \frac{1}{2})$, $(2.73205, -\frac{1}{6})$, $(2.73205, \frac{1}{2})$, $(4.73205, \frac{1}{4})$, $(2.73205, -\frac{1}{6})$, $(2.73205, -\frac{1}{6})$\}

\item $c = \frac{5}{2}$, $(d_i,\theta_i)$ = \{$(1., 0)$, $(3.73205, 0)$, $(2.73205, \frac{1}{3})$, $(1., \frac{1}{2})$, $(3.73205, \frac{1}{2})$, $(2.73205, -\frac{1}{6})$, $(3.8637, -\frac{5}{48})$, $(1.93185, -\frac{7}{16})$, $(3.34607, \frac{5}{16})$, $(1.93185, -\frac{7}{16})$, $(3.34607, \frac{5}{16})$\}

\item $c = 3$, $(d_i,\theta_i)$ = \{$(1., 0)$, $(3.73205, 0)$, $(2.73205, \frac{1}{3})$, $(1., \frac{1}{2})$, $(3.73205, \frac{1}{2})$, $(2.73205, -\frac{1}{6})$, $(2.73205, -\frac{3}{8})$, $(4.73205, \frac{3}{8})$, $(2.73205, -\frac{1}{24})$, $(2.73205, -\frac{1}{24})$\}

\item $c = \frac{7}{2}$, $(d_i,\theta_i)$ = \{$(1., 0)$, $(3.73205, 0)$, $(2.73205, \frac{1}{3})$, $(1., \frac{1}{2})$, $(3.73205, \frac{1}{2})$, $(2.73205, -\frac{1}{6})$, $(3.8637, \frac{1}{48})$, $(3.34607, \frac{7}{16})$, $(1.93185, -\frac{5}{16})$, $(3.34607, \frac{7}{16})$, $(1.93185, -\frac{5}{16})$\}

\item $c = 4$, $(d_i,\theta_i)$ = \{$(1., 0)$, $(3.73205, 0)$, $(2.73205, \frac{1}{3})$, $(1., \frac{1}{2})$, $(3.73205, \frac{1}{2})$, $(2.73205, -\frac{1}{6})$, $(2.73205, -\frac{1}{4})$, $(4.73205, \frac{1}{2})$, $(2.73205, \frac{1}{12})$, $(2.73205, \frac{1}{12})$\}

\item $c = \frac{9}{2}$, $(d_i,\theta_i)$ = \{$(1., 0)$, $(3.73205, 0)$, $(2.73205, \frac{1}{3})$, $(1., \frac{1}{2})$, $(3.73205, \frac{1}{2})$, $(2.73205, -\frac{1}{6})$, $(3.8637, \frac{7}{48})$, $(3.34607, -\frac{7}{16})$, $(1.93185, -\frac{3}{16})$, $(3.34607, -\frac{7}{16})$, $(1.93185, -\frac{3}{16})$\}

\item $c = 5$, $(d_i,\theta_i)$ = \{$(1., 0)$, $(3.73205, 0)$, $(2.73205, \frac{1}{3})$, $(1., \frac{1}{2})$, $(3.73205, \frac{1}{2})$, $(2.73205, -\frac{1}{6})$, $(2.73205, -\frac{1}{8})$, $(4.73205, -\frac{3}{8})$, $(2.73205, \frac{5}{24})$, $(2.73205, \frac{5}{24})$\}

\item $c = \frac{11}{2}$, $(d_i,\theta_i)$ = \{$(1., 0)$, $(3.73205, 0)$, $(2.73205, \frac{1}{3})$, $(1., \frac{1}{2})$, $(3.73205, \frac{1}{2})$, $(2.73205, -\frac{1}{6})$, $(3.8637, \frac{13}{48})$, $(3.34607, -\frac{5}{16})$, $(1.93185, -\frac{1}{16})$, $(3.34607, -\frac{5}{16})$, $(1.93185, -\frac{1}{16})$\}

\item $c = 6$, $(d_i,\theta_i)$ = \{$(1., 0)$, $(3.73205, \frac{1}{2})$, $(2.73205, \frac{1}{3})$, $(1., \frac{1}{2})$, $(3.73205, 0)$, $(2.73205, -\frac{1}{6})$, $(2.73205, 0)$, $(4.73205, -\frac{1}{4})$, $(2.73205, \frac{1}{3})$, $(2.73205, \frac{1}{3})$\}

\item $c = \frac{13}{2}$, $(d_i,\theta_i)$ = \{$(1., 0)$, $(3.73205, \frac{1}{2})$, $(2.73205, \frac{1}{3})$, $(1., \frac{1}{2})$, $(3.73205, 0)$, $(2.73205, -\frac{1}{6})$, $(3.8637, \frac{19}{48})$, $(1.93185, \frac{1}{16})$, $(3.34607, -\frac{3}{16})$, $(1.93185, \frac{1}{16})$, $(3.34607, -\frac{3}{16})$\}

\item $c = 7$, $(d_i,\theta_i)$ = \{$(1., 0)$, $(3.73205, \frac{1}{2})$, $(2.73205, \frac{1}{3})$, $(1., \frac{1}{2})$, $(3.73205, 0)$, $(2.73205, -\frac{1}{6})$, $(2.73205, \frac{1}{8})$, $(4.73205, -\frac{1}{8})$, $(2.73205, \frac{11}{24})$, $(2.73205, \frac{11}{24})$\}

\item $c = \frac{15}{2}$, $(d_i,\theta_i)$ = \{$(1., 0)$, $(3.73205, \frac{1}{2})$, $(2.73205, \frac{1}{3})$, $(1., \frac{1}{2})$, $(3.73205, 0)$, $(2.73205, -\frac{1}{6})$, $(3.8637, -\frac{23}{48})$, $(3.34607, -\frac{1}{16})$, $(1.93185, \frac{3}{16})$, $(3.34607, -\frac{1}{16})$, $(1.93185, \frac{3}{16})$\}

\end{enumerate}

\paragraph*{Rank 6; \#18}

\begin{enumerate}

\item $c = 0$, $(d_i,\theta_i)$ = \{$(1., 0)$, $(3.73205, 0)$, $(2.73205, \frac{1}{6})$, $(1., \frac{1}{2})$, $(3.73205, \frac{1}{2})$, $(2.73205, -\frac{1}{3})$, $(2.73205, -\frac{1}{4})$, $(4.73205, 0)$, $(2.73205, \frac{5}{12})$, $(2.73205, \frac{5}{12})$\}

\item $c = \frac{1}{2}$, $(d_i,\theta_i)$ = \{$(1., 0)$, $(3.73205, 0)$, $(2.73205, \frac{1}{6})$, $(1., \frac{1}{2})$, $(3.73205, \frac{1}{2})$, $(2.73205, -\frac{1}{3})$, $(3.8637, \frac{23}{48})$, $(1.93185, -\frac{3}{16})$, $(3.34607, \frac{1}{16})$, $(1.93185, -\frac{3}{16})$, $(3.34607, \frac{1}{16})$\}

\item $c = 1$, $(d_i,\theta_i)$ = \{$(1., 0)$, $(3.73205, 0)$, $(2.73205, \frac{1}{6})$, $(1., \frac{1}{2})$, $(3.73205, \frac{1}{2})$, $(2.73205, -\frac{1}{3})$, $(2.73205, -\frac{1}{8})$, $(4.73205, \frac{1}{8})$, $(2.73205, -\frac{11}{24})$, $(2.73205, -\frac{11}{24})$\}

\item $c = \frac{3}{2}$, $(d_i,\theta_i)$ = \{$(1., 0)$, $(3.73205, 0)$, $(2.73205, \frac{1}{6})$, $(1., \frac{1}{2})$, $(3.73205, \frac{1}{2})$, $(2.73205, -\frac{1}{3})$, $(3.8637, -\frac{19}{48})$, $(3.34607, \frac{3}{16})$, $(1.93185, -\frac{1}{16})$, $(3.34607, \frac{3}{16})$, $(1.93185, -\frac{1}{16})$\}

\item $c = 2$, $(d_i,\theta_i)$ = \{$(1., 0)$, $(3.73205, 0)$, $(2.73205, \frac{1}{6})$, $(1., \frac{1}{2})$, $(3.73205, \frac{1}{2})$, $(2.73205, -\frac{1}{3})$, $(2.73205, 0)$, $(4.73205, \frac{1}{4})$, $(2.73205, -\frac{1}{3})$, $(2.73205, -\frac{1}{3})$\}

\item $c = \frac{5}{2}$, $(d_i,\theta_i)$ = \{$(1., 0)$, $(3.73205, 0)$, $(2.73205, \frac{1}{6})$, $(1., \frac{1}{2})$, $(3.73205, \frac{1}{2})$, $(2.73205, -\frac{1}{3})$, $(3.8637, -\frac{13}{48})$, $(1.93185, \frac{1}{16})$, $(3.34607, \frac{5}{16})$, $(1.93185, \frac{1}{16})$, $(3.34607, \frac{5}{16})$\}

\item $c = 3$, $(d_i,\theta_i)$ = \{$(1., 0)$, $(3.73205, 0)$, $(2.73205, \frac{1}{6})$, $(1., \frac{1}{2})$, $(3.73205, \frac{1}{2})$, $(2.73205, -\frac{1}{3})$, $(2.73205, \frac{1}{8})$, $(4.73205, \frac{3}{8})$, $(2.73205, -\frac{5}{24})$, $(2.73205, -\frac{5}{24})$\}

\item $c = \frac{7}{2}$, $(d_i,\theta_i)$ = \{$(1., 0)$, $(3.73205, 0)$, $(2.73205, \frac{1}{6})$, $(1., \frac{1}{2})$, $(3.73205, \frac{1}{2})$, $(2.73205, -\frac{1}{3})$, $(3.8637, -\frac{7}{48})$, $(1.93185, \frac{3}{16})$, $(3.34607, \frac{7}{16})$, $(1.93185, \frac{3}{16})$, $(3.34607, \frac{7}{16})$\}

\item $c = 4$, $(d_i,\theta_i)$ = \{$(1., 0)$, $(3.73205, 0)$, $(2.73205, \frac{1}{6})$, $(1., \frac{1}{2})$, $(3.73205, \frac{1}{2})$, $(2.73205, -\frac{1}{3})$, $(2.73205, \frac{1}{4})$, $(4.73205, \frac{1}{2})$, $(2.73205, -\frac{1}{12})$, $(2.73205, -\frac{1}{12})$\}

\item $c = \frac{9}{2}$, $(d_i,\theta_i)$ = \{$(1., 0)$, $(3.73205, 0)$, $(2.73205, \frac{1}{6})$, $(1., \frac{1}{2})$, $(3.73205, \frac{1}{2})$, $(2.73205, -\frac{1}{3})$, $(3.8637, -\frac{1}{48})$, $(1.93185, \frac{5}{16})$, $(3.34607, -\frac{7}{16})$, $(1.93185, \frac{5}{16})$, $(3.34607, -\frac{7}{16})$\}

\item $c = 5$, $(d_i,\theta_i)$ = \{$(1., 0)$, $(3.73205, 0)$, $(2.73205, \frac{1}{6})$, $(1., \frac{1}{2})$, $(3.73205, \frac{1}{2})$, $(2.73205, -\frac{1}{3})$, $(2.73205, \frac{3}{8})$, $(4.73205, -\frac{3}{8})$, $(2.73205, \frac{1}{24})$, $(2.73205, \frac{1}{24})$\}

\item $c = \frac{11}{2}$, $(d_i,\theta_i)$ = \{$(1., 0)$, $(3.73205, 0)$, $(2.73205, \frac{1}{6})$, $(1., \frac{1}{2})$, $(3.73205, \frac{1}{2})$, $(2.73205, -\frac{1}{3})$, $(3.8637, \frac{5}{48})$, $(3.34607, -\frac{5}{16})$, $(1.93185, \frac{7}{16})$, $(3.34607, -\frac{5}{16})$, $(1.93185, \frac{7}{16})$\}

\item $c = 6$, $(d_i,\theta_i)$ = \{$(1., 0)$, $(3.73205, \frac{1}{2})$, $(2.73205, \frac{1}{6})$, $(1., \frac{1}{2})$, $(3.73205, 0)$, $(2.73205, -\frac{1}{3})$, $(2.73205, \frac{1}{2})$, $(4.73205, -\frac{1}{4})$, $(2.73205, \frac{1}{6})$, $(2.73205, \frac{1}{6})$\}

\item $c = \frac{13}{2}$, $(d_i,\theta_i)$ = \{$(1., 0)$, $(3.73205, \frac{1}{2})$, $(2.73205, \frac{1}{6})$, $(1., \frac{1}{2})$, $(3.73205, 0)$, $(2.73205, -\frac{1}{3})$, $(3.8637, \frac{11}{48})$, $(1.93185, -\frac{7}{16})$, $(3.34607, -\frac{3}{16})$, $(1.93185, -\frac{7}{16})$, $(3.34607, -\frac{3}{16})$\}

\item $c = 7$, $(d_i,\theta_i)$ = \{$(1., 0)$, $(3.73205, \frac{1}{2})$, $(2.73205, \frac{1}{6})$, $(1., \frac{1}{2})$, $(3.73205, 0)$, $(2.73205, -\frac{1}{3})$, $(2.73205, -\frac{3}{8})$, $(4.73205, -\frac{1}{8})$, $(2.73205, \frac{7}{24})$, $(2.73205, \frac{7}{24})$\}

\item $c = \frac{15}{2}$, $(d_i,\theta_i)$ = \{$(1., 0)$, $(3.73205, \frac{1}{2})$, $(2.73205, \frac{1}{6})$, $(1., \frac{1}{2})$, $(3.73205, 0)$, $(2.73205, -\frac{1}{3})$, $(3.8637, \frac{17}{48})$, $(1.93185, -\frac{5}{16})$, $(3.34607, -\frac{1}{16})$, $(1.93185, -\frac{5}{16})$, $(3.34607, -\frac{1}{16})$\}

\end{enumerate}

\paragraph*{Rank 6; \#19}

\begin{enumerate}

\item $c = 0$, $(d_i,\theta_i)$ = \{$(1., 0)$, $(0.267949, 0)$, $(-0.732051, -\frac{1}{6})$, $(1., \frac{1}{2})$, $(0.267949, \frac{1}{2})$, $(-0.732051, \frac{1}{3})$, $(0.732051, -\frac{1}{4})$, $(-1.26795, 0)$, $(0.732051, \frac{1}{12})$, $(0.732051, \frac{1}{12})$\}

\item $c = \frac{1}{2}$, $(d_i,\theta_i)$ = \{$(1., 0)$, $(0.267949, 0)$, $(-0.732051, -\frac{1}{6})$, $(1., \frac{1}{2})$, $(0.267949, \frac{1}{2})$, $(-0.732051, \frac{1}{3})$, $(1.03528, \frac{7}{48})$, $(0.517638, -\frac{3}{16})$, $(-0.896575, \frac{1}{16})$, $(0.517638, -\frac{3}{16})$, $(-0.896575, \frac{1}{16})$\}

\item $c = 1$, $(d_i,\theta_i)$ = \{$(1., 0)$, $(0.267949, 0)$, $(-0.732051, -\frac{1}{6})$, $(1., \frac{1}{2})$, $(0.267949, \frac{1}{2})$, $(-0.732051, \frac{1}{3})$, $(0.732051, -\frac{1}{8})$, $(-1.26795, \frac{1}{8})$, $(0.732051, \frac{5}{24})$, $(0.732051, \frac{5}{24})$\}

\item $c = \frac{3}{2}$, $(d_i,\theta_i)$ = \{$(1., 0)$, $(0.267949, 0)$, $(-0.732051, -\frac{1}{6})$, $(1., \frac{1}{2})$, $(0.267949, \frac{1}{2})$, $(-0.732051, \frac{1}{3})$, $(-1.03528, \frac{13}{48})$, $(0.896575, \frac{3}{16})$, $(-0.517638, -\frac{1}{16})$, $(0.896575, \frac{3}{16})$, $(-0.517638, -\frac{1}{16})$\}

\item $c = 2$, $(d_i,\theta_i)$ = \{$(1., 0)$, $(0.267949, 0)$, $(-0.732051, \frac{1}{3})$, $(1., \frac{1}{2})$, $(0.267949, \frac{1}{2})$, $(-0.732051, -\frac{1}{6})$, $(-0.732051, 0)$, $(1.26795, \frac{1}{4})$, $(-0.732051, \frac{1}{3})$, $(-0.732051, \frac{1}{3})$\}

\item $c = \frac{5}{2}$, $(d_i,\theta_i)$ = \{$(1., 0)$, $(0.267949, 0)$, $(-0.732051, \frac{1}{3})$, $(1., \frac{1}{2})$, $(0.267949, \frac{1}{2})$, $(-0.732051, -\frac{1}{6})$, $(-1.03528, \frac{19}{48})$, $(-0.517638, \frac{1}{16})$, $(0.896575, \frac{5}{16})$, $(-0.517638, \frac{1}{16})$, $(0.896575, \frac{5}{16})$\}

\item $c = 3$, $(d_i,\theta_i)$ = \{$(1., 0)$, $(0.267949, 0)$, $(-0.732051, \frac{1}{3})$, $(1., \frac{1}{2})$, $(0.267949, \frac{1}{2})$, $(-0.732051, -\frac{1}{6})$, $(-0.732051, \frac{1}{8})$, $(1.26795, \frac{3}{8})$, $(-0.732051, \frac{11}{24})$, $(-0.732051, \frac{11}{24})$\}

\item $c = \frac{7}{2}$, $(d_i,\theta_i)$ = \{$(1., 0)$, $(0.267949, 0)$, $(-0.732051, \frac{1}{3})$, $(1., \frac{1}{2})$, $(0.267949, \frac{1}{2})$, $(-0.732051, -\frac{1}{6})$, $(-1.03528, -\frac{23}{48})$, $(-0.517638, \frac{3}{16})$, $(0.896575, \frac{7}{16})$, $(-0.517638, \frac{3}{16})$, $(0.896575, \frac{7}{16})$\}

\item $c = 4$, $(d_i,\theta_i)$ = \{$(1., 0)$, $(0.267949, 0)$, $(-0.732051, \frac{1}{3})$, $(1., \frac{1}{2})$, $(0.267949, \frac{1}{2})$, $(-0.732051, -\frac{1}{6})$, $(-0.732051, \frac{1}{4})$, $(1.26795, \frac{1}{2})$, $(-0.732051, -\frac{5}{12})$, $(-0.732051, -\frac{5}{12})$\}

\item $c = \frac{9}{2}$, $(d_i,\theta_i)$ = \{$(1., 0)$, $(0.267949, 0)$, $(-0.732051, \frac{1}{3})$, $(1., \frac{1}{2})$, $(0.267949, \frac{1}{2})$, $(-0.732051, -\frac{1}{6})$, $(1.03528, -\frac{17}{48})$, $(0.517638, \frac{5}{16})$, $(-0.896575, -\frac{7}{16})$, $(0.517638, \frac{5}{16})$, $(-0.896575, -\frac{7}{16})$\}

\item $c = 5$, $(d_i,\theta_i)$ = \{$(1., 0)$, $(0.267949, 0)$, $(-0.732051, \frac{1}{3})$, $(1., \frac{1}{2})$, $(0.267949, \frac{1}{2})$, $(-0.732051, -\frac{1}{6})$, $(-0.732051, \frac{3}{8})$, $(1.26795, -\frac{3}{8})$, $(-0.732051, -\frac{7}{24})$, $(-0.732051, -\frac{7}{24})$\}

\item $c = \frac{11}{2}$, $(d_i,\theta_i)$ = \{$(1., 0)$, $(0.267949, 0)$, $(-0.732051, \frac{1}{3})$, $(1., \frac{1}{2})$, $(0.267949, \frac{1}{2})$, $(-0.732051, -\frac{1}{6})$, $(-1.03528, -\frac{11}{48})$, $(0.896575, -\frac{5}{16})$, $(-0.517638, \frac{7}{16})$, $(0.896575, -\frac{5}{16})$, $(-0.517638, \frac{7}{16})$\}

\item $c = 6$, $(d_i,\theta_i)$ = \{$(1., 0)$, $(0.267949, \frac{1}{2})$, $(-0.732051, \frac{1}{3})$, $(1., \frac{1}{2})$, $(0.267949, 0)$, $(-0.732051, -\frac{1}{6})$, $(0.732051, \frac{1}{2})$, $(-1.26795, -\frac{1}{4})$, $(0.732051, -\frac{1}{6})$, $(0.732051, -\frac{1}{6})$\}

\item $c = \frac{13}{2}$, $(d_i,\theta_i)$ = \{$(1., 0)$, $(0.267949, \frac{1}{2})$, $(-0.732051, \frac{1}{3})$, $(1., \frac{1}{2})$, $(0.267949, 0)$, $(-0.732051, -\frac{1}{6})$, $(1.03528, -\frac{5}{48})$, $(0.517638, -\frac{7}{16})$, $(-0.896575, -\frac{3}{16})$, $(0.517638, -\frac{7}{16})$, $(-0.896575, -\frac{3}{16})$\}

\item $c = 7$, $(d_i,\theta_i)$ = \{$(1., 0)$, $(0.267949, \frac{1}{2})$, $(-0.732051, \frac{1}{3})$, $(1., \frac{1}{2})$, $(0.267949, 0)$, $(-0.732051, -\frac{1}{6})$, $(0.732051, -\frac{3}{8})$, $(-1.26795, -\frac{1}{8})$, $(0.732051, -\frac{1}{24})$, $(0.732051, -\frac{1}{24})$\}

\item $c = \frac{15}{2}$, $(d_i,\theta_i)$ = \{$(1., 0)$, $(0.267949, \frac{1}{2})$, $(-0.732051, \frac{1}{3})$, $(1., \frac{1}{2})$, $(0.267949, 0)$, $(-0.732051, -\frac{1}{6})$, $(1.03528, \frac{1}{48})$, $(0.517638, -\frac{5}{16})$, $(-0.896575, -\frac{1}{16})$, $(0.517638, -\frac{5}{16})$, $(-0.896575, -\frac{1}{16})$\}

\end{enumerate}

\paragraph*{Rank 6; \#20}

\begin{enumerate}

\item $c = 0$, $(d_i,\theta_i)$ = \{$(1., 0)$, $(0.267949, 0)$, $(-0.732051, \frac{1}{6})$, $(1., \frac{1}{2})$, $(0.267949, \frac{1}{2})$, $(-0.732051, -\frac{1}{3})$, $(-0.732051, \frac{1}{4})$, $(1.26795, 0)$, $(-0.732051, -\frac{1}{12})$, $(-0.732051, -\frac{1}{12})$\}

\item $c = \frac{1}{2}$, $(d_i,\theta_i)$ = \{$(1., 0)$, $(0.267949, 0)$, $(-0.732051, \frac{1}{6})$, $(1., \frac{1}{2})$, $(0.267949, \frac{1}{2})$, $(-0.732051, -\frac{1}{3})$, $(-1.03528, -\frac{1}{48})$, $(0.896575, \frac{1}{16})$, $(-0.517638, \frac{5}{16})$, $(0.896575, \frac{1}{16})$, $(-0.517638, \frac{5}{16})$\}

\item $c = 1$, $(d_i,\theta_i)$ = \{$(1., 0)$, $(0.267949, 0)$, $(-0.732051, \frac{1}{6})$, $(1., \frac{1}{2})$, $(0.267949, \frac{1}{2})$, $(-0.732051, -\frac{1}{3})$, $(-0.732051, \frac{3}{8})$, $(1.26795, \frac{1}{8})$, $(-0.732051, \frac{1}{24})$, $(-0.732051, \frac{1}{24})$\}

\item $c = \frac{3}{2}$, $(d_i,\theta_i)$ = \{$(1., 0)$, $(0.267949, 0)$, $(-0.732051, \frac{1}{6})$, $(1., \frac{1}{2})$, $(0.267949, \frac{1}{2})$, $(-0.732051, -\frac{1}{3})$, $(-1.03528, \frac{5}{48})$, $(0.896575, \frac{3}{16})$, $(-0.517638, \frac{7}{16})$, $(0.896575, \frac{3}{16})$, $(-0.517638, \frac{7}{16})$\}

\item $c = 2$, $(d_i,\theta_i)$ = \{$(1., 0)$, $(0.267949, 0)$, $(-0.732051, \frac{1}{6})$, $(1., \frac{1}{2})$, $(0.267949, \frac{1}{2})$, $(-0.732051, -\frac{1}{3})$, $(1.26795, \frac{1}{4})$, $(-0.732051, \frac{1}{2})$, $(-0.732051, \frac{1}{6})$, $(-0.732051, \frac{1}{6})$\}

\item $c = \frac{5}{2}$, $(d_i,\theta_i)$ = \{$(1., 0)$, $(0.267949, 0)$, $(-0.732051, \frac{1}{6})$, $(1., \frac{1}{2})$, $(0.267949, \frac{1}{2})$, $(-0.732051, -\frac{1}{3})$, $(-1.03528, \frac{11}{48})$, $(-0.517638, -\frac{7}{16})$, $(0.896575, \frac{5}{16})$, $(-0.517638, -\frac{7}{16})$, $(0.896575, \frac{5}{16})$\}

\item $c = 3$, $(d_i,\theta_i)$ = \{$(1., 0)$, $(0.267949, 0)$, $(-0.732051, \frac{1}{6})$, $(1., \frac{1}{2})$, $(0.267949, \frac{1}{2})$, $(-0.732051, -\frac{1}{3})$, $(-0.732051, -\frac{3}{8})$, $(1.26795, \frac{3}{8})$, $(-0.732051, \frac{7}{24})$, $(-0.732051, \frac{7}{24})$\}

\item $c = \frac{7}{2}$, $(d_i,\theta_i)$ = \{$(1., 0)$, $(0.267949, 0)$, $(-0.732051, \frac{1}{6})$, $(1., \frac{1}{2})$, $(0.267949, \frac{1}{2})$, $(-0.732051, -\frac{1}{3})$, $(-1.03528, \frac{17}{48})$, $(0.896575, \frac{7}{16})$, $(-0.517638, -\frac{5}{16})$, $(0.896575, \frac{7}{16})$, $(-0.517638, -\frac{5}{16})$\}

\item $c = 4$, $(d_i,\theta_i)$ = \{$(1., 0)$, $(0.267949, 0)$, $(-0.732051, \frac{1}{6})$, $(1., \frac{1}{2})$, $(0.267949, \frac{1}{2})$, $(-0.732051, -\frac{1}{3})$, $(-0.732051, -\frac{1}{4})$, $(1.26795, \frac{1}{2})$, $(-0.732051, \frac{5}{12})$, $(-0.732051, \frac{5}{12})$\}

\item $c = \frac{9}{2}$, $(d_i,\theta_i)$ = \{$(1., 0)$, $(0.267949, 0)$, $(-0.732051, \frac{1}{6})$, $(1., \frac{1}{2})$, $(0.267949, \frac{1}{2})$, $(-0.732051, -\frac{1}{3})$, $(-1.03528, \frac{23}{48})$, $(0.896575, -\frac{7}{16})$, $(-0.517638, -\frac{3}{16})$, $(0.896575, -\frac{7}{16})$, $(-0.517638, -\frac{3}{16})$\}

\item $c = 5$, $(d_i,\theta_i)$ = \{$(1., 0)$, $(0.267949, 0)$, $(-0.732051, \frac{1}{6})$, $(1., \frac{1}{2})$, $(0.267949, \frac{1}{2})$, $(-0.732051, -\frac{1}{3})$, $(-0.732051, -\frac{1}{8})$, $(1.26795, -\frac{3}{8})$, $(-0.732051, -\frac{11}{24})$, $(-0.732051, -\frac{11}{24})$\}

\item $c = \frac{11}{2}$, $(d_i,\theta_i)$ = \{$(1., 0)$, $(0.267949, 0)$, $(-0.732051, \frac{1}{6})$, $(1., \frac{1}{2})$, $(0.267949, \frac{1}{2})$, $(-0.732051, -\frac{1}{3})$, $(-1.03528, -\frac{19}{48})$, $(0.896575, -\frac{5}{16})$, $(-0.517638, -\frac{1}{16})$, $(0.896575, -\frac{5}{16})$, $(-0.517638, -\frac{1}{16})$\}

\item $c = 6$, $(d_i,\theta_i)$ = \{$(1., 0)$, $(0.267949, \frac{1}{2})$, $(-0.732051, \frac{1}{6})$, $(1., \frac{1}{2})$, $(0.267949, 0)$, $(-0.732051, -\frac{1}{3})$, $(0.732051, 0)$, $(-1.26795, -\frac{1}{4})$, $(0.732051, -\frac{1}{3})$, $(0.732051, -\frac{1}{3})$\}

\item $c = \frac{13}{2}$, $(d_i,\theta_i)$ = \{$(1., 0)$, $(0.267949, \frac{1}{2})$, $(-0.732051, \frac{1}{6})$, $(1., \frac{1}{2})$, $(0.267949, 0)$, $(-0.732051, -\frac{1}{3})$, $(-1.03528, -\frac{13}{48})$, $(-0.517638, \frac{1}{16})$, $(0.896575, -\frac{3}{16})$, $(-0.517638, \frac{1}{16})$, $(0.896575, -\frac{3}{16})$\}

\item $c = 7$, $(d_i,\theta_i)$ = \{$(1., 0)$, $(0.267949, \frac{1}{2})$, $(-0.732051, \frac{1}{6})$, $(1., \frac{1}{2})$, $(0.267949, 0)$, $(-0.732051, -\frac{1}{3})$, $(0.732051, \frac{1}{8})$, $(-1.26795, -\frac{1}{8})$, $(0.732051, -\frac{5}{24})$, $(0.732051, -\frac{5}{24})$\}

\item $c = \frac{15}{2}$, $(d_i,\theta_i)$ = \{$(1., 0)$, $(0.267949, \frac{1}{2})$, $(-0.732051, \frac{1}{6})$, $(1., \frac{1}{2})$, $(0.267949, 0)$, $(-0.732051, -\frac{1}{3})$, $(1.03528, -\frac{7}{48})$, $(-0.896575, -\frac{1}{16})$, $(0.517638, \frac{3}{16})$, $(-0.896575, -\frac{1}{16})$, $(0.517638, \frac{3}{16})$\}

\end{enumerate}

\paragraph*{Rank 8; \#18}

\begin{enumerate}

\item $c = 0$, $(d_i,\theta_i)$ = \{$(1., 0)$, $(1., 0)$, $(2.44949, \frac{1}{16})$, $(2., \frac{1}{6})$, $(1., \frac{1}{2})$, $(1., \frac{1}{2})$, $(2.44949, -\frac{7}{16})$, $(2., -\frac{1}{3})$, $(2., \frac{1}{4})$, $(2.44949, -\frac{1}{16})$, $(2.44949, \frac{7}{16})$, $(2., -\frac{1}{12})$, $(2., -\frac{1}{12})$\}

\item $c = \frac{1}{2}$, $(d_i,\theta_i)$ = \{$(1., 0)$, $(1., 0)$, $(2.44949, \frac{1}{16})$, $(2., \frac{1}{6})$, $(1., \frac{1}{2})$, $(1., \frac{1}{2})$, $(2.44949, -\frac{7}{16})$, $(2., -\frac{1}{3})$, $(2.82843, -\frac{1}{48})$, $(1.73205, 0)$, $(1.73205, \frac{1}{2})$, $(1.41421, \frac{5}{16})$, $(1.73205, 0)$, $(1.73205, \frac{1}{2})$, $(1.41421, \frac{5}{16})$\}

\item $c = 1$, $(d_i,\theta_i)$ = \{$(1., 0)$, $(1., 0)$, $(2.44949, \frac{1}{16})$, $(2., \frac{1}{6})$, $(1., \frac{1}{2})$, $(1., \frac{1}{2})$, $(2.44949, -\frac{7}{16})$, $(2., -\frac{1}{3})$, $(2.44949, -\frac{7}{16})$, $(2.44949, \frac{1}{16})$, $(2., \frac{3}{8})$, $(2., \frac{1}{24})$, $(2., \frac{1}{24})$\}

\item $c = \frac{3}{2}$, $(d_i,\theta_i)$ = \{$(1., 0)$, $(1., 0)$, $(2.44949, \frac{1}{16})$, $(2., \frac{1}{6})$, $(1., \frac{1}{2})$, $(1., \frac{1}{2})$, $(2.44949, -\frac{7}{16})$, $(2., -\frac{1}{3})$, $(2.82843, \frac{5}{48})$, $(1.73205, \frac{1}{8})$, $(1.73205, -\frac{3}{8})$, $(1.41421, \frac{7}{16})$, $(1.73205, \frac{1}{8})$, $(1.73205, -\frac{3}{8})$, $(1.41421, \frac{7}{16})$\}

\item $c = 2$, $(d_i,\theta_i)$ = \{$(1., 0)$, $(1., 0)$, $(2.44949, \frac{1}{16})$, $(2., \frac{1}{6})$, $(1., \frac{1}{2})$, $(1., \frac{1}{2})$, $(2.44949, -\frac{7}{16})$, $(2., -\frac{1}{3})$, $(2., \frac{1}{2})$, $(2.44949, \frac{3}{16})$, $(2.44949, -\frac{5}{16})$, $(2., \frac{1}{6})$, $(2., \frac{1}{6})$\}

\item $c = \frac{5}{2}$, $(d_i,\theta_i)$ = \{$(1., 0)$, $(1., 0)$, $(2.44949, \frac{1}{16})$, $(2., \frac{1}{6})$, $(1., \frac{1}{2})$, $(1., \frac{1}{2})$, $(2.44949, -\frac{7}{16})$, $(2., -\frac{1}{3})$, $(2.82843, \frac{11}{48})$, $(1.73205, -\frac{1}{4})$, $(1.73205, \frac{1}{4})$, $(1.41421, -\frac{7}{16})$, $(1.73205, -\frac{1}{4})$, $(1.73205, \frac{1}{4})$, $(1.41421, -\frac{7}{16})$\}

\item $c = 3$, $(d_i,\theta_i)$ = \{$(1., 0)$, $(1., 0)$, $(2.44949, \frac{1}{16})$, $(2., \frac{1}{6})$, $(1., \frac{1}{2})$, $(1., \frac{1}{2})$, $(2.44949, -\frac{7}{16})$, $(2., -\frac{1}{3})$, $(2.44949, \frac{5}{16})$, $(2., -\frac{3}{8})$, $(2.44949, -\frac{3}{16})$, $(2., \frac{7}{24})$, $(2., \frac{7}{24})$\}

\item $c = \frac{7}{2}$, $(d_i,\theta_i)$ = \{$(1., 0)$, $(1., 0)$, $(2.44949, \frac{1}{16})$, $(2., \frac{1}{6})$, $(1., \frac{1}{2})$, $(1., \frac{1}{2})$, $(2.44949, -\frac{7}{16})$, $(2., -\frac{1}{3})$, $(2.82843, \frac{17}{48})$, $(1.73205, -\frac{1}{8})$, $(1.73205, \frac{3}{8})$, $(1.41421, -\frac{5}{16})$, $(1.73205, -\frac{1}{8})$, $(1.73205, \frac{3}{8})$, $(1.41421, -\frac{5}{16})$\}

\item $c = 4$, $(d_i,\theta_i)$ = \{$(1., 0)$, $(1., 0)$, $(2.44949, \frac{1}{16})$, $(2., \frac{1}{6})$, $(1., \frac{1}{2})$, $(1., \frac{1}{2})$, $(2.44949, -\frac{7}{16})$, $(2., -\frac{1}{3})$, $(2., -\frac{1}{4})$, $(2.44949, -\frac{1}{16})$, $(2.44949, \frac{7}{16})$, $(2., \frac{5}{12})$, $(2., \frac{5}{12})$\}

\item $c = \frac{9}{2}$, $(d_i,\theta_i)$ = \{$(1., 0)$, $(1., 0)$, $(2.44949, \frac{1}{16})$, $(2., \frac{1}{6})$, $(1., \frac{1}{2})$, $(1., \frac{1}{2})$, $(2.44949, -\frac{7}{16})$, $(2., -\frac{1}{3})$, $(2.82843, \frac{23}{48})$, $(1.73205, \frac{1}{2})$, $(1.73205, 0)$, $(1.41421, -\frac{3}{16})$, $(1.73205, \frac{1}{2})$, $(1.73205, 0)$, $(1.41421, -\frac{3}{16})$\}

\item $c = 5$, $(d_i,\theta_i)$ = \{$(1., 0)$, $(1., 0)$, $(2.44949, \frac{1}{16})$, $(2., \frac{1}{6})$, $(1., \frac{1}{2})$, $(1., \frac{1}{2})$, $(2.44949, -\frac{7}{16})$, $(2., -\frac{1}{3})$, $(2., -\frac{1}{8})$, $(2.44949, -\frac{7}{16})$, $(2.44949, \frac{1}{16})$, $(2., -\frac{11}{24})$, $(2., -\frac{11}{24})$\}

\item $c = \frac{11}{2}$, $(d_i,\theta_i)$ = \{$(1., 0)$, $(1., 0)$, $(2.44949, \frac{1}{16})$, $(2., \frac{1}{6})$, $(1., \frac{1}{2})$, $(1., \frac{1}{2})$, $(2.44949, -\frac{7}{16})$, $(2., -\frac{1}{3})$, $(2.82843, -\frac{19}{48})$, $(1.73205, -\frac{3}{8})$, $(1.73205, \frac{1}{8})$, $(1.41421, -\frac{1}{16})$, $(1.73205, -\frac{3}{8})$, $(1.73205, \frac{1}{8})$, $(1.41421, -\frac{1}{16})$\}

\item $c = 6$, $(d_i,\theta_i)$ = \{$(1., 0)$, $(1., \frac{1}{2})$, $(2.44949, \frac{1}{16})$, $(2., \frac{1}{6})$, $(1., \frac{1}{2})$, $(1., 0)$, $(2.44949, -\frac{7}{16})$, $(2., -\frac{1}{3})$, $(2., 0)$, $(2.44949, -\frac{5}{16})$, $(2.44949, \frac{3}{16})$, $(2., -\frac{1}{3})$, $(2., -\frac{1}{3})$\}

\item $c = \frac{13}{2}$, $(d_i,\theta_i)$ = \{$(1., 0)$, $(1., \frac{1}{2})$, $(2.44949, \frac{1}{16})$, $(2., \frac{1}{6})$, $(1., \frac{1}{2})$, $(1., 0)$, $(2.44949, -\frac{7}{16})$, $(2., -\frac{1}{3})$, $(2.82843, -\frac{13}{48})$, $(1.73205, \frac{1}{4})$, $(1.73205, -\frac{1}{4})$, $(1.41421, \frac{1}{16})$, $(1.73205, \frac{1}{4})$, $(1.73205, -\frac{1}{4})$, $(1.41421, \frac{1}{16})$\}

\item $c = 7$, $(d_i,\theta_i)$ = \{$(1., 0)$, $(1., \frac{1}{2})$, $(2.44949, \frac{1}{16})$, $(2., \frac{1}{6})$, $(1., \frac{1}{2})$, $(1., 0)$, $(2.44949, -\frac{7}{16})$, $(2., -\frac{1}{3})$, $(2.44949, \frac{5}{16})$, $(2., \frac{1}{8})$, $(2.44949, -\frac{3}{16})$, $(2., -\frac{5}{24})$, $(2., -\frac{5}{24})$\}

\item $c = \frac{15}{2}$, $(d_i,\theta_i)$ = \{$(1., 0)$, $(1., \frac{1}{2})$, $(2.44949, -\frac{7}{16})$, $(2., \frac{1}{6})$, $(1., \frac{1}{2})$, $(1., 0)$, $(2.44949, \frac{1}{16})$, $(2., -\frac{1}{3})$, $(2.82843, -\frac{7}{48})$, $(1.73205, \frac{3}{8})$, $(1.73205, -\frac{1}{8})$, $(1.41421, \frac{3}{16})$, $(1.73205, \frac{3}{8})$, $(1.73205, -\frac{1}{8})$, $(1.41421, \frac{3}{16})$\}

\end{enumerate}

\paragraph*{Rank 8; \#19}

\begin{enumerate}

\item $c = 0$, $(d_i,\theta_i)$ = \{$(1., 0)$, $(1., 0)$, $(2., \frac{1}{6})$, $(2.44949, -\frac{1}{16})$, $(1., \frac{1}{2})$, $(1., \frac{1}{2})$, $(2., -\frac{1}{3})$, $(2.44949, \frac{7}{16})$, $(2., \frac{1}{4})$, $(2.44949, \frac{1}{16})$, $(2.44949, -\frac{7}{16})$, $(2., -\frac{1}{12})$, $(2., -\frac{1}{12})$\}

\item $c = \frac{1}{2}$, $(d_i,\theta_i)$ = \{$(1., 0)$, $(1., 0)$, $(2., \frac{1}{6})$, $(2.44949, -\frac{1}{16})$, $(1., \frac{1}{2})$, $(1., \frac{1}{2})$, $(2., -\frac{1}{3})$, $(2.44949, \frac{7}{16})$, $(2.82843, -\frac{1}{48})$, $(1.73205, -\frac{3}{8})$, $(1.73205, \frac{1}{8})$, $(1.41421, \frac{5}{16})$, $(1.73205, -\frac{3}{8})$, $(1.73205, \frac{1}{8})$, $(1.41421, \frac{5}{16})$\}

\item $c = 1$, $(d_i,\theta_i)$ = \{$(1., 0)$, $(1., 0)$, $(2., \frac{1}{6})$, $(2.44949, -\frac{1}{16})$, $(1., \frac{1}{2})$, $(1., \frac{1}{2})$, $(2., -\frac{1}{3})$, $(2.44949, \frac{7}{16})$, $(2.44949, \frac{3}{16})$, $(2.44949, -\frac{5}{16})$, $(2., \frac{3}{8})$, $(2., \frac{1}{24})$, $(2., \frac{1}{24})$\}

\item $c = \frac{3}{2}$, $(d_i,\theta_i)$ = \{$(1., 0)$, $(1., 0)$, $(2., \frac{1}{6})$, $(2.44949, -\frac{1}{16})$, $(1., \frac{1}{2})$, $(1., \frac{1}{2})$, $(2., -\frac{1}{3})$, $(2.44949, \frac{7}{16})$, $(2.82843, \frac{5}{48})$, $(1.73205, \frac{1}{4})$, $(1.73205, -\frac{1}{4})$, $(1.41421, \frac{7}{16})$, $(1.73205, \frac{1}{4})$, $(1.73205, -\frac{1}{4})$, $(1.41421, \frac{7}{16})$\}

\item $c = 2$, $(d_i,\theta_i)$ = \{$(1., 0)$, $(1., 0)$, $(2., \frac{1}{6})$, $(2.44949, -\frac{1}{16})$, $(1., \frac{1}{2})$, $(1., \frac{1}{2})$, $(2., -\frac{1}{3})$, $(2.44949, \frac{7}{16})$, $(2.44949, \frac{5}{16})$, $(2.44949, -\frac{3}{16})$, $(2., \frac{1}{2})$, $(2., \frac{1}{6})$, $(2., \frac{1}{6})$\}

\item $c = \frac{5}{2}$, $(d_i,\theta_i)$ = \{$(1., 0)$, $(1., 0)$, $(2., \frac{1}{6})$, $(2.44949, -\frac{1}{16})$, $(1., \frac{1}{2})$, $(1., \frac{1}{2})$, $(2., -\frac{1}{3})$, $(2.44949, \frac{7}{16})$, $(2.82843, \frac{11}{48})$, $(1.73205, \frac{3}{8})$, $(1.73205, -\frac{1}{8})$, $(1.41421, -\frac{7}{16})$, $(1.73205, \frac{3}{8})$, $(1.73205, -\frac{1}{8})$, $(1.41421, -\frac{7}{16})$\}

\item $c = 3$, $(d_i,\theta_i)$ = \{$(1., 0)$, $(1., 0)$, $(2., \frac{1}{6})$, $(2.44949, -\frac{1}{16})$, $(1., \frac{1}{2})$, $(1., \frac{1}{2})$, $(2., -\frac{1}{3})$, $(2.44949, \frac{7}{16})$, $(2., -\frac{3}{8})$, $(2.44949, \frac{7}{16})$, $(2.44949, -\frac{1}{16})$, $(2., \frac{7}{24})$, $(2., \frac{7}{24})$\}

\item $c = \frac{7}{2}$, $(d_i,\theta_i)$ = \{$(1., 0)$, $(1., 0)$, $(2., \frac{1}{6})$, $(2.44949, -\frac{1}{16})$, $(1., \frac{1}{2})$, $(1., \frac{1}{2})$, $(2., -\frac{1}{3})$, $(2.44949, \frac{7}{16})$, $(2.82843, \frac{17}{48})$, $(1.73205, 0)$, $(1.73205, \frac{1}{2})$, $(1.41421, -\frac{5}{16})$, $(1.73205, 0)$, $(1.73205, \frac{1}{2})$, $(1.41421, -\frac{5}{16})$\}

\item $c = 4$, $(d_i,\theta_i)$ = \{$(1., 0)$, $(1., 0)$, $(2., \frac{1}{6})$, $(2.44949, -\frac{1}{16})$, $(1., \frac{1}{2})$, $(1., \frac{1}{2})$, $(2., -\frac{1}{3})$, $(2.44949, \frac{7}{16})$, $(2.44949, \frac{1}{16})$, $(2.44949, -\frac{7}{16})$, $(2., -\frac{1}{4})$, $(2., \frac{5}{12})$, $(2., \frac{5}{12})$\}

\item $c = \frac{9}{2}$, $(d_i,\theta_i)$ = \{$(1., 0)$, $(1., 0)$, $(2., \frac{1}{6})$, $(2.44949, \frac{7}{16})$, $(1., \frac{1}{2})$, $(1., \frac{1}{2})$, $(2., -\frac{1}{3})$, $(2.44949, -\frac{1}{16})$, $(2.82843, \frac{23}{48})$, $(1.73205, \frac{1}{8})$, $(1.73205, -\frac{3}{8})$, $(1.41421, -\frac{3}{16})$, $(1.73205, \frac{1}{8})$, $(1.73205, -\frac{3}{8})$, $(1.41421, -\frac{3}{16})$\}

\item $c = 5$, $(d_i,\theta_i)$ = \{$(1., 0)$, $(1., 0)$, $(2., \frac{1}{6})$, $(2.44949, \frac{7}{16})$, $(1., \frac{1}{2})$, $(1., \frac{1}{2})$, $(2., -\frac{1}{3})$, $(2.44949, -\frac{1}{16})$, $(2., -\frac{1}{8})$, $(2.44949, -\frac{5}{16})$, $(2.44949, \frac{3}{16})$, $(2., -\frac{11}{24})$, $(2., -\frac{11}{24})$\}

\item $c = \frac{11}{2}$, $(d_i,\theta_i)$ = \{$(1., 0)$, $(1., 0)$, $(2., \frac{1}{6})$, $(2.44949, \frac{7}{16})$, $(1., \frac{1}{2})$, $(1., \frac{1}{2})$, $(2., -\frac{1}{3})$, $(2.44949, -\frac{1}{16})$, $(2.82843, -\frac{19}{48})$, $(1.73205, -\frac{1}{4})$, $(1.73205, \frac{1}{4})$, $(1.41421, -\frac{1}{16})$, $(1.73205, -\frac{1}{4})$, $(1.73205, \frac{1}{4})$, $(1.41421, -\frac{1}{16})$\}

\item $c = 6$, $(d_i,\theta_i)$ = \{$(1., 0)$, $(1., \frac{1}{2})$, $(2., \frac{1}{6})$, $(2.44949, \frac{7}{16})$, $(1., \frac{1}{2})$, $(1., 0)$, $(2., -\frac{1}{3})$, $(2.44949, -\frac{1}{16})$, $(2., 0)$, $(2.44949, -\frac{3}{16})$, $(2.44949, \frac{5}{16})$, $(2., -\frac{1}{3})$, $(2., -\frac{1}{3})$\}

\item $c = \frac{13}{2}$, $(d_i,\theta_i)$ = \{$(1., 0)$, $(1., \frac{1}{2})$, $(2., \frac{1}{6})$, $(2.44949, \frac{7}{16})$, $(1., \frac{1}{2})$, $(1., 0)$, $(2., -\frac{1}{3})$, $(2.44949, -\frac{1}{16})$, $(2.82843, -\frac{13}{48})$, $(1.73205, -\frac{1}{8})$, $(1.73205, \frac{3}{8})$, $(1.41421, \frac{1}{16})$, $(1.73205, -\frac{1}{8})$, $(1.73205, \frac{3}{8})$, $(1.41421, \frac{1}{16})$\}

\item $c = 7$, $(d_i,\theta_i)$ = \{$(1., 0)$, $(1., \frac{1}{2})$, $(2., \frac{1}{6})$, $(2.44949, \frac{7}{16})$, $(1., \frac{1}{2})$, $(1., 0)$, $(2., -\frac{1}{3})$, $(2.44949, -\frac{1}{16})$, $(2.44949, -\frac{1}{16})$, $(2., \frac{1}{8})$, $(2.44949, \frac{7}{16})$, $(2., -\frac{5}{24})$, $(2., -\frac{5}{24})$\}

\item $c = \frac{15}{2}$, $(d_i,\theta_i)$ = \{$(1., 0)$, $(1., \frac{1}{2})$, $(2., \frac{1}{6})$, $(2.44949, \frac{7}{16})$, $(1., \frac{1}{2})$, $(1., 0)$, $(2., -\frac{1}{3})$, $(2.44949, -\frac{1}{16})$, $(2.82843, -\frac{7}{48})$, $(1.73205, \frac{1}{2})$, $(1.73205, 0)$, $(1.41421, \frac{3}{16})$, $(1.73205, \frac{1}{2})$, $(1.73205, 0)$, $(1.41421, \frac{3}{16})$\}

\end{enumerate}

\paragraph*{Rank 8; \#20}

\begin{enumerate}

\item $c = 0$, $(d_i,\theta_i)$ = \{$(1., 0)$, $(1., 0)$, $(2., \frac{1}{6})$, $(2.44949, \frac{3}{16})$, $(1., \frac{1}{2})$, $(1., \frac{1}{2})$, $(2., -\frac{1}{3})$, $(2.44949, -\frac{5}{16})$, $(2.44949, \frac{5}{16})$, $(2., \frac{1}{4})$, $(2.44949, -\frac{3}{16})$, $(2., -\frac{1}{12})$, $(2., -\frac{1}{12})$\}

\item $c = \frac{1}{2}$, $(d_i,\theta_i)$ = \{$(1., 0)$, $(1., 0)$, $(2., \frac{1}{6})$, $(2.44949, \frac{3}{16})$, $(1., \frac{1}{2})$, $(1., \frac{1}{2})$, $(2., -\frac{1}{3})$, $(2.44949, -\frac{5}{16})$, $(2.82843, -\frac{1}{48})$, $(1.73205, -\frac{1}{8})$, $(1.73205, \frac{3}{8})$, $(1.41421, \frac{5}{16})$, $(1.73205, -\frac{1}{8})$, $(1.73205, \frac{3}{8})$, $(1.41421, \frac{5}{16})$\}

\item $c = 1$, $(d_i,\theta_i)$ = \{$(1., 0)$, $(1., 0)$, $(2., \frac{1}{6})$, $(2.44949, \frac{3}{16})$, $(1., \frac{1}{2})$, $(1., \frac{1}{2})$, $(2., -\frac{1}{3})$, $(2.44949, -\frac{5}{16})$, $(2., \frac{3}{8})$, $(2.44949, -\frac{1}{16})$, $(2.44949, \frac{7}{16})$, $(2., \frac{1}{24})$, $(2., \frac{1}{24})$\}

\item $c = \frac{3}{2}$, $(d_i,\theta_i)$ = \{$(1., 0)$, $(1., 0)$, $(2., \frac{1}{6})$, $(2.44949, \frac{3}{16})$, $(1., \frac{1}{2})$, $(1., \frac{1}{2})$, $(2., -\frac{1}{3})$, $(2.44949, -\frac{5}{16})$, $(2.82843, \frac{5}{48})$, $(1.73205, 0)$, $(1.73205, \frac{1}{2})$, $(1.41421, \frac{7}{16})$, $(1.73205, 0)$, $(1.73205, \frac{1}{2})$, $(1.41421, \frac{7}{16})$\}

\item $c = 2$, $(d_i,\theta_i)$ = \{$(1., 0)$, $(1., 0)$, $(2., \frac{1}{6})$, $(2.44949, \frac{3}{16})$, $(1., \frac{1}{2})$, $(1., \frac{1}{2})$, $(2., -\frac{1}{3})$, $(2.44949, -\frac{5}{16})$, $(2.44949, -\frac{7}{16})$, $(2.44949, \frac{1}{16})$, $(2., \frac{1}{2})$, $(2., \frac{1}{6})$, $(2., \frac{1}{6})$\}

\item $c = \frac{5}{2}$, $(d_i,\theta_i)$ = \{$(1., 0)$, $(1., 0)$, $(2., \frac{1}{6})$, $(2.44949, \frac{3}{16})$, $(1., \frac{1}{2})$, $(1., \frac{1}{2})$, $(2., -\frac{1}{3})$, $(2.44949, -\frac{5}{16})$, $(2.82843, \frac{11}{48})$, $(1.73205, -\frac{3}{8})$, $(1.73205, \frac{1}{8})$, $(1.41421, -\frac{7}{16})$, $(1.73205, -\frac{3}{8})$, $(1.73205, \frac{1}{8})$, $(1.41421, -\frac{7}{16})$\}

\item $c = 3$, $(d_i,\theta_i)$ = \{$(1., 0)$, $(1., 0)$, $(2., \frac{1}{6})$, $(2.44949, \frac{3}{16})$, $(1., \frac{1}{2})$, $(1., \frac{1}{2})$, $(2., -\frac{1}{3})$, $(2.44949, -\frac{5}{16})$, $(2.44949, \frac{3}{16})$, $(2., -\frac{3}{8})$, $(2.44949, -\frac{5}{16})$, $(2., \frac{7}{24})$, $(2., \frac{7}{24})$\}

\item $c = \frac{7}{2}$, $(d_i,\theta_i)$ = \{$(1., 0)$, $(1., 0)$, $(2., \frac{1}{6})$, $(2.44949, \frac{3}{16})$, $(1., \frac{1}{2})$, $(1., \frac{1}{2})$, $(2., -\frac{1}{3})$, $(2.44949, -\frac{5}{16})$, $(2.82843, \frac{17}{48})$, $(1.73205, -\frac{1}{4})$, $(1.73205, \frac{1}{4})$, $(1.41421, -\frac{5}{16})$, $(1.73205, -\frac{1}{4})$, $(1.73205, \frac{1}{4})$, $(1.41421, -\frac{5}{16})$\}

\item $c = 4$, $(d_i,\theta_i)$ = \{$(1., 0)$, $(1., 0)$, $(2., \frac{1}{6})$, $(2.44949, \frac{3}{16})$, $(1., \frac{1}{2})$, $(1., \frac{1}{2})$, $(2., -\frac{1}{3})$, $(2.44949, -\frac{5}{16})$, $(2., -\frac{1}{4})$, $(2.44949, -\frac{3}{16})$, $(2.44949, \frac{5}{16})$, $(2., \frac{5}{12})$, $(2., \frac{5}{12})$\}

\item $c = \frac{9}{2}$, $(d_i,\theta_i)$ = \{$(1., 0)$, $(1., 0)$, $(2., \frac{1}{6})$, $(2.44949, \frac{3}{16})$, $(1., \frac{1}{2})$, $(1., \frac{1}{2})$, $(2., -\frac{1}{3})$, $(2.44949, -\frac{5}{16})$, $(2.82843, \frac{23}{48})$, $(1.73205, \frac{3}{8})$, $(1.73205, -\frac{1}{8})$, $(1.41421, -\frac{3}{16})$, $(1.73205, \frac{3}{8})$, $(1.73205, -\frac{1}{8})$, $(1.41421, -\frac{3}{16})$\}

\item $c = 5$, $(d_i,\theta_i)$ = \{$(1., 0)$, $(1., 0)$, $(2., \frac{1}{6})$, $(2.44949, \frac{3}{16})$, $(1., \frac{1}{2})$, $(1., \frac{1}{2})$, $(2., -\frac{1}{3})$, $(2.44949, -\frac{5}{16})$, $(2., -\frac{1}{8})$, $(2.44949, \frac{7}{16})$, $(2.44949, -\frac{1}{16})$, $(2., -\frac{11}{24})$, $(2., -\frac{11}{24})$\}

\item $c = \frac{11}{2}$, $(d_i,\theta_i)$ = \{$(1., 0)$, $(1., 0)$, $(2., \frac{1}{6})$, $(2.44949, \frac{3}{16})$, $(1., \frac{1}{2})$, $(1., \frac{1}{2})$, $(2., -\frac{1}{3})$, $(2.44949, -\frac{5}{16})$, $(2.82843, -\frac{19}{48})$, $(1.73205, \frac{1}{2})$, $(1.73205, 0)$, $(1.41421, -\frac{1}{16})$, $(1.73205, \frac{1}{2})$, $(1.73205, 0)$, $(1.41421, -\frac{1}{16})$\}

\item $c = 6$, $(d_i,\theta_i)$ = \{$(1., 0)$, $(1., \frac{1}{2})$, $(2., \frac{1}{6})$, $(2.44949, \frac{3}{16})$, $(1., \frac{1}{2})$, $(1., 0)$, $(2., -\frac{1}{3})$, $(2.44949, -\frac{5}{16})$, $(2., 0)$, $(2.44949, \frac{1}{16})$, $(2.44949, -\frac{7}{16})$, $(2., -\frac{1}{3})$, $(2., -\frac{1}{3})$\}

\item $c = \frac{13}{2}$, $(d_i,\theta_i)$ = \{$(1., 0)$, $(1., \frac{1}{2})$, $(2., \frac{1}{6})$, $(2.44949, \frac{3}{16})$, $(1., \frac{1}{2})$, $(1., 0)$, $(2., -\frac{1}{3})$, $(2.44949, -\frac{5}{16})$, $(2.82843, -\frac{13}{48})$, $(1.73205, \frac{1}{8})$, $(1.73205, -\frac{3}{8})$, $(1.41421, \frac{1}{16})$, $(1.73205, \frac{1}{8})$, $(1.73205, -\frac{3}{8})$, $(1.41421, \frac{1}{16})$\}

\item $c = 7$, $(d_i,\theta_i)$ = \{$(1., 0)$, $(1., \frac{1}{2})$, $(2., \frac{1}{6})$, $(2.44949, \frac{3}{16})$, $(1., \frac{1}{2})$, $(1., 0)$, $(2., -\frac{1}{3})$, $(2.44949, -\frac{5}{16})$, $(2.44949, -\frac{5}{16})$, $(2., \frac{1}{8})$, $(2.44949, \frac{3}{16})$, $(2., -\frac{5}{24})$, $(2., -\frac{5}{24})$\}

\item $c = \frac{15}{2}$, $(d_i,\theta_i)$ = \{$(1., 0)$, $(1., \frac{1}{2})$, $(2., \frac{1}{6})$, $(2.44949, \frac{3}{16})$, $(1., \frac{1}{2})$, $(1., 0)$, $(2., -\frac{1}{3})$, $(2.44949, -\frac{5}{16})$, $(2.82843, -\frac{7}{48})$, $(1.73205, \frac{1}{4})$, $(1.73205, -\frac{1}{4})$, $(1.41421, \frac{3}{16})$, $(1.73205, \frac{1}{4})$, $(1.73205, -\frac{1}{4})$, $(1.41421, \frac{3}{16})$\}

\end{enumerate}

\paragraph*{Rank 8; \#21}

\begin{enumerate}

\item $c = 0$, $(d_i,\theta_i)$ = \{$(1., 0)$, $(1., 0)$, $(2., \frac{1}{6})$, $(2.44949, -\frac{3}{16})$, $(1., \frac{1}{2})$, $(1., \frac{1}{2})$, $(2., -\frac{1}{3})$, $(2.44949, \frac{5}{16})$, $(2.44949, \frac{3}{16})$, $(2.44949, -\frac{5}{16})$, $(2., \frac{1}{4})$, $(2., -\frac{1}{12})$, $(2., -\frac{1}{12})$\}

\item $c = \frac{1}{2}$, $(d_i,\theta_i)$ = \{$(1., 0)$, $(1., 0)$, $(2., \frac{1}{6})$, $(2.44949, -\frac{3}{16})$, $(1., \frac{1}{2})$, $(1., \frac{1}{2})$, $(2., -\frac{1}{3})$, $(2.44949, \frac{5}{16})$, $(2.82843, -\frac{1}{48})$, $(1.73205, -\frac{1}{4})$, $(1.73205, \frac{1}{4})$, $(1.41421, \frac{5}{16})$, $(1.73205, -\frac{1}{4})$, $(1.73205, \frac{1}{4})$, $(1.41421, \frac{5}{16})$\}

\item $c = 1$, $(d_i,\theta_i)$ = \{$(1., 0)$, $(1., 0)$, $(2., \frac{1}{6})$, $(2.44949, -\frac{3}{16})$, $(1., \frac{1}{2})$, $(1., \frac{1}{2})$, $(2., -\frac{1}{3})$, $(2.44949, \frac{5}{16})$, $(2., \frac{3}{8})$, $(2.44949, -\frac{3}{16})$, $(2.44949, \frac{5}{16})$, $(2., \frac{1}{24})$, $(2., \frac{1}{24})$\}

\item $c = \frac{3}{2}$, $(d_i,\theta_i)$ = \{$(1., 0)$, $(1., 0)$, $(2., \frac{1}{6})$, $(2.44949, \frac{5}{16})$, $(1., \frac{1}{2})$, $(1., \frac{1}{2})$, $(2., -\frac{1}{3})$, $(2.44949, -\frac{3}{16})$, $(2.82843, \frac{5}{48})$, $(1.73205, \frac{3}{8})$, $(1.73205, -\frac{1}{8})$, $(1.41421, \frac{7}{16})$, $(1.73205, \frac{3}{8})$, $(1.73205, -\frac{1}{8})$, $(1.41421, \frac{7}{16})$\}

\item $c = 2$, $(d_i,\theta_i)$ = \{$(1., 0)$, $(1., 0)$, $(2., \frac{1}{6})$, $(2.44949, \frac{5}{16})$, $(1., \frac{1}{2})$, $(1., \frac{1}{2})$, $(2., -\frac{1}{3})$, $(2.44949, -\frac{3}{16})$, $(2.44949, -\frac{1}{16})$, $(2., \frac{1}{2})$, $(2.44949, \frac{7}{16})$, $(2., \frac{1}{6})$, $(2., \frac{1}{6})$\}

\item $c = \frac{5}{2}$, $(d_i,\theta_i)$ = \{$(1., 0)$, $(1., 0)$, $(2., \frac{1}{6})$, $(2.44949, \frac{5}{16})$, $(1., \frac{1}{2})$, $(1., \frac{1}{2})$, $(2., -\frac{1}{3})$, $(2.44949, -\frac{3}{16})$, $(2.82843, \frac{11}{48})$, $(1.73205, \frac{1}{2})$, $(1.73205, 0)$, $(1.41421, -\frac{7}{16})$, $(1.73205, \frac{1}{2})$, $(1.73205, 0)$, $(1.41421, -\frac{7}{16})$\}

\item $c = 3$, $(d_i,\theta_i)$ = \{$(1., 0)$, $(1., 0)$, $(2., \frac{1}{6})$, $(2.44949, \frac{5}{16})$, $(1., \frac{1}{2})$, $(1., \frac{1}{2})$, $(2., -\frac{1}{3})$, $(2.44949, -\frac{3}{16})$, $(2., -\frac{3}{8})$, $(2.44949, -\frac{7}{16})$, $(2.44949, \frac{1}{16})$, $(2., \frac{7}{24})$, $(2., \frac{7}{24})$\}

\item $c = \frac{7}{2}$, $(d_i,\theta_i)$ = \{$(1., 0)$, $(1., 0)$, $(2., \frac{1}{6})$, $(2.44949, \frac{5}{16})$, $(1., \frac{1}{2})$, $(1., \frac{1}{2})$, $(2., -\frac{1}{3})$, $(2.44949, -\frac{3}{16})$, $(2.82843, \frac{17}{48})$, $(1.73205, \frac{1}{8})$, $(1.73205, -\frac{3}{8})$, $(1.41421, -\frac{5}{16})$, $(1.73205, \frac{1}{8})$, $(1.73205, -\frac{3}{8})$, $(1.41421, -\frac{5}{16})$\}

\item $c = 4$, $(d_i,\theta_i)$ = \{$(1., 0)$, $(1., 0)$, $(2., \frac{1}{6})$, $(2.44949, \frac{5}{16})$, $(1., \frac{1}{2})$, $(1., \frac{1}{2})$, $(2., -\frac{1}{3})$, $(2.44949, -\frac{3}{16})$, $(2.44949, -\frac{5}{16})$, $(2., -\frac{1}{4})$, $(2.44949, \frac{3}{16})$, $(2., \frac{5}{12})$, $(2., \frac{5}{12})$\}

\item $c = \frac{9}{2}$, $(d_i,\theta_i)$ = \{$(1., 0)$, $(1., 0)$, $(2., \frac{1}{6})$, $(2.44949, \frac{5}{16})$, $(1., \frac{1}{2})$, $(1., \frac{1}{2})$, $(2., -\frac{1}{3})$, $(2.44949, -\frac{3}{16})$, $(2.82843, \frac{23}{48})$, $(1.73205, \frac{1}{4})$, $(1.73205, -\frac{1}{4})$, $(1.41421, -\frac{3}{16})$, $(1.73205, \frac{1}{4})$, $(1.73205, -\frac{1}{4})$, $(1.41421, -\frac{3}{16})$\}

\item $c = 5$, $(d_i,\theta_i)$ = \{$(1., 0)$, $(1., 0)$, $(2., \frac{1}{6})$, $(2.44949, \frac{5}{16})$, $(1., \frac{1}{2})$, $(1., \frac{1}{2})$, $(2., -\frac{1}{3})$, $(2.44949, -\frac{3}{16})$, $(2.44949, \frac{5}{16})$, $(2.44949, -\frac{3}{16})$, $(2., -\frac{1}{8})$, $(2., -\frac{11}{24})$, $(2., -\frac{11}{24})$\}

\item $c = \frac{11}{2}$, $(d_i,\theta_i)$ = \{$(1., 0)$, $(1., 0)$, $(2., \frac{1}{6})$, $(2.44949, \frac{5}{16})$, $(1., \frac{1}{2})$, $(1., \frac{1}{2})$, $(2., -\frac{1}{3})$, $(2.44949, -\frac{3}{16})$, $(2.82843, -\frac{19}{48})$, $(1.73205, -\frac{1}{8})$, $(1.73205, \frac{3}{8})$, $(1.41421, -\frac{1}{16})$, $(1.73205, -\frac{1}{8})$, $(1.73205, \frac{3}{8})$, $(1.41421, -\frac{1}{16})$\}

\item $c = 6$, $(d_i,\theta_i)$ = \{$(1., 0)$, $(1., \frac{1}{2})$, $(2., \frac{1}{6})$, $(2.44949, \frac{5}{16})$, $(1., \frac{1}{2})$, $(1., 0)$, $(2., -\frac{1}{3})$, $(2.44949, -\frac{3}{16})$, $(2.44949, -\frac{1}{16})$, $(2.44949, \frac{7}{16})$, $(2., 0)$, $(2., -\frac{1}{3})$, $(2., -\frac{1}{3})$\}

\item $c = \frac{13}{2}$, $(d_i,\theta_i)$ = \{$(1., 0)$, $(1., \frac{1}{2})$, $(2., \frac{1}{6})$, $(2.44949, \frac{5}{16})$, $(1., \frac{1}{2})$, $(1., 0)$, $(2., -\frac{1}{3})$, $(2.44949, -\frac{3}{16})$, $(2.82843, -\frac{13}{48})$, $(1.73205, 0)$, $(1.73205, \frac{1}{2})$, $(1.41421, \frac{1}{16})$, $(1.73205, 0)$, $(1.73205, \frac{1}{2})$, $(1.41421, \frac{1}{16})$\}

\item $c = 7$, $(d_i,\theta_i)$ = \{$(1., 0)$, $(1., \frac{1}{2})$, $(2., \frac{1}{6})$, $(2.44949, \frac{5}{16})$, $(1., \frac{1}{2})$, $(1., 0)$, $(2., -\frac{1}{3})$, $(2.44949, -\frac{3}{16})$, $(2.44949, \frac{1}{16})$, $(2., \frac{1}{8})$, $(2.44949, -\frac{7}{16})$, $(2., -\frac{5}{24})$, $(2., -\frac{5}{24})$\}

\item $c = \frac{15}{2}$, $(d_i,\theta_i)$ = \{$(1., 0)$, $(1., \frac{1}{2})$, $(2., \frac{1}{6})$, $(2.44949, \frac{5}{16})$, $(1., \frac{1}{2})$, $(1., 0)$, $(2., -\frac{1}{3})$, $(2.44949, -\frac{3}{16})$, $(2.82843, -\frac{7}{48})$, $(1.73205, -\frac{3}{8})$, $(1.73205, \frac{1}{8})$, $(1.41421, \frac{3}{16})$, $(1.73205, -\frac{3}{8})$, $(1.73205, \frac{1}{8})$, $(1.41421, \frac{3}{16})$\}

\end{enumerate}

\paragraph*{Rank 8; \#22}

\begin{enumerate}

\item $c = 0$, $(d_i,\theta_i)$ = \{$(1., 0)$, $(1., 0)$, $(2.44949, \frac{1}{16})$, $(2., -\frac{1}{6})$, $(1., \frac{1}{2})$, $(1., \frac{1}{2})$, $(2.44949, -\frac{7}{16})$, $(2., \frac{1}{3})$, $(2., -\frac{1}{4})$, $(2.44949, -\frac{1}{16})$, $(2.44949, \frac{7}{16})$, $(2., \frac{1}{12})$, $(2., \frac{1}{12})$\}

\item $c = \frac{1}{2}$, $(d_i,\theta_i)$ = \{$(1., 0)$, $(1., 0)$, $(2.44949, \frac{1}{16})$, $(2., -\frac{1}{6})$, $(1., \frac{1}{2})$, $(1., \frac{1}{2})$, $(2.44949, -\frac{7}{16})$, $(2., \frac{1}{3})$, $(2.82843, \frac{7}{48})$, $(1.73205, 0)$, $(1.73205, \frac{1}{2})$, $(1.41421, -\frac{3}{16})$, $(1.73205, 0)$, $(1.73205, \frac{1}{2})$, $(1.41421, -\frac{3}{16})$\}

\item $c = 1$, $(d_i,\theta_i)$ = \{$(1., 0)$, $(1., 0)$, $(2.44949, \frac{1}{16})$, $(2., -\frac{1}{6})$, $(1., \frac{1}{2})$, $(1., \frac{1}{2})$, $(2.44949, -\frac{7}{16})$, $(2., \frac{1}{3})$, $(2., -\frac{1}{8})$, $(2.44949, -\frac{7}{16})$, $(2.44949, \frac{1}{16})$, $(2., \frac{5}{24})$, $(2., \frac{5}{24})$\}

\item $c = \frac{3}{2}$, $(d_i,\theta_i)$ = \{$(1., 0)$, $(1., 0)$, $(2.44949, \frac{1}{16})$, $(2., -\frac{1}{6})$, $(1., \frac{1}{2})$, $(1., \frac{1}{2})$, $(2.44949, -\frac{7}{16})$, $(2., \frac{1}{3})$, $(2.82843, \frac{13}{48})$, $(1.73205, \frac{1}{8})$, $(1.73205, -\frac{3}{8})$, $(1.41421, -\frac{1}{16})$, $(1.73205, \frac{1}{8})$, $(1.73205, -\frac{3}{8})$, $(1.41421, -\frac{1}{16})$\}

\item $c = 2$, $(d_i,\theta_i)$ = \{$(1., 0)$, $(1., 0)$, $(2.44949, \frac{1}{16})$, $(2., \frac{1}{3})$, $(1., \frac{1}{2})$, $(1., \frac{1}{2})$, $(2.44949, -\frac{7}{16})$, $(2., -\frac{1}{6})$, $(2.44949, -\frac{5}{16})$, $(2.44949, \frac{3}{16})$, $(2., 0)$, $(2., \frac{1}{3})$, $(2., \frac{1}{3})$\}

\item $c = \frac{5}{2}$, $(d_i,\theta_i)$ = \{$(1., 0)$, $(1., 0)$, $(2.44949, \frac{1}{16})$, $(2., \frac{1}{3})$, $(1., \frac{1}{2})$, $(1., \frac{1}{2})$, $(2.44949, -\frac{7}{16})$, $(2., -\frac{1}{6})$, $(2.82843, \frac{19}{48})$, $(1.73205, -\frac{1}{4})$, $(1.73205, \frac{1}{4})$, $(1.41421, \frac{1}{16})$, $(1.73205, -\frac{1}{4})$, $(1.73205, \frac{1}{4})$, $(1.41421, \frac{1}{16})$\}

\item $c = 3$, $(d_i,\theta_i)$ = \{$(1., 0)$, $(1., 0)$, $(2.44949, \frac{1}{16})$, $(2., \frac{1}{3})$, $(1., \frac{1}{2})$, $(1., \frac{1}{2})$, $(2.44949, -\frac{7}{16})$, $(2., -\frac{1}{6})$, $(2., \frac{1}{8})$, $(2.44949, \frac{5}{16})$, $(2.44949, -\frac{3}{16})$, $(2., \frac{11}{24})$, $(2., \frac{11}{24})$\}

\item $c = \frac{7}{2}$, $(d_i,\theta_i)$ = \{$(1., 0)$, $(1., 0)$, $(2.44949, \frac{1}{16})$, $(2., \frac{1}{3})$, $(1., \frac{1}{2})$, $(1., \frac{1}{2})$, $(2.44949, -\frac{7}{16})$, $(2., -\frac{1}{6})$, $(2.82843, -\frac{23}{48})$, $(1.73205, -\frac{1}{8})$, $(1.73205, \frac{3}{8})$, $(1.41421, \frac{3}{16})$, $(1.73205, -\frac{1}{8})$, $(1.73205, \frac{3}{8})$, $(1.41421, \frac{3}{16})$\}

\item $c = 4$, $(d_i,\theta_i)$ = \{$(1., 0)$, $(1., 0)$, $(2.44949, \frac{1}{16})$, $(2., \frac{1}{3})$, $(1., \frac{1}{2})$, $(1., \frac{1}{2})$, $(2.44949, -\frac{7}{16})$, $(2., -\frac{1}{6})$, $(2.44949, -\frac{1}{16})$, $(2., \frac{1}{4})$, $(2.44949, \frac{7}{16})$, $(2., -\frac{5}{12})$, $(2., -\frac{5}{12})$\}

\item $c = \frac{9}{2}$, $(d_i,\theta_i)$ = \{$(1., 0)$, $(1., 0)$, $(2.44949, \frac{1}{16})$, $(2., \frac{1}{3})$, $(1., \frac{1}{2})$, $(1., \frac{1}{2})$, $(2.44949, -\frac{7}{16})$, $(2., -\frac{1}{6})$, $(2.82843, -\frac{17}{48})$, $(1.73205, \frac{1}{2})$, $(1.73205, 0)$, $(1.41421, \frac{5}{16})$, $(1.73205, \frac{1}{2})$, $(1.73205, 0)$, $(1.41421, \frac{5}{16})$\}

\item $c = 5$, $(d_i,\theta_i)$ = \{$(1., 0)$, $(1., 0)$, $(2.44949, \frac{1}{16})$, $(2., \frac{1}{3})$, $(1., \frac{1}{2})$, $(1., \frac{1}{2})$, $(2.44949, -\frac{7}{16})$, $(2., -\frac{1}{6})$, $(2., \frac{3}{8})$, $(2.44949, \frac{1}{16})$, $(2.44949, -\frac{7}{16})$, $(2., -\frac{7}{24})$, $(2., -\frac{7}{24})$\}

\item $c = \frac{11}{2}$, $(d_i,\theta_i)$ = \{$(1., 0)$, $(1., 0)$, $(2.44949, \frac{1}{16})$, $(2., \frac{1}{3})$, $(1., \frac{1}{2})$, $(1., \frac{1}{2})$, $(2.44949, -\frac{7}{16})$, $(2., -\frac{1}{6})$, $(2.82843, -\frac{11}{48})$, $(1.73205, -\frac{3}{8})$, $(1.73205, \frac{1}{8})$, $(1.41421, \frac{7}{16})$, $(1.73205, -\frac{3}{8})$, $(1.73205, \frac{1}{8})$, $(1.41421, \frac{7}{16})$\}

\item $c = 6$, $(d_i,\theta_i)$ = \{$(1., 0)$, $(1., \frac{1}{2})$, $(2.44949, \frac{1}{16})$, $(2., \frac{1}{3})$, $(1., \frac{1}{2})$, $(1., 0)$, $(2.44949, -\frac{7}{16})$, $(2., -\frac{1}{6})$, $(2., \frac{1}{2})$, $(2.44949, -\frac{5}{16})$, $(2.44949, \frac{3}{16})$, $(2., -\frac{1}{6})$, $(2., -\frac{1}{6})$\}

\item $c = \frac{13}{2}$, $(d_i,\theta_i)$ = \{$(1., 0)$, $(1., \frac{1}{2})$, $(2.44949, \frac{1}{16})$, $(2., \frac{1}{3})$, $(1., \frac{1}{2})$, $(1., 0)$, $(2.44949, -\frac{7}{16})$, $(2., -\frac{1}{6})$, $(2.82843, -\frac{5}{48})$, $(1.73205, \frac{1}{4})$, $(1.73205, -\frac{1}{4})$, $(1.41421, -\frac{7}{16})$, $(1.73205, \frac{1}{4})$, $(1.73205, -\frac{1}{4})$, $(1.41421, -\frac{7}{16})$\}

\item $c = 7$, $(d_i,\theta_i)$ = \{$(1., 0)$, $(1., \frac{1}{2})$, $(2.44949, \frac{1}{16})$, $(2., \frac{1}{3})$, $(1., \frac{1}{2})$, $(1., 0)$, $(2.44949, -\frac{7}{16})$, $(2., -\frac{1}{6})$, $(2., -\frac{3}{8})$, $(2.44949, -\frac{3}{16})$, $(2.44949, \frac{5}{16})$, $(2., -\frac{1}{24})$, $(2., -\frac{1}{24})$\}

\item $c = \frac{15}{2}$, $(d_i,\theta_i)$ = \{$(1., 0)$, $(1., \frac{1}{2})$, $(2.44949, -\frac{7}{16})$, $(2., \frac{1}{3})$, $(1., \frac{1}{2})$, $(1., 0)$, $(2.44949, \frac{1}{16})$, $(2., -\frac{1}{6})$, $(2.82843, \frac{1}{48})$, $(1.73205, \frac{3}{8})$, $(1.73205, -\frac{1}{8})$, $(1.41421, -\frac{5}{16})$, $(1.73205, \frac{3}{8})$, $(1.73205, -\frac{1}{8})$, $(1.41421, -\frac{5}{16})$\}

\end{enumerate}

\paragraph*{Rank 8; \#23}

\begin{enumerate}

\item $c = 0$, $(d_i,\theta_i)$ = \{$(1., 0)$, $(1., 0)$, $(2., -\frac{1}{6})$, $(2.44949, -\frac{1}{16})$, $(1., \frac{1}{2})$, $(1., \frac{1}{2})$, $(2., \frac{1}{3})$, $(2.44949, \frac{7}{16})$, $(2., -\frac{1}{4})$, $(2.44949, \frac{1}{16})$, $(2.44949, -\frac{7}{16})$, $(2., \frac{1}{12})$, $(2., \frac{1}{12})$\}

\item $c = \frac{1}{2}$, $(d_i,\theta_i)$ = \{$(1., 0)$, $(1., 0)$, $(2., -\frac{1}{6})$, $(2.44949, -\frac{1}{16})$, $(1., \frac{1}{2})$, $(1., \frac{1}{2})$, $(2., \frac{1}{3})$, $(2.44949, \frac{7}{16})$, $(2.82843, \frac{7}{48})$, $(1.73205, -\frac{3}{8})$, $(1.73205, \frac{1}{8})$, $(1.41421, -\frac{3}{16})$, $(1.73205, -\frac{3}{8})$, $(1.73205, \frac{1}{8})$, $(1.41421, -\frac{3}{16})$\}

\item $c = 1$, $(d_i,\theta_i)$ = \{$(1., 0)$, $(1., 0)$, $(2., -\frac{1}{6})$, $(2.44949, -\frac{1}{16})$, $(1., \frac{1}{2})$, $(1., \frac{1}{2})$, $(2., \frac{1}{3})$, $(2.44949, \frac{7}{16})$, $(2., -\frac{1}{8})$, $(2.44949, -\frac{5}{16})$, $(2.44949, \frac{3}{16})$, $(2., \frac{5}{24})$, $(2., \frac{5}{24})$\}

\item $c = \frac{3}{2}$, $(d_i,\theta_i)$ = \{$(1., 0)$, $(1., 0)$, $(2., -\frac{1}{6})$, $(2.44949, -\frac{1}{16})$, $(1., \frac{1}{2})$, $(1., \frac{1}{2})$, $(2., \frac{1}{3})$, $(2.44949, \frac{7}{16})$, $(2.82843, \frac{13}{48})$, $(1.73205, \frac{1}{4})$, $(1.73205, -\frac{1}{4})$, $(1.41421, -\frac{1}{16})$, $(1.73205, \frac{1}{4})$, $(1.73205, -\frac{1}{4})$, $(1.41421, -\frac{1}{16})$\}

\item $c = 2$, $(d_i,\theta_i)$ = \{$(1., 0)$, $(1., 0)$, $(2., \frac{1}{3})$, $(2.44949, -\frac{1}{16})$, $(1., \frac{1}{2})$, $(1., \frac{1}{2})$, $(2., -\frac{1}{6})$, $(2.44949, \frac{7}{16})$, $(2.44949, -\frac{3}{16})$, $(2., 0)$, $(2.44949, \frac{5}{16})$, $(2., \frac{1}{3})$, $(2., \frac{1}{3})$\}

\item $c = \frac{5}{2}$, $(d_i,\theta_i)$ = \{$(1., 0)$, $(1., 0)$, $(2., \frac{1}{3})$, $(2.44949, -\frac{1}{16})$, $(1., \frac{1}{2})$, $(1., \frac{1}{2})$, $(2., -\frac{1}{6})$, $(2.44949, \frac{7}{16})$, $(2.82843, \frac{19}{48})$, $(1.73205, \frac{3}{8})$, $(1.73205, -\frac{1}{8})$, $(1.41421, \frac{1}{16})$, $(1.73205, \frac{3}{8})$, $(1.73205, -\frac{1}{8})$, $(1.41421, \frac{1}{16})$\}

\item $c = 3$, $(d_i,\theta_i)$ = \{$(1., 0)$, $(1., 0)$, $(2., \frac{1}{3})$, $(2.44949, -\frac{1}{16})$, $(1., \frac{1}{2})$, $(1., \frac{1}{2})$, $(2., -\frac{1}{6})$, $(2.44949, \frac{7}{16})$, $(2.44949, -\frac{1}{16})$, $(2., \frac{1}{8})$, $(2.44949, \frac{7}{16})$, $(2., \frac{11}{24})$, $(2., \frac{11}{24})$\}

\item $c = \frac{7}{2}$, $(d_i,\theta_i)$ = \{$(1., 0)$, $(1., 0)$, $(2., \frac{1}{3})$, $(2.44949, -\frac{1}{16})$, $(1., \frac{1}{2})$, $(1., \frac{1}{2})$, $(2., -\frac{1}{6})$, $(2.44949, \frac{7}{16})$, $(2.82843, -\frac{23}{48})$, $(1.73205, 0)$, $(1.73205, \frac{1}{2})$, $(1.41421, \frac{3}{16})$, $(1.73205, 0)$, $(1.73205, \frac{1}{2})$, $(1.41421, \frac{3}{16})$\}

\item $c = 4$, $(d_i,\theta_i)$ = \{$(1., 0)$, $(1., 0)$, $(2., \frac{1}{3})$, $(2.44949, -\frac{1}{16})$, $(1., \frac{1}{2})$, $(1., \frac{1}{2})$, $(2., -\frac{1}{6})$, $(2.44949, \frac{7}{16})$, $(2., \frac{1}{4})$, $(2.44949, -\frac{7}{16})$, $(2.44949, \frac{1}{16})$, $(2., -\frac{5}{12})$, $(2., -\frac{5}{12})$\}

\item $c = \frac{9}{2}$, $(d_i,\theta_i)$ = \{$(1., 0)$, $(1., 0)$, $(2., \frac{1}{3})$, $(2.44949, \frac{7}{16})$, $(1., \frac{1}{2})$, $(1., \frac{1}{2})$, $(2., -\frac{1}{6})$, $(2.44949, -\frac{1}{16})$, $(2.82843, -\frac{17}{48})$, $(1.73205, \frac{1}{8})$, $(1.73205, -\frac{3}{8})$, $(1.41421, \frac{5}{16})$, $(1.73205, \frac{1}{8})$, $(1.73205, -\frac{3}{8})$, $(1.41421, \frac{5}{16})$\}

\item $c = 5$, $(d_i,\theta_i)$ = \{$(1., 0)$, $(1., 0)$, $(2., \frac{1}{3})$, $(2.44949, \frac{7}{16})$, $(1., \frac{1}{2})$, $(1., \frac{1}{2})$, $(2., -\frac{1}{6})$, $(2.44949, -\frac{1}{16})$, $(2., \frac{3}{8})$, $(2.44949, \frac{3}{16})$, $(2.44949, -\frac{5}{16})$, $(2., -\frac{7}{24})$, $(2., -\frac{7}{24})$\}

\item $c = \frac{11}{2}$, $(d_i,\theta_i)$ = \{$(1., 0)$, $(1., 0)$, $(2., \frac{1}{3})$, $(2.44949, \frac{7}{16})$, $(1., \frac{1}{2})$, $(1., \frac{1}{2})$, $(2., -\frac{1}{6})$, $(2.44949, -\frac{1}{16})$, $(2.82843, -\frac{11}{48})$, $(1.73205, -\frac{1}{4})$, $(1.73205, \frac{1}{4})$, $(1.41421, \frac{7}{16})$, $(1.73205, -\frac{1}{4})$, $(1.73205, \frac{1}{4})$, $(1.41421, \frac{7}{16})$\}

\item $c = 6$, $(d_i,\theta_i)$ = \{$(1., 0)$, $(1., \frac{1}{2})$, $(2., \frac{1}{3})$, $(2.44949, \frac{7}{16})$, $(1., \frac{1}{2})$, $(1., 0)$, $(2., -\frac{1}{6})$, $(2.44949, -\frac{1}{16})$, $(2., \frac{1}{2})$, $(2.44949, -\frac{3}{16})$, $(2.44949, \frac{5}{16})$, $(2., -\frac{1}{6})$, $(2., -\frac{1}{6})$\}

\item $c = \frac{13}{2}$, $(d_i,\theta_i)$ = \{$(1., 0)$, $(1., \frac{1}{2})$, $(2., \frac{1}{3})$, $(2.44949, \frac{7}{16})$, $(1., \frac{1}{2})$, $(1., 0)$, $(2., -\frac{1}{6})$, $(2.44949, -\frac{1}{16})$, $(2.82843, -\frac{5}{48})$, $(1.73205, -\frac{1}{8})$, $(1.73205, \frac{3}{8})$, $(1.41421, -\frac{7}{16})$, $(1.73205, -\frac{1}{8})$, $(1.73205, \frac{3}{8})$, $(1.41421, -\frac{7}{16})$\}

\item $c = 7$, $(d_i,\theta_i)$ = \{$(1., 0)$, $(1., \frac{1}{2})$, $(2., \frac{1}{3})$, $(2.44949, \frac{7}{16})$, $(1., \frac{1}{2})$, $(1., 0)$, $(2., -\frac{1}{6})$, $(2.44949, -\frac{1}{16})$, $(2., -\frac{3}{8})$, $(2.44949, -\frac{1}{16})$, $(2.44949, \frac{7}{16})$, $(2., -\frac{1}{24})$, $(2., -\frac{1}{24})$\}

\item $c = \frac{15}{2}$, $(d_i,\theta_i)$ = \{$(1., 0)$, $(1., \frac{1}{2})$, $(2., \frac{1}{3})$, $(2.44949, \frac{7}{16})$, $(1., \frac{1}{2})$, $(1., 0)$, $(2., -\frac{1}{6})$, $(2.44949, -\frac{1}{16})$, $(2.82843, \frac{1}{48})$, $(1.73205, \frac{1}{2})$, $(1.73205, 0)$, $(1.41421, -\frac{5}{16})$, $(1.73205, \frac{1}{2})$, $(1.73205, 0)$, $(1.41421, -\frac{5}{16})$\}

\end{enumerate}

\paragraph*{Rank 8; \#24}

\begin{enumerate}

\item $c = 0$, $(d_i,\theta_i)$ = \{$(1., 0)$, $(1., 0)$, $(2.44949, \frac{3}{16})$, $(2., -\frac{1}{6})$, $(1., \frac{1}{2})$, $(1., \frac{1}{2})$, $(2.44949, -\frac{5}{16})$, $(2., \frac{1}{3})$, $(2.44949, \frac{5}{16})$, $(2., -\frac{1}{4})$, $(2.44949, -\frac{3}{16})$, $(2., \frac{1}{12})$, $(2., \frac{1}{12})$\}

\item $c = \frac{1}{2}$, $(d_i,\theta_i)$ = \{$(1., 0)$, $(1., 0)$, $(2.44949, \frac{3}{16})$, $(2., -\frac{1}{6})$, $(1., \frac{1}{2})$, $(1., \frac{1}{2})$, $(2.44949, -\frac{5}{16})$, $(2., \frac{1}{3})$, $(2.82843, \frac{7}{48})$, $(1.73205, -\frac{1}{8})$, $(1.73205, \frac{3}{8})$, $(1.41421, -\frac{3}{16})$, $(1.73205, -\frac{1}{8})$, $(1.73205, \frac{3}{8})$, $(1.41421, -\frac{3}{16})$\}

\item $c = 1$, $(d_i,\theta_i)$ = \{$(1., 0)$, $(1., 0)$, $(2.44949, \frac{3}{16})$, $(2., -\frac{1}{6})$, $(1., \frac{1}{2})$, $(1., \frac{1}{2})$, $(2.44949, -\frac{5}{16})$, $(2., \frac{1}{3})$, $(2., -\frac{1}{8})$, $(2.44949, -\frac{1}{16})$, $(2.44949, \frac{7}{16})$, $(2., \frac{5}{24})$, $(2., \frac{5}{24})$\}

\item $c = \frac{3}{2}$, $(d_i,\theta_i)$ = \{$(1., 0)$, $(1., 0)$, $(2.44949, \frac{3}{16})$, $(2., -\frac{1}{6})$, $(1., \frac{1}{2})$, $(1., \frac{1}{2})$, $(2.44949, -\frac{5}{16})$, $(2., \frac{1}{3})$, $(2.82843, \frac{13}{48})$, $(1.73205, 0)$, $(1.73205, \frac{1}{2})$, $(1.41421, -\frac{1}{16})$, $(1.73205, 0)$, $(1.73205, \frac{1}{2})$, $(1.41421, -\frac{1}{16})$\}

\item $c = 2$, $(d_i,\theta_i)$ = \{$(1., 0)$, $(1., 0)$, $(2.44949, \frac{3}{16})$, $(2., \frac{1}{3})$, $(1., \frac{1}{2})$, $(1., \frac{1}{2})$, $(2.44949, -\frac{5}{16})$, $(2., -\frac{1}{6})$, $(2.44949, \frac{1}{16})$, $(2., 0)$, $(2.44949, -\frac{7}{16})$, $(2., \frac{1}{3})$, $(2., \frac{1}{3})$\}

\item $c = \frac{5}{2}$, $(d_i,\theta_i)$ = \{$(1., 0)$, $(1., 0)$, $(2.44949, \frac{3}{16})$, $(2., \frac{1}{3})$, $(1., \frac{1}{2})$, $(1., \frac{1}{2})$, $(2.44949, -\frac{5}{16})$, $(2., -\frac{1}{6})$, $(2.82843, \frac{19}{48})$, $(1.73205, -\frac{3}{8})$, $(1.73205, \frac{1}{8})$, $(1.41421, \frac{1}{16})$, $(1.73205, -\frac{3}{8})$, $(1.73205, \frac{1}{8})$, $(1.41421, \frac{1}{16})$\}

\item $c = 3$, $(d_i,\theta_i)$ = \{$(1., 0)$, $(1., 0)$, $(2.44949, \frac{3}{16})$, $(2., \frac{1}{3})$, $(1., \frac{1}{2})$, $(1., \frac{1}{2})$, $(2.44949, -\frac{5}{16})$, $(2., -\frac{1}{6})$, $(2., \frac{1}{8})$, $(2.44949, \frac{3}{16})$, $(2.44949, -\frac{5}{16})$, $(2., \frac{11}{24})$, $(2., \frac{11}{24})$\}

\item $c = \frac{7}{2}$, $(d_i,\theta_i)$ = \{$(1., 0)$, $(1., 0)$, $(2.44949, \frac{3}{16})$, $(2., \frac{1}{3})$, $(1., \frac{1}{2})$, $(1., \frac{1}{2})$, $(2.44949, -\frac{5}{16})$, $(2., -\frac{1}{6})$, $(2.82843, -\frac{23}{48})$, $(1.73205, -\frac{1}{4})$, $(1.73205, \frac{1}{4})$, $(1.41421, \frac{3}{16})$, $(1.73205, -\frac{1}{4})$, $(1.73205, \frac{1}{4})$, $(1.41421, \frac{3}{16})$\}

\item $c = 4$, $(d_i,\theta_i)$ = \{$(1., 0)$, $(1., 0)$, $(2.44949, \frac{3}{16})$, $(2., \frac{1}{3})$, $(1., \frac{1}{2})$, $(1., \frac{1}{2})$, $(2.44949, -\frac{5}{16})$, $(2., -\frac{1}{6})$, $(2.44949, \frac{5}{16})$, $(2.44949, -\frac{3}{16})$, $(2., \frac{1}{4})$, $(2., -\frac{5}{12})$, $(2., -\frac{5}{12})$\}

\item $c = \frac{9}{2}$, $(d_i,\theta_i)$ = \{$(1., 0)$, $(1., 0)$, $(2.44949, \frac{3}{16})$, $(2., \frac{1}{3})$, $(1., \frac{1}{2})$, $(1., \frac{1}{2})$, $(2.44949, -\frac{5}{16})$, $(2., -\frac{1}{6})$, $(2.82843, -\frac{17}{48})$, $(1.73205, \frac{3}{8})$, $(1.73205, -\frac{1}{8})$, $(1.41421, \frac{5}{16})$, $(1.73205, \frac{3}{8})$, $(1.73205, -\frac{1}{8})$, $(1.41421, \frac{5}{16})$\}

\item $c = 5$, $(d_i,\theta_i)$ = \{$(1., 0)$, $(1., 0)$, $(2.44949, \frac{3}{16})$, $(2., \frac{1}{3})$, $(1., \frac{1}{2})$, $(1., \frac{1}{2})$, $(2.44949, -\frac{5}{16})$, $(2., -\frac{1}{6})$, $(2.44949, \frac{7}{16})$, $(2., \frac{3}{8})$, $(2.44949, -\frac{1}{16})$, $(2., -\frac{7}{24})$, $(2., -\frac{7}{24})$\}

\item $c = \frac{11}{2}$, $(d_i,\theta_i)$ = \{$(1., 0)$, $(1., 0)$, $(2.44949, \frac{3}{16})$, $(2., \frac{1}{3})$, $(1., \frac{1}{2})$, $(1., \frac{1}{2})$, $(2.44949, -\frac{5}{16})$, $(2., -\frac{1}{6})$, $(2.82843, -\frac{11}{48})$, $(1.73205, \frac{1}{2})$, $(1.73205, 0)$, $(1.41421, \frac{7}{16})$, $(1.73205, \frac{1}{2})$, $(1.73205, 0)$, $(1.41421, \frac{7}{16})$\}

\item $c = 6$, $(d_i,\theta_i)$ = \{$(1., 0)$, $(1., \frac{1}{2})$, $(2.44949, \frac{3}{16})$, $(2., \frac{1}{3})$, $(1., \frac{1}{2})$, $(1., 0)$, $(2.44949, -\frac{5}{16})$, $(2., -\frac{1}{6})$, $(2.44949, \frac{1}{16})$, $(2., \frac{1}{2})$, $(2.44949, -\frac{7}{16})$, $(2., -\frac{1}{6})$, $(2., -\frac{1}{6})$\}

\item $c = \frac{13}{2}$, $(d_i,\theta_i)$ = \{$(1., 0)$, $(1., \frac{1}{2})$, $(2.44949, \frac{3}{16})$, $(2., \frac{1}{3})$, $(1., \frac{1}{2})$, $(1., 0)$, $(2.44949, -\frac{5}{16})$, $(2., -\frac{1}{6})$, $(2.82843, -\frac{5}{48})$, $(1.73205, \frac{1}{8})$, $(1.73205, -\frac{3}{8})$, $(1.41421, -\frac{7}{16})$, $(1.73205, \frac{1}{8})$, $(1.73205, -\frac{3}{8})$, $(1.41421, -\frac{7}{16})$\}

\item $c = 7$, $(d_i,\theta_i)$ = \{$(1., 0)$, $(1., \frac{1}{2})$, $(2.44949, \frac{3}{16})$, $(2., \frac{1}{3})$, $(1., \frac{1}{2})$, $(1., 0)$, $(2.44949, -\frac{5}{16})$, $(2., -\frac{1}{6})$, $(2., -\frac{3}{8})$, $(2.44949, \frac{3}{16})$, $(2.44949, -\frac{5}{16})$, $(2., -\frac{1}{24})$, $(2., -\frac{1}{24})$\}

\item $c = \frac{15}{2}$, $(d_i,\theta_i)$ = \{$(1., 0)$, $(1., \frac{1}{2})$, $(2.44949, \frac{3}{16})$, $(2., \frac{1}{3})$, $(1., \frac{1}{2})$, $(1., 0)$, $(2.44949, -\frac{5}{16})$, $(2., -\frac{1}{6})$, $(2.82843, \frac{1}{48})$, $(1.73205, \frac{1}{4})$, $(1.73205, -\frac{1}{4})$, $(1.41421, -\frac{5}{16})$, $(1.73205, \frac{1}{4})$, $(1.73205, -\frac{1}{4})$, $(1.41421, -\frac{5}{16})$\}

\end{enumerate}

\paragraph*{Rank 8; \#25}

\begin{enumerate}

\item $c = 0$, $(d_i,\theta_i)$ = \{$(1., 0)$, $(1., 0)$, $(2.44949, -\frac{3}{16})$, $(2., -\frac{1}{6})$, $(1., \frac{1}{2})$, $(1., \frac{1}{2})$, $(2.44949, \frac{5}{16})$, $(2., \frac{1}{3})$, $(2., -\frac{1}{4})$, $(2.44949, \frac{3}{16})$, $(2.44949, -\frac{5}{16})$, $(2., \frac{1}{12})$, $(2., \frac{1}{12})$\}

\item $c = \frac{1}{2}$, $(d_i,\theta_i)$ = \{$(1., 0)$, $(1., 0)$, $(2.44949, -\frac{3}{16})$, $(2., -\frac{1}{6})$, $(1., \frac{1}{2})$, $(1., \frac{1}{2})$, $(2.44949, \frac{5}{16})$, $(2., \frac{1}{3})$, $(2.82843, \frac{7}{48})$, $(1.73205, -\frac{1}{4})$, $(1.73205, \frac{1}{4})$, $(1.41421, -\frac{3}{16})$, $(1.73205, -\frac{1}{4})$, $(1.73205, \frac{1}{4})$, $(1.41421, -\frac{3}{16})$\}

\item $c = 1$, $(d_i,\theta_i)$ = \{$(1., 0)$, $(1., 0)$, $(2.44949, -\frac{3}{16})$, $(2., -\frac{1}{6})$, $(1., \frac{1}{2})$, $(1., \frac{1}{2})$, $(2.44949, \frac{5}{16})$, $(2., \frac{1}{3})$, $(2., -\frac{1}{8})$, $(2.44949, -\frac{3}{16})$, $(2.44949, \frac{5}{16})$, $(2., \frac{5}{24})$, $(2., \frac{5}{24})$\}

\item $c = \frac{3}{2}$, $(d_i,\theta_i)$ = \{$(1., 0)$, $(1., 0)$, $(2.44949, \frac{5}{16})$, $(2., -\frac{1}{6})$, $(1., \frac{1}{2})$, $(1., \frac{1}{2})$, $(2.44949, -\frac{3}{16})$, $(2., \frac{1}{3})$, $(2.82843, \frac{13}{48})$, $(1.73205, \frac{3}{8})$, $(1.73205, -\frac{1}{8})$, $(1.41421, -\frac{1}{16})$, $(1.73205, \frac{3}{8})$, $(1.73205, -\frac{1}{8})$, $(1.41421, -\frac{1}{16})$\}

\item $c = 2$, $(d_i,\theta_i)$ = \{$(1., 0)$, $(1., 0)$, $(2.44949, \frac{5}{16})$, $(2., \frac{1}{3})$, $(1., \frac{1}{2})$, $(1., \frac{1}{2})$, $(2.44949, -\frac{3}{16})$, $(2., -\frac{1}{6})$, $(2.44949, \frac{7}{16})$, $(2.44949, -\frac{1}{16})$, $(2., 0)$, $(2., \frac{1}{3})$, $(2., \frac{1}{3})$\}

\item $c = \frac{5}{2}$, $(d_i,\theta_i)$ = \{$(1., 0)$, $(1., 0)$, $(2.44949, \frac{5}{16})$, $(2., \frac{1}{3})$, $(1., \frac{1}{2})$, $(1., \frac{1}{2})$, $(2.44949, -\frac{3}{16})$, $(2., -\frac{1}{6})$, $(2.82843, \frac{19}{48})$, $(1.73205, \frac{1}{2})$, $(1.73205, 0)$, $(1.41421, \frac{1}{16})$, $(1.73205, \frac{1}{2})$, $(1.73205, 0)$, $(1.41421, \frac{1}{16})$\}

\item $c = 3$, $(d_i,\theta_i)$ = \{$(1., 0)$, $(1., 0)$, $(2.44949, \frac{5}{16})$, $(2., \frac{1}{3})$, $(1., \frac{1}{2})$, $(1., \frac{1}{2})$, $(2.44949, -\frac{3}{16})$, $(2., -\frac{1}{6})$, $(2., \frac{1}{8})$, $(2.44949, \frac{1}{16})$, $(2.44949, -\frac{7}{16})$, $(2., \frac{11}{24})$, $(2., \frac{11}{24})$\}

\item $c = \frac{7}{2}$, $(d_i,\theta_i)$ = \{$(1., 0)$, $(1., 0)$, $(2.44949, \frac{5}{16})$, $(2., \frac{1}{3})$, $(1., \frac{1}{2})$, $(1., \frac{1}{2})$, $(2.44949, -\frac{3}{16})$, $(2., -\frac{1}{6})$, $(2.82843, -\frac{23}{48})$, $(1.73205, \frac{1}{8})$, $(1.73205, -\frac{3}{8})$, $(1.41421, \frac{3}{16})$, $(1.73205, \frac{1}{8})$, $(1.73205, -\frac{3}{8})$, $(1.41421, \frac{3}{16})$\}

\item $c = 4$, $(d_i,\theta_i)$ = \{$(1., 0)$, $(1., 0)$, $(2.44949, \frac{5}{16})$, $(2., \frac{1}{3})$, $(1., \frac{1}{2})$, $(1., \frac{1}{2})$, $(2.44949, -\frac{3}{16})$, $(2., -\frac{1}{6})$, $(2., \frac{1}{4})$, $(2.44949, -\frac{5}{16})$, $(2.44949, \frac{3}{16})$, $(2., -\frac{5}{12})$, $(2., -\frac{5}{12})$\}

\item $c = \frac{9}{2}$, $(d_i,\theta_i)$ = \{$(1., 0)$, $(1., 0)$, $(2.44949, \frac{5}{16})$, $(2., \frac{1}{3})$, $(1., \frac{1}{2})$, $(1., \frac{1}{2})$, $(2.44949, -\frac{3}{16})$, $(2., -\frac{1}{6})$, $(2.82843, -\frac{17}{48})$, $(1.73205, \frac{1}{4})$, $(1.73205, -\frac{1}{4})$, $(1.41421, \frac{5}{16})$, $(1.73205, \frac{1}{4})$, $(1.73205, -\frac{1}{4})$, $(1.41421, \frac{5}{16})$\}

\item $c = 5$, $(d_i,\theta_i)$ = \{$(1., 0)$, $(1., 0)$, $(2.44949, \frac{5}{16})$, $(2., \frac{1}{3})$, $(1., \frac{1}{2})$, $(1., \frac{1}{2})$, $(2.44949, -\frac{3}{16})$, $(2., -\frac{1}{6})$, $(2.44949, -\frac{3}{16})$, $(2., \frac{3}{8})$, $(2.44949, \frac{5}{16})$, $(2., -\frac{7}{24})$, $(2., -\frac{7}{24})$\}

\item $c = \frac{11}{2}$, $(d_i,\theta_i)$ = \{$(1., 0)$, $(1., 0)$, $(2.44949, \frac{5}{16})$, $(2., \frac{1}{3})$, $(1., \frac{1}{2})$, $(1., \frac{1}{2})$, $(2.44949, -\frac{3}{16})$, $(2., -\frac{1}{6})$, $(2.82843, -\frac{11}{48})$, $(1.73205, -\frac{1}{8})$, $(1.73205, \frac{3}{8})$, $(1.41421, \frac{7}{16})$, $(1.73205, -\frac{1}{8})$, $(1.73205, \frac{3}{8})$, $(1.41421, \frac{7}{16})$\}

\item $c = 6$, $(d_i,\theta_i)$ = \{$(1., 0)$, $(1., \frac{1}{2})$, $(2.44949, \frac{5}{16})$, $(2., \frac{1}{3})$, $(1., \frac{1}{2})$, $(1., 0)$, $(2.44949, -\frac{3}{16})$, $(2., -\frac{1}{6})$, $(2., \frac{1}{2})$, $(2.44949, -\frac{1}{16})$, $(2.44949, \frac{7}{16})$, $(2., -\frac{1}{6})$, $(2., -\frac{1}{6})$\}

\item $c = \frac{13}{2}$, $(d_i,\theta_i)$ = \{$(1., 0)$, $(1., \frac{1}{2})$, $(2.44949, \frac{5}{16})$, $(2., \frac{1}{3})$, $(1., \frac{1}{2})$, $(1., 0)$, $(2.44949, -\frac{3}{16})$, $(2., -\frac{1}{6})$, $(2.82843, -\frac{5}{48})$, $(1.73205, 0)$, $(1.73205, \frac{1}{2})$, $(1.41421, -\frac{7}{16})$, $(1.73205, 0)$, $(1.73205, \frac{1}{2})$, $(1.41421, -\frac{7}{16})$\}

\item $c = 7$, $(d_i,\theta_i)$ = \{$(1., 0)$, $(1., \frac{1}{2})$, $(2.44949, \frac{5}{16})$, $(2., \frac{1}{3})$, $(1., \frac{1}{2})$, $(1., 0)$, $(2.44949, -\frac{3}{16})$, $(2., -\frac{1}{6})$, $(2., -\frac{3}{8})$, $(2.44949, -\frac{7}{16})$, $(2.44949, \frac{1}{16})$, $(2., -\frac{1}{24})$, $(2., -\frac{1}{24})$\}

\item $c = \frac{15}{2}$, $(d_i,\theta_i)$ = \{$(1., 0)$, $(1., \frac{1}{2})$, $(2.44949, \frac{5}{16})$, $(2., \frac{1}{3})$, $(1., \frac{1}{2})$, $(1., 0)$, $(2.44949, -\frac{3}{16})$, $(2., -\frac{1}{6})$, $(2.82843, \frac{1}{48})$, $(1.73205, -\frac{3}{8})$, $(1.73205, \frac{1}{8})$, $(1.41421, -\frac{5}{16})$, $(1.73205, -\frac{3}{8})$, $(1.73205, \frac{1}{8})$, $(1.41421, -\frac{5}{16})$\}

\end{enumerate}

\paragraph*{Rank 8; \#26}

\begin{enumerate}

\item $c = 0$, $(d_i,\theta_i)$ = \{$(1., 0)$, $(1., 0)$, $(2., -\frac{1}{6})$, $(-2.44949, -\frac{1}{16})$, $(1., \frac{1}{2})$, $(1., \frac{1}{2})$, $(2., \frac{1}{3})$, $(-2.44949, \frac{7}{16})$, $(2., -\frac{1}{4})$, $(-2.44949, \frac{1}{16})$, $(-2.44949, -\frac{7}{16})$, $(2., \frac{1}{12})$, $(2., \frac{1}{12})$\}

\item $c = \frac{1}{2}$, $(d_i,\theta_i)$ = \{$(1., 0)$, $(1., 0)$, $(2., -\frac{1}{6})$, $(-2.44949, -\frac{1}{16})$, $(1., \frac{1}{2})$, $(1., \frac{1}{2})$, $(2., \frac{1}{3})$, $(-2.44949, \frac{7}{16})$, $(2.82843, \frac{7}{48})$, $(-1.73205, -\frac{3}{8})$, $(-1.73205, \frac{1}{8})$, $(1.41421, -\frac{3}{16})$, $(-1.73205, -\frac{3}{8})$, $(-1.73205, \frac{1}{8})$, $(1.41421, -\frac{3}{16})$\}

\item $c = 1$, $(d_i,\theta_i)$ = \{$(1., 0)$, $(1., 0)$, $(2., -\frac{1}{6})$, $(-2.44949, -\frac{1}{16})$, $(1., \frac{1}{2})$, $(1., \frac{1}{2})$, $(2., \frac{1}{3})$, $(-2.44949, \frac{7}{16})$, $(2., -\frac{1}{8})$, $(-2.44949, -\frac{5}{16})$, $(-2.44949, \frac{3}{16})$, $(2., \frac{5}{24})$, $(2., \frac{5}{24})$\}

\item $c = \frac{3}{2}$, $(d_i,\theta_i)$ = \{$(1., 0)$, $(1., 0)$, $(2., -\frac{1}{6})$, $(-2.44949, -\frac{1}{16})$, $(1., \frac{1}{2})$, $(1., \frac{1}{2})$, $(2., \frac{1}{3})$, $(-2.44949, \frac{7}{16})$, $(2.82843, \frac{13}{48})$, $(-1.73205, \frac{1}{4})$, $(-1.73205, -\frac{1}{4})$, $(1.41421, -\frac{1}{16})$, $(-1.73205, \frac{1}{4})$, $(-1.73205, -\frac{1}{4})$, $(1.41421, -\frac{1}{16})$\}

\item $c = 2$, $(d_i,\theta_i)$ = \{$(1., 0)$, $(1., 0)$, $(2., \frac{1}{3})$, $(-2.44949, -\frac{1}{16})$, $(1., \frac{1}{2})$, $(1., \frac{1}{2})$, $(2., -\frac{1}{6})$, $(-2.44949, \frac{7}{16})$, $(-2.44949, -\frac{3}{16})$, $(2., 0)$, $(-2.44949, \frac{5}{16})$, $(2., \frac{1}{3})$, $(2., \frac{1}{3})$\}

\item $c = \frac{5}{2}$, $(d_i,\theta_i)$ = \{$(1., 0)$, $(1., 0)$, $(2., \frac{1}{3})$, $(-2.44949, -\frac{1}{16})$, $(1., \frac{1}{2})$, $(1., \frac{1}{2})$, $(2., -\frac{1}{6})$, $(-2.44949, \frac{7}{16})$, $(2.82843, \frac{19}{48})$, $(-1.73205, \frac{3}{8})$, $(-1.73205, -\frac{1}{8})$, $(1.41421, \frac{1}{16})$, $(-1.73205, \frac{3}{8})$, $(-1.73205, -\frac{1}{8})$, $(1.41421, \frac{1}{16})$\}

\item $c = 3$, $(d_i,\theta_i)$ = \{$(1., 0)$, $(1., 0)$, $(2., \frac{1}{3})$, $(-2.44949, -\frac{1}{16})$, $(1., \frac{1}{2})$, $(1., \frac{1}{2})$, $(2., -\frac{1}{6})$, $(-2.44949, \frac{7}{16})$, $(-2.44949, -\frac{1}{16})$, $(2., \frac{1}{8})$, $(-2.44949, \frac{7}{16})$, $(2., \frac{11}{24})$, $(2., \frac{11}{24})$\}

\item $c = \frac{7}{2}$, $(d_i,\theta_i)$ = \{$(1., 0)$, $(1., 0)$, $(2., \frac{1}{3})$, $(-2.44949, -\frac{1}{16})$, $(1., \frac{1}{2})$, $(1., \frac{1}{2})$, $(2., -\frac{1}{6})$, $(-2.44949, \frac{7}{16})$, $(2.82843, -\frac{23}{48})$, $(-1.73205, 0)$, $(-1.73205, \frac{1}{2})$, $(1.41421, \frac{3}{16})$, $(-1.73205, 0)$, $(-1.73205, \frac{1}{2})$, $(1.41421, \frac{3}{16})$\}

\item $c = 4$, $(d_i,\theta_i)$ = \{$(1., 0)$, $(1., 0)$, $(2., \frac{1}{3})$, $(-2.44949, -\frac{1}{16})$, $(1., \frac{1}{2})$, $(1., \frac{1}{2})$, $(2., -\frac{1}{6})$, $(-2.44949, \frac{7}{16})$, $(2., \frac{1}{4})$, $(-2.44949, -\frac{7}{16})$, $(-2.44949, \frac{1}{16})$, $(2., -\frac{5}{12})$, $(2., -\frac{5}{12})$\}

\item $c = \frac{9}{2}$, $(d_i,\theta_i)$ = \{$(1., 0)$, $(1., 0)$, $(2., \frac{1}{3})$, $(-2.44949, \frac{7}{16})$, $(1., \frac{1}{2})$, $(1., \frac{1}{2})$, $(2., -\frac{1}{6})$, $(-2.44949, -\frac{1}{16})$, $(2.82843, -\frac{17}{48})$, $(-1.73205, \frac{1}{8})$, $(-1.73205, -\frac{3}{8})$, $(1.41421, \frac{5}{16})$, $(-1.73205, \frac{1}{8})$, $(-1.73205, -\frac{3}{8})$, $(1.41421, \frac{5}{16})$\}

\item $c = 5$, $(d_i,\theta_i)$ = \{$(1., 0)$, $(1., 0)$, $(2., \frac{1}{3})$, $(-2.44949, \frac{7}{16})$, $(1., \frac{1}{2})$, $(1., \frac{1}{2})$, $(2., -\frac{1}{6})$, $(-2.44949, -\frac{1}{16})$, $(2., \frac{3}{8})$, $(-2.44949, \frac{3}{16})$, $(-2.44949, -\frac{5}{16})$, $(2., -\frac{7}{24})$, $(2., -\frac{7}{24})$\}

\item $c = \frac{11}{2}$, $(d_i,\theta_i)$ = \{$(1., 0)$, $(1., 0)$, $(2., \frac{1}{3})$, $(-2.44949, \frac{7}{16})$, $(1., \frac{1}{2})$, $(1., \frac{1}{2})$, $(2., -\frac{1}{6})$, $(-2.44949, -\frac{1}{16})$, $(2.82843, -\frac{11}{48})$, $(-1.73205, -\frac{1}{4})$, $(-1.73205, \frac{1}{4})$, $(1.41421, \frac{7}{16})$, $(-1.73205, -\frac{1}{4})$, $(-1.73205, \frac{1}{4})$, $(1.41421, \frac{7}{16})$\}

\item $c = 6$, $(d_i,\theta_i)$ = \{$(1., 0)$, $(1., \frac{1}{2})$, $(2., \frac{1}{3})$, $(-2.44949, \frac{7}{16})$, $(1., \frac{1}{2})$, $(1., 0)$, $(2., -\frac{1}{6})$, $(-2.44949, -\frac{1}{16})$, $(-2., \frac{1}{2})$, $(2.44949, -\frac{3}{16})$, $(2.44949, \frac{5}{16})$, $(-2., -\frac{1}{6})$, $(-2., -\frac{1}{6})$\}

\item $c = \frac{13}{2}$, $(d_i,\theta_i)$ = \{$(1., 0)$, $(1., \frac{1}{2})$, $(2., \frac{1}{3})$, $(-2.44949, \frac{7}{16})$, $(1., \frac{1}{2})$, $(1., 0)$, $(2., -\frac{1}{6})$, $(-2.44949, -\frac{1}{16})$, $(-2.82843, -\frac{5}{48})$, $(1.73205, -\frac{1}{8})$, $(1.73205, \frac{3}{8})$, $(-1.41421, -\frac{7}{16})$, $(1.73205, -\frac{1}{8})$, $(1.73205, \frac{3}{8})$, $(-1.41421, -\frac{7}{16})$\}

\item $c = 7$, $(d_i,\theta_i)$ = \{$(1., 0)$, $(1., \frac{1}{2})$, $(2., \frac{1}{3})$, $(-2.44949, \frac{7}{16})$, $(1., \frac{1}{2})$, $(1., 0)$, $(2., -\frac{1}{6})$, $(-2.44949, -\frac{1}{16})$, $(-2., -\frac{3}{8})$, $(2.44949, -\frac{1}{16})$, $(2.44949, \frac{7}{16})$, $(-2., -\frac{1}{24})$, $(-2., -\frac{1}{24})$\}

\item $c = \frac{15}{2}$, $(d_i,\theta_i)$ = \{$(1., 0)$, $(1., \frac{1}{2})$, $(2., \frac{1}{3})$, $(-2.44949, \frac{7}{16})$, $(1., \frac{1}{2})$, $(1., 0)$, $(2., -\frac{1}{6})$, $(-2.44949, -\frac{1}{16})$, $(-2.82843, \frac{1}{48})$, $(1.73205, \frac{1}{2})$, $(1.73205, 0)$, $(-1.41421, -\frac{5}{16})$, $(1.73205, \frac{1}{2})$, $(1.73205, 0)$, $(-1.41421, -\frac{5}{16})$\}

\end{enumerate}

\paragraph*{Rank 8; \#27}

\begin{enumerate}

\item $c = 0$, $(d_i,\theta_i)$ = \{$(1., 0)$, $(1., 0)$, $(2., \frac{1}{6})$, $(-2.44949, -\frac{1}{16})$, $(1., \frac{1}{2})$, $(1., \frac{1}{2})$, $(2., -\frac{1}{3})$, $(-2.44949, \frac{7}{16})$, $(2., \frac{1}{4})$, $(-2.44949, \frac{1}{16})$, $(-2.44949, -\frac{7}{16})$, $(2., -\frac{1}{12})$, $(2., -\frac{1}{12})$\}

\item $c = \frac{1}{2}$, $(d_i,\theta_i)$ = \{$(1., 0)$, $(1., 0)$, $(2., \frac{1}{6})$, $(-2.44949, -\frac{1}{16})$, $(1., \frac{1}{2})$, $(1., \frac{1}{2})$, $(2., -\frac{1}{3})$, $(-2.44949, \frac{7}{16})$, $(2.82843, -\frac{1}{48})$, $(-1.73205, -\frac{3}{8})$, $(-1.73205, \frac{1}{8})$, $(1.41421, \frac{5}{16})$, $(-1.73205, -\frac{3}{8})$, $(-1.73205, \frac{1}{8})$, $(1.41421, \frac{5}{16})$\}

\item $c = 1$, $(d_i,\theta_i)$ = \{$(1., 0)$, $(1., 0)$, $(2., \frac{1}{6})$, $(-2.44949, -\frac{1}{16})$, $(1., \frac{1}{2})$, $(1., \frac{1}{2})$, $(2., -\frac{1}{3})$, $(-2.44949, \frac{7}{16})$, $(-2.44949, \frac{3}{16})$, $(-2.44949, -\frac{5}{16})$, $(2., \frac{3}{8})$, $(2., \frac{1}{24})$, $(2., \frac{1}{24})$\}

\item $c = \frac{3}{2}$, $(d_i,\theta_i)$ = \{$(1., 0)$, $(1., 0)$, $(2., \frac{1}{6})$, $(-2.44949, -\frac{1}{16})$, $(1., \frac{1}{2})$, $(1., \frac{1}{2})$, $(2., -\frac{1}{3})$, $(-2.44949, \frac{7}{16})$, $(2.82843, \frac{5}{48})$, $(-1.73205, \frac{1}{4})$, $(-1.73205, -\frac{1}{4})$, $(1.41421, \frac{7}{16})$, $(-1.73205, \frac{1}{4})$, $(-1.73205, -\frac{1}{4})$, $(1.41421, \frac{7}{16})$\}

\item $c = 2$, $(d_i,\theta_i)$ = \{$(1., 0)$, $(1., 0)$, $(2., \frac{1}{6})$, $(-2.44949, -\frac{1}{16})$, $(1., \frac{1}{2})$, $(1., \frac{1}{2})$, $(2., -\frac{1}{3})$, $(-2.44949, \frac{7}{16})$, $(-2.44949, \frac{5}{16})$, $(-2.44949, -\frac{3}{16})$, $(2., \frac{1}{2})$, $(2., \frac{1}{6})$, $(2., \frac{1}{6})$\}

\item $c = \frac{5}{2}$, $(d_i,\theta_i)$ = \{$(1., 0)$, $(1., 0)$, $(2., \frac{1}{6})$, $(-2.44949, -\frac{1}{16})$, $(1., \frac{1}{2})$, $(1., \frac{1}{2})$, $(2., -\frac{1}{3})$, $(-2.44949, \frac{7}{16})$, $(2.82843, \frac{11}{48})$, $(-1.73205, \frac{3}{8})$, $(-1.73205, -\frac{1}{8})$, $(1.41421, -\frac{7}{16})$, $(-1.73205, \frac{3}{8})$, $(-1.73205, -\frac{1}{8})$, $(1.41421, -\frac{7}{16})$\}

\item $c = 3$, $(d_i,\theta_i)$ = \{$(1., 0)$, $(1., 0)$, $(2., \frac{1}{6})$, $(-2.44949, -\frac{1}{16})$, $(1., \frac{1}{2})$, $(1., \frac{1}{2})$, $(2., -\frac{1}{3})$, $(-2.44949, \frac{7}{16})$, $(2., -\frac{3}{8})$, $(-2.44949, \frac{7}{16})$, $(-2.44949, -\frac{1}{16})$, $(2., \frac{7}{24})$, $(2., \frac{7}{24})$\}

\item $c = \frac{7}{2}$, $(d_i,\theta_i)$ = \{$(1., 0)$, $(1., 0)$, $(2., \frac{1}{6})$, $(-2.44949, -\frac{1}{16})$, $(1., \frac{1}{2})$, $(1., \frac{1}{2})$, $(2., -\frac{1}{3})$, $(-2.44949, \frac{7}{16})$, $(2.82843, \frac{17}{48})$, $(-1.73205, 0)$, $(-1.73205, \frac{1}{2})$, $(1.41421, -\frac{5}{16})$, $(-1.73205, 0)$, $(-1.73205, \frac{1}{2})$, $(1.41421, -\frac{5}{16})$\}

\item $c = 4$, $(d_i,\theta_i)$ = \{$(1., 0)$, $(1., 0)$, $(2., \frac{1}{6})$, $(-2.44949, -\frac{1}{16})$, $(1., \frac{1}{2})$, $(1., \frac{1}{2})$, $(2., -\frac{1}{3})$, $(-2.44949, \frac{7}{16})$, $(2.44949, \frac{1}{16})$, $(2.44949, -\frac{7}{16})$, $(-2., -\frac{1}{4})$, $(-2., \frac{5}{12})$, $(-2., \frac{5}{12})$\}

\item $c = \frac{9}{2}$, $(d_i,\theta_i)$ = \{$(1., 0)$, $(1., 0)$, $(2., \frac{1}{6})$, $(-2.44949, \frac{7}{16})$, $(1., \frac{1}{2})$, $(1., \frac{1}{2})$, $(2., -\frac{1}{3})$, $(-2.44949, -\frac{1}{16})$, $(2.82843, \frac{23}{48})$, $(-1.73205, \frac{1}{8})$, $(-1.73205, -\frac{3}{8})$, $(1.41421, -\frac{3}{16})$, $(-1.73205, \frac{1}{8})$, $(-1.73205, -\frac{3}{8})$, $(1.41421, -\frac{3}{16})$\}

\item $c = 5$, $(d_i,\theta_i)$ = \{$(1., 0)$, $(1., 0)$, $(2., \frac{1}{6})$, $(-2.44949, \frac{7}{16})$, $(1., \frac{1}{2})$, $(1., \frac{1}{2})$, $(2., -\frac{1}{3})$, $(-2.44949, -\frac{1}{16})$, $(2., -\frac{1}{8})$, $(-2.44949, -\frac{5}{16})$, $(-2.44949, \frac{3}{16})$, $(2., -\frac{11}{24})$, $(2., -\frac{11}{24})$\}

\item $c = \frac{11}{2}$, $(d_i,\theta_i)$ = \{$(1., 0)$, $(1., 0)$, $(2., \frac{1}{6})$, $(-2.44949, \frac{7}{16})$, $(1., \frac{1}{2})$, $(1., \frac{1}{2})$, $(2., -\frac{1}{3})$, $(-2.44949, -\frac{1}{16})$, $(2.82843, -\frac{19}{48})$, $(-1.73205, -\frac{1}{4})$, $(-1.73205, \frac{1}{4})$, $(1.41421, -\frac{1}{16})$, $(-1.73205, -\frac{1}{4})$, $(-1.73205, \frac{1}{4})$, $(1.41421, -\frac{1}{16})$\}

\item $c = 6$, $(d_i,\theta_i)$ = \{$(1., 0)$, $(1., \frac{1}{2})$, $(2., \frac{1}{6})$, $(-2.44949, \frac{7}{16})$, $(1., \frac{1}{2})$, $(1., 0)$, $(2., -\frac{1}{3})$, $(-2.44949, -\frac{1}{16})$, $(-2., 0)$, $(2.44949, -\frac{3}{16})$, $(2.44949, \frac{5}{16})$, $(-2., -\frac{1}{3})$, $(-2., -\frac{1}{3})$\}

\item $c = \frac{13}{2}$, $(d_i,\theta_i)$ = \{$(1., 0)$, $(1., \frac{1}{2})$, $(2., \frac{1}{6})$, $(-2.44949, \frac{7}{16})$, $(1., \frac{1}{2})$, $(1., 0)$, $(2., -\frac{1}{3})$, $(-2.44949, -\frac{1}{16})$, $(-2.82843, -\frac{13}{48})$, $(1.73205, -\frac{1}{8})$, $(1.73205, \frac{3}{8})$, $(-1.41421, \frac{1}{16})$, $(1.73205, -\frac{1}{8})$, $(1.73205, \frac{3}{8})$, $(-1.41421, \frac{1}{16})$\}

\item $c = 7$, $(d_i,\theta_i)$ = \{$(1., 0)$, $(1., \frac{1}{2})$, $(2., \frac{1}{6})$, $(-2.44949, \frac{7}{16})$, $(1., \frac{1}{2})$, $(1., 0)$, $(2., -\frac{1}{3})$, $(-2.44949, -\frac{1}{16})$, $(-2.44949, -\frac{1}{16})$, $(2., \frac{1}{8})$, $(-2.44949, \frac{7}{16})$, $(2., -\frac{5}{24})$, $(2., -\frac{5}{24})$\}

\item $c = \frac{15}{2}$, $(d_i,\theta_i)$ = \{$(1., 0)$, $(1., \frac{1}{2})$, $(2., \frac{1}{6})$, $(-2.44949, \frac{7}{16})$, $(1., \frac{1}{2})$, $(1., 0)$, $(2., -\frac{1}{3})$, $(-2.44949, -\frac{1}{16})$, $(-2.82843, -\frac{7}{48})$, $(1.73205, \frac{1}{2})$, $(1.73205, 0)$, $(-1.41421, \frac{3}{16})$, $(1.73205, \frac{1}{2})$, $(1.73205, 0)$, $(-1.41421, \frac{3}{16})$\}

\end{enumerate}

\paragraph*{Rank 8; \#28}

\begin{enumerate}

\item $c = 0$, $(d_i,\theta_i)$ = \{$(1., 0)$, $(1., 0)$, $(2., \frac{1}{6})$, $(-2.44949, -\frac{3}{16})$, $(1., \frac{1}{2})$, $(1., \frac{1}{2})$, $(2., -\frac{1}{3})$, $(-2.44949, \frac{5}{16})$, $(-2.44949, \frac{3}{16})$, $(-2.44949, -\frac{5}{16})$, $(2., \frac{1}{4})$, $(2., -\frac{1}{12})$, $(2., -\frac{1}{12})$\}

\item $c = \frac{1}{2}$, $(d_i,\theta_i)$ = \{$(1., 0)$, $(1., 0)$, $(2., \frac{1}{6})$, $(-2.44949, -\frac{3}{16})$, $(1., \frac{1}{2})$, $(1., \frac{1}{2})$, $(2., -\frac{1}{3})$, $(-2.44949, \frac{5}{16})$, $(2.82843, -\frac{1}{48})$, $(-1.73205, -\frac{1}{4})$, $(-1.73205, \frac{1}{4})$, $(1.41421, \frac{5}{16})$, $(-1.73205, -\frac{1}{4})$, $(-1.73205, \frac{1}{4})$, $(1.41421, \frac{5}{16})$\}

\item $c = 1$, $(d_i,\theta_i)$ = \{$(1., 0)$, $(1., 0)$, $(2., \frac{1}{6})$, $(-2.44949, -\frac{3}{16})$, $(1., \frac{1}{2})$, $(1., \frac{1}{2})$, $(2., -\frac{1}{3})$, $(-2.44949, \frac{5}{16})$, $(2., \frac{3}{8})$, $(-2.44949, -\frac{3}{16})$, $(-2.44949, \frac{5}{16})$, $(2., \frac{1}{24})$, $(2., \frac{1}{24})$\}

\item $c = \frac{3}{2}$, $(d_i,\theta_i)$ = \{$(1., 0)$, $(1., 0)$, $(2., \frac{1}{6})$, $(-2.44949, \frac{5}{16})$, $(1., \frac{1}{2})$, $(1., \frac{1}{2})$, $(2., -\frac{1}{3})$, $(-2.44949, -\frac{3}{16})$, $(2.82843, \frac{5}{48})$, $(-1.73205, \frac{3}{8})$, $(-1.73205, -\frac{1}{8})$, $(1.41421, \frac{7}{16})$, $(-1.73205, \frac{3}{8})$, $(-1.73205, -\frac{1}{8})$, $(1.41421, \frac{7}{16})$\}

\item $c = 2$, $(d_i,\theta_i)$ = \{$(1., 0)$, $(1., 0)$, $(2., \frac{1}{6})$, $(-2.44949, \frac{5}{16})$, $(1., \frac{1}{2})$, $(1., \frac{1}{2})$, $(2., -\frac{1}{3})$, $(-2.44949, -\frac{3}{16})$, $(-2.44949, -\frac{1}{16})$, $(2., \frac{1}{2})$, $(-2.44949, \frac{7}{16})$, $(2., \frac{1}{6})$, $(2., \frac{1}{6})$\}

\item $c = \frac{5}{2}$, $(d_i,\theta_i)$ = \{$(1., 0)$, $(1., 0)$, $(2., \frac{1}{6})$, $(-2.44949, \frac{5}{16})$, $(1., \frac{1}{2})$, $(1., \frac{1}{2})$, $(2., -\frac{1}{3})$, $(-2.44949, -\frac{3}{16})$, $(2.82843, \frac{11}{48})$, $(-1.73205, \frac{1}{2})$, $(-1.73205, 0)$, $(1.41421, -\frac{7}{16})$, $(-1.73205, \frac{1}{2})$, $(-1.73205, 0)$, $(1.41421, -\frac{7}{16})$\}

\item $c = 3$, $(d_i,\theta_i)$ = \{$(1., 0)$, $(1., 0)$, $(2., \frac{1}{6})$, $(-2.44949, \frac{5}{16})$, $(1., \frac{1}{2})$, $(1., \frac{1}{2})$, $(2., -\frac{1}{3})$, $(-2.44949, -\frac{3}{16})$, $(2., -\frac{3}{8})$, $(-2.44949, -\frac{7}{16})$, $(-2.44949, \frac{1}{16})$, $(2., \frac{7}{24})$, $(2., \frac{7}{24})$\}

\item $c = \frac{7}{2}$, $(d_i,\theta_i)$ = \{$(1., 0)$, $(1., 0)$, $(2., \frac{1}{6})$, $(-2.44949, \frac{5}{16})$, $(1., \frac{1}{2})$, $(1., \frac{1}{2})$, $(2., -\frac{1}{3})$, $(-2.44949, -\frac{3}{16})$, $(2.82843, \frac{17}{48})$, $(-1.73205, \frac{1}{8})$, $(-1.73205, -\frac{3}{8})$, $(1.41421, -\frac{5}{16})$, $(-1.73205, \frac{1}{8})$, $(-1.73205, -\frac{3}{8})$, $(1.41421, -\frac{5}{16})$\}

\item $c = 4$, $(d_i,\theta_i)$ = \{$(1., 0)$, $(1., 0)$, $(2., \frac{1}{6})$, $(-2.44949, \frac{5}{16})$, $(1., \frac{1}{2})$, $(1., \frac{1}{2})$, $(2., -\frac{1}{3})$, $(-2.44949, -\frac{3}{16})$, $(-2.44949, -\frac{5}{16})$, $(2., -\frac{1}{4})$, $(-2.44949, \frac{3}{16})$, $(2., \frac{5}{12})$, $(2., \frac{5}{12})$\}

\item $c = \frac{9}{2}$, $(d_i,\theta_i)$ = \{$(1., 0)$, $(1., 0)$, $(2., \frac{1}{6})$, $(-2.44949, \frac{5}{16})$, $(1., \frac{1}{2})$, $(1., \frac{1}{2})$, $(2., -\frac{1}{3})$, $(-2.44949, -\frac{3}{16})$, $(2.82843, \frac{23}{48})$, $(-1.73205, -\frac{1}{4})$, $(-1.73205, \frac{1}{4})$, $(1.41421, -\frac{3}{16})$, $(-1.73205, \frac{1}{4})$, $(-1.73205, -\frac{1}{4})$, $(1.41421, -\frac{3}{16})$\}

\item $c = 5$, $(d_i,\theta_i)$ = \{$(1., 0)$, $(1., 0)$, $(2., \frac{1}{6})$, $(-2.44949, \frac{5}{16})$, $(1., \frac{1}{2})$, $(1., \frac{1}{2})$, $(2., -\frac{1}{3})$, $(-2.44949, -\frac{3}{16})$, $(-2.44949, \frac{5}{16})$, $(-2.44949, -\frac{3}{16})$, $(2., -\frac{1}{8})$, $(2., -\frac{11}{24})$, $(2., -\frac{11}{24})$\}

\item $c = \frac{11}{2}$, $(d_i,\theta_i)$ = \{$(1., 0)$, $(1., 0)$, $(2., \frac{1}{6})$, $(-2.44949, \frac{5}{16})$, $(1., \frac{1}{2})$, $(1., \frac{1}{2})$, $(2., -\frac{1}{3})$, $(-2.44949, -\frac{3}{16})$, $(2.82843, -\frac{19}{48})$, $(-1.73205, -\frac{1}{8})$, $(-1.73205, \frac{3}{8})$, $(1.41421, -\frac{1}{16})$, $(-1.73205, -\frac{1}{8})$, $(-1.73205, \frac{3}{8})$, $(1.41421, -\frac{1}{16})$\}

\item $c = 6$, $(d_i,\theta_i)$ = \{$(1., 0)$, $(1., \frac{1}{2})$, $(2., \frac{1}{6})$, $(-2.44949, \frac{5}{16})$, $(1., \frac{1}{2})$, $(1., 0)$, $(2., -\frac{1}{3})$, $(-2.44949, -\frac{3}{16})$, $(-2.44949, -\frac{1}{16})$, $(-2.44949, \frac{7}{16})$, $(2., 0)$, $(2., -\frac{1}{3})$, $(2., -\frac{1}{3})$\}

\item $c = \frac{13}{2}$, $(d_i,\theta_i)$ = \{$(1., 0)$, $(1., \frac{1}{2})$, $(2., \frac{1}{6})$, $(-2.44949, \frac{5}{16})$, $(1., \frac{1}{2})$, $(1., 0)$, $(2., -\frac{1}{3})$, $(-2.44949, -\frac{3}{16})$, $(-2.82843, -\frac{13}{48})$, $(1.73205, 0)$, $(1.73205, \frac{1}{2})$, $(-1.41421, \frac{1}{16})$, $(1.73205, 0)$, $(1.73205, \frac{1}{2})$, $(-1.41421, \frac{1}{16})$\}

\item $c = 7$, $(d_i,\theta_i)$ = \{$(1., 0)$, $(1., \frac{1}{2})$, $(2., \frac{1}{6})$, $(-2.44949, \frac{5}{16})$, $(1., \frac{1}{2})$, $(1., 0)$, $(2., -\frac{1}{3})$, $(-2.44949, -\frac{3}{16})$, $(-2.44949, \frac{1}{16})$, $(2., \frac{1}{8})$, $(-2.44949, -\frac{7}{16})$, $(2., -\frac{5}{24})$, $(2., -\frac{5}{24})$\}

\item $c = \frac{15}{2}$, $(d_i,\theta_i)$ = \{$(1., 0)$, $(1., \frac{1}{2})$, $(2., \frac{1}{6})$, $(-2.44949, \frac{5}{16})$, $(1., \frac{1}{2})$, $(1., 0)$, $(2., -\frac{1}{3})$, $(-2.44949, -\frac{3}{16})$, $(-2.82843, -\frac{7}{48})$, $(1.73205, -\frac{3}{8})$, $(1.73205, \frac{1}{8})$, $(-1.41421, \frac{3}{16})$, $(1.73205, -\frac{3}{8})$, $(1.73205, \frac{1}{8})$, $(-1.41421, \frac{3}{16})$\}

\end{enumerate}

\paragraph*{Rank 8; \#29}

\begin{enumerate}

\item $c = 0$, $(d_i,\theta_i)$ = \{$(1., 0)$, $(1., 0)$, $(2., \frac{1}{6})$, $(-2.44949, \frac{3}{16})$, $(1., \frac{1}{2})$, $(1., \frac{1}{2})$, $(2., -\frac{1}{3})$, $(-2.44949, -\frac{5}{16})$, $(-2.44949, \frac{5}{16})$, $(2., \frac{1}{4})$, $(-2.44949, -\frac{3}{16})$, $(2., -\frac{1}{12})$, $(2., -\frac{1}{12})$\}

\item $c = \frac{1}{2}$, $(d_i,\theta_i)$ = \{$(1., 0)$, $(1., 0)$, $(2., \frac{1}{6})$, $(-2.44949, \frac{3}{16})$, $(1., \frac{1}{2})$, $(1., \frac{1}{2})$, $(2., -\frac{1}{3})$, $(-2.44949, -\frac{5}{16})$, $(2.82843, -\frac{1}{48})$, $(-1.73205, -\frac{1}{8})$, $(-1.73205, \frac{3}{8})$, $(1.41421, \frac{5}{16})$, $(-1.73205, -\frac{1}{8})$, $(-1.73205, \frac{3}{8})$, $(1.41421, \frac{5}{16})$\}

\item $c = 1$, $(d_i,\theta_i)$ = \{$(1., 0)$, $(1., 0)$, $(2., \frac{1}{6})$, $(-2.44949, \frac{3}{16})$, $(1., \frac{1}{2})$, $(1., \frac{1}{2})$, $(2., -\frac{1}{3})$, $(-2.44949, -\frac{5}{16})$, $(2., \frac{3}{8})$, $(-2.44949, -\frac{1}{16})$, $(-2.44949, \frac{7}{16})$, $(2., \frac{1}{24})$, $(2., \frac{1}{24})$\}

\item $c = \frac{3}{2}$, $(d_i,\theta_i)$ = \{$(1., 0)$, $(1., 0)$, $(2., \frac{1}{6})$, $(-2.44949, \frac{3}{16})$, $(1., \frac{1}{2})$, $(1., \frac{1}{2})$, $(2., -\frac{1}{3})$, $(-2.44949, -\frac{5}{16})$, $(2.82843, \frac{5}{48})$, $(-1.73205, 0)$, $(-1.73205, \frac{1}{2})$, $(1.41421, \frac{7}{16})$, $(-1.73205, 0)$, $(-1.73205, \frac{1}{2})$, $(1.41421, \frac{7}{16})$\}

\item $c = 2$, $(d_i,\theta_i)$ = \{$(1., 0)$, $(1., 0)$, $(2., \frac{1}{6})$, $(-2.44949, \frac{3}{16})$, $(1., \frac{1}{2})$, $(1., \frac{1}{2})$, $(2., -\frac{1}{3})$, $(-2.44949, -\frac{5}{16})$, $(-2.44949, -\frac{7}{16})$, $(-2.44949, \frac{1}{16})$, $(2., \frac{1}{2})$, $(2., \frac{1}{6})$, $(2., \frac{1}{6})$\}

\item $c = \frac{5}{2}$, $(d_i,\theta_i)$ = \{$(1., 0)$, $(1., 0)$, $(2., \frac{1}{6})$, $(-2.44949, \frac{3}{16})$, $(1., \frac{1}{2})$, $(1., \frac{1}{2})$, $(2., -\frac{1}{3})$, $(-2.44949, -\frac{5}{16})$, $(2.82843, \frac{11}{48})$, $(-1.73205, -\frac{3}{8})$, $(-1.73205, \frac{1}{8})$, $(1.41421, -\frac{7}{16})$, $(-1.73205, -\frac{3}{8})$, $(-1.73205, \frac{1}{8})$, $(1.41421, -\frac{7}{16})$\}

\item $c = 3$, $(d_i,\theta_i)$ = \{$(1., 0)$, $(1., 0)$, $(2., \frac{1}{6})$, $(-2.44949, \frac{3}{16})$, $(1., \frac{1}{2})$, $(1., \frac{1}{2})$, $(2., -\frac{1}{3})$, $(-2.44949, -\frac{5}{16})$, $(-2.44949, \frac{3}{16})$, $(2., -\frac{3}{8})$, $(-2.44949, -\frac{5}{16})$, $(2., \frac{7}{24})$, $(2., \frac{7}{24})$\}

\item $c = \frac{7}{2}$, $(d_i,\theta_i)$ = \{$(1., 0)$, $(1., 0)$, $(2., \frac{1}{6})$, $(-2.44949, \frac{3}{16})$, $(1., \frac{1}{2})$, $(1., \frac{1}{2})$, $(2., -\frac{1}{3})$, $(-2.44949, -\frac{5}{16})$, $(2.82843, \frac{17}{48})$, $(-1.73205, -\frac{1}{4})$, $(-1.73205, \frac{1}{4})$, $(1.41421, -\frac{5}{16})$, $(-1.73205, -\frac{1}{4})$, $(-1.73205, \frac{1}{4})$, $(1.41421, -\frac{5}{16})$\}

\item $c = 4$, $(d_i,\theta_i)$ = \{$(1., 0)$, $(1., 0)$, $(2., \frac{1}{6})$, $(-2.44949, \frac{3}{16})$, $(1., \frac{1}{2})$, $(1., \frac{1}{2})$, $(2., -\frac{1}{3})$, $(-2.44949, -\frac{5}{16})$, $(2., -\frac{1}{4})$, $(-2.44949, -\frac{3}{16})$, $(-2.44949, \frac{5}{16})$, $(2., \frac{5}{12})$, $(2., \frac{5}{12})$\}

\item $c = \frac{9}{2}$, $(d_i,\theta_i)$ = \{$(1., 0)$, $(1., 0)$, $(2., \frac{1}{6})$, $(-2.44949, \frac{3}{16})$, $(1., \frac{1}{2})$, $(1., \frac{1}{2})$, $(2., -\frac{1}{3})$, $(-2.44949, -\frac{5}{16})$, $(2.82843, \frac{23}{48})$, $(-1.73205, \frac{3}{8})$, $(-1.73205, -\frac{1}{8})$, $(1.41421, -\frac{3}{16})$, $(-1.73205, \frac{3}{8})$, $(-1.73205, -\frac{1}{8})$, $(1.41421, -\frac{3}{16})$\}

\item $c = 5$, $(d_i,\theta_i)$ = \{$(1., 0)$, $(1., 0)$, $(2., \frac{1}{6})$, $(-2.44949, \frac{3}{16})$, $(1., \frac{1}{2})$, $(1., \frac{1}{2})$, $(2., -\frac{1}{3})$, $(-2.44949, -\frac{5}{16})$, $(-2.44949, \frac{7}{16})$, $(2., -\frac{1}{8})$, $(-2.44949, -\frac{1}{16})$, $(2., -\frac{11}{24})$, $(2., -\frac{11}{24})$\}

\item $c = \frac{11}{2}$, $(d_i,\theta_i)$ = \{$(1., 0)$, $(1., 0)$, $(2., \frac{1}{6})$, $(-2.44949, \frac{3}{16})$, $(1., \frac{1}{2})$, $(1., \frac{1}{2})$, $(2., -\frac{1}{3})$, $(-2.44949, -\frac{5}{16})$, $(2.82843, -\frac{19}{48})$, $(-1.73205, \frac{1}{2})$, $(-1.73205, 0)$, $(1.41421, -\frac{1}{16})$, $(-1.73205, \frac{1}{2})$, $(-1.73205, 0)$, $(1.41421, -\frac{1}{16})$\}

\item $c = 6$, $(d_i,\theta_i)$ = \{$(1., 0)$, $(1., \frac{1}{2})$, $(2., \frac{1}{6})$, $(-2.44949, \frac{3}{16})$, $(1., \frac{1}{2})$, $(1., 0)$, $(2., -\frac{1}{3})$, $(-2.44949, -\frac{5}{16})$, $(-2., 0)$, $(2.44949, \frac{1}{16})$, $(2.44949, -\frac{7}{16})$, $(-2., -\frac{1}{3})$, $(-2., -\frac{1}{3})$\}

\item $c = \frac{13}{2}$, $(d_i,\theta_i)$ = \{$(1., 0)$, $(1., \frac{1}{2})$, $(2., \frac{1}{6})$, $(-2.44949, \frac{3}{16})$, $(1., \frac{1}{2})$, $(1., 0)$, $(2., -\frac{1}{3})$, $(-2.44949, -\frac{5}{16})$, $(-2.82843, -\frac{13}{48})$, $(1.73205, \frac{1}{8})$, $(1.73205, -\frac{3}{8})$, $(-1.41421, \frac{1}{16})$, $(1.73205, \frac{1}{8})$, $(1.73205, -\frac{3}{8})$, $(-1.41421, \frac{1}{16})$\}

\item $c = 7$, $(d_i,\theta_i)$ = \{$(1., 0)$, $(1., \frac{1}{2})$, $(2., \frac{1}{6})$, $(-2.44949, \frac{3}{16})$, $(1., \frac{1}{2})$, $(1., 0)$, $(2., -\frac{1}{3})$, $(-2.44949, -\frac{5}{16})$, $(-2.44949, -\frac{5}{16})$, $(2., \frac{1}{8})$, $(-2.44949, \frac{3}{16})$, $(2., -\frac{5}{24})$, $(2., -\frac{5}{24})$\}

\item $c = \frac{15}{2}$, $(d_i,\theta_i)$ = \{$(1., 0)$, $(1., \frac{1}{2})$, $(2., \frac{1}{6})$, $(-2.44949, \frac{3}{16})$, $(1., \frac{1}{2})$, $(1., 0)$, $(2., -\frac{1}{3})$, $(-2.44949, -\frac{5}{16})$, $(-2.82843, -\frac{7}{48})$, $(1.73205, \frac{1}{4})$, $(1.73205, -\frac{1}{4})$, $(-1.41421, \frac{3}{16})$, $(1.73205, \frac{1}{4})$, $(1.73205, -\frac{1}{4})$, $(-1.41421, \frac{3}{16})$\}

\end{enumerate}

\paragraph*{Rank 8; \#30}

\begin{enumerate}

\item $c = 0$, $(d_i,\theta_i)$ = \{$(1., 0)$, $(1., 0)$, $(-2.44949, \frac{1}{16})$, $(2., -\frac{1}{6})$, $(1., \frac{1}{2})$, $(1., \frac{1}{2})$, $(-2.44949, -\frac{7}{16})$, $(2., \frac{1}{3})$, $(2., -\frac{1}{4})$, $(-2.44949, -\frac{1}{16})$, $(-2.44949, \frac{7}{16})$, $(2., \frac{1}{12})$, $(2., \frac{1}{12})$\}

\item $c = \frac{1}{2}$, $(d_i,\theta_i)$ = \{$(1., 0)$, $(1., 0)$, $(-2.44949, \frac{1}{16})$, $(2., -\frac{1}{6})$, $(1., \frac{1}{2})$, $(1., \frac{1}{2})$, $(-2.44949, -\frac{7}{16})$, $(2., \frac{1}{3})$, $(2.82843, \frac{7}{48})$, $(-1.73205, 0)$, $(-1.73205, \frac{1}{2})$, $(1.41421, -\frac{3}{16})$, $(-1.73205, 0)$, $(-1.73205, \frac{1}{2})$, $(1.41421, -\frac{3}{16})$\}

\item $c = 1$, $(d_i,\theta_i)$ = \{$(1., 0)$, $(1., 0)$, $(-2.44949, \frac{1}{16})$, $(2., -\frac{1}{6})$, $(1., \frac{1}{2})$, $(1., \frac{1}{2})$, $(-2.44949, -\frac{7}{16})$, $(2., \frac{1}{3})$, $(2., -\frac{1}{8})$, $(-2.44949, -\frac{7}{16})$, $(-2.44949, \frac{1}{16})$, $(2., \frac{5}{24})$, $(2., \frac{5}{24})$\}

\item $c = \frac{3}{2}$, $(d_i,\theta_i)$ = \{$(1., 0)$, $(1., 0)$, $(-2.44949, \frac{1}{16})$, $(2., -\frac{1}{6})$, $(1., \frac{1}{2})$, $(1., \frac{1}{2})$, $(-2.44949, -\frac{7}{16})$, $(2., \frac{1}{3})$, $(2.82843, \frac{13}{48})$, $(-1.73205, \frac{1}{8})$, $(-1.73205, -\frac{3}{8})$, $(1.41421, -\frac{1}{16})$, $(-1.73205, \frac{1}{8})$, $(-1.73205, -\frac{3}{8})$, $(1.41421, -\frac{1}{16})$\}

\item $c = 2$, $(d_i,\theta_i)$ = \{$(1., 0)$, $(1., 0)$, $(-2.44949, \frac{1}{16})$, $(2., \frac{1}{3})$, $(1., \frac{1}{2})$, $(1., \frac{1}{2})$, $(-2.44949, -\frac{7}{16})$, $(2., -\frac{1}{6})$, $(-2.44949, -\frac{5}{16})$, $(-2.44949, \frac{3}{16})$, $(2., 0)$, $(2., \frac{1}{3})$, $(2., \frac{1}{3})$\}

\item $c = \frac{5}{2}$, $(d_i,\theta_i)$ = \{$(1., 0)$, $(1., 0)$, $(-2.44949, \frac{1}{16})$, $(2., \frac{1}{3})$, $(1., \frac{1}{2})$, $(1., \frac{1}{2})$, $(-2.44949, -\frac{7}{16})$, $(2., -\frac{1}{6})$, $(2.82843, \frac{19}{48})$, $(-1.73205, -\frac{1}{4})$, $(-1.73205, \frac{1}{4})$, $(1.41421, \frac{1}{16})$, $(-1.73205, -\frac{1}{4})$, $(-1.73205, \frac{1}{4})$, $(1.41421, \frac{1}{16})$\}

\item $c = 3$, $(d_i,\theta_i)$ = \{$(1., 0)$, $(1., 0)$, $(-2.44949, \frac{1}{16})$, $(2., \frac{1}{3})$, $(1., \frac{1}{2})$, $(1., \frac{1}{2})$, $(-2.44949, -\frac{7}{16})$, $(2., -\frac{1}{6})$, $(2., \frac{1}{8})$, $(-2.44949, \frac{5}{16})$, $(-2.44949, -\frac{3}{16})$, $(2., \frac{11}{24})$, $(2., \frac{11}{24})$\}

\item $c = \frac{7}{2}$, $(d_i,\theta_i)$ = \{$(1., 0)$, $(1., 0)$, $(-2.44949, \frac{1}{16})$, $(2., \frac{1}{3})$, $(1., \frac{1}{2})$, $(1., \frac{1}{2})$, $(-2.44949, -\frac{7}{16})$, $(2., -\frac{1}{6})$, $(2.82843, -\frac{23}{48})$, $(-1.73205, -\frac{1}{8})$, $(-1.73205, \frac{3}{8})$, $(1.41421, \frac{3}{16})$, $(-1.73205, -\frac{1}{8})$, $(-1.73205, \frac{3}{8})$, $(1.41421, \frac{3}{16})$\}

\item $c = 4$, $(d_i,\theta_i)$ = \{$(1., 0)$, $(1., 0)$, $(-2.44949, \frac{1}{16})$, $(2., \frac{1}{3})$, $(1., \frac{1}{2})$, $(1., \frac{1}{2})$, $(-2.44949, -\frac{7}{16})$, $(2., -\frac{1}{6})$, $(-2.44949, -\frac{1}{16})$, $(2., \frac{1}{4})$, $(-2.44949, \frac{7}{16})$, $(2., -\frac{5}{12})$, $(2., -\frac{5}{12})$\}

\item $c = \frac{9}{2}$, $(d_i,\theta_i)$ = \{$(1., 0)$, $(1., 0)$, $(-2.44949, \frac{1}{16})$, $(2., \frac{1}{3})$, $(1., \frac{1}{2})$, $(1., \frac{1}{2})$, $(-2.44949, -\frac{7}{16})$, $(2., -\frac{1}{6})$, $(2.82843, -\frac{17}{48})$, $(-1.73205, \frac{1}{2})$, $(-1.73205, 0)$, $(1.41421, \frac{5}{16})$, $(-1.73205, \frac{1}{2})$, $(-1.73205, 0)$, $(1.41421, \frac{5}{16})$\}

\item $c = 5$, $(d_i,\theta_i)$ = \{$(1., 0)$, $(1., 0)$, $(-2.44949, \frac{1}{16})$, $(2., \frac{1}{3})$, $(1., \frac{1}{2})$, $(1., \frac{1}{2})$, $(-2.44949, -\frac{7}{16})$, $(2., -\frac{1}{6})$, $(2., \frac{3}{8})$, $(-2.44949, \frac{1}{16})$, $(-2.44949, -\frac{7}{16})$, $(2., -\frac{7}{24})$, $(2., -\frac{7}{24})$\}

\item $c = \frac{11}{2}$, $(d_i,\theta_i)$ = \{$(1., 0)$, $(1., 0)$, $(-2.44949, \frac{1}{16})$, $(2., \frac{1}{3})$, $(1., \frac{1}{2})$, $(1., \frac{1}{2})$, $(-2.44949, -\frac{7}{16})$, $(2., -\frac{1}{6})$, $(-2.82843, -\frac{11}{48})$, $(1.73205, -\frac{3}{8})$, $(1.73205, \frac{1}{8})$, $(-1.41421, \frac{7}{16})$, $(1.73205, -\frac{3}{8})$, $(1.73205, \frac{1}{8})$, $(-1.41421, \frac{7}{16})$\}

\item $c = 6$, $(d_i,\theta_i)$ = \{$(1., 0)$, $(1., \frac{1}{2})$, $(-2.44949, \frac{1}{16})$, $(2., \frac{1}{3})$, $(1., \frac{1}{2})$, $(1., 0)$, $(-2.44949, -\frac{7}{16})$, $(2., -\frac{1}{6})$, $(-2., \frac{1}{2})$, $(2.44949, -\frac{5}{16})$, $(2.44949, \frac{3}{16})$, $(-2., -\frac{1}{6})$, $(-2., -\frac{1}{6})$\}

\item $c = \frac{13}{2}$, $(d_i,\theta_i)$ = \{$(1., 0)$, $(1., \frac{1}{2})$, $(-2.44949, \frac{1}{16})$, $(2., \frac{1}{3})$, $(1., \frac{1}{2})$, $(1., 0)$, $(-2.44949, -\frac{7}{16})$, $(2., -\frac{1}{6})$, $(-2.82843, -\frac{5}{48})$, $(1.73205, \frac{1}{4})$, $(1.73205, -\frac{1}{4})$, $(-1.41421, -\frac{7}{16})$, $(1.73205, \frac{1}{4})$, $(1.73205, -\frac{1}{4})$, $(-1.41421, -\frac{7}{16})$\}

\item $c = 7$, $(d_i,\theta_i)$ = \{$(1., 0)$, $(1., \frac{1}{2})$, $(-2.44949, \frac{1}{16})$, $(2., \frac{1}{3})$, $(1., \frac{1}{2})$, $(1., 0)$, $(-2.44949, -\frac{7}{16})$, $(2., -\frac{1}{6})$, $(-2., -\frac{3}{8})$, $(2.44949, -\frac{3}{16})$, $(2.44949, \frac{5}{16})$, $(-2., -\frac{1}{24})$, $(-2., -\frac{1}{24})$\}

\item $c = \frac{15}{2}$, $(d_i,\theta_i)$ = \{$(1., 0)$, $(1., \frac{1}{2})$, $(-2.44949, -\frac{7}{16})$, $(2., \frac{1}{3})$, $(1., \frac{1}{2})$, $(1., 0)$, $(-2.44949, \frac{1}{16})$, $(2., -\frac{1}{6})$, $(-2.82843, \frac{1}{48})$, $(1.73205, \frac{3}{8})$, $(1.73205, -\frac{1}{8})$, $(-1.41421, -\frac{5}{16})$, $(1.73205, \frac{3}{8})$, $(1.73205, -\frac{1}{8})$, $(-1.41421, -\frac{5}{16})$\}

\end{enumerate}

\paragraph*{Rank 8; \#31}

\begin{enumerate}

\item $c = 0$, $(d_i,\theta_i)$ = \{$(1., 0)$, $(1., 0)$, $(-2.44949, \frac{1}{16})$, $(2., \frac{1}{6})$, $(1., \frac{1}{2})$, $(1., \frac{1}{2})$, $(-2.44949, -\frac{7}{16})$, $(2., -\frac{1}{3})$, $(2., \frac{1}{4})$, $(-2.44949, -\frac{1}{16})$, $(-2.44949, \frac{7}{16})$, $(2., -\frac{1}{12})$, $(2., -\frac{1}{12})$\}

\item $c = \frac{1}{2}$, $(d_i,\theta_i)$ = \{$(1., 0)$, $(1., 0)$, $(-2.44949, \frac{1}{16})$, $(2., \frac{1}{6})$, $(1., \frac{1}{2})$, $(1., \frac{1}{2})$, $(-2.44949, -\frac{7}{16})$, $(2., -\frac{1}{3})$, $(2.82843, -\frac{1}{48})$, $(-1.73205, 0)$, $(-1.73205, \frac{1}{2})$, $(1.41421, \frac{5}{16})$, $(-1.73205, 0)$, $(-1.73205, \frac{1}{2})$, $(1.41421, \frac{5}{16})$\}

\item $c = 1$, $(d_i,\theta_i)$ = \{$(1., 0)$, $(1., 0)$, $(-2.44949, \frac{1}{16})$, $(2., \frac{1}{6})$, $(1., \frac{1}{2})$, $(1., \frac{1}{2})$, $(-2.44949, -\frac{7}{16})$, $(2., -\frac{1}{3})$, $(-2.44949, -\frac{7}{16})$, $(-2.44949, \frac{1}{16})$, $(2., \frac{3}{8})$, $(2., \frac{1}{24})$, $(2., \frac{1}{24})$\}

\item $c = \frac{3}{2}$, $(d_i,\theta_i)$ = \{$(1., 0)$, $(1., 0)$, $(-2.44949, \frac{1}{16})$, $(2., \frac{1}{6})$, $(1., \frac{1}{2})$, $(1., \frac{1}{2})$, $(-2.44949, -\frac{7}{16})$, $(2., -\frac{1}{3})$, $(2.82843, \frac{5}{48})$, $(-1.73205, \frac{1}{8})$, $(-1.73205, -\frac{3}{8})$, $(1.41421, \frac{7}{16})$, $(-1.73205, \frac{1}{8})$, $(-1.73205, -\frac{3}{8})$, $(1.41421, \frac{7}{16})$\}

\item $c = 2$, $(d_i,\theta_i)$ = \{$(1., 0)$, $(1., 0)$, $(-2.44949, \frac{1}{16})$, $(2., \frac{1}{6})$, $(1., \frac{1}{2})$, $(1., \frac{1}{2})$, $(-2.44949, -\frac{7}{16})$, $(2., -\frac{1}{3})$, $(2., \frac{1}{2})$, $(-2.44949, \frac{3}{16})$, $(-2.44949, -\frac{5}{16})$, $(2., \frac{1}{6})$, $(2., \frac{1}{6})$\}

\item $c = \frac{5}{2}$, $(d_i,\theta_i)$ = \{$(1., 0)$, $(1., 0)$, $(-2.44949, \frac{1}{16})$, $(2., \frac{1}{6})$, $(1., \frac{1}{2})$, $(1., \frac{1}{2})$, $(-2.44949, -\frac{7}{16})$, $(2., -\frac{1}{3})$, $(2.82843, \frac{11}{48})$, $(-1.73205, -\frac{1}{4})$, $(-1.73205, \frac{1}{4})$, $(1.41421, -\frac{7}{16})$, $(-1.73205, -\frac{1}{4})$, $(-1.73205, \frac{1}{4})$, $(1.41421, -\frac{7}{16})$\}

\item $c = 3$, $(d_i,\theta_i)$ = \{$(1., 0)$, $(1., 0)$, $(-2.44949, \frac{1}{16})$, $(2., \frac{1}{6})$, $(1., \frac{1}{2})$, $(1., \frac{1}{2})$, $(-2.44949, -\frac{7}{16})$, $(2., -\frac{1}{3})$, $(-2.44949, \frac{5}{16})$, $(2., -\frac{3}{8})$, $(-2.44949, -\frac{3}{16})$, $(2., \frac{7}{24})$, $(2., \frac{7}{24})$\}

\item $c = \frac{7}{2}$, $(d_i,\theta_i)$ = \{$(1., 0)$, $(1., 0)$, $(-2.44949, \frac{1}{16})$, $(2., \frac{1}{6})$, $(1., \frac{1}{2})$, $(1., \frac{1}{2})$, $(-2.44949, -\frac{7}{16})$, $(2., -\frac{1}{3})$, $(2.82843, \frac{17}{48})$, $(-1.73205, -\frac{1}{8})$, $(-1.73205, \frac{3}{8})$, $(1.41421, -\frac{5}{16})$, $(-1.73205, -\frac{1}{8})$, $(-1.73205, \frac{3}{8})$, $(1.41421, -\frac{5}{16})$\}

\item $c = 4$, $(d_i,\theta_i)$ = \{$(1., 0)$, $(1., 0)$, $(-2.44949, \frac{1}{16})$, $(2., \frac{1}{6})$, $(1., \frac{1}{2})$, $(1., \frac{1}{2})$, $(-2.44949, -\frac{7}{16})$, $(2., -\frac{1}{3})$, $(2., -\frac{1}{4})$, $(-2.44949, -\frac{1}{16})$, $(-2.44949, \frac{7}{16})$, $(2., \frac{5}{12})$, $(2., \frac{5}{12})$\}

\item $c = \frac{9}{2}$, $(d_i,\theta_i)$ = \{$(1., 0)$, $(1., 0)$, $(-2.44949, \frac{1}{16})$, $(2., \frac{1}{6})$, $(1., \frac{1}{2})$, $(1., \frac{1}{2})$, $(-2.44949, -\frac{7}{16})$, $(2., -\frac{1}{3})$, $(2.82843, \frac{23}{48})$, $(-1.73205, \frac{1}{2})$, $(-1.73205, 0)$, $(1.41421, -\frac{3}{16})$, $(-1.73205, \frac{1}{2})$, $(-1.73205, 0)$, $(1.41421, -\frac{3}{16})$\}

\item $c = 5$, $(d_i,\theta_i)$ = \{$(1., 0)$, $(1., 0)$, $(-2.44949, \frac{1}{16})$, $(2., \frac{1}{6})$, $(1., \frac{1}{2})$, $(1., \frac{1}{2})$, $(-2.44949, -\frac{7}{16})$, $(2., -\frac{1}{3})$, $(2., -\frac{1}{8})$, $(-2.44949, -\frac{7}{16})$, $(-2.44949, \frac{1}{16})$, $(2., -\frac{11}{24})$, $(2., -\frac{11}{24})$\}

\item $c = \frac{11}{2}$, $(d_i,\theta_i)$ = \{$(1., 0)$, $(1., 0)$, $(-2.44949, \frac{1}{16})$, $(2., \frac{1}{6})$, $(1., \frac{1}{2})$, $(1., \frac{1}{2})$, $(-2.44949, -\frac{7}{16})$, $(2., -\frac{1}{3})$, $(2.82843, -\frac{19}{48})$, $(-1.73205, -\frac{3}{8})$, $(-1.73205, \frac{1}{8})$, $(1.41421, -\frac{1}{16})$, $(-1.73205, -\frac{3}{8})$, $(-1.73205, \frac{1}{8})$, $(1.41421, -\frac{1}{16})$\}

\item $c = 6$, $(d_i,\theta_i)$ = \{$(1., 0)$, $(1., \frac{1}{2})$, $(-2.44949, \frac{1}{16})$, $(2., \frac{1}{6})$, $(1., \frac{1}{2})$, $(1., 0)$, $(-2.44949, -\frac{7}{16})$, $(2., -\frac{1}{3})$, $(-2., 0)$, $(2.44949, -\frac{5}{16})$, $(2.44949, \frac{3}{16})$, $(-2., -\frac{1}{3})$, $(-2., -\frac{1}{3})$\}

\item $c = \frac{13}{2}$, $(d_i,\theta_i)$ = \{$(1., 0)$, $(1., \frac{1}{2})$, $(-2.44949, \frac{1}{16})$, $(2., \frac{1}{6})$, $(1., \frac{1}{2})$, $(1., 0)$, $(-2.44949, -\frac{7}{16})$, $(2., -\frac{1}{3})$, $(-2.82843, -\frac{13}{48})$, $(1.73205, \frac{1}{4})$, $(1.73205, -\frac{1}{4})$, $(-1.41421, \frac{1}{16})$, $(1.73205, \frac{1}{4})$, $(1.73205, -\frac{1}{4})$, $(-1.41421, \frac{1}{16})$\}

\item $c = 7$, $(d_i,\theta_i)$ = \{$(1., 0)$, $(1., \frac{1}{2})$, $(-2.44949, \frac{1}{16})$, $(2., \frac{1}{6})$, $(1., \frac{1}{2})$, $(1., 0)$, $(-2.44949, -\frac{7}{16})$, $(2., -\frac{1}{3})$, $(-2.44949, \frac{5}{16})$, $(2., \frac{1}{8})$, $(-2.44949, -\frac{3}{16})$, $(2., -\frac{5}{24})$, $(2., -\frac{5}{24})$\}

\item $c = \frac{15}{2}$, $(d_i,\theta_i)$ = \{$(1., 0)$, $(1., \frac{1}{2})$, $(-2.44949, -\frac{7}{16})$, $(2., \frac{1}{6})$, $(1., \frac{1}{2})$, $(1., 0)$, $(-2.44949, \frac{1}{16})$, $(2., -\frac{1}{3})$, $(-2.82843, -\frac{7}{48})$, $(1.73205, \frac{3}{8})$, $(1.73205, -\frac{1}{8})$, $(-1.41421, \frac{3}{16})$, $(1.73205, \frac{3}{8})$, $(1.73205, -\frac{1}{8})$, $(-1.41421, \frac{3}{16})$\}

\end{enumerate}

\paragraph*{Rank 8; \#32}

\begin{enumerate}

\item $c = 0$, $(d_i,\theta_i)$ = \{$(1., 0)$, $(1., 0)$, $(-2.44949, -\frac{3}{16})$, $(2., -\frac{1}{6})$, $(1., \frac{1}{2})$, $(1., \frac{1}{2})$, $(-2.44949, \frac{5}{16})$, $(2., \frac{1}{3})$, $(2., -\frac{1}{4})$, $(-2.44949, \frac{3}{16})$, $(-2.44949, -\frac{5}{16})$, $(2., \frac{1}{12})$, $(2., \frac{1}{12})$\}

\item $c = \frac{1}{2}$, $(d_i,\theta_i)$ = \{$(1., 0)$, $(1., 0)$, $(-2.44949, -\frac{3}{16})$, $(2., -\frac{1}{6})$, $(1., \frac{1}{2})$, $(1., \frac{1}{2})$, $(-2.44949, \frac{5}{16})$, $(2., \frac{1}{3})$, $(2.82843, \frac{7}{48})$, $(-1.73205, -\frac{1}{4})$, $(-1.73205, \frac{1}{4})$, $(1.41421, -\frac{3}{16})$, $(-1.73205, -\frac{1}{4})$, $(-1.73205, \frac{1}{4})$, $(1.41421, -\frac{3}{16})$\}

\item $c = 1$, $(d_i,\theta_i)$ = \{$(1., 0)$, $(1., 0)$, $(-2.44949, -\frac{3}{16})$, $(2., -\frac{1}{6})$, $(1., \frac{1}{2})$, $(1., \frac{1}{2})$, $(-2.44949, \frac{5}{16})$, $(2., \frac{1}{3})$, $(2., -\frac{1}{8})$, $(-2.44949, -\frac{3}{16})$, $(-2.44949, \frac{5}{16})$, $(2., \frac{5}{24})$, $(2., \frac{5}{24})$\}

\item $c = \frac{3}{2}$, $(d_i,\theta_i)$ = \{$(1., 0)$, $(1., 0)$, $(-2.44949, \frac{5}{16})$, $(2., -\frac{1}{6})$, $(1., \frac{1}{2})$, $(1., \frac{1}{2})$, $(-2.44949, -\frac{3}{16})$, $(2., \frac{1}{3})$, $(2.82843, \frac{13}{48})$, $(-1.73205, \frac{3}{8})$, $(-1.73205, -\frac{1}{8})$, $(1.41421, -\frac{1}{16})$, $(-1.73205, \frac{3}{8})$, $(-1.73205, -\frac{1}{8})$, $(1.41421, -\frac{1}{16})$\}

\item $c = 2$, $(d_i,\theta_i)$ = \{$(1., 0)$, $(1., 0)$, $(-2.44949, \frac{5}{16})$, $(2., \frac{1}{3})$, $(1., \frac{1}{2})$, $(1., \frac{1}{2})$, $(-2.44949, -\frac{3}{16})$, $(2., -\frac{1}{6})$, $(-2.44949, \frac{7}{16})$, $(-2.44949, -\frac{1}{16})$, $(2., 0)$, $(2., \frac{1}{3})$, $(2., \frac{1}{3})$\}

\item $c = \frac{5}{2}$, $(d_i,\theta_i)$ = \{$(1., 0)$, $(1., 0)$, $(-2.44949, \frac{5}{16})$, $(2., \frac{1}{3})$, $(1., \frac{1}{2})$, $(1., \frac{1}{2})$, $(-2.44949, -\frac{3}{16})$, $(2., -\frac{1}{6})$, $(2.82843, \frac{19}{48})$, $(-1.73205, \frac{1}{2})$, $(-1.73205, 0)$, $(1.41421, \frac{1}{16})$, $(-1.73205, \frac{1}{2})$, $(-1.73205, 0)$, $(1.41421, \frac{1}{16})$\}

\item $c = 3$, $(d_i,\theta_i)$ = \{$(1., 0)$, $(1., 0)$, $(-2.44949, \frac{5}{16})$, $(2., \frac{1}{3})$, $(1., \frac{1}{2})$, $(1., \frac{1}{2})$, $(-2.44949, -\frac{3}{16})$, $(2., -\frac{1}{6})$, $(2., \frac{1}{8})$, $(-2.44949, \frac{1}{16})$, $(-2.44949, -\frac{7}{16})$, $(2., \frac{11}{24})$, $(2., \frac{11}{24})$\}

\item $c = \frac{7}{2}$, $(d_i,\theta_i)$ = \{$(1., 0)$, $(1., 0)$, $(-2.44949, \frac{5}{16})$, $(2., \frac{1}{3})$, $(1., \frac{1}{2})$, $(1., \frac{1}{2})$, $(-2.44949, -\frac{3}{16})$, $(2., -\frac{1}{6})$, $(2.82843, -\frac{23}{48})$, $(-1.73205, \frac{1}{8})$, $(-1.73205, -\frac{3}{8})$, $(1.41421, \frac{3}{16})$, $(-1.73205, \frac{1}{8})$, $(-1.73205, -\frac{3}{8})$, $(1.41421, \frac{3}{16})$\}

\item $c = 4$, $(d_i,\theta_i)$ = \{$(1., 0)$, $(1., 0)$, $(-2.44949, \frac{5}{16})$, $(2., \frac{1}{3})$, $(1., \frac{1}{2})$, $(1., \frac{1}{2})$, $(-2.44949, -\frac{3}{16})$, $(2., -\frac{1}{6})$, $(2., \frac{1}{4})$, $(-2.44949, -\frac{5}{16})$, $(-2.44949, \frac{3}{16})$, $(2., -\frac{5}{12})$, $(2., -\frac{5}{12})$\}

\item $c = \frac{9}{2}$, $(d_i,\theta_i)$ = \{$(1., 0)$, $(1., 0)$, $(-2.44949, \frac{5}{16})$, $(2., \frac{1}{3})$, $(1., \frac{1}{2})$, $(1., \frac{1}{2})$, $(-2.44949, -\frac{3}{16})$, $(2., -\frac{1}{6})$, $(2.82843, -\frac{17}{48})$, $(-1.73205, \frac{1}{4})$, $(-1.73205, -\frac{1}{4})$, $(1.41421, \frac{5}{16})$, $(-1.73205, \frac{1}{4})$, $(-1.73205, -\frac{1}{4})$, $(1.41421, \frac{5}{16})$\}

\item $c = 5$, $(d_i,\theta_i)$ = \{$(1., 0)$, $(1., 0)$, $(-2.44949, \frac{5}{16})$, $(2., \frac{1}{3})$, $(1., \frac{1}{2})$, $(1., \frac{1}{2})$, $(-2.44949, -\frac{3}{16})$, $(2., -\frac{1}{6})$, $(2.44949, -\frac{3}{16})$, $(-2., \frac{3}{8})$, $(2.44949, \frac{5}{16})$, $(-2., -\frac{7}{24})$, $(-2., -\frac{7}{24})$\}

\item $c = \frac{11}{2}$, $(d_i,\theta_i)$ = \{$(1., 0)$, $(1., 0)$, $(-2.44949, \frac{5}{16})$, $(2., \frac{1}{3})$, $(1., \frac{1}{2})$, $(1., \frac{1}{2})$, $(-2.44949, -\frac{3}{16})$, $(2., -\frac{1}{6})$, $(2.82843, -\frac{11}{48})$, $(-1.73205, -\frac{1}{8})$, $(-1.73205, \frac{3}{8})$, $(1.41421, \frac{7}{16})$, $(-1.73205, -\frac{1}{8})$, $(-1.73205, \frac{3}{8})$, $(1.41421, \frac{7}{16})$\}

\item $c = 6$, $(d_i,\theta_i)$ = \{$(1., 0)$, $(1., \frac{1}{2})$, $(-2.44949, \frac{5}{16})$, $(2., \frac{1}{3})$, $(1., \frac{1}{2})$, $(1., 0)$, $(-2.44949, -\frac{3}{16})$, $(2., -\frac{1}{6})$, $(-2., \frac{1}{2})$, $(2.44949, -\frac{1}{16})$, $(2.44949, \frac{7}{16})$, $(-2., -\frac{1}{6})$, $(-2., -\frac{1}{6})$\}

\item $c = \frac{13}{2}$, $(d_i,\theta_i)$ = \{$(1., 0)$, $(1., \frac{1}{2})$, $(-2.44949, \frac{5}{16})$, $(2., \frac{1}{3})$, $(1., \frac{1}{2})$, $(1., 0)$, $(-2.44949, -\frac{3}{16})$, $(2., -\frac{1}{6})$, $(-2.82843, -\frac{5}{48})$, $(1.73205, 0)$, $(1.73205, \frac{1}{2})$, $(-1.41421, -\frac{7}{16})$, $(1.73205, 0)$, $(1.73205, \frac{1}{2})$, $(-1.41421, -\frac{7}{16})$\}

\item $c = 7$, $(d_i,\theta_i)$ = \{$(1., 0)$, $(1., \frac{1}{2})$, $(-2.44949, \frac{5}{16})$, $(2., \frac{1}{3})$, $(1., \frac{1}{2})$, $(1., 0)$, $(-2.44949, -\frac{3}{16})$, $(2., -\frac{1}{6})$, $(-2., -\frac{3}{8})$, $(2.44949, -\frac{7}{16})$, $(2.44949, \frac{1}{16})$, $(-2., -\frac{1}{24})$, $(-2., -\frac{1}{24})$\}

\item $c = \frac{15}{2}$, $(d_i,\theta_i)$ = \{$(1., 0)$, $(1., \frac{1}{2})$, $(-2.44949, \frac{5}{16})$, $(2., \frac{1}{3})$, $(1., \frac{1}{2})$, $(1., 0)$, $(-2.44949, -\frac{3}{16})$, $(2., -\frac{1}{6})$, $(-2.82843, \frac{1}{48})$, $(1.73205, -\frac{3}{8})$, $(1.73205, \frac{1}{8})$, $(-1.41421, -\frac{5}{16})$, $(1.73205, -\frac{3}{8})$, $(1.73205, \frac{1}{8})$, $(-1.41421, -\frac{5}{16})$\}

\end{enumerate}

\paragraph*{Rank 8; \#33}

\begin{enumerate}

\item $c = 0$, $(d_i,\theta_i)$ = \{$(1., 0)$, $(1., 0)$, $(-2.44949, \frac{3}{16})$, $(2., -\frac{1}{6})$, $(1., \frac{1}{2})$, $(1., \frac{1}{2})$, $(-2.44949, -\frac{5}{16})$, $(2., \frac{1}{3})$, $(-2.44949, \frac{5}{16})$, $(2., -\frac{1}{4})$, $(-2.44949, -\frac{3}{16})$, $(2., \frac{1}{12})$, $(2., \frac{1}{12})$\}

\item $c = \frac{1}{2}$, $(d_i,\theta_i)$ = \{$(1., 0)$, $(1., 0)$, $(-2.44949, \frac{3}{16})$, $(2., -\frac{1}{6})$, $(1., \frac{1}{2})$, $(1., \frac{1}{2})$, $(-2.44949, -\frac{5}{16})$, $(2., \frac{1}{3})$, $(2.82843, \frac{7}{48})$, $(-1.73205, -\frac{1}{8})$, $(-1.73205, \frac{3}{8})$, $(1.41421, -\frac{3}{16})$, $(-1.73205, -\frac{1}{8})$, $(-1.73205, \frac{3}{8})$, $(1.41421, -\frac{3}{16})$\}

\item $c = 1$, $(d_i,\theta_i)$ = \{$(1., 0)$, $(1., 0)$, $(-2.44949, \frac{3}{16})$, $(2., -\frac{1}{6})$, $(1., \frac{1}{2})$, $(1., \frac{1}{2})$, $(-2.44949, -\frac{5}{16})$, $(2., \frac{1}{3})$, $(2., -\frac{1}{8})$, $(-2.44949, -\frac{1}{16})$, $(-2.44949, \frac{7}{16})$, $(2., \frac{5}{24})$, $(2., \frac{5}{24})$\}

\item $c = \frac{3}{2}$, $(d_i,\theta_i)$ = \{$(1., 0)$, $(1., 0)$, $(-2.44949, \frac{3}{16})$, $(2., -\frac{1}{6})$, $(1., \frac{1}{2})$, $(1., \frac{1}{2})$, $(-2.44949, -\frac{5}{16})$, $(2., \frac{1}{3})$, $(2.82843, \frac{13}{48})$, $(-1.73205, 0)$, $(-1.73205, \frac{1}{2})$, $(1.41421, -\frac{1}{16})$, $(-1.73205, 0)$, $(-1.73205, \frac{1}{2})$, $(1.41421, -\frac{1}{16})$\}

\item $c = 2$, $(d_i,\theta_i)$ = \{$(1., 0)$, $(1., 0)$, $(-2.44949, \frac{3}{16})$, $(2., \frac{1}{3})$, $(1., \frac{1}{2})$, $(1., \frac{1}{2})$, $(-2.44949, -\frac{5}{16})$, $(2., -\frac{1}{6})$, $(-2.44949, \frac{1}{16})$, $(2., 0)$, $(-2.44949, -\frac{7}{16})$, $(2., \frac{1}{3})$, $(2., \frac{1}{3})$\}

\item $c = \frac{5}{2}$, $(d_i,\theta_i)$ = \{$(1., 0)$, $(1., 0)$, $(-2.44949, \frac{3}{16})$, $(2., \frac{1}{3})$, $(1., \frac{1}{2})$, $(1., \frac{1}{2})$, $(-2.44949, -\frac{5}{16})$, $(2., -\frac{1}{6})$, $(2.82843, \frac{19}{48})$, $(-1.73205, -\frac{3}{8})$, $(-1.73205, \frac{1}{8})$, $(1.41421, \frac{1}{16})$, $(-1.73205, -\frac{3}{8})$, $(-1.73205, \frac{1}{8})$, $(1.41421, \frac{1}{16})$\}

\item $c = 3$, $(d_i,\theta_i)$ = \{$(1., 0)$, $(1., 0)$, $(-2.44949, \frac{3}{16})$, $(2., \frac{1}{3})$, $(1., \frac{1}{2})$, $(1., \frac{1}{2})$, $(-2.44949, -\frac{5}{16})$, $(2., -\frac{1}{6})$, $(2., \frac{1}{8})$, $(-2.44949, \frac{3}{16})$, $(-2.44949, -\frac{5}{16})$, $(2., \frac{11}{24})$, $(2., \frac{11}{24})$\}

\item $c = \frac{7}{2}$, $(d_i,\theta_i)$ = \{$(1., 0)$, $(1., 0)$, $(-2.44949, \frac{3}{16})$, $(2., \frac{1}{3})$, $(1., \frac{1}{2})$, $(1., \frac{1}{2})$, $(-2.44949, -\frac{5}{16})$, $(2., -\frac{1}{6})$, $(2.82843, -\frac{23}{48})$, $(-1.73205, -\frac{1}{4})$, $(-1.73205, \frac{1}{4})$, $(1.41421, \frac{3}{16})$, $(-1.73205, -\frac{1}{4})$, $(-1.73205, \frac{1}{4})$, $(1.41421, \frac{3}{16})$\}

\item $c = 4$, $(d_i,\theta_i)$ = \{$(1., 0)$, $(1., 0)$, $(-2.44949, \frac{3}{16})$, $(2., \frac{1}{3})$, $(1., \frac{1}{2})$, $(1., \frac{1}{2})$, $(-2.44949, -\frac{5}{16})$, $(2., -\frac{1}{6})$, $(2.44949, \frac{5}{16})$, $(2.44949, -\frac{3}{16})$, $(-2., \frac{1}{4})$, $(-2., -\frac{5}{12})$, $(-2., -\frac{5}{12})$\}

\item $c = \frac{9}{2}$, $(d_i,\theta_i)$ = \{$(1., 0)$, $(1., 0)$, $(-2.44949, \frac{3}{16})$, $(2., \frac{1}{3})$, $(1., \frac{1}{2})$, $(1., \frac{1}{2})$, $(-2.44949, -\frac{5}{16})$, $(2., -\frac{1}{6})$, $(-2.82843, -\frac{17}{48})$, $(1.73205, \frac{3}{8})$, $(1.73205, -\frac{1}{8})$, $(-1.41421, \frac{5}{16})$, $(1.73205, \frac{3}{8})$, $(1.73205, -\frac{1}{8})$, $(-1.41421, \frac{5}{16})$\}

\item $c = 5$, $(d_i,\theta_i)$ = \{$(1., 0)$, $(1., 0)$, $(-2.44949, \frac{3}{16})$, $(2., \frac{1}{3})$, $(1., \frac{1}{2})$, $(1., \frac{1}{2})$, $(-2.44949, -\frac{5}{16})$, $(2., -\frac{1}{6})$, $(-2.44949, \frac{7}{16})$, $(2., \frac{3}{8})$, $(-2.44949, -\frac{1}{16})$, $(2., -\frac{7}{24})$, $(2., -\frac{7}{24})$\}

\item $c = \frac{11}{2}$, $(d_i,\theta_i)$ = \{$(1., 0)$, $(1., 0)$, $(-2.44949, \frac{3}{16})$, $(2., \frac{1}{3})$, $(1., \frac{1}{2})$, $(1., \frac{1}{2})$, $(-2.44949, -\frac{5}{16})$, $(2., -\frac{1}{6})$, $(-2.82843, -\frac{11}{48})$, $(1.73205, \frac{1}{2})$, $(1.73205, 0)$, $(-1.41421, \frac{7}{16})$, $(1.73205, \frac{1}{2})$, $(1.73205, 0)$, $(-1.41421, \frac{7}{16})$\}

\item $c = 6$, $(d_i,\theta_i)$ = \{$(1., 0)$, $(1., \frac{1}{2})$, $(-2.44949, \frac{3}{16})$, $(2., \frac{1}{3})$, $(1., \frac{1}{2})$, $(1., 0)$, $(-2.44949, -\frac{5}{16})$, $(2., -\frac{1}{6})$, $(2.44949, \frac{1}{16})$, $(-2., \frac{1}{2})$, $(2.44949, -\frac{7}{16})$, $(-2., -\frac{1}{6})$, $(-2., -\frac{1}{6})$\}

\item $c = \frac{13}{2}$, $(d_i,\theta_i)$ = \{$(1., 0)$, $(1., \frac{1}{2})$, $(-2.44949, \frac{3}{16})$, $(2., \frac{1}{3})$, $(1., \frac{1}{2})$, $(1., 0)$, $(-2.44949, -\frac{5}{16})$, $(2., -\frac{1}{6})$, $(-2.82843, -\frac{5}{48})$, $(1.73205, \frac{1}{8})$, $(1.73205, -\frac{3}{8})$, $(-1.41421, -\frac{7}{16})$, $(1.73205, \frac{1}{8})$, $(1.73205, -\frac{3}{8})$, $(-1.41421, -\frac{7}{16})$\}

\item $c = 7$, $(d_i,\theta_i)$ = \{$(1., 0)$, $(1., \frac{1}{2})$, $(-2.44949, \frac{3}{16})$, $(2., \frac{1}{3})$, $(1., \frac{1}{2})$, $(1., 0)$, $(-2.44949, -\frac{5}{16})$, $(2., -\frac{1}{6})$, $(-2., -\frac{3}{8})$, $(2.44949, \frac{3}{16})$, $(2.44949, -\frac{5}{16})$, $(-2., -\frac{1}{24})$, $(-2., -\frac{1}{24})$\}

\item $c = \frac{15}{2}$, $(d_i,\theta_i)$ = \{$(1., 0)$, $(1., \frac{1}{2})$, $(-2.44949, \frac{3}{16})$, $(2., \frac{1}{3})$, $(1., \frac{1}{2})$, $(1., 0)$, $(-2.44949, -\frac{5}{16})$, $(2., -\frac{1}{6})$, $(-2.82843, \frac{1}{48})$, $(1.73205, \frac{1}{4})$, $(1.73205, -\frac{1}{4})$, $(-1.41421, -\frac{5}{16})$, $(1.73205, \frac{1}{4})$, $(1.73205, -\frac{1}{4})$, $(-1.41421, -\frac{5}{16})$\}

\end{enumerate}

\paragraph*{Rank 8; \#62}

\begin{enumerate}

\item $c = 0$, $(d_i,\theta_i)$ = \{$(1., 0)$, $(5.82843, 0)$, $(2.41421, \frac{1}{4})$, $(2.41421, \frac{1}{4})$, $(1., \frac{1}{2})$, $(5.82843, \frac{1}{2})$, $(2.41421, -\frac{1}{4})$, $(2.41421, -\frac{1}{4})$, $(4.82843, \frac{3}{8})$, $(4.82843, -\frac{3}{8})$, $(3.41421, 0)$, $(3.41421, 0)$, $(3.41421, 0)$, $(3.41421, 0)$\}

\item $c = \frac{1}{2}$, $(d_i,\theta_i)$ = \{$(1., 0)$, $(5.82843, 0)$, $(2.41421, \frac{1}{4})$, $(2.41421, \frac{1}{4})$, $(1., \frac{1}{2})$, $(5.82843, \frac{1}{2})$, $(2.41421, -\frac{1}{4})$, $(2.41421, -\frac{1}{4})$, $(4.82843, \frac{1}{16})$, $(4.82843, \frac{1}{16})$, $(3.41421, \frac{7}{16})$, $(3.41421, -\frac{5}{16})$, $(3.41421, \frac{7}{16})$, $(3.41421, -\frac{5}{16})$\}

\item $c = 1$, $(d_i,\theta_i)$ = \{$(1., 0)$, $(5.82843, 0)$, $(2.41421, \frac{1}{4})$, $(2.41421, \frac{1}{4})$, $(1., \frac{1}{2})$, $(5.82843, \frac{1}{2})$, $(2.41421, -\frac{1}{4})$, $(2.41421, -\frac{1}{4})$, $(4.82843, -\frac{1}{4})$, $(4.82843, \frac{1}{2})$, $(3.41421, \frac{1}{8})$, $(3.41421, \frac{1}{8})$, $(3.41421, \frac{1}{8})$, $(3.41421, \frac{1}{8})$\}

\item $c = \frac{3}{2}$, $(d_i,\theta_i)$ = \{$(1., 0)$, $(5.82843, 0)$, $(2.41421, \frac{1}{4})$, $(2.41421, \frac{1}{4})$, $(1., \frac{1}{2})$, $(5.82843, \frac{1}{2})$, $(2.41421, -\frac{1}{4})$, $(2.41421, -\frac{1}{4})$, $(4.82843, \frac{3}{16})$, $(4.82843, \frac{3}{16})$, $(3.41421, -\frac{7}{16})$, $(3.41421, -\frac{3}{16})$, $(3.41421, -\frac{7}{16})$, $(3.41421, -\frac{3}{16})$\}

\item $c = 2$, $(d_i,\theta_i)$ = \{$(1., 0)$, $(5.82843, 0)$, $(2.41421, \frac{1}{4})$, $(2.41421, \frac{1}{4})$, $(1., \frac{1}{2})$, $(5.82843, \frac{1}{2})$, $(2.41421, -\frac{1}{4})$, $(2.41421, -\frac{1}{4})$, $(4.82843, -\frac{3}{8})$, $(4.82843, -\frac{1}{8})$, $(3.41421, \frac{1}{4})$, $(3.41421, \frac{1}{4})$, $(3.41421, \frac{1}{4})$, $(3.41421, \frac{1}{4})$\}

\item $c = \frac{5}{2}$, $(d_i,\theta_i)$ = \{$(1., 0)$, $(5.82843, 0)$, $(2.41421, \frac{1}{4})$, $(2.41421, \frac{1}{4})$, $(1., \frac{1}{2})$, $(5.82843, \frac{1}{2})$, $(2.41421, -\frac{1}{4})$, $(2.41421, -\frac{1}{4})$, $(4.82843, \frac{5}{16})$, $(4.82843, \frac{5}{16})$, $(3.41421, -\frac{5}{16})$, $(3.41421, -\frac{1}{16})$, $(3.41421, -\frac{5}{16})$, $(3.41421, -\frac{1}{16})$\}

\item $c = 3$, $(d_i,\theta_i)$ = \{$(1., 0)$, $(5.82843, 0)$, $(2.41421, \frac{1}{4})$, $(2.41421, \frac{1}{4})$, $(1., \frac{1}{2})$, $(5.82843, \frac{1}{2})$, $(2.41421, -\frac{1}{4})$, $(2.41421, -\frac{1}{4})$, $(4.82843, -\frac{1}{4})$, $(4.82843, 0)$, $(3.41421, \frac{3}{8})$, $(3.41421, \frac{3}{8})$, $(3.41421, \frac{3}{8})$, $(3.41421, \frac{3}{8})$\}

\item $c = \frac{7}{2}$, $(d_i,\theta_i)$ = \{$(1., 0)$, $(5.82843, 0)$, $(2.41421, \frac{1}{4})$, $(2.41421, \frac{1}{4})$, $(1., \frac{1}{2})$, $(5.82843, \frac{1}{2})$, $(2.41421, -\frac{1}{4})$, $(2.41421, -\frac{1}{4})$, $(4.82843, \frac{7}{16})$, $(4.82843, \frac{7}{16})$, $(3.41421, -\frac{3}{16})$, $(3.41421, \frac{1}{16})$, $(3.41421, -\frac{3}{16})$, $(3.41421, \frac{1}{16})$\}

\item $c = 4$, $(d_i,\theta_i)$ = \{$(1., 0)$, $(5.82843, 0)$, $(2.41421, \frac{1}{4})$, $(2.41421, \frac{1}{4})$, $(1., \frac{1}{2})$, $(5.82843, \frac{1}{2})$, $(2.41421, -\frac{1}{4})$, $(2.41421, -\frac{1}{4})$, $(4.82843, \frac{1}{8})$, $(4.82843, -\frac{1}{8})$, $(3.41421, \frac{1}{2})$, $(3.41421, \frac{1}{2})$, $(3.41421, \frac{1}{2})$, $(3.41421, \frac{1}{2})$\}

\item $c = \frac{9}{2}$, $(d_i,\theta_i)$ = \{$(1., 0)$, $(5.82843, 0)$, $(2.41421, \frac{1}{4})$, $(2.41421, \frac{1}{4})$, $(1., \frac{1}{2})$, $(5.82843, \frac{1}{2})$, $(2.41421, -\frac{1}{4})$, $(2.41421, -\frac{1}{4})$, $(4.82843, -\frac{7}{16})$, $(4.82843, -\frac{7}{16})$, $(3.41421, -\frac{1}{16})$, $(3.41421, \frac{3}{16})$, $(3.41421, -\frac{1}{16})$, $(3.41421, \frac{3}{16})$\}

\item $c = 5$, $(d_i,\theta_i)$ = \{$(1., 0)$, $(5.82843, 0)$, $(2.41421, \frac{1}{4})$, $(2.41421, \frac{1}{4})$, $(1., \frac{1}{2})$, $(5.82843, \frac{1}{2})$, $(2.41421, -\frac{1}{4})$, $(2.41421, -\frac{1}{4})$, $(4.82843, 0)$, $(4.82843, \frac{1}{4})$, $(3.41421, -\frac{3}{8})$, $(3.41421, -\frac{3}{8})$, $(3.41421, -\frac{3}{8})$, $(3.41421, -\frac{3}{8})$\}

\item $c = \frac{11}{2}$, $(d_i,\theta_i)$ = \{$(1., 0)$, $(5.82843, 0)$, $(2.41421, \frac{1}{4})$, $(2.41421, \frac{1}{4})$, $(1., \frac{1}{2})$, $(5.82843, \frac{1}{2})$, $(2.41421, -\frac{1}{4})$, $(2.41421, -\frac{1}{4})$, $(4.82843, -\frac{5}{16})$, $(4.82843, -\frac{5}{16})$, $(3.41421, \frac{1}{16})$, $(3.41421, \frac{5}{16})$, $(3.41421, \frac{1}{16})$, $(3.41421, \frac{5}{16})$\}

\item $c = 6$, $(d_i,\theta_i)$ = \{$(1., 0)$, $(5.82843, \frac{1}{2})$, $(2.41421, \frac{1}{4})$, $(2.41421, \frac{1}{4})$, $(1., \frac{1}{2})$, $(5.82843, 0)$, $(2.41421, -\frac{1}{4})$, $(2.41421, -\frac{1}{4})$, $(4.82843, \frac{3}{8})$, $(4.82843, \frac{1}{8})$, $(3.41421, -\frac{1}{4})$, $(3.41421, -\frac{1}{4})$, $(3.41421, -\frac{1}{4})$, $(3.41421, -\frac{1}{4})$\}

\item $c = \frac{13}{2}$, $(d_i,\theta_i)$ = \{$(1., 0)$, $(5.82843, \frac{1}{2})$, $(2.41421, \frac{1}{4})$, $(2.41421, \frac{1}{4})$, $(1., \frac{1}{2})$, $(5.82843, 0)$, $(2.41421, -\frac{1}{4})$, $(2.41421, -\frac{1}{4})$, $(4.82843, -\frac{3}{16})$, $(4.82843, -\frac{3}{16})$, $(3.41421, \frac{7}{16})$, $(3.41421, \frac{3}{16})$, $(3.41421, \frac{7}{16})$, $(3.41421, \frac{3}{16})$\}

\item $c = 7$, $(d_i,\theta_i)$ = \{$(1., 0)$, $(5.82843, \frac{1}{2})$, $(2.41421, \frac{1}{4})$, $(2.41421, \frac{1}{4})$, $(1., \frac{1}{2})$, $(5.82843, 0)$, $(2.41421, -\frac{1}{4})$, $(2.41421, -\frac{1}{4})$, $(4.82843, \frac{1}{2})$, $(4.82843, \frac{1}{4})$, $(3.41421, -\frac{1}{8})$, $(3.41421, -\frac{1}{8})$, $(3.41421, -\frac{1}{8})$, $(3.41421, -\frac{1}{8})$\}

\item $c = \frac{15}{2}$, $(d_i,\theta_i)$ = \{$(1., 0)$, $(5.82843, \frac{1}{2})$, $(2.41421, \frac{1}{4})$, $(2.41421, \frac{1}{4})$, $(1., \frac{1}{2})$, $(5.82843, 0)$, $(2.41421, -\frac{1}{4})$, $(2.41421, -\frac{1}{4})$, $(4.82843, -\frac{1}{16})$, $(4.82843, -\frac{1}{16})$, $(3.41421, -\frac{7}{16})$, $(3.41421, \frac{5}{16})$, $(3.41421, -\frac{7}{16})$, $(3.41421, \frac{5}{16})$\}

\end{enumerate}

\paragraph*{Rank 8; \#63}

\begin{enumerate}

\item $c = 0$, $(d_i,\theta_i)$ = \{$(1., 0)$, $(-1., 0)$, $(-0.414214, \frac{1}{4})$, $(2.41421, \frac{1}{4})$, $(1., \frac{1}{2})$, $(-1., \frac{1}{2})$, $(-0.414214, -\frac{1}{4})$, $(2.41421, -\frac{1}{4})$, $(-2., \frac{1}{8})$, $(2., -\frac{1}{8})$, $(1.41421, \frac{1}{4})$, $(-1.41421, -\frac{1}{4})$, $(1.41421, \frac{1}{4})$, $(-1.41421, -\frac{1}{4})$\}

\item $c = \frac{1}{2}$, $(d_i,\theta_i)$ = \{$(1., 0)$, $(-1., 0)$, $(-0.414214, \frac{1}{4})$, $(2.41421, \frac{1}{4})$, $(1., \frac{1}{2})$, $(-1., \frac{1}{2})$, $(-0.414214, -\frac{1}{4})$, $(2.41421, -\frac{1}{4})$, $(-2., \frac{5}{16})$, $(2., -\frac{3}{16})$, $(-1.41421, -\frac{1}{16})$, $(1.41421, \frac{3}{16})$, $(-1.41421, -\frac{1}{16})$, $(1.41421, \frac{3}{16})$\}

\item $c = 1$, $(d_i,\theta_i)$ = \{$(1., 0)$, $(-1., 0)$, $(-0.414214, \frac{1}{4})$, $(2.41421, \frac{1}{4})$, $(1., \frac{1}{2})$, $(-1., \frac{1}{2})$, $(-0.414214, -\frac{1}{4})$, $(2.41421, -\frac{1}{4})$, $(-2., \frac{1}{4})$, $(2., 0)$, $(-1.41421, -\frac{1}{8})$, $(1.41421, \frac{3}{8})$, $(-1.41421, -\frac{1}{8})$, $(1.41421, \frac{3}{8})$\}

\item $c = \frac{3}{2}$, $(d_i,\theta_i)$ = \{$(1., 0)$, $(-1., 0)$, $(-0.414214, \frac{1}{4})$, $(2.41421, \frac{1}{4})$, $(1., \frac{1}{2})$, $(-1., \frac{1}{2})$, $(-0.414214, -\frac{1}{4})$, $(2.41421, -\frac{1}{4})$, $(-2., \frac{7}{16})$, $(2., -\frac{1}{16})$, $(-1.41421, \frac{1}{16})$, $(1.41421, \frac{5}{16})$, $(-1.41421, \frac{1}{16})$, $(1.41421, \frac{5}{16})$\}

\item $c = 2$, $(d_i,\theta_i)$ = \{$(1., 0)$, $(-1., 0)$, $(-0.414214, \frac{1}{4})$, $(2.41421, \frac{1}{4})$, $(1., \frac{1}{2})$, $(-1., \frac{1}{2})$, $(-0.414214, -\frac{1}{4})$, $(2.41421, -\frac{1}{4})$, $(-2., \frac{1}{8})$, $(2., \frac{3}{8})$, $(1.41421, 0)$, $(-1.41421, \frac{1}{2})$, $(1.41421, 0)$, $(-1.41421, \frac{1}{2})$\}

\item $c = \frac{5}{2}$, $(d_i,\theta_i)$ = \{$(1., 0)$, $(-1., 0)$, $(-0.414214, \frac{1}{4})$, $(2.41421, \frac{1}{4})$, $(1., \frac{1}{2})$, $(-1., \frac{1}{2})$, $(-0.414214, -\frac{1}{4})$, $(2.41421, -\frac{1}{4})$, $(2., \frac{1}{16})$, $(-2., -\frac{7}{16})$, $(1.41421, \frac{7}{16})$, $(-1.41421, \frac{3}{16})$, $(1.41421, \frac{7}{16})$, $(-1.41421, \frac{3}{16})$\}

\item $c = 3$, $(d_i,\theta_i)$ = \{$(1., 0)$, $(-1., 0)$, $(-0.414214, \frac{1}{4})$, $(2.41421, \frac{1}{4})$, $(1., \frac{1}{2})$, $(-1., \frac{1}{2})$, $(-0.414214, -\frac{1}{4})$, $(2.41421, -\frac{1}{4})$, $(-2., \frac{1}{2})$, $(2., \frac{1}{4})$, $(-1.41421, \frac{1}{8})$, $(1.41421, -\frac{3}{8})$, $(-1.41421, \frac{1}{8})$, $(1.41421, -\frac{3}{8})$\}

\item $c = \frac{7}{2}$, $(d_i,\theta_i)$ = \{$(1., 0)$, $(-1., 0)$, $(-0.414214, \frac{1}{4})$, $(2.41421, \frac{1}{4})$, $(1., \frac{1}{2})$, $(-1., \frac{1}{2})$, $(-0.414214, -\frac{1}{4})$, $(2.41421, -\frac{1}{4})$, $(-2., -\frac{5}{16})$, $(2., \frac{3}{16})$, $(-1.41421, \frac{5}{16})$, $(1.41421, -\frac{7}{16})$, $(-1.41421, \frac{5}{16})$, $(1.41421, -\frac{7}{16})$\}

\item $c = 4$, $(d_i,\theta_i)$ = \{$(1., 0)$, $(-1., 0)$, $(-0.414214, \frac{1}{4})$, $(2.41421, \frac{1}{4})$, $(1., \frac{1}{2})$, $(-1., \frac{1}{2})$, $(-0.414214, -\frac{1}{4})$, $(2.41421, -\frac{1}{4})$, $(2., -\frac{3}{8})$, $(-2., \frac{3}{8})$, $(-1.41421, -\frac{1}{4})$, $(1.41421, \frac{1}{4})$, $(-1.41421, -\frac{1}{4})$, $(1.41421, \frac{1}{4})$\}

\item $c = \frac{9}{2}$, $(d_i,\theta_i)$ = \{$(1., 0)$, $(-1., 0)$, $(-0.414214, \frac{1}{4})$, $(2.41421, \frac{1}{4})$, $(1., \frac{1}{2})$, $(-1., \frac{1}{2})$, $(-0.414214, -\frac{1}{4})$, $(2.41421, -\frac{1}{4})$, $(-2., -\frac{3}{16})$, $(2., \frac{5}{16})$, $(1.41421, -\frac{5}{16})$, $(-1.41421, \frac{7}{16})$, $(1.41421, -\frac{5}{16})$, $(-1.41421, \frac{7}{16})$\}

\item $c = 5$, $(d_i,\theta_i)$ = \{$(1., 0)$, $(-1., 0)$, $(-0.414214, \frac{1}{4})$, $(2.41421, \frac{1}{4})$, $(1., \frac{1}{2})$, $(-1., \frac{1}{2})$, $(-0.414214, -\frac{1}{4})$, $(2.41421, -\frac{1}{4})$, $(2., \frac{1}{2})$, $(-2., -\frac{1}{4})$, $(1.41421, -\frac{1}{8})$, $(-1.41421, \frac{3}{8})$, $(1.41421, -\frac{1}{8})$, $(-1.41421, \frac{3}{8})$\}

\item $c = \frac{11}{2}$, $(d_i,\theta_i)$ = \{$(1., 0)$, $(-1., 0)$, $(-0.414214, \frac{1}{4})$, $(2.41421, \frac{1}{4})$, $(1., \frac{1}{2})$, $(-1., \frac{1}{2})$, $(-0.414214, -\frac{1}{4})$, $(2.41421, -\frac{1}{4})$, $(2., \frac{7}{16})$, $(-2., -\frac{1}{16})$, $(1.41421, -\frac{3}{16})$, $(-1.41421, -\frac{7}{16})$, $(1.41421, -\frac{3}{16})$, $(-1.41421, -\frac{7}{16})$\}

\item $c = 6$, $(d_i,\theta_i)$ = \{$(1., 0)$, $(-1., \frac{1}{2})$, $(-0.414214, \frac{1}{4})$, $(2.41421, \frac{1}{4})$, $(1., \frac{1}{2})$, $(-1., 0)$, $(-0.414214, -\frac{1}{4})$, $(2.41421, -\frac{1}{4})$, $(-2., -\frac{3}{8})$, $(2., -\frac{1}{8})$, $(1.41421, \frac{1}{2})$, $(-1.41421, 0)$, $(1.41421, \frac{1}{2})$, $(-1.41421, 0)$\}

\item $c = \frac{13}{2}$, $(d_i,\theta_i)$ = \{$(1., 0)$, $(-1., \frac{1}{2})$, $(-0.414214, \frac{1}{4})$, $(2.41421, \frac{1}{4})$, $(1., \frac{1}{2})$, $(-1., 0)$, $(-0.414214, -\frac{1}{4})$, $(2.41421, -\frac{1}{4})$, $(-2., \frac{1}{16})$, $(2., -\frac{7}{16})$, $(-1.41421, -\frac{5}{16})$, $(1.41421, -\frac{1}{16})$, $(-1.41421, -\frac{5}{16})$, $(1.41421, -\frac{1}{16})$\}

\item $c = 7$, $(d_i,\theta_i)$ = \{$(1., 0)$, $(-1., \frac{1}{2})$, $(-0.414214, \frac{1}{4})$, $(2.41421, \frac{1}{4})$, $(1., \frac{1}{2})$, $(-1., 0)$, $(-0.414214, -\frac{1}{4})$, $(2.41421, -\frac{1}{4})$, $(2., 0)$, $(-2., -\frac{1}{4})$, $(1.41421, -\frac{3}{8})$, $(-1.41421, \frac{1}{8})$, $(1.41421, -\frac{3}{8})$, $(-1.41421, \frac{1}{8})$\}

\item $c = \frac{15}{2}$, $(d_i,\theta_i)$ = \{$(1., 0)$, $(-1., \frac{1}{2})$, $(-0.414214, \frac{1}{4})$, $(2.41421, \frac{1}{4})$, $(1., \frac{1}{2})$, $(-1., 0)$, $(-0.414214, -\frac{1}{4})$, $(2.41421, -\frac{1}{4})$, $(2., -\frac{5}{16})$, $(-2., \frac{3}{16})$, $(-1.41421, -\frac{3}{16})$, $(1.41421, \frac{1}{16})$, $(-1.41421, -\frac{3}{16})$, $(1.41421, \frac{1}{16})$\}

\end{enumerate}

\paragraph*{Rank 8; \#64}

\begin{enumerate}

\item $c = 0$, $(d_i,\theta_i)$ = \{$(1., 0)$, $(0.171573, 0)$, $(-0.414214, \frac{1}{4})$, $(-0.414214, \frac{1}{4})$, $(1., \frac{1}{2})$, $(0.171573, \frac{1}{2})$, $(-0.414214, -\frac{1}{4})$, $(-0.414214, -\frac{1}{4})$, $(0.828427, -\frac{1}{8})$, $(0.828427, \frac{1}{8})$, $(-0.585786, 0)$, $(-0.585786, 0)$, $(-0.585786, 0)$, $(-0.585786, 0)$\}

\item $c = \frac{1}{2}$, $(d_i,\theta_i)$ = \{$(1., 0)$, $(0.171573, 0)$, $(-0.414214, \frac{1}{4})$, $(-0.414214, \frac{1}{4})$, $(1., \frac{1}{2})$, $(0.171573, \frac{1}{2})$, $(-0.414214, -\frac{1}{4})$, $(-0.414214, -\frac{1}{4})$, $(-0.828427, \frac{1}{16})$, $(-0.828427, \frac{1}{16})$, $(0.585786, -\frac{1}{16})$, $(0.585786, \frac{3}{16})$, $(0.585786, -\frac{1}{16})$, $(0.585786, \frac{3}{16})$\}

\item $c = 1$, $(d_i,\theta_i)$ = \{$(1., 0)$, $(0.171573, 0)$, $(-0.414214, \frac{1}{4})$, $(-0.414214, \frac{1}{4})$, $(1., \frac{1}{2})$, $(0.171573, \frac{1}{2})$, $(-0.414214, -\frac{1}{4})$, $(-0.414214, -\frac{1}{4})$, $(-0.828427, 0)$, $(-0.828427, \frac{1}{4})$, $(0.585786, \frac{1}{8})$, $(0.585786, \frac{1}{8})$, $(0.585786, \frac{1}{8})$, $(0.585786, \frac{1}{8})$\}

\item $c = \frac{3}{2}$, $(d_i,\theta_i)$ = \{$(1., 0)$, $(0.171573, 0)$, $(-0.414214, \frac{1}{4})$, $(-0.414214, \frac{1}{4})$, $(1., \frac{1}{2})$, $(0.171573, \frac{1}{2})$, $(-0.414214, -\frac{1}{4})$, $(-0.414214, -\frac{1}{4})$, $(-0.828427, \frac{3}{16})$, $(-0.828427, \frac{3}{16})$, $(0.585786, \frac{5}{16})$, $(0.585786, \frac{1}{16})$, $(0.585786, \frac{5}{16})$, $(0.585786, \frac{1}{16})$\}

\item $c = 2$, $(d_i,\theta_i)$ = \{$(1., 0)$, $(0.171573, 0)$, $(-0.414214, \frac{1}{4})$, $(-0.414214, \frac{1}{4})$, $(1., \frac{1}{2})$, $(0.171573, \frac{1}{2})$, $(-0.414214, -\frac{1}{4})$, $(-0.414214, -\frac{1}{4})$, $(-0.828427, \frac{3}{8})$, $(-0.828427, \frac{1}{8})$, $(0.585786, \frac{1}{4})$, $(0.585786, \frac{1}{4})$, $(0.585786, \frac{1}{4})$, $(0.585786, \frac{1}{4})$\}

\item $c = \frac{5}{2}$, $(d_i,\theta_i)$ = \{$(1., 0)$, $(0.171573, 0)$, $(-0.414214, \frac{1}{4})$, $(-0.414214, \frac{1}{4})$, $(1., \frac{1}{2})$, $(0.171573, \frac{1}{2})$, $(-0.414214, -\frac{1}{4})$, $(-0.414214, -\frac{1}{4})$, $(-0.828427, \frac{5}{16})$, $(-0.828427, \frac{5}{16})$, $(0.585786, \frac{3}{16})$, $(0.585786, \frac{7}{16})$, $(0.585786, \frac{3}{16})$, $(0.585786, \frac{7}{16})$\}

\item $c = 3$, $(d_i,\theta_i)$ = \{$(1., 0)$, $(0.171573, 0)$, $(-0.414214, \frac{1}{4})$, $(-0.414214, \frac{1}{4})$, $(1., \frac{1}{2})$, $(0.171573, \frac{1}{2})$, $(-0.414214, -\frac{1}{4})$, $(-0.414214, -\frac{1}{4})$, $(-0.828427, \frac{1}{4})$, $(-0.828427, \frac{1}{2})$, $(0.585786, \frac{3}{8})$, $(0.585786, \frac{3}{8})$, $(0.585786, \frac{3}{8})$, $(0.585786, \frac{3}{8})$\}

\item $c = \frac{7}{2}$, $(d_i,\theta_i)$ = \{$(1., 0)$, $(0.171573, 0)$, $(-0.414214, \frac{1}{4})$, $(-0.414214, \frac{1}{4})$, $(1., \frac{1}{2})$, $(0.171573, \frac{1}{2})$, $(-0.414214, -\frac{1}{4})$, $(-0.414214, -\frac{1}{4})$, $(-0.828427, \frac{7}{16})$, $(-0.828427, \frac{7}{16})$, $(0.585786, -\frac{7}{16})$, $(0.585786, \frac{5}{16})$, $(0.585786, -\frac{7}{16})$, $(0.585786, \frac{5}{16})$\}

\item $c = 4$, $(d_i,\theta_i)$ = \{$(1., 0)$, $(0.171573, 0)$, $(-0.414214, \frac{1}{4})$, $(-0.414214, \frac{1}{4})$, $(1., \frac{1}{2})$, $(0.171573, \frac{1}{2})$, $(-0.414214, -\frac{1}{4})$, $(-0.414214, -\frac{1}{4})$, $(-0.828427, \frac{3}{8})$, $(-0.828427, -\frac{3}{8})$, $(0.585786, \frac{1}{2})$, $(0.585786, \frac{1}{2})$, $(0.585786, \frac{1}{2})$, $(0.585786, \frac{1}{2})$\}

\item $c = \frac{9}{2}$, $(d_i,\theta_i)$ = \{$(1., 0)$, $(0.171573, 0)$, $(-0.414214, \frac{1}{4})$, $(-0.414214, \frac{1}{4})$, $(1., \frac{1}{2})$, $(0.171573, \frac{1}{2})$, $(-0.414214, -\frac{1}{4})$, $(-0.414214, -\frac{1}{4})$, $(-0.828427, -\frac{7}{16})$, $(-0.828427, -\frac{7}{16})$, $(0.585786, -\frac{5}{16})$, $(0.585786, \frac{7}{16})$, $(0.585786, -\frac{5}{16})$, $(0.585786, \frac{7}{16})$\}

\item $c = 5$, $(d_i,\theta_i)$ = \{$(1., 0)$, $(0.171573, 0)$, $(-0.414214, \frac{1}{4})$, $(-0.414214, \frac{1}{4})$, $(1., \frac{1}{2})$, $(0.171573, \frac{1}{2})$, $(-0.414214, -\frac{1}{4})$, $(-0.414214, -\frac{1}{4})$, $(-0.828427, -\frac{1}{4})$, $(-0.828427, \frac{1}{2})$, $(0.585786, -\frac{3}{8})$, $(0.585786, -\frac{3}{8})$, $(0.585786, -\frac{3}{8})$, $(0.585786, -\frac{3}{8})$\}

\item $c = \frac{11}{2}$, $(d_i,\theta_i)$ = \{$(1., 0)$, $(0.171573, 0)$, $(-0.414214, \frac{1}{4})$, $(-0.414214, \frac{1}{4})$, $(1., \frac{1}{2})$, $(0.171573, \frac{1}{2})$, $(-0.414214, -\frac{1}{4})$, $(-0.414214, -\frac{1}{4})$, $(-0.828427, -\frac{5}{16})$, $(-0.828427, -\frac{5}{16})$, $(0.585786, -\frac{7}{16})$, $(0.585786, -\frac{3}{16})$, $(0.585786, -\frac{7}{16})$, $(0.585786, -\frac{3}{16})$\}

\item $c = 6$, $(d_i,\theta_i)$ = \{$(1., 0)$, $(0.171573, \frac{1}{2})$, $(-0.414214, \frac{1}{4})$, $(-0.414214, \frac{1}{4})$, $(1., \frac{1}{2})$, $(0.171573, 0)$, $(-0.414214, -\frac{1}{4})$, $(-0.414214, -\frac{1}{4})$, $(-0.828427, -\frac{1}{8})$, $(-0.828427, -\frac{3}{8})$, $(0.585786, -\frac{1}{4})$, $(0.585786, -\frac{1}{4})$, $(0.585786, -\frac{1}{4})$, $(0.585786, -\frac{1}{4})$\}

\item $c = \frac{13}{2}$, $(d_i,\theta_i)$ = \{$(1., 0)$, $(0.171573, \frac{1}{2})$, $(-0.414214, \frac{1}{4})$, $(-0.414214, \frac{1}{4})$, $(1., \frac{1}{2})$, $(0.171573, 0)$, $(-0.414214, -\frac{1}{4})$, $(-0.414214, -\frac{1}{4})$, $(0.828427, -\frac{3}{16})$, $(0.828427, -\frac{3}{16})$, $(-0.585786, -\frac{5}{16})$, $(-0.585786, -\frac{1}{16})$, $(-0.585786, -\frac{5}{16})$, $(-0.585786, -\frac{1}{16})$\}

\item $c = 7$, $(d_i,\theta_i)$ = \{$(1., 0)$, $(0.171573, \frac{1}{2})$, $(-0.414214, \frac{1}{4})$, $(-0.414214, \frac{1}{4})$, $(1., \frac{1}{2})$, $(0.171573, 0)$, $(-0.414214, -\frac{1}{4})$, $(-0.414214, -\frac{1}{4})$, $(-0.828427, -\frac{1}{4})$, $(-0.828427, 0)$, $(0.585786, -\frac{1}{8})$, $(0.585786, -\frac{1}{8})$, $(0.585786, -\frac{1}{8})$, $(0.585786, -\frac{1}{8})$\}

\item $c = \frac{15}{2}$, $(d_i,\theta_i)$ = \{$(1., 0)$, $(0.171573, \frac{1}{2})$, $(-0.414214, \frac{1}{4})$, $(-0.414214, \frac{1}{4})$, $(1., \frac{1}{2})$, $(0.171573, 0)$, $(-0.414214, -\frac{1}{4})$, $(-0.414214, -\frac{1}{4})$, $(0.828427, -\frac{1}{16})$, $(0.828427, -\frac{1}{16})$, $(-0.585786, -\frac{3}{16})$, $(-0.585786, \frac{1}{16})$, $(-0.585786, -\frac{3}{16})$, $(-0.585786, \frac{1}{16})$\}

\end{enumerate}

\paragraph*{Rank 8; \#65}

\begin{enumerate}

\item $c = \frac{3}{8}$, $(d_i,\theta_i)$ = \{$(1., 0)$, $(4.26197, \frac{1}{8})$, $(5.02734, \frac{1}{4})$, $(2.84776, -\frac{1}{8})$, $(1., \frac{1}{2})$, $(4.26197, -\frac{3}{8})$, $(5.02734, -\frac{1}{4})$, $(2.84776, \frac{3}{8})$, $(5.12583, \frac{25}{64})$, $(1.96157, \frac{21}{64})$, $(4.73565, -\frac{11}{64})$, $(3.62451, \frac{9}{64})$, $(1.96157, \frac{21}{64})$, $(4.73565, -\frac{11}{64})$, $(3.62451, \frac{9}{64})$\}

\item $c = \frac{7}{8}$, $(d_i,\theta_i)$ = \{$(1., 0)$, $(4.26197, \frac{1}{8})$, $(5.02734, \frac{1}{4})$, $(2.84776, -\frac{1}{8})$, $(1., \frac{1}{2})$, $(4.26197, -\frac{3}{8})$, $(5.02734, -\frac{1}{4})$, $(2.84776, \frac{3}{8})$, $(2.77408, \frac{25}{64})$, $(6.69722, -\frac{7}{64})$, $(5.12583, \frac{13}{64})$, $(3.62451, \frac{29}{64})$, $(3.62451, \frac{29}{64})$\}

\item $c = \frac{11}{8}$, $(d_i,\theta_i)$ = \{$(1., 0)$, $(4.26197, \frac{1}{8})$, $(5.02734, \frac{1}{4})$, $(2.84776, -\frac{1}{8})$, $(1., \frac{1}{2})$, $(4.26197, -\frac{3}{8})$, $(5.02734, -\frac{1}{4})$, $(2.84776, \frac{3}{8})$, $(5.12583, -\frac{31}{64})$, $(1.96157, \frac{29}{64})$, $(4.73565, -\frac{3}{64})$, $(3.62451, \frac{17}{64})$, $(1.96157, \frac{29}{64})$, $(4.73565, -\frac{3}{64})$, $(3.62451, \frac{17}{64})$\}

\item $c = \frac{15}{8}$, $(d_i,\theta_i)$ = \{$(1., 0)$, $(4.26197, \frac{1}{8})$, $(5.02734, \frac{1}{4})$, $(2.84776, -\frac{1}{8})$, $(1., \frac{1}{2})$, $(4.26197, -\frac{3}{8})$, $(5.02734, -\frac{1}{4})$, $(2.84776, \frac{3}{8})$, $(2.77408, -\frac{31}{64})$, $(6.69722, \frac{1}{64})$, $(5.12583, \frac{21}{64})$, $(3.62451, -\frac{27}{64})$, $(3.62451, -\frac{27}{64})$\}

\item $c = \frac{19}{8}$, $(d_i,\theta_i)$ = \{$(1., 0)$, $(4.26197, \frac{1}{8})$, $(5.02734, \frac{1}{4})$, $(2.84776, -\frac{1}{8})$, $(1., \frac{1}{2})$, $(4.26197, -\frac{3}{8})$, $(5.02734, -\frac{1}{4})$, $(2.84776, \frac{3}{8})$, $(5.12583, -\frac{23}{64})$, $(4.73565, \frac{5}{64})$, $(1.96157, -\frac{27}{64})$, $(3.62451, \frac{25}{64})$, $(4.73565, \frac{5}{64})$, $(1.96157, -\frac{27}{64})$, $(3.62451, \frac{25}{64})$\}

\item $c = \frac{23}{8}$, $(d_i,\theta_i)$ = \{$(1., 0)$, $(4.26197, \frac{1}{8})$, $(5.02734, \frac{1}{4})$, $(2.84776, -\frac{1}{8})$, $(1., \frac{1}{2})$, $(4.26197, -\frac{3}{8})$, $(5.02734, -\frac{1}{4})$, $(2.84776, \frac{3}{8})$, $(2.77408, -\frac{23}{64})$, $(6.69722, \frac{9}{64})$, $(5.12583, \frac{29}{64})$, $(3.62451, -\frac{19}{64})$, $(3.62451, -\frac{19}{64})$\}

\item $c = \frac{27}{8}$, $(d_i,\theta_i)$ = \{$(1., 0)$, $(4.26197, \frac{1}{8})$, $(5.02734, \frac{1}{4})$, $(2.84776, \frac{3}{8})$, $(1., \frac{1}{2})$, $(4.26197, -\frac{3}{8})$, $(5.02734, -\frac{1}{4})$, $(2.84776, -\frac{1}{8})$, $(5.12583, -\frac{15}{64})$, $(4.73565, \frac{13}{64})$, $(1.96157, -\frac{19}{64})$, $(3.62451, -\frac{31}{64})$, $(4.73565, \frac{13}{64})$, $(1.96157, -\frac{19}{64})$, $(3.62451, -\frac{31}{64})$\}

\item $c = \frac{31}{8}$, $(d_i,\theta_i)$ = \{$(1., 0)$, $(4.26197, \frac{1}{8})$, $(5.02734, \frac{1}{4})$, $(2.84776, \frac{3}{8})$, $(1., \frac{1}{2})$, $(4.26197, -\frac{3}{8})$, $(5.02734, -\frac{1}{4})$, $(2.84776, -\frac{1}{8})$, $(6.69722, \frac{17}{64})$, $(2.77408, -\frac{15}{64})$, $(5.12583, -\frac{27}{64})$, $(3.62451, -\frac{11}{64})$, $(3.62451, -\frac{11}{64})$\}

\item $c = \frac{35}{8}$, $(d_i,\theta_i)$ = \{$(1., 0)$, $(4.26197, \frac{1}{8})$, $(5.02734, \frac{1}{4})$, $(2.84776, \frac{3}{8})$, $(1., \frac{1}{2})$, $(4.26197, -\frac{3}{8})$, $(5.02734, -\frac{1}{4})$, $(2.84776, -\frac{1}{8})$, $(5.12583, -\frac{7}{64})$, $(1.96157, -\frac{11}{64})$, $(4.73565, \frac{21}{64})$, $(3.62451, -\frac{23}{64})$, $(1.96157, -\frac{11}{64})$, $(4.73565, \frac{21}{64})$, $(3.62451, -\frac{23}{64})$\}

\item $c = \frac{39}{8}$, $(d_i,\theta_i)$ = \{$(1., 0)$, $(4.26197, \frac{1}{8})$, $(5.02734, \frac{1}{4})$, $(2.84776, \frac{3}{8})$, $(1., \frac{1}{2})$, $(4.26197, -\frac{3}{8})$, $(5.02734, -\frac{1}{4})$, $(2.84776, -\frac{1}{8})$, $(2.77408, -\frac{7}{64})$, $(6.69722, \frac{25}{64})$, $(5.12583, -\frac{19}{64})$, $(3.62451, -\frac{3}{64})$, $(3.62451, -\frac{3}{64})$\}

\item $c = \frac{43}{8}$, $(d_i,\theta_i)$ = \{$(1., 0)$, $(4.26197, \frac{1}{8})$, $(5.02734, \frac{1}{4})$, $(2.84776, \frac{3}{8})$, $(1., \frac{1}{2})$, $(4.26197, -\frac{3}{8})$, $(5.02734, -\frac{1}{4})$, $(2.84776, -\frac{1}{8})$, $(5.12583, \frac{1}{64})$, $(1.96157, -\frac{3}{64})$, $(4.73565, \frac{29}{64})$, $(3.62451, -\frac{15}{64})$, $(1.96157, -\frac{3}{64})$, $(4.73565, \frac{29}{64})$, $(3.62451, -\frac{15}{64})$\}

\item $c = \frac{47}{8}$, $(d_i,\theta_i)$ = \{$(1., 0)$, $(4.26197, \frac{1}{8})$, $(5.02734, \frac{1}{4})$, $(2.84776, \frac{3}{8})$, $(1., \frac{1}{2})$, $(4.26197, -\frac{3}{8})$, $(5.02734, -\frac{1}{4})$, $(2.84776, -\frac{1}{8})$, $(2.77408, \frac{1}{64})$, $(6.69722, -\frac{31}{64})$, $(5.12583, -\frac{11}{64})$, $(3.62451, \frac{5}{64})$, $(3.62451, \frac{5}{64})$\}

\item $c = \frac{51}{8}$, $(d_i,\theta_i)$ = \{$(1., 0)$, $(4.26197, \frac{1}{8})$, $(5.02734, \frac{1}{4})$, $(2.84776, \frac{3}{8})$, $(1., \frac{1}{2})$, $(4.26197, -\frac{3}{8})$, $(5.02734, -\frac{1}{4})$, $(2.84776, -\frac{1}{8})$, $(5.12583, \frac{9}{64})$, $(4.73565, -\frac{27}{64})$, $(1.96157, \frac{5}{64})$, $(3.62451, -\frac{7}{64})$, $(4.73565, -\frac{27}{64})$, $(1.96157, \frac{5}{64})$, $(3.62451, -\frac{7}{64})$\}

\item $c = \frac{55}{8}$, $(d_i,\theta_i)$ = \{$(1., 0)$, $(4.26197, \frac{1}{8})$, $(5.02734, \frac{1}{4})$, $(2.84776, \frac{3}{8})$, $(1., \frac{1}{2})$, $(4.26197, -\frac{3}{8})$, $(5.02734, -\frac{1}{4})$, $(2.84776, -\frac{1}{8})$, $(2.77408, \frac{9}{64})$, $(6.69722, -\frac{23}{64})$, $(5.12583, -\frac{3}{64})$, $(3.62451, \frac{13}{64})$, $(3.62451, \frac{13}{64})$\}

\item $c = \frac{59}{8}$, $(d_i,\theta_i)$ = \{$(1., 0)$, $(4.26197, \frac{1}{8})$, $(5.02734, \frac{1}{4})$, $(2.84776, \frac{3}{8})$, $(1., \frac{1}{2})$, $(4.26197, -\frac{3}{8})$, $(5.02734, -\frac{1}{4})$, $(2.84776, -\frac{1}{8})$, $(5.12583, \frac{17}{64})$, $(4.73565, -\frac{19}{64})$, $(1.96157, \frac{13}{64})$, $(3.62451, \frac{1}{64})$, $(4.73565, -\frac{19}{64})$, $(1.96157, \frac{13}{64})$, $(3.62451, \frac{1}{64})$\}

\item $c = \frac{63}{8}$, $(d_i,\theta_i)$ = \{$(1., 0)$, $(4.26197, \frac{1}{8})$, $(5.02734, \frac{1}{4})$, $(2.84776, \frac{3}{8})$, $(1., \frac{1}{2})$, $(4.26197, -\frac{3}{8})$, $(5.02734, -\frac{1}{4})$, $(2.84776, -\frac{1}{8})$, $(2.77408, \frac{17}{64})$, $(6.69722, -\frac{15}{64})$, $(5.12583, \frac{5}{64})$, $(3.62451, \frac{21}{64})$, $(3.62451, \frac{21}{64})$\}

\end{enumerate}

\paragraph*{Rank 8; \#66}

\begin{enumerate}

\item $c = \frac{1}{8}$, $(d_i,\theta_i)$ = \{$(1., 0)$, $(2.84776, \frac{1}{8})$, $(5.02734, \frac{1}{4})$, $(4.26197, -\frac{1}{8})$, $(1., \frac{1}{2})$, $(2.84776, -\frac{3}{8})$, $(5.02734, -\frac{1}{4})$, $(4.26197, \frac{3}{8})$, $(2.77408, -\frac{17}{64})$, $(6.69722, \frac{15}{64})$, $(5.12583, -\frac{5}{64})$, $(3.62451, -\frac{21}{64})$, $(3.62451, -\frac{21}{64})$\}

\item $c = \frac{5}{8}$, $(d_i,\theta_i)$ = \{$(1., 0)$, $(2.84776, \frac{1}{8})$, $(5.02734, \frac{1}{4})$, $(4.26197, -\frac{1}{8})$, $(1., \frac{1}{2})$, $(2.84776, -\frac{3}{8})$, $(5.02734, -\frac{1}{4})$, $(4.26197, \frac{3}{8})$, $(5.12583, -\frac{17}{64})$, $(4.73565, \frac{19}{64})$, $(1.96157, -\frac{13}{64})$, $(3.62451, -\frac{1}{64})$, $(4.73565, \frac{19}{64})$, $(1.96157, -\frac{13}{64})$, $(3.62451, -\frac{1}{64})$\}

\item $c = \frac{9}{8}$, $(d_i,\theta_i)$ = \{$(1., 0)$, $(2.84776, \frac{1}{8})$, $(5.02734, \frac{1}{4})$, $(4.26197, -\frac{1}{8})$, $(1., \frac{1}{2})$, $(2.84776, -\frac{3}{8})$, $(5.02734, -\frac{1}{4})$, $(4.26197, \frac{3}{8})$, $(2.77408, -\frac{9}{64})$, $(6.69722, \frac{23}{64})$, $(5.12583, \frac{3}{64})$, $(3.62451, -\frac{13}{64})$, $(3.62451, -\frac{13}{64})$\}

\item $c = \frac{13}{8}$, $(d_i,\theta_i)$ = \{$(1., 0)$, $(2.84776, \frac{1}{8})$, $(5.02734, \frac{1}{4})$, $(4.26197, -\frac{1}{8})$, $(1., \frac{1}{2})$, $(2.84776, -\frac{3}{8})$, $(5.02734, -\frac{1}{4})$, $(4.26197, \frac{3}{8})$, $(5.12583, -\frac{9}{64})$, $(1.96157, -\frac{5}{64})$, $(4.73565, \frac{27}{64})$, $(3.62451, \frac{7}{64})$, $(1.96157, -\frac{5}{64})$, $(4.73565, \frac{27}{64})$, $(3.62451, \frac{7}{64})$\}

\item $c = \frac{17}{8}$, $(d_i,\theta_i)$ = \{$(1., 0)$, $(2.84776, \frac{1}{8})$, $(5.02734, \frac{1}{4})$, $(4.26197, -\frac{1}{8})$, $(1., \frac{1}{2})$, $(2.84776, -\frac{3}{8})$, $(5.02734, -\frac{1}{4})$, $(4.26197, \frac{3}{8})$, $(2.77408, -\frac{1}{64})$, $(6.69722, \frac{31}{64})$, $(5.12583, \frac{11}{64})$, $(3.62451, -\frac{5}{64})$, $(3.62451, -\frac{5}{64})$\}

\item $c = \frac{21}{8}$, $(d_i,\theta_i)$ = \{$(1., 0)$, $(2.84776, \frac{1}{8})$, $(5.02734, \frac{1}{4})$, $(4.26197, -\frac{1}{8})$, $(1., \frac{1}{2})$, $(2.84776, -\frac{3}{8})$, $(5.02734, -\frac{1}{4})$, $(4.26197, \frac{3}{8})$, $(5.12583, -\frac{1}{64})$, $(1.96157, \frac{3}{64})$, $(4.73565, -\frac{29}{64})$, $(3.62451, \frac{15}{64})$, $(1.96157, \frac{3}{64})$, $(4.73565, -\frac{29}{64})$, $(3.62451, \frac{15}{64})$\}

\item $c = \frac{25}{8}$, $(d_i,\theta_i)$ = \{$(1., 0)$, $(2.84776, \frac{1}{8})$, $(5.02734, \frac{1}{4})$, $(4.26197, \frac{3}{8})$, $(1., \frac{1}{2})$, $(2.84776, -\frac{3}{8})$, $(5.02734, -\frac{1}{4})$, $(4.26197, -\frac{1}{8})$, $(6.69722, -\frac{25}{64})$, $(2.77408, \frac{7}{64})$, $(5.12583, \frac{19}{64})$, $(3.62451, \frac{3}{64})$, $(3.62451, \frac{3}{64})$\}

\item $c = \frac{29}{8}$, $(d_i,\theta_i)$ = \{$(1., 0)$, $(2.84776, \frac{1}{8})$, $(5.02734, \frac{1}{4})$, $(4.26197, \frac{3}{8})$, $(1., \frac{1}{2})$, $(2.84776, -\frac{3}{8})$, $(5.02734, -\frac{1}{4})$, $(4.26197, -\frac{1}{8})$, $(5.12583, \frac{7}{64})$, $(4.73565, -\frac{21}{64})$, $(1.96157, \frac{11}{64})$, $(3.62451, \frac{23}{64})$, $(4.73565, -\frac{21}{64})$, $(1.96157, \frac{11}{64})$, $(3.62451, \frac{23}{64})$\}

\item $c = \frac{33}{8}$, $(d_i,\theta_i)$ = \{$(1., 0)$, $(2.84776, \frac{1}{8})$, $(5.02734, \frac{1}{4})$, $(4.26197, \frac{3}{8})$, $(1., \frac{1}{2})$, $(2.84776, -\frac{3}{8})$, $(5.02734, -\frac{1}{4})$, $(4.26197, -\frac{1}{8})$, $(6.69722, -\frac{17}{64})$, $(2.77408, \frac{15}{64})$, $(5.12583, \frac{27}{64})$, $(3.62451, \frac{11}{64})$, $(3.62451, \frac{11}{64})$\}

\item $c = \frac{37}{8}$, $(d_i,\theta_i)$ = \{$(1., 0)$, $(2.84776, \frac{1}{8})$, $(5.02734, \frac{1}{4})$, $(4.26197, \frac{3}{8})$, $(1., \frac{1}{2})$, $(2.84776, -\frac{3}{8})$, $(5.02734, -\frac{1}{4})$, $(4.26197, -\frac{1}{8})$, $(5.12583, \frac{15}{64})$, $(4.73565, -\frac{13}{64})$, $(1.96157, \frac{19}{64})$, $(3.62451, \frac{31}{64})$, $(4.73565, -\frac{13}{64})$, $(1.96157, \frac{19}{64})$, $(3.62451, \frac{31}{64})$\}

\item $c = \frac{41}{8}$, $(d_i,\theta_i)$ = \{$(1., 0)$, $(2.84776, \frac{1}{8})$, $(5.02734, \frac{1}{4})$, $(4.26197, \frac{3}{8})$, $(1., \frac{1}{2})$, $(2.84776, -\frac{3}{8})$, $(5.02734, -\frac{1}{4})$, $(4.26197, -\frac{1}{8})$, $(6.69722, -\frac{9}{64})$, $(2.77408, \frac{23}{64})$, $(5.12583, -\frac{29}{64})$, $(3.62451, \frac{19}{64})$, $(3.62451, \frac{19}{64})$\}

\item $c = \frac{45}{8}$, $(d_i,\theta_i)$ = \{$(1., 0)$, $(2.84776, \frac{1}{8})$, $(5.02734, \frac{1}{4})$, $(4.26197, \frac{3}{8})$, $(1., \frac{1}{2})$, $(2.84776, -\frac{3}{8})$, $(5.02734, -\frac{1}{4})$, $(4.26197, -\frac{1}{8})$, $(5.12583, \frac{23}{64})$, $(1.96157, \frac{27}{64})$, $(4.73565, -\frac{5}{64})$, $(3.62451, -\frac{25}{64})$, $(1.96157, \frac{27}{64})$, $(4.73565, -\frac{5}{64})$, $(3.62451, -\frac{25}{64})$\}

\item $c = \frac{49}{8}$, $(d_i,\theta_i)$ = \{$(1., 0)$, $(2.84776, \frac{1}{8})$, $(5.02734, \frac{1}{4})$, $(4.26197, \frac{3}{8})$, $(1., \frac{1}{2})$, $(2.84776, -\frac{3}{8})$, $(5.02734, -\frac{1}{4})$, $(4.26197, -\frac{1}{8})$, $(6.69722, -\frac{1}{64})$, $(2.77408, \frac{31}{64})$, $(5.12583, -\frac{21}{64})$, $(3.62451, \frac{27}{64})$, $(3.62451, \frac{27}{64})$\}

\item $c = \frac{53}{8}$, $(d_i,\theta_i)$ = \{$(1., 0)$, $(2.84776, \frac{1}{8})$, $(5.02734, \frac{1}{4})$, $(4.26197, \frac{3}{8})$, $(1., \frac{1}{2})$, $(2.84776, -\frac{3}{8})$, $(5.02734, -\frac{1}{4})$, $(4.26197, -\frac{1}{8})$, $(5.12583, \frac{31}{64})$, $(1.96157, -\frac{29}{64})$, $(4.73565, \frac{3}{64})$, $(3.62451, -\frac{17}{64})$, $(1.96157, -\frac{29}{64})$, $(4.73565, \frac{3}{64})$, $(3.62451, -\frac{17}{64})$\}

\item $c = \frac{57}{8}$, $(d_i,\theta_i)$ = \{$(1., 0)$, $(2.84776, \frac{1}{8})$, $(5.02734, \frac{1}{4})$, $(4.26197, \frac{3}{8})$, $(1., \frac{1}{2})$, $(2.84776, -\frac{3}{8})$, $(5.02734, -\frac{1}{4})$, $(4.26197, -\frac{1}{8})$, $(2.77408, -\frac{25}{64})$, $(6.69722, \frac{7}{64})$, $(5.12583, -\frac{13}{64})$, $(3.62451, -\frac{29}{64})$, $(3.62451, -\frac{29}{64})$\}

\item $c = \frac{61}{8}$, $(d_i,\theta_i)$ = \{$(1., 0)$, $(2.84776, \frac{1}{8})$, $(5.02734, \frac{1}{4})$, $(4.26197, \frac{3}{8})$, $(1., \frac{1}{2})$, $(2.84776, -\frac{3}{8})$, $(5.02734, -\frac{1}{4})$, $(4.26197, -\frac{1}{8})$, $(5.12583, -\frac{25}{64})$, $(4.73565, \frac{11}{64})$, $(1.96157, -\frac{21}{64})$, $(3.62451, -\frac{9}{64})$, $(4.73565, \frac{11}{64})$, $(1.96157, -\frac{21}{64})$, $(3.62451, -\frac{9}{64})$\}

\end{enumerate}

\paragraph*{Rank 8; \#67}

\begin{enumerate}

\item $c = \frac{1}{8}$, $(d_i,\theta_i)$ = \{$(1., 0)$, $(1.76537, \frac{1}{8})$, $(-1.49661, \frac{1}{4})$, $(0.351153, -\frac{1}{8})$, $(1., \frac{1}{2})$, $(1.76537, -\frac{3}{8})$, $(-1.49661, -\frac{1}{4})$, $(0.351153, \frac{3}{8})$, $(1.79995, \frac{3}{64})$, $(-1.66294, -\frac{9}{64})$, $(0.688812, \frac{23}{64})$, $(-1.27276, \frac{19}{64})$, $(-1.66294, -\frac{9}{64})$, $(0.688812, \frac{23}{64})$, $(-1.27276, \frac{19}{64})$\}

\item $c = \frac{5}{8}$, $(d_i,\theta_i)$ = \{$(1., 0)$, $(1.76537, \frac{1}{8})$, $(-1.49661, \frac{1}{4})$, $(0.351153, -\frac{1}{8})$, $(1., \frac{1}{2})$, $(1.76537, -\frac{3}{8})$, $(-1.49661, -\frac{1}{4})$, $(0.351153, \frac{3}{8})$, $(-1.79995, \frac{23}{64})$, $(0.974127, \frac{27}{64})$, $(-2.35175, -\frac{5}{64})$, $(1.27276, \frac{7}{64})$, $(1.27276, \frac{7}{64})$\}

\item $c = \frac{9}{8}$, $(d_i,\theta_i)$ = \{$(1., 0)$, $(1.76537, \frac{1}{8})$, $(-1.49661, \frac{1}{4})$, $(0.351153, -\frac{1}{8})$, $(1., \frac{1}{2})$, $(1.76537, -\frac{3}{8})$, $(-1.49661, -\frac{1}{4})$, $(0.351153, \frac{3}{8})$, $(1.79995, \frac{11}{64})$, $(-1.66294, -\frac{1}{64})$, $(0.688812, \frac{31}{64})$, $(-1.27276, \frac{27}{64})$, $(-1.66294, -\frac{1}{64})$, $(0.688812, \frac{31}{64})$, $(-1.27276, \frac{27}{64})$\}

\item $c = \frac{13}{8}$, $(d_i,\theta_i)$ = \{$(1., 0)$, $(1.76537, \frac{1}{8})$, $(-1.49661, \frac{1}{4})$, $(0.351153, -\frac{1}{8})$, $(1., \frac{1}{2})$, $(1.76537, -\frac{3}{8})$, $(-1.49661, -\frac{1}{4})$, $(0.351153, \frac{3}{8})$, $(-1.79995, \frac{31}{64})$, $(0.974127, -\frac{29}{64})$, $(-2.35175, \frac{3}{64})$, $(1.27276, \frac{15}{64})$, $(1.27276, \frac{15}{64})$\}

\item $c = \frac{17}{8}$, $(d_i,\theta_i)$ = \{$(1., 0)$, $(1.76537, \frac{1}{8})$, $(-1.49661, \frac{1}{4})$, $(0.351153, -\frac{1}{8})$, $(1., \frac{1}{2})$, $(1.76537, -\frac{3}{8})$, $(-1.49661, -\frac{1}{4})$, $(0.351153, \frac{3}{8})$, $(1.79995, \frac{19}{64})$, $(0.688812, -\frac{25}{64})$, $(-1.66294, \frac{7}{64})$, $(-1.27276, -\frac{29}{64})$, $(0.688812, -\frac{25}{64})$, $(-1.66294, \frac{7}{64})$, $(-1.27276, -\frac{29}{64})$\}

\item $c = \frac{21}{8}$, $(d_i,\theta_i)$ = \{$(1., 0)$, $(1.76537, \frac{1}{8})$, $(-1.49661, \frac{1}{4})$, $(0.351153, -\frac{1}{8})$, $(1., \frac{1}{2})$, $(1.76537, -\frac{3}{8})$, $(-1.49661, -\frac{1}{4})$, $(0.351153, \frac{3}{8})$, $(1.79995, -\frac{25}{64})$, $(2.35175, \frac{11}{64})$, $(-0.974127, -\frac{21}{64})$, $(-1.27276, \frac{23}{64})$, $(-1.27276, \frac{23}{64})$\}

\item $c = \frac{25}{8}$, $(d_i,\theta_i)$ = \{$(1., 0)$, $(1.76537, \frac{1}{8})$, $(-1.49661, \frac{1}{4})$, $(0.351153, \frac{3}{8})$, $(1., \frac{1}{2})$, $(1.76537, -\frac{3}{8})$, $(-1.49661, -\frac{1}{4})$, $(0.351153, -\frac{1}{8})$, $(1.79995, \frac{27}{64})$, $(0.688812, -\frac{17}{64})$, $(-1.66294, \frac{15}{64})$, $(-1.27276, -\frac{21}{64})$, $(0.688812, -\frac{17}{64})$, $(-1.66294, \frac{15}{64})$, $(-1.27276, -\frac{21}{64})$\}

\item $c = \frac{29}{8}$, $(d_i,\theta_i)$ = \{$(1., 0)$, $(1.76537, \frac{1}{8})$, $(-1.49661, \frac{1}{4})$, $(0.351153, \frac{3}{8})$, $(1., \frac{1}{2})$, $(1.76537, -\frac{3}{8})$, $(-1.49661, -\frac{1}{4})$, $(0.351153, -\frac{1}{8})$, $(1.79995, -\frac{17}{64})$, $(2.35175, \frac{19}{64})$, $(-0.974127, -\frac{13}{64})$, $(-1.27276, \frac{31}{64})$, $(-1.27276, \frac{31}{64})$\}

\item $c = \frac{33}{8}$, $(d_i,\theta_i)$ = \{$(1., 0)$, $(1.76537, \frac{1}{8})$, $(-1.49661, \frac{1}{4})$, $(0.351153, \frac{3}{8})$, $(1., \frac{1}{2})$, $(1.76537, -\frac{3}{8})$, $(-1.49661, -\frac{1}{4})$, $(0.351153, -\frac{1}{8})$, $(-1.79995, -\frac{29}{64})$, $(1.66294, \frac{23}{64})$, $(-0.688812, -\frac{9}{64})$, $(1.27276, -\frac{13}{64})$, $(1.66294, \frac{23}{64})$, $(-0.688812, -\frac{9}{64})$, $(1.27276, -\frac{13}{64})$\}

\item $c = \frac{37}{8}$, $(d_i,\theta_i)$ = \{$(1., 0)$, $(1.76537, \frac{1}{8})$, $(-1.49661, \frac{1}{4})$, $(0.351153, \frac{3}{8})$, $(1., \frac{1}{2})$, $(1.76537, -\frac{3}{8})$, $(-1.49661, -\frac{1}{4})$, $(0.351153, -\frac{1}{8})$, $(-1.79995, -\frac{9}{64})$, $(-2.35175, \frac{27}{64})$, $(0.974127, -\frac{5}{64})$, $(1.27276, -\frac{25}{64})$, $(1.27276, -\frac{25}{64})$\}

\item $c = \frac{41}{8}$, $(d_i,\theta_i)$ = \{$(1., 0)$, $(1.76537, \frac{1}{8})$, $(-1.49661, \frac{1}{4})$, $(0.351153, \frac{3}{8})$, $(1., \frac{1}{2})$, $(1.76537, -\frac{3}{8})$, $(-1.49661, -\frac{1}{4})$, $(0.351153, -\frac{1}{8})$, $(-1.79995, -\frac{21}{64})$, $(1.66294, \frac{31}{64})$, $(-0.688812, -\frac{1}{64})$, $(1.27276, -\frac{5}{64})$, $(1.66294, \frac{31}{64})$, $(-0.688812, -\frac{1}{64})$, $(1.27276, -\frac{5}{64})$\}

\item $c = \frac{45}{8}$, $(d_i,\theta_i)$ = \{$(1., 0)$, $(1.76537, \frac{1}{8})$, $(-1.49661, \frac{1}{4})$, $(0.351153, \frac{3}{8})$, $(1., \frac{1}{2})$, $(1.76537, -\frac{3}{8})$, $(-1.49661, -\frac{1}{4})$, $(0.351153, -\frac{1}{8})$, $(1.79995, -\frac{1}{64})$, $(2.35175, -\frac{29}{64})$, $(-0.974127, \frac{3}{64})$, $(-1.27276, -\frac{17}{64})$, $(-1.27276, -\frac{17}{64})$\}

\item $c = \frac{49}{8}$, $(d_i,\theta_i)$ = \{$(1., 0)$, $(1.76537, \frac{1}{8})$, $(-1.49661, \frac{1}{4})$, $(0.351153, \frac{3}{8})$, $(1., \frac{1}{2})$, $(1.76537, -\frac{3}{8})$, $(-1.49661, -\frac{1}{4})$, $(0.351153, -\frac{1}{8})$, $(1.79995, -\frac{13}{64})$, $(0.688812, \frac{7}{64})$, $(-1.66294, -\frac{25}{64})$, $(-1.27276, \frac{3}{64})$, $(0.688812, \frac{7}{64})$, $(-1.66294, -\frac{25}{64})$, $(-1.27276, \frac{3}{64})$\}

\item $c = \frac{53}{8}$, $(d_i,\theta_i)$ = \{$(1., 0)$, $(1.76537, \frac{1}{8})$, $(-1.49661, \frac{1}{4})$, $(0.351153, \frac{3}{8})$, $(1., \frac{1}{2})$, $(1.76537, -\frac{3}{8})$, $(-1.49661, -\frac{1}{4})$, $(0.351153, -\frac{1}{8})$, $(1.79995, \frac{7}{64})$, $(-0.974127, \frac{11}{64})$, $(2.35175, -\frac{21}{64})$, $(-1.27276, -\frac{9}{64})$, $(-1.27276, -\frac{9}{64})$\}

\item $c = \frac{57}{8}$, $(d_i,\theta_i)$ = \{$(1., 0)$, $(1.76537, \frac{1}{8})$, $(-1.49661, \frac{1}{4})$, $(0.351153, \frac{3}{8})$, $(1., \frac{1}{2})$, $(1.76537, -\frac{3}{8})$, $(-1.49661, -\frac{1}{4})$, $(0.351153, -\frac{1}{8})$, $(-1.79995, -\frac{5}{64})$, $(-0.688812, \frac{15}{64})$, $(1.66294, -\frac{17}{64})$, $(1.27276, \frac{11}{64})$, $(-0.688812, \frac{15}{64})$, $(1.66294, -\frac{17}{64})$, $(1.27276, \frac{11}{64})$\}

\item $c = \frac{61}{8}$, $(d_i,\theta_i)$ = \{$(1., 0)$, $(1.76537, \frac{1}{8})$, $(-1.49661, \frac{1}{4})$, $(0.351153, \frac{3}{8})$, $(1., \frac{1}{2})$, $(1.76537, -\frac{3}{8})$, $(-1.49661, -\frac{1}{4})$, $(0.351153, -\frac{1}{8})$, $(1.79995, \frac{15}{64})$, $(-0.974127, \frac{19}{64})$, $(2.35175, -\frac{13}{64})$, $(-1.27276, -\frac{1}{64})$, $(-1.27276, -\frac{1}{64})$\}

\end{enumerate}

\paragraph*{Rank 8; \#68}

\begin{enumerate}

\item $c = \frac{3}{8}$, $(d_i,\theta_i)$ = \{$(1., 0)$, $(0.351153, \frac{1}{8})$, $(-1.49661, \frac{1}{4})$, $(1.76537, -\frac{1}{8})$, $(1., \frac{1}{2})$, $(0.351153, -\frac{3}{8})$, $(-1.49661, -\frac{1}{4})$, $(1.76537, \frac{3}{8})$, $(-1.79995, -\frac{15}{64})$, $(-2.35175, \frac{13}{64})$, $(0.974127, -\frac{19}{64})$, $(1.27276, \frac{1}{64})$, $(1.27276, \frac{1}{64})$\}

\item $c = \frac{7}{8}$, $(d_i,\theta_i)$ = \{$(1., 0)$, $(0.351153, \frac{1}{8})$, $(-1.49661, \frac{1}{4})$, $(1.76537, -\frac{1}{8})$, $(1., \frac{1}{2})$, $(0.351153, -\frac{3}{8})$, $(-1.49661, -\frac{1}{4})$, $(1.76537, \frac{3}{8})$, $(1.79995, \frac{5}{64})$, $(0.688812, -\frac{15}{64})$, $(-1.66294, \frac{17}{64})$, $(-1.27276, -\frac{11}{64})$, $(0.688812, -\frac{15}{64})$, $(-1.66294, \frac{17}{64})$, $(-1.27276, -\frac{11}{64})$\}

\item $c = \frac{11}{8}$, $(d_i,\theta_i)$ = \{$(1., 0)$, $(0.351153, \frac{1}{8})$, $(-1.49661, \frac{1}{4})$, $(1.76537, -\frac{1}{8})$, $(1., \frac{1}{2})$, $(0.351153, -\frac{3}{8})$, $(-1.49661, -\frac{1}{4})$, $(1.76537, \frac{3}{8})$, $(1.79995, -\frac{7}{64})$, $(2.35175, \frac{21}{64})$, $(-0.974127, -\frac{11}{64})$, $(-1.27276, \frac{9}{64})$, $(-1.27276, \frac{9}{64})$\}

\item $c = \frac{15}{8}$, $(d_i,\theta_i)$ = \{$(1., 0)$, $(0.351153, \frac{1}{8})$, $(-1.49661, \frac{1}{4})$, $(1.76537, -\frac{1}{8})$, $(1., \frac{1}{2})$, $(0.351153, -\frac{3}{8})$, $(-1.49661, -\frac{1}{4})$, $(1.76537, \frac{3}{8})$, $(1.79995, \frac{13}{64})$, $(-1.66294, \frac{25}{64})$, $(0.688812, -\frac{7}{64})$, $(-1.27276, -\frac{3}{64})$, $(-1.66294, \frac{25}{64})$, $(0.688812, -\frac{7}{64})$, $(-1.27276, -\frac{3}{64})$\}

\item $c = \frac{19}{8}$, $(d_i,\theta_i)$ = \{$(1., 0)$, $(0.351153, \frac{1}{8})$, $(-1.49661, \frac{1}{4})$, $(1.76537, -\frac{1}{8})$, $(1., \frac{1}{2})$, $(0.351153, -\frac{3}{8})$, $(-1.49661, -\frac{1}{4})$, $(1.76537, \frac{3}{8})$, $(-1.79995, \frac{1}{64})$, $(-2.35175, \frac{29}{64})$, $(0.974127, -\frac{3}{64})$, $(1.27276, \frac{17}{64})$, $(1.27276, \frac{17}{64})$\}

\item $c = \frac{23}{8}$, $(d_i,\theta_i)$ = \{$(1., 0)$, $(0.351153, \frac{1}{8})$, $(-1.49661, \frac{1}{4})$, $(1.76537, -\frac{1}{8})$, $(1., \frac{1}{2})$, $(0.351153, -\frac{3}{8})$, $(-1.49661, -\frac{1}{4})$, $(1.76537, \frac{3}{8})$, $(1.79995, \frac{21}{64})$, $(-1.66294, -\frac{31}{64})$, $(0.688812, \frac{1}{64})$, $(-1.27276, \frac{5}{64})$, $(-1.66294, -\frac{31}{64})$, $(0.688812, \frac{1}{64})$, $(-1.27276, \frac{5}{64})$\}

\item $c = \frac{27}{8}$, $(d_i,\theta_i)$ = \{$(1., 0)$, $(0.351153, \frac{1}{8})$, $(-1.49661, \frac{1}{4})$, $(1.76537, \frac{3}{8})$, $(1., \frac{1}{2})$, $(0.351153, -\frac{3}{8})$, $(-1.49661, -\frac{1}{4})$, $(1.76537, -\frac{1}{8})$, $(-1.79995, \frac{9}{64})$, $(0.974127, \frac{5}{64})$, $(-2.35175, -\frac{27}{64})$, $(1.27276, \frac{25}{64})$, $(1.27276, \frac{25}{64})$\}

\item $c = \frac{31}{8}$, $(d_i,\theta_i)$ = \{$(1., 0)$, $(0.351153, \frac{1}{8})$, $(-1.49661, \frac{1}{4})$, $(1.76537, \frac{3}{8})$, $(1., \frac{1}{2})$, $(0.351153, -\frac{3}{8})$, $(-1.49661, -\frac{1}{4})$, $(1.76537, -\frac{1}{8})$, $(1.79995, \frac{29}{64})$, $(0.688812, \frac{9}{64})$, $(-1.66294, -\frac{23}{64})$, $(-1.27276, \frac{13}{64})$, $(0.688812, \frac{9}{64})$, $(-1.66294, -\frac{23}{64})$, $(-1.27276, \frac{13}{64})$\}

\item $c = \frac{35}{8}$, $(d_i,\theta_i)$ = \{$(1., 0)$, $(0.351153, \frac{1}{8})$, $(-1.49661, \frac{1}{4})$, $(1.76537, \frac{3}{8})$, $(1., \frac{1}{2})$, $(0.351153, -\frac{3}{8})$, $(-1.49661, -\frac{1}{4})$, $(1.76537, -\frac{1}{8})$, $(-1.79995, \frac{17}{64})$, $(0.974127, \frac{13}{64})$, $(-2.35175, -\frac{19}{64})$, $(1.27276, -\frac{31}{64})$, $(1.27276, -\frac{31}{64})$\}

\item $c = \frac{39}{8}$, $(d_i,\theta_i)$ = \{$(1., 0)$, $(0.351153, \frac{1}{8})$, $(-1.49661, \frac{1}{4})$, $(1.76537, \frac{3}{8})$, $(1., \frac{1}{2})$, $(0.351153, -\frac{3}{8})$, $(-1.49661, -\frac{1}{4})$, $(1.76537, -\frac{1}{8})$, $(1.79995, -\frac{27}{64})$, $(0.688812, \frac{17}{64})$, $(-1.66294, -\frac{15}{64})$, $(-1.27276, \frac{21}{64})$, $(0.688812, \frac{17}{64})$, $(-1.66294, -\frac{15}{64})$, $(-1.27276, \frac{21}{64})$\}

\item $c = \frac{43}{8}$, $(d_i,\theta_i)$ = \{$(1., 0)$, $(0.351153, \frac{1}{8})$, $(-1.49661, \frac{1}{4})$, $(1.76537, \frac{3}{8})$, $(1., \frac{1}{2})$, $(0.351153, -\frac{3}{8})$, $(-1.49661, -\frac{1}{4})$, $(1.76537, -\frac{1}{8})$, $(-1.79995, \frac{25}{64})$, $(0.974127, \frac{21}{64})$, $(-2.35175, -\frac{11}{64})$, $(1.27276, -\frac{23}{64})$, $(1.27276, -\frac{23}{64})$\}

\item $c = \frac{47}{8}$, $(d_i,\theta_i)$ = \{$(1., 0)$, $(0.351153, \frac{1}{8})$, $(-1.49661, \frac{1}{4})$, $(1.76537, \frac{3}{8})$, $(1., \frac{1}{2})$, $(0.351153, -\frac{3}{8})$, $(-1.49661, -\frac{1}{4})$, $(1.76537, -\frac{1}{8})$, $(-1.79995, -\frac{19}{64})$, $(1.66294, -\frac{7}{64})$, $(-0.688812, \frac{25}{64})$, $(1.27276, \frac{29}{64})$, $(1.66294, -\frac{7}{64})$, $(-0.688812, \frac{25}{64})$, $(1.27276, \frac{29}{64})$\}

\item $c = \frac{51}{8}$, $(d_i,\theta_i)$ = \{$(1., 0)$, $(0.351153, \frac{1}{8})$, $(-1.49661, \frac{1}{4})$, $(1.76537, \frac{3}{8})$, $(1., \frac{1}{2})$, $(0.351153, -\frac{3}{8})$, $(-1.49661, -\frac{1}{4})$, $(1.76537, -\frac{1}{8})$, $(-1.79995, -\frac{31}{64})$, $(-2.35175, -\frac{3}{64})$, $(0.974127, \frac{29}{64})$, $(1.27276, -\frac{15}{64})$, $(1.27276, -\frac{15}{64})$\}

\item $c = \frac{55}{8}$, $(d_i,\theta_i)$ = \{$(1., 0)$, $(0.351153, \frac{1}{8})$, $(-1.49661, \frac{1}{4})$, $(1.76537, \frac{3}{8})$, $(1., \frac{1}{2})$, $(0.351153, -\frac{3}{8})$, $(-1.49661, -\frac{1}{4})$, $(1.76537, -\frac{1}{8})$, $(1.79995, -\frac{11}{64})$, $(-1.66294, \frac{1}{64})$, $(0.688812, -\frac{31}{64})$, $(-1.27276, -\frac{27}{64})$, $(-1.66294, \frac{1}{64})$, $(0.688812, -\frac{31}{64})$, $(-1.27276, -\frac{27}{64})$\}

\item $c = \frac{59}{8}$, $(d_i,\theta_i)$ = \{$(1., 0)$, $(0.351153, \frac{1}{8})$, $(-1.49661, \frac{1}{4})$, $(1.76537, \frac{3}{8})$, $(1., \frac{1}{2})$, $(0.351153, -\frac{3}{8})$, $(-1.49661, -\frac{1}{4})$, $(1.76537, -\frac{1}{8})$, $(-1.79995, -\frac{23}{64})$, $(0.974127, -\frac{27}{64})$, $(-2.35175, \frac{5}{64})$, $(1.27276, -\frac{7}{64})$, $(1.27276, -\frac{7}{64})$\}

\item $c = \frac{63}{8}$, $(d_i,\theta_i)$ = \{$(1., 0)$, $(0.351153, \frac{1}{8})$, $(-1.49661, \frac{1}{4})$, $(1.76537, \frac{3}{8})$, $(1., \frac{1}{2})$, $(0.351153, -\frac{3}{8})$, $(-1.49661, -\frac{1}{4})$, $(1.76537, -\frac{1}{8})$, $(-1.79995, -\frac{3}{64})$, $(-0.688812, -\frac{23}{64})$, $(1.66294, \frac{9}{64})$, $(1.27276, -\frac{19}{64})$, $(-0.688812, -\frac{23}{64})$, $(1.66294, \frac{9}{64})$, $(1.27276, -\frac{19}{64})$\}

\end{enumerate}

\paragraph*{Rank 8; \#69}

\begin{enumerate}

\item $c = \frac{1}{8}$, $(d_i,\theta_i)$ = \{$(1., 0)$, $(0.234633, \frac{1}{8})$, $(0.668179, \frac{1}{4})$, $(-1.17958, -\frac{1}{8})$, $(1., \frac{1}{2})$, $(0.234633, -\frac{3}{8})$, $(0.668179, -\frac{1}{4})$, $(-1.17958, \frac{3}{8})$, $(1.20269, -\frac{13}{64})$, $(-0.460249, -\frac{25}{64})$, $(1.11114, \frac{7}{64})$, $(-0.85043, \frac{3}{64})$, $(-0.460249, -\frac{25}{64})$, $(1.11114, \frac{7}{64})$, $(-0.85043, \frac{3}{64})$\}

\item $c = \frac{5}{8}$, $(d_i,\theta_i)$ = \{$(1., 0)$, $(0.234633, \frac{1}{8})$, $(0.668179, \frac{1}{4})$, $(-1.17958, -\frac{1}{8})$, $(1., \frac{1}{2})$, $(0.234633, -\frac{3}{8})$, $(0.668179, -\frac{1}{4})$, $(-1.17958, \frac{3}{8})$, $(0.650891, -\frac{21}{64})$, $(1.20269, \frac{7}{64})$, $(-1.57139, \frac{11}{64})$, $(-0.85043, -\frac{9}{64})$, $(-0.85043, -\frac{9}{64})$\}

\item $c = \frac{9}{8}$, $(d_i,\theta_i)$ = \{$(1., 0)$, $(0.234633, \frac{1}{8})$, $(0.668179, \frac{1}{4})$, $(-1.17958, -\frac{1}{8})$, $(1., \frac{1}{2})$, $(0.234633, -\frac{3}{8})$, $(0.668179, -\frac{1}{4})$, $(-1.17958, \frac{3}{8})$, $(1.20269, -\frac{5}{64})$, $(1.11114, \frac{15}{64})$, $(-0.460249, -\frac{17}{64})$, $(-0.85043, \frac{11}{64})$, $(1.11114, \frac{15}{64})$, $(-0.460249, -\frac{17}{64})$, $(-0.85043, \frac{11}{64})$\}

\item $c = \frac{13}{8}$, $(d_i,\theta_i)$ = \{$(1., 0)$, $(0.234633, \frac{1}{8})$, $(0.668179, \frac{1}{4})$, $(-1.17958, -\frac{1}{8})$, $(1., \frac{1}{2})$, $(0.234633, -\frac{3}{8})$, $(0.668179, -\frac{1}{4})$, $(-1.17958, \frac{3}{8})$, $(-0.650891, -\frac{13}{64})$, $(-1.20269, \frac{15}{64})$, $(1.57139, \frac{19}{64})$, $(0.85043, -\frac{1}{64})$, $(0.85043, -\frac{1}{64})$\}

\item $c = \frac{17}{8}$, $(d_i,\theta_i)$ = \{$(1., 0)$, $(0.234633, \frac{1}{8})$, $(0.668179, \frac{1}{4})$, $(-1.17958, -\frac{1}{8})$, $(1., \frac{1}{2})$, $(0.234633, -\frac{3}{8})$, $(0.668179, -\frac{1}{4})$, $(-1.17958, \frac{3}{8})$, $(1.20269, \frac{3}{64})$, $(1.11114, \frac{23}{64})$, $(-0.460249, -\frac{9}{64})$, $(-0.85043, \frac{19}{64})$, $(1.11114, \frac{23}{64})$, $(-0.460249, -\frac{9}{64})$, $(-0.85043, \frac{19}{64})$\}

\item $c = \frac{21}{8}$, $(d_i,\theta_i)$ = \{$(1., 0)$, $(0.234633, \frac{1}{8})$, $(0.668179, \frac{1}{4})$, $(-1.17958, -\frac{1}{8})$, $(1., \frac{1}{2})$, $(0.234633, -\frac{3}{8})$, $(0.668179, -\frac{1}{4})$, $(-1.17958, \frac{3}{8})$, $(-0.650891, -\frac{5}{64})$, $(-1.20269, \frac{23}{64})$, $(1.57139, \frac{27}{64})$, $(0.85043, \frac{7}{64})$, $(0.85043, \frac{7}{64})$\}

\item $c = \frac{25}{8}$, $(d_i,\theta_i)$ = \{$(1., 0)$, $(0.234633, \frac{1}{8})$, $(0.668179, \frac{1}{4})$, $(-1.17958, \frac{3}{8})$, $(1., \frac{1}{2})$, $(0.234633, -\frac{3}{8})$, $(0.668179, -\frac{1}{4})$, $(-1.17958, -\frac{1}{8})$, $(1.20269, \frac{11}{64})$, $(-0.460249, -\frac{1}{64})$, $(1.11114, \frac{31}{64})$, $(-0.85043, \frac{27}{64})$, $(-0.460249, -\frac{1}{64})$, $(1.11114, \frac{31}{64})$, $(-0.85043, \frac{27}{64})$\}

\item $c = \frac{29}{8}$, $(d_i,\theta_i)$ = \{$(1., 0)$, $(0.234633, \frac{1}{8})$, $(0.668179, \frac{1}{4})$, $(-1.17958, \frac{3}{8})$, $(1., \frac{1}{2})$, $(0.234633, -\frac{3}{8})$, $(0.668179, -\frac{1}{4})$, $(-1.17958, -\frac{1}{8})$, $(-1.20269, \frac{31}{64})$, $(-0.650891, \frac{3}{64})$, $(1.57139, -\frac{29}{64})$, $(0.85043, \frac{15}{64})$, $(0.85043, \frac{15}{64})$\}

\item $c = \frac{33}{8}$, $(d_i,\theta_i)$ = \{$(1., 0)$, $(0.234633, \frac{1}{8})$, $(0.668179, \frac{1}{4})$, $(-1.17958, \frac{3}{8})$, $(1., \frac{1}{2})$, $(0.234633, -\frac{3}{8})$, $(0.668179, -\frac{1}{4})$, $(-1.17958, -\frac{1}{8})$, $(-1.20269, \frac{19}{64})$, $(0.460249, \frac{7}{64})$, $(-1.11114, -\frac{25}{64})$, $(0.85043, -\frac{29}{64})$, $(0.460249, \frac{7}{64})$, $(-1.11114, -\frac{25}{64})$, $(0.85043, -\frac{29}{64})$\}

\item $c = \frac{37}{8}$, $(d_i,\theta_i)$ = \{$(1., 0)$, $(0.234633, \frac{1}{8})$, $(0.668179, \frac{1}{4})$, $(-1.17958, \frac{3}{8})$, $(1., \frac{1}{2})$, $(0.234633, -\frac{3}{8})$, $(0.668179, -\frac{1}{4})$, $(-1.17958, -\frac{1}{8})$, $(-0.650891, \frac{11}{64})$, $(1.57139, -\frac{21}{64})$, $(-1.20269, -\frac{25}{64})$, $(0.85043, \frac{23}{64})$, $(0.85043, \frac{23}{64})$\}

\item $c = \frac{41}{8}$, $(d_i,\theta_i)$ = \{$(1., 0)$, $(0.234633, \frac{1}{8})$, $(0.668179, \frac{1}{4})$, $(-1.17958, \frac{3}{8})$, $(1., \frac{1}{2})$, $(0.234633, -\frac{3}{8})$, $(0.668179, -\frac{1}{4})$, $(-1.17958, -\frac{1}{8})$, $(1.20269, \frac{27}{64})$, $(1.11114, -\frac{17}{64})$, $(-0.460249, \frac{15}{64})$, $(-0.85043, -\frac{21}{64})$, $(1.11114, -\frac{17}{64})$, $(-0.460249, \frac{15}{64})$, $(-0.85043, -\frac{21}{64})$\}

\item $c = \frac{45}{8}$, $(d_i,\theta_i)$ = \{$(1., 0)$, $(0.234633, \frac{1}{8})$, $(0.668179, \frac{1}{4})$, $(-1.17958, \frac{3}{8})$, $(1., \frac{1}{2})$, $(0.234633, -\frac{3}{8})$, $(0.668179, -\frac{1}{4})$, $(-1.17958, -\frac{1}{8})$, $(1.57139, -\frac{13}{64})$, $(-1.20269, -\frac{17}{64})$, $(-0.650891, \frac{19}{64})$, $(0.85043, \frac{31}{64})$, $(0.85043, \frac{31}{64})$\}

\item $c = \frac{49}{8}$, $(d_i,\theta_i)$ = \{$(1., 0)$, $(0.234633, \frac{1}{8})$, $(0.668179, \frac{1}{4})$, $(-1.17958, \frac{3}{8})$, $(1., \frac{1}{2})$, $(0.234633, -\frac{3}{8})$, $(0.668179, -\frac{1}{4})$, $(-1.17958, -\frac{1}{8})$, $(1.20269, -\frac{29}{64})$, $(1.11114, -\frac{9}{64})$, $(-0.460249, \frac{23}{64})$, $(-0.85043, -\frac{13}{64})$, $(1.11114, -\frac{9}{64})$, $(-0.460249, \frac{23}{64})$, $(-0.85043, -\frac{13}{64})$\}

\item $c = \frac{53}{8}$, $(d_i,\theta_i)$ = \{$(1., 0)$, $(0.234633, \frac{1}{8})$, $(0.668179, \frac{1}{4})$, $(-1.17958, \frac{3}{8})$, $(1., \frac{1}{2})$, $(0.234633, -\frac{3}{8})$, $(0.668179, -\frac{1}{4})$, $(-1.17958, -\frac{1}{8})$, $(-0.650891, \frac{27}{64})$, $(1.57139, -\frac{5}{64})$, $(-1.20269, -\frac{9}{64})$, $(0.85043, -\frac{25}{64})$, $(0.85043, -\frac{25}{64})$\}

\item $c = \frac{57}{8}$, $(d_i,\theta_i)$ = \{$(1., 0)$, $(0.234633, \frac{1}{8})$, $(0.668179, \frac{1}{4})$, $(-1.17958, \frac{3}{8})$, $(1., \frac{1}{2})$, $(0.234633, -\frac{3}{8})$, $(0.668179, -\frac{1}{4})$, $(-1.17958, -\frac{1}{8})$, $(1.20269, -\frac{21}{64})$, $(-0.460249, \frac{31}{64})$, $(1.11114, -\frac{1}{64})$, $(-0.85043, -\frac{5}{64})$, $(-0.460249, \frac{31}{64})$, $(1.11114, -\frac{1}{64})$, $(-0.85043, -\frac{5}{64})$\}

\item $c = \frac{61}{8}$, $(d_i,\theta_i)$ = \{$(1., 0)$, $(0.234633, \frac{1}{8})$, $(0.668179, \frac{1}{4})$, $(-1.17958, \frac{3}{8})$, $(1., \frac{1}{2})$, $(0.234633, -\frac{3}{8})$, $(0.668179, -\frac{1}{4})$, $(-1.17958, -\frac{1}{8})$, $(-1.20269, -\frac{1}{64})$, $(1.57139, \frac{3}{64})$, $(-0.650891, -\frac{29}{64})$, $(0.85043, -\frac{17}{64})$, $(0.85043, -\frac{17}{64})$\}

\end{enumerate}

\paragraph*{Rank 8; \#70}

\begin{enumerate}

\item $c = \frac{3}{8}$, $(d_i,\theta_i)$ = \{$(1., 0)$, $(-1.17958, \frac{1}{8})$, $(0.668179, \frac{1}{4})$, $(0.234633, -\frac{1}{8})$, $(1., \frac{1}{2})$, $(-1.17958, -\frac{3}{8})$, $(0.668179, -\frac{1}{4})$, $(0.234633, \frac{3}{8})$, $(1.20269, \frac{1}{64})$, $(-1.57139, -\frac{3}{64})$, $(0.650891, \frac{29}{64})$, $(-0.85043, \frac{17}{64})$, $(-0.85043, \frac{17}{64})$\}

\item $c = \frac{7}{8}$, $(d_i,\theta_i)$ = \{$(1., 0)$, $(-1.17958, \frac{1}{8})$, $(0.668179, \frac{1}{4})$, $(0.234633, -\frac{1}{8})$, $(1., \frac{1}{2})$, $(-1.17958, -\frac{3}{8})$, $(0.668179, -\frac{1}{4})$, $(0.234633, \frac{3}{8})$, $(-1.20269, \frac{21}{64})$, $(-1.11114, \frac{1}{64})$, $(0.460249, -\frac{31}{64})$, $(0.85043, \frac{5}{64})$, $(-1.11114, \frac{1}{64})$, $(0.460249, -\frac{31}{64})$, $(0.85043, \frac{5}{64})$\}

\item $c = \frac{11}{8}$, $(d_i,\theta_i)$ = \{$(1., 0)$, $(-1.17958, \frac{1}{8})$, $(0.668179, \frac{1}{4})$, $(0.234633, -\frac{1}{8})$, $(1., \frac{1}{2})$, $(-1.17958, -\frac{3}{8})$, $(0.668179, -\frac{1}{4})$, $(0.234633, \frac{3}{8})$, $(0.650891, -\frac{27}{64})$, $(1.20269, \frac{9}{64})$, $(-1.57139, \frac{5}{64})$, $(-0.85043, \frac{25}{64})$, $(-0.85043, \frac{25}{64})$\}

\item $c = \frac{15}{8}$, $(d_i,\theta_i)$ = \{$(1., 0)$, $(-1.17958, \frac{1}{8})$, $(0.668179, \frac{1}{4})$, $(0.234633, -\frac{1}{8})$, $(1., \frac{1}{2})$, $(-1.17958, -\frac{3}{8})$, $(0.668179, -\frac{1}{4})$, $(0.234633, \frac{3}{8})$, $(1.20269, \frac{29}{64})$, $(1.11114, \frac{9}{64})$, $(-0.460249, -\frac{23}{64})$, $(-0.85043, \frac{13}{64})$, $(1.11114, \frac{9}{64})$, $(-0.460249, -\frac{23}{64})$, $(-0.85043, \frac{13}{64})$\}

\item $c = \frac{19}{8}$, $(d_i,\theta_i)$ = \{$(1., 0)$, $(-1.17958, \frac{1}{8})$, $(0.668179, \frac{1}{4})$, $(0.234633, -\frac{1}{8})$, $(1., \frac{1}{2})$, $(-1.17958, -\frac{3}{8})$, $(0.668179, -\frac{1}{4})$, $(0.234633, \frac{3}{8})$, $(0.650891, -\frac{19}{64})$, $(1.20269, \frac{17}{64})$, $(-1.57139, \frac{13}{64})$, $(-0.85043, -\frac{31}{64})$, $(-0.85043, -\frac{31}{64})$\}

\item $c = \frac{23}{8}$, $(d_i,\theta_i)$ = \{$(1., 0)$, $(-1.17958, \frac{1}{8})$, $(0.668179, \frac{1}{4})$, $(0.234633, -\frac{1}{8})$, $(1., \frac{1}{2})$, $(-1.17958, -\frac{3}{8})$, $(0.668179, -\frac{1}{4})$, $(0.234633, \frac{3}{8})$, $(-1.20269, -\frac{27}{64})$, $(0.460249, -\frac{15}{64})$, $(-1.11114, \frac{17}{64})$, $(0.85043, \frac{21}{64})$, $(0.460249, -\frac{15}{64})$, $(-1.11114, \frac{17}{64})$, $(0.85043, \frac{21}{64})$\}

\item $c = \frac{27}{8}$, $(d_i,\theta_i)$ = \{$(1., 0)$, $(-1.17958, \frac{1}{8})$, $(0.668179, \frac{1}{4})$, $(0.234633, \frac{3}{8})$, $(1., \frac{1}{2})$, $(-1.17958, -\frac{3}{8})$, $(0.668179, -\frac{1}{4})$, $(0.234633, -\frac{1}{8})$, $(0.650891, -\frac{11}{64})$, $(-1.57139, \frac{21}{64})$, $(1.20269, \frac{25}{64})$, $(-0.85043, -\frac{23}{64})$, $(-0.85043, -\frac{23}{64})$\}

\item $c = \frac{31}{8}$, $(d_i,\theta_i)$ = \{$(1., 0)$, $(-1.17958, \frac{1}{8})$, $(0.668179, \frac{1}{4})$, $(0.234633, \frac{3}{8})$, $(1., \frac{1}{2})$, $(-1.17958, -\frac{3}{8})$, $(0.668179, -\frac{1}{4})$, $(0.234633, -\frac{1}{8})$, $(-1.20269, -\frac{19}{64})$, $(0.460249, -\frac{7}{64})$, $(-1.11114, \frac{25}{64})$, $(0.85043, \frac{29}{64})$, $(0.460249, -\frac{7}{64})$, $(-1.11114, \frac{25}{64})$, $(0.85043, \frac{29}{64})$\}

\item $c = \frac{35}{8}$, $(d_i,\theta_i)$ = \{$(1., 0)$, $(-1.17958, \frac{1}{8})$, $(0.668179, \frac{1}{4})$, $(0.234633, \frac{3}{8})$, $(1., \frac{1}{2})$, $(-1.17958, -\frac{3}{8})$, $(0.668179, -\frac{1}{4})$, $(0.234633, -\frac{1}{8})$, $(1.20269, -\frac{31}{64})$, $(-1.57139, \frac{29}{64})$, $(0.650891, -\frac{3}{64})$, $(-0.85043, -\frac{15}{64})$, $(-0.85043, -\frac{15}{64})$\}

\item $c = \frac{39}{8}$, $(d_i,\theta_i)$ = \{$(1., 0)$, $(-1.17958, \frac{1}{8})$, $(0.668179, \frac{1}{4})$, $(0.234633, \frac{3}{8})$, $(1., \frac{1}{2})$, $(-1.17958, -\frac{3}{8})$, $(0.668179, -\frac{1}{4})$, $(0.234633, -\frac{1}{8})$, $(1.20269, -\frac{11}{64})$, $(1.11114, -\frac{31}{64})$, $(-0.460249, \frac{1}{64})$, $(-0.85043, -\frac{27}{64})$, $(1.11114, -\frac{31}{64})$, $(-0.460249, \frac{1}{64})$, $(-0.85043, -\frac{27}{64})$\}

\item $c = \frac{43}{8}$, $(d_i,\theta_i)$ = \{$(1., 0)$, $(-1.17958, \frac{1}{8})$, $(0.668179, \frac{1}{4})$, $(0.234633, \frac{3}{8})$, $(1., \frac{1}{2})$, $(-1.17958, -\frac{3}{8})$, $(0.668179, -\frac{1}{4})$, $(0.234633, -\frac{1}{8})$, $(0.650891, \frac{5}{64})$, $(1.20269, -\frac{23}{64})$, $(-1.57139, -\frac{27}{64})$, $(-0.85043, -\frac{7}{64})$, $(-0.85043, -\frac{7}{64})$\}

\item $c = \frac{47}{8}$, $(d_i,\theta_i)$ = \{$(1., 0)$, $(-1.17958, \frac{1}{8})$, $(0.668179, \frac{1}{4})$, $(0.234633, \frac{3}{8})$, $(1., \frac{1}{2})$, $(-1.17958, -\frac{3}{8})$, $(0.668179, -\frac{1}{4})$, $(0.234633, -\frac{1}{8})$, $(1.20269, -\frac{3}{64})$, $(1.11114, -\frac{23}{64})$, $(-0.460249, \frac{9}{64})$, $(-0.85043, -\frac{19}{64})$, $(1.11114, -\frac{23}{64})$, $(-0.460249, \frac{9}{64})$, $(-0.85043, -\frac{19}{64})$\}

\item $c = \frac{51}{8}$, $(d_i,\theta_i)$ = \{$(1., 0)$, $(-1.17958, \frac{1}{8})$, $(0.668179, \frac{1}{4})$, $(0.234633, \frac{3}{8})$, $(1., \frac{1}{2})$, $(-1.17958, -\frac{3}{8})$, $(0.668179, -\frac{1}{4})$, $(0.234633, -\frac{1}{8})$, $(0.650891, \frac{13}{64})$, $(1.20269, -\frac{15}{64})$, $(-1.57139, -\frac{19}{64})$, $(-0.85043, \frac{1}{64})$, $(-0.85043, \frac{1}{64})$\}

\item $c = \frac{55}{8}$, $(d_i,\theta_i)$ = \{$(1., 0)$, $(-1.17958, \frac{1}{8})$, $(0.668179, \frac{1}{4})$, $(0.234633, \frac{3}{8})$, $(1., \frac{1}{2})$, $(-1.17958, -\frac{3}{8})$, $(0.668179, -\frac{1}{4})$, $(0.234633, -\frac{1}{8})$, $(-1.20269, \frac{5}{64})$, $(0.460249, \frac{17}{64})$, $(-1.11114, -\frac{15}{64})$, $(0.85043, -\frac{11}{64})$, $(0.460249, \frac{17}{64})$, $(-1.11114, -\frac{15}{64})$, $(0.85043, -\frac{11}{64})$\}

\item $c = \frac{59}{8}$, $(d_i,\theta_i)$ = \{$(1., 0)$, $(-1.17958, \frac{1}{8})$, $(0.668179, \frac{1}{4})$, $(0.234633, \frac{3}{8})$, $(1., \frac{1}{2})$, $(-1.17958, -\frac{3}{8})$, $(0.668179, -\frac{1}{4})$, $(0.234633, -\frac{1}{8})$, $(-1.57139, -\frac{11}{64})$, $(1.20269, -\frac{7}{64})$, $(0.650891, \frac{21}{64})$, $(-0.85043, \frac{9}{64})$, $(-0.85043, \frac{9}{64})$\}

\item $c = \frac{63}{8}$, $(d_i,\theta_i)$ = \{$(1., 0)$, $(-1.17958, \frac{1}{8})$, $(0.668179, \frac{1}{4})$, $(0.234633, \frac{3}{8})$, $(1., \frac{1}{2})$, $(-1.17958, -\frac{3}{8})$, $(0.668179, -\frac{1}{4})$, $(0.234633, -\frac{1}{8})$, $(-1.20269, \frac{13}{64})$, $(0.460249, \frac{25}{64})$, $(-1.11114, -\frac{7}{64})$, $(0.85043, -\frac{3}{64})$, $(0.460249, \frac{25}{64})$, $(-1.11114, -\frac{7}{64})$, $(0.85043, -\frac{3}{64})$\}

\end{enumerate}

\paragraph*{Rank 8; \#71}

\begin{enumerate}

\item $c = \frac{3}{8}$, $(d_i,\theta_i)$ = \{$(1., 0)$, $(0.566454, \frac{1}{8})$, $(-0.198912, \frac{1}{4})$, $(-0.847759, -\frac{1}{8})$, $(1., \frac{1}{2})$, $(0.566454, -\frac{3}{8})$, $(-0.198912, -\frac{1}{4})$, $(-0.847759, \frac{3}{8})$, $(-1.01959, \frac{9}{64})$, $(0.941979, \frac{5}{64})$, $(0.390181, -\frac{27}{64})$, $(-0.72096, -\frac{7}{64})$, $(0.941979, \frac{5}{64})$, $(0.390181, -\frac{27}{64})$, $(-0.72096, -\frac{7}{64})$\}

\item $c = \frac{7}{8}$, $(d_i,\theta_i)$ = \{$(1., 0)$, $(0.566454, \frac{1}{8})$, $(-0.198912, \frac{1}{4})$, $(-0.847759, -\frac{1}{8})$, $(1., \frac{1}{2})$, $(0.566454, -\frac{3}{8})$, $(-0.198912, -\frac{1}{4})$, $(-0.847759, \frac{3}{8})$, $(1.01959, -\frac{3}{64})$, $(-0.551799, -\frac{23}{64})$, $(-1.33216, \frac{9}{64})$, $(0.72096, \frac{13}{64})$, $(0.72096, \frac{13}{64})$\}

\item $c = \frac{11}{8}$, $(d_i,\theta_i)$ = \{$(1., 0)$, $(0.566454, \frac{1}{8})$, $(-0.198912, \frac{1}{4})$, $(-0.847759, -\frac{1}{8})$, $(1., \frac{1}{2})$, $(0.566454, -\frac{3}{8})$, $(-0.198912, -\frac{1}{4})$, $(-0.847759, \frac{3}{8})$, $(-1.01959, \frac{17}{64})$, $(0.390181, -\frac{19}{64})$, $(0.941979, \frac{13}{64})$, $(-0.72096, \frac{1}{64})$, $(0.390181, -\frac{19}{64})$, $(0.941979, \frac{13}{64})$, $(-0.72096, \frac{1}{64})$\}

\item $c = \frac{15}{8}$, $(d_i,\theta_i)$ = \{$(1., 0)$, $(0.566454, \frac{1}{8})$, $(-0.198912, \frac{1}{4})$, $(-0.847759, -\frac{1}{8})$, $(1., \frac{1}{2})$, $(0.566454, -\frac{3}{8})$, $(-0.198912, -\frac{1}{4})$, $(-0.847759, \frac{3}{8})$, $(-1.33216, \frac{17}{64})$, $(1.01959, \frac{5}{64})$, $(-0.551799, -\frac{15}{64})$, $(0.72096, \frac{21}{64})$, $(0.72096, \frac{21}{64})$\}

\item $c = \frac{19}{8}$, $(d_i,\theta_i)$ = \{$(1., 0)$, $(0.566454, \frac{1}{8})$, $(-0.198912, \frac{1}{4})$, $(-0.847759, -\frac{1}{8})$, $(1., \frac{1}{2})$, $(0.566454, -\frac{3}{8})$, $(-0.198912, -\frac{1}{4})$, $(-0.847759, \frac{3}{8})$, $(1.01959, \frac{25}{64})$, $(-0.390181, -\frac{11}{64})$, $(-0.941979, \frac{21}{64})$, $(0.72096, \frac{9}{64})$, $(-0.390181, -\frac{11}{64})$, $(-0.941979, \frac{21}{64})$, $(0.72096, \frac{9}{64})$\}

\item $c = \frac{23}{8}$, $(d_i,\theta_i)$ = \{$(1., 0)$, $(0.566454, \frac{1}{8})$, $(-0.198912, \frac{1}{4})$, $(-0.847759, -\frac{1}{8})$, $(1., \frac{1}{2})$, $(0.566454, -\frac{3}{8})$, $(-0.198912, -\frac{1}{4})$, $(-0.847759, \frac{3}{8})$, $(-0.551799, -\frac{7}{64})$, $(1.01959, \frac{13}{64})$, $(-1.33216, \frac{25}{64})$, $(0.72096, \frac{29}{64})$, $(0.72096, \frac{29}{64})$\}

\item $c = \frac{27}{8}$, $(d_i,\theta_i)$ = \{$(1., 0)$, $(0.566454, \frac{1}{8})$, $(-0.198912, \frac{1}{4})$, $(-0.847759, \frac{3}{8})$, $(1., \frac{1}{2})$, $(0.566454, -\frac{3}{8})$, $(-0.198912, -\frac{1}{4})$, $(-0.847759, -\frac{1}{8})$, $(1.01959, -\frac{31}{64})$, $(-0.941979, \frac{29}{64})$, $(-0.390181, -\frac{3}{64})$, $(0.72096, \frac{17}{64})$, $(-0.941979, \frac{29}{64})$, $(-0.390181, -\frac{3}{64})$, $(0.72096, \frac{17}{64})$\}

\item $c = \frac{31}{8}$, $(d_i,\theta_i)$ = \{$(1., 0)$, $(0.566454, \frac{1}{8})$, $(-0.198912, \frac{1}{4})$, $(-0.847759, \frac{3}{8})$, $(1., \frac{1}{2})$, $(0.566454, -\frac{3}{8})$, $(-0.198912, -\frac{1}{4})$, $(-0.847759, -\frac{1}{8})$, $(1.01959, \frac{21}{64})$, $(-1.33216, -\frac{31}{64})$, $(-0.551799, \frac{1}{64})$, $(0.72096, -\frac{27}{64})$, $(0.72096, -\frac{27}{64})$\}

\item $c = \frac{35}{8}$, $(d_i,\theta_i)$ = \{$(1., 0)$, $(0.566454, \frac{1}{8})$, $(-0.198912, \frac{1}{4})$, $(-0.847759, \frac{3}{8})$, $(1., \frac{1}{2})$, $(0.566454, -\frac{3}{8})$, $(-0.198912, -\frac{1}{4})$, $(-0.847759, -\frac{1}{8})$, $(1.01959, -\frac{23}{64})$, $(-0.941979, -\frac{27}{64})$, $(-0.390181, \frac{5}{64})$, $(0.72096, \frac{25}{64})$, $(-0.941979, -\frac{27}{64})$, $(-0.390181, \frac{5}{64})$, $(0.72096, \frac{25}{64})$\}

\item $c = \frac{39}{8}$, $(d_i,\theta_i)$ = \{$(1., 0)$, $(0.566454, \frac{1}{8})$, $(-0.198912, \frac{1}{4})$, $(-0.847759, \frac{3}{8})$, $(1., \frac{1}{2})$, $(0.566454, -\frac{3}{8})$, $(-0.198912, -\frac{1}{4})$, $(-0.847759, -\frac{1}{8})$, $(1.01959, \frac{29}{64})$, $(-0.551799, \frac{9}{64})$, $(-1.33216, -\frac{23}{64})$, $(0.72096, -\frac{19}{64})$, $(0.72096, -\frac{19}{64})$\}

\item $c = \frac{43}{8}$, $(d_i,\theta_i)$ = \{$(1., 0)$, $(0.566454, \frac{1}{8})$, $(-0.198912, \frac{1}{4})$, $(-0.847759, \frac{3}{8})$, $(1., \frac{1}{2})$, $(0.566454, -\frac{3}{8})$, $(-0.198912, -\frac{1}{4})$, $(-0.847759, -\frac{1}{8})$, $(1.01959, -\frac{15}{64})$, $(-0.390181, \frac{13}{64})$, $(-0.941979, -\frac{19}{64})$, $(0.72096, -\frac{31}{64})$, $(-0.390181, \frac{13}{64})$, $(-0.941979, -\frac{19}{64})$, $(0.72096, -\frac{31}{64})$\}

\item $c = \frac{47}{8}$, $(d_i,\theta_i)$ = \{$(1., 0)$, $(0.566454, \frac{1}{8})$, $(-0.198912, \frac{1}{4})$, $(-0.847759, \frac{3}{8})$, $(1., \frac{1}{2})$, $(0.566454, -\frac{3}{8})$, $(-0.198912, -\frac{1}{4})$, $(-0.847759, -\frac{1}{8})$, $(-1.33216, -\frac{15}{64})$, $(1.01959, -\frac{27}{64})$, $(-0.551799, \frac{17}{64})$, $(0.72096, -\frac{11}{64})$, $(0.72096, -\frac{11}{64})$\}

\item $c = \frac{51}{8}$, $(d_i,\theta_i)$ = \{$(1., 0)$, $(0.566454, \frac{1}{8})$, $(-0.198912, \frac{1}{4})$, $(-0.847759, \frac{3}{8})$, $(1., \frac{1}{2})$, $(0.566454, -\frac{3}{8})$, $(-0.198912, -\frac{1}{4})$, $(-0.847759, -\frac{1}{8})$, $(1.01959, -\frac{7}{64})$, $(-0.390181, \frac{21}{64})$, $(-0.941979, -\frac{11}{64})$, $(0.72096, -\frac{23}{64})$, $(-0.390181, \frac{21}{64})$, $(-0.941979, -\frac{11}{64})$, $(0.72096, -\frac{23}{64})$\}

\item $c = \frac{55}{8}$, $(d_i,\theta_i)$ = \{$(1., 0)$, $(0.566454, \frac{1}{8})$, $(-0.198912, \frac{1}{4})$, $(-0.847759, \frac{3}{8})$, $(1., \frac{1}{2})$, $(0.566454, -\frac{3}{8})$, $(-0.198912, -\frac{1}{4})$, $(-0.847759, -\frac{1}{8})$, $(1.01959, -\frac{19}{64})$, $(-0.551799, \frac{25}{64})$, $(-1.33216, -\frac{7}{64})$, $(0.72096, -\frac{3}{64})$, $(0.72096, -\frac{3}{64})$\}

\item $c = \frac{59}{8}$, $(d_i,\theta_i)$ = \{$(1., 0)$, $(0.566454, \frac{1}{8})$, $(-0.198912, \frac{1}{4})$, $(-0.847759, \frac{3}{8})$, $(1., \frac{1}{2})$, $(0.566454, -\frac{3}{8})$, $(-0.198912, -\frac{1}{4})$, $(-0.847759, -\frac{1}{8})$, $(1.01959, \frac{1}{64})$, $(-0.941979, -\frac{3}{64})$, $(-0.390181, \frac{29}{64})$, $(0.72096, -\frac{15}{64})$, $(-0.941979, -\frac{3}{64})$, $(-0.390181, \frac{29}{64})$, $(0.72096, -\frac{15}{64})$\}

\item $c = \frac{63}{8}$, $(d_i,\theta_i)$ = \{$(1., 0)$, $(0.566454, \frac{1}{8})$, $(-0.198912, \frac{1}{4})$, $(-0.847759, \frac{3}{8})$, $(1., \frac{1}{2})$, $(0.566454, -\frac{3}{8})$, $(-0.198912, -\frac{1}{4})$, $(-0.847759, -\frac{1}{8})$, $(1.01959, -\frac{11}{64})$, $(-0.551799, -\frac{31}{64})$, $(-1.33216, \frac{1}{64})$, $(0.72096, \frac{5}{64})$, $(0.72096, \frac{5}{64})$\}

\end{enumerate}

\paragraph*{Rank 8; \#72}

\begin{enumerate}

\item $c = \frac{1}{8}$, $(d_i,\theta_i)$ = \{$(1., 0)$, $(-0.847759, \frac{1}{8})$, $(-0.198912, \frac{1}{4})$, $(0.566454, -\frac{1}{8})$, $(1., \frac{1}{2})$, $(-0.847759, -\frac{3}{8})$, $(-0.198912, -\frac{1}{4})$, $(0.566454, \frac{3}{8})$, $(1.33216, -\frac{1}{64})$, $(-1.01959, \frac{11}{64})$, $(0.551799, \frac{31}{64})$, $(-0.72096, -\frac{5}{64})$, $(-0.72096, -\frac{5}{64})$\}

\item $c = \frac{5}{8}$, $(d_i,\theta_i)$ = \{$(1., 0)$, $(-0.847759, \frac{1}{8})$, $(-0.198912, \frac{1}{4})$, $(0.566454, -\frac{1}{8})$, $(1., \frac{1}{2})$, $(-0.847759, -\frac{3}{8})$, $(-0.198912, -\frac{1}{4})$, $(0.566454, \frac{3}{8})$, $(-1.01959, -\frac{1}{64})$, $(0.390181, -\frac{29}{64})$, $(0.941979, \frac{3}{64})$, $(-0.72096, \frac{15}{64})$, $(0.390181, -\frac{29}{64})$, $(0.941979, \frac{3}{64})$, $(-0.72096, \frac{15}{64})$\}

\item $c = \frac{9}{8}$, $(d_i,\theta_i)$ = \{$(1., 0)$, $(-0.847759, \frac{1}{8})$, $(-0.198912, \frac{1}{4})$, $(0.566454, -\frac{1}{8})$, $(1., \frac{1}{2})$, $(-0.847759, -\frac{3}{8})$, $(-0.198912, -\frac{1}{4})$, $(0.566454, \frac{3}{8})$, $(-1.33216, \frac{7}{64})$, $(1.01959, \frac{19}{64})$, $(-0.551799, -\frac{25}{64})$, $(0.72096, \frac{3}{64})$, $(0.72096, \frac{3}{64})$\}

\item $c = \frac{13}{8}$, $(d_i,\theta_i)$ = \{$(1., 0)$, $(-0.847759, \frac{1}{8})$, $(-0.198912, \frac{1}{4})$, $(0.566454, -\frac{1}{8})$, $(1., \frac{1}{2})$, $(-0.847759, -\frac{3}{8})$, $(-0.198912, -\frac{1}{4})$, $(0.566454, \frac{3}{8})$, $(-1.01959, \frac{7}{64})$, $(0.390181, -\frac{21}{64})$, $(0.941979, \frac{11}{64})$, $(-0.72096, \frac{23}{64})$, $(0.390181, -\frac{21}{64})$, $(0.941979, \frac{11}{64})$, $(-0.72096, \frac{23}{64})$\}

\item $c = \frac{17}{8}$, $(d_i,\theta_i)$ = \{$(1., 0)$, $(-0.847759, \frac{1}{8})$, $(-0.198912, \frac{1}{4})$, $(0.566454, -\frac{1}{8})$, $(1., \frac{1}{2})$, $(-0.847759, -\frac{3}{8})$, $(-0.198912, -\frac{1}{4})$, $(0.566454, \frac{3}{8})$, $(1.01959, \frac{27}{64})$, $(-0.551799, -\frac{17}{64})$, $(-1.33216, \frac{15}{64})$, $(0.72096, \frac{11}{64})$, $(0.72096, \frac{11}{64})$\}

\item $c = \frac{21}{8}$, $(d_i,\theta_i)$ = \{$(1., 0)$, $(-0.847759, \frac{1}{8})$, $(-0.198912, \frac{1}{4})$, $(0.566454, -\frac{1}{8})$, $(1., \frac{1}{2})$, $(-0.847759, -\frac{3}{8})$, $(-0.198912, -\frac{1}{4})$, $(0.566454, \frac{3}{8})$, $(1.01959, \frac{15}{64})$, $(-0.941979, \frac{19}{64})$, $(-0.390181, -\frac{13}{64})$, $(0.72096, \frac{31}{64})$, $(-0.941979, \frac{19}{64})$, $(-0.390181, -\frac{13}{64})$, $(0.72096, \frac{31}{64})$\}

\item $c = \frac{25}{8}$, $(d_i,\theta_i)$ = \{$(1., 0)$, $(-0.847759, \frac{1}{8})$, $(-0.198912, \frac{1}{4})$, $(0.566454, \frac{3}{8})$, $(1., \frac{1}{2})$, $(-0.847759, -\frac{3}{8})$, $(-0.198912, -\frac{1}{4})$, $(0.566454, -\frac{1}{8})$, $(-1.01959, -\frac{29}{64})$, $(0.551799, -\frac{9}{64})$, $(1.33216, \frac{23}{64})$, $(-0.72096, \frac{19}{64})$, $(-0.72096, \frac{19}{64})$\}

\item $c = \frac{29}{8}$, $(d_i,\theta_i)$ = \{$(1., 0)$, $(-0.847759, \frac{1}{8})$, $(-0.198912, \frac{1}{4})$, $(0.566454, \frac{3}{8})$, $(1., \frac{1}{2})$, $(-0.847759, -\frac{3}{8})$, $(-0.198912, -\frac{1}{4})$, $(0.566454, -\frac{1}{8})$, $(-1.01959, \frac{23}{64})$, $(0.941979, \frac{27}{64})$, $(0.390181, -\frac{5}{64})$, $(-0.72096, -\frac{25}{64})$, $(0.941979, \frac{27}{64})$, $(0.390181, -\frac{5}{64})$, $(-0.72096, -\frac{25}{64})$\}

\item $c = \frac{33}{8}$, $(d_i,\theta_i)$ = \{$(1., 0)$, $(-0.847759, \frac{1}{8})$, $(-0.198912, \frac{1}{4})$, $(0.566454, \frac{3}{8})$, $(1., \frac{1}{2})$, $(-0.847759, -\frac{3}{8})$, $(-0.198912, -\frac{1}{4})$, $(0.566454, -\frac{1}{8})$, $(1.33216, \frac{31}{64})$, $(-1.01959, -\frac{21}{64})$, $(0.551799, -\frac{1}{64})$, $(-0.72096, \frac{27}{64})$, $(-0.72096, \frac{27}{64})$\}

\item $c = \frac{37}{8}$, $(d_i,\theta_i)$ = \{$(1., 0)$, $(-0.847759, \frac{1}{8})$, $(-0.198912, \frac{1}{4})$, $(0.566454, \frac{3}{8})$, $(1., \frac{1}{2})$, $(-0.847759, -\frac{3}{8})$, $(-0.198912, -\frac{1}{4})$, $(0.566454, -\frac{1}{8})$, $(-1.01959, \frac{31}{64})$, $(0.390181, \frac{3}{64})$, $(0.941979, -\frac{29}{64})$, $(-0.72096, -\frac{17}{64})$, $(0.390181, \frac{3}{64})$, $(0.941979, -\frac{29}{64})$, $(-0.72096, -\frac{17}{64})$\}

\item $c = \frac{41}{8}$, $(d_i,\theta_i)$ = \{$(1., 0)$, $(-0.847759, \frac{1}{8})$, $(-0.198912, \frac{1}{4})$, $(0.566454, \frac{3}{8})$, $(1., \frac{1}{2})$, $(-0.847759, -\frac{3}{8})$, $(-0.198912, -\frac{1}{4})$, $(0.566454, -\frac{1}{8})$, $(1.33216, -\frac{25}{64})$, $(0.551799, \frac{7}{64})$, $(-1.01959, -\frac{13}{64})$, $(-0.72096, -\frac{29}{64})$, $(-0.72096, -\frac{29}{64})$\}

\item $c = \frac{45}{8}$, $(d_i,\theta_i)$ = \{$(1., 0)$, $(-0.847759, \frac{1}{8})$, $(-0.198912, \frac{1}{4})$, $(0.566454, \frac{3}{8})$, $(1., \frac{1}{2})$, $(-0.847759, -\frac{3}{8})$, $(-0.198912, -\frac{1}{4})$, $(0.566454, -\frac{1}{8})$, $(-1.01959, -\frac{25}{64})$, $(0.390181, \frac{11}{64})$, $(0.941979, -\frac{21}{64})$, $(-0.72096, -\frac{9}{64})$, $(0.390181, \frac{11}{64})$, $(0.941979, -\frac{21}{64})$, $(-0.72096, -\frac{9}{64})$\}

\item $c = \frac{49}{8}$, $(d_i,\theta_i)$ = \{$(1., 0)$, $(-0.847759, \frac{1}{8})$, $(-0.198912, \frac{1}{4})$, $(0.566454, \frac{3}{8})$, $(1., \frac{1}{2})$, $(-0.847759, -\frac{3}{8})$, $(-0.198912, -\frac{1}{4})$, $(0.566454, -\frac{1}{8})$, $(1.33216, -\frac{17}{64})$, $(0.551799, \frac{15}{64})$, $(-1.01959, -\frac{5}{64})$, $(-0.72096, -\frac{21}{64})$, $(-0.72096, -\frac{21}{64})$\}

\item $c = \frac{53}{8}$, $(d_i,\theta_i)$ = \{$(1., 0)$, $(-0.847759, \frac{1}{8})$, $(-0.198912, \frac{1}{4})$, $(0.566454, \frac{3}{8})$, $(1., \frac{1}{2})$, $(-0.847759, -\frac{3}{8})$, $(-0.198912, -\frac{1}{4})$, $(0.566454, -\frac{1}{8})$, $(-1.01959, -\frac{17}{64})$, $(0.941979, -\frac{13}{64})$, $(0.390181, \frac{19}{64})$, $(-0.72096, -\frac{1}{64})$, $(0.941979, -\frac{13}{64})$, $(0.390181, \frac{19}{64})$, $(-0.72096, -\frac{1}{64})$\}

\item $c = \frac{57}{8}$, $(d_i,\theta_i)$ = \{$(1., 0)$, $(-0.847759, \frac{1}{8})$, $(-0.198912, \frac{1}{4})$, $(0.566454, \frac{3}{8})$, $(1., \frac{1}{2})$, $(-0.847759, -\frac{3}{8})$, $(-0.198912, -\frac{1}{4})$, $(0.566454, -\frac{1}{8})$, $(-1.01959, \frac{3}{64})$, $(0.551799, \frac{23}{64})$, $(1.33216, -\frac{9}{64})$, $(-0.72096, -\frac{13}{64})$, $(-0.72096, -\frac{13}{64})$\}

\item $c = \frac{61}{8}$, $(d_i,\theta_i)$ = \{$(1., 0)$, $(-0.847759, \frac{1}{8})$, $(-0.198912, \frac{1}{4})$, $(0.566454, \frac{3}{8})$, $(1., \frac{1}{2})$, $(-0.847759, -\frac{3}{8})$, $(-0.198912, -\frac{1}{4})$, $(0.566454, -\frac{1}{8})$, $(-1.01959, -\frac{9}{64})$, $(0.941979, -\frac{5}{64})$, $(0.390181, \frac{27}{64})$, $(-0.72096, \frac{7}{64})$, $(0.941979, -\frac{5}{64})$, $(0.390181, \frac{27}{64})$, $(-0.72096, \frac{7}{64})$\}

\end{enumerate}

\paragraph*{Rank 10; \#3}

\begin{enumerate}

\item $c = 0$, $(d_i,\theta_i)$ = \{$(1., 0)$, $(1.73205, 0)$, $(1.73205, 0)$, $(1., 0)$, $(2., \frac{1}{6})$, $(1., \frac{1}{2})$, $(1.73205, \frac{1}{2})$, $(1.73205, \frac{1}{2})$, $(1., \frac{1}{2})$, $(2., -\frac{1}{3})$, $(1., \frac{1}{4})$, $(1.73205, 0)$, $(1.73205, \frac{1}{2})$, $(2., -\frac{1}{12})$, $(1., \frac{1}{4})$, $(1., \frac{1}{4})$, $(1.73205, 0)$, $(1.73205, \frac{1}{2})$, $(2., -\frac{1}{12})$, $(1., \frac{1}{4})$\}

\item $c = \frac{1}{2}$, $(d_i,\theta_i)$ = \{$(1., 0)$, $(1.73205, 0)$, $(1.73205, 0)$, $(1., 0)$, $(2., \frac{1}{6})$, $(1., \frac{1}{2})$, $(1.73205, \frac{1}{2})$, $(1.73205, \frac{1}{2})$, $(1., \frac{1}{2})$, $(2., -\frac{1}{3})$, $(2.44949, -\frac{7}{16})$, $(2.44949, \frac{1}{16})$, $(2.82843, -\frac{1}{48})$, $(1.41421, \frac{5}{16})$, $(1.41421, \frac{5}{16})$\}

\item $c = 1$, $(d_i,\theta_i)$ = \{$(1., 0)$, $(1.73205, 0)$, $(1.73205, 0)$, $(1., 0)$, $(2., \frac{1}{6})$, $(1., \frac{1}{2})$, $(1.73205, \frac{1}{2})$, $(1.73205, \frac{1}{2})$, $(1., \frac{1}{2})$, $(2., -\frac{1}{3})$, $(1.73205, -\frac{3}{8})$, $(1.73205, \frac{1}{8})$, $(1., \frac{3}{8})$, $(1., \frac{3}{8})$, $(2., \frac{1}{24})$, $(1.73205, -\frac{3}{8})$, $(1.73205, \frac{1}{8})$, $(1., \frac{3}{8})$, $(1., \frac{3}{8})$, $(2., \frac{1}{24})$\}

\item $c = \frac{3}{2}$, $(d_i,\theta_i)$ = \{$(1., 0)$, $(1.73205, 0)$, $(1.73205, 0)$, $(1., 0)$, $(2., \frac{1}{6})$, $(1., \frac{1}{2})$, $(1.73205, \frac{1}{2})$, $(1.73205, \frac{1}{2})$, $(1., \frac{1}{2})$, $(2., -\frac{1}{3})$, $(2.82843, \frac{5}{48})$, $(2.44949, \frac{3}{16})$, $(2.44949, -\frac{5}{16})$, $(1.41421, \frac{7}{16})$, $(1.41421, \frac{7}{16})$\}

\item $c = 2$, $(d_i,\theta_i)$ = \{$(1., 0)$, $(1.73205, 0)$, $(1.73205, 0)$, $(1., 0)$, $(2., \frac{1}{6})$, $(1., \frac{1}{2})$, $(1.73205, \frac{1}{2})$, $(1.73205, \frac{1}{2})$, $(1., \frac{1}{2})$, $(2., -\frac{1}{3})$, $(1., \frac{1}{2})$, $(1.73205, -\frac{1}{4})$, $(1.73205, \frac{1}{4})$, $(2., \frac{1}{6})$, $(1., \frac{1}{2})$, $(1., \frac{1}{2})$, $(1.73205, -\frac{1}{4})$, $(1.73205, \frac{1}{4})$, $(2., \frac{1}{6})$, $(1., \frac{1}{2})$\}

\item $c = \frac{5}{2}$, $(d_i,\theta_i)$ = \{$(1., 0)$, $(1.73205, 0)$, $(1.73205, 0)$, $(1., 0)$, $(2., \frac{1}{6})$, $(1., \frac{1}{2})$, $(1.73205, \frac{1}{2})$, $(1.73205, \frac{1}{2})$, $(1., \frac{1}{2})$, $(2., -\frac{1}{3})$, $(2.82843, \frac{11}{48})$, $(2.44949, -\frac{3}{16})$, $(2.44949, \frac{5}{16})$, $(1.41421, -\frac{7}{16})$, $(1.41421, -\frac{7}{16})$\}

\item $c = 3$, $(d_i,\theta_i)$ = \{$(1., 0)$, $(1.73205, 0)$, $(1.73205, 0)$, $(1., 0)$, $(2., \frac{1}{6})$, $(1., \frac{1}{2})$, $(1.73205, \frac{1}{2})$, $(1.73205, \frac{1}{2})$, $(1., \frac{1}{2})$, $(2., -\frac{1}{3})$, $(1.73205, -\frac{1}{8})$, $(1.73205, \frac{3}{8})$, $(1., -\frac{3}{8})$, $(2., \frac{7}{24})$, $(1., -\frac{3}{8})$, $(1.73205, -\frac{1}{8})$, $(1.73205, \frac{3}{8})$, $(1., -\frac{3}{8})$, $(2., \frac{7}{24})$, $(1., -\frac{3}{8})$\}

\item $c = \frac{7}{2}$, $(d_i,\theta_i)$ = \{$(1., 0)$, $(1.73205, 0)$, $(1.73205, 0)$, $(1., 0)$, $(2., \frac{1}{6})$, $(1., \frac{1}{2})$, $(1.73205, \frac{1}{2})$, $(1.73205, \frac{1}{2})$, $(1., \frac{1}{2})$, $(2., -\frac{1}{3})$, $(2.44949, \frac{7}{16})$, $(2.44949, -\frac{1}{16})$, $(2.82843, \frac{17}{48})$, $(1.41421, -\frac{5}{16})$, $(1.41421, -\frac{5}{16})$\}

\item $c = 4$, $(d_i,\theta_i)$ = \{$(1., 0)$, $(1.73205, 0)$, $(1.73205, 0)$, $(1., 0)$, $(2., \frac{1}{6})$, $(1., \frac{1}{2})$, $(1.73205, \frac{1}{2})$, $(1.73205, \frac{1}{2})$, $(1., \frac{1}{2})$, $(2., -\frac{1}{3})$, $(1., -\frac{1}{4})$, $(1.73205, \frac{1}{2})$, $(1.73205, 0)$, $(1., -\frac{1}{4})$, $(2., \frac{5}{12})$, $(1., -\frac{1}{4})$, $(1.73205, \frac{1}{2})$, $(1.73205, 0)$, $(1., -\frac{1}{4})$, $(2., \frac{5}{12})$\}

\item $c = \frac{9}{2}$, $(d_i,\theta_i)$ = \{$(1., 0)$, $(1.73205, 0)$, $(1.73205, 0)$, $(1., 0)$, $(2., \frac{1}{6})$, $(1., \frac{1}{2})$, $(1.73205, \frac{1}{2})$, $(1.73205, \frac{1}{2})$, $(1., \frac{1}{2})$, $(2., -\frac{1}{3})$, $(2.82843, \frac{23}{48})$, $(2.44949, \frac{1}{16})$, $(2.44949, -\frac{7}{16})$, $(1.41421, -\frac{3}{16})$, $(1.41421, -\frac{3}{16})$\}

\item $c = 5$, $(d_i,\theta_i)$ = \{$(1., 0)$, $(1.73205, 0)$, $(1.73205, 0)$, $(1., 0)$, $(2., \frac{1}{6})$, $(1., \frac{1}{2})$, $(1.73205, \frac{1}{2})$, $(1.73205, \frac{1}{2})$, $(1., \frac{1}{2})$, $(2., -\frac{1}{3})$, $(1.73205, \frac{1}{8})$, $(1.73205, -\frac{3}{8})$, $(1., -\frac{1}{8})$, $(2., -\frac{11}{24})$, $(1., -\frac{1}{8})$, $(1.73205, \frac{1}{8})$, $(1.73205, -\frac{3}{8})$, $(1., -\frac{1}{8})$, $(2., -\frac{11}{24})$, $(1., -\frac{1}{8})$\}

\item $c = \frac{11}{2}$, $(d_i,\theta_i)$ = \{$(1., 0)$, $(1.73205, 0)$, $(1.73205, 0)$, $(1., 0)$, $(2., \frac{1}{6})$, $(1., \frac{1}{2})$, $(1.73205, \frac{1}{2})$, $(1.73205, \frac{1}{2})$, $(1., \frac{1}{2})$, $(2., -\frac{1}{3})$, $(2.44949, -\frac{5}{16})$, $(2.82843, -\frac{19}{48})$, $(2.44949, \frac{3}{16})$, $(1.41421, -\frac{1}{16})$, $(1.41421, -\frac{1}{16})$\}

\item $c = 6$, $(d_i,\theta_i)$ = \{$(1., 0)$, $(1.73205, \frac{1}{2})$, $(1.73205, \frac{1}{2})$, $(1., \frac{1}{2})$, $(2., \frac{1}{6})$, $(1., \frac{1}{2})$, $(1.73205, 0)$, $(1.73205, 0)$, $(1., 0)$, $(2., -\frac{1}{3})$, $(1., 0)$, $(1.73205, \frac{1}{4})$, $(1.73205, -\frac{1}{4})$, $(2., -\frac{1}{3})$, $(1., 0)$, $(1., 0)$, $(1.73205, \frac{1}{4})$, $(1.73205, -\frac{1}{4})$, $(2., -\frac{1}{3})$, $(1., 0)$\}

\item $c = \frac{13}{2}$, $(d_i,\theta_i)$ = \{$(1., 0)$, $(1.73205, \frac{1}{2})$, $(1.73205, \frac{1}{2})$, $(1., \frac{1}{2})$, $(2., \frac{1}{6})$, $(1., \frac{1}{2})$, $(1.73205, 0)$, $(1.73205, 0)$, $(1., 0)$, $(2., -\frac{1}{3})$, $(2.82843, -\frac{13}{48})$, $(2.44949, \frac{5}{16})$, $(2.44949, -\frac{3}{16})$, $(1.41421, \frac{1}{16})$, $(1.41421, \frac{1}{16})$\}

\item $c = 7$, $(d_i,\theta_i)$ = \{$(1., 0)$, $(1.73205, \frac{1}{2})$, $(1.73205, \frac{1}{2})$, $(1., \frac{1}{2})$, $(2., \frac{1}{6})$, $(1., \frac{1}{2})$, $(1.73205, 0)$, $(1.73205, 0)$, $(1., 0)$, $(2., -\frac{1}{3})$, $(1.73205, \frac{3}{8})$, $(1.73205, -\frac{1}{8})$, $(1., \frac{1}{8})$, $(1., \frac{1}{8})$, $(2., -\frac{5}{24})$, $(1.73205, \frac{3}{8})$, $(1.73205, -\frac{1}{8})$, $(1., \frac{1}{8})$, $(1., \frac{1}{8})$, $(2., -\frac{5}{24})$\}

\item $c = \frac{15}{2}$, $(d_i,\theta_i)$ = \{$(1., 0)$, $(1.73205, \frac{1}{2})$, $(1.73205, \frac{1}{2})$, $(1., \frac{1}{2})$, $(2., \frac{1}{6})$, $(1., \frac{1}{2})$, $(1.73205, 0)$, $(1.73205, 0)$, $(1., 0)$, $(2., -\frac{1}{3})$, $(2.82843, -\frac{7}{48})$, $(2.44949, \frac{7}{16})$, $(2.44949, -\frac{1}{16})$, $(1.41421, \frac{3}{16})$, $(1.41421, \frac{3}{16})$\}

\end{enumerate}

\paragraph*{Rank 10; \#4}

\begin{enumerate}

\item $c = 0$, $(d_i,\theta_i)$ = \{$(1., 0)$, $(1.73205, 0)$, $(1.73205, 0)$, $(1., 0)$, $(2., -\frac{1}{6})$, $(1., \frac{1}{2})$, $(1.73205, \frac{1}{2})$, $(1.73205, \frac{1}{2})$, $(1., \frac{1}{2})$, $(2., \frac{1}{3})$, $(1., -\frac{1}{4})$, $(1.73205, 0)$, $(1.73205, \frac{1}{2})$, $(2., \frac{1}{12})$, $(1., -\frac{1}{4})$, $(1., -\frac{1}{4})$, $(1.73205, 0)$, $(1.73205, \frac{1}{2})$, $(2., \frac{1}{12})$, $(1., -\frac{1}{4})$\}

\item $c = \frac{1}{2}$, $(d_i,\theta_i)$ = \{$(1., 0)$, $(1.73205, 0)$, $(1.73205, 0)$, $(1., 0)$, $(2., -\frac{1}{6})$, $(1., \frac{1}{2})$, $(1.73205, \frac{1}{2})$, $(1.73205, \frac{1}{2})$, $(1., \frac{1}{2})$, $(2., \frac{1}{3})$, $(2.82843, \frac{7}{48})$, $(2.44949, \frac{1}{16})$, $(2.44949, -\frac{7}{16})$, $(1.41421, -\frac{3}{16})$, $(1.41421, -\frac{3}{16})$\}

\item $c = 1$, $(d_i,\theta_i)$ = \{$(1., 0)$, $(1.73205, 0)$, $(1.73205, 0)$, $(1., 0)$, $(2., -\frac{1}{6})$, $(1., \frac{1}{2})$, $(1.73205, \frac{1}{2})$, $(1.73205, \frac{1}{2})$, $(1., \frac{1}{2})$, $(2., \frac{1}{3})$, $(1.73205, \frac{1}{8})$, $(1.73205, -\frac{3}{8})$, $(1., -\frac{1}{8})$, $(1., -\frac{1}{8})$, $(2., \frac{5}{24})$, $(1.73205, \frac{1}{8})$, $(1.73205, -\frac{3}{8})$, $(1., -\frac{1}{8})$, $(1., -\frac{1}{8})$, $(2., \frac{5}{24})$\}

\item $c = \frac{3}{2}$, $(d_i,\theta_i)$ = \{$(1., 0)$, $(1.73205, 0)$, $(1.73205, 0)$, $(1., 0)$, $(2., -\frac{1}{6})$, $(1., \frac{1}{2})$, $(1.73205, \frac{1}{2})$, $(1.73205, \frac{1}{2})$, $(1., \frac{1}{2})$, $(2., \frac{1}{3})$, $(2.82843, \frac{13}{48})$, $(2.44949, \frac{3}{16})$, $(2.44949, -\frac{5}{16})$, $(1.41421, -\frac{1}{16})$, $(1.41421, -\frac{1}{16})$\}

\item $c = 2$, $(d_i,\theta_i)$ = \{$(1., 0)$, $(1.73205, 0)$, $(1.73205, 0)$, $(1., 0)$, $(2., \frac{1}{3})$, $(1., \frac{1}{2})$, $(1.73205, \frac{1}{2})$, $(1.73205, \frac{1}{2})$, $(1., \frac{1}{2})$, $(2., -\frac{1}{6})$, $(1., 0)$, $(1.73205, -\frac{1}{4})$, $(1.73205, \frac{1}{4})$, $(1., 0)$, $(2., \frac{1}{3})$, $(1., 0)$, $(1.73205, -\frac{1}{4})$, $(1.73205, \frac{1}{4})$, $(1., 0)$, $(2., \frac{1}{3})$\}

\item $c = \frac{5}{2}$, $(d_i,\theta_i)$ = \{$(1., 0)$, $(1.73205, 0)$, $(1.73205, 0)$, $(1., 0)$, $(2., \frac{1}{3})$, $(1., \frac{1}{2})$, $(1.73205, \frac{1}{2})$, $(1.73205, \frac{1}{2})$, $(1., \frac{1}{2})$, $(2., -\frac{1}{6})$, $(2.44949, -\frac{3}{16})$, $(2.44949, \frac{5}{16})$, $(2.82843, \frac{19}{48})$, $(1.41421, \frac{1}{16})$, $(1.41421, \frac{1}{16})$\}

\item $c = 3$, $(d_i,\theta_i)$ = \{$(1., 0)$, $(1.73205, 0)$, $(1.73205, 0)$, $(1., 0)$, $(2., \frac{1}{3})$, $(1., \frac{1}{2})$, $(1.73205, \frac{1}{2})$, $(1.73205, \frac{1}{2})$, $(1., \frac{1}{2})$, $(2., -\frac{1}{6})$, $(1.73205, \frac{3}{8})$, $(1.73205, -\frac{1}{8})$, $(1., \frac{1}{8})$, $(2., \frac{11}{24})$, $(1., \frac{1}{8})$, $(1.73205, \frac{3}{8})$, $(1.73205, -\frac{1}{8})$, $(1., \frac{1}{8})$, $(2., \frac{11}{24})$, $(1., \frac{1}{8})$\}

\item $c = \frac{7}{2}$, $(d_i,\theta_i)$ = \{$(1., 0)$, $(1.73205, 0)$, $(1.73205, 0)$, $(1., 0)$, $(2., \frac{1}{3})$, $(1., \frac{1}{2})$, $(1.73205, \frac{1}{2})$, $(1.73205, \frac{1}{2})$, $(1., \frac{1}{2})$, $(2., -\frac{1}{6})$, $(2.82843, -\frac{23}{48})$, $(2.44949, \frac{7}{16})$, $(2.44949, -\frac{1}{16})$, $(1.41421, \frac{3}{16})$, $(1.41421, \frac{3}{16})$\}

\item $c = 4$, $(d_i,\theta_i)$ = \{$(1., 0)$, $(1.73205, 0)$, $(1.73205, 0)$, $(1., 0)$, $(2., \frac{1}{3})$, $(1., \frac{1}{2})$, $(1.73205, \frac{1}{2})$, $(1.73205, \frac{1}{2})$, $(1., \frac{1}{2})$, $(2., -\frac{1}{6})$, $(1., \frac{1}{4})$, $(1.73205, \frac{1}{2})$, $(1.73205, 0)$, $(1., \frac{1}{4})$, $(2., -\frac{5}{12})$, $(1., \frac{1}{4})$, $(1.73205, \frac{1}{2})$, $(1.73205, 0)$, $(1., \frac{1}{4})$, $(2., -\frac{5}{12})$\}

\item $c = \frac{9}{2}$, $(d_i,\theta_i)$ = \{$(1., 0)$, $(1.73205, 0)$, $(1.73205, 0)$, $(1., 0)$, $(2., \frac{1}{3})$, $(1., \frac{1}{2})$, $(1.73205, \frac{1}{2})$, $(1.73205, \frac{1}{2})$, $(1., \frac{1}{2})$, $(2., -\frac{1}{6})$, $(2.82843, -\frac{17}{48})$, $(2.44949, \frac{1}{16})$, $(2.44949, -\frac{7}{16})$, $(1.41421, \frac{5}{16})$, $(1.41421, \frac{5}{16})$\}

\item $c = 5$, $(d_i,\theta_i)$ = \{$(1., 0)$, $(1.73205, 0)$, $(1.73205, 0)$, $(1., 0)$, $(2., \frac{1}{3})$, $(1., \frac{1}{2})$, $(1.73205, \frac{1}{2})$, $(1.73205, \frac{1}{2})$, $(1., \frac{1}{2})$, $(2., -\frac{1}{6})$, $(1.73205, -\frac{3}{8})$, $(1.73205, \frac{1}{8})$, $(1., \frac{3}{8})$, $(1., \frac{3}{8})$, $(2., -\frac{7}{24})$, $(1.73205, -\frac{3}{8})$, $(1.73205, \frac{1}{8})$, $(1., \frac{3}{8})$, $(1., \frac{3}{8})$, $(2., -\frac{7}{24})$\}

\item $c = \frac{11}{2}$, $(d_i,\theta_i)$ = \{$(1., 0)$, $(1.73205, 0)$, $(1.73205, 0)$, $(1., 0)$, $(2., \frac{1}{3})$, $(1., \frac{1}{2})$, $(1.73205, \frac{1}{2})$, $(1.73205, \frac{1}{2})$, $(1., \frac{1}{2})$, $(2., -\frac{1}{6})$, $(2.82843, -\frac{11}{48})$, $(2.44949, \frac{3}{16})$, $(2.44949, -\frac{5}{16})$, $(1.41421, \frac{7}{16})$, $(1.41421, \frac{7}{16})$\}

\item $c = 6$, $(d_i,\theta_i)$ = \{$(1., 0)$, $(1.73205, \frac{1}{2})$, $(1.73205, \frac{1}{2})$, $(1., \frac{1}{2})$, $(2., \frac{1}{3})$, $(1., \frac{1}{2})$, $(1.73205, 0)$, $(1.73205, 0)$, $(1., 0)$, $(2., -\frac{1}{6})$, $(1., \frac{1}{2})$, $(1.73205, \frac{1}{4})$, $(1.73205, -\frac{1}{4})$, $(2., -\frac{1}{6})$, $(1., \frac{1}{2})$, $(1., \frac{1}{2})$, $(1.73205, \frac{1}{4})$, $(1.73205, -\frac{1}{4})$, $(2., -\frac{1}{6})$, $(1., \frac{1}{2})$\}

\item $c = \frac{13}{2}$, $(d_i,\theta_i)$ = \{$(1., 0)$, $(1.73205, \frac{1}{2})$, $(1.73205, \frac{1}{2})$, $(1., \frac{1}{2})$, $(2., \frac{1}{3})$, $(1., \frac{1}{2})$, $(1.73205, 0)$, $(1.73205, 0)$, $(1., 0)$, $(2., -\frac{1}{6})$, $(2.44949, -\frac{3}{16})$, $(2.82843, -\frac{5}{48})$, $(2.44949, \frac{5}{16})$, $(1.41421, -\frac{7}{16})$, $(1.41421, -\frac{7}{16})$\}

\item $c = 7$, $(d_i,\theta_i)$ = \{$(1., 0)$, $(1.73205, \frac{1}{2})$, $(1.73205, \frac{1}{2})$, $(1., \frac{1}{2})$, $(2., \frac{1}{3})$, $(1., \frac{1}{2})$, $(1.73205, 0)$, $(1.73205, 0)$, $(1., 0)$, $(2., -\frac{1}{6})$, $(1.73205, -\frac{1}{8})$, $(1.73205, \frac{3}{8})$, $(1., -\frac{3}{8})$, $(1., -\frac{3}{8})$, $(2., -\frac{1}{24})$, $(1.73205, -\frac{1}{8})$, $(1.73205, \frac{3}{8})$, $(1., -\frac{3}{8})$, $(1., -\frac{3}{8})$, $(2., -\frac{1}{24})$\}

\item $c = \frac{15}{2}$, $(d_i,\theta_i)$ = \{$(1., 0)$, $(1.73205, \frac{1}{2})$, $(1.73205, \frac{1}{2})$, $(1., \frac{1}{2})$, $(2., \frac{1}{3})$, $(1., \frac{1}{2})$, $(1.73205, 0)$, $(1.73205, 0)$, $(1., 0)$, $(2., -\frac{1}{6})$, $(2.82843, \frac{1}{48})$, $(2.44949, \frac{7}{16})$, $(2.44949, -\frac{1}{16})$, $(1.41421, -\frac{5}{16})$, $(1.41421, -\frac{5}{16})$\}

\end{enumerate}

\paragraph*{Rank 10; \#5}

\begin{enumerate}

\item $c = 0$, $(d_i,\theta_i)$ = \{$(1., 0)$, $(-1.73205, 0)$, $(-1.73205, 0)$, $(1., 0)$, $(2., \frac{1}{6})$, $(1., \frac{1}{2})$, $(-1.73205, \frac{1}{2})$, $(-1.73205, \frac{1}{2})$, $(1., \frac{1}{2})$, $(2., -\frac{1}{3})$, $(-1., \frac{1}{4})$, $(1.73205, 0)$, $(1.73205, \frac{1}{2})$, $(-2., -\frac{1}{12})$, $(-1., \frac{1}{4})$, $(-1., \frac{1}{4})$, $(1.73205, 0)$, $(1.73205, \frac{1}{2})$, $(-2., -\frac{1}{12})$, $(-1., \frac{1}{4})$\}

\item $c = \frac{1}{2}$, $(d_i,\theta_i)$ = \{$(1., 0)$, $(-1.73205, 0)$, $(-1.73205, 0)$, $(1., 0)$, $(2., \frac{1}{6})$, $(1., \frac{1}{2})$, $(-1.73205, \frac{1}{2})$, $(-1.73205, \frac{1}{2})$, $(1., \frac{1}{2})$, $(2., -\frac{1}{3})$, $(-2.44949, -\frac{7}{16})$, $(-2.44949, \frac{1}{16})$, $(2.82843, -\frac{1}{48})$, $(1.41421, \frac{5}{16})$, $(1.41421, \frac{5}{16})$\}

\item $c = 1$, $(d_i,\theta_i)$ = \{$(1., 0)$, $(-1.73205, 0)$, $(-1.73205, 0)$, $(1., 0)$, $(2., \frac{1}{6})$, $(1., \frac{1}{2})$, $(-1.73205, \frac{1}{2})$, $(-1.73205, \frac{1}{2})$, $(1., \frac{1}{2})$, $(2., -\frac{1}{3})$, $(-1.73205, -\frac{3}{8})$, $(-1.73205, \frac{1}{8})$, $(1., \frac{3}{8})$, $(1., \frac{3}{8})$, $(2., \frac{1}{24})$, $(-1.73205, -\frac{3}{8})$, $(-1.73205, \frac{1}{8})$, $(1., \frac{3}{8})$, $(1., \frac{3}{8})$, $(2., \frac{1}{24})$\}

\item $c = \frac{3}{2}$, $(d_i,\theta_i)$ = \{$(1., 0)$, $(-1.73205, 0)$, $(-1.73205, 0)$, $(1., 0)$, $(2., \frac{1}{6})$, $(1., \frac{1}{2})$, $(-1.73205, \frac{1}{2})$, $(-1.73205, \frac{1}{2})$, $(1., \frac{1}{2})$, $(2., -\frac{1}{3})$, $(2.82843, \frac{5}{48})$, $(-2.44949, \frac{3}{16})$, $(-2.44949, -\frac{5}{16})$, $(1.41421, \frac{7}{16})$, $(1.41421, \frac{7}{16})$\}

\item $c = 2$, $(d_i,\theta_i)$ = \{$(1., 0)$, $(-1.73205, 0)$, $(-1.73205, 0)$, $(1., 0)$, $(2., \frac{1}{6})$, $(1., \frac{1}{2})$, $(-1.73205, \frac{1}{2})$, $(-1.73205, \frac{1}{2})$, $(1., \frac{1}{2})$, $(2., -\frac{1}{3})$, $(1., \frac{1}{2})$, $(-1.73205, -\frac{1}{4})$, $(-1.73205, \frac{1}{4})$, $(2., \frac{1}{6})$, $(1., \frac{1}{2})$, $(1., \frac{1}{2})$, $(-1.73205, -\frac{1}{4})$, $(-1.73205, \frac{1}{4})$, $(2., \frac{1}{6})$, $(1., \frac{1}{2})$\}

\item $c = \frac{5}{2}$, $(d_i,\theta_i)$ = \{$(1., 0)$, $(-1.73205, 0)$, $(-1.73205, 0)$, $(1., 0)$, $(2., \frac{1}{6})$, $(1., \frac{1}{2})$, $(-1.73205, \frac{1}{2})$, $(-1.73205, \frac{1}{2})$, $(1., \frac{1}{2})$, $(2., -\frac{1}{3})$, $(2.82843, \frac{11}{48})$, $(-2.44949, -\frac{3}{16})$, $(-2.44949, \frac{5}{16})$, $(1.41421, -\frac{7}{16})$, $(1.41421, -\frac{7}{16})$\}

\item $c = 3$, $(d_i,\theta_i)$ = \{$(1., 0)$, $(-1.73205, 0)$, $(-1.73205, 0)$, $(1., 0)$, $(2., \frac{1}{6})$, $(1., \frac{1}{2})$, $(-1.73205, \frac{1}{2})$, $(-1.73205, \frac{1}{2})$, $(1., \frac{1}{2})$, $(2., -\frac{1}{3})$, $(-1.73205, -\frac{1}{8})$, $(-1.73205, \frac{3}{8})$, $(1., -\frac{3}{8})$, $(2., \frac{7}{24})$, $(1., -\frac{3}{8})$, $(-1.73205, -\frac{1}{8})$, $(-1.73205, \frac{3}{8})$, $(1., -\frac{3}{8})$, $(2., \frac{7}{24})$, $(1., -\frac{3}{8})$\}

\item $c = \frac{7}{2}$, $(d_i,\theta_i)$ = \{$(1., 0)$, $(-1.73205, 0)$, $(-1.73205, 0)$, $(1., 0)$, $(2., \frac{1}{6})$, $(1., \frac{1}{2})$, $(-1.73205, \frac{1}{2})$, $(-1.73205, \frac{1}{2})$, $(1., \frac{1}{2})$, $(2., -\frac{1}{3})$, $(-2.44949, \frac{7}{16})$, $(-2.44949, -\frac{1}{16})$, $(2.82843, \frac{17}{48})$, $(1.41421, -\frac{5}{16})$, $(1.41421, -\frac{5}{16})$\}

\item $c = 4$, $(d_i,\theta_i)$ = \{$(1., 0)$, $(-1.73205, 0)$, $(-1.73205, 0)$, $(1., 0)$, $(2., \frac{1}{6})$, $(1., \frac{1}{2})$, $(-1.73205, \frac{1}{2})$, $(-1.73205, \frac{1}{2})$, $(1., \frac{1}{2})$, $(2., -\frac{1}{3})$, $(1., -\frac{1}{4})$, $(-1.73205, \frac{1}{2})$, $(-1.73205, 0)$, $(1., -\frac{1}{4})$, $(2., \frac{5}{12})$, $(1., -\frac{1}{4})$, $(-1.73205, \frac{1}{2})$, $(-1.73205, 0)$, $(1., -\frac{1}{4})$, $(2., \frac{5}{12})$\}

\item $c = \frac{9}{2}$, $(d_i,\theta_i)$ = \{$(1., 0)$, $(-1.73205, 0)$, $(-1.73205, 0)$, $(1., 0)$, $(2., \frac{1}{6})$, $(1., \frac{1}{2})$, $(-1.73205, \frac{1}{2})$, $(-1.73205, \frac{1}{2})$, $(1., \frac{1}{2})$, $(2., -\frac{1}{3})$, $(2.82843, \frac{23}{48})$, $(-2.44949, \frac{1}{16})$, $(-2.44949, -\frac{7}{16})$, $(1.41421, -\frac{3}{16})$, $(1.41421, -\frac{3}{16})$\}

\item $c = 5$, $(d_i,\theta_i)$ = \{$(1., 0)$, $(-1.73205, 0)$, $(-1.73205, 0)$, $(1., 0)$, $(2., \frac{1}{6})$, $(1., \frac{1}{2})$, $(-1.73205, \frac{1}{2})$, $(-1.73205, \frac{1}{2})$, $(1., \frac{1}{2})$, $(2., -\frac{1}{3})$, $(-1.73205, \frac{1}{8})$, $(-1.73205, -\frac{3}{8})$, $(1., -\frac{1}{8})$, $(2., -\frac{11}{24})$, $(1., -\frac{1}{8})$, $(-1.73205, \frac{1}{8})$, $(-1.73205, -\frac{3}{8})$, $(1., -\frac{1}{8})$, $(2., -\frac{11}{24})$, $(1., -\frac{1}{8})$\}

\item $c = \frac{11}{2}$, $(d_i,\theta_i)$ = \{$(1., 0)$, $(-1.73205, 0)$, $(-1.73205, 0)$, $(1., 0)$, $(2., \frac{1}{6})$, $(1., \frac{1}{2})$, $(-1.73205, \frac{1}{2})$, $(-1.73205, \frac{1}{2})$, $(1., \frac{1}{2})$, $(2., -\frac{1}{3})$, $(-2.44949, -\frac{5}{16})$, $(2.82843, -\frac{19}{48})$, $(-2.44949, \frac{3}{16})$, $(1.41421, -\frac{1}{16})$, $(1.41421, -\frac{1}{16})$\}

\item $c = 6$, $(d_i,\theta_i)$ = \{$(1., 0)$, $(-1.73205, \frac{1}{2})$, $(-1.73205, \frac{1}{2})$, $(1., \frac{1}{2})$, $(2., \frac{1}{6})$, $(1., \frac{1}{2})$, $(-1.73205, 0)$, $(-1.73205, 0)$, $(1., 0)$, $(2., -\frac{1}{3})$, $(1., 0)$, $(-1.73205, \frac{1}{4})$, $(-1.73205, -\frac{1}{4})$, $(2., -\frac{1}{3})$, $(1., 0)$, $(1., 0)$, $(-1.73205, \frac{1}{4})$, $(-1.73205, -\frac{1}{4})$, $(2., -\frac{1}{3})$, $(1., 0)$\}

\item $c = \frac{13}{2}$, $(d_i,\theta_i)$ = \{$(1., 0)$, $(-1.73205, \frac{1}{2})$, $(-1.73205, \frac{1}{2})$, $(1., \frac{1}{2})$, $(2., \frac{1}{6})$, $(1., \frac{1}{2})$, $(-1.73205, 0)$, $(-1.73205, 0)$, $(1., 0)$, $(2., -\frac{1}{3})$, $(2.82843, -\frac{13}{48})$, $(-2.44949, \frac{5}{16})$, $(-2.44949, -\frac{3}{16})$, $(1.41421, \frac{1}{16})$, $(1.41421, \frac{1}{16})$\}

\item $c = 7$, $(d_i,\theta_i)$ = \{$(1., 0)$, $(-1.73205, \frac{1}{2})$, $(-1.73205, \frac{1}{2})$, $(1., \frac{1}{2})$, $(2., \frac{1}{6})$, $(1., \frac{1}{2})$, $(-1.73205, 0)$, $(-1.73205, 0)$, $(1., 0)$, $(2., -\frac{1}{3})$, $(-1.73205, \frac{3}{8})$, $(-1.73205, -\frac{1}{8})$, $(1., \frac{1}{8})$, $(1., \frac{1}{8})$, $(2., -\frac{5}{24})$, $(-1.73205, \frac{3}{8})$, $(-1.73205, -\frac{1}{8})$, $(1., \frac{1}{8})$, $(1., \frac{1}{8})$, $(2., -\frac{5}{24})$\}

\item $c = \frac{15}{2}$, $(d_i,\theta_i)$ = \{$(1., 0)$, $(-1.73205, \frac{1}{2})$, $(-1.73205, \frac{1}{2})$, $(1., \frac{1}{2})$, $(2., \frac{1}{6})$, $(1., \frac{1}{2})$, $(-1.73205, 0)$, $(-1.73205, 0)$, $(1., 0)$, $(2., -\frac{1}{3})$, $(2.82843, -\frac{7}{48})$, $(-2.44949, \frac{7}{16})$, $(-2.44949, -\frac{1}{16})$, $(1.41421, \frac{3}{16})$, $(1.41421, \frac{3}{16})$\}

\end{enumerate}

\paragraph*{Rank 10; \#6}

\begin{enumerate}

\item $c = 0$, $(d_i,\theta_i)$ = \{$(1., 0)$, $(-1.73205, 0)$, $(-1.73205, 0)$, $(1., 0)$, $(2., -\frac{1}{6})$, $(1., \frac{1}{2})$, $(-1.73205, \frac{1}{2})$, $(-1.73205, \frac{1}{2})$, $(1., \frac{1}{2})$, $(2., \frac{1}{3})$, $(1., -\frac{1}{4})$, $(-1.73205, 0)$, $(-1.73205, \frac{1}{2})$, $(2., \frac{1}{12})$, $(1., -\frac{1}{4})$, $(1., -\frac{1}{4})$, $(-1.73205, 0)$, $(-1.73205, \frac{1}{2})$, $(2., \frac{1}{12})$, $(1., -\frac{1}{4})$\}

\item $c = \frac{1}{2}$, $(d_i,\theta_i)$ = \{$(1., 0)$, $(-1.73205, 0)$, $(-1.73205, 0)$, $(1., 0)$, $(2., -\frac{1}{6})$, $(1., \frac{1}{2})$, $(-1.73205, \frac{1}{2})$, $(-1.73205, \frac{1}{2})$, $(1., \frac{1}{2})$, $(2., \frac{1}{3})$, $(2.82843, \frac{7}{48})$, $(-2.44949, \frac{1}{16})$, $(-2.44949, -\frac{7}{16})$, $(1.41421, -\frac{3}{16})$, $(1.41421, -\frac{3}{16})$\}

\item $c = 1$, $(d_i,\theta_i)$ = \{$(1., 0)$, $(-1.73205, 0)$, $(-1.73205, 0)$, $(1., 0)$, $(2., -\frac{1}{6})$, $(1., \frac{1}{2})$, $(-1.73205, \frac{1}{2})$, $(-1.73205, \frac{1}{2})$, $(1., \frac{1}{2})$, $(2., \frac{1}{3})$, $(-1.73205, \frac{1}{8})$, $(-1.73205, -\frac{3}{8})$, $(1., -\frac{1}{8})$, $(1., -\frac{1}{8})$, $(2., \frac{5}{24})$, $(-1.73205, \frac{1}{8})$, $(-1.73205, -\frac{3}{8})$, $(1., -\frac{1}{8})$, $(1., -\frac{1}{8})$, $(2., \frac{5}{24})$\}

\item $c = \frac{3}{2}$, $(d_i,\theta_i)$ = \{$(1., 0)$, $(-1.73205, 0)$, $(-1.73205, 0)$, $(1., 0)$, $(2., -\frac{1}{6})$, $(1., \frac{1}{2})$, $(-1.73205, \frac{1}{2})$, $(-1.73205, \frac{1}{2})$, $(1., \frac{1}{2})$, $(2., \frac{1}{3})$, $(2.82843, \frac{13}{48})$, $(-2.44949, \frac{3}{16})$, $(-2.44949, -\frac{5}{16})$, $(1.41421, -\frac{1}{16})$, $(1.41421, -\frac{1}{16})$\}

\item $c = 2$, $(d_i,\theta_i)$ = \{$(1., 0)$, $(-1.73205, 0)$, $(-1.73205, 0)$, $(1., 0)$, $(2., \frac{1}{3})$, $(1., \frac{1}{2})$, $(-1.73205, \frac{1}{2})$, $(-1.73205, \frac{1}{2})$, $(1., \frac{1}{2})$, $(2., -\frac{1}{6})$, $(1., 0)$, $(-1.73205, -\frac{1}{4})$, $(-1.73205, \frac{1}{4})$, $(1., 0)$, $(2., \frac{1}{3})$, $(1., 0)$, $(-1.73205, -\frac{1}{4})$, $(-1.73205, \frac{1}{4})$, $(1., 0)$, $(2., \frac{1}{3})$\}

\item $c = \frac{5}{2}$, $(d_i,\theta_i)$ = \{$(1., 0)$, $(-1.73205, 0)$, $(-1.73205, 0)$, $(1., 0)$, $(2., \frac{1}{3})$, $(1., \frac{1}{2})$, $(-1.73205, \frac{1}{2})$, $(-1.73205, \frac{1}{2})$, $(1., \frac{1}{2})$, $(2., -\frac{1}{6})$, $(-2.44949, -\frac{3}{16})$, $(-2.44949, \frac{5}{16})$, $(2.82843, \frac{19}{48})$, $(1.41421, \frac{1}{16})$, $(1.41421, \frac{1}{16})$\}

\item $c = 3$, $(d_i,\theta_i)$ = \{$(1., 0)$, $(-1.73205, 0)$, $(-1.73205, 0)$, $(1., 0)$, $(2., \frac{1}{3})$, $(1., \frac{1}{2})$, $(-1.73205, \frac{1}{2})$, $(-1.73205, \frac{1}{2})$, $(1., \frac{1}{2})$, $(2., -\frac{1}{6})$, $(-1.73205, \frac{3}{8})$, $(-1.73205, -\frac{1}{8})$, $(1., \frac{1}{8})$, $(2., \frac{11}{24})$, $(1., \frac{1}{8})$, $(-1.73205, \frac{3}{8})$, $(-1.73205, -\frac{1}{8})$, $(1., \frac{1}{8})$, $(2., \frac{11}{24})$, $(1., \frac{1}{8})$\}

\item $c = \frac{7}{2}$, $(d_i,\theta_i)$ = \{$(1., 0)$, $(-1.73205, 0)$, $(-1.73205, 0)$, $(1., 0)$, $(2., \frac{1}{3})$, $(1., \frac{1}{2})$, $(-1.73205, \frac{1}{2})$, $(-1.73205, \frac{1}{2})$, $(1., \frac{1}{2})$, $(2., -\frac{1}{6})$, $(2.82843, -\frac{23}{48})$, $(-2.44949, \frac{7}{16})$, $(-2.44949, -\frac{1}{16})$, $(1.41421, \frac{3}{16})$, $(1.41421, \frac{3}{16})$\}

\item $c = 4$, $(d_i,\theta_i)$ = \{$(1., 0)$, $(-1.73205, 0)$, $(-1.73205, 0)$, $(1., 0)$, $(2., \frac{1}{3})$, $(1., \frac{1}{2})$, $(-1.73205, \frac{1}{2})$, $(-1.73205, \frac{1}{2})$, $(1., \frac{1}{2})$, $(2., -\frac{1}{6})$, $(1., \frac{1}{4})$, $(-1.73205, \frac{1}{2})$, $(-1.73205, 0)$, $(1., \frac{1}{4})$, $(2., -\frac{5}{12})$, $(1., \frac{1}{4})$, $(-1.73205, \frac{1}{2})$, $(-1.73205, 0)$, $(1., \frac{1}{4})$, $(2., -\frac{5}{12})$\}

\item $c = \frac{9}{2}$, $(d_i,\theta_i)$ = \{$(1., 0)$, $(-1.73205, 0)$, $(-1.73205, 0)$, $(1., 0)$, $(2., \frac{1}{3})$, $(1., \frac{1}{2})$, $(-1.73205, \frac{1}{2})$, $(-1.73205, \frac{1}{2})$, $(1., \frac{1}{2})$, $(2., -\frac{1}{6})$, $(2.82843, -\frac{17}{48})$, $(-2.44949, \frac{1}{16})$, $(-2.44949, -\frac{7}{16})$, $(1.41421, \frac{5}{16})$, $(1.41421, \frac{5}{16})$\}

\item $c = 5$, $(d_i,\theta_i)$ = \{$(1., 0)$, $(-1.73205, 0)$, $(-1.73205, 0)$, $(1., 0)$, $(2., \frac{1}{3})$, $(1., \frac{1}{2})$, $(-1.73205, \frac{1}{2})$, $(-1.73205, \frac{1}{2})$, $(1., \frac{1}{2})$, $(2., -\frac{1}{6})$, $(-1.73205, -\frac{3}{8})$, $(-1.73205, \frac{1}{8})$, $(1., \frac{3}{8})$, $(1., \frac{3}{8})$, $(2., -\frac{7}{24})$, $(-1.73205, -\frac{3}{8})$, $(-1.73205, \frac{1}{8})$, $(1., \frac{3}{8})$, $(1., \frac{3}{8})$, $(2., -\frac{7}{24})$\}

\item $c = \frac{11}{2}$, $(d_i,\theta_i)$ = \{$(1., 0)$, $(-1.73205, 0)$, $(-1.73205, 0)$, $(1., 0)$, $(2., \frac{1}{3})$, $(1., \frac{1}{2})$, $(-1.73205, \frac{1}{2})$, $(-1.73205, \frac{1}{2})$, $(1., \frac{1}{2})$, $(2., -\frac{1}{6})$, $(2.82843, -\frac{11}{48})$, $(-2.44949, \frac{3}{16})$, $(-2.44949, -\frac{5}{16})$, $(1.41421, \frac{7}{16})$, $(1.41421, \frac{7}{16})$\}

\item $c = 6$, $(d_i,\theta_i)$ = \{$(1., 0)$, $(-1.73205, \frac{1}{2})$, $(-1.73205, \frac{1}{2})$, $(1., \frac{1}{2})$, $(2., \frac{1}{3})$, $(1., \frac{1}{2})$, $(-1.73205, 0)$, $(-1.73205, 0)$, $(1., 0)$, $(2., -\frac{1}{6})$, $(1., \frac{1}{2})$, $(-1.73205, \frac{1}{4})$, $(-1.73205, -\frac{1}{4})$, $(2., -\frac{1}{6})$, $(1., \frac{1}{2})$, $(1., \frac{1}{2})$, $(-1.73205, \frac{1}{4})$, $(-1.73205, -\frac{1}{4})$, $(2., -\frac{1}{6})$, $(1., \frac{1}{2})$\}

\item $c = \frac{13}{2}$, $(d_i,\theta_i)$ = \{$(1., 0)$, $(-1.73205, \frac{1}{2})$, $(-1.73205, \frac{1}{2})$, $(1., \frac{1}{2})$, $(2., \frac{1}{3})$, $(1., \frac{1}{2})$, $(-1.73205, 0)$, $(-1.73205, 0)$, $(1., 0)$, $(2., -\frac{1}{6})$, $(-2.44949, -\frac{3}{16})$, $(2.82843, -\frac{5}{48})$, $(-2.44949, \frac{5}{16})$, $(1.41421, -\frac{7}{16})$, $(1.41421, -\frac{7}{16})$\}

\item $c = 7$, $(d_i,\theta_i)$ = \{$(1., 0)$, $(-1.73205, \frac{1}{2})$, $(-1.73205, \frac{1}{2})$, $(1., \frac{1}{2})$, $(2., \frac{1}{3})$, $(1., \frac{1}{2})$, $(-1.73205, 0)$, $(-1.73205, 0)$, $(1., 0)$, $(2., -\frac{1}{6})$, $(-1.73205, -\frac{1}{8})$, $(-1.73205, \frac{3}{8})$, $(1., -\frac{3}{8})$, $(1., -\frac{3}{8})$, $(2., -\frac{1}{24})$, $(-1.73205, -\frac{1}{8})$, $(-1.73205, \frac{3}{8})$, $(1., -\frac{3}{8})$, $(1., -\frac{3}{8})$, $(2., -\frac{1}{24})$\}

\item $c = \frac{15}{2}$, $(d_i,\theta_i)$ = \{$(1., 0)$, $(-1.73205, \frac{1}{2})$, $(-1.73205, \frac{1}{2})$, $(1., \frac{1}{2})$, $(2., \frac{1}{3})$, $(1., \frac{1}{2})$, $(-1.73205, 0)$, $(-1.73205, 0)$, $(1., 0)$, $(2., -\frac{1}{6})$, $(2.82843, \frac{1}{48})$, $(-2.44949, \frac{7}{16})$, $(-2.44949, -\frac{1}{16})$, $(1.41421, -\frac{5}{16})$, $(1.41421, -\frac{5}{16})$\}

\end{enumerate}

\paragraph*{Rank 10; \#7}

\begin{enumerate}

\item $c = 0$, $(d_i,\theta_i)$ = \{$(1., 0)$, $(1., 0)$, $(2., -\frac{1}{3})$, $(1.73205, \frac{1}{4})$, $(1.73205, -\frac{1}{4})$, $(1., \frac{1}{2})$, $(1., \frac{1}{2})$, $(2., \frac{1}{6})$, $(1.73205, \frac{1}{4})$, $(1.73205, -\frac{1}{4})$, $(1., \frac{1}{4})$, $(1., \frac{1}{4})$, $(2., -\frac{1}{12})$, $(1.73205, -\frac{1}{4})$, $(1.73205, \frac{1}{4})$, $(1., \frac{1}{4})$, $(1., \frac{1}{4})$, $(2., -\frac{1}{12})$, $(1.73205, \frac{1}{4})$, $(1.73205, -\frac{1}{4})$\}

\item $c = \frac{1}{2}$, $(d_i,\theta_i)$ = \{$(1., 0)$, $(1., 0)$, $(2., -\frac{1}{3})$, $(1.73205, \frac{1}{4})$, $(1.73205, -\frac{1}{4})$, $(1., \frac{1}{2})$, $(1., \frac{1}{2})$, $(2., \frac{1}{6})$, $(1.73205, \frac{1}{4})$, $(1.73205, -\frac{1}{4})$, $(1.41421, \frac{5}{16})$, $(1.41421, \frac{5}{16})$, $(2.82843, -\frac{1}{48})$, $(2.44949, -\frac{3}{16})$, $(2.44949, \frac{5}{16})$\}

\item $c = 1$, $(d_i,\theta_i)$ = \{$(1., 0)$, $(1., 0)$, $(2., -\frac{1}{3})$, $(1.73205, \frac{1}{4})$, $(1.73205, -\frac{1}{4})$, $(1., \frac{1}{2})$, $(1., \frac{1}{2})$, $(2., \frac{1}{6})$, $(1.73205, \frac{1}{4})$, $(1.73205, -\frac{1}{4})$, $(1., \frac{3}{8})$, $(1., \frac{3}{8})$, $(2., \frac{1}{24})$, $(1.73205, -\frac{1}{8})$, $(1.73205, \frac{3}{8})$, $(1., \frac{3}{8})$, $(1., \frac{3}{8})$, $(2., \frac{1}{24})$, $(1.73205, \frac{3}{8})$, $(1.73205, -\frac{1}{8})$\}

\item $c = \frac{3}{2}$, $(d_i,\theta_i)$ = \{$(1., 0)$, $(1., 0)$, $(2., -\frac{1}{3})$, $(1.73205, \frac{1}{4})$, $(1.73205, -\frac{1}{4})$, $(1., \frac{1}{2})$, $(1., \frac{1}{2})$, $(2., \frac{1}{6})$, $(1.73205, \frac{1}{4})$, $(1.73205, -\frac{1}{4})$, $(1.41421, \frac{7}{16})$, $(1.41421, \frac{7}{16})$, $(2.82843, \frac{5}{48})$, $(2.44949, -\frac{1}{16})$, $(2.44949, \frac{7}{16})$\}

\item $c = 2$, $(d_i,\theta_i)$ = \{$(1., 0)$, $(1., 0)$, $(2., -\frac{1}{3})$, $(1.73205, \frac{1}{4})$, $(1.73205, -\frac{1}{4})$, $(1., \frac{1}{2})$, $(1., \frac{1}{2})$, $(2., \frac{1}{6})$, $(1.73205, \frac{1}{4})$, $(1.73205, -\frac{1}{4})$, $(1., \frac{1}{2})$, $(1., \frac{1}{2})$, $(2., \frac{1}{6})$, $(1.73205, 0)$, $(1.73205, \frac{1}{2})$, $(1., \frac{1}{2})$, $(1., \frac{1}{2})$, $(2., \frac{1}{6})$, $(1.73205, \frac{1}{2})$, $(1.73205, 0)$\}

\item $c = \frac{5}{2}$, $(d_i,\theta_i)$ = \{$(1., 0)$, $(1., 0)$, $(2., -\frac{1}{3})$, $(1.73205, \frac{1}{4})$, $(1.73205, -\frac{1}{4})$, $(1., \frac{1}{2})$, $(1., \frac{1}{2})$, $(2., \frac{1}{6})$, $(1.73205, \frac{1}{4})$, $(1.73205, -\frac{1}{4})$, $(1.41421, -\frac{7}{16})$, $(1.41421, -\frac{7}{16})$, $(2.82843, \frac{11}{48})$, $(2.44949, \frac{1}{16})$, $(2.44949, -\frac{7}{16})$\}

\item $c = 3$, $(d_i,\theta_i)$ = \{$(1., 0)$, $(1., 0)$, $(2., -\frac{1}{3})$, $(1.73205, \frac{1}{4})$, $(1.73205, -\frac{1}{4})$, $(1., \frac{1}{2})$, $(1., \frac{1}{2})$, $(2., \frac{1}{6})$, $(1.73205, \frac{1}{4})$, $(1.73205, -\frac{1}{4})$, $(1., -\frac{3}{8})$, $(1., -\frac{3}{8})$, $(2., \frac{7}{24})$, $(1.73205, \frac{1}{8})$, $(1.73205, -\frac{3}{8})$, $(1., -\frac{3}{8})$, $(1., -\frac{3}{8})$, $(2., \frac{7}{24})$, $(1.73205, -\frac{3}{8})$, $(1.73205, \frac{1}{8})$\}

\item $c = \frac{7}{2}$, $(d_i,\theta_i)$ = \{$(1., 0)$, $(1., 0)$, $(2., -\frac{1}{3})$, $(1.73205, \frac{1}{4})$, $(1.73205, -\frac{1}{4})$, $(1., \frac{1}{2})$, $(1., \frac{1}{2})$, $(2., \frac{1}{6})$, $(1.73205, \frac{1}{4})$, $(1.73205, -\frac{1}{4})$, $(1.41421, -\frac{5}{16})$, $(1.41421, -\frac{5}{16})$, $(2.82843, \frac{17}{48})$, $(2.44949, \frac{3}{16})$, $(2.44949, -\frac{5}{16})$\}

\item $c = 4$, $(d_i,\theta_i)$ = \{$(1., 0)$, $(1., 0)$, $(2., -\frac{1}{3})$, $(1.73205, \frac{1}{4})$, $(1.73205, -\frac{1}{4})$, $(1., \frac{1}{2})$, $(1., \frac{1}{2})$, $(2., \frac{1}{6})$, $(1.73205, \frac{1}{4})$, $(1.73205, -\frac{1}{4})$, $(1., -\frac{1}{4})$, $(1., -\frac{1}{4})$, $(2., \frac{5}{12})$, $(1.73205, \frac{1}{4})$, $(1.73205, -\frac{1}{4})$, $(1., -\frac{1}{4})$, $(1., -\frac{1}{4})$, $(2., \frac{5}{12})$, $(1.73205, -\frac{1}{4})$, $(1.73205, \frac{1}{4})$\}

\item $c = \frac{9}{2}$, $(d_i,\theta_i)$ = \{$(1., 0)$, $(1., 0)$, $(2., -\frac{1}{3})$, $(1.73205, \frac{1}{4})$, $(1.73205, -\frac{1}{4})$, $(1., \frac{1}{2})$, $(1., \frac{1}{2})$, $(2., \frac{1}{6})$, $(1.73205, \frac{1}{4})$, $(1.73205, -\frac{1}{4})$, $(1.41421, -\frac{3}{16})$, $(1.41421, -\frac{3}{16})$, $(2.82843, \frac{23}{48})$, $(2.44949, \frac{5}{16})$, $(2.44949, -\frac{3}{16})$\}

\item $c = 5$, $(d_i,\theta_i)$ = \{$(1., 0)$, $(1., 0)$, $(2., -\frac{1}{3})$, $(1.73205, \frac{1}{4})$, $(1.73205, -\frac{1}{4})$, $(1., \frac{1}{2})$, $(1., \frac{1}{2})$, $(2., \frac{1}{6})$, $(1.73205, \frac{1}{4})$, $(1.73205, -\frac{1}{4})$, $(1., -\frac{1}{8})$, $(1., -\frac{1}{8})$, $(2., -\frac{11}{24})$, $(1.73205, \frac{3}{8})$, $(1.73205, -\frac{1}{8})$, $(1., -\frac{1}{8})$, $(1., -\frac{1}{8})$, $(2., -\frac{11}{24})$, $(1.73205, -\frac{1}{8})$, $(1.73205, \frac{3}{8})$\}

\item $c = \frac{11}{2}$, $(d_i,\theta_i)$ = \{$(1., 0)$, $(1., 0)$, $(2., -\frac{1}{3})$, $(1.73205, \frac{1}{4})$, $(1.73205, -\frac{1}{4})$, $(1., \frac{1}{2})$, $(1., \frac{1}{2})$, $(2., \frac{1}{6})$, $(1.73205, \frac{1}{4})$, $(1.73205, -\frac{1}{4})$, $(1.41421, -\frac{1}{16})$, $(1.41421, -\frac{1}{16})$, $(2.82843, -\frac{19}{48})$, $(2.44949, \frac{7}{16})$, $(2.44949, -\frac{1}{16})$\}

\item $c = 6$, $(d_i,\theta_i)$ = \{$(1., 0)$, $(1., 0)$, $(2., -\frac{1}{3})$, $(1.73205, \frac{1}{4})$, $(1.73205, -\frac{1}{4})$, $(1., \frac{1}{2})$, $(1., \frac{1}{2})$, $(2., \frac{1}{6})$, $(1.73205, \frac{1}{4})$, $(1.73205, -\frac{1}{4})$, $(1., 0)$, $(1., 0)$, $(2., -\frac{1}{3})$, $(1.73205, \frac{1}{2})$, $(1.73205, 0)$, $(1., 0)$, $(1., 0)$, $(2., -\frac{1}{3})$, $(1.73205, 0)$, $(1.73205, \frac{1}{2})$\}

\item $c = \frac{13}{2}$, $(d_i,\theta_i)$ = \{$(1., 0)$, $(1., 0)$, $(2., -\frac{1}{3})$, $(1.73205, \frac{1}{4})$, $(1.73205, -\frac{1}{4})$, $(1., \frac{1}{2})$, $(1., \frac{1}{2})$, $(2., \frac{1}{6})$, $(1.73205, \frac{1}{4})$, $(1.73205, -\frac{1}{4})$, $(1.41421, \frac{1}{16})$, $(1.41421, \frac{1}{16})$, $(2.82843, -\frac{13}{48})$, $(2.44949, -\frac{7}{16})$, $(2.44949, \frac{1}{16})$\}

\item $c = 7$, $(d_i,\theta_i)$ = \{$(1., 0)$, $(1., 0)$, $(2., -\frac{1}{3})$, $(1.73205, \frac{1}{4})$, $(1.73205, -\frac{1}{4})$, $(1., \frac{1}{2})$, $(1., \frac{1}{2})$, $(2., \frac{1}{6})$, $(1.73205, \frac{1}{4})$, $(1.73205, -\frac{1}{4})$, $(1., \frac{1}{8})$, $(1., \frac{1}{8})$, $(2., -\frac{5}{24})$, $(1.73205, -\frac{3}{8})$, $(1.73205, \frac{1}{8})$, $(1., \frac{1}{8})$, $(1., \frac{1}{8})$, $(2., -\frac{5}{24})$, $(1.73205, \frac{1}{8})$, $(1.73205, -\frac{3}{8})$\}

\item $c = \frac{15}{2}$, $(d_i,\theta_i)$ = \{$(1., 0)$, $(1., 0)$, $(2., -\frac{1}{3})$, $(1.73205, \frac{1}{4})$, $(1.73205, -\frac{1}{4})$, $(1., \frac{1}{2})$, $(1., \frac{1}{2})$, $(2., \frac{1}{6})$, $(1.73205, \frac{1}{4})$, $(1.73205, -\frac{1}{4})$, $(1.41421, \frac{3}{16})$, $(1.41421, \frac{3}{16})$, $(2.82843, -\frac{7}{48})$, $(2.44949, -\frac{5}{16})$, $(2.44949, \frac{3}{16})$\}

\end{enumerate}

\paragraph*{Rank 10; \#8}

\begin{enumerate}

\item $c = 0$, $(d_i,\theta_i)$ = \{$(1., 0)$, $(1., 0)$, $(2., \frac{1}{3})$, $(1.73205, -\frac{1}{4})$, $(1.73205, \frac{1}{4})$, $(1., \frac{1}{2})$, $(1., \frac{1}{2})$, $(2., -\frac{1}{6})$, $(1.73205, -\frac{1}{4})$, $(1.73205, \frac{1}{4})$, $(1., -\frac{1}{4})$, $(1., -\frac{1}{4})$, $(2., \frac{1}{12})$, $(1.73205, \frac{1}{4})$, $(1.73205, -\frac{1}{4})$, $(1., -\frac{1}{4})$, $(1., -\frac{1}{4})$, $(2., \frac{1}{12})$, $(1.73205, -\frac{1}{4})$, $(1.73205, \frac{1}{4})$\}

\item $c = \frac{1}{2}$, $(d_i,\theta_i)$ = \{$(1., 0)$, $(1., 0)$, $(2., \frac{1}{3})$, $(1.73205, -\frac{1}{4})$, $(1.73205, \frac{1}{4})$, $(1., \frac{1}{2})$, $(1., \frac{1}{2})$, $(2., -\frac{1}{6})$, $(1.73205, -\frac{1}{4})$, $(1.73205, \frac{1}{4})$, $(1.41421, -\frac{3}{16})$, $(1.41421, -\frac{3}{16})$, $(2.82843, \frac{7}{48})$, $(2.44949, \frac{5}{16})$, $(2.44949, -\frac{3}{16})$\}

\item $c = 1$, $(d_i,\theta_i)$ = \{$(1., 0)$, $(1., 0)$, $(2., \frac{1}{3})$, $(1.73205, -\frac{1}{4})$, $(1.73205, \frac{1}{4})$, $(1., \frac{1}{2})$, $(1., \frac{1}{2})$, $(2., -\frac{1}{6})$, $(1.73205, -\frac{1}{4})$, $(1.73205, \frac{1}{4})$, $(1., -\frac{1}{8})$, $(1., -\frac{1}{8})$, $(2., \frac{5}{24})$, $(1.73205, \frac{3}{8})$, $(1.73205, -\frac{1}{8})$, $(1., -\frac{1}{8})$, $(1., -\frac{1}{8})$, $(2., \frac{5}{24})$, $(1.73205, -\frac{1}{8})$, $(1.73205, \frac{3}{8})$\}

\item $c = \frac{3}{2}$, $(d_i,\theta_i)$ = \{$(1., 0)$, $(1., 0)$, $(2., \frac{1}{3})$, $(1.73205, -\frac{1}{4})$, $(1.73205, \frac{1}{4})$, $(1., \frac{1}{2})$, $(1., \frac{1}{2})$, $(2., -\frac{1}{6})$, $(1.73205, -\frac{1}{4})$, $(1.73205, \frac{1}{4})$, $(1.41421, -\frac{1}{16})$, $(1.41421, -\frac{1}{16})$, $(2.82843, \frac{13}{48})$, $(2.44949, \frac{7}{16})$, $(2.44949, -\frac{1}{16})$\}

\item $c = 2$, $(d_i,\theta_i)$ = \{$(1., 0)$, $(1., 0)$, $(2., \frac{1}{3})$, $(1.73205, -\frac{1}{4})$, $(1.73205, \frac{1}{4})$, $(1., \frac{1}{2})$, $(1., \frac{1}{2})$, $(2., -\frac{1}{6})$, $(1.73205, -\frac{1}{4})$, $(1.73205, \frac{1}{4})$, $(1., 0)$, $(1., 0)$, $(2., \frac{1}{3})$, $(1.73205, \frac{1}{2})$, $(1.73205, 0)$, $(1., 0)$, $(1., 0)$, $(2., \frac{1}{3})$, $(1.73205, 0)$, $(1.73205, \frac{1}{2})$\}

\item $c = \frac{5}{2}$, $(d_i,\theta_i)$ = \{$(1., 0)$, $(1., 0)$, $(2., \frac{1}{3})$, $(1.73205, -\frac{1}{4})$, $(1.73205, \frac{1}{4})$, $(1., \frac{1}{2})$, $(1., \frac{1}{2})$, $(2., -\frac{1}{6})$, $(1.73205, -\frac{1}{4})$, $(1.73205, \frac{1}{4})$, $(1.41421, \frac{1}{16})$, $(1.41421, \frac{1}{16})$, $(2.82843, \frac{19}{48})$, $(2.44949, -\frac{7}{16})$, $(2.44949, \frac{1}{16})$\}

\item $c = 3$, $(d_i,\theta_i)$ = \{$(1., 0)$, $(1., 0)$, $(2., \frac{1}{3})$, $(1.73205, -\frac{1}{4})$, $(1.73205, \frac{1}{4})$, $(1., \frac{1}{2})$, $(1., \frac{1}{2})$, $(2., -\frac{1}{6})$, $(1.73205, -\frac{1}{4})$, $(1.73205, \frac{1}{4})$, $(1., \frac{1}{8})$, $(1., \frac{1}{8})$, $(2., \frac{11}{24})$, $(1.73205, -\frac{3}{8})$, $(1.73205, \frac{1}{8})$, $(1., \frac{1}{8})$, $(1., \frac{1}{8})$, $(2., \frac{11}{24})$, $(1.73205, \frac{1}{8})$, $(1.73205, -\frac{3}{8})$\}

\item $c = \frac{7}{2}$, $(d_i,\theta_i)$ = \{$(1., 0)$, $(1., 0)$, $(2., \frac{1}{3})$, $(1.73205, -\frac{1}{4})$, $(1.73205, \frac{1}{4})$, $(1., \frac{1}{2})$, $(1., \frac{1}{2})$, $(2., -\frac{1}{6})$, $(1.73205, -\frac{1}{4})$, $(1.73205, \frac{1}{4})$, $(1.41421, \frac{3}{16})$, $(1.41421, \frac{3}{16})$, $(2.82843, -\frac{23}{48})$, $(2.44949, -\frac{5}{16})$, $(2.44949, \frac{3}{16})$\}

\item $c = 4$, $(d_i,\theta_i)$ = \{$(1., 0)$, $(1., 0)$, $(2., \frac{1}{3})$, $(1.73205, -\frac{1}{4})$, $(1.73205, \frac{1}{4})$, $(1., \frac{1}{2})$, $(1., \frac{1}{2})$, $(2., -\frac{1}{6})$, $(1.73205, -\frac{1}{4})$, $(1.73205, \frac{1}{4})$, $(1., \frac{1}{4})$, $(1., \frac{1}{4})$, $(2., -\frac{5}{12})$, $(1.73205, -\frac{1}{4})$, $(1.73205, \frac{1}{4})$, $(1., \frac{1}{4})$, $(1., \frac{1}{4})$, $(2., -\frac{5}{12})$, $(1.73205, \frac{1}{4})$, $(1.73205, -\frac{1}{4})$\}

\item $c = \frac{9}{2}$, $(d_i,\theta_i)$ = \{$(1., 0)$, $(1., 0)$, $(2., \frac{1}{3})$, $(1.73205, -\frac{1}{4})$, $(1.73205, \frac{1}{4})$, $(1., \frac{1}{2})$, $(1., \frac{1}{2})$, $(2., -\frac{1}{6})$, $(1.73205, -\frac{1}{4})$, $(1.73205, \frac{1}{4})$, $(1.41421, \frac{5}{16})$, $(1.41421, \frac{5}{16})$, $(2.82843, -\frac{17}{48})$, $(2.44949, -\frac{3}{16})$, $(2.44949, \frac{5}{16})$\}

\item $c = 5$, $(d_i,\theta_i)$ = \{$(1., 0)$, $(1., 0)$, $(2., \frac{1}{3})$, $(1.73205, -\frac{1}{4})$, $(1.73205, \frac{1}{4})$, $(1., \frac{1}{2})$, $(1., \frac{1}{2})$, $(2., -\frac{1}{6})$, $(1.73205, -\frac{1}{4})$, $(1.73205, \frac{1}{4})$, $(1., \frac{3}{8})$, $(1., \frac{3}{8})$, $(2., -\frac{7}{24})$, $(1.73205, -\frac{1}{8})$, $(1.73205, \frac{3}{8})$, $(1., \frac{3}{8})$, $(1., \frac{3}{8})$, $(2., -\frac{7}{24})$, $(1.73205, \frac{3}{8})$, $(1.73205, -\frac{1}{8})$\}

\item $c = \frac{11}{2}$, $(d_i,\theta_i)$ = \{$(1., 0)$, $(1., 0)$, $(2., \frac{1}{3})$, $(1.73205, -\frac{1}{4})$, $(1.73205, \frac{1}{4})$, $(1., \frac{1}{2})$, $(1., \frac{1}{2})$, $(2., -\frac{1}{6})$, $(1.73205, -\frac{1}{4})$, $(1.73205, \frac{1}{4})$, $(1.41421, \frac{7}{16})$, $(1.41421, \frac{7}{16})$, $(2.82843, -\frac{11}{48})$, $(2.44949, -\frac{1}{16})$, $(2.44949, \frac{7}{16})$\}

\item $c = 6$, $(d_i,\theta_i)$ = \{$(1., 0)$, $(1., 0)$, $(2., \frac{1}{3})$, $(1.73205, -\frac{1}{4})$, $(1.73205, \frac{1}{4})$, $(1., \frac{1}{2})$, $(1., \frac{1}{2})$, $(2., -\frac{1}{6})$, $(1.73205, -\frac{1}{4})$, $(1.73205, \frac{1}{4})$, , $(1., \frac{1}{2})$, $(1., \frac{1}{2})$, $(2., -\frac{1}{6})$, $(1.73205, 0)$, $(1.73205, \frac{1}{2})$, $(1., \frac{1}{2})$, $(1., \frac{1}{2})$, $(2., -\frac{1}{6})$, $(1.73205, \frac{1}{2})$, $(1.73205, 0)$\}

\item $c = \frac{13}{2}$, $(d_i,\theta_i)$ = \{$(1., 0)$, $(1., 0)$, $(2., \frac{1}{3})$, $(1.73205, -\frac{1}{4})$, $(1.73205, \frac{1}{4})$, $(1., \frac{1}{2})$, $(1., \frac{1}{2})$, $(2., -\frac{1}{6})$, $(1.73205, -\frac{1}{4})$, $(1.73205, \frac{1}{4})$, $(1.41421, -\frac{7}{16})$, $(1.41421, -\frac{7}{16})$, $(2.82843, -\frac{5}{48})$, $(2.44949, \frac{1}{16})$, $(2.44949, -\frac{7}{16})$\}

\item $c = 7$, $(d_i,\theta_i)$ = \{$(1., 0)$, $(1., 0)$, $(2., \frac{1}{3})$, $(1.73205, -\frac{1}{4})$, $(1.73205, \frac{1}{4})$, $(1., \frac{1}{2})$, $(1., \frac{1}{2})$, $(2., -\frac{1}{6})$, $(1.73205, -\frac{1}{4})$, $(1.73205, \frac{1}{4})$, $(1., -\frac{3}{8})$, $(1., -\frac{3}{8})$, $(2., -\frac{1}{24})$, $(1.73205, \frac{1}{8})$, $(1.73205, -\frac{3}{8})$, $(1., -\frac{3}{8})$, $(1., -\frac{3}{8})$, $(2., -\frac{1}{24})$, $(1.73205, -\frac{3}{8})$, $(1.73205, \frac{1}{8})$\}

\item $c = \frac{15}{2}$, $(d_i,\theta_i)$ = \{$(1., 0)$, $(1., 0)$, $(2., \frac{1}{3})$, $(1.73205, -\frac{1}{4})$, $(1.73205, \frac{1}{4})$, $(1., \frac{1}{2})$, $(1., \frac{1}{2})$, $(2., -\frac{1}{6})$, $(1.73205, -\frac{1}{4})$, $(1.73205, \frac{1}{4})$, $(1.41421, -\frac{5}{16})$, $(1.41421, -\frac{5}{16})$, $(2.82843, \frac{1}{48})$, $(2.44949, \frac{3}{16})$, $(2.44949, -\frac{5}{16})$\}

\end{enumerate}

\paragraph*{Rank 10; \#9}

\begin{enumerate}

\item $c = 0$, $(d_i,\theta_i)$ = \{$(1., 0)$, $(1., 0)$, $(2., -\frac{1}{3})$, $(-1.73205, \frac{1}{4})$, $(-1.73205, -\frac{1}{4})$, $(1., \frac{1}{2})$, $(1., \frac{1}{2})$, $(2., \frac{1}{6})$, $(-1.73205, \frac{1}{4})$, $(-1.73205, -\frac{1}{4})$, $(1., \frac{1}{4})$, $(1., \frac{1}{4})$, $(2., -\frac{1}{12})$, $(-1.73205, -\frac{1}{4})$, $(-1.73205, \frac{1}{4})$, $(1., \frac{1}{4})$, $(1., \frac{1}{4})$, $(2., -\frac{1}{12})$, $(-1.73205, \frac{1}{4})$, $(-1.73205, -\frac{1}{4})$\}

\item $c = \frac{1}{2}$, $(d_i,\theta_i)$ = \{$(1., 0)$, $(1., 0)$, $(2., -\frac{1}{3})$, $(-1.73205, \frac{1}{4})$, $(-1.73205, -\frac{1}{4})$, $(1., \frac{1}{2})$, $(1., \frac{1}{2})$, $(2., \frac{1}{6})$, $(-1.73205, \frac{1}{4})$, $(-1.73205, -\frac{1}{4})$, $(1.41421, \frac{5}{16})$, $(1.41421, \frac{5}{16})$, $(2.82843, -\frac{1}{48})$, $(-2.44949, -\frac{3}{16})$, $(-2.44949, \frac{5}{16})$\}

\item $c = 1$, $(d_i,\theta_i)$ = \{$(1., 0)$, $(1., 0)$, $(2., -\frac{1}{3})$, $(-1.73205, \frac{1}{4})$, $(-1.73205, -\frac{1}{4})$, $(1., \frac{1}{2})$, $(1., \frac{1}{2})$, $(2., \frac{1}{6})$, $(-1.73205, \frac{1}{4})$, $(-1.73205, -\frac{1}{4})$, $(1., \frac{3}{8})$, $(1., \frac{3}{8})$, $(2., \frac{1}{24})$, $(-1.73205, -\frac{1}{8})$, $(-1.73205, \frac{3}{8})$, $(1., \frac{3}{8})$, $(1., \frac{3}{8})$, $(2., \frac{1}{24})$, $(-1.73205, \frac{3}{8})$, $(-1.73205, -\frac{1}{8})$\}

\item $c = \frac{3}{2}$, $(d_i,\theta_i)$ = \{$(1., 0)$, $(1., 0)$, $(2., -\frac{1}{3})$, $(-1.73205, \frac{1}{4})$, $(-1.73205, -\frac{1}{4})$, $(1., \frac{1}{2})$, $(1., \frac{1}{2})$, $(2., \frac{1}{6})$, $(-1.73205, \frac{1}{4})$, $(-1.73205, -\frac{1}{4})$, $(1.41421, \frac{7}{16})$, $(1.41421, \frac{7}{16})$, $(2.82843, \frac{5}{48})$, $(-2.44949, -\frac{1}{16})$, $(-2.44949, \frac{7}{16})$\}

\item $c = 2$, $(d_i,\theta_i)$ = \{$(1., 0)$, $(1., 0)$, $(2., -\frac{1}{3})$, $(-1.73205, \frac{1}{4})$, $(-1.73205, -\frac{1}{4})$, $(1., \frac{1}{2})$, $(1., \frac{1}{2})$, $(2., \frac{1}{6})$, $(-1.73205, \frac{1}{4})$, $(-1.73205, -\frac{1}{4})$, $(1., \frac{1}{2})$, $(1., \frac{1}{2})$, $(2., \frac{1}{6})$, $(-1.73205, 0)$, $(-1.73205, \frac{1}{2})$, $(1., \frac{1}{2})$, $(1., \frac{1}{2})$, $(2., \frac{1}{6})$, $(-1.73205, \frac{1}{2})$, $(-1.73205, 0)$\}

\item $c = \frac{5}{2}$, $(d_i,\theta_i)$ = \{$(1., 0)$, $(1., 0)$, $(2., -\frac{1}{3})$, $(-1.73205, \frac{1}{4})$, $(-1.73205, -\frac{1}{4})$, $(1., \frac{1}{2})$, $(1., \frac{1}{2})$, $(2., \frac{1}{6})$, $(-1.73205, \frac{1}{4})$, $(-1.73205, -\frac{1}{4})$, $(1.41421, -\frac{7}{16})$, $(1.41421, -\frac{7}{16})$, $(2.82843, \frac{11}{48})$, $(-2.44949, \frac{1}{16})$, $(-2.44949, -\frac{7}{16})$\}

\item $c = 3$, $(d_i,\theta_i)$ = \{$(1., 0)$, $(1., 0)$, $(2., -\frac{1}{3})$, $(-1.73205, \frac{1}{4})$, $(-1.73205, -\frac{1}{4})$, $(1., \frac{1}{2})$, $(1., \frac{1}{2})$, $(2., \frac{1}{6})$, $(-1.73205, \frac{1}{4})$, $(-1.73205, -\frac{1}{4})$, $(1., -\frac{3}{8})$, $(1., -\frac{3}{8})$, $(2., \frac{7}{24})$, $(-1.73205, \frac{1}{8})$, $(-1.73205, -\frac{3}{8})$, $(1., -\frac{3}{8})$, $(1., -\frac{3}{8})$, $(2., \frac{7}{24})$, $(-1.73205, -\frac{3}{8})$, $(-1.73205, \frac{1}{8})$\}

\item $c = \frac{7}{2}$, $(d_i,\theta_i)$ = \{$(1., 0)$, $(1., 0)$, $(2., -\frac{1}{3})$, $(-1.73205, \frac{1}{4})$, $(-1.73205, -\frac{1}{4})$, $(1., \frac{1}{2})$, $(1., \frac{1}{2})$, $(2., \frac{1}{6})$, $(-1.73205, \frac{1}{4})$, $(-1.73205, -\frac{1}{4})$, $(1.41421, -\frac{5}{16})$, $(1.41421, -\frac{5}{16})$, $(2.82843, \frac{17}{48})$, $(-2.44949, \frac{3}{16})$, $(-2.44949, -\frac{5}{16})$\}

\item $c = 4$, $(d_i,\theta_i)$ = \{$(1., 0)$, $(1., 0)$, $(2., -\frac{1}{3})$, $(-1.73205, \frac{1}{4})$, $(-1.73205, -\frac{1}{4})$, $(1., \frac{1}{2})$, $(1., \frac{1}{2})$, $(2., \frac{1}{6})$, $(-1.73205, \frac{1}{4})$, $(-1.73205, -\frac{1}{4})$, $(1., -\frac{1}{4})$, $(1., -\frac{1}{4})$, $(2., \frac{5}{12})$, $(-1.73205, \frac{1}{4})$, $(-1.73205, -\frac{1}{4})$, $(1., -\frac{1}{4})$, $(1., -\frac{1}{4})$, $(2., \frac{5}{12})$, $(-1.73205, -\frac{1}{4})$, $(-1.73205, \frac{1}{4})$\}

\item $c = \frac{9}{2}$, $(d_i,\theta_i)$ = \{$(1., 0)$, $(1., 0)$, $(2., -\frac{1}{3})$, $(-1.73205, \frac{1}{4})$, $(-1.73205, -\frac{1}{4})$, $(1., \frac{1}{2})$, $(1., \frac{1}{2})$, $(2., \frac{1}{6})$, $(-1.73205, \frac{1}{4})$, $(-1.73205, -\frac{1}{4})$, $(1.41421, -\frac{3}{16})$, $(1.41421, -\frac{3}{16})$, $(2.82843, \frac{23}{48})$, $(-2.44949, \frac{5}{16})$, $(-2.44949, -\frac{3}{16})$\}

\item $c = 5$, $(d_i,\theta_i)$ = \{$(1., 0)$, $(1., 0)$, $(2., -\frac{1}{3})$, $(-1.73205, \frac{1}{4})$, $(-1.73205, -\frac{1}{4})$, $(1., \frac{1}{2})$, $(1., \frac{1}{2})$, $(2., \frac{1}{6})$, $(-1.73205, \frac{1}{4})$, $(-1.73205, -\frac{1}{4})$, $(1., -\frac{1}{8})$, $(1., -\frac{1}{8})$, $(2., -\frac{11}{24})$, $(-1.73205, \frac{3}{8})$, $(-1.73205, -\frac{1}{8})$, $(1., -\frac{1}{8})$, $(1., -\frac{1}{8})$, $(2., -\frac{11}{24})$, $(-1.73205, -\frac{1}{8})$, $(-1.73205, \frac{3}{8})$\}

\item $c = \frac{11}{2}$, $(d_i,\theta_i)$ = \{ $(1., 0)$, $(1., 0)$, $(2., -\frac{1}{3})$, $(-1.73205, \frac{1}{4})$, $(-1.73205, -\frac{1}{4})$, $(1., \frac{1}{2})$, $(1., \frac{1}{2})$, $(2., \frac{1}{6})$, $(-1.73205, \frac{1}{4})$, $(-1.73205, -\frac{1}{4})$, $(1.41421, -\frac{1}{16})$, $(1.41421, -\frac{1}{16})$, $(2.82843, -\frac{19}{48})$, $(-2.44949, \frac{7}{16})$, $(-2.44949, -\frac{1}{16})$\}

\item $c = 6$, $(d_i,\theta_i)$ = \{$(1., 0)$, $(1., 0)$, $(2., -\frac{1}{3})$, $(-1.73205, \frac{1}{4})$, $(-1.73205, -\frac{1}{4})$, $(1., \frac{1}{2})$, $(1., \frac{1}{2})$, $(2., \frac{1}{6})$, $(-1.73205, \frac{1}{4})$, $(-1.73205, -\frac{1}{4})$, $(1., 0)$, $(1., 0)$, $(2., -\frac{1}{3})$, $(-1.73205, \frac{1}{2})$, $(-1.73205, 0)$, $(1., 0)$, $(1., 0)$, $(2., -\frac{1}{3})$, $(-1.73205, 0)$, $(-1.73205, \frac{1}{2})$\}

\item $c = \frac{13}{2}$, $(d_i,\theta_i)$ = \{$(1., 0)$, $(1., 0)$, $(2., -\frac{1}{3})$, $(-1.73205, \frac{1}{4})$, $(-1.73205, -\frac{1}{4})$, $(1., \frac{1}{2})$, $(1., \frac{1}{2})$, $(2., \frac{1}{6})$, $(-1.73205, \frac{1}{4})$, $(-1.73205, -\frac{1}{4})$, $(1.41421, \frac{1}{16})$, $(1.41421, \frac{1}{16})$, $(2.82843, -\frac{13}{48})$, $(-2.44949, -\frac{7}{16})$, $(-2.44949, \frac{1}{16})$\}

\item $c = 7$, $(d_i,\theta_i)$ = \{$(1., 0)$, $(1., 0)$, $(2., -\frac{1}{3})$, $(-1.73205, \frac{1}{4})$, $(-1.73205, -\frac{1}{4})$, $(1., \frac{1}{2})$, $(1., \frac{1}{2})$, $(2., \frac{1}{6})$, $(-1.73205, \frac{1}{4})$, $(-1.73205, -\frac{1}{4})$, $(1., \frac{1}{8})$, $(1., \frac{1}{8})$, $(2., -\frac{5}{24})$, $(-1.73205, -\frac{3}{8})$, $(-1.73205, \frac{1}{8})$, $(1., \frac{1}{8})$, $(1., \frac{1}{8})$, $(2., -\frac{5}{24})$, $(-1.73205, \frac{1}{8})$, $(-1.73205, -\frac{3}{8})$\}

\item $c = \frac{15}{2}$, $(d_i,\theta_i)$ = \{$(1., 0)$, $(1., 0)$, $(2., -\frac{1}{3})$, $(-1.73205, \frac{1}{4})$, $(-1.73205, -\frac{1}{4})$, $(1., \frac{1}{2})$, $(1., \frac{1}{2})$, $(2., \frac{1}{6})$, $(-1.73205, \frac{1}{4})$, $(-1.73205, -\frac{1}{4})$, $(1.41421, \frac{3}{16})$, $(1.41421, \frac{3}{16})$, $(2.82843, -\frac{7}{48})$, $(-2.44949, -\frac{5}{16})$, $(-2.44949, \frac{3}{16})$\}

\end{enumerate}

\paragraph*{Rank 10; \#10}

\begin{enumerate}

\item $c = 0$, $(d_i,\theta_i)$ = \{$(1., 0)$, $(1., 0)$, $(2., \frac{1}{3})$, $(-1.73205, -\frac{1}{4})$, $(-1.73205, \frac{1}{4})$, $(1., \frac{1}{2})$, $(1., \frac{1}{2})$, $(2., -\frac{1}{6})$, $(-1.73205, -\frac{1}{4})$, $(-1.73205, \frac{1}{4})$, $(1., -\frac{1}{4})$, $(1., -\frac{1}{4})$, $(2., \frac{1}{12})$, $(-1.73205, \frac{1}{4})$, $(-1.73205, -\frac{1}{4})$, $(1., -\frac{1}{4})$, $(1., -\frac{1}{4})$, $(2., \frac{1}{12})$, $(-1.73205, -\frac{1}{4})$, $(-1.73205, \frac{1}{4})$\}

\item $c = \frac{1}{2}$, $(d_i,\theta_i)$ = \{$(1., 0)$, $(1., 0)$, $(2., \frac{1}{3})$, $(-1.73205, -\frac{1}{4})$, $(-1.73205, \frac{1}{4})$, $(1., \frac{1}{2})$, $(1., \frac{1}{2})$, $(2., -\frac{1}{6})$, $(-1.73205, -\frac{1}{4})$, $(-1.73205, \frac{1}{4})$, $(1.41421, -\frac{3}{16})$, $(1.41421, -\frac{3}{16})$, $(2.82843, \frac{7}{48})$, $(-2.44949, \frac{5}{16})$, $(-2.44949, -\frac{3}{16})$\}

\item $c = 1$, $(d_i,\theta_i)$ = \{$(1., 0)$, $(1., 0)$, $(2., \frac{1}{3})$, $(-1.73205, -\frac{1}{4})$, $(-1.73205, \frac{1}{4})$, $(1., \frac{1}{2})$, $(1., \frac{1}{2})$, $(2., -\frac{1}{6})$, $(-1.73205, -\frac{1}{4})$, $(-1.73205, \frac{1}{4})$, $(1., -\frac{1}{8})$, $(1., -\frac{1}{8})$, $(2., \frac{5}{24})$, $(-1.73205, \frac{3}{8})$, $(-1.73205, -\frac{1}{8})$, $(1., -\frac{1}{8})$, $(1., -\frac{1}{8})$, $(2., \frac{5}{24})$, $(-1.73205, -\frac{1}{8})$, $(-1.73205, \frac{3}{8})$\}

\item $c = \frac{3}{2}$, $(d_i,\theta_i)$ = \{$(1., 0)$, $(1., 0)$, $(2., \frac{1}{3})$, $(-1.73205, -\frac{1}{4})$, $(-1.73205, \frac{1}{4})$, $(1., \frac{1}{2})$, $(1., \frac{1}{2})$, $(2., -\frac{1}{6})$, $(-1.73205, -\frac{1}{4})$, $(-1.73205, \frac{1}{4})$, $(1.41421, -\frac{1}{16})$, $(1.41421, -\frac{1}{16})$, $(2.82843, \frac{13}{48})$, $(-2.44949, \frac{7}{16})$, $(-2.44949, -\frac{1}{16})$\}

\item $c = 2$, $(d_i,\theta_i)$ = \{$(1., 0)$, $(1., 0)$, $(2., \frac{1}{3})$, $(-1.73205, -\frac{1}{4})$, $(-1.73205, \frac{1}{4})$, $(1., \frac{1}{2})$, $(1., \frac{1}{2})$, $(2., -\frac{1}{6})$, $(-1.73205, -\frac{1}{4})$, $(-1.73205, \frac{1}{4})$, $(1., 0)$, $(1., 0)$, $(2., \frac{1}{3})$, $(-1.73205, \frac{1}{2})$, $(-1.73205, 0)$, $(1., 0)$, $(1., 0)$, $(2., \frac{1}{3})$, $(-1.73205, 0)$, $(-1.73205, \frac{1}{2})$\}

\item $c = \frac{5}{2}$, $(d_i,\theta_i)$ = \{$(1., 0)$, $(1., 0)$, $(2., \frac{1}{3})$, $(-1.73205, -\frac{1}{4})$, $(-1.73205, \frac{1}{4})$, $(1., \frac{1}{2})$, $(1., \frac{1}{2})$, $(2., -\frac{1}{6})$, $(-1.73205, -\frac{1}{4})$, $(-1.73205, \frac{1}{4})$, $(1.41421, \frac{1}{16})$, $(1.41421, \frac{1}{16})$, $(2.82843, \frac{19}{48})$, $(-2.44949, -\frac{7}{16})$, $(-2.44949, \frac{1}{16})$\}

\item $c = 3$, $(d_i,\theta_i)$ = \{$(1., 0)$, $(1., 0)$, $(2., \frac{1}{3})$, $(-1.73205, -\frac{1}{4})$, $(-1.73205, \frac{1}{4})$, $(1., \frac{1}{2})$, $(1., \frac{1}{2})$, $(2., -\frac{1}{6})$, $(-1.73205, -\frac{1}{4})$, $(-1.73205, \frac{1}{4})$, $(1., \frac{1}{8})$, $(1., \frac{1}{8})$, $(2., \frac{11}{24})$, $(-1.73205, -\frac{3}{8})$, $(-1.73205, \frac{1}{8})$, $(1., \frac{1}{8})$, $(1., \frac{1}{8})$, $(2., \frac{11}{24})$, $(-1.73205, \frac{1}{8})$, $(-1.73205, -\frac{3}{8})$\}

\item $c = \frac{7}{2}$, $(d_i,\theta_i)$ = \{$(1., 0)$, $(1., 0)$, $(2., \frac{1}{3})$, $(-1.73205, -\frac{1}{4})$, $(-1.73205, \frac{1}{4})$, $(1., \frac{1}{2})$, $(1., \frac{1}{2})$, $(2., -\frac{1}{6})$, $(-1.73205, -\frac{1}{4})$, $(-1.73205, \frac{1}{4})$, $(1.41421, \frac{3}{16})$, $(1.41421, \frac{3}{16})$, $(2.82843, -\frac{23}{48})$, $(-2.44949, -\frac{5}{16})$, $(-2.44949, \frac{3}{16})$\}

\item $c = 4$, $(d_i,\theta_i)$ = \{$(1., 0)$, $(1., 0)$, $(2., \frac{1}{3})$, $(-1.73205, -\frac{1}{4})$, $(-1.73205, \frac{1}{4})$, $(1., \frac{1}{2})$, $(1., \frac{1}{2})$, $(2., -\frac{1}{6})$, $(-1.73205, -\frac{1}{4})$, $(-1.73205, \frac{1}{4})$, $(1., \frac{1}{4})$, $(1., \frac{1}{4})$, $(2., -\frac{5}{12})$, $(-1.73205, -\frac{1}{4})$, $(-1.73205, \frac{1}{4})$, $(1., \frac{1}{4})$, $(1., \frac{1}{4})$, $(2., -\frac{5}{12})$, $(-1.73205, \frac{1}{4})$, $(-1.73205, -\frac{1}{4})$\}

\item $c = \frac{9}{2}$, $(d_i,\theta_i)$ = \{$(1., 0)$, $(1., 0)$, $(2., \frac{1}{3})$, $(-1.73205, -\frac{1}{4})$, $(-1.73205, \frac{1}{4})$, $(1., \frac{1}{2})$, $(1., \frac{1}{2})$, $(2., -\frac{1}{6})$, $(-1.73205, -\frac{1}{4})$, $(-1.73205, \frac{1}{4})$, $(1.41421, \frac{5}{16})$, $(1.41421, \frac{5}{16})$, $(2.82843, -\frac{17}{48})$, $(-2.44949, -\frac{3}{16})$, $(-2.44949, \frac{5}{16})$\}

\item $c = 5$, $(d_i,\theta_i)$ = \{$(1., 0)$, $(1., 0)$, $(2., \frac{1}{3})$, $(-1.73205, -\frac{1}{4})$, $(-1.73205, \frac{1}{4})$, $(1., \frac{1}{2})$, $(1., \frac{1}{2})$, $(2., -\frac{1}{6})$, $(-1.73205, -\frac{1}{4})$, $(-1.73205, \frac{1}{4})$, $(1., \frac{3}{8})$, $(1., \frac{3}{8})$, $(2., -\frac{7}{24})$, $(-1.73205, -\frac{1}{8})$, $(-1.73205, \frac{3}{8})$, $(1., \frac{3}{8})$, $(1., \frac{3}{8})$, $(2., -\frac{7}{24})$, $(-1.73205, \frac{3}{8})$, $(-1.73205, -\frac{1}{8})$\}

\item $c = \frac{11}{2}$, $(d_i,\theta_i)$ = \{$(1., 0)$, $(1., 0)$, $(2., \frac{1}{3})$, $(-1.73205, -\frac{1}{4})$, $(-1.73205, \frac{1}{4})$, $(1., \frac{1}{2})$, $(1., \frac{1}{2})$, $(2., -\frac{1}{6})$, $(-1.73205, -\frac{1}{4})$, $(-1.73205, \frac{1}{4})$, $(1.41421, \frac{7}{16})$, $(1.41421, \frac{7}{16})$, $(2.82843, -\frac{11}{48})$, $(-2.44949, -\frac{1}{16})$, $(-2.44949, \frac{7}{16})$\}

\item $c = 6$, $(d_i,\theta_i)$ = \{$(1., 0)$, $(1., 0)$, $(2., \frac{1}{3})$, $(-1.73205, -\frac{1}{4})$, $(-1.73205, \frac{1}{4})$, $(1., \frac{1}{2})$, $(1., \frac{1}{2})$, $(2., -\frac{1}{6})$, $(-1.73205, -\frac{1}{4})$, $(-1.73205, \frac{1}{4})$, $(1., \frac{1}{2})$, $(1., \frac{1}{2})$, $(2., -\frac{1}{6})$, $(-1.73205, 0)$, $(-1.73205, \frac{1}{2})$, $(1., \frac{1}{2})$, $(1., \frac{1}{2})$, $(2., -\frac{1}{6})$, $(-1.73205, \frac{1}{2})$, $(-1.73205, 0)$\}

\item $c = \frac{13}{2}$, $(d_i,\theta_i)$ = \{$(1., 0)$, $(1., 0)$, $(2., \frac{1}{3})$, $(-1.73205, -\frac{1}{4})$, $(-1.73205, \frac{1}{4})$, $(1., \frac{1}{2})$, $(1., \frac{1}{2})$, $(2., -\frac{1}{6})$, $(-1.73205, -\frac{1}{4})$, $(-1.73205, \frac{1}{4})$, $(1.41421, -\frac{7}{16})$, $(1.41421, -\frac{7}{16})$, $(2.82843, -\frac{5}{48})$, $(-2.44949, \frac{1}{16})$, $(-2.44949, -\frac{7}{16})$\}

\item $c = 7$, $(d_i,\theta_i)$ = \{$(1., 0)$, $(1., 0)$, $(2., \frac{1}{3})$, $(-1.73205, -\frac{1}{4})$, $(-1.73205, \frac{1}{4})$, $(1., \frac{1}{2})$, $(1., \frac{1}{2})$, $(2., -\frac{1}{6})$, $(-1.73205, -\frac{1}{4})$, $(-1.73205, \frac{1}{4})$, $(1., -\frac{3}{8})$, $(1., -\frac{3}{8})$, $(2., -\frac{1}{24})$, $(-1.73205, \frac{1}{8})$, $(-1.73205, -\frac{3}{8})$, $(1., -\frac{3}{8})$, $(1., -\frac{3}{8})$, $(2., -\frac{1}{24})$, $(-1.73205, -\frac{3}{8})$, $(-1.73205, \frac{1}{8})$\}

\item $c = \frac{15}{2}$, $(d_i,\theta_i)$ = \{$(1., 0)$, $(1., 0)$, $(2., \frac{1}{3})$, $(-1.73205, -\frac{1}{4})$, $(-1.73205, \frac{1}{4})$, $(1., \frac{1}{2})$, $(1., \frac{1}{2})$, $(2., -\frac{1}{6})$, $(-1.73205, -\frac{1}{4})$, $(-1.73205, \frac{1}{4})$, $(1.41421, -\frac{5}{16})$, $(1.41421, -\frac{5}{16})$, $(2.82843, \frac{1}{48})$, $(-2.44949, \frac{3}{16})$, $(-2.44949, -\frac{5}{16})$\}

\end{enumerate}

\paragraph*{Rank 10; \#19}

\begin{enumerate}

\item $c = 0$, $(d_i,\theta_i)$ = \{$(1., 0)$, $(1., 0)$, $(3.16228, \frac{1}{16})$, $(2., \frac{1}{10})$, $(2., -\frac{1}{10})$, $(1., \frac{1}{2})$, $(1., \frac{1}{2})$, $(3.16228, -\frac{7}{16})$, $(2., -\frac{2}{5})$, $(2., \frac{2}{5})$, $(2., \frac{1}{2})$, $(3.16228, \frac{7}{16})$, $(3.16228, -\frac{1}{16})$, $(2., -\frac{1}{10})$, $(2., \frac{1}{10})$, $(2., -\frac{1}{10})$, $(2., \frac{1}{10})$\}

\item $c = \frac{1}{2}$, $(d_i,\theta_i)$ = \{$(1., 0)$, $(1., 0)$, $(3.16228, \frac{1}{16})$, $(2., \frac{1}{10})$, $(2., -\frac{1}{10})$, $(1., \frac{1}{2})$, $(1., \frac{1}{2})$, $(3.16228, -\frac{7}{16})$, $(2., -\frac{2}{5})$, $(2., \frac{2}{5})$, $(2.82843, \frac{13}{80})$, $(2.82843, -\frac{3}{80})$, $(2.23607, 0)$, $(2.23607, \frac{1}{2})$, $(1.41421, -\frac{7}{16})$, $(2.23607, 0)$, $(2.23607, \frac{1}{2})$, $(1.41421, -\frac{7}{16})$\}

\item $c = 1$, $(d_i,\theta_i)$ = \{$(1., 0)$, $(1., 0)$, $(3.16228, \frac{1}{16})$, $(2., \frac{1}{10})$, $(2., -\frac{1}{10})$, $(1., \frac{1}{2})$, $(1., \frac{1}{2})$, $(3.16228, -\frac{7}{16})$, $(2., -\frac{2}{5})$, $(2., \frac{2}{5})$, $(3.16228, -\frac{7}{16})$, $(3.16228, \frac{1}{16})$, $(2., -\frac{3}{8})$, $(2., \frac{1}{40})$, $(2., \frac{9}{40})$, $(2., \frac{1}{40})$, $(2., \frac{9}{40})$\}

\item $c = \frac{3}{2}$, $(d_i,\theta_i)$ = \{$(1., 0)$, $(1., 0)$, $(3.16228, \frac{1}{16})$, $(2., \frac{1}{10})$, $(2., -\frac{1}{10})$, $(1., \frac{1}{2})$, $(1., \frac{1}{2})$, $(3.16228, -\frac{7}{16})$, $(2., -\frac{2}{5})$, $(2., \frac{2}{5})$, $(2.82843, \frac{23}{80})$, $(2.82843, \frac{7}{80})$, $(2.23607, \frac{1}{8})$, $(2.23607, -\frac{3}{8})$, $(1.41421, -\frac{5}{16})$, $(2.23607, \frac{1}{8})$, $(2.23607, -\frac{3}{8})$, $(1.41421, -\frac{5}{16})$\}

\item $c = 2$, $(d_i,\theta_i)$ = \{$(1., 0)$, $(1., 0)$, $(3.16228, \frac{1}{16})$, $(2., \frac{1}{10})$, $(2., -\frac{1}{10})$, $(1., \frac{1}{2})$, $(1., \frac{1}{2})$, $(3.16228, -\frac{7}{16})$, $(2., -\frac{2}{5})$, $(2., \frac{2}{5})$, $(2., -\frac{1}{4})$, $(3.16228, -\frac{5}{16})$, $(3.16228, \frac{3}{16})$, $(2., \frac{3}{20})$, $(2., \frac{7}{20})$, $(2., \frac{3}{20})$, $(2., \frac{7}{20})$\}

\item $c = \frac{5}{2}$, $(d_i,\theta_i)$ = \{$(1., 0)$, $(1., 0)$, $(3.16228, \frac{1}{16})$, $(2., \frac{1}{10})$, $(2., -\frac{1}{10})$, $(1., \frac{1}{2})$, $(1., \frac{1}{2})$, $(3.16228, -\frac{7}{16})$, $(2., -\frac{2}{5})$, $(2., \frac{2}{5})$, $(2.82843, \frac{33}{80})$, $(2.82843, \frac{17}{80})$, $(2.23607, -\frac{1}{4})$, $(2.23607, \frac{1}{4})$, $(1.41421, -\frac{3}{16})$, $(2.23607, -\frac{1}{4})$, $(2.23607, \frac{1}{4})$, $(1.41421, -\frac{3}{16})$\}

\item $c = 3$, $(d_i,\theta_i)$ = \{$(1., 0)$, $(1., 0)$, $(3.16228, \frac{1}{16})$, $(2., \frac{1}{10})$, $(2., -\frac{1}{10})$, $(1., \frac{1}{2})$, $(1., \frac{1}{2})$, $(3.16228, -\frac{7}{16})$, $(2., -\frac{2}{5})$, $(2., \frac{2}{5})$, $(3.16228, -\frac{3}{16})$, $(3.16228, \frac{5}{16})$, $(2., -\frac{1}{8})$, $(2., \frac{11}{40})$, $(2., \frac{19}{40})$, $(2., \frac{11}{40})$, $(2., \frac{19}{40})$\}

\item $c = \frac{7}{2}$, $(d_i,\theta_i)$ = \{$(1., 0)$, $(1., 0)$, $(3.16228, \frac{1}{16})$, $(2., \frac{1}{10})$, $(2., -\frac{1}{10})$, $(1., \frac{1}{2})$, $(1., \frac{1}{2})$, $(3.16228, -\frac{7}{16})$, $(2., -\frac{2}{5})$, $(2., \frac{2}{5})$, $(2.82843, \frac{27}{80})$, $(2.82843, -\frac{37}{80})$, $(2.23607, -\frac{1}{8})$, $(2.23607, \frac{3}{8})$, $(1.41421, -\frac{1}{16})$, $(2.23607, -\frac{1}{8})$, $(2.23607, \frac{3}{8})$, $(1.41421, -\frac{1}{16})$\}

\item $c = 4$, $(d_i,\theta_i)$ = \{$(1., 0)$, $(1., 0)$, $(3.16228, \frac{1}{16})$, $(2., \frac{1}{10})$, $(2., \frac{2}{5})$, $(1., \frac{1}{2})$, $(1., \frac{1}{2})$, $(3.16228, -\frac{7}{16})$, $(2., -\frac{2}{5})$, $(2., -\frac{1}{10})$, $(3.16228, -\frac{1}{16})$, $(2., 0)$, $(3.16228, \frac{7}{16})$, $(2., \frac{2}{5})$, $(2., -\frac{2}{5})$, $(2., \frac{2}{5})$, $(2., -\frac{2}{5})$\}

\item $c = \frac{9}{2}$, $(d_i,\theta_i)$ = \{$(1., 0)$, $(1., 0)$, $(3.16228, \frac{1}{16})$, $(2., \frac{1}{10})$, $(2., \frac{2}{5})$, $(1., \frac{1}{2})$, $(1., \frac{1}{2})$, $(3.16228, -\frac{7}{16})$, $(2., -\frac{2}{5})$, $(2., -\frac{1}{10})$, $(2.82843, -\frac{27}{80})$, $(2.82843, \frac{37}{80})$, $(2.23607, \frac{1}{2})$, $(2.23607, 0)$, $(1.41421, \frac{1}{16})$, $(2.23607, \frac{1}{2})$, $(2.23607, 0)$, $(1.41421, \frac{1}{16})$\}

\item $c = 5$, $(d_i,\theta_i)$ = \{$(1., 0)$, $(1., 0)$, $(3.16228, \frac{1}{16})$, $(2., \frac{1}{10})$, $(2., \frac{2}{5})$, $(1., \frac{1}{2})$, $(1., \frac{1}{2})$, $(3.16228, -\frac{7}{16})$, $(2., -\frac{2}{5})$, $(2., -\frac{1}{10})$, $(3.16228, \frac{1}{16})$, $(2., \frac{1}{8})$, $(3.16228, -\frac{7}{16})$, $(2., -\frac{19}{40})$, $(2., -\frac{11}{40})$, $(2., -\frac{19}{40})$, $(2., -\frac{11}{40})$\}

\item $c = \frac{11}{2}$, $(d_i,\theta_i)$ = \{$(1., 0)$, $(1., 0)$, $(3.16228, \frac{1}{16})$, $(2., \frac{1}{10})$, $(2., \frac{2}{5})$, $(1., \frac{1}{2})$, $(1., \frac{1}{2})$, $(3.16228, -\frac{7}{16})$, $(2., -\frac{2}{5})$, $(2., -\frac{1}{10})$, $(2.82843, -\frac{17}{80})$, $(2.82843, -\frac{33}{80})$, $(2.23607, -\frac{3}{8})$, $(2.23607, \frac{1}{8})$, $(1.41421, \frac{3}{16})$, $(2.23607, -\frac{3}{8})$, $(2.23607, \frac{1}{8})$, $(1.41421, \frac{3}{16})$\}

\item $c = 6$, $(d_i,\theta_i)$ = \{$(1., 0)$, $(1., \frac{1}{2})$, $(3.16228, \frac{1}{16})$, $(2., \frac{1}{10})$, $(2., \frac{2}{5})$, $(1., \frac{1}{2})$, $(1., 0)$, $(3.16228, -\frac{7}{16})$, $(2., -\frac{2}{5})$, $(2., -\frac{1}{10})$, $(2., \frac{1}{4})$, $(3.16228, \frac{3}{16})$, $(3.16228, -\frac{5}{16})$, $(2., -\frac{7}{20})$, $(2., -\frac{3}{20})$, $(2., -\frac{7}{20})$, $(2., -\frac{3}{20})$\}

\item $c = \frac{13}{2}$, $(d_i,\theta_i)$ = \{$(1., 0)$, $(1., \frac{1}{2})$, $(3.16228, \frac{1}{16})$, $(2., \frac{1}{10})$, $(2., \frac{2}{5})$, $(1., \frac{1}{2})$, $(1., 0)$, $(3.16228, -\frac{7}{16})$, $(2., -\frac{2}{5})$, $(2., -\frac{1}{10})$, $(2.82843, -\frac{23}{80})$, $(2.82843, -\frac{7}{80})$, $(2.23607, \frac{1}{4})$, $(2.23607, -\frac{1}{4})$, $(1.41421, \frac{5}{16})$, $(2.23607, \frac{1}{4})$, $(2.23607, -\frac{1}{4})$, $(1.41421, \frac{5}{16})$\}

\item $c = 7$, $(d_i,\theta_i)$ = \{$(1., 0)$, $(1., \frac{1}{2})$, $(3.16228, \frac{1}{16})$, $(2., \frac{1}{10})$, $(2., \frac{2}{5})$, $(1., \frac{1}{2})$, $(1., 0)$, $(3.16228, -\frac{7}{16})$, $(2., -\frac{2}{5})$, $(2., -\frac{1}{10})$, $(3.16228, \frac{5}{16})$, $(2., \frac{3}{8})$, $(3.16228, -\frac{3}{16})$, $(2., -\frac{9}{40})$, $(2., -\frac{1}{40})$, $(2., -\frac{9}{40})$, $(2., -\frac{1}{40})$\}

\item $c = \frac{15}{2}$, $(d_i,\theta_i)$ = \{$(1., 0)$, $(1., \frac{1}{2})$, $(3.16228, -\frac{7}{16})$, $(2., \frac{1}{10})$, $(2., \frac{2}{5})$, $(1., \frac{1}{2})$, $(1., 0)$, $(3.16228, \frac{1}{16})$, $(2., -\frac{2}{5})$, $(2., -\frac{1}{10})$, $(2.82843, \frac{3}{80})$, $(2.82843, -\frac{13}{80})$, $(2.23607, \frac{3}{8})$, $(2.23607, -\frac{1}{8})$, $(1.41421, \frac{7}{16})$, $(2.23607, \frac{3}{8})$, $(2.23607, -\frac{1}{8})$, $(1.41421, \frac{7}{16})$\}

\end{enumerate}

\paragraph*{Rank 10; \#20}

\begin{enumerate}

\item $c = 0$, $(d_i,\theta_i)$ = \{$(1., 0)$, $(1., 0)$, $(2., \frac{1}{10})$, $(3.16228, \frac{3}{16})$, $(2., -\frac{1}{10})$, $(1., \frac{1}{2})$, $(1., \frac{1}{2})$, $(2., -\frac{2}{5})$, $(3.16228, -\frac{5}{16})$, $(2., \frac{2}{5})$, $(3.16228, -\frac{3}{16})$, $(3.16228, \frac{5}{16})$, $(2., \frac{1}{2})$, $(2., -\frac{1}{10})$, $(2., \frac{1}{10})$, $(2., -\frac{1}{10})$, $(2., \frac{1}{10})$\}

\item $c = \frac{1}{2}$, $(d_i,\theta_i)$ = \{$(1., 0)$, $(1., 0)$, $(2., \frac{1}{10})$, $(3.16228, \frac{3}{16})$, $(2., -\frac{1}{10})$, $(1., \frac{1}{2})$, $(1., \frac{1}{2})$, $(2., -\frac{2}{5})$, $(3.16228, -\frac{5}{16})$, $(2., \frac{2}{5})$, $(2.82843, \frac{13}{80})$, $(2.82843, -\frac{3}{80})$, $(2.23607, -\frac{1}{8})$, $(2.23607, \frac{3}{8})$, $(1.41421, -\frac{7}{16})$, $(2.23607, -\frac{1}{8})$, $(2.23607, \frac{3}{8})$, $(1.41421, -\frac{7}{16})$\}

\item $c = 1$, $(d_i,\theta_i)$ = \{$(1., 0)$, $(1., 0)$, $(2., \frac{1}{10})$, $(3.16228, \frac{3}{16})$, $(2., -\frac{1}{10})$, $(1., \frac{1}{2})$, $(1., \frac{1}{2})$, $(2., -\frac{2}{5})$, $(3.16228, -\frac{5}{16})$, $(2., \frac{2}{5})$, $(3.16228, -\frac{1}{16})$, $(2., -\frac{3}{8})$, $(3.16228, \frac{7}{16})$, $(2., \frac{1}{40})$, $(2., \frac{9}{40})$, $(2., \frac{1}{40})$, $(2., \frac{9}{40})$\}

\item $c = \frac{3}{2}$, $(d_i,\theta_i)$ = \{$(1., 0)$, $(1., 0)$, $(2., \frac{1}{10})$, $(3.16228, \frac{3}{16})$, $(2., -\frac{1}{10})$, $(1., \frac{1}{2})$, $(1., \frac{1}{2})$, $(2., -\frac{2}{5})$, $(3.16228, -\frac{5}{16})$, $(2., \frac{2}{5})$, $(2.82843, \frac{23}{80})$, $(2.82843, \frac{7}{80})$, $(2.23607, 0)$, $(2.23607, \frac{1}{2})$, $(1.41421, -\frac{5}{16})$, $(2.23607, 0)$, $(2.23607, \frac{1}{2})$, $(1.41421, -\frac{5}{16})$\}

\item $c = 2$, $(d_i,\theta_i)$ = \{$(1., 0)$, $(1., 0)$, $(2., \frac{1}{10})$, $(3.16228, \frac{3}{16})$, $(2., -\frac{1}{10})$, $(1., \frac{1}{2})$, $(1., \frac{1}{2})$, $(2., -\frac{2}{5})$, $(3.16228, -\frac{5}{16})$, $(2., \frac{2}{5})$, $(2., -\frac{1}{4})$, $(3.16228, -\frac{7}{16})$, $(3.16228, \frac{1}{16})$, $(2., \frac{3}{20})$, $(2., \frac{7}{20})$, $(2., \frac{3}{20})$, $(2., \frac{7}{20})$\}

\item $c = \frac{5}{2}$, $(d_i,\theta_i)$ = \{$(1., 0)$, $(1., 0)$, $(2., \frac{1}{10})$, $(3.16228, \frac{3}{16})$, $(2., -\frac{1}{10})$, $(1., \frac{1}{2})$, $(1., \frac{1}{2})$, $(2., -\frac{2}{5})$, $(3.16228, -\frac{5}{16})$, $(2., \frac{2}{5})$, $(2.82843, \frac{17}{80})$, $(2.82843, \frac{33}{80})$, $(2.23607, -\frac{3}{8})$, $(2.23607, \frac{1}{8})$, $(1.41421, -\frac{3}{16})$, $(2.23607, -\frac{3}{8})$, $(2.23607, \frac{1}{8})$, $(1.41421, -\frac{3}{16})$\}

\item $c = 3$, $(d_i,\theta_i)$ = \{$(1., 0)$, $(1., 0)$, $(2., \frac{1}{10})$, $(3.16228, \frac{3}{16})$, $(2., -\frac{1}{10})$, $(1., \frac{1}{2})$, $(1., \frac{1}{2})$, $(2., -\frac{2}{5})$, $(3.16228, -\frac{5}{16})$, $(2., \frac{2}{5})$, $(2., -\frac{1}{8})$, $(3.16228, \frac{3}{16})$, $(3.16228, -\frac{5}{16})$, $(2., \frac{11}{40})$, $(2., \frac{19}{40})$, $(2., \frac{11}{40})$, $(2., \frac{19}{40})$\}

\item $c = \frac{7}{2}$, $(d_i,\theta_i)$ = \{$(1., 0)$, $(1., 0)$, $(2., \frac{1}{10})$, $(3.16228, \frac{3}{16})$, $(2., -\frac{1}{10})$, $(1., \frac{1}{2})$, $(1., \frac{1}{2})$, $(2., -\frac{2}{5})$, $(3.16228, -\frac{5}{16})$, $(2., \frac{2}{5})$, $(2.82843, \frac{27}{80})$, $(2.82843, -\frac{37}{80})$, $(2.23607, -\frac{1}{4})$, $(2.23607, \frac{1}{4})$, $(1.41421, -\frac{1}{16})$, $(2.23607, -\frac{1}{4})$, $(2.23607, \frac{1}{4})$, $(1.41421, -\frac{1}{16})$\}

\item $c = 4$, $(d_i,\theta_i)$ = \{$(1., 0)$, $(1., 0)$, $(2., \frac{1}{10})$, $(3.16228, \frac{3}{16})$, $(2., \frac{2}{5})$, $(1., \frac{1}{2})$, $(1., \frac{1}{2})$, $(2., -\frac{2}{5})$, $(3.16228, -\frac{5}{16})$, $(2., -\frac{1}{10})$, $(3.16228, -\frac{3}{16})$, $(2., 0)$, $(3.16228, \frac{5}{16})$, $(2., \frac{2}{5})$, $(2., -\frac{2}{5})$, $(2., \frac{2}{5})$, $(2., -\frac{2}{5})$\}

\item $c = \frac{9}{2}$, $(d_i,\theta_i)$ = \{$(1., 0)$, $(1., 0)$, $(2., \frac{1}{10})$, $(3.16228, \frac{3}{16})$, $(2., \frac{2}{5})$, $(1., \frac{1}{2})$, $(1., \frac{1}{2})$, $(2., -\frac{2}{5})$, $(3.16228, -\frac{5}{16})$, $(2., -\frac{1}{10})$, $(2.82843, -\frac{27}{80})$, $(2.82843, \frac{37}{80})$, $(2.23607, \frac{3}{8})$, $(2.23607, -\frac{1}{8})$, $(1.41421, \frac{1}{16})$, $(2.23607, \frac{3}{8})$, $(2.23607, -\frac{1}{8})$, $(1.41421, \frac{1}{16})$\}

\item $c = 5$, $(d_i,\theta_i)$ = \{$(1., 0)$, $(1., 0)$, $(2., \frac{1}{10})$, $(3.16228, \frac{3}{16})$, $(2., \frac{2}{5})$, $(1., \frac{1}{2})$, $(1., \frac{1}{2})$, $(2., -\frac{2}{5})$, $(3.16228, -\frac{5}{16})$, $(2., -\frac{1}{10})$, $(2., \frac{1}{8})$, $(3.16228, -\frac{1}{16})$, $(3.16228, \frac{7}{16})$, $(2., -\frac{19}{40})$, $(2., -\frac{11}{40})$, $(2., -\frac{19}{40})$, $(2., -\frac{11}{40})$\}

\item $c = \frac{11}{2}$, $(d_i,\theta_i)$ = \{$(1., 0)$, $(1., 0)$, $(2., \frac{1}{10})$, $(3.16228, \frac{3}{16})$, $(2., \frac{2}{5})$, $(1., \frac{1}{2})$, $(1., \frac{1}{2})$, $(2., -\frac{2}{5})$, $(3.16228, -\frac{5}{16})$, $(2., -\frac{1}{10})$, $(2.82843, -\frac{17}{80})$, $(2.82843, -\frac{33}{80})$, $(2.23607, \frac{1}{2})$, $(2.23607, 0)$, $(1.41421, \frac{3}{16})$, $(2.23607, \frac{1}{2})$, $(2.23607, 0)$, $(1.41421, \frac{3}{16})$\}

\item $c = 6$, $(d_i,\theta_i)$ = \{$(1., 0)$, $(1., \frac{1}{2})$, $(2., \frac{1}{10})$, $(3.16228, \frac{3}{16})$, $(2., \frac{2}{5})$, $(1., \frac{1}{2})$, $(1., 0)$, $(2., -\frac{2}{5})$, $(3.16228, -\frac{5}{16})$, $(2., -\frac{1}{10})$, $(3.16228, -\frac{7}{16})$, $(3.16228, \frac{1}{16})$, $(2., \frac{1}{4})$, $(2., -\frac{7}{20})$, $(2., -\frac{3}{20})$, $(2., -\frac{7}{20})$, $(2., -\frac{3}{20})$\}

\item $c = \frac{13}{2}$, $(d_i,\theta_i)$ = \{$(1., 0)$, $(1., \frac{1}{2})$, $(2., \frac{1}{10})$, $(3.16228, \frac{3}{16})$, $(2., \frac{2}{5})$, $(1., \frac{1}{2})$, $(1., 0)$, $(2., -\frac{2}{5})$, $(3.16228, -\frac{5}{16})$, $(2., -\frac{1}{10})$, $(2.82843, -\frac{7}{80})$, $(2.82843, -\frac{23}{80})$, $(2.23607, \frac{1}{8})$, $(2.23607, -\frac{3}{8})$, $(1.41421, \frac{5}{16})$, $(2.23607, \frac{1}{8})$, $(2.23607, -\frac{3}{8})$, $(1.41421, \frac{5}{16})$\}

\item $c = 7$, $(d_i,\theta_i)$ = \{$(1., 0)$, $(1., \frac{1}{2})$, $(2., \frac{1}{10})$, $(3.16228, \frac{3}{16})$, $(2., \frac{2}{5})$, $(1., \frac{1}{2})$, $(1., 0)$, $(2., -\frac{2}{5})$, $(3.16228, -\frac{5}{16})$, $(2., -\frac{1}{10})$, $(2., \frac{3}{8})$, $(3.16228, -\frac{5}{16})$, $(3.16228, \frac{3}{16})$, $(2., -\frac{9}{40})$, $(2., -\frac{1}{40})$, $(2., -\frac{9}{40})$, $(2., -\frac{1}{40})$\}

\item $c = \frac{15}{2}$, $(d_i,\theta_i)$ = \{$(1., 0)$, $(1., \frac{1}{2})$, $(2., \frac{1}{10})$, $(3.16228, \frac{3}{16})$, $(2., \frac{2}{5})$, $(1., \frac{1}{2})$, $(1., 0)$, $(2., -\frac{2}{5})$, $(3.16228, -\frac{5}{16})$, $(2., -\frac{1}{10})$, $(2.82843, \frac{3}{80})$, $(2.82843, -\frac{13}{80})$, $(2.23607, \frac{1}{4})$, $(2.23607, -\frac{1}{4})$, $(1.41421, \frac{7}{16})$, $(2.23607, \frac{1}{4})$, $(2.23607, -\frac{1}{4})$, $(1.41421, \frac{7}{16})$\}

\end{enumerate}

\paragraph*{Rank 10; \#21}

\begin{enumerate}

\item $c = 0$, $(d_i,\theta_i)$ = \{$(1., 0)$, $(1., 0)$, $(2., \frac{1}{10})$, $(3.16228, -\frac{3}{16})$, $(2., -\frac{1}{10})$, $(1., \frac{1}{2})$, $(1., \frac{1}{2})$, $(2., -\frac{2}{5})$, $(3.16228, \frac{5}{16})$, $(2., \frac{2}{5})$, $(3.16228, \frac{3}{16})$, $(2., \frac{1}{2})$, $(3.16228, -\frac{5}{16})$, $(2., -\frac{1}{10})$, $(2., \frac{1}{10})$, $(2., -\frac{1}{10})$, $(2., \frac{1}{10})$\}

\item $c = \frac{1}{2}$, $(d_i,\theta_i)$ = \{$(1., 0)$, $(1., 0)$, $(2., \frac{1}{10})$, $(3.16228, -\frac{3}{16})$, $(2., -\frac{1}{10})$, $(1., \frac{1}{2})$, $(1., \frac{1}{2})$, $(2., -\frac{2}{5})$, $(3.16228, \frac{5}{16})$, $(2., \frac{2}{5})$, $(2.82843, \frac{13}{80})$, $(2.82843, -\frac{3}{80})$, $(2.23607, -\frac{1}{4})$, $(2.23607, \frac{1}{4})$, $(1.41421, -\frac{7}{16})$, $(2.23607, -\frac{1}{4})$, $(2.23607, \frac{1}{4})$, $(1.41421, -\frac{7}{16})$\}

\item $c = 1$, $(d_i,\theta_i)$ = \{$(1., 0)$, $(1., 0)$, $(2., \frac{1}{10})$, $(3.16228, -\frac{3}{16})$, $(2., -\frac{1}{10})$, $(1., \frac{1}{2})$, $(1., \frac{1}{2})$, $(2., -\frac{2}{5})$, $(3.16228, \frac{5}{16})$, $(2., \frac{2}{5})$, $(2., -\frac{3}{8})$, $(3.16228, -\frac{3}{16})$, $(3.16228, \frac{5}{16})$, $(2., \frac{1}{40})$, $(2., \frac{9}{40})$, $(2., \frac{1}{40})$, $(2., \frac{9}{40})$\}

\item $c = \frac{3}{2}$, $(d_i,\theta_i)$ = \{$(1., 0)$, $(1., 0)$, $(2., \frac{1}{10})$, $(3.16228, \frac{5}{16})$, $(2., -\frac{1}{10})$, $(1., \frac{1}{2})$, $(1., \frac{1}{2})$, $(2., -\frac{2}{5})$, $(3.16228, -\frac{3}{16})$, $(2., \frac{2}{5})$, $(2.82843, \frac{7}{80})$, $(2.82843, \frac{23}{80})$, $(2.23607, \frac{3}{8})$, $(2.23607, -\frac{1}{8})$, $(1.41421, -\frac{5}{16})$, $(2.23607, \frac{3}{8})$, $(2.23607, -\frac{1}{8})$, $(1.41421, -\frac{5}{16})$\}

\item $c = 2$, $(d_i,\theta_i)$ = \{$(1., 0)$, $(1., 0)$, $(2., \frac{1}{10})$, $(3.16228, \frac{5}{16})$, $(2., -\frac{1}{10})$, $(1., \frac{1}{2})$, $(1., \frac{1}{2})$, $(2., -\frac{2}{5})$, $(3.16228, -\frac{3}{16})$, $(2., \frac{2}{5})$, $(2., -\frac{1}{4})$, $(3.16228, -\frac{1}{16})$, $(3.16228, \frac{7}{16})$, $(2., \frac{3}{20})$, $(2., \frac{7}{20})$, $(2., \frac{3}{20})$, $(2., \frac{7}{20})$\}

\item $c = \frac{5}{2}$, $(d_i,\theta_i)$ = \{$(1., 0)$, $(1., 0)$, $(2., \frac{1}{10})$, $(3.16228, \frac{5}{16})$, $(2., -\frac{1}{10})$, $(1., \frac{1}{2})$, $(1., \frac{1}{2})$, $(2., -\frac{2}{5})$, $(3.16228, -\frac{3}{16})$, $(2., \frac{2}{5})$, $(2.82843, \frac{33}{80})$, $(2.82843, \frac{17}{80})$, $(2.23607, \frac{1}{2})$, $(2.23607, 0)$, $(1.41421, -\frac{3}{16})$, $(2.23607, \frac{1}{2})$, $(2.23607, 0)$, $(1.41421, -\frac{3}{16})$\}

\item $c = 3$, $(d_i,\theta_i)$ = \{$(1., 0)$, $(1., 0)$, $(2., \frac{1}{10})$, $(3.16228, \frac{5}{16})$, $(2., -\frac{1}{10})$, $(1., \frac{1}{2})$, $(1., \frac{1}{2})$, $(2., -\frac{2}{5})$, $(3.16228, -\frac{3}{16})$, $(2., \frac{2}{5})$, $(2., -\frac{1}{8})$, $(3.16228, \frac{1}{16})$, $(3.16228, -\frac{7}{16})$, $(2., \frac{11}{40})$, $(2., \frac{19}{40})$, $(2., \frac{11}{40})$, $(2., \frac{19}{40})$\}

\item $c = \frac{7}{2}$, $(d_i,\theta_i)$ = \{$(1., 0)$, $(1., 0)$, $(2., \frac{1}{10})$, $(3.16228, \frac{5}{16})$, $(2., -\frac{1}{10})$, $(1., \frac{1}{2})$, $(1., \frac{1}{2})$, $(2., -\frac{2}{5})$, $(3.16228, -\frac{3}{16})$, $(2., \frac{2}{5})$, $(2.82843, \frac{27}{80})$, $(2.82843, -\frac{37}{80})$, $(2.23607, \frac{1}{8})$, $(2.23607, -\frac{3}{8})$, $(1.41421, -\frac{1}{16})$, $(2.23607, \frac{1}{8})$, $(2.23607, -\frac{3}{8})$, $(1.41421, -\frac{1}{16})$\}

\item $c = 4$, $(d_i,\theta_i)$ = \{$(1., 0)$, $(1., 0)$, $(2., \frac{1}{10})$, $(3.16228, \frac{5}{16})$, $(2., \frac{2}{5})$, $(1., \frac{1}{2})$, $(1., \frac{1}{2})$, $(2., -\frac{2}{5})$, $(3.16228, -\frac{3}{16})$, $(2., -\frac{1}{10})$, $(2., 0)$, $(3.16228, \frac{3}{16})$, $(3.16228, -\frac{5}{16})$, $(2., \frac{2}{5})$, $(2., -\frac{2}{5})$, $(2., \frac{2}{5})$, $(2., -\frac{2}{5})$\}

\item $c = \frac{9}{2}$, $(d_i,\theta_i)$ = \{$(1., 0)$, $(1., 0)$, $(2., \frac{1}{10})$, $(3.16228, \frac{5}{16})$, $(2., \frac{2}{5})$, $(1., \frac{1}{2})$, $(1., \frac{1}{2})$, $(2., -\frac{2}{5})$, $(3.16228, -\frac{3}{16})$, $(2., -\frac{1}{10})$, $(2.82843, \frac{37}{80})$, $(2.82843, -\frac{27}{80})$, $(2.23607, \frac{1}{4})$, $(2.23607, -\frac{1}{4})$, $(1.41421, \frac{1}{16})$, $(2.23607, \frac{1}{4})$, $(2.23607, -\frac{1}{4})$, $(1.41421, \frac{1}{16})$\}

\item $c = 5$, $(d_i,\theta_i)$ = \{$(1., 0)$, $(1., 0)$, $(2., \frac{1}{10})$, $(3.16228, \frac{5}{16})$, $(2., \frac{2}{5})$, $(1., \frac{1}{2})$, $(1., \frac{1}{2})$, $(2., -\frac{2}{5})$, $(3.16228, -\frac{3}{16})$, $(2., -\frac{1}{10})$, $(3.16228, \frac{5}{16})$, $(2., \frac{1}{8})$, $(3.16228, -\frac{3}{16})$, $(2., -\frac{19}{40})$, $(2., -\frac{11}{40})$, $(2., -\frac{19}{40})$, $(2., -\frac{11}{40})$\}

\item $c = \frac{11}{2}$, $(d_i,\theta_i)$ = \{$(1., 0)$, $(1., 0)$, $(2., \frac{1}{10})$, $(3.16228, \frac{5}{16})$, $(2., \frac{2}{5})$, $(1., \frac{1}{2})$, $(1., \frac{1}{2})$, $(2., -\frac{2}{5})$, $(3.16228, -\frac{3}{16})$, $(2., -\frac{1}{10})$, $(2.82843, -\frac{17}{80})$, $(2.82843, -\frac{33}{80})$, $(2.23607, -\frac{1}{8})$, $(2.23607, \frac{3}{8})$, $(1.41421, \frac{3}{16})$, $(2.23607, -\frac{1}{8})$, $(2.23607, \frac{3}{8})$, $(1.41421, \frac{3}{16})$\}

\item $c = 6$, $(d_i,\theta_i)$ = \{$(1., 0)$, $(1., \frac{1}{2})$, $(2., \frac{1}{10})$, $(3.16228, \frac{5}{16})$, $(2., \frac{2}{5})$, $(1., \frac{1}{2})$, $(1., 0)$, $(2., -\frac{2}{5})$, $(3.16228, -\frac{3}{16})$, $(2., -\frac{1}{10})$, $(3.16228, -\frac{1}{16})$, $(2., \frac{1}{4})$, $(3.16228, \frac{7}{16})$, $(2., -\frac{7}{20})$, $(2., -\frac{3}{20})$, $(2., -\frac{7}{20})$, $(2., -\frac{3}{20})$\}

\item $c = \frac{13}{2}$, $(d_i,\theta_i)$ = \{$(1., 0)$, $(1., \frac{1}{2})$, $(2., \frac{1}{10})$, $(3.16228, \frac{5}{16})$, $(2., \frac{2}{5})$, $(1., \frac{1}{2})$, $(1., 0)$, $(2., -\frac{2}{5})$, $(3.16228, -\frac{3}{16})$, $(2., -\frac{1}{10})$, $(2.82843, -\frac{23}{80})$, $(2.82843, -\frac{7}{80})$, $(2.23607, 0)$, $(2.23607, \frac{1}{2})$, $(1.41421, \frac{5}{16})$, $(2.23607, 0)$, $(2.23607, \frac{1}{2})$, $(1.41421, \frac{5}{16})$\}

\item $c = 7$, $(d_i,\theta_i)$ = \{$(1., 0)$, $(1., \frac{1}{2})$, $(2., \frac{1}{10})$, $(3.16228, \frac{5}{16})$, $(2., \frac{2}{5})$, $(1., \frac{1}{2})$, $(1., 0)$, $(2., -\frac{2}{5})$, $(3.16228, -\frac{3}{16})$, $(2., -\frac{1}{10})$, $(2., \frac{3}{8})$, $(3.16228, \frac{1}{16})$, $(3.16228, -\frac{7}{16})$, $(2., -\frac{9}{40})$, $(2., -\frac{1}{40})$, $(2., -\frac{9}{40})$, $(2., -\frac{1}{40})$\}

\item $c = \frac{15}{2}$, $(d_i,\theta_i)$ = \{$(1., 0)$, $(1., \frac{1}{2})$, $(2., \frac{1}{10})$, $(3.16228, \frac{5}{16})$, $(2., \frac{2}{5})$, $(1., \frac{1}{2})$, $(1., 0)$, $(2., -\frac{2}{5})$, $(3.16228, -\frac{3}{16})$, $(2., -\frac{1}{10})$, $(2.82843, -\frac{13}{80})$, $(2.82843, \frac{3}{80})$, $(2.23607, -\frac{3}{8})$, $(2.23607, \frac{1}{8})$, $(1.41421, \frac{7}{16})$, $(2.23607, -\frac{3}{8})$, $(2.23607, \frac{1}{8})$, $(1.41421, \frac{7}{16})$\}

\end{enumerate}

\paragraph*{Rank 10; \#22}

\begin{enumerate}

\item $c = 0$, $(d_i,\theta_i)$ = \{$(1., 0)$, $(1., 0)$, $(2., \frac{1}{10})$, $(2., -\frac{1}{10})$, $(3.16228, -\frac{1}{16})$, $(1., \frac{1}{2})$, $(1., \frac{1}{2})$, $(2., -\frac{2}{5})$, $(2., \frac{2}{5})$, $(3.16228, \frac{7}{16})$, $(2., \frac{1}{2})$, $(3.16228, -\frac{7}{16})$, $(3.16228, \frac{1}{16})$, $(2., -\frac{1}{10})$, $(2., \frac{1}{10})$, $(2., -\frac{1}{10})$, $(2., \frac{1}{10})$\}

\item $c = \frac{1}{2}$, $(d_i,\theta_i)$ = \{$(1., 0)$, $(1., 0)$, $(2., \frac{1}{10})$, $(2., -\frac{1}{10})$, $(3.16228, -\frac{1}{16})$, $(1., \frac{1}{2})$, $(1., \frac{1}{2})$, $(2., -\frac{2}{5})$, $(2., \frac{2}{5})$, $(3.16228, \frac{7}{16})$, $(2.82843, \frac{13}{80})$, $(2.82843, -\frac{3}{80})$, $(2.23607, -\frac{3}{8})$, $(2.23607, \frac{1}{8})$, $(1.41421, -\frac{7}{16})$, $(2.23607, -\frac{3}{8})$, $(2.23607, \frac{1}{8})$, $(1.41421, -\frac{7}{16})$\}

\item $c = 1$, $(d_i,\theta_i)$ = \{$(1., 0)$, $(1., 0)$, $(2., \frac{1}{10})$, $(2., -\frac{1}{10})$, $(3.16228, -\frac{1}{16})$, $(1., \frac{1}{2})$, $(1., \frac{1}{2})$, $(2., -\frac{2}{5})$, $(2., \frac{2}{5})$, $(3.16228, \frac{7}{16})$, $(3.16228, -\frac{5}{16})$, $(2., -\frac{3}{8})$, $(3.16228, \frac{3}{16})$, $(2., \frac{1}{40})$, $(2., \frac{9}{40})$, $(2., \frac{1}{40})$, $(2., \frac{9}{40})$\}

\item $c = \frac{3}{2}$, $(d_i,\theta_i)$ = \{$(1., 0)$, $(1., 0)$, $(2., \frac{1}{10})$, $(2., -\frac{1}{10})$, $(3.16228, -\frac{1}{16})$, $(1., \frac{1}{2})$, $(1., \frac{1}{2})$, $(2., -\frac{2}{5})$, $(2., \frac{2}{5})$, $(3.16228, \frac{7}{16})$, $(2.82843, \frac{23}{80})$, $(2.82843, \frac{7}{80})$, $(2.23607, \frac{1}{4})$, $(2.23607, -\frac{1}{4})$, $(1.41421, -\frac{5}{16})$, $(2.23607, \frac{1}{4})$, $(2.23607, -\frac{1}{4})$, $(1.41421, -\frac{5}{16})$\}

\item $c = 2$, $(d_i,\theta_i)$ = \{$(1., 0)$, $(1., 0)$, $(2., \frac{1}{10})$, $(2., -\frac{1}{10})$, $(3.16228, -\frac{1}{16})$, $(1., \frac{1}{2})$, $(1., \frac{1}{2})$, $(2., -\frac{2}{5})$, $(2., \frac{2}{5})$, $(3.16228, \frac{7}{16})$, $(2., -\frac{1}{4})$, $(3.16228, \frac{5}{16})$, $(3.16228, -\frac{3}{16})$, $(2., \frac{3}{20})$, $(2., \frac{7}{20})$, $(2., \frac{3}{20})$, $(2., \frac{7}{20})$\}

\item $c = \frac{5}{2}$, $(d_i,\theta_i)$ = \{$(1., 0)$, $(1., 0)$, $(2., \frac{1}{10})$, $(2., -\frac{1}{10})$, $(3.16228, -\frac{1}{16})$, $(1., \frac{1}{2})$, $(1., \frac{1}{2})$, $(2., -\frac{2}{5})$, $(2., \frac{2}{5})$, $(3.16228, \frac{7}{16})$, $(2.82843, \frac{17}{80})$, $(2.82843, \frac{33}{80})$, $(2.23607, \frac{3}{8})$, $(2.23607, -\frac{1}{8})$, $(1.41421, -\frac{3}{16})$, $(2.23607, \frac{3}{8})$, $(2.23607, -\frac{1}{8})$, $(1.41421, -\frac{3}{16})$\}

\item $c = 3$, $(d_i,\theta_i)$ = \{$(1., 0)$, $(1., 0)$, $(2., \frac{1}{10})$, $(2., -\frac{1}{10})$, $(3.16228, -\frac{1}{16})$, $(1., \frac{1}{2})$, $(1., \frac{1}{2})$, $(2., -\frac{2}{5})$, $(2., \frac{2}{5})$, $(3.16228, \frac{7}{16})$, $(2., -\frac{1}{8})$, $(3.16228, \frac{7}{16})$, $(3.16228, -\frac{1}{16})$, $(2., \frac{11}{40})$, $(2., \frac{19}{40})$, $(2., \frac{11}{40})$, $(2., \frac{19}{40})$\}

\item $c = \frac{7}{2}$, $(d_i,\theta_i)$ = \{$(1., 0)$, $(1., 0)$, $(2., \frac{1}{10})$, $(2., -\frac{1}{10})$, $(3.16228, -\frac{1}{16})$, $(1., \frac{1}{2})$, $(1., \frac{1}{2})$, $(2., -\frac{2}{5})$, $(2., \frac{2}{5})$, $(3.16228, \frac{7}{16})$, $(2.82843, \frac{27}{80})$, $(2.82843, -\frac{37}{80})$, $(2.23607, 0)$, $(2.23607, \frac{1}{2})$, $(1.41421, -\frac{1}{16})$, $(2.23607, 0)$, $(2.23607, \frac{1}{2})$, $(1.41421, -\frac{1}{16})$\}

\item $c = 4$, $(d_i,\theta_i)$ = \{$(1., 0)$, $(1., 0)$, $(2., \frac{1}{10})$, $(2., \frac{2}{5})$, $(3.16228, -\frac{1}{16})$, $(1., \frac{1}{2})$, $(1., \frac{1}{2})$, $(2., -\frac{2}{5})$, $(2., -\frac{1}{10})$, $(3.16228, \frac{7}{16})$, $(2., 0)$, $(3.16228, \frac{1}{16})$, $(3.16228, -\frac{7}{16})$, $(2., \frac{2}{5})$, $(2., -\frac{2}{5})$, $(2., \frac{2}{5})$, $(2., -\frac{2}{5})$\}

\item $c = \frac{9}{2}$, $(d_i,\theta_i)$ = \{$(1., 0)$, $(1., 0)$, $(2., \frac{1}{10})$, $(2., \frac{2}{5})$, $(3.16228, \frac{7}{16})$, $(1., \frac{1}{2})$, $(1., \frac{1}{2})$, $(2., -\frac{2}{5})$, $(2., -\frac{1}{10})$, $(3.16228, -\frac{1}{16})$, $(2.82843, \frac{37}{80})$, $(2.82843, -\frac{27}{80})$, $(2.23607, \frac{1}{8})$, $(2.23607, -\frac{3}{8})$, $(1.41421, \frac{1}{16})$, $(2.23607, \frac{1}{8})$, $(2.23607, -\frac{3}{8})$, $(1.41421, \frac{1}{16})$\}

\item $c = 5$, $(d_i,\theta_i)$ = \{$(1., 0)$, $(1., 0)$, $(2., \frac{1}{10})$, $(2., \frac{2}{5})$, $(3.16228, \frac{7}{16})$, $(1., \frac{1}{2})$, $(1., \frac{1}{2})$, $(2., -\frac{2}{5})$, $(2., -\frac{1}{10})$, $(3.16228, -\frac{1}{16})$, $(3.16228, \frac{3}{16})$, $(3.16228, -\frac{5}{16})$, $(2., \frac{1}{8})$, $(2., -\frac{19}{40})$, $(2., -\frac{11}{40})$, $(2., -\frac{19}{40})$, $(2., -\frac{11}{40})$\}

\item $c = \frac{11}{2}$, $(d_i,\theta_i)$ = \{$(1., 0)$, $(1., 0)$, $(2., \frac{1}{10})$, $(2., \frac{2}{5})$, $(3.16228, \frac{7}{16})$, $(1., \frac{1}{2})$, $(1., \frac{1}{2})$, $(2., -\frac{2}{5})$, $(2., -\frac{1}{10})$, $(3.16228, -\frac{1}{16})$, $(2.82843, -\frac{17}{80})$, $(2.82843, -\frac{33}{80})$, $(2.23607, -\frac{1}{4})$, $(2.23607, \frac{1}{4})$, $(1.41421, \frac{3}{16})$, $(2.23607, -\frac{1}{4})$, $(2.23607, \frac{1}{4})$, $(1.41421, \frac{3}{16})$\}

\item $c = 6$, $(d_i,\theta_i)$ = \{$(1., 0)$, $(1., \frac{1}{2})$, $(2., \frac{1}{10})$, $(2., \frac{2}{5})$, $(3.16228, \frac{7}{16})$, $(1., \frac{1}{2})$, $(1., 0)$, $(2., -\frac{2}{5})$, $(2., -\frac{1}{10})$, $(3.16228, -\frac{1}{16})$, $(2., \frac{1}{4})$, $(3.16228, \frac{5}{16})$, $(3.16228, -\frac{3}{16})$, $(2., -\frac{7}{20})$, $(2., -\frac{3}{20})$, $(2., -\frac{7}{20})$, $(2., -\frac{3}{20})$\}

\item $c = \frac{13}{2}$, $(d_i,\theta_i)$ = \{$(1., 0)$, $(1., \frac{1}{2})$, $(2., \frac{1}{10})$, $(2., \frac{2}{5})$, $(3.16228, \frac{7}{16})$, $(1., \frac{1}{2})$, $(1., 0)$, $(2., -\frac{2}{5})$, $(2., -\frac{1}{10})$, $(3.16228, -\frac{1}{16})$, $(2.82843, -\frac{7}{80})$, $(2.82843, -\frac{23}{80})$, $(2.23607, -\frac{1}{8})$, $(2.23607, \frac{3}{8})$, $(1.41421, \frac{5}{16})$, $(2.23607, -\frac{1}{8})$, $(2.23607, \frac{3}{8})$, $(1.41421, \frac{5}{16})$\}

\item $c = 7$, $(d_i,\theta_i)$ = \{$(1., 0)$, $(1., \frac{1}{2})$, $(2., \frac{1}{10})$, $(2., \frac{2}{5})$, $(3.16228, \frac{7}{16})$, $(1., \frac{1}{2})$, $(1., 0)$, $(2., -\frac{2}{5})$, $(2., -\frac{1}{10})$, $(3.16228, -\frac{1}{16})$, $(2., \frac{3}{8})$, $(3.16228, \frac{7}{16})$, $(3.16228, -\frac{1}{16})$, $(2., -\frac{9}{40})$, $(2., -\frac{1}{40})$, $(2., -\frac{9}{40})$, $(2., -\frac{1}{40})$\}

\item $c = \frac{15}{2}$, $(d_i,\theta_i)$ = \{$(1., 0)$, $(1., \frac{1}{2})$, $(2., \frac{1}{10})$, $(2., \frac{2}{5})$, $(3.16228, \frac{7}{16})$, $(1., \frac{1}{2})$, $(1., 0)$, $(2., -\frac{2}{5})$, $(2., -\frac{1}{10})$, $(3.16228, -\frac{1}{16})$, $(2.82843, \frac{3}{80})$, $(2.82843, -\frac{13}{80})$, $(2.23607, \frac{1}{2})$, $(2.23607, 0)$, $(1.41421, \frac{7}{16})$, $(2.23607, \frac{1}{2})$, $(2.23607, 0)$, $(1.41421, \frac{7}{16})$\}

\end{enumerate}

\paragraph*{Rank 10; \#23}

\begin{enumerate}

\item $c = 0$, $(d_i,\theta_i)$ = \{$(1., 0)$, $(1., 0)$, $(3.16228, \frac{1}{16})$, $(2., \frac{1}{5})$, $(2., -\frac{1}{5})$, $(1., \frac{1}{2})$, $(1., \frac{1}{2})$, $(3.16228, -\frac{7}{16})$, $(2., -\frac{3}{10})$, $(2., \frac{3}{10})$, $(3.16228, \frac{7}{16})$, $(3.16228, -\frac{1}{16})$, $(2., 0)$, $(2., \frac{1}{5})$, $(2., -\frac{1}{5})$, $(2., \frac{1}{5})$, $(2., -\frac{1}{5})$\}

\item $c = \frac{1}{2}$, $(d_i,\theta_i)$ = \{$(1., 0)$, $(1., 0)$, $(3.16228, \frac{1}{16})$, $(2., \frac{1}{5})$, $(2., -\frac{1}{5})$, $(1., \frac{1}{2})$, $(1., \frac{1}{2})$, $(3.16228, -\frac{7}{16})$, $(2., -\frac{3}{10})$, $(2., \frac{3}{10})$, $(2.82843, -\frac{11}{80})$, $(2.82843, \frac{21}{80})$, $(2.23607, 0)$, $(2.23607, \frac{1}{2})$, $(1.41421, \frac{1}{16})$, $(2.23607, 0)$, $(2.23607, \frac{1}{2})$, $(1.41421, \frac{1}{16})$\}

\item $c = 1$, $(d_i,\theta_i)$ = \{$(1., 0)$, $(1., 0)$, $(3.16228, \frac{1}{16})$, $(2., \frac{1}{5})$, $(2., -\frac{1}{5})$, $(1., \frac{1}{2})$, $(1., \frac{1}{2})$, $(3.16228, -\frac{7}{16})$, $(2., -\frac{3}{10})$, $(2., \frac{3}{10})$, $(3.16228, \frac{1}{16})$, $(2., \frac{1}{8})$, $(3.16228, -\frac{7}{16})$, $(2., \frac{13}{40})$, $(2., -\frac{3}{40})$, $(2., \frac{13}{40})$, $(2., -\frac{3}{40})$\}

\item $c = \frac{3}{2}$, $(d_i,\theta_i)$ = \{$(1., 0)$, $(1., 0)$, $(3.16228, \frac{1}{16})$, $(2., \frac{1}{5})$, $(2., \frac{3}{10})$, $(1., \frac{1}{2})$, $(1., \frac{1}{2})$, $(3.16228, -\frac{7}{16})$, $(2., -\frac{3}{10})$, $(2., -\frac{1}{5})$, $(2.82843, \frac{31}{80})$, $(2.82843, -\frac{1}{80})$, $(2.23607, \frac{1}{8})$, $(2.23607, -\frac{3}{8})$, $(1.41421, \frac{3}{16})$, $(2.23607, \frac{1}{8})$, $(2.23607, -\frac{3}{8})$, $(1.41421, \frac{3}{16})$\}

\item $c = 2$, $(d_i,\theta_i)$ = \{$(1., 0)$, $(1., 0)$, $(3.16228, \frac{1}{16})$, $(2., \frac{1}{5})$, $(2., \frac{3}{10})$, $(1., \frac{1}{2})$, $(1., \frac{1}{2})$, $(3.16228, -\frac{7}{16})$, $(2., -\frac{3}{10})$, $(2., -\frac{1}{5})$, $(3.16228, -\frac{5}{16})$, $(2., \frac{1}{4})$, $(3.16228, \frac{3}{16})$, $(2., \frac{9}{20})$, $(2., \frac{1}{20})$, $(2., \frac{9}{20})$, $(2., \frac{1}{20})$\}

\item $c = \frac{5}{2}$, $(d_i,\theta_i)$ = \{$(1., 0)$, $(1., 0)$, $(3.16228, \frac{1}{16})$, $(2., \frac{1}{5})$, $(2., \frac{3}{10})$, $(1., \frac{1}{2})$, $(1., \frac{1}{2})$, $(3.16228, -\frac{7}{16})$, $(2., -\frac{3}{10})$, $(2., -\frac{1}{5})$, $(2.82843, \frac{9}{80})$, $(2.82843, -\frac{39}{80})$, $(2.23607, -\frac{1}{4})$, $(2.23607, \frac{1}{4})$, $(1.41421, \frac{5}{16})$, $(2.23607, -\frac{1}{4})$, $(2.23607, \frac{1}{4})$, $(1.41421, \frac{5}{16})$\}

\item $c = 3$, $(d_i,\theta_i)$ = \{$(1., 0)$, $(1., 0)$, $(3.16228, \frac{1}{16})$, $(2., \frac{1}{5})$, $(2., \frac{3}{10})$, $(1., \frac{1}{2})$, $(1., \frac{1}{2})$, $(3.16228, -\frac{7}{16})$, $(2., -\frac{3}{10})$, $(2., -\frac{1}{5})$, $(2., \frac{3}{8})$, $(3.16228, \frac{5}{16})$, $(3.16228, -\frac{3}{16})$, $(2., -\frac{17}{40})$, $(2., \frac{7}{40})$, $(2., -\frac{17}{40})$, $(2., \frac{7}{40})$\}

\item $c = \frac{7}{2}$, $(d_i,\theta_i)$ = \{$(1., 0)$, $(1., 0)$, $(3.16228, \frac{1}{16})$, $(2., \frac{1}{5})$, $(2., \frac{3}{10})$, $(1., \frac{1}{2})$, $(1., \frac{1}{2})$, $(3.16228, -\frac{7}{16})$, $(2., -\frac{3}{10})$, $(2., -\frac{1}{5})$, $(2.82843, -\frac{29}{80})$, $(2.82843, \frac{19}{80})$, $(2.23607, -\frac{1}{8})$, $(2.23607, \frac{3}{8})$, $(1.41421, \frac{7}{16})$, $(2.23607, -\frac{1}{8})$, $(2.23607, \frac{3}{8})$, $(1.41421, \frac{7}{16})$\}

\item $c = 4$, $(d_i,\theta_i)$ = \{$(1., 0)$, $(1., 0)$, $(3.16228, \frac{1}{16})$, $(2., \frac{1}{5})$, $(2., \frac{3}{10})$, $(1., \frac{1}{2})$, $(1., \frac{1}{2})$, $(3.16228, -\frac{7}{16})$, $(2., -\frac{3}{10})$, $(2., -\frac{1}{5})$, $(3.16228, -\frac{1}{16})$, $(2., \frac{1}{2})$, $(3.16228, \frac{7}{16})$, $(2., -\frac{3}{10})$, $(2., \frac{3}{10})$, $(2., -\frac{3}{10})$, $(2., \frac{3}{10})$\}

\item $c = \frac{9}{2}$, $(d_i,\theta_i)$ = \{$(1., 0)$, $(1., 0)$, $(3.16228, \frac{1}{16})$, $(2., \frac{1}{5})$, $(2., \frac{3}{10})$, $(1., \frac{1}{2})$, $(1., \frac{1}{2})$, $(3.16228, -\frac{7}{16})$, $(2., -\frac{3}{10})$, $(2., -\frac{1}{5})$, $(2.82843, \frac{29}{80})$, $(2.82843, -\frac{19}{80})$, $(2.23607, \frac{1}{2})$, $(2.23607, 0)$, $(1.41421, -\frac{7}{16})$, $(2.23607, \frac{1}{2})$, $(2.23607, 0)$, $(1.41421, -\frac{7}{16})$\}

\item $c = 5$, $(d_i,\theta_i)$ = \{$(1., 0)$, $(1., 0)$, $(3.16228, \frac{1}{16})$, $(2., \frac{1}{5})$, $(2., \frac{3}{10})$, $(1., \frac{1}{2})$, $(1., \frac{1}{2})$, $(3.16228, -\frac{7}{16})$, $(2., -\frac{3}{10})$, $(2., -\frac{1}{5})$, $(2., -\frac{3}{8})$, $(3.16228, -\frac{7}{16})$, $(3.16228, \frac{1}{16})$, $(2., -\frac{7}{40})$, $(2., \frac{17}{40})$, $(2., -\frac{7}{40})$, $(2., \frac{17}{40})$\}

\item $c = \frac{11}{2}$, $(d_i,\theta_i)$ = \{$(1., 0)$, $(1., 0)$, $(3.16228, \frac{1}{16})$, $(2., \frac{1}{5})$, $(2., \frac{3}{10})$, $(1., \frac{1}{2})$, $(1., \frac{1}{2})$, $(3.16228, -\frac{7}{16})$, $(2., -\frac{3}{10})$, $(2., -\frac{1}{5})$, $(2.82843, \frac{39}{80})$, $(2.82843, -\frac{9}{80})$, $(2.23607, -\frac{3}{8})$, $(2.23607, \frac{1}{8})$, $(1.41421, -\frac{5}{16})$, $(2.23607, -\frac{3}{8})$, $(2.23607, \frac{1}{8})$, $(1.41421, -\frac{5}{16})$\}

\item $c = 6$, $(d_i,\theta_i)$ = \{$(1., 0)$, $(1., \frac{1}{2})$, $(3.16228, \frac{1}{16})$, $(2., \frac{1}{5})$, $(2., \frac{3}{10})$, $(1., \frac{1}{2})$, $(1., 0)$, $(3.16228, -\frac{7}{16})$, $(2., -\frac{3}{10})$, $(2., -\frac{1}{5})$, $(2., -\frac{1}{4})$, $(3.16228, -\frac{5}{16})$, $(3.16228, \frac{3}{16})$, $(2., -\frac{1}{20})$, $(2., -\frac{9}{20})$, $(2., -\frac{1}{20})$, $(2., -\frac{9}{20})$\}

\item $c = \frac{13}{2}$, $(d_i,\theta_i)$ = \{$(1., 0)$, $(1., \frac{1}{2})$, $(3.16228, \frac{1}{16})$, $(2., \frac{1}{5})$, $(2., \frac{3}{10})$, $(1., \frac{1}{2})$, $(1., 0)$, $(3.16228, -\frac{7}{16})$, $(2., -\frac{3}{10})$, $(2., -\frac{1}{5})$, $(2.82843, \frac{1}{80})$, $(2.82843, -\frac{31}{80})$, $(2.23607, \frac{1}{4})$, $(2.23607, -\frac{1}{4})$, $(1.41421, -\frac{3}{16})$, $(2.23607, \frac{1}{4})$, $(2.23607, -\frac{1}{4})$, $(1.41421, -\frac{3}{16})$\}

\item $c = 7$, $(d_i,\theta_i)$ = \{$(1., 0)$, $(1., \frac{1}{2})$, $(3.16228, \frac{1}{16})$, $(2., \frac{1}{5})$, $(2., \frac{3}{10})$, $(1., \frac{1}{2})$, $(1., 0)$, $(3.16228, -\frac{7}{16})$, $(2., -\frac{3}{10})$, $(2., -\frac{1}{5})$, $(3.16228, \frac{5}{16})$, $(3.16228, -\frac{3}{16})$, $(2., -\frac{1}{8})$, $(2., \frac{3}{40})$, $(2., -\frac{13}{40})$, $(2., \frac{3}{40})$, $(2., -\frac{13}{40})$\}

\item $c = \frac{15}{2}$, $(d_i,\theta_i)$ = \{$(1., 0)$, $(1., \frac{1}{2})$, $(3.16228, -\frac{7}{16})$, $(2., \frac{1}{5})$, $(2., \frac{3}{10})$, $(1., \frac{1}{2})$, $(1., 0)$, $(3.16228, \frac{1}{16})$, $(2., -\frac{3}{10})$, $(2., -\frac{1}{5})$, $(2.82843, -\frac{21}{80})$, $(2.82843, \frac{11}{80})$, $(2.23607, \frac{3}{8})$, $(2.23607, -\frac{1}{8})$, $(1.41421, -\frac{1}{16})$, $(2.23607, \frac{3}{8})$, $(2.23607, -\frac{1}{8})$, $(1.41421, -\frac{1}{16})$\}

\end{enumerate}

\paragraph*{Rank 10; \#24}

\begin{enumerate}

\item $c = 0$, $(d_i,\theta_i)$ = \{$(1., 0)$, $(1., 0)$, $(3.16228, \frac{3}{16})$, $(2., \frac{1}{5})$, $(2., -\frac{1}{5})$, $(1., \frac{1}{2})$, $(1., \frac{1}{2})$, $(3.16228, -\frac{5}{16})$, $(2., -\frac{3}{10})$, $(2., \frac{3}{10})$, $(3.16228, \frac{5}{16})$, $(2., 0)$, $(3.16228, -\frac{3}{16})$, $(2., \frac{1}{5})$, $(2., -\frac{1}{5})$, $(2., \frac{1}{5})$, $(2., -\frac{1}{5})$\}

\item $c = \frac{1}{2}$, $(d_i,\theta_i)$ = \{$(1., 0)$, $(1., 0)$, $(3.16228, \frac{3}{16})$, $(2., \frac{1}{5})$, $(2., -\frac{1}{5})$, $(1., \frac{1}{2})$, $(1., \frac{1}{2})$, $(3.16228, -\frac{5}{16})$, $(2., -\frac{3}{10})$, $(2., \frac{3}{10})$, $(2.82843, -\frac{11}{80})$, $(2.82843, \frac{21}{80})$, $(2.23607, -\frac{1}{8})$, $(2.23607, \frac{3}{8})$, $(1.41421, \frac{1}{16})$, $(2.23607, -\frac{1}{8})$, $(2.23607, \frac{3}{8})$, $(1.41421, \frac{1}{16})$\}

\item $c = 1$, $(d_i,\theta_i)$ = \{$(1., 0)$, $(1., 0)$, $(3.16228, \frac{3}{16})$, $(2., \frac{1}{5})$, $(2., -\frac{1}{5})$, $(1., \frac{1}{2})$, $(1., \frac{1}{2})$, $(3.16228, -\frac{5}{16})$, $(2., -\frac{3}{10})$, $(2., \frac{3}{10})$, $(3.16228, -\frac{1}{16})$, $(3.16228, \frac{7}{16})$, $(2., \frac{1}{8})$, $(2., \frac{13}{40})$, $(2., -\frac{3}{40})$, $(2., \frac{13}{40})$, $(2., -\frac{3}{40})$\}

\item $c = \frac{3}{2}$, $(d_i,\theta_i)$ = \{$(1., 0)$, $(1., 0)$, $(3.16228, \frac{3}{16})$, $(2., \frac{1}{5})$, $(2., \frac{3}{10})$, $(1., \frac{1}{2})$, $(1., \frac{1}{2})$, $(3.16228, -\frac{5}{16})$, $(2., -\frac{3}{10})$, $(2., -\frac{1}{5})$, $(2.82843, \frac{31}{80})$, $(2.82843, -\frac{1}{80})$, $(2.23607, 0)$, $(2.23607, \frac{1}{2})$, $(1.41421, \frac{3}{16})$, $(2.23607, 0)$, $(2.23607, \frac{1}{2})$, $(1.41421, \frac{3}{16})$\}

\item $c = 2$, $(d_i,\theta_i)$ = \{$(1., 0)$, $(1., 0)$, $(3.16228, \frac{3}{16})$, $(2., \frac{1}{5})$, $(2., \frac{3}{10})$, $(1., \frac{1}{2})$, $(1., \frac{1}{2})$, $(3.16228, -\frac{5}{16})$, $(2., -\frac{3}{10})$, $(2., -\frac{1}{5})$, $(3.16228, \frac{1}{16})$, $(3.16228, -\frac{7}{16})$, $(2., \frac{1}{4})$, $(2., \frac{9}{20})$, $(2., \frac{1}{20})$, $(2., \frac{9}{20})$, $(2., \frac{1}{20})$\}

\item $c = \frac{5}{2}$, $(d_i,\theta_i)$ = \{$(1., 0)$, $(1., 0)$, $(3.16228, \frac{3}{16})$, $(2., \frac{1}{5})$, $(2., \frac{3}{10})$, $(1., \frac{1}{2})$, $(1., \frac{1}{2})$, $(3.16228, -\frac{5}{16})$, $(2., -\frac{3}{10})$, $(2., -\frac{1}{5})$, $(2.82843, \frac{9}{80})$, $(2.82843, -\frac{39}{80})$, $(2.23607, -\frac{3}{8})$, $(2.23607, \frac{1}{8})$, $(1.41421, \frac{5}{16})$, $(2.23607, -\frac{3}{8})$, $(2.23607, \frac{1}{8})$, $(1.41421, \frac{5}{16})$\}

\item $c = 3$, $(d_i,\theta_i)$ = \{$(1., 0)$, $(1., 0)$, $(3.16228, \frac{3}{16})$, $(2., \frac{1}{5})$, $(2., \frac{3}{10})$, $(1., \frac{1}{2})$, $(1., \frac{1}{2})$, $(3.16228, -\frac{5}{16})$, $(2., -\frac{3}{10})$, $(2., -\frac{1}{5})$, $(3.16228, -\frac{5}{16})$, $(3.16228, \frac{3}{16})$, $(2., \frac{3}{8})$, $(2., -\frac{17}{40})$, $(2., \frac{7}{40})$, $(2., -\frac{17}{40})$, $(2., \frac{7}{40})$\}

\item $c = \frac{7}{2}$, $(d_i,\theta_i)$ = \{$(1., 0)$, $(1., 0)$, $(3.16228, \frac{3}{16})$, $(2., \frac{1}{5})$, $(2., \frac{3}{10})$, $(1., \frac{1}{2})$, $(1., \frac{1}{2})$, $(3.16228, -\frac{5}{16})$, $(2., -\frac{3}{10})$, $(2., -\frac{1}{5})$, $(2.82843, \frac{19}{80})$, $(2.82843, -\frac{29}{80})$, $(2.23607, -\frac{1}{4})$, $(2.23607, \frac{1}{4})$, $(1.41421, \frac{7}{16})$, $(2.23607, -\frac{1}{4})$, $(2.23607, \frac{1}{4})$, $(1.41421, \frac{7}{16})$\}

\item $c = 4$, $(d_i,\theta_i)$ = \{$(1., 0)$, $(1., 0)$, $(3.16228, \frac{3}{16})$, $(2., \frac{1}{5})$, $(2., \frac{3}{10})$, $(1., \frac{1}{2})$, $(1., \frac{1}{2})$, $(3.16228, -\frac{5}{16})$, $(2., -\frac{3}{10})$, $(2., -\frac{1}{5})$, $(2., \frac{1}{2})$, $(3.16228, \frac{5}{16})$, $(3.16228, -\frac{3}{16})$, $(2., -\frac{3}{10})$, $(2., \frac{3}{10})$, $(2., -\frac{3}{10})$, $(2., \frac{3}{10})$\}

\item $c = \frac{9}{2}$, $(d_i,\theta_i)$ = \{$(1., 0)$, $(1., 0)$, $(3.16228, \frac{3}{16})$, $(2., \frac{1}{5})$, $(2., \frac{3}{10})$, $(1., \frac{1}{2})$, $(1., \frac{1}{2})$, $(3.16228, -\frac{5}{16})$, $(2., -\frac{3}{10})$, $(2., -\frac{1}{5})$, $(2.82843, -\frac{19}{80})$, $(2.82843, \frac{29}{80})$, $(2.23607, \frac{3}{8})$, $(2.23607, -\frac{1}{8})$, $(1.41421, -\frac{7}{16})$, $(2.23607, \frac{3}{8})$, $(2.23607, -\frac{1}{8})$, $(1.41421, -\frac{7}{16})$\}

\item $c = 5$, $(d_i,\theta_i)$ = \{$(1., 0)$, $(1., 0)$, $(3.16228, \frac{3}{16})$, $(2., \frac{1}{5})$, $(2., \frac{3}{10})$, $(1., \frac{1}{2})$, $(1., \frac{1}{2})$, $(3.16228, -\frac{5}{16})$, $(2., -\frac{3}{10})$, $(2., -\frac{1}{5})$, $(3.16228, -\frac{1}{16})$, $(3.16228, \frac{7}{16})$, $(2., -\frac{3}{8})$, $(2., -\frac{7}{40})$, $(2., \frac{17}{40})$, $(2., -\frac{7}{40})$, $(2., \frac{17}{40})$\}

\item $c = \frac{11}{2}$, $(d_i,\theta_i)$ = \{$(1., 0)$, $(1., 0)$, $(3.16228, \frac{3}{16})$, $(2., \frac{1}{5})$, $(2., \frac{3}{10})$, $(1., \frac{1}{2})$, $(1., \frac{1}{2})$, $(3.16228, -\frac{5}{16})$, $(2., -\frac{3}{10})$, $(2., -\frac{1}{5})$, $(2.82843, -\frac{9}{80})$, $(2.82843, \frac{39}{80})$, $(2.23607, \frac{1}{2})$, $(2.23607, 0)$, $(1.41421, -\frac{5}{16})$, $(2.23607, \frac{1}{2})$, $(2.23607, 0)$, $(1.41421, -\frac{5}{16})$\}

\item $c = 6$, $(d_i,\theta_i)$ = \{$(1., 0)$, $(1., \frac{1}{2})$, $(3.16228, \frac{3}{16})$, $(2., \frac{1}{5})$, $(2., \frac{3}{10})$, $(1., \frac{1}{2})$, $(1., 0)$, $(3.16228, -\frac{5}{16})$, $(2., -\frac{3}{10})$, $(2., -\frac{1}{5})$, $(3.16228, \frac{1}{16})$, $(2., -\frac{1}{4})$, $(3.16228, -\frac{7}{16})$, $(2., -\frac{1}{20})$, $(2., -\frac{9}{20})$, $(2., -\frac{1}{20})$, $(2., -\frac{9}{20})$\}

\item $c = \frac{13}{2}$, $(d_i,\theta_i)$ = \{$(1., 0)$, $(1., \frac{1}{2})$, $(3.16228, \frac{3}{16})$, $(2., \frac{1}{5})$, $(2., \frac{3}{10})$, $(1., \frac{1}{2})$, $(1., 0)$, $(3.16228, -\frac{5}{16})$, $(2., -\frac{3}{10})$, $(2., -\frac{1}{5})$, $(2.82843, -\frac{31}{80})$, $(2.82843, \frac{1}{80})$, $(2.23607, \frac{1}{8})$, $(2.23607, -\frac{3}{8})$, $(1.41421, -\frac{3}{16})$, $(2.23607, \frac{1}{8})$, $(2.23607, -\frac{3}{8})$, $(1.41421, -\frac{3}{16})$\}

\item $c = 7$, $(d_i,\theta_i)$ = \{$(1., 0)$, $(1., \frac{1}{2})$, $(3.16228, \frac{3}{16})$, $(2., \frac{1}{5})$, $(2., \frac{3}{10})$, $(1., \frac{1}{2})$, $(1., 0)$, $(3.16228, -\frac{5}{16})$, $(2., -\frac{3}{10})$, $(2., -\frac{1}{5})$, $(3.16228, -\frac{5}{16})$, $(3.16228, \frac{3}{16})$, $(2., -\frac{1}{8})$, $(2., \frac{3}{40})$, $(2., -\frac{13}{40})$, $(2., \frac{3}{40})$, $(2., -\frac{13}{40})$\}

\item $c = \frac{15}{2}$, $(d_i,\theta_i)$ = \{$(1., 0)$, $(1., \frac{1}{2})$, $(3.16228, \frac{3}{16})$, $(2., \frac{1}{5})$, $(2., \frac{3}{10})$, $(1., \frac{1}{2})$, $(1., 0)$, $(3.16228, -\frac{5}{16})$, $(2., -\frac{3}{10})$, $(2., -\frac{1}{5})$, $(2.82843, -\frac{21}{80})$, $(2.82843, \frac{11}{80})$, $(2.23607, \frac{1}{4})$, $(2.23607, -\frac{1}{4})$, $(1.41421, -\frac{1}{16})$, $(2.23607, \frac{1}{4})$, $(2.23607, -\frac{1}{4})$, $(1.41421, -\frac{1}{16})$\}

\end{enumerate}

\paragraph*{Rank 10; \#25}

\begin{enumerate}

\item $c = 0$, $(d_i,\theta_i)$ = \{$(1., 0)$, $(1., 0)$, $(2., \frac{1}{5})$, $(2., -\frac{1}{5})$, $(3.16228, -\frac{3}{16})$, $(1., \frac{1}{2})$, $(1., \frac{1}{2})$, $(2., -\frac{3}{10})$, $(2., \frac{3}{10})$, $(3.16228, \frac{5}{16})$, $(3.16228, -\frac{5}{16})$, $(2., 0)$, $(3.16228, \frac{3}{16})$, $(2., \frac{1}{5})$, $(2., -\frac{1}{5})$, $(2., \frac{1}{5})$, $(2., -\frac{1}{5})$\}

\item $c = \frac{1}{2}$, $(d_i,\theta_i)$ = \{$(1., 0)$, $(1., 0)$, $(2., \frac{1}{5})$, $(2., -\frac{1}{5})$, $(3.16228, -\frac{3}{16})$, $(1., \frac{1}{2})$, $(1., \frac{1}{2})$, $(2., -\frac{3}{10})$, $(2., \frac{3}{10})$, $(3.16228, \frac{5}{16})$, $(2.82843, \frac{21}{80})$, $(2.82843, -\frac{11}{80})$, $(2.23607, -\frac{1}{4})$, $(2.23607, \frac{1}{4})$, $(1.41421, \frac{1}{16})$, $(2.23607, -\frac{1}{4})$, $(2.23607, \frac{1}{4})$, $(1.41421, \frac{1}{16})$\}

\item $c = 1$, $(d_i,\theta_i)$ = \{$(1., 0)$, $(1., 0)$, $(2., \frac{1}{5})$, $(2., -\frac{1}{5})$, $(3.16228, -\frac{3}{16})$, $(1., \frac{1}{2})$, $(1., \frac{1}{2})$, $(2., -\frac{3}{10})$, $(2., \frac{3}{10})$, $(3.16228, \frac{5}{16})$, $(3.16228, -\frac{3}{16})$, $(3.16228, \frac{5}{16})$, $(2., \frac{1}{8})$, $(2., \frac{13}{40})$, $(2., -\frac{3}{40})$, $(2., \frac{13}{40})$, $(2., -\frac{3}{40})$\}

\item $c = \frac{3}{2}$, $(d_i,\theta_i)$ = \{$(1., 0)$, $(1., 0)$, $(2., \frac{1}{5})$, $(2., \frac{3}{10})$, $(3.16228, \frac{5}{16})$, $(1., \frac{1}{2})$, $(1., \frac{1}{2})$, $(2., -\frac{3}{10})$, $(2., -\frac{1}{5})$, $(3.16228, -\frac{3}{16})$, $(2.82843, -\frac{1}{80})$, $(2.82843, \frac{31}{80})$, $(2.23607, \frac{3}{8})$, $(2.23607, -\frac{1}{8})$, $(1.41421, \frac{3}{16})$, $(2.23607, \frac{3}{8})$, $(2.23607, -\frac{1}{8})$, $(1.41421, \frac{3}{16})$\}

\item $c = 2$, $(d_i,\theta_i)$ = \{$(1., 0)$, $(1., 0)$, $(2., \frac{1}{5})$, $(2., \frac{3}{10})$, $(3.16228, \frac{5}{16})$, $(1., \frac{1}{2})$, $(1., \frac{1}{2})$, $(2., -\frac{3}{10})$, $(2., -\frac{1}{5})$, $(3.16228, -\frac{3}{16})$, $(2., \frac{1}{4})$, $(3.16228, -\frac{1}{16})$, $(3.16228, \frac{7}{16})$, $(2., \frac{9}{20})$, $(2., \frac{1}{20})$, $(2., \frac{9}{20})$, $(2., \frac{1}{20})$\}

\item $c = \frac{5}{2}$, $(d_i,\theta_i)$ = \{$(1., 0)$, $(1., 0)$, $(2., \frac{1}{5})$, $(2., \frac{3}{10})$, $(3.16228, \frac{5}{16})$, $(1., \frac{1}{2})$, $(1., \frac{1}{2})$, $(2., -\frac{3}{10})$, $(2., -\frac{1}{5})$, $(3.16228, -\frac{3}{16})$, $(2.82843, -\frac{39}{80})$, $(2.82843, \frac{9}{80})$, $(2.23607, \frac{1}{2})$, $(2.23607, 0)$, $(1.41421, \frac{5}{16})$, $(2.23607, \frac{1}{2})$, $(2.23607, 0)$, $(1.41421, \frac{5}{16})$\}

\item $c = 3$, $(d_i,\theta_i)$ = \{$(1., 0)$, $(1., 0)$, $(2., \frac{1}{5})$, $(2., \frac{3}{10})$, $(3.16228, \frac{5}{16})$, $(1., \frac{1}{2})$, $(1., \frac{1}{2})$, $(2., -\frac{3}{10})$, $(2., -\frac{1}{5})$, $(3.16228, -\frac{3}{16})$, $(2., \frac{3}{8})$, $(3.16228, \frac{1}{16})$, $(3.16228, -\frac{7}{16})$, $(2., -\frac{17}{40})$, $(2., \frac{7}{40})$, $(2., -\frac{17}{40})$, $(2., \frac{7}{40})$\}

\item $c = \frac{7}{2}$, $(d_i,\theta_i)$ = \{$(1., 0)$, $(1., 0)$, $(2., \frac{1}{5})$, $(2., \frac{3}{10})$, $(3.16228, \frac{5}{16})$, $(1., \frac{1}{2})$, $(1., \frac{1}{2})$, $(2., -\frac{3}{10})$, $(2., -\frac{1}{5})$, $(3.16228, -\frac{3}{16})$, $(2.82843, \frac{19}{80})$, $(2.82843, -\frac{29}{80})$, $(2.23607, \frac{1}{8})$, $(2.23607, -\frac{3}{8})$, $(1.41421, \frac{7}{16})$, $(2.23607, \frac{1}{8})$, $(2.23607, -\frac{3}{8})$, $(1.41421, \frac{7}{16})$\}

\item $c = 4$, $(d_i,\theta_i)$ = \{$(1., 0)$, $(1., 0)$, $(2., \frac{1}{5})$, $(2., \frac{3}{10})$, $(3.16228, \frac{5}{16})$, $(1., \frac{1}{2})$, $(1., \frac{1}{2})$, $(2., -\frac{3}{10})$, $(2., -\frac{1}{5})$, $(3.16228, -\frac{3}{16})$, $(3.16228, -\frac{5}{16})$, $(2., \frac{1}{2})$, $(3.16228, \frac{3}{16})$, $(2., -\frac{3}{10})$, $(2., \frac{3}{10})$, $(2., -\frac{3}{10})$, $(2., \frac{3}{10})$\}

\item $c = \frac{9}{2}$, $(d_i,\theta_i)$ = \{$(1., 0)$, $(1., 0)$, $(2., \frac{1}{5})$, $(2., \frac{3}{10})$, $(3.16228, \frac{5}{16})$, $(1., \frac{1}{2})$, $(1., \frac{1}{2})$, $(2., -\frac{3}{10})$, $(2., -\frac{1}{5})$, $(3.16228, -\frac{3}{16})$, $(2.82843, \frac{29}{80})$, $(2.82843, -\frac{19}{80})$, $(2.23607, \frac{1}{4})$, $(2.23607, -\frac{1}{4})$, $(1.41421, -\frac{7}{16})$, $(2.23607, \frac{1}{4})$, $(2.23607, -\frac{1}{4})$, $(1.41421, -\frac{7}{16})$\}

\item $c = 5$, $(d_i,\theta_i)$ = \{$(1., 0)$, $(1., 0)$, $(2., \frac{1}{5})$, $(2., \frac{3}{10})$, $(3.16228, \frac{5}{16})$, $(1., \frac{1}{2})$, $(1., \frac{1}{2})$, $(2., -\frac{3}{10})$, $(2., -\frac{1}{5})$, $(3.16228, -\frac{3}{16})$, $(3.16228, \frac{5}{16})$, $(2., -\frac{3}{8})$, $(3.16228, -\frac{3}{16})$, $(2., -\frac{7}{40})$, $(2., \frac{17}{40})$, $(2., -\frac{7}{40})$, $(2., \frac{17}{40})$\}

\item $c = \frac{11}{2}$, $(d_i,\theta_i)$ = \{$(1., 0)$, $(1., 0)$, $(2., \frac{1}{5})$, $(2., \frac{3}{10})$, $(3.16228, \frac{5}{16})$, $(1., \frac{1}{2})$, $(1., \frac{1}{2})$, $(2., -\frac{3}{10})$, $(2., -\frac{1}{5})$, $(3.16228, -\frac{3}{16})$, $(2.82843, \frac{39}{80})$, $(2.82843, -\frac{9}{80})$, $(2.23607, -\frac{1}{8})$, $(2.23607, \frac{3}{8})$, $(1.41421, -\frac{5}{16})$, $(2.23607, -\frac{1}{8})$, $(2.23607, \frac{3}{8})$, $(1.41421, -\frac{5}{16})$\}

\item $c = 6$, $(d_i,\theta_i)$ = \{$(1., 0)$, $(1., \frac{1}{2})$, $(2., \frac{1}{5})$, $(2., \frac{3}{10})$, $(3.16228, \frac{5}{16})$, $(1., \frac{1}{2})$, $(1., 0)$, $(2., -\frac{3}{10})$, $(2., -\frac{1}{5})$, $(3.16228, -\frac{3}{16})$, $(3.16228, \frac{7}{16})$, $(2., -\frac{1}{4})$, $(3.16228, -\frac{1}{16})$, $(2., -\frac{1}{20})$, $(2., -\frac{9}{20})$, $(2., -\frac{1}{20})$, $(2., -\frac{9}{20})$\}

\item $c = \frac{13}{2}$, $(d_i,\theta_i)$ = \{$(1., 0)$, $(1., \frac{1}{2})$, $(2., \frac{1}{5})$, $(2., \frac{3}{10})$, $(3.16228, \frac{5}{16})$, $(1., \frac{1}{2})$, $(1., 0)$, $(2., -\frac{3}{10})$, $(2., -\frac{1}{5})$, $(3.16228, -\frac{3}{16})$, $(2.82843, \frac{1}{80})$, $(2.82843, -\frac{31}{80})$, $(2.23607, 0)$, $(2.23607, \frac{1}{2})$, $(1.41421, -\frac{3}{16})$, $(2.23607, 0)$, $(2.23607, \frac{1}{2})$, $(1.41421, -\frac{3}{16})$\}

\item $c = 7$, $(d_i,\theta_i)$ = \{$(1., 0)$, $(1., \frac{1}{2})$, $(2., \frac{1}{5})$, $(2., \frac{3}{10})$, $(3.16228, \frac{5}{16})$, $(1., \frac{1}{2})$, $(1., 0)$, $(2., -\frac{3}{10})$, $(2., -\frac{1}{5})$, $(3.16228, -\frac{3}{16})$, $(2., -\frac{1}{8})$, $(3.16228, -\frac{7}{16})$, $(3.16228, \frac{1}{16})$, $(2., \frac{3}{40})$, $(2., -\frac{13}{40})$, $(2., \frac{3}{40})$, $(2., -\frac{13}{40})$\}

\item $c = \frac{15}{2}$, $(d_i,\theta_i)$ = \{$(1., 0)$, $(1., \frac{1}{2})$, $(2., \frac{1}{5})$, $(2., \frac{3}{10})$, $(3.16228, \frac{5}{16})$, $(1., \frac{1}{2})$, $(1., 0)$, $(2., -\frac{3}{10})$, $(2., -\frac{1}{5})$, $(3.16228, -\frac{3}{16})$, $(2.82843, -\frac{21}{80})$, $(2.82843, \frac{11}{80})$, $(2.23607, -\frac{3}{8})$, $(2.23607, \frac{1}{8})$, $(1.41421, -\frac{1}{16})$, $(2.23607, -\frac{3}{8})$, $(2.23607, \frac{1}{8})$, $(1.41421, -\frac{1}{16})$\}

\end{enumerate}

\paragraph*{Rank 10; \#26}

\begin{enumerate}

\item $c = 0$, $(d_i,\theta_i)$ = \{$(1., 0)$, $(1., 0)$, $(2., \frac{1}{5})$, $(2., -\frac{1}{5})$, $(3.16228, -\frac{1}{16})$, $(1., \frac{1}{2})$, $(1., \frac{1}{2})$, $(2., -\frac{3}{10})$, $(2., \frac{3}{10})$, $(3.16228, \frac{7}{16})$, $(3.16228, \frac{1}{16})$, $(2., 0)$, $(3.16228, -\frac{7}{16})$, $(2., \frac{1}{5})$, $(2., -\frac{1}{5})$, $(2., \frac{1}{5})$, $(2., -\frac{1}{5})$\}

\item $c = \frac{1}{2}$, $(d_i,\theta_i)$ = \{$(1., 0)$, $(1., 0)$, $(2., \frac{1}{5})$, $(2., -\frac{1}{5})$, $(3.16228, -\frac{1}{16})$, $(1., \frac{1}{2})$, $(1., \frac{1}{2})$, $(2., -\frac{3}{10})$, $(2., \frac{3}{10})$, $(3.16228, \frac{7}{16})$, $(2.82843, -\frac{11}{80})$, $(2.82843, \frac{21}{80})$, $(2.23607, -\frac{3}{8})$, $(2.23607, \frac{1}{8})$, $(1.41421, \frac{1}{16})$, $(2.23607, -\frac{3}{8})$, $(2.23607, \frac{1}{8})$, $(1.41421, \frac{1}{16})$\}

\item $c = 1$, $(d_i,\theta_i)$ = \{$(1., 0)$, $(1., 0)$, $(2., \frac{1}{5})$, $(2., -\frac{1}{5})$, $(3.16228, -\frac{1}{16})$, $(1., \frac{1}{2})$, $(1., \frac{1}{2})$, $(2., -\frac{3}{10})$, $(2., \frac{3}{10})$, $(3.16228, \frac{7}{16})$, $(2., \frac{1}{8})$, $(3.16228, -\frac{5}{16})$, $(3.16228, \frac{3}{16})$, $(2., \frac{13}{40})$, $(2., -\frac{3}{40})$, $(2., \frac{13}{40})$, $(2., -\frac{3}{40})$\}

\item $c = \frac{3}{2}$, $(d_i,\theta_i)$ = \{$(1., 0)$, $(1., 0)$, $(2., \frac{1}{5})$, $(2., \frac{3}{10})$, $(3.16228, -\frac{1}{16})$, $(1., \frac{1}{2})$, $(1., \frac{1}{2})$, $(2., -\frac{3}{10})$, $(2., -\frac{1}{5})$, $(3.16228, \frac{7}{16})$, $(2.82843, \frac{31}{80})$, $(2.82843, -\frac{1}{80})$, $(2.23607, \frac{1}{4})$, $(2.23607, -\frac{1}{4})$, $(1.41421, \frac{3}{16})$, $(2.23607, \frac{1}{4})$, $(2.23607, -\frac{1}{4})$, $(1.41421, \frac{3}{16})$\}

\item $c = 2$, $(d_i,\theta_i)$ = \{$(1., 0)$, $(1., 0)$, $(2., \frac{1}{5})$, $(2., \frac{3}{10})$, $(3.16228, -\frac{1}{16})$, $(1., \frac{1}{2})$, $(1., \frac{1}{2})$, $(2., -\frac{3}{10})$, $(2., -\frac{1}{5})$, $(3.16228, \frac{7}{16})$, $(3.16228, \frac{5}{16})$, $(3.16228, -\frac{3}{16})$, $(2., \frac{1}{4})$, $(2., \frac{9}{20})$, $(2., \frac{1}{20})$, $(2., \frac{9}{20})$, $(2., \frac{1}{20})$\}

\item $c = \frac{5}{2}$, $(d_i,\theta_i)$ = \{$(1., 0)$, $(1., 0)$, $(2., \frac{1}{5})$, $(2., \frac{3}{10})$, $(3.16228, -\frac{1}{16})$, $(1., \frac{1}{2})$, $(1., \frac{1}{2})$, $(2., -\frac{3}{10})$, $(2., -\frac{1}{5})$, $(3.16228, \frac{7}{16})$, $(2.82843, -\frac{39}{80})$, $(2.82843, \frac{9}{80})$, $(2.23607, \frac{3}{8})$, $(2.23607, -\frac{1}{8})$, $(1.41421, \frac{5}{16})$, $(2.23607, \frac{3}{8})$, $(2.23607, -\frac{1}{8})$, $(1.41421, \frac{5}{16})$\}

\item $c = 3$, $(d_i,\theta_i)$ = \{$(1., 0)$, $(1., 0)$, $(2., \frac{1}{5})$, $(2., \frac{3}{10})$, $(3.16228, -\frac{1}{16})$, $(1., \frac{1}{2})$, $(1., \frac{1}{2})$, $(2., -\frac{3}{10})$, $(2., -\frac{1}{5})$, $(3.16228, \frac{7}{16})$, $(3.16228, -\frac{1}{16})$, $(3.16228, \frac{7}{16})$, $(2., \frac{3}{8})$, $(2., -\frac{17}{40})$, $(2., \frac{7}{40})$, $(2., -\frac{17}{40})$, $(2., \frac{7}{40})$\}

\item $c = \frac{7}{2}$, $(d_i,\theta_i)$ = \{$(1., 0)$, $(1., 0)$, $(2., \frac{1}{5})$, $(2., \frac{3}{10})$, $(3.16228, -\frac{1}{16})$, $(1., \frac{1}{2})$, $(1., \frac{1}{2})$, $(2., -\frac{3}{10})$, $(2., -\frac{1}{5})$, $(3.16228, \frac{7}{16})$, $(2.82843, \frac{19}{80})$, $(2.82843, -\frac{29}{80})$, $(2.23607, 0)$, $(2.23607, \frac{1}{2})$, $(1.41421, \frac{7}{16})$, $(2.23607, 0)$, $(2.23607, \frac{1}{2})$, $(1.41421, \frac{7}{16})$\}

\item $c = 4$, $(d_i,\theta_i)$ = \{$(1., 0)$, $(1., 0)$, $(2., \frac{1}{5})$, $(2., \frac{3}{10})$, $(3.16228, -\frac{1}{16})$, $(1., \frac{1}{2})$, $(1., \frac{1}{2})$, $(2., -\frac{3}{10})$, $(2., -\frac{1}{5})$, $(3.16228, \frac{7}{16})$, $(2., \frac{1}{2})$, $(3.16228, -\frac{7}{16})$, $(3.16228, \frac{1}{16})$, $(2., -\frac{3}{10})$, $(2., \frac{3}{10})$, $(2., -\frac{3}{10})$, $(2., \frac{3}{10})$\}

\item $c = \frac{9}{2}$, $(d_i,\theta_i)$ = \{$(1., 0)$, $(1., 0)$, $(2., \frac{1}{5})$, $(2., \frac{3}{10})$, $(3.16228, \frac{7}{16})$, $(1., \frac{1}{2})$, $(1., \frac{1}{2})$, $(2., -\frac{3}{10})$, $(2., -\frac{1}{5})$, $(3.16228, -\frac{1}{16})$, $(2.82843, \frac{29}{80})$, $(2.82843, -\frac{19}{80})$, $(2.23607, \frac{1}{8})$, $(2.23607, -\frac{3}{8})$, $(1.41421, -\frac{7}{16})$, $(2.23607, \frac{1}{8})$, $(2.23607, -\frac{3}{8})$, $(1.41421, -\frac{7}{16})$\}

\item $c = 5$, $(d_i,\theta_i)$ = \{$(1., 0)$, $(1., 0)$, $(2., \frac{1}{5})$, $(2., \frac{3}{10})$, $(3.16228, \frac{7}{16})$, $(1., \frac{1}{2})$, $(1., \frac{1}{2})$, $(2., -\frac{3}{10})$, $(2., -\frac{1}{5})$, $(3.16228, -\frac{1}{16})$, $(2., -\frac{3}{8})$, $(3.16228, \frac{3}{16})$, $(3.16228, -\frac{5}{16})$, $(2., -\frac{7}{40})$, $(2., \frac{17}{40})$, $(2., -\frac{7}{40})$, $(2., \frac{17}{40})$\}

\item $c = \frac{11}{2}$, $(d_i,\theta_i)$ = \{$(1., 0)$, $(1., 0)$, $(2., \frac{1}{5})$, $(2., \frac{3}{10})$, $(3.16228, \frac{7}{16})$, $(1., \frac{1}{2})$, $(1., \frac{1}{2})$, $(2., -\frac{3}{10})$, $(2., -\frac{1}{5})$, $(3.16228, -\frac{1}{16})$, $(2.82843, -\frac{9}{80})$, $(2.82843, \frac{39}{80})$, $(2.23607, -\frac{1}{4})$, $(2.23607, \frac{1}{4})$, $(1.41421, -\frac{5}{16})$, $(2.23607, -\frac{1}{4})$, $(2.23607, \frac{1}{4})$, $(1.41421, -\frac{5}{16})$\}

\item $c = 6$, $(d_i,\theta_i)$ = \{$(1., 0)$, $(1., \frac{1}{2})$, $(2., \frac{1}{5})$, $(2., \frac{3}{10})$, $(3.16228, \frac{7}{16})$, $(1., \frac{1}{2})$, $(1., 0)$, $(2., -\frac{3}{10})$, $(2., -\frac{1}{5})$, $(3.16228, -\frac{1}{16})$, $(3.16228, -\frac{3}{16})$, $(2., -\frac{1}{4})$, $(3.16228, \frac{5}{16})$, $(2., -\frac{1}{20})$, $(2., -\frac{9}{20})$, $(2., -\frac{1}{20})$, $(2., -\frac{9}{20})$\}

\item $c = \frac{13}{2}$, $(d_i,\theta_i)$ = \{$(1., 0)$, $(1., \frac{1}{2})$, $(2., \frac{1}{5})$, $(2., \frac{3}{10})$, $(3.16228, \frac{7}{16})$, $(1., \frac{1}{2})$, $(1., 0)$, $(2., -\frac{3}{10})$, $(2., -\frac{1}{5})$, $(3.16228, -\frac{1}{16})$, $(2.82843, \frac{1}{80})$, $(2.82843, -\frac{31}{80})$, $(2.23607, -\frac{1}{8})$, $(2.23607, \frac{3}{8})$, $(1.41421, -\frac{3}{16})$, $(2.23607, -\frac{1}{8})$, $(2.23607, \frac{3}{8})$, $(1.41421, -\frac{3}{16})$\}

\item $c = 7$, $(d_i,\theta_i)$ = \{$(1., 0)$, $(1., \frac{1}{2})$, $(2., \frac{1}{5})$, $(2., \frac{3}{10})$, $(3.16228, \frac{7}{16})$, $(1., \frac{1}{2})$, $(1., 0)$, $(2., -\frac{3}{10})$, $(2., -\frac{1}{5})$, $(3.16228, -\frac{1}{16})$, $(2., -\frac{1}{8})$, $(3.16228, -\frac{1}{16})$, $(3.16228, \frac{7}{16})$, $(2., \frac{3}{40})$, $(2., -\frac{13}{40})$, $(2., \frac{3}{40})$, $(2., -\frac{13}{40})$\}

\item $c = \frac{15}{2}$, $(d_i,\theta_i)$ = \{$(1., 0)$, $(1., \frac{1}{2})$, $(2., \frac{1}{5})$, $(2., \frac{3}{10})$, $(3.16228, \frac{7}{16})$, $(1., \frac{1}{2})$, $(1., 0)$, $(2., -\frac{3}{10})$, $(2., -\frac{1}{5})$, $(3.16228, -\frac{1}{16})$, $(2.82843, -\frac{21}{80})$, $(2.82843, \frac{11}{80})$, $(2.23607, \frac{1}{2})$, $(2.23607, 0)$, $(1.41421, -\frac{1}{16})$, $(2.23607, \frac{1}{2})$, $(2.23607, 0)$, $(1.41421, -\frac{1}{16})$\}

\end{enumerate}

\paragraph*{Rank 10; \#27}

\begin{enumerate}

\item $c = 0$, $(d_i,\theta_i)$ = \{$(1., 0)$, $(1., 0)$, $(-3.16228, \frac{1}{16})$, $(2., \frac{1}{10})$, $(2., -\frac{1}{10})$, $(1., \frac{1}{2})$, $(1., \frac{1}{2})$, $(-3.16228, -\frac{7}{16})$, $(2., -\frac{2}{5})$, $(2., \frac{2}{5})$, $(2., \frac{1}{2})$, $(-3.16228, \frac{7}{16})$, $(-3.16228, -\frac{1}{16})$, $(2., -\frac{1}{10})$, $(2., \frac{1}{10})$, $(2., -\frac{1}{10})$, $(2., \frac{1}{10})$\}

\item $c = \frac{1}{2}$, $(d_i,\theta_i)$ = \{$(1., 0)$, $(1., 0)$, $(-3.16228, \frac{1}{16})$, $(2., \frac{1}{10})$, $(2., -\frac{1}{10})$, $(1., \frac{1}{2})$, $(1., \frac{1}{2})$, $(-3.16228, -\frac{7}{16})$, $(2., -\frac{2}{5})$, $(2., \frac{2}{5})$, $(2.82843, \frac{13}{80})$, $(2.82843, -\frac{3}{80})$, $(-2.23607, 0)$, $(-2.23607, \frac{1}{2})$, $(1.41421, -\frac{7}{16})$, $(-2.23607, 0)$, $(-2.23607, \frac{1}{2})$, $(1.41421, -\frac{7}{16})$\}

\item $c = 1$, $(d_i,\theta_i)$ = \{$(1., 0)$, $(1., 0)$, $(-3.16228, \frac{1}{16})$, $(2., \frac{1}{10})$, $(2., -\frac{1}{10})$, $(1., \frac{1}{2})$, $(1., \frac{1}{2})$, $(-3.16228, -\frac{7}{16})$, $(2., -\frac{2}{5})$, $(2., \frac{2}{5})$, $(-3.16228, -\frac{7}{16})$, $(-3.16228, \frac{1}{16})$, $(2., -\frac{3}{8})$, $(2., \frac{1}{40})$, $(2., \frac{9}{40})$, $(2., \frac{1}{40})$, $(2., \frac{9}{40})$\}

\item $c = \frac{3}{2}$, $(d_i,\theta_i)$ = \{$(1., 0)$, $(1., 0)$, $(-3.16228, \frac{1}{16})$, $(2., \frac{1}{10})$, $(2., -\frac{1}{10})$, $(1., \frac{1}{2})$, $(1., \frac{1}{2})$, $(-3.16228, -\frac{7}{16})$, $(2., -\frac{2}{5})$, $(2., \frac{2}{5})$, $(2.82843, \frac{23}{80})$, $(2.82843, \frac{7}{80})$, $(-2.23607, \frac{1}{8})$, $(-2.23607, -\frac{3}{8})$, $(1.41421, -\frac{5}{16})$, $(-2.23607, \frac{1}{8})$, $(-2.23607, -\frac{3}{8})$, $(1.41421, -\frac{5}{16})$\}

\item $c = 2$, $(d_i,\theta_i)$ = \{$(1., 0)$, $(1., 0)$, $(-3.16228, \frac{1}{16})$, $(2., \frac{1}{10})$, $(2., -\frac{1}{10})$, $(1., \frac{1}{2})$, $(1., \frac{1}{2})$, $(-3.16228, -\frac{7}{16})$, $(2., -\frac{2}{5})$, $(2., \frac{2}{5})$, $(2., -\frac{1}{4})$, $(-3.16228, -\frac{5}{16})$, $(-3.16228, \frac{3}{16})$, $(2., \frac{3}{20})$, $(2., \frac{7}{20})$, $(2., \frac{3}{20})$, $(2., \frac{7}{20})$\}

\item $c = \frac{5}{2}$, $(d_i,\theta_i)$ = \{$(1., 0)$, $(1., 0)$, $(-3.16228, \frac{1}{16})$, $(2., \frac{1}{10})$, $(2., -\frac{1}{10})$, $(1., \frac{1}{2})$, $(1., \frac{1}{2})$, $(-3.16228, -\frac{7}{16})$, $(2., -\frac{2}{5})$, $(2., \frac{2}{5})$, $(2.82843, \frac{33}{80})$, $(2.82843, \frac{17}{80})$, $(-2.23607, -\frac{1}{4})$, $(-2.23607, \frac{1}{4})$, $(1.41421, -\frac{3}{16})$, $(-2.23607, -\frac{1}{4})$, $(-2.23607, \frac{1}{4})$, $(1.41421, -\frac{3}{16})$\}

\item $c = 3$, $(d_i,\theta_i)$ = \{$(1., 0)$, $(1., 0)$, $(-3.16228, \frac{1}{16})$, $(2., \frac{1}{10})$, $(2., -\frac{1}{10})$, $(1., \frac{1}{2})$, $(1., \frac{1}{2})$, $(-3.16228, -\frac{7}{16})$, $(2., -\frac{2}{5})$, $(2., \frac{2}{5})$, $(-3.16228, -\frac{3}{16})$, $(-3.16228, \frac{5}{16})$, $(2., -\frac{1}{8})$, $(2., \frac{11}{40})$, $(2., \frac{19}{40})$, $(2., \frac{11}{40})$, $(2., \frac{19}{40})$\}

\item $c = \frac{7}{2}$, $(d_i,\theta_i)$ = \{$(1., 0)$, $(1., 0)$, $(-3.16228, \frac{1}{16})$, $(2., \frac{1}{10})$, $(2., -\frac{1}{10})$, $(1., \frac{1}{2})$, $(1., \frac{1}{2})$, $(-3.16228, -\frac{7}{16})$, $(2., -\frac{2}{5})$, $(2., \frac{2}{5})$, $(2.82843, \frac{27}{80})$, $(2.82843, -\frac{37}{80})$, $(-2.23607, -\frac{1}{8})$, $(-2.23607, \frac{3}{8})$, $(1.41421, -\frac{1}{16})$, $(-2.23607, -\frac{1}{8})$, $(-2.23607, \frac{3}{8})$, $(1.41421, -\frac{1}{16})$\}

\item $c = 4$, $(d_i,\theta_i)$ = \{$(1., 0)$, $(1., 0)$, $(-3.16228, \frac{1}{16})$, $(2., \frac{1}{10})$, $(2., \frac{2}{5})$, $(1., \frac{1}{2})$, $(1., \frac{1}{2})$, $(-3.16228, -\frac{7}{16})$, $(2., -\frac{2}{5})$, $(2., -\frac{1}{10})$, $(-3.16228, -\frac{1}{16})$, $(2., 0)$, $(-3.16228, \frac{7}{16})$, $(2., \frac{2}{5})$, $(2., -\frac{2}{5})$, $(2., \frac{2}{5})$, $(2., -\frac{2}{5})$\}

\item $c = \frac{9}{2}$, $(d_i,\theta_i)$ = \{$(1., 0)$, $(1., 0)$, $(-3.16228, \frac{1}{16})$, $(2., \frac{1}{10})$, $(2., \frac{2}{5})$, $(1., \frac{1}{2})$, $(1., \frac{1}{2})$, $(-3.16228, -\frac{7}{16})$, $(2., -\frac{2}{5})$, $(2., -\frac{1}{10})$, $(2.82843, -\frac{27}{80})$, $(2.82843, \frac{37}{80})$, $(-2.23607, \frac{1}{2})$, $(-2.23607, 0)$, $(1.41421, \frac{1}{16})$, $(-2.23607, \frac{1}{2})$, $(-2.23607, 0)$, $(1.41421, \frac{1}{16})$\}

\item $c = 5$, $(d_i,\theta_i)$ = \{$(1., 0)$, $(1., 0)$, $(-3.16228, \frac{1}{16})$, $(2., \frac{1}{10})$, $(2., \frac{2}{5})$, $(1., \frac{1}{2})$, $(1., \frac{1}{2})$, $(-3.16228, -\frac{7}{16})$, $(2., -\frac{2}{5})$, $(2., -\frac{1}{10})$, $(-3.16228, \frac{1}{16})$, $(2., \frac{1}{8})$, $(-3.16228, -\frac{7}{16})$, $(2., -\frac{19}{40})$, $(2., -\frac{11}{40})$, $(2., -\frac{19}{40})$, $(2., -\frac{11}{40})$\}

\item $c = \frac{11}{2}$, $(d_i,\theta_i)$ = \{$(1., 0)$, $(1., 0)$, $(-3.16228, \frac{1}{16})$, $(2., \frac{1}{10})$, $(2., \frac{2}{5})$, $(1., \frac{1}{2})$, $(1., \frac{1}{2})$, $(-3.16228, -\frac{7}{16})$, $(2., -\frac{2}{5})$, $(2., -\frac{1}{10})$, $(2.82843, -\frac{17}{80})$, $(2.82843, -\frac{33}{80})$, $(-2.23607, -\frac{3}{8})$, $(-2.23607, \frac{1}{8})$, $(1.41421, \frac{3}{16})$, $(-2.23607, -\frac{3}{8})$, $(-2.23607, \frac{1}{8})$, $(1.41421, \frac{3}{16})$\}

\item $c = 6$, $(d_i,\theta_i)$ = \{$(1., 0)$, $(1., \frac{1}{2})$, $(-3.16228, \frac{1}{16})$, $(2., \frac{1}{10})$, $(2., \frac{2}{5})$, $(1., \frac{1}{2})$, $(1., 0)$, $(-3.16228, -\frac{7}{16})$, $(2., -\frac{2}{5})$, $(2., -\frac{1}{10})$, $(-2., \frac{1}{4})$, $(3.16228, \frac{3}{16})$, $(3.16228, -\frac{5}{16})$, $(-2., -\frac{7}{20})$, $(-2., -\frac{3}{20})$, $(-2., -\frac{7}{20})$, $(-2., -\frac{3}{20})$\}

\item $c = \frac{13}{2}$, $(d_i,\theta_i)$ = \{$(1., 0)$, $(1., \frac{1}{2})$, $(-3.16228, \frac{1}{16})$, $(2., \frac{1}{10})$, $(2., \frac{2}{5})$, $(1., \frac{1}{2})$, $(1., 0)$, $(-3.16228, -\frac{7}{16})$, $(2., -\frac{2}{5})$, $(2., -\frac{1}{10})$, $(2.82843, -\frac{23}{80})$, $(2.82843, -\frac{7}{80})$, $(-2.23607, \frac{1}{4})$, $(-2.23607, -\frac{1}{4})$, $(1.41421, \frac{5}{16})$, $(-2.23607, \frac{1}{4})$, $(-2.23607, -\frac{1}{4})$, $(1.41421, \frac{5}{16})$\}

\item $c = 7$, $(d_i,\theta_i)$ = \{$(1., 0)$, $(1., \frac{1}{2})$, $(-3.16228, \frac{1}{16})$, $(2., \frac{1}{10})$, $(2., \frac{2}{5})$, $(1., \frac{1}{2})$, $(1., 0)$, $(-3.16228, -\frac{7}{16})$, $(2., -\frac{2}{5})$, $(2., -\frac{1}{10})$, $(3.16228, \frac{5}{16})$, $(-2., \frac{3}{8})$, $(3.16228, -\frac{3}{16})$, $(-2., -\frac{9}{40})$, $(-2., -\frac{1}{40})$, $(-2., -\frac{9}{40})$, $(-2., -\frac{1}{40})$\}

\item $c = \frac{15}{2}$, $(d_i,\theta_i)$ = \{$(1., 0)$, $(1., \frac{1}{2})$, $(-3.16228, -\frac{7}{16})$, $(2., \frac{1}{10})$, $(2., \frac{2}{5})$, $(1., \frac{1}{2})$, $(1., 0)$, $(-3.16228, \frac{1}{16})$, $(2., -\frac{2}{5})$, $(2., -\frac{1}{10})$, $(-2.82843, \frac{3}{80})$, $(-2.82843, -\frac{13}{80})$, $(2.23607, \frac{3}{8})$, $(2.23607, -\frac{1}{8})$, $(-1.41421, \frac{7}{16})$, $(2.23607, \frac{3}{8})$, $(2.23607, -\frac{1}{8})$, $(-1.41421, \frac{7}{16})$\}

\end{enumerate}

\paragraph*{Rank 10; \#28}

\begin{enumerate}

\item $c = 0$, $(d_i,\theta_i)$ = \{$(1., 0)$, $(1., 0)$, $(2., \frac{1}{10})$, $(-3.16228, \frac{3}{16})$, $(2., -\frac{1}{10})$, $(1., \frac{1}{2})$, $(1., \frac{1}{2})$, $(2., -\frac{2}{5})$, $(-3.16228, -\frac{5}{16})$, $(2., \frac{2}{5})$, $(-3.16228, -\frac{3}{16})$, $(-3.16228, \frac{5}{16})$, $(2., \frac{1}{2})$, $(2., -\frac{1}{10})$, $(2., \frac{1}{10})$, $(2., -\frac{1}{10})$, $(2., \frac{1}{10})$\}

\item $c = \frac{1}{2}$, $(d_i,\theta_i)$ = \{$(1., 0)$, $(1., 0)$, $(2., \frac{1}{10})$, $(-3.16228, \frac{3}{16})$, $(2., -\frac{1}{10})$, $(1., \frac{1}{2})$, $(1., \frac{1}{2})$, $(2., -\frac{2}{5})$, $(-3.16228, -\frac{5}{16})$, $(2., \frac{2}{5})$, $(2.82843, \frac{13}{80})$, $(2.82843, -\frac{3}{80})$, $(-2.23607, -\frac{1}{8})$, $(-2.23607, \frac{3}{8})$, $(1.41421, -\frac{7}{16})$, $(-2.23607, -\frac{1}{8})$, $(-2.23607, \frac{3}{8})$, $(1.41421, -\frac{7}{16})$\}

\item $c = 1$, $(d_i,\theta_i)$ = \{$(1., 0)$, $(1., 0)$, $(2., \frac{1}{10})$, $(-3.16228, \frac{3}{16})$, $(2., -\frac{1}{10})$, $(1., \frac{1}{2})$, $(1., \frac{1}{2})$, $(2., -\frac{2}{5})$, $(-3.16228, -\frac{5}{16})$, $(2., \frac{2}{5})$, $(-3.16228, -\frac{1}{16})$, $(2., -\frac{3}{8})$, $(-3.16228, \frac{7}{16})$, $(2., \frac{1}{40})$, $(2., \frac{9}{40})$, $(2., \frac{1}{40})$, $(2., \frac{9}{40})$\}

\item $c = \frac{3}{2}$, $(d_i,\theta_i)$ = \{$(1., 0)$, $(1., 0)$, $(2., \frac{1}{10})$, $(-3.16228, \frac{3}{16})$, $(2., -\frac{1}{10})$, $(1., \frac{1}{2})$, $(1., \frac{1}{2})$, $(2., -\frac{2}{5})$, $(-3.16228, -\frac{5}{16})$, $(2., \frac{2}{5})$, $(2.82843, \frac{23}{80})$, $(2.82843, \frac{7}{80})$, $(-2.23607, 0)$, $(-2.23607, \frac{1}{2})$, $(1.41421, -\frac{5}{16})$, $(-2.23607, 0)$, $(-2.23607, \frac{1}{2})$, $(1.41421, -\frac{5}{16})$\}

\item $c = 2$, $(d_i,\theta_i)$ = \{$(1., 0)$, $(1., 0)$, $(2., \frac{1}{10})$, $(-3.16228, \frac{3}{16})$, $(2., -\frac{1}{10})$, $(1., \frac{1}{2})$, $(1., \frac{1}{2})$, $(2., -\frac{2}{5})$, $(-3.16228, -\frac{5}{16})$, $(2., \frac{2}{5})$, $(2., -\frac{1}{4})$, $(-3.16228, -\frac{7}{16})$, $(-3.16228, \frac{1}{16})$, $(2., \frac{3}{20})$, $(2., \frac{7}{20})$, $(2., \frac{3}{20})$, $(2., \frac{7}{20})$\}

\item $c = \frac{5}{2}$, $(d_i,\theta_i)$ = \{$(1., 0)$, $(1., 0)$, $(2., \frac{1}{10})$, $(-3.16228, \frac{3}{16})$, $(2., -\frac{1}{10})$, $(1., \frac{1}{2})$, $(1., \frac{1}{2})$, $(2., -\frac{2}{5})$, $(-3.16228, -\frac{5}{16})$, $(2., \frac{2}{5})$, $(2.82843, \frac{17}{80})$, $(2.82843, \frac{33}{80})$, $(-2.23607, -\frac{3}{8})$, $(-2.23607, \frac{1}{8})$, $(1.41421, -\frac{3}{16})$, $(-2.23607, -\frac{3}{8})$, $(-2.23607, \frac{1}{8})$, $(1.41421, -\frac{3}{16})$\}

\item $c = 3$, $(d_i,\theta_i)$ = \{$(1., 0)$, $(1., 0)$, $(2., \frac{1}{10})$, $(-3.16228, \frac{3}{16})$, $(2., -\frac{1}{10})$, $(1., \frac{1}{2})$, $(1., \frac{1}{2})$, $(2., -\frac{2}{5})$, $(-3.16228, -\frac{5}{16})$, $(2., \frac{2}{5})$, $(2., -\frac{1}{8})$, $(-3.16228, \frac{3}{16})$, $(-3.16228, -\frac{5}{16})$, $(2., \frac{11}{40})$, $(2., \frac{19}{40})$, $(2., \frac{11}{40})$, $(2., \frac{19}{40})$\}

\item $c = \frac{7}{2}$, $(d_i,\theta_i)$ = \{$(1., 0)$, $(1., 0)$, $(2., \frac{1}{10})$, $(-3.16228, \frac{3}{16})$, $(2., -\frac{1}{10})$, $(1., \frac{1}{2})$, $(1., \frac{1}{2})$, $(2., -\frac{2}{5})$, $(-3.16228, -\frac{5}{16})$, $(2., \frac{2}{5})$, $(2.82843, \frac{27}{80})$, $(2.82843, -\frac{37}{80})$, $(-2.23607, -\frac{1}{4})$, $(-2.23607, \frac{1}{4})$, $(1.41421, -\frac{1}{16})$, $(-2.23607, -\frac{1}{4})$, $(-2.23607, \frac{1}{4})$, $(1.41421, -\frac{1}{16})$\}

\item $c = 4$, $(d_i,\theta_i)$ = \{$(1., 0)$, $(1., 0)$, $(2., \frac{1}{10})$, $(-3.16228, \frac{3}{16})$, $(2., \frac{2}{5})$, $(1., \frac{1}{2})$, $(1., \frac{1}{2})$, $(2., -\frac{2}{5})$, $(-3.16228, -\frac{5}{16})$, $(2., -\frac{1}{10})$, $(-3.16228, -\frac{3}{16})$, $(2., 0)$, $(-3.16228, \frac{5}{16})$, $(2., \frac{2}{5})$, $(2., -\frac{2}{5})$, $(2., \frac{2}{5})$, $(2., -\frac{2}{5})$\}

\item $c = \frac{9}{2}$, $(d_i,\theta_i)$ = \{$(1., 0)$, $(1., 0)$, $(2., \frac{1}{10})$, $(-3.16228, \frac{3}{16})$, $(2., \frac{2}{5})$, $(1., \frac{1}{2})$, $(1., \frac{1}{2})$, $(2., -\frac{2}{5})$, $(-3.16228, -\frac{5}{16})$, $(2., -\frac{1}{10})$, $(2.82843, -\frac{27}{80})$, $(2.82843, \frac{37}{80})$, $(-2.23607, \frac{3}{8})$, $(-2.23607, -\frac{1}{8})$, $(1.41421, \frac{1}{16})$, $(-2.23607, \frac{3}{8})$, $(-2.23607, -\frac{1}{8})$, $(1.41421, \frac{1}{16})$\}

\item $c = 5$, $(d_i,\theta_i)$ = \{$(1., 0)$, $(1., 0)$, $(2., \frac{1}{10})$, $(-3.16228, \frac{3}{16})$, $(2., \frac{2}{5})$, $(1., \frac{1}{2})$, $(1., \frac{1}{2})$, $(2., -\frac{2}{5})$, $(-3.16228, -\frac{5}{16})$, $(2., -\frac{1}{10})$, $(2., \frac{1}{8})$, $(-3.16228, -\frac{1}{16})$, $(-3.16228, \frac{7}{16})$, $(2., -\frac{19}{40})$, $(2., -\frac{11}{40})$, $(2., -\frac{19}{40})$, $(2., -\frac{11}{40})$\}

\item $c = \frac{11}{2}$, $(d_i,\theta_i)$ = \{$(1., 0)$, $(1., 0)$, $(2., \frac{1}{10})$, $(-3.16228, \frac{3}{16})$, $(2., \frac{2}{5})$, $(1., \frac{1}{2})$, $(1., \frac{1}{2})$, $(2., -\frac{2}{5})$, $(-3.16228, -\frac{5}{16})$, $(2., -\frac{1}{10})$, $(2.82843, -\frac{17}{80})$, $(2.82843, -\frac{33}{80})$, $(-2.23607, \frac{1}{2})$, $(-2.23607, 0)$, $(1.41421, \frac{3}{16})$, $(-2.23607, \frac{1}{2})$, $(-2.23607, 0)$, $(1.41421, \frac{3}{16})$\}

\item $c = 6$, $(d_i,\theta_i)$ = \{$(1., 0)$, $(1., \frac{1}{2})$, $(2., \frac{1}{10})$, $(-3.16228, \frac{3}{16})$, $(2., \frac{2}{5})$, $(1., \frac{1}{2})$, $(1., 0)$, $(2., -\frac{2}{5})$, $(-3.16228, -\frac{5}{16})$, $(2., -\frac{1}{10})$, $(3.16228, -\frac{7}{16})$, $(3.16228, \frac{1}{16})$, $(-2., \frac{1}{4})$, $(-2., -\frac{7}{20})$, $(-2., -\frac{3}{20})$, $(-2., -\frac{7}{20})$, $(-2., -\frac{3}{20})$\}

\item $c = \frac{13}{2}$, $(d_i,\theta_i)$ = \{$(1., 0)$, $(1., \frac{1}{2})$, $(2., \frac{1}{10})$, $(-3.16228, \frac{3}{16})$, $(2., \frac{2}{5})$, $(1., \frac{1}{2})$, $(1., 0)$, $(2., -\frac{2}{5})$, $(-3.16228, -\frac{5}{16})$, $(2., -\frac{1}{10})$, $(-2.82843, -\frac{7}{80})$, $(-2.82843, -\frac{23}{80})$, $(2.23607, \frac{1}{8})$, $(2.23607, -\frac{3}{8})$, $(-1.41421, \frac{5}{16})$, $(2.23607, \frac{1}{8})$, $(2.23607, -\frac{3}{8})$, $(-1.41421, \frac{5}{16})$\}

\item $c = 7$, $(d_i,\theta_i)$ = \{$(1., 0)$, $(1., \frac{1}{2})$, $(2., \frac{1}{10})$, $(-3.16228, \frac{3}{16})$, $(2., \frac{2}{5})$, $(1., \frac{1}{2})$, $(1., 0)$, $(2., -\frac{2}{5})$, $(-3.16228, -\frac{5}{16})$, $(2., -\frac{1}{10})$, $(-2., \frac{3}{8})$, $(3.16228, -\frac{5}{16})$, $(3.16228, \frac{3}{16})$, $(-2., -\frac{9}{40})$, $(-2., -\frac{1}{40})$, $(-2., -\frac{9}{40})$, $(-2., -\frac{1}{40})$\}

\item $c = \frac{15}{2}$, $(d_i,\theta_i)$ = \{$(1., 0)$, $(1., \frac{1}{2})$, $(2., \frac{1}{10})$, $(-3.16228, \frac{3}{16})$, $(2., \frac{2}{5})$, $(1., \frac{1}{2})$, $(1., 0)$, $(2., -\frac{2}{5})$, $(-3.16228, -\frac{5}{16})$, $(2., -\frac{1}{10})$, $(-2.82843, \frac{3}{80})$, $(-2.82843, -\frac{13}{80})$, $(2.23607, \frac{1}{4})$, $(2.23607, -\frac{1}{4})$, $(-1.41421, \frac{7}{16})$, $(2.23607, \frac{1}{4})$, $(2.23607, -\frac{1}{4})$, $(-1.41421, \frac{7}{16})$\}

\end{enumerate}

\paragraph*{Rank 10; \#29}

\begin{enumerate}

\item $c = 0$, $(d_i,\theta_i)$ = \{$(1., 0)$, $(1., 0)$, $(2., \frac{1}{10})$, $(-3.16228, -\frac{3}{16})$, $(2., -\frac{1}{10})$, $(1., \frac{1}{2})$, $(1., \frac{1}{2})$, $(2., -\frac{2}{5})$, $(-3.16228, \frac{5}{16})$, $(2., \frac{2}{5})$, $(-3.16228, \frac{3}{16})$, $(2., \frac{1}{2})$, $(-3.16228, -\frac{5}{16})$, $(2., -\frac{1}{10})$, $(2., \frac{1}{10})$, $(2., -\frac{1}{10})$, $(2., \frac{1}{10})$\}

\item $c = \frac{1}{2}$, $(d_i,\theta_i)$ = \{$(1., 0)$, $(1., 0)$, $(2., \frac{1}{10})$, $(-3.16228, -\frac{3}{16})$, $(2., -\frac{1}{10})$, $(1., \frac{1}{2})$, $(1., \frac{1}{2})$, $(2., -\frac{2}{5})$, $(-3.16228, \frac{5}{16})$, $(2., \frac{2}{5})$, $(2.82843, \frac{13}{80})$, $(2.82843, -\frac{3}{80})$, $(-2.23607, -\frac{1}{4})$, $(-2.23607, \frac{1}{4})$, $(1.41421, -\frac{7}{16})$, $(-2.23607, -\frac{1}{4})$, $(-2.23607, \frac{1}{4})$, $(1.41421, -\frac{7}{16})$\}

\item $c = 1$, $(d_i,\theta_i)$ = \{$(1., 0)$, $(1., 0)$, $(2., \frac{1}{10})$, $(-3.16228, -\frac{3}{16})$, $(2., -\frac{1}{10})$, $(1., \frac{1}{2})$, $(1., \frac{1}{2})$, $(2., -\frac{2}{5})$, $(-3.16228, \frac{5}{16})$, $(2., \frac{2}{5})$, $(2., -\frac{3}{8})$, $(-3.16228, -\frac{3}{16})$, $(-3.16228, \frac{5}{16})$, $(2., \frac{1}{40})$, $(2., \frac{9}{40})$, $(2., \frac{1}{40})$, $(2., \frac{9}{40})$\}

\item $c = \frac{3}{2}$, $(d_i,\theta_i)$ = \{$(1., 0)$, $(1., 0)$, $(2., \frac{1}{10})$, $(-3.16228, \frac{5}{16})$, $(2., -\frac{1}{10})$, $(1., \frac{1}{2})$, $(1., \frac{1}{2})$, $(2., -\frac{2}{5})$, $(-3.16228, -\frac{3}{16})$, $(2., \frac{2}{5})$, $(2.82843, \frac{7}{80})$, $(2.82843, \frac{23}{80})$, $(-2.23607, \frac{3}{8})$, $(-2.23607, -\frac{1}{8})$, $(1.41421, -\frac{5}{16})$, $(-2.23607, \frac{3}{8})$, $(-2.23607, -\frac{1}{8})$, $(1.41421, -\frac{5}{16})$\}

\item $c = 2$, $(d_i,\theta_i)$ = \{$(1., 0)$, $(1., 0)$, $(2., \frac{1}{10})$, $(-3.16228, \frac{5}{16})$, $(2., -\frac{1}{10})$, $(1., \frac{1}{2})$, $(1., \frac{1}{2})$, $(2., -\frac{2}{5})$, $(-3.16228, -\frac{3}{16})$, $(2., \frac{2}{5})$, $(2., -\frac{1}{4})$, $(-3.16228, -\frac{1}{16})$, $(-3.16228, \frac{7}{16})$, $(2., \frac{3}{20})$, $(2., \frac{7}{20})$, $(2., \frac{3}{20})$, $(2., \frac{7}{20})$\}

\item $c = \frac{5}{2}$, $(d_i,\theta_i)$ = \{$(1., 0)$, $(1., 0)$, $(2., \frac{1}{10})$, $(-3.16228, \frac{5}{16})$, $(2., -\frac{1}{10})$, $(1., \frac{1}{2})$, $(1., \frac{1}{2})$, $(2., -\frac{2}{5})$, $(-3.16228, -\frac{3}{16})$, $(2., \frac{2}{5})$, $(2.82843, \frac{33}{80})$, $(2.82843, \frac{17}{80})$, $(-2.23607, \frac{1}{2})$, $(-2.23607, 0)$, $(1.41421, -\frac{3}{16})$, $(-2.23607, \frac{1}{2})$, $(-2.23607, 0)$, $(1.41421, -\frac{3}{16})$\}

\item $c = 3$, $(d_i,\theta_i)$ = \{$(1., 0)$, $(1., 0)$, $(2., \frac{1}{10})$, $(-3.16228, \frac{5}{16})$, $(2., -\frac{1}{10})$, $(1., \frac{1}{2})$, $(1., \frac{1}{2})$, $(2., -\frac{2}{5})$, $(-3.16228, -\frac{3}{16})$, $(2., \frac{2}{5})$, $(2., -\frac{1}{8})$, $(-3.16228, \frac{1}{16})$, $(-3.16228, -\frac{7}{16})$, $(2., \frac{11}{40})$, $(2., \frac{19}{40})$, $(2., \frac{11}{40})$, $(2., \frac{19}{40})$\}

\item $c = \frac{7}{2}$, $(d_i,\theta_i)$ = \{$(1., 0)$, $(1., 0)$, $(2., \frac{1}{10})$, $(-3.16228, \frac{5}{16})$, $(2., -\frac{1}{10})$, $(1., \frac{1}{2})$, $(1., \frac{1}{2})$, $(2., -\frac{2}{5})$, $(-3.16228, -\frac{3}{16})$, $(2., \frac{2}{5})$, $(2.82843, \frac{27}{80})$, $(2.82843, -\frac{37}{80})$, $(-2.23607, \frac{1}{8})$, $(-2.23607, -\frac{3}{8})$, $(1.41421, -\frac{1}{16})$, $(-2.23607, \frac{1}{8})$, $(-2.23607, -\frac{3}{8})$, $(1.41421, -\frac{1}{16})$\}

\item $c = 4$, $(d_i,\theta_i)$ = \{$(1., 0)$, $(1., 0)$, $(2., \frac{1}{10})$, $(-3.16228, \frac{5}{16})$, $(2., \frac{2}{5})$, $(1., \frac{1}{2})$, $(1., \frac{1}{2})$, $(2., -\frac{2}{5})$, $(-3.16228, -\frac{3}{16})$, $(2., -\frac{1}{10})$, $(2., 0)$, $(-3.16228, \frac{3}{16})$, $(-3.16228, -\frac{5}{16})$, $(2., \frac{2}{5})$, $(2., -\frac{2}{5})$, $(2., \frac{2}{5})$, $(2., -\frac{2}{5})$\}

\item $c = \frac{9}{2}$, $(d_i,\theta_i)$ = \{$(1., 0)$, $(1., 0)$, $(2., \frac{1}{10})$, $(-3.16228, \frac{5}{16})$, $(2., \frac{2}{5})$, $(1., \frac{1}{2})$, $(1., \frac{1}{2})$, $(2., -\frac{2}{5})$, $(-3.16228, -\frac{3}{16})$, $(2., -\frac{1}{10})$, $(2.82843, \frac{37}{80})$, $(2.82843, -\frac{27}{80})$, $(-2.23607, \frac{1}{4})$, $(-2.23607, -\frac{1}{4})$, $(1.41421, \frac{1}{16})$, $(-2.23607, \frac{1}{4})$, $(-2.23607, -\frac{1}{4})$, $(1.41421, \frac{1}{16})$\}

\item $c = 5$, $(d_i,\theta_i)$ = \{$(1., 0)$, $(1., 0)$, $(2., \frac{1}{10})$, $(-3.16228, \frac{5}{16})$, $(2., \frac{2}{5})$, $(1., \frac{1}{2})$, $(1., \frac{1}{2})$, $(2., -\frac{2}{5})$, $(-3.16228, -\frac{3}{16})$, $(2., -\frac{1}{10})$, $(-3.16228, \frac{5}{16})$, $(2., \frac{1}{8})$, $(-3.16228, -\frac{3}{16})$, $(2., -\frac{19}{40})$, $(2., -\frac{11}{40})$, $(2., -\frac{19}{40})$, $(2., -\frac{11}{40})$\}

\item $c = \frac{11}{2}$, $(d_i,\theta_i)$ = \{$(1., 0)$, $(1., 0)$, $(2., \frac{1}{10})$, $(-3.16228, \frac{5}{16})$, $(2., \frac{2}{5})$, $(1., \frac{1}{2})$, $(1., \frac{1}{2})$, $(2., -\frac{2}{5})$, $(-3.16228, -\frac{3}{16})$, $(2., -\frac{1}{10})$, $(2.82843, -\frac{17}{80})$, $(2.82843, -\frac{33}{80})$, $(-2.23607, -\frac{1}{8})$, $(-2.23607, \frac{3}{8})$, $(1.41421, \frac{3}{16})$, $(-2.23607, -\frac{1}{8})$, $(-2.23607, \frac{3}{8})$, $(1.41421, \frac{3}{16})$\}

\item $c = 6$, $(d_i,\theta_i)$ = \{$(1., 0)$, $(1., \frac{1}{2})$, $(2., \frac{1}{10})$, $(-3.16228, \frac{5}{16})$, $(2., \frac{2}{5})$, $(1., \frac{1}{2})$, $(1., 0)$, $(2., -\frac{2}{5})$, $(-3.16228, -\frac{3}{16})$, $(2., -\frac{1}{10})$, $(3.16228, -\frac{1}{16})$, $(-2., \frac{1}{4})$, $(3.16228, \frac{7}{16})$, $(-2., -\frac{7}{20})$, $(-2., -\frac{3}{20})$, $(-2., -\frac{7}{20})$, $(-2., -\frac{3}{20})$\}

\item $c = \frac{13}{2}$, $(d_i,\theta_i)$ = \{$(1., 0)$, $(1., \frac{1}{2})$, $(2., \frac{1}{10})$, $(-3.16228, \frac{5}{16})$, $(2., \frac{2}{5})$, $(1., \frac{1}{2})$, $(1., 0)$, $(2., -\frac{2}{5})$, $(-3.16228, -\frac{3}{16})$, $(2., -\frac{1}{10})$, $(-2.82843, -\frac{23}{80})$, $(-2.82843, -\frac{7}{80})$, $(2.23607, 0)$, $(2.23607, \frac{1}{2})$, $(-1.41421, \frac{5}{16})$, $(2.23607, 0)$, $(2.23607, \frac{1}{2})$, $(-1.41421, \frac{5}{16})$\}

\item $c = 7$, $(d_i,\theta_i)$ = \{$(1., 0)$, $(1., \frac{1}{2})$, $(2., \frac{1}{10})$, $(-3.16228, \frac{5}{16})$, $(2., \frac{2}{5})$, $(1., \frac{1}{2})$, $(1., 0)$, $(2., -\frac{2}{5})$, $(-3.16228, -\frac{3}{16})$, $(2., -\frac{1}{10})$, $(-2., \frac{3}{8})$, $(3.16228, \frac{1}{16})$, $(3.16228, -\frac{7}{16})$, $(-2., -\frac{9}{40})$, $(-2., -\frac{1}{40})$, $(-2., -\frac{9}{40})$, $(-2., -\frac{1}{40})$\}

\item $c = \frac{15}{2}$, $(d_i,\theta_i)$ = \{$(1., 0)$, $(1., \frac{1}{2})$, $(2., \frac{1}{10})$, $(-3.16228, \frac{5}{16})$, $(2., \frac{2}{5})$, $(1., \frac{1}{2})$, $(1., 0)$, $(2., -\frac{2}{5})$, $(-3.16228, -\frac{3}{16})$, $(2., -\frac{1}{10})$, $(-2.82843, -\frac{13}{80})$, $(-2.82843, \frac{3}{80})$, $(2.23607, -\frac{3}{8})$, $(2.23607, \frac{1}{8})$, $(-1.41421, \frac{7}{16})$, $(2.23607, -\frac{3}{8})$, $(2.23607, \frac{1}{8})$, $(-1.41421, \frac{7}{16})$\}

\end{enumerate}

\paragraph*{Rank 10; \#30}

\begin{enumerate}

\item $c = 0$, $(d_i,\theta_i)$ = \{$(1., 0)$, $(1., 0)$, $(2., \frac{1}{10})$, $(2., -\frac{1}{10})$, $(-3.16228, -\frac{1}{16})$, $(1., \frac{1}{2})$, $(1., \frac{1}{2})$, $(2., -\frac{2}{5})$, $(2., \frac{2}{5})$, $(-3.16228, \frac{7}{16})$, $(2., \frac{1}{2})$, $(-3.16228, -\frac{7}{16})$, $(-3.16228, \frac{1}{16})$, $(2., -\frac{1}{10})$, $(2., \frac{1}{10})$, $(2., -\frac{1}{10})$, $(2., \frac{1}{10})$\}

\item $c = \frac{1}{2}$, $(d_i,\theta_i)$ = \{$(1., 0)$, $(1., 0)$, $(2., \frac{1}{10})$, $(2., -\frac{1}{10})$, $(-3.16228, -\frac{1}{16})$, $(1., \frac{1}{2})$, $(1., \frac{1}{2})$, $(2., -\frac{2}{5})$, $(2., \frac{2}{5})$, $(-3.16228, \frac{7}{16})$, $(2.82843, \frac{13}{80})$, $(2.82843, -\frac{3}{80})$, $(-2.23607, -\frac{3}{8})$, $(-2.23607, \frac{1}{8})$, $(1.41421, -\frac{7}{16})$, $(-2.23607, -\frac{3}{8})$, $(-2.23607, \frac{1}{8})$, $(1.41421, -\frac{7}{16})$\}

\item $c = 1$, $(d_i,\theta_i)$ = \{$(1., 0)$, $(1., 0)$, $(2., \frac{1}{10})$, $(2., -\frac{1}{10})$, $(-3.16228, -\frac{1}{16})$, $(1., \frac{1}{2})$, $(1., \frac{1}{2})$, $(2., -\frac{2}{5})$, $(2., \frac{2}{5})$, $(-3.16228, \frac{7}{16})$, $(-3.16228, -\frac{5}{16})$, $(2., -\frac{3}{8})$, $(-3.16228, \frac{3}{16})$, $(2., \frac{1}{40})$, $(2., \frac{9}{40})$, $(2., \frac{1}{40})$, $(2., \frac{9}{40})$\}

\item $c = \frac{3}{2}$, $(d_i,\theta_i)$ = \{$(1., 0)$, $(1., 0)$, $(2., \frac{1}{10})$, $(2., -\frac{1}{10})$, $(-3.16228, -\frac{1}{16})$, $(1., \frac{1}{2})$, $(1., \frac{1}{2})$, $(2., -\frac{2}{5})$, $(2., \frac{2}{5})$, $(-3.16228, \frac{7}{16})$, $(2.82843, \frac{23}{80})$, $(2.82843, \frac{7}{80})$, $(-2.23607, \frac{1}{4})$, $(-2.23607, -\frac{1}{4})$, $(1.41421, -\frac{5}{16})$, $(-2.23607, \frac{1}{4})$, $(-2.23607, -\frac{1}{4})$, $(1.41421, -\frac{5}{16})$\}

\item $c = 2$, $(d_i,\theta_i)$ = \{$(1., 0)$, $(1., 0)$, $(2., \frac{1}{10})$, $(2., -\frac{1}{10})$, $(-3.16228, -\frac{1}{16})$, $(1., \frac{1}{2})$, $(1., \frac{1}{2})$, $(2., -\frac{2}{5})$, $(2., \frac{2}{5})$, $(-3.16228, \frac{7}{16})$, $(2., -\frac{1}{4})$, $(-3.16228, \frac{5}{16})$, $(-3.16228, -\frac{3}{16})$, $(2., \frac{3}{20})$, $(2., \frac{7}{20})$, $(2., \frac{3}{20})$, $(2., \frac{7}{20})$\}

\item $c = \frac{5}{2}$, $(d_i,\theta_i)$ = \{$(1., 0)$, $(1., 0)$, $(2., \frac{1}{10})$, $(2., -\frac{1}{10})$, $(-3.16228, -\frac{1}{16})$, $(1., \frac{1}{2})$, $(1., \frac{1}{2})$, $(2., -\frac{2}{5})$, $(2., \frac{2}{5})$, $(-3.16228, \frac{7}{16})$, $(2.82843, \frac{17}{80})$, $(2.82843, \frac{33}{80})$, $(-2.23607, \frac{3}{8})$, $(-2.23607, -\frac{1}{8})$, $(1.41421, -\frac{3}{16})$, $(-2.23607, \frac{3}{8})$, $(-2.23607, -\frac{1}{8})$, $(1.41421, -\frac{3}{16})$\}

\item $c = 3$, $(d_i,\theta_i)$ = \{$(1., 0)$, $(1., 0)$, $(2., \frac{1}{10})$, $(2., -\frac{1}{10})$, $(-3.16228, -\frac{1}{16})$, $(1., \frac{1}{2})$, $(1., \frac{1}{2})$, $(2., -\frac{2}{5})$, $(2., \frac{2}{5})$, $(-3.16228, \frac{7}{16})$, $(2., -\frac{1}{8})$, $(-3.16228, \frac{7}{16})$, $(-3.16228, -\frac{1}{16})$, $(2., \frac{11}{40})$, $(2., \frac{19}{40})$, $(2., \frac{11}{40})$, $(2., \frac{19}{40})$\}

\item $c = \frac{7}{2}$, $(d_i,\theta_i)$ = \{$(1., 0)$, $(1., 0)$, $(2., \frac{1}{10})$, $(2., -\frac{1}{10})$, $(-3.16228, -\frac{1}{16})$, $(1., \frac{1}{2})$, $(1., \frac{1}{2})$, $(2., -\frac{2}{5})$, $(2., \frac{2}{5})$, $(-3.16228, \frac{7}{16})$, $(2.82843, \frac{27}{80})$, $(2.82843, -\frac{37}{80})$, $(-2.23607, 0)$, $(-2.23607, \frac{1}{2})$, $(1.41421, -\frac{1}{16})$, $(-2.23607, 0)$, $(-2.23607, \frac{1}{2})$, $(1.41421, -\frac{1}{16})$\}

\item $c = 4$, $(d_i,\theta_i)$ = \{$(1., 0)$, $(1., 0)$, $(2., \frac{1}{10})$, $(2., \frac{2}{5})$, $(-3.16228, -\frac{1}{16})$, $(1., \frac{1}{2})$, $(1., \frac{1}{2})$, $(2., -\frac{2}{5})$, $(2., -\frac{1}{10})$, $(-3.16228, \frac{7}{16})$, $(2., 0)$, $(-3.16228, \frac{1}{16})$, $(-3.16228, -\frac{7}{16})$, $(2., \frac{2}{5})$, $(2., -\frac{2}{5})$, $(2., \frac{2}{5})$, $(2., -\frac{2}{5})$\}

\item $c = \frac{9}{2}$, $(d_i,\theta_i)$ = \{$(1., 0)$, $(1., 0)$, $(2., \frac{1}{10})$, $(2., \frac{2}{5})$, $(-3.16228, \frac{7}{16})$, $(1., \frac{1}{2})$, $(1., \frac{1}{2})$, $(2., -\frac{2}{5})$, $(2., -\frac{1}{10})$, $(-3.16228, -\frac{1}{16})$, $(2.82843, \frac{37}{80})$, $(2.82843, -\frac{27}{80})$, $(-2.23607, \frac{1}{8})$, $(-2.23607, -\frac{3}{8})$, $(1.41421, \frac{1}{16})$, $(-2.23607, \frac{1}{8})$, $(-2.23607, -\frac{3}{8})$, $(1.41421, \frac{1}{16})$\}

\item $c = 5$, $(d_i,\theta_i)$ = \{$(1., 0)$, $(1., 0)$, $(2., \frac{1}{10})$, $(2., \frac{2}{5})$, $(-3.16228, \frac{7}{16})$, $(1., \frac{1}{2})$, $(1., \frac{1}{2})$, $(2., -\frac{2}{5})$, $(2., -\frac{1}{10})$, $(-3.16228, -\frac{1}{16})$, $(-3.16228, \frac{3}{16})$, $(-3.16228, -\frac{5}{16})$, $(2., \frac{1}{8})$, $(2., -\frac{19}{40})$, $(2., -\frac{11}{40})$, $(2., -\frac{19}{40})$, $(2., -\frac{11}{40})$\}

\item $c = \frac{11}{2}$, $(d_i,\theta_i)$ = \{$(1., 0)$, $(1., 0)$, $(2., \frac{1}{10})$, $(2., \frac{2}{5})$, $(-3.16228, \frac{7}{16})$, $(1., \frac{1}{2})$, $(1., \frac{1}{2})$, $(2., -\frac{2}{5})$, $(2., -\frac{1}{10})$, $(-3.16228, -\frac{1}{16})$, $(2.82843, -\frac{17}{80})$, $(2.82843, -\frac{33}{80})$, $(-2.23607, -\frac{1}{4})$, $(-2.23607, \frac{1}{4})$, $(1.41421, \frac{3}{16})$, $(-2.23607, -\frac{1}{4})$, $(-2.23607, \frac{1}{4})$, $(1.41421, \frac{3}{16})$\}

\item $c = 6$, $(d_i,\theta_i)$ = \{$(1., 0)$, $(1., \frac{1}{2})$, $(2., \frac{1}{10})$, $(2., \frac{2}{5})$, $(-3.16228, \frac{7}{16})$, $(1., \frac{1}{2})$, $(1., 0)$, $(2., -\frac{2}{5})$, $(2., -\frac{1}{10})$, $(-3.16228, -\frac{1}{16})$, $(-2., \frac{1}{4})$, $(3.16228, \frac{5}{16})$, $(3.16228, -\frac{3}{16})$, $(-2., -\frac{7}{20})$, $(-2., -\frac{3}{20})$, $(-2., -\frac{7}{20})$, $(-2., -\frac{3}{20})$\}

\item $c = \frac{13}{2}$, $(d_i,\theta_i)$ = \{$(1., 0)$, $(1., \frac{1}{2})$, $(2., \frac{1}{10})$, $(2., \frac{2}{5})$, $(-3.16228, \frac{7}{16})$, $(1., \frac{1}{2})$, $(1., 0)$, $(2., -\frac{2}{5})$, $(2., -\frac{1}{10})$, $(-3.16228, -\frac{1}{16})$, $(-2.82843, -\frac{7}{80})$, $(-2.82843, -\frac{23}{80})$, $(2.23607, -\frac{1}{8})$, $(2.23607, \frac{3}{8})$, $(-1.41421, \frac{5}{16})$, $(2.23607, -\frac{1}{8})$, $(2.23607, \frac{3}{8})$, $(-1.41421, \frac{5}{16})$\}

\item $c = 7$, $(d_i,\theta_i)$ = \{$(1., 0)$, $(1., \frac{1}{2})$, $(2., \frac{1}{10})$, $(2., \frac{2}{5})$, $(-3.16228, \frac{7}{16})$, $(1., \frac{1}{2})$, $(1., 0)$, $(2., -\frac{2}{5})$, $(2., -\frac{1}{10})$, $(-3.16228, -\frac{1}{16})$, $(-2., \frac{3}{8})$, $(3.16228, \frac{7}{16})$, $(3.16228, -\frac{1}{16})$, $(-2., -\frac{9}{40})$, $(-2., -\frac{1}{40})$, $(-2., -\frac{9}{40})$, $(-2., -\frac{1}{40})$\}

\item $c = \frac{15}{2}$, $(d_i,\theta_i)$ = \{$(1., 0)$, $(1., \frac{1}{2})$, $(2., \frac{1}{10})$, $(2., \frac{2}{5})$, $(-3.16228, \frac{7}{16})$, $(1., \frac{1}{2})$, $(1., 0)$, $(2., -\frac{2}{5})$, $(2., -\frac{1}{10})$, $(-3.16228, -\frac{1}{16})$, $(-2.82843, \frac{3}{80})$, $(-2.82843, -\frac{13}{80})$, $(2.23607, \frac{1}{2})$, $(2.23607, 0)$, $(-1.41421, \frac{7}{16})$, $(2.23607, \frac{1}{2})$, $(2.23607, 0)$, $(-1.41421, \frac{7}{16})$\}

\end{enumerate}

\paragraph*{Rank 10; \#31}

\begin{enumerate}

\item $c = 0$, $(d_i,\theta_i)$ = \{$(1., 0)$, $(1., 0)$, $(-3.16228, \frac{1}{16})$, $(2., \frac{1}{5})$, $(2., -\frac{1}{5})$, $(1., \frac{1}{2})$, $(1., \frac{1}{2})$, $(-3.16228, -\frac{7}{16})$, $(2., -\frac{3}{10})$, $(2., \frac{3}{10})$, $(-3.16228, \frac{7}{16})$, $(-3.16228, -\frac{1}{16})$, $(2., 0)$, $(2., \frac{1}{5})$, $(2., -\frac{1}{5})$, $(2., \frac{1}{5})$, $(2., -\frac{1}{5})$\}

\item $c = \frac{1}{2}$, $(d_i,\theta_i)$ = \{$(1., 0)$, $(1., 0)$, $(-3.16228, \frac{1}{16})$, $(2., \frac{1}{5})$, $(2., -\frac{1}{5})$, $(1., \frac{1}{2})$, $(1., \frac{1}{2})$, $(-3.16228, -\frac{7}{16})$, $(2., -\frac{3}{10})$, $(2., \frac{3}{10})$, $(2.82843, -\frac{11}{80})$, $(2.82843, \frac{21}{80})$, $(-2.23607, 0)$, $(-2.23607, \frac{1}{2})$, $(1.41421, \frac{1}{16})$, $(-2.23607, 0)$, $(-2.23607, \frac{1}{2})$, $(1.41421, \frac{1}{16})$\}

\item $c = 1$, $(d_i,\theta_i)$ = \{$(1., 0)$, $(1., 0)$, $(-3.16228, \frac{1}{16})$, $(2., \frac{1}{5})$, $(2., -\frac{1}{5})$, $(1., \frac{1}{2})$, $(1., \frac{1}{2})$, $(-3.16228, -\frac{7}{16})$, $(2., -\frac{3}{10})$, $(2., \frac{3}{10})$, $(-3.16228, \frac{1}{16})$, $(2., \frac{1}{8})$, $(-3.16228, -\frac{7}{16})$, $(2., \frac{13}{40})$, $(2., -\frac{3}{40})$, $(2., \frac{13}{40})$, $(2., -\frac{3}{40})$\}

\item $c = \frac{3}{2}$, $(d_i,\theta_i)$ = \{$(1., 0)$, $(1., 0)$, $(-3.16228, \frac{1}{16})$, $(2., \frac{1}{5})$, $(2., \frac{3}{10})$, $(1., \frac{1}{2})$, $(1., \frac{1}{2})$, $(-3.16228, -\frac{7}{16})$, $(2., -\frac{3}{10})$, $(2., -\frac{1}{5})$, $(2.82843, \frac{31}{80})$, $(2.82843, -\frac{1}{80})$, $(-2.23607, \frac{1}{8})$, $(-2.23607, -\frac{3}{8})$, $(1.41421, \frac{3}{16})$, $(-2.23607, \frac{1}{8})$, $(-2.23607, -\frac{3}{8})$, $(1.41421, \frac{3}{16})$\}

\item $c = 2$, $(d_i,\theta_i)$ = \{$(1., 0)$, $(1., 0)$, $(-3.16228, \frac{1}{16})$, $(2., \frac{1}{5})$, $(2., \frac{3}{10})$, $(1., \frac{1}{2})$, $(1., \frac{1}{2})$, $(-3.16228, -\frac{7}{16})$, $(2., -\frac{3}{10})$, $(2., -\frac{1}{5})$, $(-3.16228, -\frac{5}{16})$, $(2., \frac{1}{4})$, $(-3.16228, \frac{3}{16})$, $(2., \frac{9}{20})$, $(2., \frac{1}{20})$, $(2., \frac{9}{20})$, $(2., \frac{1}{20})$\}

\item $c = \frac{5}{2}$, $(d_i,\theta_i)$ = \{$(1., 0)$, $(1., 0)$, $(-3.16228, \frac{1}{16})$, $(2., \frac{1}{5})$, $(2., \frac{3}{10})$, $(1., \frac{1}{2})$, $(1., \frac{1}{2})$, $(-3.16228, -\frac{7}{16})$, $(2., -\frac{3}{10})$, $(2., -\frac{1}{5})$, $(2.82843, \frac{9}{80})$, $(2.82843, -\frac{39}{80})$, $(-2.23607, -\frac{1}{4})$, $(-2.23607, \frac{1}{4})$, $(1.41421, \frac{5}{16})$, $(-2.23607, -\frac{1}{4})$, $(-2.23607, \frac{1}{4})$, $(1.41421, \frac{5}{16})$\}

\item $c = 3$, $(d_i,\theta_i)$ = \{$(1., 0)$, $(1., 0)$, $(-3.16228, \frac{1}{16})$, $(2., \frac{1}{5})$, $(2., \frac{3}{10})$, $(1., \frac{1}{2})$, $(1., \frac{1}{2})$, $(-3.16228, -\frac{7}{16})$, $(2., -\frac{3}{10})$, $(2., -\frac{1}{5})$, $(2., \frac{3}{8})$, $(-3.16228, \frac{5}{16})$, $(-3.16228, -\frac{3}{16})$, $(2., -\frac{17}{40})$, $(2., \frac{7}{40})$, $(2., -\frac{17}{40})$, $(2., \frac{7}{40})$\}

\item $c = \frac{7}{2}$, $(d_i,\theta_i)$ = \{$(1., 0)$, $(1., 0)$, $(-3.16228, \frac{1}{16})$, $(2., \frac{1}{5})$, $(2., \frac{3}{10})$, $(1., \frac{1}{2})$, $(1., \frac{1}{2})$, $(-3.16228, -\frac{7}{16})$, $(2., -\frac{3}{10})$, $(2., -\frac{1}{5})$, $(2.82843, -\frac{29}{80})$, $(2.82843, \frac{19}{80})$, $(-2.23607, -\frac{1}{8})$, $(-2.23607, \frac{3}{8})$, $(1.41421, \frac{7}{16})$, $(-2.23607, -\frac{1}{8})$, $(-2.23607, \frac{3}{8})$, $(1.41421, \frac{7}{16})$\}

\item $c = 4$, $(d_i,\theta_i)$ = \{$(1., 0)$, $(1., 0)$, $(-3.16228, \frac{1}{16})$, $(2., \frac{1}{5})$, $(2., \frac{3}{10})$, $(1., \frac{1}{2})$, $(1., \frac{1}{2})$, $(-3.16228, -\frac{7}{16})$, $(2., -\frac{3}{10})$, $(2., -\frac{1}{5})$, $(-3.16228, -\frac{1}{16})$, $(2., \frac{1}{2})$, $(-3.16228, \frac{7}{16})$, $(2., -\frac{3}{10})$, $(2., \frac{3}{10})$, $(2., -\frac{3}{10})$, $(2., \frac{3}{10})$\}

\item $c = \frac{9}{2}$, $(d_i,\theta_i)$ = \{$(1., 0)$, $(1., 0)$, $(-3.16228, \frac{1}{16})$, $(2., \frac{1}{5})$, $(2., \frac{3}{10})$, $(1., \frac{1}{2})$, $(1., \frac{1}{2})$, $(-3.16228, -\frac{7}{16})$, $(2., -\frac{3}{10})$, $(2., -\frac{1}{5})$, $(2.82843, \frac{29}{80})$, $(2.82843, -\frac{19}{80})$, $(-2.23607, \frac{1}{2})$, $(-2.23607, 0)$, $(1.41421, -\frac{7}{16})$, $(-2.23607, \frac{1}{2})$, $(-2.23607, 0)$, $(1.41421, -\frac{7}{16})$\}

\item $c = 5$, $(d_i,\theta_i)$ = \{$(1., 0)$, $(1., 0)$, $(-3.16228, \frac{1}{16})$, $(2., \frac{1}{5})$, $(2., \frac{3}{10})$, $(1., \frac{1}{2})$, $(1., \frac{1}{2})$, $(-3.16228, -\frac{7}{16})$, $(2., -\frac{3}{10})$, $(2., -\frac{1}{5})$, $(2., -\frac{3}{8})$, $(-3.16228, -\frac{7}{16})$, $(-3.16228, \frac{1}{16})$, $(2., -\frac{7}{40})$, $(2., \frac{17}{40})$, $(2., -\frac{7}{40})$, $(2., \frac{17}{40})$\}

\item $c = \frac{11}{2}$, $(d_i,\theta_i)$ = \{$(1., 0)$, $(1., 0)$, $(-3.16228, \frac{1}{16})$, $(2., \frac{1}{5})$, $(2., \frac{3}{10})$, $(1., \frac{1}{2})$, $(1., \frac{1}{2})$, $(-3.16228, -\frac{7}{16})$, $(2., -\frac{3}{10})$, $(2., -\frac{1}{5})$, $(2.82843, \frac{39}{80})$, $(2.82843, -\frac{9}{80})$, $(-2.23607, -\frac{3}{8})$, $(-2.23607, \frac{1}{8})$, $(1.41421, -\frac{5}{16})$, $(-2.23607, -\frac{3}{8})$, $(-2.23607, \frac{1}{8})$, $(1.41421, -\frac{5}{16})$\}

\item $c = 6$, $(d_i,\theta_i)$ = \{$(1., 0)$, $(1., \frac{1}{2})$, $(-3.16228, \frac{1}{16})$, $(2., \frac{1}{5})$, $(2., \frac{3}{10})$, $(1., \frac{1}{2})$, $(1., 0)$, $(-3.16228, -\frac{7}{16})$, $(2., -\frac{3}{10})$, $(2., -\frac{1}{5})$, $(-2., -\frac{1}{4})$, $(3.16228, -\frac{5}{16})$, $(3.16228, \frac{3}{16})$, $(-2., -\frac{1}{20})$, $(-2., -\frac{9}{20})$, $(-2., -\frac{1}{20})$, $(-2., -\frac{9}{20})$\}

\item $c = \frac{13}{2}$, $(d_i,\theta_i)$ = \{$(1., 0)$, $(1., \frac{1}{2})$, $(-3.16228, \frac{1}{16})$, $(2., \frac{1}{5})$, $(2., \frac{3}{10})$, $(1., \frac{1}{2})$, $(1., 0)$, $(-3.16228, -\frac{7}{16})$, $(2., -\frac{3}{10})$, $(2., -\frac{1}{5})$, $(-2.82843, \frac{1}{80})$, $(-2.82843, -\frac{31}{80})$, $(2.23607, \frac{1}{4})$, $(2.23607, -\frac{1}{4})$, $(-1.41421, -\frac{3}{16})$, $(2.23607, \frac{1}{4})$, $(2.23607, -\frac{1}{4})$, $(-1.41421, -\frac{3}{16})$\}

\item $c = 7$, $(d_i,\theta_i)$ = \{$(1., 0)$, $(1., \frac{1}{2})$, $(-3.16228, \frac{1}{16})$, $(2., \frac{1}{5})$, $(2., \frac{3}{10})$, $(1., \frac{1}{2})$, $(1., 0)$, $(-3.16228, -\frac{7}{16})$, $(2., -\frac{3}{10})$, $(2., -\frac{1}{5})$, $(3.16228, \frac{5}{16})$, $(3.16228, -\frac{3}{16})$, $(-2., -\frac{1}{8})$, $(-2., \frac{3}{40})$, $(-2., -\frac{13}{40})$, $(-2., \frac{3}{40})$, $(-2., -\frac{13}{40})$\}

\item $c = \frac{15}{2}$, $(d_i,\theta_i)$ = \{$(1., 0)$, $(1., \frac{1}{2})$, $(-3.16228, -\frac{7}{16})$, $(2., \frac{1}{5})$, $(2., \frac{3}{10})$, $(1., \frac{1}{2})$, $(1., 0)$, $(-3.16228, \frac{1}{16})$, $(2., -\frac{3}{10})$, $(2., -\frac{1}{5})$, $(-2.82843, -\frac{21}{80})$, $(-2.82843, \frac{11}{80})$, $(2.23607, \frac{3}{8})$, $(2.23607, -\frac{1}{8})$, $(-1.41421, -\frac{1}{16})$, $(2.23607, \frac{3}{8})$, $(2.23607, -\frac{1}{8})$, $(-1.41421, -\frac{1}{16})$\}

\end{enumerate}

\paragraph*{Rank 10; \#32}

\begin{enumerate}

\item $c = 0$, $(d_i,\theta_i)$ = \{$(1., 0)$, $(1., 0)$, $(-3.16228, \frac{3}{16})$, $(2., \frac{1}{5})$, $(2., -\frac{1}{5})$, $(1., \frac{1}{2})$, $(1., \frac{1}{2})$, $(-3.16228, -\frac{5}{16})$, $(2., -\frac{3}{10})$, $(2., \frac{3}{10})$, $(-3.16228, \frac{5}{16})$, $(2., 0)$, $(-3.16228, -\frac{3}{16})$, $(2., \frac{1}{5})$, $(2., -\frac{1}{5})$, $(2., \frac{1}{5})$, $(2., -\frac{1}{5})$\}

\item $c = \frac{1}{2}$, $(d_i,\theta_i)$ = \{$(1., 0)$, $(1., 0)$, $(-3.16228, \frac{3}{16})$, $(2., \frac{1}{5})$, $(2., -\frac{1}{5})$, $(1., \frac{1}{2})$, $(1., \frac{1}{2})$, $(-3.16228, -\frac{5}{16})$, $(2., -\frac{3}{10})$, $(2., \frac{3}{10})$, $(2.82843, -\frac{11}{80})$, $(2.82843, \frac{21}{80})$, $(-2.23607, -\frac{1}{8})$, $(-2.23607, \frac{3}{8})$, $(1.41421, \frac{1}{16})$, $(-2.23607, -\frac{1}{8})$, $(-2.23607, \frac{3}{8})$, $(1.41421, \frac{1}{16})$\}

\item $c = 1$, $(d_i,\theta_i)$ = \{$(1., 0)$, $(1., 0)$, $(-3.16228, \frac{3}{16})$, $(2., \frac{1}{5})$, $(2., -\frac{1}{5})$, $(1., \frac{1}{2})$, $(1., \frac{1}{2})$, $(-3.16228, -\frac{5}{16})$, $(2., -\frac{3}{10})$, $(2., \frac{3}{10})$, $(-3.16228, -\frac{1}{16})$, $(-3.16228, \frac{7}{16})$, $(2., \frac{1}{8})$, $(2., \frac{13}{40})$, $(2., -\frac{3}{40})$, $(2., \frac{13}{40})$, $(2., -\frac{3}{40})$\}

\item $c = \frac{3}{2}$, $(d_i,\theta_i)$ = \{$(1., 0)$, $(1., 0)$, $(-3.16228, \frac{3}{16})$, $(2., \frac{1}{5})$, $(2., \frac{3}{10})$, $(1., \frac{1}{2})$, $(1., \frac{1}{2})$, $(-3.16228, -\frac{5}{16})$, $(2., -\frac{3}{10})$, $(2., -\frac{1}{5})$, $(2.82843, \frac{31}{80})$, $(2.82843, -\frac{1}{80})$, $(-2.23607, 0)$, $(-2.23607, \frac{1}{2})$, $(1.41421, \frac{3}{16})$, $(-2.23607, 0)$, $(-2.23607, \frac{1}{2})$, $(1.41421, \frac{3}{16})$\}

\item $c = 2$, $(d_i,\theta_i)$ = \{$(1., 0)$, $(1., 0)$, $(-3.16228, \frac{3}{16})$, $(2., \frac{1}{5})$, $(2., \frac{3}{10})$, $(1., \frac{1}{2})$, $(1., \frac{1}{2})$, $(-3.16228, -\frac{5}{16})$, $(2., -\frac{3}{10})$, $(2., -\frac{1}{5})$, $(-3.16228, \frac{1}{16})$, $(-3.16228, -\frac{7}{16})$, $(2., \frac{1}{4})$, $(2., \frac{9}{20})$, $(2., \frac{1}{20})$, $(2., \frac{9}{20})$, $(2., \frac{1}{20})$\}

\item $c = \frac{5}{2}$, $(d_i,\theta_i)$ = \{$(1., 0)$, $(1., 0)$, $(-3.16228, \frac{3}{16})$, $(2., \frac{1}{5})$, $(2., \frac{3}{10})$, $(1., \frac{1}{2})$, $(1., \frac{1}{2})$, $(-3.16228, -\frac{5}{16})$, $(2., -\frac{3}{10})$, $(2., -\frac{1}{5})$, $(2.82843, \frac{9}{80})$, $(2.82843, -\frac{39}{80})$, $(-2.23607, -\frac{3}{8})$, $(-2.23607, \frac{1}{8})$, $(1.41421, \frac{5}{16})$, $(-2.23607, -\frac{3}{8})$, $(-2.23607, \frac{1}{8})$, $(1.41421, \frac{5}{16})$\}

\item $c = 3$, $(d_i,\theta_i)$ = \{$(1., 0)$, $(1., 0)$, $(-3.16228, \frac{3}{16})$, $(2., \frac{1}{5})$, $(2., \frac{3}{10})$, $(1., \frac{1}{2})$, $(1., \frac{1}{2})$, $(-3.16228, -\frac{5}{16})$, $(2., -\frac{3}{10})$, $(2., -\frac{1}{5})$, $(-3.16228, -\frac{5}{16})$, $(-3.16228, \frac{3}{16})$, $(2., \frac{3}{8})$, $(2., -\frac{17}{40})$, $(2., \frac{7}{40})$, $(2., -\frac{17}{40})$, $(2., \frac{7}{40})$\}

\item $c = \frac{7}{2}$, $(d_i,\theta_i)$ = \{$(1., 0)$, $(1., 0)$, $(-3.16228, \frac{3}{16})$, $(2., \frac{1}{5})$, $(2., \frac{3}{10})$, $(1., \frac{1}{2})$, $(1., \frac{1}{2})$, $(-3.16228, -\frac{5}{16})$, $(2., -\frac{3}{10})$, $(2., -\frac{1}{5})$, $(2.82843, \frac{19}{80})$, $(2.82843, -\frac{29}{80})$, $(-2.23607, -\frac{1}{4})$, $(-2.23607, \frac{1}{4})$, $(1.41421, \frac{7}{16})$, $(-2.23607, -\frac{1}{4})$, $(-2.23607, \frac{1}{4})$, $(1.41421, \frac{7}{16})$\}

\item $c = 4$, $(d_i,\theta_i)$ = \{$(1., 0)$, $(1., 0)$, $(-3.16228, \frac{3}{16})$, $(2., \frac{1}{5})$, $(2., \frac{3}{10})$, $(1., \frac{1}{2})$, $(1., \frac{1}{2})$, $(-3.16228, -\frac{5}{16})$, $(2., -\frac{3}{10})$, $(2., -\frac{1}{5})$, $(2., \frac{1}{2})$, $(-3.16228, \frac{5}{16})$, $(-3.16228, -\frac{3}{16})$, $(2., -\frac{3}{10})$, $(2., \frac{3}{10})$, $(2., -\frac{3}{10})$, $(2., \frac{3}{10})$\}

\item $c = \frac{9}{2}$, $(d_i,\theta_i)$ = \{$(1., 0)$, $(1., 0)$, $(-3.16228, \frac{3}{16})$, $(2., \frac{1}{5})$, $(2., \frac{3}{10})$, $(1., \frac{1}{2})$, $(1., \frac{1}{2})$, $(-3.16228, -\frac{5}{16})$, $(2., -\frac{3}{10})$, $(2., -\frac{1}{5})$, $(2.82843, -\frac{19}{80})$, $(2.82843, \frac{29}{80})$, $(-2.23607, \frac{3}{8})$, $(-2.23607, -\frac{1}{8})$, $(1.41421, -\frac{7}{16})$, $(-2.23607, \frac{3}{8})$, $(-2.23607, -\frac{1}{8})$, $(1.41421, -\frac{7}{16})$\}

\item $c = 5$, $(d_i,\theta_i)$ = \{$(1., 0)$, $(1., 0)$, $(-3.16228, \frac{3}{16})$, $(2., \frac{1}{5})$, $(2., \frac{3}{10})$, $(1., \frac{1}{2})$, $(1., \frac{1}{2})$, $(-3.16228, -\frac{5}{16})$, $(2., -\frac{3}{10})$, $(2., -\frac{1}{5})$, $(-3.16228, -\frac{1}{16})$, $(-3.16228, \frac{7}{16})$, $(2., -\frac{3}{8})$, $(2., -\frac{7}{40})$, $(2., \frac{17}{40})$, $(2., -\frac{7}{40})$, $(2., \frac{17}{40})$\}

\item $c = \frac{11}{2}$, $(d_i,\theta_i)$ = \{$(1., 0)$, $(1., 0)$, $(-3.16228, \frac{3}{16})$, $(2., \frac{1}{5})$, $(2., \frac{3}{10})$, $(1., \frac{1}{2})$, $(1., \frac{1}{2})$, $(-3.16228, -\frac{5}{16})$, $(2., -\frac{3}{10})$, $(2., -\frac{1}{5})$, $(2.82843, -\frac{9}{80})$, $(2.82843, \frac{39}{80})$, $(-2.23607, \frac{1}{2})$, $(-2.23607, 0)$, $(1.41421, -\frac{5}{16})$, $(-2.23607, \frac{1}{2})$, $(-2.23607, 0)$, $(1.41421, -\frac{5}{16})$\}

\item $c = 6$, $(d_i,\theta_i)$ = \{$(1., 0)$, $(1., \frac{1}{2})$, $(-3.16228, \frac{3}{16})$, $(2., \frac{1}{5})$, $(2., \frac{3}{10})$, $(1., \frac{1}{2})$, $(1., 0)$, $(-3.16228, -\frac{5}{16})$, $(2., -\frac{3}{10})$, $(2., -\frac{1}{5})$, $(3.16228, \frac{1}{16})$, $(-2., -\frac{1}{4})$, $(3.16228, -\frac{7}{16})$, $(-2., -\frac{1}{20})$, $(-2., -\frac{9}{20})$, $(-2., -\frac{1}{20})$, $(-2., -\frac{9}{20})$\}

\item $c = \frac{13}{2}$, $(d_i,\theta_i)$ = \{$(1., 0)$, $(1., \frac{1}{2})$, $(-3.16228, \frac{3}{16})$, $(2., \frac{1}{5})$, $(2., \frac{3}{10})$, $(1., \frac{1}{2})$, $(1., 0)$, $(-3.16228, -\frac{5}{16})$, $(2., -\frac{3}{10})$, $(2., -\frac{1}{5})$, $(-2.82843, -\frac{31}{80})$, $(-2.82843, \frac{1}{80})$, $(2.23607, \frac{1}{8})$, $(2.23607, -\frac{3}{8})$, $(-1.41421, -\frac{3}{16})$, $(2.23607, \frac{1}{8})$, $(2.23607, -\frac{3}{8})$, $(-1.41421, -\frac{3}{16})$\}

\item $c = 7$, $(d_i,\theta_i)$ = \{$(1., 0)$, $(1., \frac{1}{2})$, $(-3.16228, \frac{3}{16})$, $(2., \frac{1}{5})$, $(2., \frac{3}{10})$, $(1., \frac{1}{2})$, $(1., 0)$, $(-3.16228, -\frac{5}{16})$, $(2., -\frac{3}{10})$, $(2., -\frac{1}{5})$, $(3.16228, -\frac{5}{16})$, $(3.16228, \frac{3}{16})$, $(-2., -\frac{1}{8})$, $(-2., \frac{3}{40})$, $(-2., -\frac{13}{40})$, $(-2., \frac{3}{40})$, $(-2., -\frac{13}{40})$\}

\item $c = \frac{15}{2}$, $(d_i,\theta_i)$ = \{$(1., 0)$, $(1., \frac{1}{2})$, $(-3.16228, \frac{3}{16})$, $(2., \frac{1}{5})$, $(2., \frac{3}{10})$, $(1., \frac{1}{2})$, $(1., 0)$, $(-3.16228, -\frac{5}{16})$, $(2., -\frac{3}{10})$, $(2., -\frac{1}{5})$, $(-2.82843, -\frac{21}{80})$, $(-2.82843, \frac{11}{80})$, $(2.23607, \frac{1}{4})$, $(2.23607, -\frac{1}{4})$, $(-1.41421, -\frac{1}{16})$, $(2.23607, \frac{1}{4})$, $(2.23607, -\frac{1}{4})$, $(-1.41421, -\frac{1}{16})$\}

\end{enumerate}

\paragraph*{Rank 10; \#33}

\begin{enumerate}

\item $c = 0$, $(d_i,\theta_i)$ = \{$(1., 0)$, $(1., 0)$, $(2., \frac{1}{5})$, $(2., -\frac{1}{5})$, $(-3.16228, -\frac{3}{16})$, $(1., \frac{1}{2})$, $(1., \frac{1}{2})$, $(2., -\frac{3}{10})$, $(2., \frac{3}{10})$, $(-3.16228, \frac{5}{16})$, $(-3.16228, -\frac{5}{16})$, $(2., 0)$, $(-3.16228, \frac{3}{16})$, $(2., \frac{1}{5})$, $(2., -\frac{1}{5})$, $(2., \frac{1}{5})$, $(2., -\frac{1}{5})$\}

\item $c = \frac{1}{2}$, $(d_i,\theta_i)$ = \{$(1., 0)$, $(1., 0)$, $(2., \frac{1}{5})$, $(2., -\frac{1}{5})$, $(-3.16228, -\frac{3}{16})$, $(1., \frac{1}{2})$, $(1., \frac{1}{2})$, $(2., -\frac{3}{10})$, $(2., \frac{3}{10})$, $(-3.16228, \frac{5}{16})$, $(2.82843, \frac{21}{80})$, $(2.82843, -\frac{11}{80})$, $(-2.23607, -\frac{1}{4})$, $(-2.23607, \frac{1}{4})$, $(1.41421, \frac{1}{16})$, $(-2.23607, -\frac{1}{4})$, $(-2.23607, \frac{1}{4})$, $(1.41421, \frac{1}{16})$\}

\item $c = 1$, $(d_i,\theta_i)$ = \{$(1., 0)$, $(1., 0)$, $(2., \frac{1}{5})$, $(2., -\frac{1}{5})$, $(-3.16228, -\frac{3}{16})$, $(1., \frac{1}{2})$, $(1., \frac{1}{2})$, $(2., -\frac{3}{10})$, $(2., \frac{3}{10})$, $(-3.16228, \frac{5}{16})$, $(-3.16228, -\frac{3}{16})$, $(-3.16228, \frac{5}{16})$, $(2., \frac{1}{8})$, $(2., \frac{13}{40})$, $(2., -\frac{3}{40})$, $(2., \frac{13}{40})$, $(2., -\frac{3}{40})$\}

\item $c = \frac{3}{2}$, $(d_i,\theta_i)$ = \{$(1., 0)$, $(1., 0)$, $(2., \frac{1}{5})$, $(2., \frac{3}{10})$, $(-3.16228, \frac{5}{16})$, $(1., \frac{1}{2})$, $(1., \frac{1}{2})$, $(2., -\frac{3}{10})$, $(2., -\frac{1}{5})$, $(-3.16228, -\frac{3}{16})$, $(2.82843, -\frac{1}{80})$, $(2.82843, \frac{31}{80})$, $(-2.23607, \frac{3}{8})$, $(-2.23607, -\frac{1}{8})$, $(1.41421, \frac{3}{16})$, $(-2.23607, \frac{3}{8})$, $(-2.23607, -\frac{1}{8})$, $(1.41421, \frac{3}{16})$\}

\item $c = 2$, $(d_i,\theta_i)$ = \{$(1., 0)$, $(1., 0)$, $(2., \frac{1}{5})$, $(2., \frac{3}{10})$, $(-3.16228, \frac{5}{16})$, $(1., \frac{1}{2})$, $(1., \frac{1}{2})$, $(2., -\frac{3}{10})$, $(2., -\frac{1}{5})$, $(-3.16228, -\frac{3}{16})$, $(2., \frac{1}{4})$, $(-3.16228, -\frac{1}{16})$, $(-3.16228, \frac{7}{16})$, $(2., \frac{9}{20})$, $(2., \frac{1}{20})$, $(2., \frac{9}{20})$, $(2., \frac{1}{20})$\}

\item $c = \frac{5}{2}$, $(d_i,\theta_i)$ = \{$(1., 0)$, $(1., 0)$, $(2., \frac{1}{5})$, $(2., \frac{3}{10})$, $(-3.16228, \frac{5}{16})$, $(1., \frac{1}{2})$, $(1., \frac{1}{2})$, $(2., -\frac{3}{10})$, $(2., -\frac{1}{5})$, $(-3.16228, -\frac{3}{16})$, $(2.82843, -\frac{39}{80})$, $(2.82843, \frac{9}{80})$, $(-2.23607, \frac{1}{2})$, $(-2.23607, 0)$, $(1.41421, \frac{5}{16})$, $(-2.23607, \frac{1}{2})$, $(-2.23607, 0)$, $(1.41421, \frac{5}{16})$\}

\item $c = 3$, $(d_i,\theta_i)$ = \{$(1., 0)$, $(1., 0)$, $(2., \frac{1}{5})$, $(2., \frac{3}{10})$, $(-3.16228, \frac{5}{16})$, $(1., \frac{1}{2})$, $(1., \frac{1}{2})$, $(2., -\frac{3}{10})$, $(2., -\frac{1}{5})$, $(-3.16228, -\frac{3}{16})$, $(2., \frac{3}{8})$, $(-3.16228, \frac{1}{16})$, $(-3.16228, -\frac{7}{16})$, $(2., -\frac{17}{40})$, $(2., \frac{7}{40})$, $(2., -\frac{17}{40})$, $(2., \frac{7}{40})$\}

\item $c = \frac{7}{2}$, $(d_i,\theta_i)$ = \{$(1., 0)$, $(1., 0)$, $(2., \frac{1}{5})$, $(2., \frac{3}{10})$, $(-3.16228, \frac{5}{16})$, $(1., \frac{1}{2})$, $(1., \frac{1}{2})$, $(2., -\frac{3}{10})$, $(2., -\frac{1}{5})$, $(-3.16228, -\frac{3}{16})$, $(2.82843, \frac{19}{80})$, $(2.82843, -\frac{29}{80})$, $(-2.23607, \frac{1}{8})$, $(-2.23607, -\frac{3}{8})$, $(1.41421, \frac{7}{16})$, $(-2.23607, \frac{1}{8})$, $(-2.23607, -\frac{3}{8})$, $(1.41421, \frac{7}{16})$\}

\item $c = 4$, $(d_i,\theta_i)$ = \{$(1., 0)$, $(1., 0)$, $(2., \frac{1}{5})$, $(2., \frac{3}{10})$, $(-3.16228, \frac{5}{16})$, $(1., \frac{1}{2})$, $(1., \frac{1}{2})$, $(2., -\frac{3}{10})$, $(2., -\frac{1}{5})$, $(-3.16228, -\frac{3}{16})$, $(-3.16228, -\frac{5}{16})$, $(2., \frac{1}{2})$, $(-3.16228, \frac{3}{16})$, $(2., -\frac{3}{10})$, $(2., \frac{3}{10})$, $(2., -\frac{3}{10})$, $(2., \frac{3}{10})$\}

\item $c = \frac{9}{2}$, $(d_i,\theta_i)$ = \{$(1., 0)$, $(1., 0)$, $(2., \frac{1}{5})$, $(2., \frac{3}{10})$, $(-3.16228, \frac{5}{16})$, $(1., \frac{1}{2})$, $(1., \frac{1}{2})$, $(2., -\frac{3}{10})$, $(2., -\frac{1}{5})$, $(-3.16228, -\frac{3}{16})$, $(2.82843, \frac{29}{80})$, $(2.82843, -\frac{19}{80})$, $(-2.23607, \frac{1}{4})$, $(-2.23607, -\frac{1}{4})$, $(1.41421, -\frac{7}{16})$, $(-2.23607, \frac{1}{4})$, $(-2.23607, -\frac{1}{4})$, $(1.41421, -\frac{7}{16})$\}

\item $c = 5$, $(d_i,\theta_i)$ = \{$(1., 0)$, $(1., 0)$, $(2., \frac{1}{5})$, $(2., \frac{3}{10})$, $(-3.16228, \frac{5}{16})$, $(1., \frac{1}{2})$, $(1., \frac{1}{2})$, $(2., -\frac{3}{10})$, $(2., -\frac{1}{5})$, $(-3.16228, -\frac{3}{16})$, $(-3.16228, \frac{5}{16})$, $(2., -\frac{3}{8})$, $(-3.16228, -\frac{3}{16})$, $(2., -\frac{7}{40})$, $(2., \frac{17}{40})$, $(2., -\frac{7}{40})$, $(2., \frac{17}{40})$\}

\item $c = \frac{11}{2}$, $(d_i,\theta_i)$ = \{$(1., 0)$, $(1., 0)$, $(2., \frac{1}{5})$, $(2., \frac{3}{10})$, $(-3.16228, \frac{5}{16})$, $(1., \frac{1}{2})$, $(1., \frac{1}{2})$, $(2., -\frac{3}{10})$, $(2., -\frac{1}{5})$, $(-3.16228, -\frac{3}{16})$, $(2.82843, \frac{39}{80})$, $(2.82843, -\frac{9}{80})$, $(-2.23607, -\frac{1}{8})$, $(-2.23607, \frac{3}{8})$, $(1.41421, -\frac{5}{16})$, $(-2.23607, -\frac{1}{8})$, $(-2.23607, \frac{3}{8})$, $(1.41421, -\frac{5}{16})$\}

\item $c = 6$, $(d_i,\theta_i)$ = \{$(1., 0)$, $(1., \frac{1}{2})$, $(2., \frac{1}{5})$, $(2., \frac{3}{10})$, $(-3.16228, \frac{5}{16})$, $(1., \frac{1}{2})$, $(1., 0)$, $(2., -\frac{3}{10})$, $(2., -\frac{1}{5})$, $(-3.16228, -\frac{3}{16})$, $(3.16228, \frac{7}{16})$, $(-2., -\frac{1}{4})$, $(3.16228, -\frac{1}{16})$, $(-2., -\frac{1}{20})$, $(-2., -\frac{9}{20})$, $(-2., -\frac{1}{20})$, $(-2., -\frac{9}{20})$\}

\item $c = \frac{13}{2}$, $(d_i,\theta_i)$ = \{$(1., 0)$, $(1., \frac{1}{2})$, $(2., \frac{1}{5})$, $(2., \frac{3}{10})$, $(-3.16228, \frac{5}{16})$, $(1., \frac{1}{2})$, $(1., 0)$, $(2., -\frac{3}{10})$, $(2., -\frac{1}{5})$, $(-3.16228, -\frac{3}{16})$, $(-2.82843, \frac{1}{80})$, $(-2.82843, -\frac{31}{80})$, $(2.23607, 0)$, $(2.23607, \frac{1}{2})$, $(-1.41421, -\frac{3}{16})$, $(2.23607, 0)$, $(2.23607, \frac{1}{2})$, $(-1.41421, -\frac{3}{16})$\}

\item $c = 7$, $(d_i,\theta_i)$ = \{$(1., 0)$, $(1., \frac{1}{2})$, $(2., \frac{1}{5})$, $(2., \frac{3}{10})$, $(-3.16228, \frac{5}{16})$, $(1., \frac{1}{2})$, $(1., 0)$, $(2., -\frac{3}{10})$, $(2., -\frac{1}{5})$, $(-3.16228, -\frac{3}{16})$, $(-2., -\frac{1}{8})$, $(3.16228, -\frac{7}{16})$, $(3.16228, \frac{1}{16})$, $(-2., \frac{3}{40})$, $(-2., -\frac{13}{40})$, $(-2., \frac{3}{40})$, $(-2., -\frac{13}{40})$\}

\item $c = \frac{15}{2}$, $(d_i,\theta_i)$ = \{$(1., 0)$, $(1., \frac{1}{2})$, $(2., \frac{1}{5})$, $(2., \frac{3}{10})$, $(-3.16228, \frac{5}{16})$, $(1., \frac{1}{2})$, $(1., 0)$, $(2., -\frac{3}{10})$, $(2., -\frac{1}{5})$, $(-3.16228, -\frac{3}{16})$, $(-2.82843, -\frac{21}{80})$, $(-2.82843, \frac{11}{80})$, $(2.23607, -\frac{3}{8})$, $(2.23607, \frac{1}{8})$, $(-1.41421, -\frac{1}{16})$, $(2.23607, -\frac{3}{8})$, $(2.23607, \frac{1}{8})$, $(-1.41421, -\frac{1}{16})$\}

\end{enumerate}

\paragraph*{Rank 10; \#34}

\begin{enumerate}

\item $c = 0$, $(d_i,\theta_i)$ = \{$(1., 0)$, $(1., 0)$, $(2., \frac{1}{5})$, $(2., -\frac{1}{5})$, $(-3.16228, -\frac{1}{16})$, $(1., \frac{1}{2})$, $(1., \frac{1}{2})$, $(2., -\frac{3}{10})$, $(2., \frac{3}{10})$, $(-3.16228, \frac{7}{16})$, $(-3.16228, \frac{1}{16})$, $(2., 0)$, $(-3.16228, -\frac{7}{16})$, $(2., \frac{1}{5})$, $(2., -\frac{1}{5})$, $(2., \frac{1}{5})$, $(2., -\frac{1}{5})$\}

\item $c = \frac{1}{2}$, $(d_i,\theta_i)$ = \{$(1., 0)$, $(1., 0)$, $(2., \frac{1}{5})$, $(2., -\frac{1}{5})$, $(-3.16228, -\frac{1}{16})$, $(1., \frac{1}{2})$, $(1., \frac{1}{2})$, $(2., -\frac{3}{10})$, $(2., \frac{3}{10})$, $(-3.16228, \frac{7}{16})$, $(2.82843, -\frac{11}{80})$, $(2.82843, \frac{21}{80})$, $(-2.23607, -\frac{3}{8})$, $(-2.23607, \frac{1}{8})$, $(1.41421, \frac{1}{16})$, $(-2.23607, -\frac{3}{8})$, $(-2.23607, \frac{1}{8})$, $(1.41421, \frac{1}{16})$\}

\item $c = 1$, $(d_i,\theta_i)$ = \{$(1., 0)$, $(1., 0)$, $(2., \frac{1}{5})$, $(2., -\frac{1}{5})$, $(-3.16228, -\frac{1}{16})$, $(1., \frac{1}{2})$, $(1., \frac{1}{2})$, $(2., -\frac{3}{10})$, $(2., \frac{3}{10})$, $(-3.16228, \frac{7}{16})$, $(2., \frac{1}{8})$, $(-3.16228, -\frac{5}{16})$, $(-3.16228, \frac{3}{16})$, $(2., \frac{13}{40})$, $(2., -\frac{3}{40})$, $(2., \frac{13}{40})$, $(2., -\frac{3}{40})$\}

\item $c = \frac{3}{2}$, $(d_i,\theta_i)$ = \{$(1., 0)$, $(1., 0)$, $(2., \frac{1}{5})$, $(2., \frac{3}{10})$, $(-3.16228, -\frac{1}{16})$, $(1., \frac{1}{2})$, $(1., \frac{1}{2})$, $(2., -\frac{3}{10})$, $(2., -\frac{1}{5})$, $(-3.16228, \frac{7}{16})$, $(2.82843, \frac{31}{80})$, $(2.82843, -\frac{1}{80})$, $(-2.23607, \frac{1}{4})$, $(-2.23607, -\frac{1}{4})$, $(1.41421, \frac{3}{16})$, $(-2.23607, \frac{1}{4})$, $(-2.23607, -\frac{1}{4})$, $(1.41421, \frac{3}{16})$\}

\item $c = 2$, $(d_i,\theta_i)$ = \{$(1., 0)$, $(1., 0)$, $(2., \frac{1}{5})$, $(2., \frac{3}{10})$, $(-3.16228, -\frac{1}{16})$, $(1., \frac{1}{2})$, $(1., \frac{1}{2})$, $(2., -\frac{3}{10})$, $(2., -\frac{1}{5})$, $(-3.16228, \frac{7}{16})$, $(-3.16228, \frac{5}{16})$, $(-3.16228, -\frac{3}{16})$, $(2., \frac{1}{4})$, $(2., \frac{9}{20})$, $(2., \frac{1}{20})$, $(2., \frac{9}{20})$, $(2., \frac{1}{20})$\}

\item $c = \frac{5}{2}$, $(d_i,\theta_i)$ = \{$(1., 0)$, $(1., 0)$, $(2., \frac{1}{5})$, $(2., \frac{3}{10})$, $(-3.16228, -\frac{1}{16})$, $(1., \frac{1}{2})$, $(1., \frac{1}{2})$, $(2., -\frac{3}{10})$, $(2., -\frac{1}{5})$, $(-3.16228, \frac{7}{16})$, $(2.82843, -\frac{39}{80})$, $(2.82843, \frac{9}{80})$, $(-2.23607, \frac{3}{8})$, $(-2.23607, -\frac{1}{8})$, $(1.41421, \frac{5}{16})$, $(-2.23607, \frac{3}{8})$, $(-2.23607, -\frac{1}{8})$, $(1.41421, \frac{5}{16})$\}

\item $c = 3$, $(d_i,\theta_i)$ = \{$(1., 0)$, $(1., 0)$, $(2., \frac{1}{5})$, $(2., \frac{3}{10})$, $(-3.16228, -\frac{1}{16})$, $(1., \frac{1}{2})$, $(1., \frac{1}{2})$, $(2., -\frac{3}{10})$, $(2., -\frac{1}{5})$, $(-3.16228, \frac{7}{16})$, $(-3.16228, -\frac{1}{16})$, $(-3.16228, \frac{7}{16})$, $(2., \frac{3}{8})$, $(2., -\frac{17}{40})$, $(2., \frac{7}{40})$, $(2., -\frac{17}{40})$, $(2., \frac{7}{40})$\}

\item $c = \frac{7}{2}$, $(d_i,\theta_i)$ = \{$(1., 0)$, $(1., 0)$, $(2., \frac{1}{5})$, $(2., \frac{3}{10})$, $(-3.16228, -\frac{1}{16})$, $(1., \frac{1}{2})$, $(1., \frac{1}{2})$, $(2., -\frac{3}{10})$, $(2., -\frac{1}{5})$, $(-3.16228, \frac{7}{16})$, $(2.82843, \frac{19}{80})$, $(2.82843, -\frac{29}{80})$, $(-2.23607, 0)$, $(-2.23607, \frac{1}{2})$, $(1.41421, \frac{7}{16})$, $(-2.23607, 0)$, $(-2.23607, \frac{1}{2})$, $(1.41421, \frac{7}{16})$\}

\item $c = 4$, $(d_i,\theta_i)$ = \{$(1., 0)$, $(1., 0)$, $(2., \frac{1}{5})$, $(2., \frac{3}{10})$, $(-3.16228, -\frac{1}{16})$, $(1., \frac{1}{2})$, $(1., \frac{1}{2})$, $(2., -\frac{3}{10})$, $(2., -\frac{1}{5})$, $(-3.16228, \frac{7}{16})$, $(2., \frac{1}{2})$, $(-3.16228, -\frac{7}{16})$, $(-3.16228, \frac{1}{16})$, $(2., -\frac{3}{10})$, $(2., \frac{3}{10})$, $(2., -\frac{3}{10})$, $(2., \frac{3}{10})$\}

\item $c = \frac{9}{2}$, $(d_i,\theta_i)$ = \{$(1., 0)$, $(1., 0)$, $(2., \frac{1}{5})$, $(2., \frac{3}{10})$, $(-3.16228, \frac{7}{16})$, $(1., \frac{1}{2})$, $(1., \frac{1}{2})$, $(2., -\frac{3}{10})$, $(2., -\frac{1}{5})$, $(-3.16228, -\frac{1}{16})$, $(2.82843, \frac{29}{80})$, $(2.82843, -\frac{19}{80})$, $(-2.23607, \frac{1}{8})$, $(-2.23607, -\frac{3}{8})$, $(1.41421, -\frac{7}{16})$, $(-2.23607, \frac{1}{8})$, $(-2.23607, -\frac{3}{8})$, $(1.41421, -\frac{7}{16})$\}

\item $c = 5$, $(d_i,\theta_i)$ = \{$(1., 0)$, $(1., 0)$, $(2., \frac{1}{5})$, $(2., \frac{3}{10})$, $(-3.16228, \frac{7}{16})$, $(1., \frac{1}{2})$, $(1., \frac{1}{2})$, $(2., -\frac{3}{10})$, $(2., -\frac{1}{5})$, $(-3.16228, -\frac{1}{16})$, $(2., -\frac{3}{8})$, $(-3.16228, \frac{3}{16})$, $(-3.16228, -\frac{5}{16})$, $(2., -\frac{7}{40})$, $(2., \frac{17}{40})$, $(2., -\frac{7}{40})$, $(2., \frac{17}{40})$\}

\item $c = \frac{11}{2}$, $(d_i,\theta_i)$ = \{$(1., 0)$, $(1., 0)$, $(2., \frac{1}{5})$, $(2., \frac{3}{10})$, $(-3.16228, \frac{7}{16})$, $(1., \frac{1}{2})$, $(1., \frac{1}{2})$, $(2., -\frac{3}{10})$, $(2., -\frac{1}{5})$, $(-3.16228, -\frac{1}{16})$, $(2.82843, -\frac{9}{80})$, $(2.82843, \frac{39}{80})$, $(-2.23607, -\frac{1}{4})$, $(-2.23607, \frac{1}{4})$, $(1.41421, -\frac{5}{16})$, $(-2.23607, -\frac{1}{4})$, $(-2.23607, \frac{1}{4})$, $(1.41421, -\frac{5}{16})$\}

\item $c = 6$, $(d_i,\theta_i)$ = \{$(1., 0)$, $(1., \frac{1}{2})$, $(2., \frac{1}{5})$, $(2., \frac{3}{10})$, $(-3.16228, \frac{7}{16})$, $(1., \frac{1}{2})$, $(1., 0)$, $(2., -\frac{3}{10})$, $(2., -\frac{1}{5})$, $(-3.16228, -\frac{1}{16})$, $(3.16228, -\frac{3}{16})$, $(-2., -\frac{1}{4})$, $(3.16228, \frac{5}{16})$, $(-2., -\frac{1}{20})$, $(-2., -\frac{9}{20})$, $(-2., -\frac{1}{20})$, $(-2., -\frac{9}{20})$\}

\item $c = \frac{13}{2}$, $(d_i,\theta_i)$ = \{$(1., 0)$, $(1., \frac{1}{2})$, $(2., \frac{1}{5})$, $(2., \frac{3}{10})$, $(-3.16228, \frac{7}{16})$, $(1., \frac{1}{2})$, $(1., 0)$, $(2., -\frac{3}{10})$, $(2., -\frac{1}{5})$, $(-3.16228, -\frac{1}{16})$, $(-2.82843, \frac{1}{80})$, $(-2.82843, -\frac{31}{80})$, $(2.23607, -\frac{1}{8})$, $(2.23607, \frac{3}{8})$, $(-1.41421, -\frac{3}{16})$, $(2.23607, -\frac{1}{8})$, $(2.23607, \frac{3}{8})$, $(-1.41421, -\frac{3}{16})$\}

\item $c = 7$, $(d_i,\theta_i)$ = \{$(1., 0)$, $(1., \frac{1}{2})$, $(2., \frac{1}{5})$, $(2., \frac{3}{10})$, $(-3.16228, \frac{7}{16})$, $(1., \frac{1}{2})$, $(1., 0)$, $(2., -\frac{3}{10})$, $(2., -\frac{1}{5})$, $(-3.16228, -\frac{1}{16})$, $(-2., -\frac{1}{8})$, $(3.16228, -\frac{1}{16})$, $(3.16228, \frac{7}{16})$, $(-2., \frac{3}{40})$, $(-2., -\frac{13}{40})$, $(-2., \frac{3}{40})$, $(-2., -\frac{13}{40})$\}

\item $c = \frac{15}{2}$, $(d_i,\theta_i)$ = \{$(1., 0)$, $(1., \frac{1}{2})$, $(2., \frac{1}{5})$, $(2., \frac{3}{10})$, $(-3.16228, \frac{7}{16})$, $(1., \frac{1}{2})$, $(1., 0)$, $(2., -\frac{3}{10})$, $(2., -\frac{1}{5})$, $(-3.16228, -\frac{1}{16})$, $(-2.82843, -\frac{21}{80})$, $(-2.82843, \frac{11}{80})$, $(2.23607, \frac{1}{2})$, $(2.23607, 0)$, $(-1.41421, -\frac{1}{16})$, $(2.23607, \frac{1}{2})$, $(2.23607, 0)$, $(-1.41421, -\frac{1}{16})$\}

\end{enumerate}

\paragraph*{Rank 10; \#51}

\begin{enumerate}

\item $c = \frac{1}{5}$, $(d_i,\theta_i)$ = \{$(1., 0)$, $(6.31375, 0)$, $(5.69572, \frac{1}{10})$, $(2.90211, \frac{1}{10})$, $(4.52015, -\frac{1}{5})$, $(1., \frac{1}{2})$, $(6.31375, \frac{1}{2})$, $(5.69572, -\frac{2}{5})$, $(2.90211, -\frac{2}{5})$, $(4.52015, \frac{3}{10})$, $(8.59783, \frac{19}{40})$, $(7.31375, \frac{1}{8})$, $(2.7936, -\frac{11}{40})$, $(5.31375, -\frac{1}{8})$, $(4.52015, -\frac{3}{40})$, $(4.52015, -\frac{3}{40})$\}

\item $c = \frac{7}{10}$, $(d_i,\theta_i)$ = \{$(1., 0)$, $(6.31375, 0)$, $(5.69572, \frac{1}{10})$, $(2.90211, \frac{1}{10})$, $(4.52015, -\frac{1}{5})$, $(1., \frac{1}{2})$, $(6.31375, \frac{1}{2})$, $(5.69572, -\frac{2}{5})$, $(2.90211, -\frac{2}{5})$, $(4.52015, \frac{3}{10})$, $(6.39245, -\frac{1}{80})$, $(3.75739, -\frac{1}{16})$, $(5.1716, \frac{3}{16})$, $(6.07958, -\frac{37}{80})$, $(1.97538, -\frac{17}{80})$, $(3.75739, -\frac{1}{16})$, $(5.1716, \frac{3}{16})$, $(6.07958, -\frac{37}{80})$, $(1.97538, -\frac{17}{80})$\}

\item $c = \frac{6}{5}$, $(d_i,\theta_i)$ = \{$(1., 0)$, $(6.31375, 0)$, $(5.69572, \frac{1}{10})$, $(2.90211, \frac{1}{10})$, $(4.52015, \frac{3}{10})$, $(1., \frac{1}{2})$, $(6.31375, \frac{1}{2})$, $(5.69572, -\frac{2}{5})$, $(2.90211, -\frac{2}{5})$, $(4.52015, -\frac{1}{5})$, $(2.7936, -\frac{3}{20})$, $(7.31375, \frac{1}{4})$, $(5.31375, 0)$, $(8.59783, -\frac{2}{5})$, $(4.52015, \frac{1}{20})$, $(4.52015, \frac{1}{20})$\}

\item $c = \frac{17}{10}$, $(d_i,\theta_i)$ = \{$(1., 0)$, $(6.31375, 0)$, $(5.69572, \frac{1}{10})$, $(2.90211, \frac{1}{10})$, $(4.52015, \frac{3}{10})$, $(1., \frac{1}{2})$, $(6.31375, \frac{1}{2})$, $(5.69572, -\frac{2}{5})$, $(2.90211, -\frac{2}{5})$, $(4.52015, -\frac{1}{5})$, $(6.39245, \frac{9}{80})$, $(3.75739, \frac{1}{16})$, $(5.1716, \frac{5}{16})$, $(6.07958, -\frac{27}{80})$, $(1.97538, -\frac{7}{80})$, $(3.75739, \frac{1}{16})$, $(5.1716, \frac{5}{16})$, $(6.07958, -\frac{27}{80})$, $(1.97538, -\frac{7}{80})$\}

\item $c = \frac{11}{5}$, $(d_i,\theta_i)$ = \{$(1., 0)$, $(6.31375, 0)$, $(5.69572, \frac{1}{10})$, $(2.90211, \frac{1}{10})$, $(4.52015, \frac{3}{10})$, $(1., \frac{1}{2})$, $(6.31375, \frac{1}{2})$, $(5.69572, -\frac{2}{5})$, $(2.90211, -\frac{2}{5})$, $(4.52015, -\frac{1}{5})$, $(5.31375, \frac{1}{8})$, $(8.59783, -\frac{11}{40})$, $(2.7936, -\frac{1}{40})$, $(7.31375, \frac{3}{8})$, $(4.52015, \frac{7}{40})$, $(4.52015, \frac{7}{40})$\}

\item $c = \frac{27}{10}$, $(d_i,\theta_i)$ = \{$(1., 0)$, $(6.31375, 0)$, $(5.69572, \frac{1}{10})$, $(2.90211, \frac{1}{10})$, $(4.52015, \frac{3}{10})$, $(1., \frac{1}{2})$, $(6.31375, \frac{1}{2})$, $(5.69572, -\frac{2}{5})$, $(2.90211, -\frac{2}{5})$, $(4.52015, -\frac{1}{5})$, $(6.39245, \frac{19}{80})$, $(5.1716, \frac{7}{16})$, $(3.75739, \frac{3}{16})$, $(1.97538, \frac{3}{80})$, $(6.07958, -\frac{17}{80})$, $(5.1716, \frac{7}{16})$, $(3.75739, \frac{3}{16})$, $(1.97538, \frac{3}{80})$, $(6.07958, -\frac{17}{80})$\}

\item $c = \frac{16}{5}$, $(d_i,\theta_i)$ = \{$(1., 0)$, $(6.31375, 0)$, $(5.69572, \frac{1}{10})$, $(2.90211, \frac{1}{10})$, $(4.52015, \frac{3}{10})$, $(1., \frac{1}{2})$, $(6.31375, \frac{1}{2})$, $(5.69572, -\frac{2}{5})$, $(2.90211, -\frac{2}{5})$, $(4.52015, -\frac{1}{5})$, $(7.31375, \frac{1}{2})$, $(5.31375, \frac{1}{4})$, $(2.7936, \frac{1}{10})$, $(8.59783, -\frac{3}{20})$, $(4.52015, \frac{3}{10})$, $(4.52015, \frac{3}{10})$\}

\item $c = \frac{37}{10}$, $(d_i,\theta_i)$ = \{$(1., 0)$, $(6.31375, 0)$, $(5.69572, \frac{1}{10})$, $(2.90211, \frac{1}{10})$, $(4.52015, \frac{3}{10})$, $(1., \frac{1}{2})$, $(6.31375, \frac{1}{2})$, $(5.69572, -\frac{2}{5})$, $(2.90211, -\frac{2}{5})$, $(4.52015, -\frac{1}{5})$, $(6.39245, \frac{29}{80})$, $(3.75739, \frac{5}{16})$, $(5.1716, -\frac{7}{16})$, $(6.07958, -\frac{7}{80})$, $(1.97538, \frac{13}{80})$, $(3.75739, \frac{5}{16})$, $(5.1716, -\frac{7}{16})$, $(6.07958, -\frac{7}{80})$, $(1.97538, \frac{13}{80})$\}

\item $c = \frac{21}{5}$, $(d_i,\theta_i)$ = \{$(1., 0)$, $(6.31375, 0)$, $(5.69572, \frac{1}{10})$, $(2.90211, \frac{1}{10})$, $(4.52015, \frac{3}{10})$, $(1., \frac{1}{2})$, $(6.31375, \frac{1}{2})$, $(5.69572, -\frac{2}{5})$, $(2.90211, -\frac{2}{5})$, $(4.52015, -\frac{1}{5})$, $(5.31375, \frac{3}{8})$, $(7.31375, -\frac{3}{8})$, $(2.7936, \frac{9}{40})$, $(8.59783, -\frac{1}{40})$, $(4.52015, \frac{17}{40})$, $(4.52015, \frac{17}{40})$\}

\item $c = \frac{47}{10}$, $(d_i,\theta_i)$ = \{$(1., 0)$, $(6.31375, 0)$, $(5.69572, \frac{1}{10})$, $(2.90211, \frac{1}{10})$, $(4.52015, \frac{3}{10})$, $(1., \frac{1}{2})$, $(6.31375, \frac{1}{2})$, $(5.69572, -\frac{2}{5})$, $(2.90211, -\frac{2}{5})$, $(4.52015, -\frac{1}{5})$, $(6.39245, \frac{39}{80})$, $(3.75739, \frac{7}{16})$, $(5.1716, -\frac{5}{16})$, $(6.07958, \frac{3}{80})$, $(1.97538, \frac{23}{80})$, $(3.75739, \frac{7}{16})$, $(5.1716, -\frac{5}{16})$, $(6.07958, \frac{3}{80})$, $(1.97538, \frac{23}{80})$\}

\item $c = \frac{26}{5}$, $(d_i,\theta_i)$ = \{$(1., 0)$, $(6.31375, 0)$, $(5.69572, \frac{1}{10})$, $(2.90211, \frac{1}{10})$, $(4.52015, \frac{3}{10})$, $(1., \frac{1}{2})$, $(6.31375, \frac{1}{2})$, $(5.69572, -\frac{2}{5})$, $(2.90211, -\frac{2}{5})$, $(4.52015, -\frac{1}{5})$, $(5.31375, \frac{1}{2})$, $(7.31375, -\frac{1}{4})$, $(2.7936, \frac{7}{20})$, $(8.59783, \frac{1}{10})$, $(4.52015, -\frac{9}{20})$, $(4.52015, -\frac{9}{20})$\}

\item $c = \frac{57}{10}$, $(d_i,\theta_i)$ = \{$(1., 0)$, $(6.31375, 0)$, $(5.69572, \frac{1}{10})$, $(2.90211, \frac{1}{10})$, $(4.52015, \frac{3}{10})$, $(1., \frac{1}{2})$, $(6.31375, \frac{1}{2})$, $(5.69572, -\frac{2}{5})$, $(2.90211, -\frac{2}{5})$, $(4.52015, -\frac{1}{5})$, $(6.39245, -\frac{31}{80})$, $(3.75739, -\frac{7}{16})$, $(5.1716, -\frac{3}{16})$, $(6.07958, \frac{13}{80})$, $(1.97538, \frac{33}{80})$, $(3.75739, -\frac{7}{16})$, $(5.1716, -\frac{3}{16})$, $(6.07958, \frac{13}{80})$, $(1.97538, \frac{33}{80})$\}

\item $c = \frac{31}{5}$, $(d_i,\theta_i)$ = \{$(1., 0)$, $(6.31375, \frac{1}{2})$, $(5.69572, \frac{1}{10})$, $(2.90211, \frac{1}{10})$, $(4.52015, \frac{3}{10})$, $(1., \frac{1}{2})$, $(6.31375, 0)$, $(5.69572, -\frac{2}{5})$, $(2.90211, -\frac{2}{5})$, $(4.52015, -\frac{1}{5})$, $(7.31375, -\frac{1}{8})$, $(8.59783, \frac{9}{40})$, $(2.7936, \frac{19}{40})$, $(5.31375, -\frac{3}{8})$, $(4.52015, -\frac{13}{40})$, $(4.52015, -\frac{13}{40})$\}

\item $c = \frac{67}{10}$, $(d_i,\theta_i)$ = \{$(1., 0)$, $(6.31375, \frac{1}{2})$, $(5.69572, \frac{1}{10})$, $(2.90211, \frac{1}{10})$, $(4.52015, \frac{3}{10})$, $(1., \frac{1}{2})$, $(6.31375, 0)$, $(5.69572, -\frac{2}{5})$, $(2.90211, -\frac{2}{5})$, $(4.52015, -\frac{1}{5})$, $(6.39245, -\frac{21}{80})$, $(5.1716, -\frac{1}{16})$, $(3.75739, -\frac{5}{16})$, $(1.97538, -\frac{37}{80})$, $(6.07958, \frac{23}{80})$, $(5.1716, -\frac{1}{16})$, $(3.75739, -\frac{5}{16})$, $(1.97538, -\frac{37}{80})$, $(6.07958, \frac{23}{80})$\}

\item $c = \frac{36}{5}$, $(d_i,\theta_i)$ = \{$(1., 0)$, $(6.31375, \frac{1}{2})$, $(5.69572, \frac{1}{10})$, $(2.90211, \frac{1}{10})$, $(4.52015, \frac{3}{10})$, $(1., \frac{1}{2})$, $(6.31375, 0)$, $(5.69572, -\frac{2}{5})$, $(2.90211, -\frac{2}{5})$, $(4.52015, -\frac{1}{5})$, $(2.7936, -\frac{2}{5})$, $(8.59783, \frac{7}{20})$, $(5.31375, -\frac{1}{4})$, $(7.31375, 0)$, $(4.52015, -\frac{1}{5})$, $(4.52015, -\frac{1}{5})$\}

\item $c = \frac{77}{10}$, $(d_i,\theta_i)$ = \{$(1., 0)$, $(6.31375, \frac{1}{2})$, $(5.69572, \frac{1}{10})$, $(2.90211, \frac{1}{10})$, $(4.52015, \frac{3}{10})$, $(1., \frac{1}{2})$, $(6.31375, 0)$, $(5.69572, -\frac{2}{5})$, $(2.90211, -\frac{2}{5})$, $(4.52015, -\frac{1}{5})$, $(6.39245, -\frac{11}{80})$, $(3.75739, -\frac{3}{16})$, $(5.1716, \frac{1}{16})$, $(6.07958, \frac{33}{80})$, $(1.97538, -\frac{27}{80})$, $(3.75739, -\frac{3}{16})$, $(5.1716, \frac{1}{16})$, $(6.07958, \frac{33}{80})$, $(1.97538, -\frac{27}{80})$\}

\end{enumerate}

\paragraph*{Rank 10; \#52}

\begin{enumerate}

\item $c = \frac{3}{10}$, $(d_i,\theta_i)$ = \{$(1., 0)$, $(6.31375, 0)$, $(4.52015, \frac{1}{5})$, $(5.69572, -\frac{1}{10})$, $(2.90211, -\frac{1}{10})$, $(1., \frac{1}{2})$, $(6.31375, \frac{1}{2})$, $(4.52015, -\frac{3}{10})$, $(5.69572, \frac{2}{5})$, $(2.90211, \frac{2}{5})$, $(6.39245, \frac{11}{80})$, $(1.97538, \frac{27}{80})$, $(6.07958, -\frac{33}{80})$, $(5.1716, -\frac{1}{16})$, $(3.75739, \frac{3}{16})$, $(1.97538, \frac{27}{80})$, $(6.07958, -\frac{33}{80})$, $(5.1716, -\frac{1}{16})$, $(3.75739, \frac{3}{16})$\}

\item $c = \frac{4}{5}$, $(d_i,\theta_i)$ = \{$(1., 0)$, $(6.31375, 0)$, $(4.52015, \frac{1}{5})$, $(5.69572, -\frac{1}{10})$, $(2.90211, -\frac{1}{10})$, $(1., \frac{1}{2})$, $(6.31375, \frac{1}{2})$, $(4.52015, -\frac{3}{10})$, $(5.69572, \frac{2}{5})$, $(2.90211, \frac{2}{5})$, $(5.31375, \frac{1}{4})$, $(7.31375, 0)$, $(8.59783, -\frac{7}{20})$, $(2.7936, \frac{2}{5})$, $(4.52015, \frac{1}{5})$, $(4.52015, \frac{1}{5})$\}

\item $c = \frac{13}{10}$, $(d_i,\theta_i)$ = \{$(1., 0)$, $(6.31375, 0)$, $(4.52015, \frac{1}{5})$, $(5.69572, -\frac{1}{10})$, $(2.90211, -\frac{1}{10})$, $(1., \frac{1}{2})$, $(6.31375, \frac{1}{2})$, $(4.52015, -\frac{3}{10})$, $(5.69572, \frac{2}{5})$, $(2.90211, \frac{2}{5})$, $(6.39245, \frac{21}{80})$, $(6.07958, -\frac{23}{80})$, $(1.97538, \frac{37}{80})$, $(3.75739, \frac{5}{16})$, $(5.1716, \frac{1}{16})$, $(6.07958, -\frac{23}{80})$, $(1.97538, \frac{37}{80})$, $(3.75739, \frac{5}{16})$, $(5.1716, \frac{1}{16})$\}

\item $c = \frac{9}{5}$, $(d_i,\theta_i)$ = \{$(1., 0)$, $(6.31375, 0)$, $(4.52015, \frac{1}{5})$, $(5.69572, -\frac{1}{10})$, $(2.90211, -\frac{1}{10})$, $(1., \frac{1}{2})$, $(6.31375, \frac{1}{2})$, $(4.52015, -\frac{3}{10})$, $(5.69572, \frac{2}{5})$, $(2.90211, \frac{2}{5})$, $(7.31375, \frac{1}{8})$, $(2.7936, -\frac{19}{40})$, $(8.59783, -\frac{9}{40})$, $(5.31375, \frac{3}{8})$, $(4.52015, \frac{13}{40})$, $(4.52015, \frac{13}{40})$\}

\item $c = \frac{23}{10}$, $(d_i,\theta_i)$ = \{$(1., 0)$, $(6.31375, 0)$, $(4.52015, \frac{1}{5})$, $(5.69572, -\frac{1}{10})$, $(2.90211, -\frac{1}{10})$, $(1., \frac{1}{2})$, $(6.31375, \frac{1}{2})$, $(4.52015, -\frac{3}{10})$, $(5.69572, \frac{2}{5})$, $(2.90211, \frac{2}{5})$, $(6.39245, \frac{31}{80})$, $(1.97538, -\frac{33}{80})$, $(6.07958, -\frac{13}{80})$, $(5.1716, \frac{3}{16})$, $(3.75739, \frac{7}{16})$, $(1.97538, -\frac{33}{80})$, $(6.07958, -\frac{13}{80})$, $(5.1716, \frac{3}{16})$, $(3.75739, \frac{7}{16})$\}

\item $c = \frac{14}{5}$, $(d_i,\theta_i)$ = \{$(1., 0)$, $(6.31375, 0)$, $(4.52015, \frac{1}{5})$, $(5.69572, -\frac{1}{10})$, $(2.90211, -\frac{1}{10})$, $(1., \frac{1}{2})$, $(6.31375, \frac{1}{2})$, $(4.52015, -\frac{3}{10})$, $(5.69572, \frac{2}{5})$, $(2.90211, \frac{2}{5})$, $(8.59783, -\frac{1}{10})$, $(2.7936, -\frac{7}{20})$, $(7.31375, \frac{1}{4})$, $(5.31375, \frac{1}{2})$, $(4.52015, \frac{9}{20})$, $(4.52015, \frac{9}{20})$\}

\item $c = \frac{33}{10}$, $(d_i,\theta_i)$ = \{$(1., 0)$, $(6.31375, 0)$, $(4.52015, \frac{1}{5})$, $(5.69572, -\frac{1}{10})$, $(2.90211, -\frac{1}{10})$, $(1., \frac{1}{2})$, $(6.31375, \frac{1}{2})$, $(4.52015, -\frac{3}{10})$, $(5.69572, \frac{2}{5})$, $(2.90211, \frac{2}{5})$, $(6.39245, -\frac{39}{80})$, $(1.97538, -\frac{23}{80})$, $(6.07958, -\frac{3}{80})$, $(5.1716, \frac{5}{16})$, $(3.75739, -\frac{7}{16})$, $(1.97538, -\frac{23}{80})$, $(6.07958, -\frac{3}{80})$, $(5.1716, \frac{5}{16})$, $(3.75739, -\frac{7}{16})$\}

\item $c = \frac{19}{5}$, $(d_i,\theta_i)$ = \{$(1., 0)$, $(6.31375, 0)$, $(4.52015, \frac{1}{5})$, $(5.69572, \frac{2}{5})$, $(2.90211, \frac{2}{5})$, $(1., \frac{1}{2})$, $(6.31375, \frac{1}{2})$, $(4.52015, -\frac{3}{10})$, $(5.69572, -\frac{1}{10})$, $(2.90211, -\frac{1}{10})$, $(7.31375, \frac{3}{8})$, $(5.31375, -\frac{3}{8})$, $(8.59783, \frac{1}{40})$, $(2.7936, -\frac{9}{40})$, $(4.52015, -\frac{17}{40})$, $(4.52015, -\frac{17}{40})$\}

\item $c = \frac{43}{10}$, $(d_i,\theta_i)$ = \{$(1., 0)$, $(6.31375, 0)$, $(4.52015, \frac{1}{5})$, $(5.69572, \frac{2}{5})$, $(2.90211, \frac{2}{5})$, $(1., \frac{1}{2})$, $(6.31375, \frac{1}{2})$, $(4.52015, -\frac{3}{10})$, $(5.69572, -\frac{1}{10})$, $(2.90211, -\frac{1}{10})$, $(6.39245, -\frac{29}{80})$, $(1.97538, -\frac{13}{80})$, $(6.07958, \frac{7}{80})$, $(5.1716, \frac{7}{16})$, $(3.75739, -\frac{5}{16})$, $(1.97538, -\frac{13}{80})$, $(6.07958, \frac{7}{80})$, $(5.1716, \frac{7}{16})$, $(3.75739, -\frac{5}{16})$\}

\item $c = \frac{24}{5}$, $(d_i,\theta_i)$ = \{$(1., 0)$, $(6.31375, 0)$, $(4.52015, \frac{1}{5})$, $(5.69572, \frac{2}{5})$, $(2.90211, \frac{2}{5})$, $(1., \frac{1}{2})$, $(6.31375, \frac{1}{2})$, $(4.52015, -\frac{3}{10})$, $(5.69572, -\frac{1}{10})$, $(2.90211, -\frac{1}{10})$, $(8.59783, \frac{3}{20})$, $(2.7936, -\frac{1}{10})$, $(7.31375, \frac{1}{2})$, $(5.31375, -\frac{1}{4})$, $(4.52015, -\frac{3}{10})$, $(4.52015, -\frac{3}{10})$\}

\item $c = \frac{53}{10}$, $(d_i,\theta_i)$ = \{$(1., 0)$, $(6.31375, 0)$, $(4.52015, \frac{1}{5})$, $(5.69572, \frac{2}{5})$, $(2.90211, \frac{2}{5})$, $(1., \frac{1}{2})$, $(6.31375, \frac{1}{2})$, $(4.52015, -\frac{3}{10})$, $(5.69572, -\frac{1}{10})$, $(2.90211, -\frac{1}{10})$, $(6.39245, -\frac{19}{80})$, $(6.07958, \frac{17}{80})$, $(1.97538, -\frac{3}{80})$, $(3.75739, -\frac{3}{16})$, $(5.1716, -\frac{7}{16})$, $(6.07958, \frac{17}{80})$, $(1.97538, -\frac{3}{80})$, $(3.75739, -\frac{3}{16})$, $(5.1716, -\frac{7}{16})$\}

\item $c = \frac{29}{5}$, $(d_i,\theta_i)$ = \{$(1., 0)$, $(6.31375, 0)$, $(4.52015, \frac{1}{5})$, $(5.69572, \frac{2}{5})$, $(2.90211, \frac{2}{5})$, $(1., \frac{1}{2})$, $(6.31375, \frac{1}{2})$, $(4.52015, -\frac{3}{10})$, $(5.69572, -\frac{1}{10})$, $(2.90211, -\frac{1}{10})$, $(5.31375, -\frac{1}{8})$, $(2.7936, \frac{1}{40})$, $(8.59783, \frac{11}{40})$, $(7.31375, -\frac{3}{8})$, $(4.52015, -\frac{7}{40})$, $(4.52015, -\frac{7}{40})$\}

\item $c = \frac{63}{10}$, $(d_i,\theta_i)$ = \{$(1., 0)$, $(6.31375, \frac{1}{2})$, $(4.52015, \frac{1}{5})$, $(5.69572, \frac{2}{5})$, $(2.90211, \frac{2}{5})$, $(1., \frac{1}{2})$, $(6.31375, 0)$, $(4.52015, -\frac{3}{10})$, $(5.69572, -\frac{1}{10})$, $(2.90211, -\frac{1}{10})$, $(6.39245, -\frac{9}{80})$, $(1.97538, \frac{7}{80})$, $(6.07958, \frac{27}{80})$, $(5.1716, -\frac{5}{16})$, $(3.75739, -\frac{1}{16})$, $(1.97538, \frac{7}{80})$, $(6.07958, \frac{27}{80})$, $(5.1716, -\frac{5}{16})$, $(3.75739, -\frac{1}{16})$\}

\item $c = \frac{34}{5}$, $(d_i,\theta_i)$ = \{$(1., 0)$, $(6.31375, \frac{1}{2})$, $(4.52015, \frac{1}{5})$, $(5.69572, \frac{2}{5})$, $(2.90211, \frac{2}{5})$, $(1., \frac{1}{2})$, $(6.31375, 0)$, $(4.52015, -\frac{3}{10})$, $(5.69572, -\frac{1}{10})$, $(2.90211, -\frac{1}{10})$, $(2.7936, \frac{3}{20})$, $(7.31375, -\frac{1}{4})$, $(5.31375, 0)$, $(8.59783, \frac{2}{5})$, $(4.52015, -\frac{1}{20})$, $(4.52015, -\frac{1}{20})$\}

\item $c = \frac{73}{10}$, $(d_i,\theta_i)$ = \{$(1., 0)$, $(6.31375, \frac{1}{2})$, $(4.52015, \frac{1}{5})$, $(5.69572, \frac{2}{5})$, $(2.90211, \frac{2}{5})$, $(1., \frac{1}{2})$, $(6.31375, 0)$, $(4.52015, -\frac{3}{10})$, $(5.69572, -\frac{1}{10})$, $(2.90211, -\frac{1}{10})$, $(6.39245, \frac{1}{80})$, $(1.97538, \frac{17}{80})$, $(6.07958, \frac{37}{80})$, $(5.1716, -\frac{3}{16})$, $(3.75739, \frac{1}{16})$, $(1.97538, \frac{17}{80})$, $(6.07958, \frac{37}{80})$, $(5.1716, -\frac{3}{16})$, $(3.75739, \frac{1}{16})$\}

\item $c = \frac{39}{5}$, $(d_i,\theta_i)$ = \{$(1., 0)$, $(6.31375, \frac{1}{2})$, $(4.52015, \frac{1}{5})$, $(5.69572, \frac{2}{5})$, $(2.90211, \frac{2}{5})$, $(1., \frac{1}{2})$, $(6.31375, 0)$, $(4.52015, -\frac{3}{10})$, $(5.69572, -\frac{1}{10})$, $(2.90211, -\frac{1}{10})$, $(2.7936, \frac{11}{40})$, $(7.31375, -\frac{1}{8})$, $(5.31375, \frac{1}{8})$, $(8.59783, -\frac{19}{40})$, $(4.52015, \frac{3}{40})$, $(4.52015, \frac{3}{40})$\}

\end{enumerate}

\paragraph*{Rank 10; \#53}

\begin{enumerate}

\item $c = \frac{1}{5}$, $(d_i,\theta_i)$ = \{$(1., 0)$, $(0.158384, 0)$, $(-0.45965, \frac{1}{10})$, $(-0.902113, \frac{1}{10})$, $(0.715921, -\frac{1}{5})$, $(1., \frac{1}{2})$, $(0.158384, \frac{1}{2})$, $(-0.45965, -\frac{2}{5})$, $(-0.902113, -\frac{2}{5})$, $(0.715921, \frac{3}{10})$, $(0.442463, -\frac{11}{40})$, $(-0.841616, \frac{3}{8})$, $(1.15838, \frac{1}{8})$, $(-1.36176, -\frac{1}{40})$, $(0.715921, -\frac{3}{40})$, $(0.715921, -\frac{3}{40})$\}

\item $c = \frac{7}{10}$, $(d_i,\theta_i)$ = \{$(1., 0)$, $(0.158384, 0)$, $(-0.45965, \frac{1}{10})$, $(-0.902113, \frac{1}{10})$, $(0.715921, -\frac{1}{5})$, $(1., \frac{1}{2})$, $(0.158384, \frac{1}{2})$, $(-0.45965, -\frac{2}{5})$, $(-0.902113, -\frac{2}{5})$, $(0.715921, \frac{3}{10})$, $(1.01247, -\frac{1}{80})$, $(0.819101, \frac{3}{16})$, $(-0.595112, \frac{7}{16})$, $(0.312869, -\frac{17}{80})$, $(-0.962912, \frac{3}{80})$, $(0.819101, \frac{3}{16})$, $(-0.595112, \frac{7}{16})$, $(0.312869, -\frac{17}{80})$, $(-0.962912, \frac{3}{80})$\}

\item $c = \frac{6}{5}$, $(d_i,\theta_i)$ = \{$(1., 0)$, $(0.158384, 0)$, $(-0.45965, \frac{1}{10})$, $(-0.902113, \frac{1}{10})$, $(0.715921, \frac{3}{10})$, $(1., \frac{1}{2})$, $(0.158384, \frac{1}{2})$, $(-0.45965, -\frac{2}{5})$, $(-0.902113, -\frac{2}{5})$, $(0.715921, -\frac{1}{5})$, $(0.442463, -\frac{3}{20})$, $(1.15838, \frac{1}{4})$, $(-1.36176, \frac{1}{10})$, $(-0.841616, \frac{1}{2})$, $(0.715921, \frac{1}{20})$, $(0.715921, \frac{1}{20})$\}

\item $c = \frac{17}{10}$, $(d_i,\theta_i)$ = \{$(1., 0)$, $(0.158384, 0)$, $(-0.45965, \frac{1}{10})$, $(-0.902113, \frac{1}{10})$, $(0.715921, \frac{3}{10})$, $(1., \frac{1}{2})$, $(0.158384, \frac{1}{2})$, $(-0.45965, -\frac{2}{5})$, $(-0.902113, -\frac{2}{5})$, $(0.715921, -\frac{1}{5})$, $(1.01247, \frac{9}{80})$, $(0.819101, \frac{5}{16})$, $(-0.595112, -\frac{7}{16})$, $(0.312869, -\frac{7}{80})$, $(-0.962912, \frac{13}{80})$, $(0.819101, \frac{5}{16})$, $(-0.595112, -\frac{7}{16})$, $(0.312869, -\frac{7}{80})$, $(-0.962912, \frac{13}{80})$\}

\item $c = \frac{11}{5}$, $(d_i,\theta_i)$ = \{$(1., 0)$, $(0.158384, 0)$, $(-0.45965, \frac{1}{10})$, $(-0.902113, \frac{1}{10})$, $(0.715921, \frac{3}{10})$, $(1., \frac{1}{2})$, $(0.158384, \frac{1}{2})$, $(-0.45965, -\frac{2}{5})$, $(-0.902113, -\frac{2}{5})$, $(0.715921, -\frac{1}{5})$, $(-0.841616, -\frac{3}{8})$, $(1.15838, \frac{3}{8})$, $(0.442463, -\frac{1}{40})$, $(-1.36176, \frac{9}{40})$, $(0.715921, \frac{7}{40})$, $(0.715921, \frac{7}{40})$\}

\item $c = \frac{27}{10}$, $(d_i,\theta_i)$ = \{$(1., 0)$, $(0.158384, 0)$, $(-0.45965, \frac{1}{10})$, $(-0.902113, \frac{1}{10})$, $(0.715921, \frac{3}{10})$, $(1., \frac{1}{2})$, $(0.158384, \frac{1}{2})$, $(-0.45965, -\frac{2}{5})$, $(-0.902113, -\frac{2}{5})$, $(0.715921, -\frac{1}{5})$, $(1.01247, \frac{19}{80})$, $(0.819101, \frac{7}{16})$, $(-0.595112, -\frac{5}{16})$, $(0.312869, \frac{3}{80})$, $(-0.962912, \frac{23}{80})$, $(0.819101, \frac{7}{16})$, $(-0.595112, -\frac{5}{16})$, $(0.312869, \frac{3}{80})$, $(-0.962912, \frac{23}{80})$\}

\item $c = \frac{16}{5}$, $(d_i,\theta_i)$ = \{$(1., 0)$, $(0.158384, 0)$, $(-0.45965, \frac{1}{10})$, $(-0.902113, \frac{1}{10})$, $(0.715921, \frac{3}{10})$, $(1., \frac{1}{2})$, $(0.158384, \frac{1}{2})$, $(-0.45965, -\frac{2}{5})$, $(-0.902113, -\frac{2}{5})$, $(0.715921, -\frac{1}{5})$, $(0.442463, \frac{1}{10})$, $(-1.36176, \frac{7}{20})$, $(1.15838, \frac{1}{2})$, $(-0.841616, -\frac{1}{4})$, $(0.715921, \frac{3}{10})$, $(0.715921, \frac{3}{10})$\}

\item $c = \frac{37}{10}$, $(d_i,\theta_i)$ = \{$(1., 0)$, $(0.158384, 0)$, $(-0.45965, \frac{1}{10})$, $(-0.902113, \frac{1}{10})$, $(0.715921, \frac{3}{10})$, $(1., \frac{1}{2})$, $(0.158384, \frac{1}{2})$, $(-0.45965, -\frac{2}{5})$, $(-0.902113, -\frac{2}{5})$, $(0.715921, -\frac{1}{5})$, $(1.01247, \frac{29}{80})$, $(-0.595112, -\frac{3}{16})$, $(0.819101, -\frac{7}{16})$, $(-0.962912, \frac{33}{80})$, $(0.312869, \frac{13}{80})$, $(-0.595112, -\frac{3}{16})$, $(0.819101, -\frac{7}{16})$, $(-0.962912, \frac{33}{80})$, $(0.312869, \frac{13}{80})$\}

\item $c = \frac{21}{5}$, $(d_i,\theta_i)$ = \{$(1., 0)$, $(0.158384, 0)$, $(-0.45965, \frac{1}{10})$, $(-0.902113, \frac{1}{10})$, $(0.715921, \frac{3}{10})$, $(1., \frac{1}{2})$, $(0.158384, \frac{1}{2})$, $(-0.45965, -\frac{2}{5})$, $(-0.902113, -\frac{2}{5})$, $(0.715921, -\frac{1}{5})$, $(-0.841616, -\frac{1}{8})$, $(-1.36176, \frac{19}{40})$, $(1.15838, -\frac{3}{8})$, $(0.442463, \frac{9}{40})$, $(0.715921, \frac{17}{40})$, $(0.715921, \frac{17}{40})$\}

\item $c = \frac{47}{10}$, $(d_i,\theta_i)$ = \{$(1., 0)$, $(0.158384, 0)$, $(-0.45965, \frac{1}{10})$, $(-0.902113, \frac{1}{10})$, $(0.715921, \frac{3}{10})$, $(1., \frac{1}{2})$, $(0.158384, \frac{1}{2})$, $(-0.45965, -\frac{2}{5})$, $(-0.902113, -\frac{2}{5})$, $(0.715921, -\frac{1}{5})$, $(1.01247, \frac{39}{80})$, $(0.819101, -\frac{5}{16})$, $(-0.595112, -\frac{1}{16})$, $(0.312869, \frac{23}{80})$, $(-0.962912, -\frac{37}{80})$, $(0.819101, -\frac{5}{16})$, $(-0.595112, -\frac{1}{16})$, $(0.312869, \frac{23}{80})$, $(-0.962912, -\frac{37}{80})$\}

\item $c = \frac{26}{5}$, $(d_i,\theta_i)$ = \{$(1., 0)$, $(0.158384, 0)$, $(-0.45965, \frac{1}{10})$, $(-0.902113, \frac{1}{10})$, $(0.715921, \frac{3}{10})$, $(1., \frac{1}{2})$, $(0.158384, \frac{1}{2})$, $(-0.45965, -\frac{2}{5})$, $(-0.902113, -\frac{2}{5})$, $(0.715921, -\frac{1}{5})$, $(0.442463, \frac{7}{20})$, $(1.15838, -\frac{1}{4})$, $(-1.36176, -\frac{2}{5})$, $(-0.841616, 0)$, $(0.715921, -\frac{9}{20})$, $(0.715921, -\frac{9}{20})$\}

\item $c = \frac{57}{10}$, $(d_i,\theta_i)$ = \{$(1., 0)$, $(0.158384, 0)$, $(-0.45965, \frac{1}{10})$, $(-0.902113, \frac{1}{10})$, $(0.715921, \frac{3}{10})$, $(1., \frac{1}{2})$, $(0.158384, \frac{1}{2})$, $(-0.45965, -\frac{2}{5})$, $(-0.902113, -\frac{2}{5})$, $(0.715921, -\frac{1}{5})$, $(1.01247, -\frac{31}{80})$, $(0.819101, -\frac{3}{16})$, $(-0.595112, \frac{1}{16})$, $(0.312869, \frac{33}{80})$, $(-0.962912, -\frac{27}{80})$, $(0.819101, -\frac{3}{16})$, $(-0.595112, \frac{1}{16})$, $(0.312869, \frac{33}{80})$, $(-0.962912, -\frac{27}{80})$\}

\item $c = \frac{31}{5}$, $(d_i,\theta_i)$ = \{$(1., 0)$, $(0.158384, \frac{1}{2})$, $(-0.45965, \frac{1}{10})$, $(-0.902113, \frac{1}{10})$, $(0.715921, \frac{3}{10})$, $(1., \frac{1}{2})$, $(0.158384, 0)$, $(-0.45965, -\frac{2}{5})$, $(-0.902113, -\frac{2}{5})$, $(0.715921, -\frac{1}{5})$, $(1.15838, -\frac{1}{8})$, $(0.442463, \frac{19}{40})$, $(-1.36176, -\frac{11}{40})$, $(-0.841616, \frac{1}{8})$, $(0.715921, -\frac{13}{40})$, $(0.715921, -\frac{13}{40})$\}

\item $c = \frac{67}{10}$, $(d_i,\theta_i)$ = \{$(1., 0)$, $(0.158384, \frac{1}{2})$, $(-0.45965, \frac{1}{10})$, $(-0.902113, \frac{1}{10})$, $(0.715921, \frac{3}{10})$, $(1., \frac{1}{2})$, $(0.158384, 0)$, $(-0.45965, -\frac{2}{5})$, $(-0.902113, -\frac{2}{5})$, $(0.715921, -\frac{1}{5})$, $(1.01247, -\frac{21}{80})$, $(0.819101, -\frac{1}{16})$, $(-0.595112, \frac{3}{16})$, $(0.312869, -\frac{37}{80})$, $(-0.962912, -\frac{17}{80})$, $(0.819101, -\frac{1}{16})$, $(-0.595112, \frac{3}{16})$, $(0.312869, -\frac{37}{80})$, $(-0.962912, -\frac{17}{80})$\}

\item $c = \frac{36}{5}$, $(d_i,\theta_i)$ = \{$(1., 0)$, $(0.158384, \frac{1}{2})$, $(-0.45965, \frac{1}{10})$, $(-0.902113, \frac{1}{10})$, $(0.715921, \frac{3}{10})$, $(1., \frac{1}{2})$, $(0.158384, 0)$, $(-0.45965, -\frac{2}{5})$, $(-0.902113, -\frac{2}{5})$, $(0.715921, -\frac{1}{5})$, $(1.15838, 0)$, $(-1.36176, -\frac{3}{20})$, $(0.442463, -\frac{2}{5})$, $(-0.841616, \frac{1}{4})$, $(0.715921, -\frac{1}{5})$, $(0.715921, -\frac{1}{5})$\}

\item $c = \frac{77}{10}$, $(d_i,\theta_i)$ = \{$(1., 0)$, $(0.158384, \frac{1}{2})$, $(-0.45965, \frac{1}{10})$, $(-0.902113, \frac{1}{10})$, $(0.715921, \frac{3}{10})$, $(1., \frac{1}{2})$, $(0.158384, 0)$, $(-0.45965, -\frac{2}{5})$, $(-0.902113, -\frac{2}{5})$, $(0.715921, -\frac{1}{5})$, $(1.01247, -\frac{11}{80})$, $(-0.595112, \frac{5}{16})$, $(0.819101, \frac{1}{16})$, $(-0.962912, -\frac{7}{80})$, $(0.312869, -\frac{27}{80})$, $(-0.595112, \frac{5}{16})$, $(0.819101, \frac{1}{16})$, $(-0.962912, -\frac{7}{80})$, $(0.312869, -\frac{27}{80})$\}

\end{enumerate}

\paragraph*{Rank 10; \#54}

\begin{enumerate}

\item $c = \frac{3}{10}$, $(d_i,\theta_i)$ = \{$(1., 0)$, $(0.158384, 0)$, $(0.715921, \frac{1}{5})$, $(-0.45965, -\frac{1}{10})$, $(-0.902113, -\frac{1}{10})$, $(1., \frac{1}{2})$, $(0.158384, \frac{1}{2})$, $(0.715921, -\frac{3}{10})$, $(-0.45965, \frac{2}{5})$, $(-0.902113, \frac{2}{5})$, $(1.01247, \frac{11}{80})$, $(0.312869, \frac{27}{80})$, $(-0.962912, \frac{7}{80})$, $(0.819101, -\frac{1}{16})$, $(-0.595112, -\frac{5}{16})$, $(0.312869, \frac{27}{80})$, $(-0.962912, \frac{7}{80})$, $(0.819101, -\frac{1}{16})$, $(-0.595112, -\frac{5}{16})$\}

\item $c = \frac{4}{5}$, $(d_i,\theta_i)$ = \{$(1., 0)$, $(0.158384, 0)$, $(0.715921, \frac{1}{5})$, $(-0.45965, -\frac{1}{10})$, $(-0.902113, -\frac{1}{10})$, $(1., \frac{1}{2})$, $(0.158384, \frac{1}{2})$, $(0.715921, -\frac{3}{10})$, $(-0.45965, \frac{2}{5})$, $(-0.902113, \frac{2}{5})$, $(0.442463, \frac{2}{5})$, $(-1.36176, \frac{3}{20})$, $(1.15838, 0)$, $(-0.841616, -\frac{1}{4})$, $(0.715921, \frac{1}{5})$, $(0.715921, \frac{1}{5})$\}

\item $c = \frac{13}{10}$, $(d_i,\theta_i)$ = \{$(1., 0)$, $(0.158384, 0)$, $(0.715921, \frac{1}{5})$, $(-0.45965, -\frac{1}{10})$, $(-0.902113, -\frac{1}{10})$, $(1., \frac{1}{2})$, $(0.158384, \frac{1}{2})$, $(0.715921, -\frac{3}{10})$, $(-0.45965, \frac{2}{5})$, $(-0.902113, \frac{2}{5})$, $(1.01247, \frac{21}{80})$, $(-0.962912, \frac{17}{80})$, $(0.312869, \frac{37}{80})$, $(-0.595112, -\frac{3}{16})$, $(0.819101, \frac{1}{16})$, $(-0.962912, \frac{17}{80})$, $(0.312869, \frac{37}{80})$, $(-0.595112, -\frac{3}{16})$, $(0.819101, \frac{1}{16})$\}

\item $c = \frac{9}{5}$, $(d_i,\theta_i)$ = \{$(1., 0)$, $(0.158384, 0)$, $(0.715921, \frac{1}{5})$, $(-0.45965, -\frac{1}{10})$, $(-0.902113, -\frac{1}{10})$, $(1., \frac{1}{2})$, $(0.158384, \frac{1}{2})$, $(0.715921, -\frac{3}{10})$, $(-0.45965, \frac{2}{5})$, $(-0.902113, \frac{2}{5})$, $(0.442463, -\frac{19}{40})$, $(-0.841616, -\frac{1}{8})$, $(-1.36176, \frac{11}{40})$, $(1.15838, \frac{1}{8})$, $(0.715921, \frac{13}{40})$, $(0.715921, \frac{13}{40})$\}

\item $c = \frac{23}{10}$, $(d_i,\theta_i)$ = \{$(1., 0)$, $(0.158384, 0)$, $(0.715921, \frac{1}{5})$, $(-0.45965, -\frac{1}{10})$, $(-0.902113, -\frac{1}{10})$, $(1., \frac{1}{2})$, $(0.158384, \frac{1}{2})$, $(0.715921, -\frac{3}{10})$, $(-0.45965, \frac{2}{5})$, $(-0.902113, \frac{2}{5})$, $(1.01247, \frac{31}{80})$, $(-0.962912, \frac{27}{80})$, $(0.312869, -\frac{33}{80})$, $(-0.595112, -\frac{1}{16})$, $(0.819101, \frac{3}{16})$, $(-0.962912, \frac{27}{80})$, $(0.312869, -\frac{33}{80})$, $(-0.595112, -\frac{1}{16})$, $(0.819101, \frac{3}{16})$\}

\item $c = \frac{14}{5}$, $(d_i,\theta_i)$ = \{$(1., 0)$, $(0.158384, 0)$, $(0.715921, \frac{1}{5})$, $(-0.45965, -\frac{1}{10})$, $(-0.902113, -\frac{1}{10})$, $(1., \frac{1}{2})$, $(0.158384, \frac{1}{2})$, $(0.715921, -\frac{3}{10})$, $(-0.45965, \frac{2}{5})$, $(-0.902113, \frac{2}{5})$, $(1.15838, \frac{1}{4})$, $(-0.841616, 0)$, $(-1.36176, \frac{2}{5})$, $(0.442463, -\frac{7}{20})$, $(0.715921, \frac{9}{20})$, $(0.715921, \frac{9}{20})$\}

\item $c = \frac{33}{10}$, $(d_i,\theta_i)$ = \{$(1., 0)$, $(0.158384, 0)$, $(0.715921, \frac{1}{5})$, $(-0.45965, -\frac{1}{10})$, $(-0.902113, -\frac{1}{10})$, $(1., \frac{1}{2})$, $(0.158384, \frac{1}{2})$, $(0.715921, -\frac{3}{10})$, $(-0.45965, \frac{2}{5})$, $(-0.902113, \frac{2}{5})$, $(1.01247, -\frac{39}{80})$, $(-0.962912, \frac{37}{80})$, $(0.312869, -\frac{23}{80})$, $(-0.595112, \frac{1}{16})$, $(0.819101, \frac{5}{16})$, $(-0.962912, \frac{37}{80})$, $(0.312869, -\frac{23}{80})$, $(-0.595112, \frac{1}{16})$, $(0.819101, \frac{5}{16})$\}

\item $c = \frac{19}{5}$, $(d_i,\theta_i)$ = \{$(1., 0)$, $(0.158384, 0)$, $(0.715921, \frac{1}{5})$, $(-0.45965, \frac{2}{5})$, $(-0.902113, \frac{2}{5})$, $(1., \frac{1}{2})$, $(0.158384, \frac{1}{2})$, $(0.715921, -\frac{3}{10})$, $(-0.45965, -\frac{1}{10})$, $(-0.902113, -\frac{1}{10})$, $(1.15838, \frac{3}{8})$, $(0.442463, -\frac{9}{40})$, $(-1.36176, -\frac{19}{40})$, $(-0.841616, \frac{1}{8})$, $(0.715921, -\frac{17}{40})$, $(0.715921, -\frac{17}{40})$\}

\item $c = \frac{43}{10}$, $(d_i,\theta_i)$ = \{$(1., 0)$, $(0.158384, 0)$, $(0.715921, \frac{1}{5})$, $(-0.45965, \frac{2}{5})$, $(-0.902113, \frac{2}{5})$, $(1., \frac{1}{2})$, $(0.158384, \frac{1}{2})$, $(0.715921, -\frac{3}{10})$, $(-0.45965, -\frac{1}{10})$, $(-0.902113, -\frac{1}{10})$, $(1.01247, -\frac{29}{80})$, $(0.312869, -\frac{13}{80})$, $(-0.962912, -\frac{33}{80})$, $(0.819101, \frac{7}{16})$, $(-0.595112, \frac{3}{16})$, $(0.312869, -\frac{13}{80})$, $(-0.962912, -\frac{33}{80})$, $(0.819101, \frac{7}{16})$, $(-0.595112, \frac{3}{16})$\}

\item $c = \frac{24}{5}$, $(d_i,\theta_i)$ = \{$(1., 0)$, $(0.158384, 0)$, $(0.715921, \frac{1}{5})$, $(-0.45965, \frac{2}{5})$, $(-0.902113, \frac{2}{5})$, $(1., \frac{1}{2})$, $(0.158384, \frac{1}{2})$, $(0.715921, -\frac{3}{10})$, $(-0.45965, -\frac{1}{10})$, $(-0.902113, -\frac{1}{10})$, $(-1.36176, -\frac{7}{20})$, $(1.15838, \frac{1}{2})$, $(-0.841616, \frac{1}{4})$, $(0.442463, -\frac{1}{10})$, $(0.715921, -\frac{3}{10})$, $(0.715921, -\frac{3}{10})$\}

\item $c = \frac{53}{10}$, $(d_i,\theta_i)$ = \{$(1., 0)$, $(0.158384, 0)$, $(0.715921, \frac{1}{5})$, $(-0.45965, \frac{2}{5})$, $(-0.902113, \frac{2}{5})$, $(1., \frac{1}{2})$, $(0.158384, \frac{1}{2})$, $(0.715921, -\frac{3}{10})$, $(-0.45965, -\frac{1}{10})$, $(-0.902113, -\frac{1}{10})$, $(1.01247, -\frac{19}{80})$, $(-0.962912, -\frac{23}{80})$, $(0.312869, -\frac{3}{80})$, $(-0.595112, \frac{5}{16})$, $(0.819101, -\frac{7}{16})$, $(-0.962912, -\frac{23}{80})$, $(0.312869, -\frac{3}{80})$, $(-0.595112, \frac{5}{16})$, $(0.819101, -\frac{7}{16})$\}

\item $c = \frac{29}{5}$, $(d_i,\theta_i)$ = \{$(1., 0)$, $(0.158384, 0)$, $(0.715921, \frac{1}{5})$, $(-0.45965, \frac{2}{5})$, $(-0.902113, \frac{2}{5})$, $(1., \frac{1}{2})$, $(0.158384, \frac{1}{2})$, $(0.715921, -\frac{3}{10})$, $(-0.45965, -\frac{1}{10})$, $(-0.902113, -\frac{1}{10})$, $(-0.841616, \frac{3}{8})$, $(1.15838, -\frac{3}{8})$, $(0.442463, \frac{1}{40})$, $(-1.36176, -\frac{9}{40})$, $(0.715921, -\frac{7}{40})$, $(0.715921, -\frac{7}{40})$\}

\item $c = \frac{63}{10}$, $(d_i,\theta_i)$ = \{$(1., 0)$, $(0.158384, \frac{1}{2})$, $(0.715921, \frac{1}{5})$, $(-0.45965, \frac{2}{5})$, $(-0.902113, \frac{2}{5})$, $(1., \frac{1}{2})$, $(0.158384, 0)$, $(0.715921, -\frac{3}{10})$, $(-0.45965, -\frac{1}{10})$, $(-0.902113, -\frac{1}{10})$, $(1.01247, -\frac{9}{80})$, $(-0.962912, -\frac{13}{80})$, $(0.312869, \frac{7}{80})$, $(-0.595112, \frac{7}{16})$, $(0.819101, -\frac{5}{16})$, $(-0.962912, -\frac{13}{80})$, $(0.312869, \frac{7}{80})$, $(-0.595112, \frac{7}{16})$, $(0.819101, -\frac{5}{16})$\}

\item $c = \frac{34}{5}$, $(d_i,\theta_i)$ = \{$(1., 0)$, $(0.158384, \frac{1}{2})$, $(0.715921, \frac{1}{5})$, $(-0.45965, \frac{2}{5})$, $(-0.902113, \frac{2}{5})$, $(1., \frac{1}{2})$, $(0.158384, 0)$, $(0.715921, -\frac{3}{10})$, $(-0.45965, -\frac{1}{10})$, $(-0.902113, -\frac{1}{10})$, $(0.442463, \frac{3}{20})$, $(1.15838, -\frac{1}{4})$, $(-0.841616, \frac{1}{2})$, $(-1.36176, -\frac{1}{10})$, $(0.715921, -\frac{1}{20})$, $(0.715921, -\frac{1}{20})$\}

\item $c = \frac{73}{10}$, $(d_i,\theta_i)$ = \{$(1., 0)$, $(0.158384, \frac{1}{2})$, $(0.715921, \frac{1}{5})$, $(-0.45965, \frac{2}{5})$, $(-0.902113, \frac{2}{5})$, $(1., \frac{1}{2})$, $(0.158384, 0)$, $(0.715921, -\frac{3}{10})$, $(-0.45965, -\frac{1}{10})$, $(-0.902113, -\frac{1}{10})$, $(1.01247, \frac{1}{80})$, $(-0.962912, -\frac{3}{80})$, $(0.312869, \frac{17}{80})$, $(-0.595112, -\frac{7}{16})$, $(0.819101, -\frac{3}{16})$, $(-0.962912, -\frac{3}{80})$, $(0.312869, \frac{17}{80})$, $(-0.595112, -\frac{7}{16})$, $(0.819101, -\frac{3}{16})$\}

\item $c = \frac{39}{5}$, $(d_i,\theta_i)$ = \{$(1., 0)$, $(0.158384, \frac{1}{2})$, $(0.715921, \frac{1}{5})$, $(-0.45965, \frac{2}{5})$, $(-0.902113, \frac{2}{5})$, $(1., \frac{1}{2})$, $(0.158384, 0)$, $(0.715921, -\frac{3}{10})$, $(-0.45965, -\frac{1}{10})$, $(-0.902113, -\frac{1}{10})$, $(0.442463, \frac{11}{40})$, $(-1.36176, \frac{1}{40})$, $(-0.841616, -\frac{3}{8})$, $(1.15838, -\frac{1}{8})$, $(0.715921, \frac{3}{40})$, $(0.715921, \frac{3}{40})$\}

\end{enumerate}

\paragraph*{Rank 10; \#55}

\begin{enumerate}

\item $c = \frac{1}{10}$, $(d_i,\theta_i)$ = \{$(1., 0)$, $(-0.509525, 0)$, $(1.10851, -\frac{1}{5})$, $(-0.175571, -\frac{1}{5})$, $(-0.793604, -\frac{1}{10})$, $(1., \frac{1}{2})$, $(-0.509525, \frac{1}{2})$, $(1.10851, \frac{3}{10})$, $(-0.175571, \frac{3}{10})$, $(-0.793604, \frac{2}{5})$, $(1.12233, -\frac{23}{80})$, $(-0.659687, -\frac{11}{80})$, $(-0.907981, \frac{9}{80})$, $(1.0674, \frac{1}{16})$, $(-0.346818, \frac{5}{16})$, $(-0.659687, -\frac{11}{80})$, $(-0.907981, \frac{9}{80})$, $(1.0674, \frac{1}{16})$, $(-0.346818, \frac{5}{16})$\}

\item $c = \frac{3}{5}$, $(d_i,\theta_i)$ = \{$(1., 0)$, $(-0.509525, 0)$, $(1.10851, -\frac{1}{5})$, $(-0.175571, -\frac{1}{5})$, $(-0.793604, -\frac{1}{10})$, $(1., \frac{1}{2})$, $(-0.509525, \frac{1}{2})$, $(1.10851, \frac{3}{10})$, $(-0.175571, \frac{3}{10})$, $(-0.793604, \frac{2}{5})$, $(-1.28408, \frac{7}{40})$, $(1.50953, \frac{1}{8})$, $(-0.490475, \frac{3}{8})$, $(-0.932938, -\frac{3}{40})$, $(0.793604, -\frac{9}{40})$, $(0.793604, -\frac{9}{40})$\}

\item $c = \frac{11}{10}$, $(d_i,\theta_i)$ = \{$(1., 0)$, $(-0.509525, 0)$, $(1.10851, -\frac{1}{5})$, $(-0.175571, -\frac{1}{5})$, $(-0.793604, -\frac{1}{10})$, $(1., \frac{1}{2})$, $(-0.509525, \frac{1}{2})$, $(1.10851, \frac{3}{10})$, $(-0.175571, \frac{3}{10})$, $(-0.793604, \frac{2}{5})$, $(1.12233, -\frac{13}{80})$, $(-0.907981, \frac{19}{80})$, $(-0.659687, -\frac{1}{80})$, $(-0.346818, \frac{7}{16})$, $(1.0674, \frac{3}{16})$, $(-0.907981, \frac{19}{80})$, $(-0.659687, -\frac{1}{80})$, $(-0.346818, \frac{7}{16})$, $(1.0674, \frac{3}{16})$\}

\item $c = \frac{8}{5}$, $(d_i,\theta_i)$ = \{$(1., 0)$, $(-0.509525, 0)$, $(1.10851, \frac{3}{10})$, $(-0.175571, \frac{3}{10})$, $(-0.793604, -\frac{1}{10})$, $(1., \frac{1}{2})$, $(-0.509525, \frac{1}{2})$, $(1.10851, -\frac{1}{5})$, $(-0.175571, -\frac{1}{5})$, $(-0.793604, \frac{2}{5})$, $(1.50953, \frac{1}{4})$, $(-1.28408, \frac{3}{10})$, $(-0.932938, \frac{1}{20})$, $(-0.490475, \frac{1}{2})$, $(0.793604, -\frac{1}{10})$, $(0.793604, -\frac{1}{10})$\}

\item $c = \frac{21}{10}$, $(d_i,\theta_i)$ = \{$(1., 0)$, $(-0.509525, 0)$, $(1.10851, \frac{3}{10})$, $(-0.175571, \frac{3}{10})$, $(-0.793604, -\frac{1}{10})$, $(1., \frac{1}{2})$, $(-0.509525, \frac{1}{2})$, $(1.10851, -\frac{1}{5})$, $(-0.175571, -\frac{1}{5})$, $(-0.793604, \frac{2}{5})$, $(1.12233, -\frac{3}{80})$, $(-0.659687, \frac{9}{80})$, $(-0.907981, \frac{29}{80})$, $(1.0674, \frac{5}{16})$, $(-0.346818, -\frac{7}{16})$, $(-0.659687, \frac{9}{80})$, $(-0.907981, \frac{29}{80})$, $(1.0674, \frac{5}{16})$, $(-0.346818, -\frac{7}{16})$\}

\item $c = \frac{13}{5}$, $(d_i,\theta_i)$ = \{$(1., 0)$, $(-0.509525, 0)$, $(1.10851, \frac{3}{10})$, $(-0.175571, \frac{3}{10})$, $(-0.793604, -\frac{1}{10})$, $(1., \frac{1}{2})$, $(-0.509525, \frac{1}{2})$, $(1.10851, -\frac{1}{5})$, $(-0.175571, -\frac{1}{5})$, $(-0.793604, \frac{2}{5})$, $(-0.932938, \frac{7}{40})$, $(-0.490475, -\frac{3}{8})$, $(-1.28408, \frac{17}{40})$, $(1.50953, \frac{3}{8})$, $(0.793604, \frac{1}{40})$, $(0.793604, \frac{1}{40})$\}

\item $c = \frac{31}{10}$, $(d_i,\theta_i)$ = \{$(1., 0)$, $(-0.509525, 0)$, $(1.10851, \frac{3}{10})$, $(-0.175571, \frac{3}{10})$, $(-0.793604, -\frac{1}{10})$, $(1., \frac{1}{2})$, $(-0.509525, \frac{1}{2})$, $(1.10851, -\frac{1}{5})$, $(-0.175571, -\frac{1}{5})$, $(-0.793604, \frac{2}{5})$, $(1.12233, \frac{7}{80})$, $(-0.659687, \frac{19}{80})$, $(-0.907981, \frac{39}{80})$, $(1.0674, \frac{7}{16})$, $(-0.346818, -\frac{5}{16})$, $(-0.659687, \frac{19}{80})$, $(-0.907981, \frac{39}{80})$, $(1.0674, \frac{7}{16})$, $(-0.346818, -\frac{5}{16})$\}

\item $c = \frac{18}{5}$, $(d_i,\theta_i)$ = \{$(1., 0)$, $(-0.509525, 0)$, $(1.10851, \frac{3}{10})$, $(-0.175571, \frac{3}{10})$, $(-0.793604, \frac{2}{5})$, $(1., \frac{1}{2})$, $(-0.509525, \frac{1}{2})$, $(1.10851, -\frac{1}{5})$, $(-0.175571, -\frac{1}{5})$, $(-0.793604, -\frac{1}{10})$, $(-0.490475, -\frac{1}{4})$, $(-0.932938, \frac{3}{10})$, $(1.50953, \frac{1}{2})$, $(-1.28408, -\frac{9}{20})$, $(0.793604, \frac{3}{20})$, $(0.793604, \frac{3}{20})$\}

\item $c = \frac{41}{10}$, $(d_i,\theta_i)$ = \{$(1., 0)$, $(-0.509525, 0)$, $(1.10851, \frac{3}{10})$, $(-0.175571, \frac{3}{10})$, $(-0.793604, \frac{2}{5})$, $(1., \frac{1}{2})$, $(-0.509525, \frac{1}{2})$, $(1.10851, -\frac{1}{5})$, $(-0.175571, -\frac{1}{5})$, $(-0.793604, -\frac{1}{10})$, $(1.12233, \frac{17}{80})$, $(-0.659687, \frac{29}{80})$, $(-0.907981, -\frac{31}{80})$, $(1.0674, -\frac{7}{16})$, $(-0.346818, -\frac{3}{16})$, $(-0.659687, \frac{29}{80})$, $(-0.907981, -\frac{31}{80})$, $(1.0674, -\frac{7}{16})$, $(-0.346818, -\frac{3}{16})$\}

\item $c = \frac{23}{5}$, $(d_i,\theta_i)$ = \{$(1., 0)$, $(-0.509525, 0)$, $(1.10851, \frac{3}{10})$, $(-0.175571, \frac{3}{10})$, $(-0.793604, \frac{2}{5})$, $(1., \frac{1}{2})$, $(-0.509525, \frac{1}{2})$, $(1.10851, -\frac{1}{5})$, $(-0.175571, -\frac{1}{5})$, $(-0.793604, -\frac{1}{10})$, $(-1.28408, -\frac{13}{40})$, $(-0.932938, \frac{17}{40})$, $(-0.490475, -\frac{1}{8})$, $(1.50953, -\frac{3}{8})$, $(0.793604, \frac{11}{40})$, $(0.793604, \frac{11}{40})$\}

\item $c = \frac{51}{10}$, $(d_i,\theta_i)$ = \{$(1., 0)$, $(-0.509525, 0)$, $(1.10851, \frac{3}{10})$, $(-0.175571, \frac{3}{10})$, $(-0.793604, \frac{2}{5})$, $(1., \frac{1}{2})$, $(-0.509525, \frac{1}{2})$, $(1.10851, -\frac{1}{5})$, $(-0.175571, -\frac{1}{5})$, $(-0.793604, -\frac{1}{10})$, $(1.12233, \frac{27}{80})$, $(-0.907981, -\frac{21}{80})$, $(-0.659687, \frac{39}{80})$, $(-0.346818, -\frac{1}{16})$, $(1.0674, -\frac{5}{16})$, $(-0.907981, -\frac{21}{80})$, $(-0.659687, \frac{39}{80})$, $(-0.346818, -\frac{1}{16})$, $(1.0674, -\frac{5}{16})$\}

\item $c = \frac{28}{5}$, $(d_i,\theta_i)$ = \{$(1., 0)$, $(-0.509525, 0)$, $(1.10851, \frac{3}{10})$, $(-0.175571, \frac{3}{10})$, $(-0.793604, \frac{2}{5})$, $(1., \frac{1}{2})$, $(-0.509525, \frac{1}{2})$, $(1.10851, -\frac{1}{5})$, $(-0.175571, -\frac{1}{5})$, $(-0.793604, -\frac{1}{10})$, $(1.50953, -\frac{1}{4})$, $(-0.932938, -\frac{9}{20})$, $(-1.28408, -\frac{1}{5})$, $(-0.490475, 0)$, $(0.793604, \frac{2}{5})$, $(0.793604, \frac{2}{5})$\}

\item $c = \frac{61}{10}$, $(d_i,\theta_i)$ = \{$(1., 0)$, $(-0.509525, \frac{1}{2})$, $(1.10851, \frac{3}{10})$, $(-0.175571, \frac{3}{10})$, $(-0.793604, \frac{2}{5})$, $(1., \frac{1}{2})$, $(-0.509525, 0)$, $(1.10851, -\frac{1}{5})$, $(-0.175571, -\frac{1}{5})$, $(-0.793604, -\frac{1}{10})$, $(1.12233, \frac{37}{80})$, $(-0.659687, -\frac{31}{80})$, $(-0.907981, -\frac{11}{80})$, $(1.0674, -\frac{3}{16})$, $(-0.346818, \frac{1}{16})$, $(-0.659687, -\frac{31}{80})$, $(-0.907981, -\frac{11}{80})$, $(1.0674, -\frac{3}{16})$, $(-0.346818, \frac{1}{16})$\}

\item $c = \frac{33}{5}$, $(d_i,\theta_i)$ = \{$(1., 0)$, $(-0.509525, \frac{1}{2})$, $(1.10851, \frac{3}{10})$, $(-0.175571, \frac{3}{10})$, $(-0.793604, \frac{2}{5})$, $(1., \frac{1}{2})$, $(-0.509525, 0)$, $(1.10851, -\frac{1}{5})$, $(-0.175571, -\frac{1}{5})$, $(-0.793604, -\frac{1}{10})$, $(-0.932938, -\frac{13}{40})$, $(-1.28408, -\frac{3}{40})$, $(-0.490475, \frac{1}{8})$, $(1.50953, -\frac{1}{8})$, $(0.793604, -\frac{19}{40})$, $(0.793604, -\frac{19}{40})$\}

\item $c = \frac{71}{10}$, $(d_i,\theta_i)$ = \{$(1., 0)$, $(-0.509525, \frac{1}{2})$, $(1.10851, \frac{3}{10})$, $(-0.175571, \frac{3}{10})$, $(-0.793604, \frac{2}{5})$, $(1., \frac{1}{2})$, $(-0.509525, 0)$, $(1.10851, -\frac{1}{5})$, $(-0.175571, -\frac{1}{5})$, $(-0.793604, -\frac{1}{10})$, $(1.12233, -\frac{33}{80})$, $(-0.659687, -\frac{21}{80})$, $(-0.907981, -\frac{1}{80})$, $(1.0674, -\frac{1}{16})$, $(-0.346818, \frac{3}{16})$, $(-0.659687, -\frac{21}{80})$, $(-0.907981, -\frac{1}{80})$, $(1.0674, -\frac{1}{16})$, $(-0.346818, \frac{3}{16})$\}

\item $c = \frac{38}{5}$, $(d_i,\theta_i)$ = \{$(1., 0)$, $(-0.509525, \frac{1}{2})$, $(1.10851, \frac{3}{10})$, $(-0.175571, \frac{3}{10})$, $(-0.793604, \frac{2}{5})$, $(1., \frac{1}{2})$, $(-0.509525, 0)$, $(1.10851, -\frac{1}{5})$, $(-0.175571, -\frac{1}{5})$, $(-0.793604, -\frac{1}{10})$, $(1.50953, 0)$, $(-0.932938, -\frac{1}{5})$, $(-1.28408, \frac{1}{20})$, $(-0.490475, \frac{1}{4})$, $(0.793604, -\frac{7}{20})$, $(0.793604, -\frac{7}{20})$\}

\end{enumerate}

\paragraph*{Rank 10; \#56}

\begin{enumerate}

\item $c = \frac{2}{5}$, $(d_i,\theta_i)$ = \{$(1., 0)$, $(-0.509525, 0)$, $(-0.793604, \frac{1}{10})$, $(1.10851, \frac{1}{5})$, $(-0.175571, \frac{1}{5})$, $(1., \frac{1}{2})$, $(-0.509525, \frac{1}{2})$, $(-0.793604, -\frac{2}{5})$, $(1.10851, -\frac{3}{10})$, $(-0.175571, -\frac{3}{10})$, $(1.50953, 0)$, $(-0.932938, \frac{1}{5})$, $(-1.28408, -\frac{1}{20})$, $(-0.490475, -\frac{1}{4})$, $(0.793604, \frac{7}{20})$, $(0.793604, \frac{7}{20})$\}

\item $c = \frac{9}{10}$, $(d_i,\theta_i)$ = \{$(1., 0)$, $(-0.509525, 0)$, $(-0.793604, \frac{1}{10})$, $(1.10851, \frac{1}{5})$, $(-0.175571, \frac{1}{5})$, $(1., \frac{1}{2})$, $(-0.509525, \frac{1}{2})$, $(-0.793604, -\frac{2}{5})$, $(1.10851, -\frac{3}{10})$, $(-0.175571, -\frac{3}{10})$, $(1.12233, \frac{33}{80})$, $(-0.346818, -\frac{3}{16})$, $(1.0674, \frac{1}{16})$, $(-0.907981, \frac{1}{80})$, $(-0.659687, \frac{21}{80})$, $(-0.346818, -\frac{3}{16})$, $(1.0674, \frac{1}{16})$, $(-0.907981, \frac{1}{80})$, $(-0.659687, \frac{21}{80})$\}

\item $c = \frac{7}{5}$, $(d_i,\theta_i)$ = \{$(1., 0)$, $(-0.509525, 0)$, $(-0.793604, \frac{1}{10})$, $(1.10851, \frac{1}{5})$, $(-0.175571, \frac{1}{5})$, $(1., \frac{1}{2})$, $(-0.509525, \frac{1}{2})$, $(-0.793604, -\frac{2}{5})$, $(1.10851, -\frac{3}{10})$, $(-0.175571, -\frac{3}{10})$, $(-1.28408, \frac{3}{40})$, $(1.50953, \frac{1}{8})$, $(-0.932938, \frac{13}{40})$, $(-0.490475, -\frac{1}{8})$, $(0.793604, \frac{19}{40})$, $(0.793604, \frac{19}{40})$\}

\item $c = \frac{19}{10}$, $(d_i,\theta_i)$ = \{$(1., 0)$, $(-0.509525, 0)$, $(-0.793604, \frac{1}{10})$, $(1.10851, \frac{1}{5})$, $(-0.175571, \frac{1}{5})$, $(1., \frac{1}{2})$, $(-0.509525, \frac{1}{2})$, $(-0.793604, -\frac{2}{5})$, $(1.10851, -\frac{3}{10})$, $(-0.175571, -\frac{3}{10})$, $(1.12233, -\frac{37}{80})$, $(-0.346818, -\frac{1}{16})$, $(1.0674, \frac{3}{16})$, $(-0.907981, \frac{11}{80})$, $(-0.659687, \frac{31}{80})$, $(-0.346818, -\frac{1}{16})$, $(1.0674, \frac{3}{16})$, $(-0.907981, \frac{11}{80})$, $(-0.659687, \frac{31}{80})$\}

\item $c = \frac{12}{5}$, $(d_i,\theta_i)$ = \{$(1., 0)$, $(-0.509525, 0)$, $(-0.793604, \frac{1}{10})$, $(1.10851, \frac{1}{5})$, $(-0.175571, \frac{1}{5})$, $(1., \frac{1}{2})$, $(-0.509525, \frac{1}{2})$, $(-0.793604, -\frac{2}{5})$, $(1.10851, -\frac{3}{10})$, $(-0.175571, -\frac{3}{10})$, $(-0.932938, \frac{9}{20})$, $(1.50953, \frac{1}{4})$, $(-0.490475, 0)$, $(-1.28408, \frac{1}{5})$, $(0.793604, -\frac{2}{5})$, $(0.793604, -\frac{2}{5})$\}

\item $c = \frac{29}{10}$, $(d_i,\theta_i)$ = \{$(1., 0)$, $(-0.509525, 0)$, $(-0.793604, \frac{1}{10})$, $(1.10851, \frac{1}{5})$, $(-0.175571, \frac{1}{5})$, $(1., \frac{1}{2})$, $(-0.509525, \frac{1}{2})$, $(-0.793604, -\frac{2}{5})$, $(1.10851, -\frac{3}{10})$, $(-0.175571, -\frac{3}{10})$, $(1.12233, -\frac{27}{80})$, $(1.0674, \frac{5}{16})$, $(-0.346818, \frac{1}{16})$, $(-0.659687, -\frac{39}{80})$, $(-0.907981, \frac{21}{80})$, $(1.0674, \frac{5}{16})$, $(-0.346818, \frac{1}{16})$, $(-0.659687, -\frac{39}{80})$, $(-0.907981, \frac{21}{80})$\}

\item $c = \frac{17}{5}$, $(d_i,\theta_i)$ = \{$(1., 0)$, $(-0.509525, 0)$, $(-0.793604, \frac{1}{10})$, $(1.10851, \frac{1}{5})$, $(-0.175571, \frac{1}{5})$, $(1., \frac{1}{2})$, $(-0.509525, \frac{1}{2})$, $(-0.793604, -\frac{2}{5})$, $(1.10851, -\frac{3}{10})$, $(-0.175571, -\frac{3}{10})$, $(-0.490475, \frac{1}{8})$, $(-1.28408, \frac{13}{40})$, $(1.50953, \frac{3}{8})$, $(-0.932938, -\frac{17}{40})$, $(0.793604, -\frac{11}{40})$, $(0.793604, -\frac{11}{40})$\}

\item $c = \frac{39}{10}$, $(d_i,\theta_i)$ = \{$(1., 0)$, $(-0.509525, 0)$, $(-0.793604, \frac{1}{10})$, $(1.10851, \frac{1}{5})$, $(-0.175571, \frac{1}{5})$, $(1., \frac{1}{2})$, $(-0.509525, \frac{1}{2})$, $(-0.793604, -\frac{2}{5})$, $(1.10851, -\frac{3}{10})$, $(-0.175571, -\frac{3}{10})$, $(1.12233, -\frac{17}{80})$, $(-0.346818, \frac{3}{16})$, $(1.0674, \frac{7}{16})$, $(-0.907981, \frac{31}{80})$, $(-0.659687, -\frac{29}{80})$, $(-0.346818, \frac{3}{16})$, $(1.0674, \frac{7}{16})$, $(-0.907981, \frac{31}{80})$, $(-0.659687, -\frac{29}{80})$\}

\item $c = \frac{22}{5}$, $(d_i,\theta_i)$ = \{$(1., 0)$, $(-0.509525, 0)$, $(-0.793604, \frac{1}{10})$, $(1.10851, \frac{1}{5})$, $(-0.175571, \frac{1}{5})$, $(1., \frac{1}{2})$, $(-0.509525, \frac{1}{2})$, $(-0.793604, -\frac{2}{5})$, $(1.10851, -\frac{3}{10})$, $(-0.175571, -\frac{3}{10})$, $(-0.932938, -\frac{3}{10})$, $(-1.28408, \frac{9}{20})$, $(1.50953, \frac{1}{2})$, $(-0.490475, \frac{1}{4})$, $(0.793604, -\frac{3}{20})$, $(0.793604, -\frac{3}{20})$\}

\item $c = \frac{49}{10}$, $(d_i,\theta_i)$ = \{$(1., 0)$, $(-0.509525, 0)$, $(-0.793604, \frac{1}{10})$, $(1.10851, \frac{1}{5})$, $(-0.175571, \frac{1}{5})$, $(1., \frac{1}{2})$, $(-0.509525, \frac{1}{2})$, $(-0.793604, -\frac{2}{5})$, $(1.10851, -\frac{3}{10})$, $(-0.175571, -\frac{3}{10})$, $(1.12233, -\frac{7}{80})$, $(-0.346818, \frac{5}{16})$, $(1.0674, -\frac{7}{16})$, $(-0.907981, -\frac{39}{80})$, $(-0.659687, -\frac{19}{80})$, $(-0.346818, \frac{5}{16})$, $(1.0674, -\frac{7}{16})$, $(-0.907981, -\frac{39}{80})$, $(-0.659687, -\frac{19}{80})$\}

\item $c = \frac{27}{5}$, $(d_i,\theta_i)$ = \{$(1., 0)$, $(-0.509525, 0)$, $(-0.793604, \frac{1}{10})$, $(1.10851, \frac{1}{5})$, $(-0.175571, \frac{1}{5})$, $(1., \frac{1}{2})$, $(-0.509525, \frac{1}{2})$, $(-0.793604, -\frac{2}{5})$, $(1.10851, -\frac{3}{10})$, $(-0.175571, -\frac{3}{10})$, $(-0.932938, -\frac{7}{40})$, $(1.50953, -\frac{3}{8})$, $(-1.28408, -\frac{17}{40})$, $(-0.490475, \frac{3}{8})$, $(0.793604, -\frac{1}{40})$, $(0.793604, -\frac{1}{40})$\}

\item $c = \frac{59}{10}$, $(d_i,\theta_i)$ = \{$(1., 0)$, $(-0.509525, 0)$, $(-0.793604, \frac{1}{10})$, $(1.10851, \frac{1}{5})$, $(-0.175571, \frac{1}{5})$, $(1., \frac{1}{2})$, $(-0.509525, \frac{1}{2})$, $(-0.793604, -\frac{2}{5})$, $(1.10851, -\frac{3}{10})$, $(-0.175571, -\frac{3}{10})$, $(1.12233, \frac{3}{80})$, $(-0.346818, \frac{7}{16})$, $(1.0674, -\frac{5}{16})$, $(-0.907981, -\frac{29}{80})$, $(-0.659687, -\frac{9}{80})$, $(-0.346818, \frac{7}{16})$, $(1.0674, -\frac{5}{16})$, $(-0.907981, -\frac{29}{80})$, $(-0.659687, -\frac{9}{80})$\}

\item $c = \frac{32}{5}$, $(d_i,\theta_i)$ = \{$(1., 0)$, $(-0.509525, \frac{1}{2})$, $(-0.793604, \frac{1}{10})$, $(1.10851, \frac{1}{5})$, $(-0.175571, \frac{1}{5})$, $(1., \frac{1}{2})$, $(-0.509525, 0)$, $(-0.793604, -\frac{2}{5})$, $(1.10851, -\frac{3}{10})$, $(-0.175571, -\frac{3}{10})$, $(-0.932938, -\frac{1}{20})$, $(1.50953, -\frac{1}{4})$, $(-1.28408, -\frac{3}{10})$, $(-0.490475, \frac{1}{2})$, $(0.793604, \frac{1}{10})$, $(0.793604, \frac{1}{10})$\}

\item $c = \frac{69}{10}$, $(d_i,\theta_i)$ = \{$(1., 0)$, $(-0.509525, \frac{1}{2})$, $(-0.793604, \frac{1}{10})$, $(1.10851, \frac{1}{5})$, $(-0.175571, \frac{1}{5})$, $(1., \frac{1}{2})$, $(-0.509525, 0)$, $(-0.793604, -\frac{2}{5})$, $(1.10851, -\frac{3}{10})$, $(-0.175571, -\frac{3}{10})$, $(1.12233, \frac{13}{80})$, $(1.0674, -\frac{3}{16})$, $(-0.346818, -\frac{7}{16})$, $(-0.659687, \frac{1}{80})$, $(-0.907981, -\frac{19}{80})$, $(1.0674, -\frac{3}{16})$, $(-0.346818, -\frac{7}{16})$, $(-0.659687, \frac{1}{80})$, $(-0.907981, -\frac{19}{80})$\}

\item $c = \frac{37}{5}$, $(d_i,\theta_i)$ = \{$(1., 0)$, $(-0.509525, \frac{1}{2})$, $(-0.793604, \frac{1}{10})$, $(1.10851, \frac{1}{5})$, $(-0.175571, \frac{1}{5})$, $(1., \frac{1}{2})$, $(-0.509525, 0)$, $(-0.793604, -\frac{2}{5})$, $(1.10851, -\frac{3}{10})$, $(-0.175571, -\frac{3}{10})$, $(-0.932938, \frac{3}{40})$, $(1.50953, -\frac{1}{8})$, $(-1.28408, -\frac{7}{40})$, $(-0.490475, -\frac{3}{8})$, $(0.793604, \frac{9}{40})$, $(0.793604, \frac{9}{40})$\}

\item $c = \frac{79}{10}$, $(d_i,\theta_i)$ = \{$(1., 0)$, $(-0.509525, \frac{1}{2})$, $(-0.793604, \frac{1}{10})$, $(1.10851, \frac{1}{5})$, $(-0.175571, \frac{1}{5})$, $(1., \frac{1}{2})$, $(-0.509525, 0)$, $(-0.793604, -\frac{2}{5})$, $(1.10851, -\frac{3}{10})$, $(-0.175571, -\frac{3}{10})$, $(1.12233, \frac{23}{80})$, $(-0.346818, -\frac{5}{16})$, $(1.0674, -\frac{1}{16})$, $(-0.907981, -\frac{9}{80})$, $(-0.659687, \frac{11}{80})$, $(-0.346818, -\frac{5}{16})$, $(1.0674, -\frac{1}{16})$, $(-0.907981, -\frac{9}{80})$, $(-0.659687, \frac{11}{80})$\}

\end{enumerate}

\paragraph*{Rank 10; \#57}

\begin{enumerate}

\item $c = \frac{1}{10}$, $(d_i,\theta_i)$ = \{$(1., 0)$, $(-1.96261, 0)$, $(-0.344577, -\frac{1}{5})$, $(2.17557, -\frac{1}{5})$, $(1.55754, -\frac{1}{10})$, $(1., \frac{1}{2})$, $(-1.96261, \frac{1}{2})$, $(-0.344577, \frac{3}{10})$, $(2.17557, \frac{3}{10})$, $(1.55754, \frac{2}{5})$, $(2.20269, \frac{17}{80})$, $(1.29471, -\frac{11}{80})$, $(-1.78201, -\frac{31}{80})$, $(-2.09488, \frac{1}{16})$, $(-0.680668, -\frac{3}{16})$, $(1.29471, -\frac{11}{80})$, $(-1.78201, -\frac{31}{80})$, $(-2.09488, \frac{1}{16})$, $(-0.680668, -\frac{3}{16})$\}

\item $c = \frac{3}{5}$, $(d_i,\theta_i)$ = \{$(1., 0)$, $(-1.96261, 0)$, $(-0.344577, -\frac{1}{5})$, $(2.17557, -\frac{1}{5})$, $(1.55754, -\frac{1}{10})$, $(1., \frac{1}{2})$, $(-1.96261, \frac{1}{2})$, $(-0.344577, \frac{3}{10})$, $(2.17557, \frac{3}{10})$, $(1.55754, \frac{2}{5})$, $(-2.96261, \frac{1}{8})$, $(-0.962611, -\frac{1}{8})$, $(1.83099, -\frac{3}{40})$, $(-2.52015, -\frac{13}{40})$, $(1.55754, \frac{11}{40})$, $(1.55754, \frac{11}{40})$\}

\item $c = \frac{11}{10}$, $(d_i,\theta_i)$ = \{$(1., 0)$, $(-1.96261, 0)$, $(-0.344577, -\frac{1}{5})$, $(2.17557, -\frac{1}{5})$, $(1.55754, -\frac{1}{10})$, $(1., \frac{1}{2})$, $(-1.96261, \frac{1}{2})$, $(-0.344577, \frac{3}{10})$, $(2.17557, \frac{3}{10})$, $(1.55754, \frac{2}{5})$, $(2.20269, \frac{27}{80})$, $(-1.78201, -\frac{21}{80})$, $(1.29471, -\frac{1}{80})$, $(-0.680668, -\frac{1}{16})$, $(-2.09488, \frac{3}{16})$, $(-1.78201, -\frac{21}{80})$, $(1.29471, -\frac{1}{80})$, $(-0.680668, -\frac{1}{16})$, $(-2.09488, \frac{3}{16})$\}

\item $c = \frac{8}{5}$, $(d_i,\theta_i)$ = \{$(1., 0)$, $(-1.96261, 0)$, $(-0.344577, \frac{3}{10})$, $(2.17557, \frac{3}{10})$, $(1.55754, -\frac{1}{10})$, $(1., \frac{1}{2})$, $(-1.96261, \frac{1}{2})$, $(-0.344577, -\frac{1}{5})$, $(2.17557, -\frac{1}{5})$, $(1.55754, \frac{2}{5})$, $(-2.52015, -\frac{1}{5})$, $(-0.962611, 0)$, $(-2.96261, \frac{1}{4})$, $(1.83099, \frac{1}{20})$, $(1.55754, \frac{2}{5})$, $(1.55754, \frac{2}{5})$\}

\item $c = \frac{21}{10}$, $(d_i,\theta_i)$ = \{$(1., 0)$, $(-1.96261, 0)$, $(-0.344577, \frac{3}{10})$, $(2.17557, \frac{3}{10})$, $(1.55754, -\frac{1}{10})$, $(1., \frac{1}{2})$, $(-1.96261, \frac{1}{2})$, $(-0.344577, -\frac{1}{5})$, $(2.17557, -\frac{1}{5})$, $(1.55754, \frac{2}{5})$, $(2.20269, \frac{37}{80})$, $(-1.78201, -\frac{11}{80})$, $(1.29471, \frac{9}{80})$, $(-0.680668, \frac{1}{16})$, $(-2.09488, \frac{5}{16})$, $(-1.78201, -\frac{11}{80})$, $(1.29471, \frac{9}{80})$, $(-0.680668, \frac{1}{16})$, $(-2.09488, \frac{5}{16})$\}

\item $c = \frac{13}{5}$, $(d_i,\theta_i)$ = \{$(1., 0)$, $(-1.96261, 0)$, $(-0.344577, \frac{3}{10})$, $(2.17557, \frac{3}{10})$, $(1.55754, -\frac{1}{10})$, $(1., \frac{1}{2})$, $(-1.96261, \frac{1}{2})$, $(-0.344577, -\frac{1}{5})$, $(2.17557, -\frac{1}{5})$, $(1.55754, \frac{2}{5})$, $(-2.96261, \frac{3}{8})$, $(-0.962611, \frac{1}{8})$, $(-2.52015, -\frac{3}{40})$, $(1.83099, \frac{7}{40})$, $(1.55754, -\frac{19}{40})$, $(1.55754, -\frac{19}{40})$\}

\item $c = \frac{31}{10}$, $(d_i,\theta_i)$ = \{$(1., 0)$, $(-1.96261, 0)$, $(-0.344577, \frac{3}{10})$, $(2.17557, \frac{3}{10})$, $(1.55754, -\frac{1}{10})$, $(1., \frac{1}{2})$, $(-1.96261, \frac{1}{2})$, $(-0.344577, -\frac{1}{5})$, $(2.17557, -\frac{1}{5})$, $(1.55754, \frac{2}{5})$, $(2.20269, -\frac{33}{80})$, $(-1.78201, -\frac{1}{80})$, $(1.29471, \frac{19}{80})$, $(-0.680668, \frac{3}{16})$, $(-2.09488, \frac{7}{16})$, $(-1.78201, -\frac{1}{80})$, $(1.29471, \frac{19}{80})$, $(-0.680668, \frac{3}{16})$, $(-2.09488, \frac{7}{16})$\}

\item $c = \frac{18}{5}$, $(d_i,\theta_i)$ = \{$(1., 0)$, $(-1.96261, 0)$, $(-0.344577, \frac{3}{10})$, $(2.17557, \frac{3}{10})$, $(1.55754, \frac{2}{5})$, $(1., \frac{1}{2})$, $(-1.96261, \frac{1}{2})$, $(-0.344577, -\frac{1}{5})$, $(2.17557, -\frac{1}{5})$, $(1.55754, -\frac{1}{10})$, $(-2.96261, \frac{1}{2})$, $(1.83099, \frac{3}{10})$, $(-2.52015, \frac{1}{20})$, $(-0.962611, \frac{1}{4})$, $(1.55754, -\frac{7}{20})$, $(1.55754, -\frac{7}{20})$\}

\item $c = \frac{41}{10}$, $(d_i,\theta_i)$ = \{$(1., 0)$, $(-1.96261, 0)$, $(-0.344577, \frac{3}{10})$, $(2.17557, \frac{3}{10})$, $(1.55754, \frac{2}{5})$, $(1., \frac{1}{2})$, $(-1.96261, \frac{1}{2})$, $(-0.344577, -\frac{1}{5})$, $(2.17557, -\frac{1}{5})$, $(1.55754, -\frac{1}{10})$, $(2.20269, -\frac{23}{80})$, $(1.29471, \frac{29}{80})$, $(-1.78201, \frac{9}{80})$, $(-2.09488, -\frac{7}{16})$, $(-0.680668, \frac{5}{16})$, $(1.29471, \frac{29}{80})$, $(-1.78201, \frac{9}{80})$, $(-2.09488, -\frac{7}{16})$, $(-0.680668, \frac{5}{16})$\}

\item $c = \frac{23}{5}$, $(d_i,\theta_i)$ = \{$(1., 0)$, $(-1.96261, 0)$, $(-0.344577, \frac{3}{10})$, $(2.17557, \frac{3}{10})$, $(1.55754, \frac{2}{5})$, $(1., \frac{1}{2})$, $(-1.96261, \frac{1}{2})$, $(-0.344577, -\frac{1}{5})$, $(2.17557, -\frac{1}{5})$, $(1.55754, -\frac{1}{10})$, $(-2.96261, -\frac{3}{8})$, $(-0.962611, \frac{3}{8})$, $(-2.52015, \frac{7}{40})$, $(1.83099, \frac{17}{40})$, $(1.55754, -\frac{9}{40})$, $(1.55754, -\frac{9}{40})$\}

\item $c = \frac{51}{10}$, $(d_i,\theta_i)$ = \{$(1., 0)$, $(-1.96261, 0)$, $(-0.344577, \frac{3}{10})$, $(2.17557, \frac{3}{10})$, $(1.55754, \frac{2}{5})$, $(1., \frac{1}{2})$, $(-1.96261, \frac{1}{2})$, $(-0.344577, -\frac{1}{5})$, $(2.17557, -\frac{1}{5})$, $(1.55754, -\frac{1}{10})$, $(2.20269, -\frac{13}{80})$, $(-1.78201, \frac{19}{80})$, $(1.29471, \frac{39}{80})$, $(-0.680668, \frac{7}{16})$, $(-2.09488, -\frac{5}{16})$, $(-1.78201, \frac{19}{80})$, $(1.29471, \frac{39}{80})$, $(-0.680668, \frac{7}{16})$, $(-2.09488, -\frac{5}{16})$\}

\item $c = \frac{28}{5}$, $(d_i,\theta_i)$ = \{$(1., 0)$, $(-1.96261, 0)$, $(-0.344577, \frac{3}{10})$, $(2.17557, \frac{3}{10})$, $(1.55754, \frac{2}{5})$, $(1., \frac{1}{2})$, $(-1.96261, \frac{1}{2})$, $(-0.344577, -\frac{1}{5})$, $(2.17557, -\frac{1}{5})$, $(1.55754, -\frac{1}{10})$, $(1.83099, -\frac{9}{20})$, $(-2.96261, -\frac{1}{4})$, $(-0.962611, \frac{1}{2})$, $(-2.52015, \frac{3}{10})$, $(1.55754, -\frac{1}{10})$, $(1.55754, -\frac{1}{10})$\}

\item $c = \frac{61}{10}$, $(d_i,\theta_i)$ = \{$(1., 0)$, $(-1.96261, \frac{1}{2})$, $(-0.344577, \frac{3}{10})$, $(2.17557, \frac{3}{10})$, $(1.55754, \frac{2}{5})$, $(1., \frac{1}{2})$, $(-1.96261, 0)$, $(-0.344577, -\frac{1}{5})$, $(2.17557, -\frac{1}{5})$, $(1.55754, -\frac{1}{10})$, $(2.20269, -\frac{3}{80})$, $(-1.78201, \frac{29}{80})$, $(1.29471, -\frac{31}{80})$, $(-0.680668, -\frac{7}{16})$, $(-2.09488, -\frac{3}{16})$, $(-1.78201, \frac{29}{80})$, $(1.29471, -\frac{31}{80})$, $(-0.680668, -\frac{7}{16})$, $(-2.09488, -\frac{3}{16})$\}

\item $c = \frac{33}{5}$, $(d_i,\theta_i)$ = \{$(1., 0)$, $(-1.96261, \frac{1}{2})$, $(-0.344577, \frac{3}{10})$, $(2.17557, \frac{3}{10})$, $(1.55754, \frac{2}{5})$, $(1., \frac{1}{2})$, $(-1.96261, 0)$, $(-0.344577, -\frac{1}{5})$, $(2.17557, -\frac{1}{5})$, $(1.55754, -\frac{1}{10})$, $(1.83099, -\frac{13}{40})$, $(-2.96261, -\frac{1}{8})$, $(-0.962611, -\frac{3}{8})$, $(-2.52015, \frac{17}{40})$, $(1.55754, \frac{1}{40})$, $(1.55754, \frac{1}{40})$\}

\item $c = \frac{71}{10}$, $(d_i,\theta_i)$ = \{$(1., 0)$, $(-1.96261, \frac{1}{2})$, $(-0.344577, \frac{3}{10})$, $(2.17557, \frac{3}{10})$, $(1.55754, \frac{2}{5})$, $(1., \frac{1}{2})$, $(-1.96261, 0)$, $(-0.344577, -\frac{1}{5})$, $(2.17557, -\frac{1}{5})$, $(1.55754, -\frac{1}{10})$, $(2.20269, \frac{7}{80})$, $(-1.78201, \frac{39}{80})$, $(1.29471, -\frac{21}{80})$, $(-0.680668, -\frac{5}{16})$, $(-2.09488, -\frac{1}{16})$, $(-1.78201, \frac{39}{80})$, $(1.29471, -\frac{21}{80})$, $(-0.680668, -\frac{5}{16})$, $(-2.09488, -\frac{1}{16})$\}

\item $c = \frac{38}{5}$, $(d_i,\theta_i)$ = \{$(1., 0)$, $(-1.96261, \frac{1}{2})$, $(-0.344577, \frac{3}{10})$, $(2.17557, \frac{3}{10})$, $(1.55754, \frac{2}{5})$, $(1., \frac{1}{2})$, $(-1.96261, 0)$, $(-0.344577, -\frac{1}{5})$, $(2.17557, -\frac{1}{5})$, $(1.55754, -\frac{1}{10})$, $(-2.96261, 0)$, $(1.83099, -\frac{1}{5})$, $(-0.962611, -\frac{1}{4})$, $(-2.52015, -\frac{9}{20})$, $(1.55754, \frac{3}{20})$, $(1.55754, \frac{3}{20})$\}

\end{enumerate}

\paragraph*{Rank 10; \#58}

\begin{enumerate}

\item $c = \frac{2}{5}$, $(d_i,\theta_i)$ = \{$(1., 0)$, $(-1.96261, 0)$, $(1.55754, \frac{1}{10})$, $(-0.344577, \frac{1}{5})$, $(2.17557, \frac{1}{5})$, $(1., \frac{1}{2})$, $(-1.96261, \frac{1}{2})$, $(1.55754, -\frac{2}{5})$, $(-0.344577, -\frac{3}{10})$, $(2.17557, -\frac{3}{10})$, $(1.83099, \frac{1}{5})$, $(-0.962611, \frac{1}{4})$, $(-2.52015, \frac{9}{20})$, $(-2.96261, 0)$, $(1.55754, -\frac{3}{20})$, $(1.55754, -\frac{3}{20})$\}

\item $c = \frac{9}{10}$, $(d_i,\theta_i)$ = \{$(1., 0)$, $(-1.96261, 0)$, $(1.55754, \frac{1}{10})$, $(-0.344577, \frac{1}{5})$, $(2.17557, \frac{1}{5})$, $(1., \frac{1}{2})$, $(-1.96261, \frac{1}{2})$, $(1.55754, -\frac{2}{5})$, $(-0.344577, -\frac{3}{10})$, $(2.17557, -\frac{3}{10})$, $(2.20269, -\frac{7}{80})$, $(-2.09488, \frac{1}{16})$, $(-0.680668, \frac{5}{16})$, $(1.29471, \frac{21}{80})$, $(-1.78201, -\frac{39}{80})$, $(-2.09488, \frac{1}{16})$, $(-0.680668, \frac{5}{16})$, $(1.29471, \frac{21}{80})$, $(-1.78201, -\frac{39}{80})$\}

\item $c = \frac{7}{5}$, $(d_i,\theta_i)$ = \{$(1., 0)$, $(-1.96261, 0)$, $(1.55754, \frac{1}{10})$, $(-0.344577, \frac{1}{5})$, $(2.17557, \frac{1}{5})$, $(1., \frac{1}{2})$, $(-1.96261, \frac{1}{2})$, $(1.55754, -\frac{2}{5})$, $(-0.344577, -\frac{3}{10})$, $(2.17557, -\frac{3}{10})$, $(-0.962611, \frac{3}{8})$, $(-2.96261, \frac{1}{8})$, $(-2.52015, -\frac{17}{40})$, $(1.83099, \frac{13}{40})$, $(1.55754, -\frac{1}{40})$, $(1.55754, -\frac{1}{40})$\}

\item $c = \frac{19}{10}$, $(d_i,\theta_i)$ = \{$(1., 0)$, $(-1.96261, 0)$, $(1.55754, \frac{1}{10})$, $(-0.344577, \frac{1}{5})$, $(2.17557, \frac{1}{5})$, $(1., \frac{1}{2})$, $(-1.96261, \frac{1}{2})$, $(1.55754, -\frac{2}{5})$, $(-0.344577, -\frac{3}{10})$, $(2.17557, -\frac{3}{10})$, $(2.20269, \frac{3}{80})$, $(-2.09488, \frac{3}{16})$, $(-0.680668, \frac{7}{16})$, $(1.29471, \frac{31}{80})$, $(-1.78201, -\frac{29}{80})$, $(-2.09488, \frac{3}{16})$, $(-0.680668, \frac{7}{16})$, $(1.29471, \frac{31}{80})$, $(-1.78201, -\frac{29}{80})$\}

\item $c = \frac{12}{5}$, $(d_i,\theta_i)$ = \{$(1., 0)$, $(-1.96261, 0)$, $(1.55754, \frac{1}{10})$, $(-0.344577, \frac{1}{5})$, $(2.17557, \frac{1}{5})$, $(1., \frac{1}{2})$, $(-1.96261, \frac{1}{2})$, $(1.55754, -\frac{2}{5})$, $(-0.344577, -\frac{3}{10})$, $(2.17557, -\frac{3}{10})$, $(1.83099, \frac{9}{20})$, $(-0.962611, \frac{1}{2})$, $(-2.96261, \frac{1}{4})$, $(-2.52015, -\frac{3}{10})$, $(1.55754, \frac{1}{10})$, $(1.55754, \frac{1}{10})$\}

\item $c = \frac{29}{10}$, $(d_i,\theta_i)$ = \{$(1., 0)$, $(-1.96261, 0)$, $(1.55754, \frac{1}{10})$, $(-0.344577, \frac{1}{5})$, $(2.17557, \frac{1}{5})$, $(1., \frac{1}{2})$, $(-1.96261, \frac{1}{2})$, $(1.55754, -\frac{2}{5})$, $(-0.344577, -\frac{3}{10})$, $(2.17557, -\frac{3}{10})$, $(2.20269, \frac{13}{80})$, $(-2.09488, \frac{5}{16})$, $(-0.680668, -\frac{7}{16})$, $(1.29471, -\frac{39}{80})$, $(-1.78201, -\frac{19}{80})$, $(-2.09488, \frac{5}{16})$, $(-0.680668, -\frac{7}{16})$, $(1.29471, -\frac{39}{80})$, $(-1.78201, -\frac{19}{80})$\}

\item $c = \frac{17}{5}$, $(d_i,\theta_i)$ = \{$(1., 0)$, $(-1.96261, 0)$, $(1.55754, \frac{1}{10})$, $(-0.344577, \frac{1}{5})$, $(2.17557, \frac{1}{5})$, $(1., \frac{1}{2})$, $(-1.96261, \frac{1}{2})$, $(1.55754, -\frac{2}{5})$, $(-0.344577, -\frac{3}{10})$, $(2.17557, -\frac{3}{10})$, $(-2.96261, \frac{3}{8})$, $(-2.52015, -\frac{7}{40})$, $(-0.962611, -\frac{3}{8})$, $(1.83099, -\frac{17}{40})$, $(1.55754, \frac{9}{40})$, $(1.55754, \frac{9}{40})$\}

\item $c = \frac{39}{10}$, $(d_i,\theta_i)$ = \{$(1., 0)$, $(-1.96261, 0)$, $(1.55754, \frac{1}{10})$, $(-0.344577, \frac{1}{5})$, $(2.17557, \frac{1}{5})$, $(1., \frac{1}{2})$, $(-1.96261, \frac{1}{2})$, $(1.55754, -\frac{2}{5})$, $(-0.344577, -\frac{3}{10})$, $(2.17557, -\frac{3}{10})$, $(2.20269, \frac{23}{80})$, $(-0.680668, -\frac{5}{16})$, $(-2.09488, \frac{7}{16})$, $(-1.78201, -\frac{9}{80})$, $(1.29471, -\frac{29}{80})$, $(-0.680668, -\frac{5}{16})$, $(-2.09488, \frac{7}{16})$, $(-1.78201, -\frac{9}{80})$, $(1.29471, -\frac{29}{80})$\}

\item $c = \frac{22}{5}$, $(d_i,\theta_i)$ = \{$(1., 0)$, $(-1.96261, 0)$, $(1.55754, \frac{1}{10})$, $(-0.344577, \frac{1}{5})$, $(2.17557, \frac{1}{5})$, $(1., \frac{1}{2})$, $(-1.96261, \frac{1}{2})$, $(1.55754, -\frac{2}{5})$, $(-0.344577, -\frac{3}{10})$, $(2.17557, -\frac{3}{10})$, $(-2.96261, \frac{1}{2})$, $(-2.52015, -\frac{1}{20})$, $(1.83099, -\frac{3}{10})$, $(-0.962611, -\frac{1}{4})$, $(1.55754, \frac{7}{20})$, $(1.55754, \frac{7}{20})$\}

\item $c = \frac{49}{10}$, $(d_i,\theta_i)$ = \{$(1., 0)$, $(-1.96261, 0)$, $(1.55754, \frac{1}{10})$, $(-0.344577, \frac{1}{5})$, $(2.17557, \frac{1}{5})$, $(1., \frac{1}{2})$, $(-1.96261, \frac{1}{2})$, $(1.55754, -\frac{2}{5})$, $(-0.344577, -\frac{3}{10})$, $(2.17557, -\frac{3}{10})$, $(2.20269, \frac{33}{80})$, $(-2.09488, -\frac{7}{16})$, $(-0.680668, -\frac{3}{16})$, $(1.29471, -\frac{19}{80})$, $(-1.78201, \frac{1}{80})$, $(-2.09488, -\frac{7}{16})$, $(-0.680668, -\frac{3}{16})$, $(1.29471, -\frac{19}{80})$, $(-1.78201, \frac{1}{80})$\}

\item $c = \frac{27}{5}$, $(d_i,\theta_i)$ = \{$(1., 0)$, $(-1.96261, 0)$, $(1.55754, \frac{1}{10})$, $(-0.344577, \frac{1}{5})$, $(2.17557, \frac{1}{5})$, $(1., \frac{1}{2})$, $(-1.96261, \frac{1}{2})$, $(1.55754, -\frac{2}{5})$, $(-0.344577, -\frac{3}{10})$, $(2.17557, -\frac{3}{10})$, $(1.83099, -\frac{7}{40})$, $(-0.962611, -\frac{1}{8})$, $(-2.96261, -\frac{3}{8})$, $(-2.52015, \frac{3}{40})$, $(1.55754, \frac{19}{40})$, $(1.55754, \frac{19}{40})$\}

\item $c = \frac{59}{10}$, $(d_i,\theta_i)$ = \{$(1., 0)$, $(-1.96261, 0)$, $(1.55754, \frac{1}{10})$, $(-0.344577, \frac{1}{5})$, $(2.17557, \frac{1}{5})$, $(1., \frac{1}{2})$, $(-1.96261, \frac{1}{2})$, $(1.55754, -\frac{2}{5})$, $(-0.344577, -\frac{3}{10})$, $(2.17557, -\frac{3}{10})$, $(2.20269, -\frac{37}{80})$, $(-2.09488, -\frac{5}{16})$, $(-0.680668, -\frac{1}{16})$, $(1.29471, -\frac{9}{80})$, $(-1.78201, \frac{11}{80})$, $(-2.09488, -\frac{5}{16})$, $(-0.680668, -\frac{1}{16})$, $(1.29471, -\frac{9}{80})$, $(-1.78201, \frac{11}{80})$\}

\item $c = \frac{32}{5}$, $(d_i,\theta_i)$ = \{$(1., 0)$, $(-1.96261, \frac{1}{2})$, $(1.55754, \frac{1}{10})$, $(-0.344577, \frac{1}{5})$, $(2.17557, \frac{1}{5})$, $(1., \frac{1}{2})$, $(-1.96261, 0)$, $(1.55754, -\frac{2}{5})$, $(-0.344577, -\frac{3}{10})$, $(2.17557, -\frac{3}{10})$, $(-2.96261, -\frac{1}{4})$, $(-2.52015, \frac{1}{5})$, $(-0.962611, 0)$, $(1.83099, -\frac{1}{20})$, $(1.55754, -\frac{2}{5})$, $(1.55754, -\frac{2}{5})$\}

\item $c = \frac{69}{10}$, $(d_i,\theta_i)$ = \{$(1., 0)$, $(-1.96261, \frac{1}{2})$, $(1.55754, \frac{1}{10})$, $(-0.344577, \frac{1}{5})$, $(2.17557, \frac{1}{5})$, $(1., \frac{1}{2})$, $(-1.96261, 0)$, $(1.55754, -\frac{2}{5})$, $(-0.344577, -\frac{3}{10})$, $(2.17557, -\frac{3}{10})$, $(2.20269, -\frac{27}{80})$, $(-2.09488, -\frac{3}{16})$, $(-0.680668, \frac{1}{16})$, $(1.29471, \frac{1}{80})$, $(-1.78201, \frac{21}{80})$, $(-2.09488, -\frac{3}{16})$, $(-0.680668, \frac{1}{16})$, $(1.29471, \frac{1}{80})$, $(-1.78201, \frac{21}{80})$\}

\item $c = \frac{37}{5}$, $(d_i,\theta_i)$ = \{$(1., 0)$, $(-1.96261, \frac{1}{2})$, $(1.55754, \frac{1}{10})$, $(-0.344577, \frac{1}{5})$, $(2.17557, \frac{1}{5})$, $(1., \frac{1}{2})$, $(-1.96261, 0)$, $(1.55754, -\frac{2}{5})$, $(-0.344577, -\frac{3}{10})$, $(2.17557, -\frac{3}{10})$, $(-2.96261, -\frac{1}{8})$, $(-0.962611, \frac{1}{8})$, $(1.83099, \frac{3}{40})$, $(-2.52015, \frac{13}{40})$, $(1.55754, -\frac{11}{40})$, $(1.55754, -\frac{11}{40})$\}

\item $c = \frac{79}{10}$, $(d_i,\theta_i)$ = \{$(1., 0)$, $(-1.96261, 0)$, $(-0.344577, \frac{1}{5})$, $(2.17557, \frac{1}{5})$, $(1.55754, \frac{1}{10})$, $(1., \frac{1}{2})$, $(-1.96261, \frac{1}{2})$, $(-0.344577, -\frac{3}{10})$, $(2.17557, -\frac{3}{10})$, $(1.55754, -\frac{2}{5})$, $(2.20269, -\frac{17}{80})$, $(1.29471, \frac{11}{80})$, $(-1.78201, \frac{31}{80})$, $(-2.09488, -\frac{1}{16})$, $(-0.680668, \frac{3}{16})$, $(1.29471, \frac{11}{80})$, $(-1.78201, \frac{31}{80})$, $(-2.09488, -\frac{1}{16})$, $(-0.680668, \frac{3}{16})$\}

\end{enumerate}

\paragraph*{Rank 10; \#59}

\begin{enumerate}

\item $c = 0$, $(d_i,\theta_i)$ = \{$(1., 0)$, $(7.87298, 0)$, $(6.87298, \frac{1}{10})$, $(8.87298, \frac{1}{6})$, $(6.87298, -\frac{1}{10})$, $(1., \frac{1}{2})$, $(7.87298, \frac{1}{2})$, $(6.87298, -\frac{2}{5})$, $(8.87298, -\frac{1}{3})$, $(6.87298, \frac{2}{5})$, $(6.87298, 0)$, $(8.87298, \frac{1}{4})$, $(8.87298, -\frac{1}{12})$, $(6.87298, \frac{2}{5})$, $(6.87298, -\frac{2}{5})$, $(8.87298, -\frac{1}{12})$, $(6.87298, \frac{2}{5})$, $(6.87298, -\frac{2}{5})$\}

\item $c = \frac{1}{2}$, $(d_i,\theta_i)$ = \{$(1., 0)$, $(7.87298, 0)$, $(6.87298, \frac{1}{10})$, $(8.87298, \frac{1}{6})$, $(6.87298, -\frac{1}{10})$, $(1., \frac{1}{2})$, $(7.87298, \frac{1}{2})$, $(6.87298, -\frac{2}{5})$, $(8.87298, -\frac{1}{3})$, $(6.87298, \frac{2}{5})$, $(12.5483, -\frac{1}{48})$, $(9.71987, \frac{37}{80})$, $(9.71987, -\frac{27}{80})$, $(4.85993, \frac{1}{16})$, $(6.27415, \frac{5}{16})$, $(4.85993, \frac{1}{16})$, $(6.27415, \frac{5}{16})$\}

\item $c = 1$, $(d_i,\theta_i)$ = \{$(1., 0)$, $(7.87298, 0)$, $(6.87298, \frac{1}{10})$, $(8.87298, \frac{1}{6})$, $(6.87298, -\frac{1}{10})$, $(1., \frac{1}{2})$, $(7.87298, \frac{1}{2})$, $(6.87298, -\frac{2}{5})$, $(8.87298, -\frac{1}{3})$, $(6.87298, \frac{2}{5})$, $(8.87298, \frac{3}{8})$, $(6.87298, \frac{1}{8})$, $(8.87298, \frac{1}{24})$, $(6.87298, -\frac{19}{40})$, $(6.87298, -\frac{11}{40})$, $(8.87298, \frac{1}{24})$, $(6.87298, -\frac{19}{40})$, $(6.87298, -\frac{11}{40})$\}

\item $c = \frac{3}{2}$, $(d_i,\theta_i)$ = \{$(1., 0)$, $(7.87298, 0)$, $(6.87298, \frac{1}{10})$, $(8.87298, \frac{1}{6})$, $(6.87298, -\frac{1}{10})$, $(1., \frac{1}{2})$, $(7.87298, \frac{1}{2})$, $(6.87298, -\frac{2}{5})$, $(8.87298, -\frac{1}{3})$, $(6.87298, \frac{2}{5})$, $(12.5483, \frac{5}{48})$, $(9.71987, -\frac{17}{80})$, $(9.71987, -\frac{33}{80})$, $(4.85993, \frac{3}{16})$, $(6.27415, \frac{7}{16})$, $(4.85993, \frac{3}{16})$, $(6.27415, \frac{7}{16})$\}

\item $c = 2$, $(d_i,\theta_i)$ = \{$(1., 0)$, $(7.87298, 0)$, $(6.87298, \frac{1}{10})$, $(8.87298, \frac{1}{6})$, $(6.87298, -\frac{1}{10})$, $(1., \frac{1}{2})$, $(7.87298, \frac{1}{2})$, $(6.87298, -\frac{2}{5})$, $(8.87298, -\frac{1}{3})$, $(6.87298, \frac{2}{5})$, $(6.87298, \frac{1}{4})$, $(8.87298, \frac{1}{2})$, $(8.87298, \frac{1}{6})$, $(6.87298, -\frac{7}{20})$, $(6.87298, -\frac{3}{20})$, $(8.87298, \frac{1}{6})$, $(6.87298, -\frac{7}{20})$, $(6.87298, -\frac{3}{20})$\}

\item $c = \frac{5}{2}$, $(d_i,\theta_i)$ = \{$(1., 0)$, $(7.87298, 0)$, $(6.87298, \frac{1}{10})$, $(8.87298, \frac{1}{6})$, $(6.87298, -\frac{1}{10})$, $(1., \frac{1}{2})$, $(7.87298, \frac{1}{2})$, $(6.87298, -\frac{2}{5})$, $(8.87298, -\frac{1}{3})$, $(6.87298, \frac{2}{5})$, $(12.5483, \frac{11}{48})$, $(9.71987, -\frac{23}{80})$, $(9.71987, -\frac{7}{80})$, $(6.27415, -\frac{7}{16})$, $(4.85993, \frac{5}{16})$, $(6.27415, -\frac{7}{16})$, $(4.85993, \frac{5}{16})$\}

\item $c = 3$, $(d_i,\theta_i)$ = \{$(1., 0)$, $(7.87298, 0)$, $(6.87298, \frac{1}{10})$, $(8.87298, \frac{1}{6})$, $(6.87298, -\frac{1}{10})$, $(1., \frac{1}{2})$, $(7.87298, \frac{1}{2})$, $(6.87298, -\frac{2}{5})$, $(8.87298, -\frac{1}{3})$, $(6.87298, \frac{2}{5})$, $(6.87298, \frac{3}{8})$, $(8.87298, -\frac{3}{8})$, $(8.87298, \frac{7}{24})$, $(6.87298, -\frac{9}{40})$, $(6.87298, -\frac{1}{40})$, $(8.87298, \frac{7}{24})$, $(6.87298, -\frac{9}{40})$, $(6.87298, -\frac{1}{40})$\}

\item $c = \frac{7}{2}$, $(d_i,\theta_i)$ = \{$(1., 0)$, $(7.87298, 0)$, $(6.87298, \frac{1}{10})$, $(8.87298, \frac{1}{6})$, $(6.87298, -\frac{1}{10})$, $(1., \frac{1}{2})$, $(7.87298, \frac{1}{2})$, $(6.87298, -\frac{2}{5})$, $(8.87298, -\frac{1}{3})$, $(6.87298, \frac{2}{5})$, $(12.5483, \frac{17}{48})$, $(9.71987, \frac{3}{80})$, $(9.71987, -\frac{13}{80})$, $(4.85993, \frac{7}{16})$, $(6.27415, -\frac{5}{16})$, $(4.85993, \frac{7}{16})$, $(6.27415, -\frac{5}{16})$\}

\item $c = 4$, $(d_i,\theta_i)$ = \{$(1., 0)$, $(7.87298, 0)$, $(6.87298, \frac{1}{10})$, $(8.87298, \frac{1}{6})$, $(6.87298, \frac{2}{5})$, $(1., \frac{1}{2})$, $(7.87298, \frac{1}{2})$, $(6.87298, -\frac{2}{5})$, $(8.87298, -\frac{1}{3})$, $(6.87298, -\frac{1}{10})$, $(6.87298, \frac{1}{2})$, $(8.87298, -\frac{1}{4})$, $(8.87298, \frac{5}{12})$, $(6.87298, -\frac{1}{10})$, $(6.87298, \frac{1}{10})$, $(8.87298, \frac{5}{12})$, $(6.87298, -\frac{1}{10})$, $(6.87298, \frac{1}{10})$\}

\item $c = \frac{9}{2}$, $(d_i,\theta_i)$ = \{$(1., 0)$, $(7.87298, 0)$, $(6.87298, \frac{1}{10})$, $(8.87298, \frac{1}{6})$, $(6.87298, \frac{2}{5})$, $(1., \frac{1}{2})$, $(7.87298, \frac{1}{2})$, $(6.87298, -\frac{2}{5})$, $(8.87298, -\frac{1}{3})$, $(6.87298, -\frac{1}{10})$, $(12.5483, \frac{23}{48})$, $(9.71987, \frac{13}{80})$, $(9.71987, -\frac{3}{80})$, $(4.85993, -\frac{7}{16})$, $(6.27415, -\frac{3}{16})$, $(4.85993, -\frac{7}{16})$, $(6.27415, -\frac{3}{16})$\}

\item $c = 5$, $(d_i,\theta_i)$ = \{$(1., 0)$, $(7.87298, 0)$, $(6.87298, \frac{1}{10})$, $(8.87298, \frac{1}{6})$, $(6.87298, \frac{2}{5})$, $(1., \frac{1}{2})$, $(7.87298, \frac{1}{2})$, $(6.87298, -\frac{2}{5})$, $(8.87298, -\frac{1}{3})$, $(6.87298, -\frac{1}{10})$, $(6.87298, -\frac{3}{8})$, $(8.87298, -\frac{1}{8})$, $(8.87298, -\frac{11}{24})$, $(6.87298, \frac{1}{40})$, $(6.87298, \frac{9}{40})$, $(8.87298, -\frac{11}{24})$, $(6.87298, \frac{1}{40})$, $(6.87298, \frac{9}{40})$\}

\item $c = \frac{11}{2}$, $(d_i,\theta_i)$ = \{$(1., 0)$, $(7.87298, 0)$, $(6.87298, \frac{1}{10})$, $(8.87298, \frac{1}{6})$, $(6.87298, \frac{2}{5})$, $(1., \frac{1}{2})$, $(7.87298, \frac{1}{2})$, $(6.87298, -\frac{2}{5})$, $(8.87298, -\frac{1}{3})$, $(6.87298, -\frac{1}{10})$, $(12.5483, -\frac{19}{48})$, $(9.71987, \frac{23}{80})$, $(9.71987, \frac{7}{80})$, $(4.85993, -\frac{5}{16})$, $(6.27415, -\frac{1}{16})$, $(4.85993, -\frac{5}{16})$, $(6.27415, -\frac{1}{16})$\}

\item $c = 6$, $(d_i,\theta_i)$ = \{$(1., 0)$, $(7.87298, \frac{1}{2})$, $(6.87298, \frac{1}{10})$, $(8.87298, \frac{1}{6})$, $(6.87298, \frac{2}{5})$, $(1., \frac{1}{2})$, $(7.87298, 0)$, $(6.87298, -\frac{2}{5})$, $(8.87298, -\frac{1}{3})$, $(6.87298, -\frac{1}{10})$, $(6.87298, -\frac{1}{4})$, $(8.87298, 0)$, $(8.87298, -\frac{1}{3})$, $(6.87298, \frac{3}{20})$, $(6.87298, \frac{7}{20})$, $(8.87298, -\frac{1}{3})$, $(6.87298, \frac{3}{20})$, $(6.87298, \frac{7}{20})$\}

\item $c = \frac{13}{2}$, $(d_i,\theta_i)$ = \{$(1., 0)$, $(7.87298, \frac{1}{2})$, $(6.87298, \frac{1}{10})$, $(8.87298, \frac{1}{6})$, $(6.87298, \frac{2}{5})$, $(1., \frac{1}{2})$, $(7.87298, 0)$, $(6.87298, -\frac{2}{5})$, $(8.87298, -\frac{1}{3})$, $(6.87298, -\frac{1}{10})$, $(12.5483, -\frac{13}{48})$, $(9.71987, \frac{17}{80})$, $(9.71987, \frac{33}{80})$, $(6.27415, \frac{1}{16})$, $(4.85993, -\frac{3}{16})$, $(6.27415, \frac{1}{16})$, $(4.85993, -\frac{3}{16})$\}

\item $c = 7$, $(d_i,\theta_i)$ = \{$(1., 0)$, $(7.87298, \frac{1}{2})$, $(6.87298, \frac{1}{10})$, $(8.87298, \frac{1}{6})$, $(6.87298, \frac{2}{5})$, $(1., \frac{1}{2})$, $(7.87298, 0)$, $(6.87298, -\frac{2}{5})$, $(8.87298, -\frac{1}{3})$, $(6.87298, -\frac{1}{10})$, $(6.87298, -\frac{1}{8})$, $(8.87298, \frac{1}{8})$, $(8.87298, -\frac{5}{24})$, $(6.87298, \frac{11}{40})$, $(6.87298, \frac{19}{40})$, $(8.87298, -\frac{5}{24})$, $(6.87298, \frac{11}{40})$, $(6.87298, \frac{19}{40})$\}

\item $c = \frac{15}{2}$, $(d_i,\theta_i)$ = \{$(1., 0)$, $(7.87298, \frac{1}{2})$, $(6.87298, \frac{1}{10})$, $(8.87298, \frac{1}{6})$, $(6.87298, \frac{2}{5})$, $(1., \frac{1}{2})$, $(7.87298, 0)$, $(6.87298, -\frac{2}{5})$, $(8.87298, -\frac{1}{3})$, $(6.87298, -\frac{1}{10})$, $(12.5483, -\frac{7}{48})$, $(9.71987, -\frac{37}{80})$, $(9.71987, \frac{27}{80})$, $(4.85993, -\frac{1}{16})$, $(6.27415, \frac{3}{16})$, $(4.85993, -\frac{1}{16})$, $(6.27415, \frac{3}{16})$\}

\end{enumerate}

\paragraph*{Rank 10; \#60}

\begin{enumerate}

\item $c = 0$, $(d_i,\theta_i)$ = \{$(1., 0)$, $(7.87298, 0)$, $(8.87298, \frac{1}{6})$, $(6.87298, \frac{1}{5})$, $(6.87298, -\frac{1}{5})$, $(1., \frac{1}{2})$, $(7.87298, \frac{1}{2})$, $(8.87298, -\frac{1}{3})$, $(6.87298, -\frac{3}{10})$, $(6.87298, \frac{3}{10})$, $(8.87298, \frac{1}{4})$, $(6.87298, \frac{1}{2})$, $(8.87298, -\frac{1}{12})$, $(6.87298, -\frac{3}{10})$, $(6.87298, \frac{3}{10})$, $(8.87298, -\frac{1}{12})$, $(6.87298, -\frac{3}{10})$, $(6.87298, \frac{3}{10})$\}

\item $c = \frac{1}{2}$, $(d_i,\theta_i)$ = \{$(1., 0)$, $(7.87298, 0)$, $(8.87298, \frac{1}{6})$, $(6.87298, \frac{1}{5})$, $(6.87298, -\frac{1}{5})$, $(1., \frac{1}{2})$, $(7.87298, \frac{1}{2})$, $(8.87298, -\frac{1}{3})$, $(6.87298, -\frac{3}{10})$, $(6.87298, \frac{3}{10})$, $(12.5483, -\frac{1}{48})$, $(9.71987, -\frac{19}{80})$, $(9.71987, \frac{29}{80})$, $(6.27415, \frac{5}{16})$, $(4.85993, -\frac{7}{16})$, $(6.27415, \frac{5}{16})$, $(4.85993, -\frac{7}{16})$\}

\item $c = 1$, $(d_i,\theta_i)$ = \{$(1., 0)$, $(7.87298, 0)$, $(8.87298, \frac{1}{6})$, $(6.87298, \frac{1}{5})$, $(6.87298, -\frac{1}{5})$, $(1., \frac{1}{2})$, $(7.87298, \frac{1}{2})$, $(8.87298, -\frac{1}{3})$, $(6.87298, -\frac{3}{10})$, $(6.87298, \frac{3}{10})$, $(8.87298, \frac{3}{8})$, $(6.87298, -\frac{3}{8})$, $(8.87298, \frac{1}{24})$, $(6.87298, -\frac{7}{40})$, $(6.87298, \frac{17}{40})$, $(8.87298, \frac{1}{24})$, $(6.87298, -\frac{7}{40})$, $(6.87298, \frac{17}{40})$\}

\item $c = \frac{3}{2}$, $(d_i,\theta_i)$ = \{$(1., 0)$, $(7.87298, 0)$, $(8.87298, \frac{1}{6})$, $(6.87298, \frac{1}{5})$, $(6.87298, \frac{3}{10})$, $(1., \frac{1}{2})$, $(7.87298, \frac{1}{2})$, $(8.87298, -\frac{1}{3})$, $(6.87298, -\frac{3}{10})$, $(6.87298, -\frac{1}{5})$, $(12.5483, \frac{5}{48})$, $(9.71987, -\frac{9}{80})$, $(9.71987, \frac{39}{80})$, $(6.27415, \frac{7}{16})$, $(4.85993, -\frac{5}{16})$, $(6.27415, \frac{7}{16})$, $(4.85993, -\frac{5}{16})$\}

\item $c = 2$, $(d_i,\theta_i)$ = \{$(1., 0)$, $(7.87298, 0)$, $(8.87298, \frac{1}{6})$, $(6.87298, \frac{1}{5})$, $(6.87298, \frac{3}{10})$, $(1., \frac{1}{2})$, $(7.87298, \frac{1}{2})$, $(8.87298, -\frac{1}{3})$, $(6.87298, -\frac{3}{10})$, $(6.87298, -\frac{1}{5})$, $(8.87298, \frac{1}{2})$, $(6.87298, -\frac{1}{4})$, $(8.87298, \frac{1}{6})$, $(6.87298, -\frac{1}{20})$, $(6.87298, -\frac{9}{20})$, $(8.87298, \frac{1}{6})$, $(6.87298, -\frac{1}{20})$, $(6.87298, -\frac{9}{20})$\}

\item $c = \frac{5}{2}$, $(d_i,\theta_i)$ = \{$(1., 0)$, $(7.87298, 0)$, $(8.87298, \frac{1}{6})$, $(6.87298, \frac{1}{5})$, $(6.87298, \frac{3}{10})$, $(1., \frac{1}{2})$, $(7.87298, \frac{1}{2})$, $(8.87298, -\frac{1}{3})$, $(6.87298, -\frac{3}{10})$, $(6.87298, -\frac{1}{5})$, $(12.5483, \frac{11}{48})$, $(9.71987, \frac{1}{80})$, $(9.71987, -\frac{31}{80})$, $(6.27415, -\frac{7}{16})$, $(4.85993, -\frac{3}{16})$, $(6.27415, -\frac{7}{16})$, $(4.85993, -\frac{3}{16})$\}

\item $c = 3$, $(d_i,\theta_i)$ = \{$(1., 0)$, $(7.87298, 0)$, $(8.87298, \frac{1}{6})$, $(6.87298, \frac{1}{5})$, $(6.87298, \frac{3}{10})$, $(1., \frac{1}{2})$, $(7.87298, \frac{1}{2})$, $(8.87298, -\frac{1}{3})$, $(6.87298, -\frac{3}{10})$, $(6.87298, -\frac{1}{5})$, $(8.87298, -\frac{3}{8})$, $(6.87298, -\frac{1}{8})$, $(8.87298, \frac{7}{24})$, $(6.87298, \frac{3}{40})$, $(6.87298, -\frac{13}{40})$, $(8.87298, \frac{7}{24})$, $(6.87298, \frac{3}{40})$, $(6.87298, -\frac{13}{40})$\}

\item $c = \frac{7}{2}$, $(d_i,\theta_i)$ = \{$(1., 0)$, $(7.87298, 0)$, $(8.87298, \frac{1}{6})$, $(6.87298, \frac{1}{5})$, $(6.87298, \frac{3}{10})$, $(1., \frac{1}{2})$, $(7.87298, \frac{1}{2})$, $(8.87298, -\frac{1}{3})$, $(6.87298, -\frac{3}{10})$, $(6.87298, -\frac{1}{5})$, $(12.5483, \frac{17}{48})$, $(9.71987, \frac{11}{80})$, $(9.71987, -\frac{21}{80})$, $(4.85993, -\frac{1}{16})$, $(6.27415, -\frac{5}{16})$, $(4.85993, -\frac{1}{16})$, $(6.27415, -\frac{5}{16})$\}

\item $c = 4$, $(d_i,\theta_i)$ = \{$(1., 0)$, $(7.87298, 0)$, $(8.87298, \frac{1}{6})$, $(6.87298, \frac{1}{5})$, $(6.87298, \frac{3}{10})$, $(1., \frac{1}{2})$, $(7.87298, \frac{1}{2})$, $(8.87298, -\frac{1}{3})$, $(6.87298, -\frac{3}{10})$, $(6.87298, -\frac{1}{5})$, $(8.87298, -\frac{1}{4})$, $(6.87298, 0)$, $(8.87298, \frac{5}{12})$, $(6.87298, \frac{1}{5})$, $(6.87298, -\frac{1}{5})$, $(8.87298, \frac{5}{12})$, $(6.87298, \frac{1}{5})$, $(6.87298, -\frac{1}{5})$\}

\item $c = \frac{9}{2}$, $(d_i,\theta_i)$ = \{$(1., 0)$, $(7.87298, 0)$, $(8.87298, \frac{1}{6})$, $(6.87298, \frac{1}{5})$, $(6.87298, \frac{3}{10})$, $(1., \frac{1}{2})$, $(7.87298, \frac{1}{2})$, $(8.87298, -\frac{1}{3})$, $(6.87298, -\frac{3}{10})$, $(6.87298, -\frac{1}{5})$, $(12.5483, \frac{23}{48})$, $(9.71987, \frac{21}{80})$, $(9.71987, -\frac{11}{80})$, $(6.27415, -\frac{3}{16})$, $(4.85993, \frac{1}{16})$, $(6.27415, -\frac{3}{16})$, $(4.85993, \frac{1}{16})$\}

\item $c = 5$, $(d_i,\theta_i)$ = \{$(1., 0)$, $(7.87298, 0)$, $(8.87298, \frac{1}{6})$, $(6.87298, \frac{1}{5})$, $(6.87298, \frac{3}{10})$, $(1., \frac{1}{2})$, $(7.87298, \frac{1}{2})$, $(8.87298, -\frac{1}{3})$, $(6.87298, -\frac{3}{10})$, $(6.87298, -\frac{1}{5})$, $(8.87298, -\frac{1}{8})$, $(6.87298, \frac{1}{8})$, $(8.87298, -\frac{11}{24})$, $(6.87298, \frac{13}{40})$, $(6.87298, -\frac{3}{40})$, $(8.87298, -\frac{11}{24})$, $(6.87298, \frac{13}{40})$, $(6.87298, -\frac{3}{40})$\}

\item $c = \frac{11}{2}$, $(d_i,\theta_i)$ = \{$(1., 0)$, $(7.87298, 0)$, $(8.87298, \frac{1}{6})$, $(6.87298, \frac{1}{5})$, $(6.87298, \frac{3}{10})$, $(1., \frac{1}{2})$, $(7.87298, \frac{1}{2})$, $(8.87298, -\frac{1}{3})$, $(6.87298, -\frac{3}{10})$, $(6.87298, -\frac{1}{5})$, $(12.5483, -\frac{19}{48})$, $(9.71987, \frac{31}{80})$, $(9.71987, -\frac{1}{80})$, $(6.27415, -\frac{1}{16})$, $(4.85993, \frac{3}{16})$, $(6.27415, -\frac{1}{16})$, $(4.85993, \frac{3}{16})$\}

\item $c = 6$, $(d_i,\theta_i)$ = \{$(1., 0)$, $(7.87298, \frac{1}{2})$, $(8.87298, \frac{1}{6})$, $(6.87298, \frac{1}{5})$, $(6.87298, \frac{3}{10})$, $(1., \frac{1}{2})$, $(7.87298, 0)$, $(8.87298, -\frac{1}{3})$, $(6.87298, -\frac{3}{10})$, $(6.87298, -\frac{1}{5})$, $(8.87298, 0)$, $(6.87298, \frac{1}{4})$, $(8.87298, -\frac{1}{3})$, $(6.87298, \frac{9}{20})$, $(6.87298, \frac{1}{20})$, $(8.87298, -\frac{1}{3})$, $(6.87298, \frac{9}{20})$, $(6.87298, \frac{1}{20})$\}

\item $c = \frac{13}{2}$, $(d_i,\theta_i)$ = \{$(1., 0)$, $(7.87298, \frac{1}{2})$, $(8.87298, \frac{1}{6})$, $(6.87298, \frac{1}{5})$, $(6.87298, \frac{3}{10})$, $(1., \frac{1}{2})$, $(7.87298, 0)$, $(8.87298, -\frac{1}{3})$, $(6.87298, -\frac{3}{10})$, $(6.87298, -\frac{1}{5})$, $(12.5483, -\frac{13}{48})$, $(9.71987, \frac{9}{80})$, $(9.71987, -\frac{39}{80})$, $(6.27415, \frac{1}{16})$, $(4.85993, \frac{5}{16})$, $(6.27415, \frac{1}{16})$, $(4.85993, \frac{5}{16})$\}

\item $c = 7$, $(d_i,\theta_i)$ = \{$(1., 0)$, $(7.87298, \frac{1}{2})$, $(8.87298, \frac{1}{6})$, $(6.87298, \frac{1}{5})$, $(6.87298, \frac{3}{10})$, $(1., \frac{1}{2})$, $(7.87298, 0)$, $(8.87298, -\frac{1}{3})$, $(6.87298, -\frac{3}{10})$, $(6.87298, -\frac{1}{5})$, $(8.87298, \frac{1}{8})$, $(6.87298, \frac{3}{8})$, $(8.87298, -\frac{5}{24})$, $(6.87298, -\frac{17}{40})$, $(6.87298, \frac{7}{40})$, $(8.87298, -\frac{5}{24})$, $(6.87298, -\frac{17}{40})$, $(6.87298, \frac{7}{40})$\}

\item $c = \frac{15}{2}$, $(d_i,\theta_i)$ = \{$(1., 0)$, $(7.87298, \frac{1}{2})$, $(8.87298, \frac{1}{6})$, $(6.87298, \frac{1}{5})$, $(6.87298, \frac{3}{10})$, $(1., \frac{1}{2})$, $(7.87298, 0)$, $(8.87298, -\frac{1}{3})$, $(6.87298, -\frac{3}{10})$, $(6.87298, -\frac{1}{5})$, $(12.5483, -\frac{7}{48})$, $(9.71987, -\frac{29}{80})$, $(9.71987, \frac{19}{80})$, $(4.85993, \frac{7}{16})$, $(6.27415, \frac{3}{16})$, $(4.85993, \frac{7}{16})$, $(6.27415, \frac{3}{16})$\}

\end{enumerate}

\paragraph*{Rank 10; \#61}

\begin{enumerate}

\item $c = 0$, $(d_i,\theta_i)$ = \{$(1., 0)$, $(7.87298, 0)$, $(6.87298, \frac{1}{10})$, $(8.87298, -\frac{1}{6})$, $(6.87298, -\frac{1}{10})$, $(1., \frac{1}{2})$, $(7.87298, \frac{1}{2})$, $(6.87298, -\frac{2}{5})$, $(8.87298, \frac{1}{3})$, $(6.87298, \frac{2}{5})$, $(8.87298, -\frac{1}{4})$, $(6.87298, 0)$, $(8.87298, \frac{1}{12})$, $(6.87298, \frac{2}{5})$, $(6.87298, -\frac{2}{5})$, $(8.87298, \frac{1}{12})$, $(6.87298, \frac{2}{5})$, $(6.87298, -\frac{2}{5})$\}

\item $c = \frac{1}{2}$, $(d_i,\theta_i)$ = \{$(1., 0)$, $(7.87298, 0)$, $(6.87298, \frac{1}{10})$, $(8.87298, -\frac{1}{6})$, $(6.87298, -\frac{1}{10})$, $(1., \frac{1}{2})$, $(7.87298, \frac{1}{2})$, $(6.87298, -\frac{2}{5})$, $(8.87298, \frac{1}{3})$, $(6.87298, \frac{2}{5})$, $(12.5483, \frac{7}{48})$, $(9.71987, -\frac{27}{80})$, $(9.71987, \frac{37}{80})$, $(6.27415, -\frac{3}{16})$, $(4.85993, \frac{1}{16})$, $(6.27415, -\frac{3}{16})$, $(4.85993, \frac{1}{16})$\}

\item $c = 1$, $(d_i,\theta_i)$ = \{$(1., 0)$, $(7.87298, 0)$, $(6.87298, \frac{1}{10})$, $(8.87298, -\frac{1}{6})$, $(6.87298, -\frac{1}{10})$, $(1., \frac{1}{2})$, $(7.87298, \frac{1}{2})$, $(6.87298, -\frac{2}{5})$, $(8.87298, \frac{1}{3})$, $(6.87298, \frac{2}{5})$, $(6.87298, \frac{1}{8})$, $(8.87298, -\frac{1}{8})$, $(8.87298, \frac{5}{24})$, $(6.87298, -\frac{19}{40})$, $(6.87298, -\frac{11}{40})$, $(8.87298, \frac{5}{24})$, $(6.87298, -\frac{19}{40})$, $(6.87298, -\frac{11}{40})$\}

\item $c = \frac{3}{2}$, $(d_i,\theta_i)$ = \{$(1., 0)$, $(7.87298, 0)$, $(6.87298, \frac{1}{10})$, $(8.87298, -\frac{1}{6})$, $(6.87298, -\frac{1}{10})$, $(1., \frac{1}{2})$, $(7.87298, \frac{1}{2})$, $(6.87298, -\frac{2}{5})$, $(8.87298, \frac{1}{3})$, $(6.87298, \frac{2}{5})$, $(12.5483, \frac{13}{48})$, $(9.71987, -\frac{33}{80})$, $(9.71987, -\frac{17}{80})$, $(4.85993, \frac{3}{16})$, $(6.27415, -\frac{1}{16})$, $(4.85993, \frac{3}{16})$, $(6.27415, -\frac{1}{16})$\}

\item $c = 2$, $(d_i,\theta_i)$ = \{$(1., 0)$, $(7.87298, 0)$, $(6.87298, \frac{1}{10})$, $(8.87298, \frac{1}{3})$, $(6.87298, -\frac{1}{10})$, $(1., \frac{1}{2})$, $(7.87298, \frac{1}{2})$, $(6.87298, -\frac{2}{5})$, $(8.87298, -\frac{1}{6})$, $(6.87298, \frac{2}{5})$, $(6.87298, \frac{1}{4})$, $(8.87298, 0)$, $(8.87298, \frac{1}{3})$, $(6.87298, -\frac{7}{20})$, $(6.87298, -\frac{3}{20})$, $(8.87298, \frac{1}{3})$, $(6.87298, -\frac{7}{20})$, $(6.87298, -\frac{3}{20})$\}

\item $c = \frac{5}{2}$, $(d_i,\theta_i)$ = \{$(1., 0)$, $(7.87298, 0)$, $(6.87298, \frac{1}{10})$, $(8.87298, \frac{1}{3})$, $(6.87298, -\frac{1}{10})$, $(1., \frac{1}{2})$, $(7.87298, \frac{1}{2})$, $(6.87298, -\frac{2}{5})$, $(8.87298, -\frac{1}{6})$, $(6.87298, \frac{2}{5})$, $(12.5483, \frac{19}{48})$, $(9.71987, -\frac{23}{80})$, $(9.71987, -\frac{7}{80})$, $(6.27415, \frac{1}{16})$, $(4.85993, \frac{5}{16})$, $(6.27415, \frac{1}{16})$, $(4.85993, \frac{5}{16})$\}

\item $c = 3$, $(d_i,\theta_i)$ = \{$(1., 0)$, $(7.87298, 0)$, $(6.87298, \frac{1}{10})$, $(8.87298, \frac{1}{3})$, $(6.87298, -\frac{1}{10})$, $(1., \frac{1}{2})$, $(7.87298, \frac{1}{2})$, $(6.87298, -\frac{2}{5})$, $(8.87298, -\frac{1}{6})$, $(6.87298, \frac{2}{5})$, $(6.87298, \frac{3}{8})$, $(8.87298, \frac{1}{8})$, $(8.87298, \frac{11}{24})$, $(6.87298, -\frac{9}{40})$, $(6.87298, -\frac{1}{40})$, $(8.87298, \frac{11}{24})$, $(6.87298, -\frac{9}{40})$, $(6.87298, -\frac{1}{40})$\}

\item $c = \frac{7}{2}$, $(d_i,\theta_i)$ = \{$(1., 0)$, $(7.87298, 0)$, $(6.87298, \frac{1}{10})$, $(8.87298, \frac{1}{3})$, $(6.87298, -\frac{1}{10})$, $(1., \frac{1}{2})$, $(7.87298, \frac{1}{2})$, $(6.87298, -\frac{2}{5})$, $(8.87298, -\frac{1}{6})$, $(6.87298, \frac{2}{5})$, $(12.5483, -\frac{23}{48})$, $(9.71987, \frac{3}{80})$, $(9.71987, -\frac{13}{80})$, $(6.27415, \frac{3}{16})$, $(4.85993, \frac{7}{16})$, $(6.27415, \frac{3}{16})$, $(4.85993, \frac{7}{16})$\}

\item $c = 4$, $(d_i,\theta_i)$ = \{$(1., 0)$, $(7.87298, 0)$, $(6.87298, \frac{1}{10})$, $(8.87298, \frac{1}{3})$, $(6.87298, \frac{2}{5})$, $(1., \frac{1}{2})$, $(7.87298, \frac{1}{2})$, $(6.87298, -\frac{2}{5})$, $(8.87298, -\frac{1}{6})$, $(6.87298, -\frac{1}{10})$, $(6.87298, \frac{1}{2})$, $(8.87298, \frac{1}{4})$, $(8.87298, -\frac{5}{12})$, $(6.87298, -\frac{1}{10})$, $(6.87298, \frac{1}{10})$, $(8.87298, -\frac{5}{12})$, $(6.87298, -\frac{1}{10})$, $(6.87298, \frac{1}{10})$\}

\item $c = \frac{9}{2}$, $(d_i,\theta_i)$ = \{$(1., 0)$, $(7.87298, 0)$, $(6.87298, \frac{1}{10})$, $(8.87298, \frac{1}{3})$, $(6.87298, \frac{2}{5})$, $(1., \frac{1}{2})$, $(7.87298, \frac{1}{2})$, $(6.87298, -\frac{2}{5})$, $(8.87298, -\frac{1}{6})$, $(6.87298, -\frac{1}{10})$, $(12.5483, -\frac{17}{48})$, $(9.71987, -\frac{3}{80})$, $(9.71987, \frac{13}{80})$, $(6.27415, \frac{5}{16})$, $(4.85993, -\frac{7}{16})$, $(6.27415, \frac{5}{16})$, $(4.85993, -\frac{7}{16})$\}

\item $c = 5$, $(d_i,\theta_i)$ = \{$(1., 0)$, $(7.87298, 0)$, $(6.87298, \frac{1}{10})$, $(8.87298, \frac{1}{3})$, $(6.87298, \frac{2}{5})$, $(1., \frac{1}{2})$, $(7.87298, \frac{1}{2})$, $(6.87298, -\frac{2}{5})$, $(8.87298, -\frac{1}{6})$, $(6.87298, -\frac{1}{10})$, $(6.87298, -\frac{3}{8})$, $(8.87298, \frac{3}{8})$, $(8.87298, -\frac{7}{24})$, $(6.87298, \frac{1}{40})$, $(6.87298, \frac{9}{40})$, $(8.87298, -\frac{7}{24})$, $(6.87298, \frac{1}{40})$, $(6.87298, \frac{9}{40})$\}

\item $c = \frac{11}{2}$, $(d_i,\theta_i)$ = \{$(1., 0)$, $(7.87298, 0)$, $(6.87298, \frac{1}{10})$, $(8.87298, \frac{1}{3})$, $(6.87298, \frac{2}{5})$, $(1., \frac{1}{2})$, $(7.87298, \frac{1}{2})$, $(6.87298, -\frac{2}{5})$, $(8.87298, -\frac{1}{6})$, $(6.87298, -\frac{1}{10})$, $(12.5483, -\frac{11}{48})$, $(9.71987, \frac{23}{80})$, $(9.71987, \frac{7}{80})$, $(4.85993, -\frac{5}{16})$, $(6.27415, \frac{7}{16})$, $(4.85993, -\frac{5}{16})$, $(6.27415, \frac{7}{16})$\}

\item $c = 6$, $(d_i,\theta_i)$ = \{$(1., 0)$, $(7.87298, \frac{1}{2})$, $(6.87298, \frac{1}{10})$, $(8.87298, \frac{1}{3})$, $(6.87298, \frac{2}{5})$, $(1., \frac{1}{2})$, $(7.87298, 0)$, $(6.87298, -\frac{2}{5})$, $(8.87298, -\frac{1}{6})$, $(6.87298, -\frac{1}{10})$, $(8.87298, \frac{1}{2})$, $(6.87298, -\frac{1}{4})$, $(8.87298, -\frac{1}{6})$, $(6.87298, \frac{3}{20})$, $(6.87298, \frac{7}{20})$, $(8.87298, -\frac{1}{6})$, $(6.87298, \frac{3}{20})$, $(6.87298, \frac{7}{20})$\}

\item $c = \frac{13}{2}$, $(d_i,\theta_i)$ = \{$(1., 0)$, $(7.87298, \frac{1}{2})$, $(6.87298, \frac{1}{10})$, $(8.87298, \frac{1}{3})$, $(6.87298, \frac{2}{5})$, $(1., \frac{1}{2})$, $(7.87298, 0)$, $(6.87298, -\frac{2}{5})$, $(8.87298, -\frac{1}{6})$, $(6.87298, -\frac{1}{10})$, $(12.5483, -\frac{5}{48})$, $(9.71987, \frac{33}{80})$, $(9.71987, \frac{17}{80})$, $(6.27415, -\frac{7}{16})$, $(4.85993, -\frac{3}{16})$, $(6.27415, -\frac{7}{16})$, $(4.85993, -\frac{3}{16})$\}

\item $c = 7$, $(d_i,\theta_i)$ = \{$(1., 0)$, $(7.87298, \frac{1}{2})$, $(6.87298, \frac{1}{10})$, $(8.87298, \frac{1}{3})$, $(6.87298, \frac{2}{5})$, $(1., \frac{1}{2})$, $(7.87298, 0)$, $(6.87298, -\frac{2}{5})$, $(8.87298, -\frac{1}{6})$, $(6.87298, -\frac{1}{10})$, $(6.87298, -\frac{1}{8})$, $(8.87298, -\frac{3}{8})$, $(8.87298, -\frac{1}{24})$, $(6.87298, \frac{11}{40})$, $(6.87298, \frac{19}{40})$, $(8.87298, -\frac{1}{24})$, $(6.87298, \frac{11}{40})$, $(6.87298, \frac{19}{40})$\}

\item $c = \frac{15}{2}$, $(d_i,\theta_i)$ = \{$(1., 0)$, $(7.87298, \frac{1}{2})$, $(6.87298, \frac{1}{10})$, $(8.87298, \frac{1}{3})$, $(6.87298, \frac{2}{5})$, $(1., \frac{1}{2})$, $(7.87298, 0)$, $(6.87298, -\frac{2}{5})$, $(8.87298, -\frac{1}{6})$, $(6.87298, -\frac{1}{10})$, $(12.5483, \frac{1}{48})$, $(9.71987, -\frac{37}{80})$, $(9.71987, \frac{27}{80})$, $(6.27415, -\frac{5}{16})$, $(4.85993, -\frac{1}{16})$, $(6.27415, -\frac{5}{16})$, $(4.85993, -\frac{1}{16})$\}

\end{enumerate}

\paragraph*{Rank 10; \#62}

\begin{enumerate}

\item $c = 0$, $(d_i,\theta_i)$ = \{$(1., 0)$, $(7.87298, 0)$, $(6.87298, \frac{1}{5})$, $(6.87298, -\frac{1}{5})$, $(8.87298, -\frac{1}{6})$, $(1., \frac{1}{2})$, $(7.87298, \frac{1}{2})$, $(6.87298, -\frac{3}{10})$, $(6.87298, \frac{3}{10})$, $(8.87298, \frac{1}{3})$, $(8.87298, -\frac{1}{4})$, $(6.87298, \frac{1}{2})$, $(8.87298, \frac{1}{12})$, $(6.87298, -\frac{3}{10})$, $(6.87298, \frac{3}{10})$, $(8.87298, \frac{1}{12})$, $(6.87298, -\frac{3}{10})$, $(6.87298, \frac{3}{10})$\}

\item $c = \frac{1}{2}$, $(d_i,\theta_i)$ = \{$(1., 0)$, $(7.87298, 0)$, $(6.87298, \frac{1}{5})$, $(6.87298, -\frac{1}{5})$, $(8.87298, -\frac{1}{6})$, $(1., \frac{1}{2})$, $(7.87298, \frac{1}{2})$, $(6.87298, -\frac{3}{10})$, $(6.87298, \frac{3}{10})$, $(8.87298, \frac{1}{3})$, $(12.5483, \frac{7}{48})$, $(9.71987, -\frac{19}{80})$, $(9.71987, \frac{29}{80})$, $(6.27415, -\frac{3}{16})$, $(4.85993, -\frac{7}{16})$, $(6.27415, -\frac{3}{16})$, $(4.85993, -\frac{7}{16})$\}

\item $c = 1$, $(d_i,\theta_i)$ = \{$(1., 0)$, $(7.87298, 0)$, $(6.87298, \frac{1}{5})$, $(6.87298, -\frac{1}{5})$, $(8.87298, -\frac{1}{6})$, $(1., \frac{1}{2})$, $(7.87298, \frac{1}{2})$, $(6.87298, -\frac{3}{10})$, $(6.87298, \frac{3}{10})$, $(8.87298, \frac{1}{3})$, $(8.87298, -\frac{1}{8})$, $(6.87298, -\frac{3}{8})$, $(8.87298, \frac{5}{24})$, $(6.87298, -\frac{7}{40})$, $(6.87298, \frac{17}{40})$, $(8.87298, \frac{5}{24})$, $(6.87298, -\frac{7}{40})$, $(6.87298, \frac{17}{40})$\}

\item $c = \frac{3}{2}$, $(d_i,\theta_i)$ = \{$(1., 0)$, $(7.87298, 0)$, $(6.87298, \frac{1}{5})$, $(6.87298, \frac{3}{10})$, $(8.87298, -\frac{1}{6})$, $(1., \frac{1}{2})$, $(7.87298, \frac{1}{2})$, $(6.87298, -\frac{3}{10})$, $(6.87298, -\frac{1}{5})$, $(8.87298, \frac{1}{3})$, $(12.5483, \frac{13}{48})$, $(9.71987, -\frac{9}{80})$, $(9.71987, \frac{39}{80})$, $(4.85993, -\frac{5}{16})$, $(6.27415, -\frac{1}{16})$, $(4.85993, -\frac{5}{16})$, $(6.27415, -\frac{1}{16})$\}

\item $c = 2$, $(d_i,\theta_i)$ = \{$(1., 0)$, $(7.87298, 0)$, $(6.87298, \frac{1}{5})$, $(6.87298, \frac{3}{10})$, $(8.87298, \frac{1}{3})$, $(1., \frac{1}{2})$, $(7.87298, \frac{1}{2})$, $(6.87298, -\frac{3}{10})$, $(6.87298, -\frac{1}{5})$, $(8.87298, -\frac{1}{6})$, $(8.87298, 0)$, $(6.87298, -\frac{1}{4})$, $(8.87298, \frac{1}{3})$, $(6.87298, -\frac{1}{20})$, $(6.87298, -\frac{9}{20})$, $(8.87298, \frac{1}{3})$, $(6.87298, -\frac{1}{20})$, $(6.87298, -\frac{9}{20})$\}

\item $c = \frac{5}{2}$, $(d_i,\theta_i)$ = \{$(1., 0)$, $(7.87298, 0)$, $(6.87298, \frac{1}{5})$, $(6.87298, \frac{3}{10})$, $(8.87298, \frac{1}{3})$, $(1., \frac{1}{2})$, $(7.87298, \frac{1}{2})$, $(6.87298, -\frac{3}{10})$, $(6.87298, -\frac{1}{5})$, $(8.87298, -\frac{1}{6})$, $(12.5483, \frac{19}{48})$, $(9.71987, \frac{1}{80})$, $(9.71987, -\frac{31}{80})$, $(4.85993, -\frac{3}{16})$, $(6.27415, \frac{1}{16})$, $(4.85993, -\frac{3}{16})$, $(6.27415, \frac{1}{16})$\}

\item $c = 3$, $(d_i,\theta_i)$ = \{$(1., 0)$, $(7.87298, 0)$, $(6.87298, \frac{1}{5})$, $(6.87298, \frac{3}{10})$, $(8.87298, \frac{1}{3})$, $(1., \frac{1}{2})$, $(7.87298, \frac{1}{2})$, $(6.87298, -\frac{3}{10})$, $(6.87298, -\frac{1}{5})$, $(8.87298, -\frac{1}{6})$, $(8.87298, \frac{1}{8})$, $(6.87298, -\frac{1}{8})$, $(8.87298, \frac{11}{24})$, $(6.87298, \frac{3}{40})$, $(6.87298, -\frac{13}{40})$, $(8.87298, \frac{11}{24})$, $(6.87298, \frac{3}{40})$, $(6.87298, -\frac{13}{40})$\}

\item $c = \frac{7}{2}$, $(d_i,\theta_i)$ = \{$(1., 0)$, $(7.87298, 0)$, $(6.87298, \frac{1}{5})$, $(6.87298, \frac{3}{10})$, $(8.87298, \frac{1}{3})$, $(1., \frac{1}{2})$, $(7.87298, \frac{1}{2})$, $(6.87298, -\frac{3}{10})$, $(6.87298, -\frac{1}{5})$, $(8.87298, -\frac{1}{6})$, $(12.5483, -\frac{23}{48})$, $(9.71987, -\frac{21}{80})$, $(9.71987, \frac{11}{80})$, $(4.85993, -\frac{1}{16})$, $(6.27415, \frac{3}{16})$, $(4.85993, -\frac{1}{16})$, $(6.27415, \frac{3}{16})$\}

\item $c = 4$, $(d_i,\theta_i)$ = \{$(1., 0)$, $(7.87298, 0)$, $(6.87298, \frac{1}{5})$, $(6.87298, \frac{3}{10})$, $(8.87298, \frac{1}{3})$, $(1., \frac{1}{2})$, $(7.87298, \frac{1}{2})$, $(6.87298, -\frac{3}{10})$, $(6.87298, -\frac{1}{5})$, $(8.87298, -\frac{1}{6})$, $(8.87298, \frac{1}{4})$, $(6.87298, 0)$, $(8.87298, -\frac{5}{12})$, $(6.87298, \frac{1}{5})$, $(6.87298, -\frac{1}{5})$, $(8.87298, -\frac{5}{12})$, $(6.87298, \frac{1}{5})$, $(6.87298, -\frac{1}{5})$\}

\item $c = \frac{9}{2}$, $(d_i,\theta_i)$ = \{$(1., 0)$, $(7.87298, 0)$, $(6.87298, \frac{1}{5})$, $(6.87298, \frac{3}{10})$, $(8.87298, \frac{1}{3})$, $(1., \frac{1}{2})$, $(7.87298, \frac{1}{2})$, $(6.87298, -\frac{3}{10})$, $(6.87298, -\frac{1}{5})$, $(8.87298, -\frac{1}{6})$, $(12.5483, -\frac{17}{48})$, $(9.71987, \frac{21}{80})$, $(9.71987, -\frac{11}{80})$, $(6.27415, \frac{5}{16})$, $(4.85993, \frac{1}{16})$, $(6.27415, \frac{5}{16})$, $(4.85993, \frac{1}{16})$\}

\item $c = 5$, $(d_i,\theta_i)$ = \{$(1., 0)$, $(7.87298, 0)$, $(6.87298, \frac{1}{5})$, $(6.87298, \frac{3}{10})$, $(8.87298, \frac{1}{3})$, $(1., \frac{1}{2})$, $(7.87298, \frac{1}{2})$, $(6.87298, -\frac{3}{10})$, $(6.87298, -\frac{1}{5})$, $(8.87298, -\frac{1}{6})$, $(8.87298, \frac{3}{8})$, $(6.87298, \frac{1}{8})$, $(8.87298, -\frac{7}{24})$, $(6.87298, \frac{13}{40})$, $(6.87298, -\frac{3}{40})$, $(8.87298, -\frac{7}{24})$, $(6.87298, \frac{13}{40})$, $(6.87298, -\frac{3}{40})$\}

\item $c = \frac{11}{2}$, $(d_i,\theta_i)$ = \{$(1., 0)$, $(7.87298, 0)$, $(6.87298, \frac{1}{5})$, $(6.87298, \frac{3}{10})$, $(8.87298, \frac{1}{3})$, $(1., \frac{1}{2})$, $(7.87298, \frac{1}{2})$, $(6.87298, -\frac{3}{10})$, $(6.87298, -\frac{1}{5})$, $(8.87298, -\frac{1}{6})$, $(12.5483, -\frac{11}{48})$, $(9.71987, -\frac{1}{80})$, $(9.71987, \frac{31}{80})$, $(4.85993, \frac{3}{16})$, $(6.27415, \frac{7}{16})$, $(4.85993, \frac{3}{16})$, $(6.27415, \frac{7}{16})$\}

\item $c = 6$, $(d_i,\theta_i)$ = \{$(1., 0)$, $(7.87298, \frac{1}{2})$, $(6.87298, \frac{1}{5})$, $(6.87298, \frac{3}{10})$, $(8.87298, \frac{1}{3})$, $(1., \frac{1}{2})$, $(7.87298, 0)$, $(6.87298, -\frac{3}{10})$, $(6.87298, -\frac{1}{5})$, $(8.87298, -\frac{1}{6})$, $(8.87298, \frac{1}{2})$, $(6.87298, \frac{1}{4})$, $(8.87298, -\frac{1}{6})$, $(6.87298, \frac{9}{20})$, $(6.87298, \frac{1}{20})$, $(8.87298, -\frac{1}{6})$, $(6.87298, \frac{9}{20})$, $(6.87298, \frac{1}{20})$\}

\item $c = \frac{13}{2}$, $(d_i,\theta_i)$ = \{$(1., 0)$, $(7.87298, \frac{1}{2})$, $(6.87298, \frac{1}{5})$, $(6.87298, \frac{3}{10})$, $(8.87298, \frac{1}{3})$, $(1., \frac{1}{2})$, $(7.87298, 0)$, $(6.87298, -\frac{3}{10})$, $(6.87298, -\frac{1}{5})$, $(8.87298, -\frac{1}{6})$, $(12.5483, -\frac{5}{48})$, $(9.71987, \frac{9}{80})$, $(9.71987, -\frac{39}{80})$, $(4.85993, \frac{5}{16})$, $(6.27415, -\frac{7}{16})$, $(4.85993, \frac{5}{16})$, $(6.27415, -\frac{7}{16})$\}

\item $c = 7$, $(d_i,\theta_i)$ = \{$(1., 0)$, $(7.87298, \frac{1}{2})$, $(6.87298, \frac{1}{5})$, $(6.87298, \frac{3}{10})$, $(8.87298, \frac{1}{3})$, $(1., \frac{1}{2})$, $(7.87298, 0)$, $(6.87298, -\frac{3}{10})$, $(6.87298, -\frac{1}{5})$, $(8.87298, -\frac{1}{6})$, $(8.87298, -\frac{3}{8})$, $(6.87298, \frac{3}{8})$, $(8.87298, -\frac{1}{24})$, $(6.87298, -\frac{17}{40})$, $(6.87298, \frac{7}{40})$, $(8.87298, -\frac{1}{24})$, $(6.87298, -\frac{17}{40})$, $(6.87298, \frac{7}{40})$\}

\item $c = \frac{15}{2}$, $(d_i,\theta_i)$ = \{$(1., 0)$, $(7.87298, \frac{1}{2})$, $(6.87298, \frac{1}{5})$, $(6.87298, \frac{3}{10})$, $(8.87298, \frac{1}{3})$, $(1., \frac{1}{2})$, $(7.87298, 0)$, $(6.87298, -\frac{3}{10})$, $(6.87298, -\frac{1}{5})$, $(8.87298, -\frac{1}{6})$, $(12.5483, \frac{1}{48})$, $(9.71987, \frac{19}{80})$, $(9.71987, -\frac{29}{80})$, $(4.85993, \frac{7}{16})$, $(6.27415, -\frac{5}{16})$, $(4.85993, \frac{7}{16})$, $(6.27415, -\frac{5}{16})$\}

\end{enumerate}

\paragraph*{Rank 10; \#63}

\begin{enumerate}

\item $c = 0$, $(d_i,\theta_i)$ = \{$(1., 0)$, $(0.127017, 0)$, $(-0.872983, \frac{1}{10})$, $(1.12702, \frac{1}{6})$, $(-0.872983, -\frac{1}{10})$, $(1., \frac{1}{2})$, $(0.127017, \frac{1}{2})$, $(-0.872983, -\frac{2}{5})$, $(1.12702, -\frac{1}{3})$, $(-0.872983, \frac{2}{5})$, $(-0.872983, \frac{1}{2})$, $(1.12702, \frac{1}{4})$, $(1.12702, -\frac{1}{12})$, $(-0.872983, -\frac{1}{10})$, $(-0.872983, \frac{1}{10})$, $(1.12702, -\frac{1}{12})$, $(-0.872983, -\frac{1}{10})$, $(-0.872983, \frac{1}{10})$\}

\item $c = \frac{1}{2}$, $(d_i,\theta_i)$ = \{$(1., 0)$, $(0.127017, 0)$, $(-0.872983, \frac{1}{10})$, $(1.12702, \frac{1}{6})$, $(-0.872983, -\frac{1}{10})$, $(1., \frac{1}{2})$, $(0.127017, \frac{1}{2})$, $(-0.872983, -\frac{2}{5})$, $(1.12702, -\frac{1}{3})$, $(-0.872983, \frac{2}{5})$, $(1.59384, -\frac{1}{48})$, $(-1.23458, -\frac{3}{80})$, $(-1.23458, \frac{13}{80})$, $(0.796921, \frac{5}{16})$, $(-0.617292, -\frac{7}{16})$, $(0.796921, \frac{5}{16})$, $(-0.617292, -\frac{7}{16})$\}

\item $c = 1$, $(d_i,\theta_i)$ = \{$(1., 0)$, $(0.127017, 0)$, $(-0.872983, \frac{1}{10})$, $(1.12702, \frac{1}{6})$, $(-0.872983, -\frac{1}{10})$, $(1., \frac{1}{2})$, $(0.127017, \frac{1}{2})$, $(-0.872983, -\frac{2}{5})$, $(1.12702, -\frac{1}{3})$, $(-0.872983, \frac{2}{5})$, $(-0.872983, -\frac{3}{8})$, $(1.12702, \frac{3}{8})$, $(1.12702, \frac{1}{24})$, $(-0.872983, \frac{1}{40})$, $(-0.872983, \frac{9}{40})$, $(1.12702, \frac{1}{24})$, $(-0.872983, \frac{1}{40})$, $(-0.872983, \frac{9}{40})$\}

\item $c = \frac{3}{2}$, $(d_i,\theta_i)$ = \{$(1., 0)$, $(0.127017, 0)$, $(-0.872983, \frac{1}{10})$, $(1.12702, \frac{1}{6})$, $(-0.872983, -\frac{1}{10})$, $(1., \frac{1}{2})$, $(0.127017, \frac{1}{2})$, $(-0.872983, -\frac{2}{5})$, $(1.12702, -\frac{1}{3})$, $(-0.872983, \frac{2}{5})$, $(1.59384, \frac{5}{48})$, $(-1.23458, \frac{23}{80})$, $(-1.23458, \frac{7}{80})$, $(0.796921, \frac{7}{16})$, $(-0.617292, -\frac{5}{16})$, $(0.796921, \frac{7}{16})$, $(-0.617292, -\frac{5}{16})$\}

\item $c = 2$, $(d_i,\theta_i)$ = \{$(1., 0)$, $(0.127017, 0)$, $(-0.872983, \frac{1}{10})$, $(1.12702, \frac{1}{6})$, $(-0.872983, -\frac{1}{10})$, $(1., \frac{1}{2})$, $(0.127017, \frac{1}{2})$, $(-0.872983, -\frac{2}{5})$, $(1.12702, -\frac{1}{3})$, $(-0.872983, \frac{2}{5})$, $(-0.872983, -\frac{1}{4})$, $(1.12702, \frac{1}{2})$, $(1.12702, \frac{1}{6})$, $(-0.872983, \frac{3}{20})$, $(-0.872983, \frac{7}{20})$, $(1.12702, \frac{1}{6})$, $(-0.872983, \frac{3}{20})$, $(-0.872983, \frac{7}{20})$\}

\item $c = \frac{5}{2}$, $(d_i,\theta_i)$ = \{$(1., 0)$, $(0.127017, 0)$, $(-0.872983, \frac{1}{10})$, $(1.12702, \frac{1}{6})$, $(-0.872983, -\frac{1}{10})$, $(1., \frac{1}{2})$, $(0.127017, \frac{1}{2})$, $(-0.872983, -\frac{2}{5})$, $(1.12702, -\frac{1}{3})$, $(-0.872983, \frac{2}{5})$, $(1.59384, \frac{11}{48})$, $(-1.23458, \frac{33}{80})$, $(-1.23458, \frac{17}{80})$, $(0.796921, -\frac{7}{16})$, $(-0.617292, -\frac{3}{16})$, $(0.796921, -\frac{7}{16})$, $(-0.617292, -\frac{3}{16})$\}

\item $c = 3$, $(d_i,\theta_i)$ = \{$(1., 0)$, $(0.127017, 0)$, $(-0.872983, \frac{1}{10})$, $(1.12702, \frac{1}{6})$, $(-0.872983, -\frac{1}{10})$, $(1., \frac{1}{2})$, $(0.127017, \frac{1}{2})$, $(-0.872983, -\frac{2}{5})$, $(1.12702, -\frac{1}{3})$, $(-0.872983, \frac{2}{5})$, $(-0.872983, -\frac{1}{8})$, $(1.12702, -\frac{3}{8})$, $(1.12702, \frac{7}{24})$, $(-0.872983, \frac{11}{40})$, $(-0.872983, \frac{19}{40})$, $(1.12702, \frac{7}{24})$, $(-0.872983, \frac{11}{40})$, $(-0.872983, \frac{19}{40})$\}

\item $c = \frac{7}{2}$, $(d_i,\theta_i)$ = \{$(1., 0)$, $(0.127017, 0)$, $(-0.872983, \frac{1}{10})$, $(1.12702, \frac{1}{6})$, $(-0.872983, -\frac{1}{10})$, $(1., \frac{1}{2})$, $(0.127017, \frac{1}{2})$, $(-0.872983, -\frac{2}{5})$, $(1.12702, -\frac{1}{3})$, $(-0.872983, \frac{2}{5})$, $(1.59384, \frac{17}{48})$, $(-1.23458, -\frac{37}{80})$, $(-1.23458, \frac{27}{80})$, $(-0.617292, -\frac{1}{16})$, $(0.796921, -\frac{5}{16})$, $(-0.617292, -\frac{1}{16})$, $(0.796921, -\frac{5}{16})$\}

\item $c = 4$, $(d_i,\theta_i)$ = \{$(1., 0)$, $(0.127017, 0)$, $(-0.872983, \frac{1}{10})$, $(1.12702, \frac{1}{6})$, $(-0.872983, \frac{2}{5})$, $(1., \frac{1}{2})$, $(0.127017, \frac{1}{2})$, $(-0.872983, -\frac{2}{5})$, $(1.12702, -\frac{1}{3})$, $(-0.872983, -\frac{1}{10})$, $(-0.872983, 0)$, $(1.12702, -\frac{1}{4})$, $(1.12702, \frac{5}{12})$, $(-0.872983, \frac{2}{5})$, $(-0.872983, -\frac{2}{5})$, $(1.12702, \frac{5}{12})$, $(-0.872983, \frac{2}{5})$, $(-0.872983, -\frac{2}{5})$\}

\item $c = \frac{9}{2}$, $(d_i,\theta_i)$ = \{$(1., 0)$, $(0.127017, 0)$, $(-0.872983, \frac{1}{10})$, $(1.12702, \frac{1}{6})$, $(-0.872983, \frac{2}{5})$, $(1., \frac{1}{2})$, $(0.127017, \frac{1}{2})$, $(-0.872983, -\frac{2}{5})$, $(1.12702, -\frac{1}{3})$, $(-0.872983, -\frac{1}{10})$, $(1.59384, \frac{23}{48})$, $(-1.23458, \frac{37}{80})$, $(-1.23458, -\frac{27}{80})$, $(0.796921, -\frac{3}{16})$, $(-0.617292, \frac{1}{16})$, $(0.796921, -\frac{3}{16})$, $(-0.617292, \frac{1}{16})$\}

\item $c = 5$, $(d_i,\theta_i)$ = \{$(1., 0)$, $(0.127017, 0)$, $(-0.872983, \frac{1}{10})$, $(1.12702, \frac{1}{6})$, $(-0.872983, \frac{2}{5})$, $(1., \frac{1}{2})$, $(0.127017, \frac{1}{2})$, $(-0.872983, -\frac{2}{5})$, $(1.12702, -\frac{1}{3})$, $(-0.872983, -\frac{1}{10})$, $(1.12702, -\frac{1}{8})$, $(-0.872983, \frac{1}{8})$, $(1.12702, -\frac{11}{24})$, $(-0.872983, -\frac{19}{40})$, $(-0.872983, -\frac{11}{40})$, $(1.12702, -\frac{11}{24})$, $(-0.872983, -\frac{19}{40})$, $(-0.872983, -\frac{11}{40})$\}

\item $c = \frac{11}{2}$, $(d_i,\theta_i)$ = \{$(1., 0)$, $(0.127017, 0)$, $(-0.872983, \frac{1}{10})$, $(1.12702, \frac{1}{6})$, $(-0.872983, \frac{2}{5})$, $(1., \frac{1}{2})$, $(0.127017, \frac{1}{2})$, $(-0.872983, -\frac{2}{5})$, $(1.12702, -\frac{1}{3})$, $(-0.872983, -\frac{1}{10})$, $(1.59384, -\frac{19}{48})$, $(-1.23458, -\frac{33}{80})$, $(-1.23458, -\frac{17}{80})$, $(0.796921, -\frac{1}{16})$, $(-0.617292, \frac{3}{16})$, $(0.796921, -\frac{1}{16})$, $(-0.617292, \frac{3}{16})$\}

\item $c = 6$, $(d_i,\theta_i)$ = \{$(1., 0)$, $(0.127017, \frac{1}{2})$, $(-0.872983, \frac{1}{10})$, $(1.12702, \frac{1}{6})$, $(-0.872983, \frac{2}{5})$, $(1., \frac{1}{2})$, $(0.127017, 0)$, $(-0.872983, -\frac{2}{5})$, $(1.12702, -\frac{1}{3})$, $(-0.872983, -\frac{1}{10})$, $(-0.872983, \frac{1}{4})$, $(1.12702, 0)$, $(1.12702, -\frac{1}{3})$, $(-0.872983, -\frac{7}{20})$, $(-0.872983, -\frac{3}{20})$, $(1.12702, -\frac{1}{3})$, $(-0.872983, -\frac{7}{20})$, $(-0.872983, -\frac{3}{20})$\}

\item $c = \frac{13}{2}$, $(d_i,\theta_i)$ = \{$(1., 0)$, $(0.127017, \frac{1}{2})$, $(-0.872983, \frac{1}{10})$, $(1.12702, \frac{1}{6})$, $(-0.872983, \frac{2}{5})$, $(1., \frac{1}{2})$, $(0.127017, 0)$, $(-0.872983, -\frac{2}{5})$, $(1.12702, -\frac{1}{3})$, $(-0.872983, -\frac{1}{10})$, $(1.59384, -\frac{13}{48})$, $(-1.23458, -\frac{23}{80})$, $(-1.23458, -\frac{7}{80})$, $(0.796921, \frac{1}{16})$, $(-0.617292, \frac{5}{16})$, $(0.796921, \frac{1}{16})$, $(-0.617292, \frac{5}{16})$\}

\item $c = 7$, $(d_i,\theta_i)$ = \{$(1., 0)$, $(0.127017, \frac{1}{2})$, $(-0.872983, \frac{1}{10})$, $(1.12702, \frac{1}{6})$, $(-0.872983, \frac{2}{5})$, $(1., \frac{1}{2})$, $(0.127017, 0)$, $(-0.872983, -\frac{2}{5})$, $(1.12702, -\frac{1}{3})$, $(-0.872983, -\frac{1}{10})$, $(-0.872983, \frac{3}{8})$, $(1.12702, \frac{1}{8})$, $(1.12702, -\frac{5}{24})$, $(-0.872983, -\frac{9}{40})$, $(-0.872983, -\frac{1}{40})$, $(1.12702, -\frac{5}{24})$, $(-0.872983, -\frac{9}{40})$, $(-0.872983, -\frac{1}{40})$\}

\item $c = \frac{15}{2}$, $(d_i,\theta_i)$ = \{$(1., 0)$, $(0.127017, \frac{1}{2})$, $(-0.872983, \frac{1}{10})$, $(1.12702, \frac{1}{6})$, $(-0.872983, \frac{2}{5})$, $(1., \frac{1}{2})$, $(0.127017, 0)$, $(-0.872983, -\frac{2}{5})$, $(1.12702, -\frac{1}{3})$, $(-0.872983, -\frac{1}{10})$, $(1.59384, -\frac{7}{48})$, $(-1.23458, \frac{3}{80})$, $(-1.23458, -\frac{13}{80})$, $(-0.617292, \frac{7}{16})$, $(0.796921, \frac{3}{16})$, $(-0.617292, \frac{7}{16})$, $(0.796921, \frac{3}{16})$\}

\end{enumerate}

\paragraph*{Rank 10; \#64}

\begin{enumerate}

\item $c = 0$, $(d_i,\theta_i)$ = \{$(1., 0)$, $(0.127017, 0)$, $(1.12702, \frac{1}{6})$, $(-0.872983, \frac{1}{5})$, $(-0.872983, -\frac{1}{5})$, $(1., \frac{1}{2})$, $(0.127017, \frac{1}{2})$, $(1.12702, -\frac{1}{3})$, $(-0.872983, -\frac{3}{10})$, $(-0.872983, \frac{3}{10})$, $(1.12702, \frac{1}{4})$, $(-0.872983, 0)$, $(1.12702, -\frac{1}{12})$, $(-0.872983, \frac{1}{5})$, $(-0.872983, -\frac{1}{5})$, $(1.12702, -\frac{1}{12})$, $(-0.872983, \frac{1}{5})$, $(-0.872983, -\frac{1}{5})$\}

\item $c = \frac{1}{2}$, $(d_i,\theta_i)$ = \{$(1., 0)$, $(0.127017, 0)$, $(1.12702, \frac{1}{6})$, $(-0.872983, \frac{1}{5})$, $(-0.872983, -\frac{1}{5})$, $(1., \frac{1}{2})$, $(0.127017, \frac{1}{2})$, $(1.12702, -\frac{1}{3})$, $(-0.872983, -\frac{3}{10})$, $(-0.872983, \frac{3}{10})$, $(1.59384, -\frac{1}{48})$, $(-1.23458, -\frac{11}{80})$, $(-1.23458, \frac{21}{80})$, $(-0.617292, \frac{1}{16})$, $(0.796921, \frac{5}{16})$, $(-0.617292, \frac{1}{16})$, $(0.796921, \frac{5}{16})$\}

\item $c = 1$, $(d_i,\theta_i)$ = \{$(1., 0)$, $(0.127017, 0)$, $(1.12702, \frac{1}{6})$, $(-0.872983, \frac{1}{5})$, $(-0.872983, -\frac{1}{5})$, $(1., \frac{1}{2})$, $(0.127017, \frac{1}{2})$, $(1.12702, -\frac{1}{3})$, $(-0.872983, -\frac{3}{10})$, $(-0.872983, \frac{3}{10})$, $(1.12702, \frac{3}{8})$, $(-0.872983, \frac{1}{8})$, $(1.12702, \frac{1}{24})$, $(-0.872983, \frac{13}{40})$, $(-0.872983, -\frac{3}{40})$, $(1.12702, \frac{1}{24})$, $(-0.872983, \frac{13}{40})$, $(-0.872983, -\frac{3}{40})$\}

\item $c = \frac{3}{2}$, $(d_i,\theta_i)$ = \{$(1., 0)$, $(0.127017, 0)$, $(1.12702, \frac{1}{6})$, $(-0.872983, \frac{1}{5})$, $(-0.872983, \frac{3}{10})$, $(1., \frac{1}{2})$, $(0.127017, \frac{1}{2})$, $(1.12702, -\frac{1}{3})$, $(-0.872983, -\frac{3}{10})$, $(-0.872983, -\frac{1}{5})$, $(1.59384, \frac{5}{48})$, $(-1.23458, -\frac{1}{80})$, $(-1.23458, \frac{31}{80})$, $(-0.617292, \frac{3}{16})$, $(0.796921, \frac{7}{16})$, $(-0.617292, \frac{3}{16})$, $(0.796921, \frac{7}{16})$\}

\item $c = 2$, $(d_i,\theta_i)$ = \{$(1., 0)$, $(0.127017, 0)$, $(1.12702, \frac{1}{6})$, $(-0.872983, \frac{1}{5})$, $(-0.872983, \frac{3}{10})$, $(1., \frac{1}{2})$, $(0.127017, \frac{1}{2})$, $(1.12702, -\frac{1}{3})$, $(-0.872983, -\frac{3}{10})$, $(-0.872983, -\frac{1}{5})$, $(1.12702, \frac{1}{2})$, $(-0.872983, \frac{1}{4})$, $(1.12702, \frac{1}{6})$, $(-0.872983, \frac{9}{20})$, $(-0.872983, \frac{1}{20})$, $(1.12702, \frac{1}{6})$, $(-0.872983, \frac{9}{20})$, $(-0.872983, \frac{1}{20})$\}

\item $c = \frac{5}{2}$, $(d_i,\theta_i)$ = \{$(1., 0)$, $(0.127017, 0)$, $(1.12702, \frac{1}{6})$, $(-0.872983, \frac{1}{5})$, $(-0.872983, \frac{3}{10})$, $(1., \frac{1}{2})$, $(0.127017, \frac{1}{2})$, $(1.12702, -\frac{1}{3})$, $(-0.872983, -\frac{3}{10})$, $(-0.872983, -\frac{1}{5})$, $(1.59384, \frac{11}{48})$, $(-1.23458, -\frac{39}{80})$, $(-1.23458, \frac{9}{80})$, $(0.796921, -\frac{7}{16})$, $(-0.617292, \frac{5}{16})$, $(0.796921, -\frac{7}{16})$, $(-0.617292, \frac{5}{16})$\}

\item $c = 3$, $(d_i,\theta_i)$ = \{$(1., 0)$, $(0.127017, 0)$, $(1.12702, \frac{1}{6})$, $(-0.872983, \frac{1}{5})$, $(-0.872983, \frac{3}{10})$, $(1., \frac{1}{2})$, $(0.127017, \frac{1}{2})$, $(1.12702, -\frac{1}{3})$, $(-0.872983, -\frac{3}{10})$, $(-0.872983, -\frac{1}{5})$, $(1.12702, -\frac{3}{8})$, $(-0.872983, \frac{3}{8})$, $(1.12702, \frac{7}{24})$, $(-0.872983, -\frac{17}{40})$, $(-0.872983, \frac{7}{40})$, $(1.12702, \frac{7}{24})$, $(-0.872983, -\frac{17}{40})$, $(-0.872983, \frac{7}{40})$\}

\item $c = \frac{7}{2}$, $(d_i,\theta_i)$ = \{$(1., 0)$, $(0.127017, 0)$, $(1.12702, \frac{1}{6})$, $(-0.872983, \frac{1}{5})$, $(-0.872983, \frac{3}{10})$, $(1., \frac{1}{2})$, $(0.127017, \frac{1}{2})$, $(1.12702, -\frac{1}{3})$, $(-0.872983, -\frac{3}{10})$, $(-0.872983, -\frac{1}{5})$, $(1.59384, \frac{17}{48})$, $(-1.23458, -\frac{29}{80})$, $(-1.23458, \frac{19}{80})$, $(-0.617292, \frac{7}{16})$, $(0.796921, -\frac{5}{16})$, $(-0.617292, \frac{7}{16})$, $(0.796921, -\frac{5}{16})$\}

\item $c = 4$, $(d_i,\theta_i)$ = \{$(1., 0)$, $(0.127017, 0)$, $(1.12702, \frac{1}{6})$, $(-0.872983, \frac{1}{5})$, $(-0.872983, \frac{3}{10})$, $(1., \frac{1}{2})$, $(0.127017, \frac{1}{2})$, $(1.12702, -\frac{1}{3})$, $(-0.872983, -\frac{3}{10})$, $(-0.872983, -\frac{1}{5})$, $(1.12702, -\frac{1}{4})$, $(-0.872983, \frac{1}{2})$, $(1.12702, \frac{5}{12})$, $(-0.872983, -\frac{3}{10})$, $(-0.872983, \frac{3}{10})$, $(1.12702, \frac{5}{12})$, $(-0.872983, -\frac{3}{10})$, $(-0.872983, \frac{3}{10})$\}

\item $c = \frac{9}{2}$, $(d_i,\theta_i)$ = \{$(1., 0)$, $(0.127017, 0)$, $(1.12702, \frac{1}{6})$, $(-0.872983, \frac{1}{5})$, $(-0.872983, \frac{3}{10})$, $(1., \frac{1}{2})$, $(0.127017, \frac{1}{2})$, $(1.12702, -\frac{1}{3})$, $(-0.872983, -\frac{3}{10})$, $(-0.872983, -\frac{1}{5})$, $(1.59384, \frac{23}{48})$, $(-1.23458, -\frac{19}{80})$, $(-1.23458, \frac{29}{80})$, $(-0.617292, -\frac{7}{16})$, $(0.796921, -\frac{3}{16})$, $(-0.617292, -\frac{7}{16})$, $(0.796921, -\frac{3}{16})$\}

\item $c = 5$, $(d_i,\theta_i)$ = \{$(1., 0)$, $(0.127017, 0)$, $(1.12702, \frac{1}{6})$, $(-0.872983, \frac{1}{5})$, $(-0.872983, \frac{3}{10})$, $(1., \frac{1}{2})$, $(0.127017, \frac{1}{2})$, $(1.12702, -\frac{1}{3})$, $(-0.872983, -\frac{3}{10})$, $(-0.872983, -\frac{1}{5})$, $(1.12702, -\frac{1}{8})$, $(-0.872983, -\frac{3}{8})$, $(1.12702, -\frac{11}{24})$, $(-0.872983, -\frac{7}{40})$, $(-0.872983, \frac{17}{40})$, $(1.12702, -\frac{11}{24})$, $(-0.872983, -\frac{7}{40})$, $(-0.872983, \frac{17}{40})$\}

\item $c = \frac{11}{2}$, $(d_i,\theta_i)$ = \{$(1., 0)$, $(0.127017, 0)$, $(1.12702, \frac{1}{6})$, $(-0.872983, \frac{1}{5})$, $(-0.872983, \frac{3}{10})$, $(1., \frac{1}{2})$, $(0.127017, \frac{1}{2})$, $(1.12702, -\frac{1}{3})$, $(-0.872983, -\frac{3}{10})$, $(-0.872983, -\frac{1}{5})$, $(1.59384, -\frac{19}{48})$, $(-1.23458, -\frac{9}{80})$, $(-1.23458, \frac{39}{80})$, $(-0.617292, -\frac{5}{16})$, $(0.796921, -\frac{1}{16})$, $(-0.617292, -\frac{5}{16})$, $(0.796921, -\frac{1}{16})$\}

\item $c = 6$, $(d_i,\theta_i)$ = \{$(1., 0)$, $(0.127017, \frac{1}{2})$, $(1.12702, \frac{1}{6})$, $(-0.872983, \frac{1}{5})$, $(-0.872983, \frac{3}{10})$, $(1., \frac{1}{2})$, $(0.127017, 0)$, $(1.12702, -\frac{1}{3})$, $(-0.872983, -\frac{3}{10})$, $(-0.872983, -\frac{1}{5})$, $(1.12702, 0)$, $(-0.872983, -\frac{1}{4})$, $(1.12702, -\frac{1}{3})$, $(-0.872983, -\frac{1}{20})$, $(-0.872983, -\frac{9}{20})$, $(1.12702, -\frac{1}{3})$, $(-0.872983, -\frac{1}{20})$, $(-0.872983, -\frac{9}{20})$\}

\item $c = \frac{13}{2}$, $(d_i,\theta_i)$ = \{$(1., 0)$, $(0.127017, \frac{1}{2})$, $(1.12702, \frac{1}{6})$, $(-0.872983, \frac{1}{5})$, $(-0.872983, \frac{3}{10})$, $(1., \frac{1}{2})$, $(0.127017, 0)$, $(1.12702, -\frac{1}{3})$, $(-0.872983, -\frac{3}{10})$, $(-0.872983, -\frac{1}{5})$, $(1.59384, -\frac{13}{48})$, $(-1.23458, -\frac{31}{80})$, $(-1.23458, \frac{1}{80})$, $(0.796921, \frac{1}{16})$, $(-0.617292, -\frac{3}{16})$, $(0.796921, \frac{1}{16})$, $(-0.617292, -\frac{3}{16})$\}

\item $c = 7$, $(d_i,\theta_i)$ = \{$(1., 0)$, $(0.127017, \frac{1}{2})$, $(1.12702, \frac{1}{6})$, $(-0.872983, \frac{1}{5})$, $(-0.872983, \frac{3}{10})$, $(1., \frac{1}{2})$, $(0.127017, 0)$, $(1.12702, -\frac{1}{3})$, $(-0.872983, -\frac{3}{10})$, $(-0.872983, -\frac{1}{5})$, $(-0.872983, -\frac{1}{8})$, $(1.12702, \frac{1}{8})$, $(1.12702, -\frac{5}{24})$, $(-0.872983, \frac{3}{40})$, $(-0.872983, -\frac{13}{40})$, $(1.12702, -\frac{5}{24})$, $(-0.872983, \frac{3}{40})$, $(-0.872983, -\frac{13}{40})$\}

\item $c = \frac{15}{2}$, $(d_i,\theta_i)$ = \{$(1., 0)$, $(0.127017, \frac{1}{2})$, $(1.12702, \frac{1}{6})$, $(-0.872983, \frac{1}{5})$, $(-0.872983, \frac{3}{10})$, $(1., \frac{1}{2})$, $(0.127017, 0)$, $(1.12702, -\frac{1}{3})$, $(-0.872983, -\frac{3}{10})$, $(-0.872983, -\frac{1}{5})$, $(1.59384, -\frac{7}{48})$, $(-1.23458, -\frac{21}{80})$, $(-1.23458, \frac{11}{80})$, $(-0.617292, -\frac{1}{16})$, $(0.796921, \frac{3}{16})$, $(-0.617292, -\frac{1}{16})$, $(0.796921, \frac{3}{16})$\}

\end{enumerate}

\paragraph*{Rank 10; \#65}

\begin{enumerate}

\item $c = 0$, $(d_i,\theta_i)$ = \{$(1., 0)$, $(0.127017, 0)$, $(-0.872983, \frac{1}{10})$, $(1.12702, -\frac{1}{6})$, $(-0.872983, -\frac{1}{10})$, $(1., \frac{1}{2})$, $(0.127017, \frac{1}{2})$, $(-0.872983, -\frac{2}{5})$, $(1.12702, \frac{1}{3})$, $(-0.872983, \frac{2}{5})$, $(-0.872983, \frac{1}{2})$, $(1.12702, -\frac{1}{4})$, $(1.12702, \frac{1}{12})$, $(-0.872983, -\frac{1}{10})$, $(-0.872983, \frac{1}{10})$, $(1.12702, \frac{1}{12})$, $(-0.872983, -\frac{1}{10})$, $(-0.872983, \frac{1}{10})$\}

\item $c = \frac{1}{2}$, $(d_i,\theta_i)$ = \{$(1., 0)$, $(0.127017, 0)$, $(-0.872983, \frac{1}{10})$, $(1.12702, -\frac{1}{6})$, $(-0.872983, -\frac{1}{10})$, $(1., \frac{1}{2})$, $(0.127017, \frac{1}{2})$, $(-0.872983, -\frac{2}{5})$, $(1.12702, \frac{1}{3})$, $(-0.872983, \frac{2}{5})$, $(1.59384, \frac{7}{48})$, $(-1.23458, \frac{13}{80})$, $(-1.23458, -\frac{3}{80})$, $(0.796921, -\frac{3}{16})$, $(-0.617292, -\frac{7}{16})$, $(0.796921, -\frac{3}{16})$, $(-0.617292, -\frac{7}{16})$\}

\item $c = 1$, $(d_i,\theta_i)$ = \{$(1., 0)$, $(0.127017, 0)$, $(-0.872983, \frac{1}{10})$, $(1.12702, -\frac{1}{6})$, $(-0.872983, -\frac{1}{10})$, $(1., \frac{1}{2})$, $(0.127017, \frac{1}{2})$, $(-0.872983, -\frac{2}{5})$, $(1.12702, \frac{1}{3})$, $(-0.872983, \frac{2}{5})$, $(-0.872983, -\frac{3}{8})$, $(1.12702, -\frac{1}{8})$, $(1.12702, \frac{5}{24})$, $(-0.872983, \frac{1}{40})$, $(-0.872983, \frac{9}{40})$, $(1.12702, \frac{5}{24})$, $(-0.872983, \frac{1}{40})$, $(-0.872983, \frac{9}{40})$\}

\item $c = \frac{3}{2}$, $(d_i,\theta_i)$ = \{$(1., 0)$, $(0.127017, 0)$, $(-0.872983, \frac{1}{10})$, $(1.12702, -\frac{1}{6})$, $(-0.872983, -\frac{1}{10})$, $(1., \frac{1}{2})$, $(0.127017, \frac{1}{2})$, $(-0.872983, -\frac{2}{5})$, $(1.12702, \frac{1}{3})$, $(-0.872983, \frac{2}{5})$, $(1.59384, \frac{13}{48})$, $(-1.23458, \frac{23}{80})$, $(-1.23458, \frac{7}{80})$, $(-0.617292, -\frac{5}{16})$, $(0.796921, -\frac{1}{16})$, $(-0.617292, -\frac{5}{16})$, $(0.796921, -\frac{1}{16})$\}

\item $c = 2$, $(d_i,\theta_i)$ = \{$(1., 0)$, $(0.127017, 0)$, $(-0.872983, \frac{1}{10})$, $(1.12702, \frac{1}{3})$, $(-0.872983, -\frac{1}{10})$, $(1., \frac{1}{2})$, $(0.127017, \frac{1}{2})$, $(-0.872983, -\frac{2}{5})$, $(1.12702, -\frac{1}{6})$, $(-0.872983, \frac{2}{5})$, $(-0.872983, -\frac{1}{4})$, $(1.12702, 0)$, $(1.12702, \frac{1}{3})$, $(-0.872983, \frac{3}{20})$, $(-0.872983, \frac{7}{20})$, $(1.12702, \frac{1}{3})$, $(-0.872983, \frac{3}{20})$, $(-0.872983, \frac{7}{20})$\}

\item $c = \frac{5}{2}$, $(d_i,\theta_i)$ = \{$(1., 0)$, $(0.127017, 0)$, $(-0.872983, \frac{1}{10})$, $(1.12702, \frac{1}{3})$, $(-0.872983, -\frac{1}{10})$, $(1., \frac{1}{2})$, $(0.127017, \frac{1}{2})$, $(-0.872983, -\frac{2}{5})$, $(1.12702, -\frac{1}{6})$, $(-0.872983, \frac{2}{5})$, $(1.59384, \frac{19}{48})$, $(-1.23458, \frac{17}{80})$, $(-1.23458, \frac{33}{80})$, $(-0.617292, -\frac{3}{16})$, $(0.796921, \frac{1}{16})$, $(-0.617292, -\frac{3}{16})$, $(0.796921, \frac{1}{16})$\}

\item $c = 3$, $(d_i,\theta_i)$ = \{$(1., 0)$, $(0.127017, 0)$, $(-0.872983, \frac{1}{10})$, $(1.12702, \frac{1}{3})$, $(-0.872983, -\frac{1}{10})$, $(1., \frac{1}{2})$, $(0.127017, \frac{1}{2})$, $(-0.872983, -\frac{2}{5})$, $(1.12702, -\frac{1}{6})$, $(-0.872983, \frac{2}{5})$, $(1.12702, \frac{1}{8})$, $(-0.872983, -\frac{1}{8})$, $(1.12702, \frac{11}{24})$, $(-0.872983, \frac{11}{40})$, $(-0.872983, \frac{19}{40})$, $(1.12702, \frac{11}{24})$, $(-0.872983, \frac{11}{40})$, $(-0.872983, \frac{19}{40})$\}

\item $c = \frac{7}{2}$, $(d_i,\theta_i)$ = \{$(1., 0)$, $(0.127017, 0)$, $(-0.872983, \frac{1}{10})$, $(1.12702, \frac{1}{3})$, $(-0.872983, -\frac{1}{10})$, $(1., \frac{1}{2})$, $(0.127017, \frac{1}{2})$, $(-0.872983, -\frac{2}{5})$, $(1.12702, -\frac{1}{6})$, $(-0.872983, \frac{2}{5})$, $(1.59384, -\frac{23}{48})$, $(-1.23458, -\frac{37}{80})$, $(-1.23458, \frac{27}{80})$, $(-0.617292, -\frac{1}{16})$, $(0.796921, \frac{3}{16})$, $(-0.617292, -\frac{1}{16})$, $(0.796921, \frac{3}{16})$\}

\item $c = 4$, $(d_i,\theta_i)$ = \{$(1., 0)$, $(0.127017, 0)$, $(-0.872983, \frac{1}{10})$, $(1.12702, \frac{1}{3})$, $(-0.872983, \frac{2}{5})$, $(1., \frac{1}{2})$, $(0.127017, \frac{1}{2})$, $(-0.872983, -\frac{2}{5})$, $(1.12702, -\frac{1}{6})$, $(-0.872983, -\frac{1}{10})$, $(-0.872983, 0)$, $(1.12702, \frac{1}{4})$, $(1.12702, -\frac{5}{12})$, $(-0.872983, \frac{2}{5})$, $(-0.872983, -\frac{2}{5})$, $(1.12702, -\frac{5}{12})$, $(-0.872983, \frac{2}{5})$, $(-0.872983, -\frac{2}{5})$\}

\item $c = \frac{9}{2}$, $(d_i,\theta_i)$ = \{$(1., 0)$, $(0.127017, 0)$, $(-0.872983, \frac{1}{10})$, $(1.12702, \frac{1}{3})$, $(-0.872983, \frac{2}{5})$, $(1., \frac{1}{2})$, $(0.127017, \frac{1}{2})$, $(-0.872983, -\frac{2}{5})$, $(1.12702, -\frac{1}{6})$, $(-0.872983, -\frac{1}{10})$, $(1.59384, -\frac{17}{48})$, $(-1.23458, \frac{37}{80})$, $(-1.23458, -\frac{27}{80})$, $(0.796921, \frac{5}{16})$, $(-0.617292, \frac{1}{16})$, $(0.796921, \frac{5}{16})$, $(-0.617292, \frac{1}{16})$\}

\item $c = 5$, $(d_i,\theta_i)$ = \{$(1., 0)$, $(0.127017, 0)$, $(-0.872983, \frac{1}{10})$, $(1.12702, \frac{1}{3})$, $(-0.872983, \frac{2}{5})$, $(1., \frac{1}{2})$, $(0.127017, \frac{1}{2})$, $(-0.872983, -\frac{2}{5})$, $(1.12702, -\frac{1}{6})$, $(-0.872983, -\frac{1}{10})$, $(-0.872983, \frac{1}{8})$, $(1.12702, \frac{3}{8})$, $(1.12702, -\frac{7}{24})$, $(-0.872983, -\frac{19}{40})$, $(-0.872983, -\frac{11}{40})$, $(1.12702, -\frac{7}{24})$, $(-0.872983, -\frac{19}{40})$, $(-0.872983, -\frac{11}{40})$\}

\item $c = \frac{11}{2}$, $(d_i,\theta_i)$ = \{$(1., 0)$, $(0.127017, 0)$, $(-0.872983, \frac{1}{10})$, $(1.12702, \frac{1}{3})$, $(-0.872983, \frac{2}{5})$, $(1., \frac{1}{2})$, $(0.127017, \frac{1}{2})$, $(-0.872983, -\frac{2}{5})$, $(1.12702, -\frac{1}{6})$, $(-0.872983, -\frac{1}{10})$, $(1.59384, -\frac{11}{48})$, $(-1.23458, -\frac{17}{80})$, $(-1.23458, -\frac{33}{80})$, $(-0.617292, \frac{3}{16})$, $(0.796921, \frac{7}{16})$, $(-0.617292, \frac{3}{16})$, $(0.796921, \frac{7}{16})$\}

\item $c = 6$, $(d_i,\theta_i)$ = \{$(1., 0)$, $(0.127017, \frac{1}{2})$, $(-0.872983, \frac{1}{10})$, $(1.12702, \frac{1}{3})$, $(-0.872983, \frac{2}{5})$, $(1., \frac{1}{2})$, $(0.127017, 0)$, $(-0.872983, -\frac{2}{5})$, $(1.12702, -\frac{1}{6})$, $(-0.872983, -\frac{1}{10})$, $(1.12702, \frac{1}{2})$, $(-0.872983, \frac{1}{4})$, $(1.12702, -\frac{1}{6})$, $(-0.872983, -\frac{7}{20})$, $(-0.872983, -\frac{3}{20})$, $(1.12702, -\frac{1}{6})$, $(-0.872983, -\frac{7}{20})$, $(-0.872983, -\frac{3}{20})$\}

\item $c = \frac{13}{2}$, $(d_i,\theta_i)$ = \{$(1., 0)$, $(0.127017, \frac{1}{2})$, $(-0.872983, \frac{1}{10})$, $(1.12702, \frac{1}{3})$, $(-0.872983, \frac{2}{5})$, $(1., \frac{1}{2})$, $(0.127017, 0)$, $(-0.872983, -\frac{2}{5})$, $(1.12702, -\frac{1}{6})$, $(-0.872983, -\frac{1}{10})$, $(1.59384, -\frac{5}{48})$, $(-1.23458, -\frac{23}{80})$, $(-1.23458, -\frac{7}{80})$, $(-0.617292, \frac{5}{16})$, $(0.796921, -\frac{7}{16})$, $(-0.617292, \frac{5}{16})$, $(0.796921, -\frac{7}{16})$\}

\item $c = 7$, $(d_i,\theta_i)$ = \{$(1., 0)$, $(0.127017, \frac{1}{2})$, $(-0.872983, \frac{1}{10})$, $(1.12702, \frac{1}{3})$, $(-0.872983, \frac{2}{5})$, $(1., \frac{1}{2})$, $(0.127017, 0)$, $(-0.872983, -\frac{2}{5})$, $(1.12702, -\frac{1}{6})$, $(-0.872983, -\frac{1}{10})$, $(-0.872983, \frac{3}{8})$, $(1.12702, -\frac{3}{8})$, $(1.12702, -\frac{1}{24})$, $(-0.872983, -\frac{9}{40})$, $(-0.872983, -\frac{1}{40})$, $(1.12702, -\frac{1}{24})$, $(-0.872983, -\frac{9}{40})$, $(-0.872983, -\frac{1}{40})$\}

\item $c = \frac{15}{2}$, $(d_i,\theta_i)$ = \{$(1., 0)$, $(0.127017, \frac{1}{2})$, $(-0.872983, \frac{1}{10})$, $(1.12702, \frac{1}{3})$, $(-0.872983, \frac{2}{5})$, $(1., \frac{1}{2})$, $(0.127017, 0)$, $(-0.872983, -\frac{2}{5})$, $(1.12702, -\frac{1}{6})$, $(-0.872983, -\frac{1}{10})$, $(1.59384, \frac{1}{48})$, $(-1.23458, \frac{3}{80})$, $(-1.23458, -\frac{13}{80})$, $(-0.617292, \frac{7}{16})$, $(0.796921, -\frac{5}{16})$, $(-0.617292, \frac{7}{16})$, $(0.796921, -\frac{5}{16})$\}

\end{enumerate}

\paragraph*{Rank 10; \#66}

\begin{enumerate}

\item $c = 0$, $(d_i,\theta_i)$ = \{$(1., 0)$, $(0.127017, 0)$, $(-0.872983, \frac{1}{5})$, $(-0.872983, -\frac{1}{5})$, $(1.12702, -\frac{1}{6})$, $(1., \frac{1}{2})$, $(0.127017, \frac{1}{2})$, $(-0.872983, -\frac{3}{10})$, $(-0.872983, \frac{3}{10})$, $(1.12702, \frac{1}{3})$, $(1.12702, -\frac{1}{4})$, $(-0.872983, 0)$, $(1.12702, \frac{1}{12})$, $(-0.872983, \frac{1}{5})$, $(-0.872983, -\frac{1}{5})$, $(1.12702, \frac{1}{12})$, $(-0.872983, \frac{1}{5})$, $(-0.872983, -\frac{1}{5})$\}

\item $c = \frac{1}{2}$, $(d_i,\theta_i)$ = \{$(1., 0)$, $(0.127017, 0)$, $(-0.872983, \frac{1}{5})$, $(-0.872983, -\frac{1}{5})$, $(1.12702, -\frac{1}{6})$, $(1., \frac{1}{2})$, $(0.127017, \frac{1}{2})$, $(-0.872983, -\frac{3}{10})$, $(-0.872983, \frac{3}{10})$, $(1.12702, \frac{1}{3})$, $(1.59384, \frac{7}{48})$, $(-1.23458, -\frac{11}{80})$, $(-1.23458, \frac{21}{80})$, $(0.796921, -\frac{3}{16})$, $(-0.617292, \frac{1}{16})$, $(0.796921, -\frac{3}{16})$, $(-0.617292, \frac{1}{16})$\}

\item $c = 1$, $(d_i,\theta_i)$ = \{$(1., 0)$, $(0.127017, 0)$, $(-0.872983, \frac{1}{5})$, $(-0.872983, -\frac{1}{5})$, $(1.12702, -\frac{1}{6})$, $(1., \frac{1}{2})$, $(0.127017, \frac{1}{2})$, $(-0.872983, -\frac{3}{10})$, $(-0.872983, \frac{3}{10})$, $(1.12702, \frac{1}{3})$, $(1.12702, -\frac{1}{8})$, $(-0.872983, \frac{1}{8})$, $(1.12702, \frac{5}{24})$, $(-0.872983, \frac{13}{40})$, $(-0.872983, -\frac{3}{40})$, $(1.12702, \frac{5}{24})$, $(-0.872983, \frac{13}{40})$, $(-0.872983, -\frac{3}{40})$\}

\item $c = \frac{3}{2}$, $(d_i,\theta_i)$ = \{$(1., 0)$, $(0.127017, 0)$, $(-0.872983, \frac{1}{5})$, $(-0.872983, \frac{3}{10})$, $(1.12702, -\frac{1}{6})$, $(1., \frac{1}{2})$, $(0.127017, \frac{1}{2})$, $(-0.872983, -\frac{3}{10})$, $(-0.872983, -\frac{1}{5})$, $(1.12702, \frac{1}{3})$, $(1.59384, \frac{13}{48})$, $(-1.23458, -\frac{1}{80})$, $(-1.23458, \frac{31}{80})$, $(-0.617292, \frac{3}{16})$, $(0.796921, -\frac{1}{16})$, $(-0.617292, \frac{3}{16})$, $(0.796921, -\frac{1}{16})$\}

\item $c = 2$, $(d_i,\theta_i)$ = \{$(1., 0)$, $(0.127017, 0)$, $(-0.872983, \frac{1}{5})$, $(-0.872983, \frac{3}{10})$, $(1.12702, \frac{1}{3})$, $(1., \frac{1}{2})$, $(0.127017, \frac{1}{2})$, $(-0.872983, -\frac{3}{10})$, $(-0.872983, -\frac{1}{5})$, $(1.12702, -\frac{1}{6})$, $(1.12702, 0)$, $(-0.872983, \frac{1}{4})$, $(1.12702, \frac{1}{3})$, $(-0.872983, \frac{9}{20})$, $(-0.872983, \frac{1}{20})$, $(1.12702, \frac{1}{3})$, $(-0.872983, \frac{9}{20})$, $(-0.872983, \frac{1}{20})$\}

\item $c = \frac{5}{2}$, $(d_i,\theta_i)$ = \{$(1., 0)$, $(0.127017, 0)$, $(-0.872983, \frac{1}{5})$, $(-0.872983, \frac{3}{10})$, $(1.12702, \frac{1}{3})$, $(1., \frac{1}{2})$, $(0.127017, \frac{1}{2})$, $(-0.872983, -\frac{3}{10})$, $(-0.872983, -\frac{1}{5})$, $(1.12702, -\frac{1}{6})$, $(1.59384, \frac{19}{48})$, $(-1.23458, \frac{9}{80})$, $(-1.23458, -\frac{39}{80})$, $(0.796921, \frac{1}{16})$, $(-0.617292, \frac{5}{16})$, $(0.796921, \frac{1}{16})$, $(-0.617292, \frac{5}{16})$\}

\item $c = 3$, $(d_i,\theta_i)$ = \{$(1., 0)$, $(0.127017, 0)$, $(-0.872983, \frac{1}{5})$, $(-0.872983, \frac{3}{10})$, $(1.12702, \frac{1}{3})$, $(1., \frac{1}{2})$, $(0.127017, \frac{1}{2})$, $(-0.872983, -\frac{3}{10})$, $(-0.872983, -\frac{1}{5})$, $(1.12702, -\frac{1}{6})$, $(1.12702, \frac{1}{8})$, $(-0.872983, \frac{3}{8})$, $(1.12702, \frac{11}{24})$, $(-0.872983, -\frac{17}{40})$, $(-0.872983, \frac{7}{40})$, $(1.12702, \frac{11}{24})$, $(-0.872983, -\frac{17}{40})$, $(-0.872983, \frac{7}{40})$\}

\item $c = \frac{7}{2}$, $(d_i,\theta_i)$ = \{$(1., 0)$, $(0.127017, 0)$, $(-0.872983, \frac{1}{5})$, $(-0.872983, \frac{3}{10})$, $(1.12702, \frac{1}{3})$, $(1., \frac{1}{2})$, $(0.127017, \frac{1}{2})$, $(-0.872983, -\frac{3}{10})$, $(-0.872983, -\frac{1}{5})$, $(1.12702, -\frac{1}{6})$, $(1.59384, -\frac{23}{48})$, $(-1.23458, \frac{19}{80})$, $(-1.23458, -\frac{29}{80})$, $(0.796921, \frac{3}{16})$, $(-0.617292, \frac{7}{16})$, $(0.796921, \frac{3}{16})$, $(-0.617292, \frac{7}{16})$\}

\item $c = 4$, $(d_i,\theta_i)$ = \{$(1., 0)$, $(0.127017, 0)$, $(-0.872983, \frac{1}{5})$, $(-0.872983, \frac{3}{10})$, $(1.12702, \frac{1}{3})$, $(1., \frac{1}{2})$, $(0.127017, \frac{1}{2})$, $(-0.872983, -\frac{3}{10})$, $(-0.872983, -\frac{1}{5})$, $(1.12702, -\frac{1}{6})$, $(1.12702, \frac{1}{4})$, $(-0.872983, \frac{1}{2})$, $(1.12702, -\frac{5}{12})$, $(-0.872983, -\frac{3}{10})$, $(-0.872983, \frac{3}{10})$, $(1.12702, -\frac{5}{12})$, $(-0.872983, -\frac{3}{10})$, $(-0.872983, \frac{3}{10})$\}

\item $c = \frac{9}{2}$, $(d_i,\theta_i)$ = \{$(1., 0)$, $(0.127017, 0)$, $(-0.872983, \frac{1}{5})$, $(-0.872983, \frac{3}{10})$, $(1.12702, \frac{1}{3})$, $(1., \frac{1}{2})$, $(0.127017, \frac{1}{2})$, $(-0.872983, -\frac{3}{10})$, $(-0.872983, -\frac{1}{5})$, $(1.12702, -\frac{1}{6})$, $(1.59384, -\frac{17}{48})$, $(-1.23458, \frac{29}{80})$, $(-1.23458, -\frac{19}{80})$, $(0.796921, \frac{5}{16})$, $(-0.617292, -\frac{7}{16})$, $(0.796921, \frac{5}{16})$, $(-0.617292, -\frac{7}{16})$\}

\item $c = 5$, $(d_i,\theta_i)$ = \{$(1., 0)$, $(0.127017, 0)$, $(-0.872983, \frac{1}{5})$, $(-0.872983, \frac{3}{10})$, $(1.12702, \frac{1}{3})$, $(1., \frac{1}{2})$, $(0.127017, \frac{1}{2})$, $(-0.872983, -\frac{3}{10})$, $(-0.872983, -\frac{1}{5})$, $(1.12702, -\frac{1}{6})$, $(1.12702, \frac{3}{8})$, $(-0.872983, -\frac{3}{8})$, $(1.12702, -\frac{7}{24})$, $(-0.872983, -\frac{7}{40})$, $(-0.872983, \frac{17}{40})$, $(1.12702, -\frac{7}{24})$, $(-0.872983, -\frac{7}{40})$, $(-0.872983, \frac{17}{40})$\}

\item $c = \frac{11}{2}$, $(d_i,\theta_i)$ = \{$(1., 0)$, $(0.127017, 0)$, $(-0.872983, \frac{1}{5})$, $(-0.872983, \frac{3}{10})$, $(1.12702, \frac{1}{3})$, $(1., \frac{1}{2})$, $(0.127017, \frac{1}{2})$, $(-0.872983, -\frac{3}{10})$, $(-0.872983, -\frac{1}{5})$, $(1.12702, -\frac{1}{6})$, $(1.59384, -\frac{11}{48})$, $(-1.23458, \frac{39}{80})$, $(-1.23458, -\frac{9}{80})$, $(-0.617292, -\frac{5}{16})$, $(0.796921, \frac{7}{16})$, $(-0.617292, -\frac{5}{16})$, $(0.796921, \frac{7}{16})$\}

\item $c = 6$, $(d_i,\theta_i)$ = \{$(1., 0)$, $(0.127017, \frac{1}{2})$, $(-0.872983, \frac{1}{5})$, $(-0.872983, \frac{3}{10})$, $(1.12702, \frac{1}{3})$, $(1., \frac{1}{2})$, $(0.127017, 0)$, $(-0.872983, -\frac{3}{10})$, $(-0.872983, -\frac{1}{5})$, $(1.12702, -\frac{1}{6})$, $(1.12702, \frac{1}{2})$, $(-0.872983, -\frac{1}{4})$, $(1.12702, -\frac{1}{6})$, $(-0.872983, -\frac{1}{20})$, $(-0.872983, -\frac{9}{20})$, $(1.12702, -\frac{1}{6})$, $(-0.872983, -\frac{1}{20})$, $(-0.872983, -\frac{9}{20})$\}

\item $c = \frac{13}{2}$, $(d_i,\theta_i)$ = \{$(1., 0)$, $(0.127017, \frac{1}{2})$, $(-0.872983, \frac{1}{5})$, $(-0.872983, \frac{3}{10})$, $(1.12702, \frac{1}{3})$, $(1., \frac{1}{2})$, $(0.127017, 0)$, $(-0.872983, -\frac{3}{10})$, $(-0.872983, -\frac{1}{5})$, $(1.12702, -\frac{1}{6})$, $(1.59384, -\frac{5}{48})$, $(-1.23458, \frac{1}{80})$, $(-1.23458, -\frac{31}{80})$, $(0.796921, -\frac{7}{16})$, $(-0.617292, -\frac{3}{16})$, $(0.796921, -\frac{7}{16})$, $(-0.617292, -\frac{3}{16})$\}

\item $c = 7$, $(d_i,\theta_i)$ = \{$(1., 0)$, $(0.127017, \frac{1}{2})$, $(-0.872983, \frac{1}{5})$, $(-0.872983, \frac{3}{10})$, $(1.12702, \frac{1}{3})$, $(1., \frac{1}{2})$, $(0.127017, 0)$, $(-0.872983, -\frac{3}{10})$, $(-0.872983, -\frac{1}{5})$, $(1.12702, -\frac{1}{6})$, $(1.12702, -\frac{3}{8})$, $(-0.872983, -\frac{1}{8})$, $(1.12702, -\frac{1}{24})$, $(-0.872983, \frac{3}{40})$, $(-0.872983, -\frac{13}{40})$, $(1.12702, -\frac{1}{24})$, $(-0.872983, \frac{3}{40})$, $(-0.872983, -\frac{13}{40})$\}

\item $c = \frac{15}{2}$, $(d_i,\theta_i)$ = \{$(1., 0)$, $(0.127017, \frac{1}{2})$, $(-0.872983, \frac{1}{5})$, $(-0.872983, \frac{3}{10})$, $(1.12702, \frac{1}{3})$, $(1., \frac{1}{2})$, $(0.127017, 0)$, $(-0.872983, -\frac{3}{10})$, $(-0.872983, -\frac{1}{5})$, $(1.12702, -\frac{1}{6})$, $(1.59384, \frac{1}{48})$, $(-1.23458, -\frac{21}{80})$, $(-1.23458, \frac{11}{80})$, $(0.796921, -\frac{5}{16})$, $(-0.617292, -\frac{1}{16})$, $(0.796921, -\frac{5}{16})$, $(-0.617292, -\frac{1}{16})$\}

\end{enumerate}

\paragraph*{Rank 10; \#67}

\begin{enumerate}

\item $c = 0$, $(d_i,\theta_i)$ = \{$(1., 0)$, $(9.89898, 0)$, $(8.89898, \frac{1}{6})$, $(5.44949, \frac{1}{4})$, $(5.44949, \frac{1}{4})$, $(1., \frac{1}{2})$, $(9.89898, \frac{1}{2})$, $(8.89898, -\frac{1}{3})$, $(5.44949, -\frac{1}{4})$, $(5.44949, -\frac{1}{4})$, $(10.899, -\frac{1}{8})$, $(10.899, \frac{1}{8})$, $(8.89898, \frac{5}{12})$, $(4.44949, -\frac{1}{4})$, $(4.44949, -\frac{1}{4})$, $(8.89898, \frac{5}{12})$, $(4.44949, -\frac{1}{4})$, $(4.44949, -\frac{1}{4})$\}

\item $c = \frac{1}{2}$, $(d_i,\theta_i)$ = \{$(1., 0)$, $(9.89898, 0)$, $(8.89898, \frac{1}{6})$, $(5.44949, \frac{1}{4})$, $(5.44949, \frac{1}{4})$, $(1., \frac{1}{2})$, $(9.89898, \frac{1}{2})$, $(8.89898, -\frac{1}{3})$, $(5.44949, -\frac{1}{4})$, $(5.44949, -\frac{1}{4})$, $(12.5851, \frac{23}{48})$, $(6.29253, -\frac{3}{16})$, $(6.29253, -\frac{3}{16})$, $(7.70674, \frac{3}{16})$, $(7.70674, -\frac{1}{16})$, $(7.70674, \frac{3}{16})$, $(7.70674, -\frac{1}{16})$\}

\item $c = 1$, $(d_i,\theta_i)$ = \{$(1., 0)$, $(9.89898, 0)$, $(8.89898, \frac{1}{6})$, $(5.44949, \frac{1}{4})$, $(5.44949, \frac{1}{4})$, $(1., \frac{1}{2})$, $(9.89898, \frac{1}{2})$, $(8.89898, -\frac{1}{3})$, $(5.44949, -\frac{1}{4})$, $(5.44949, -\frac{1}{4})$, $(10.899, \frac{1}{4})$, $(10.899, 0)$, $(8.89898, -\frac{11}{24})$, $(4.44949, -\frac{1}{8})$, $(4.44949, -\frac{1}{8})$, $(8.89898, -\frac{11}{24})$, $(4.44949, -\frac{1}{8})$, $(4.44949, -\frac{1}{8})$\}

\item $c = \frac{3}{2}$, $(d_i,\theta_i)$ = \{$(1., 0)$, $(9.89898, 0)$, $(8.89898, \frac{1}{6})$, $(5.44949, \frac{1}{4})$, $(5.44949, \frac{1}{4})$, $(1., \frac{1}{2})$, $(9.89898, \frac{1}{2})$, $(8.89898, -\frac{1}{3})$, $(5.44949, -\frac{1}{4})$, $(5.44949, -\frac{1}{4})$, $(12.5851, -\frac{19}{48})$, $(6.29253, -\frac{1}{16})$, $(6.29253, -\frac{1}{16})$, $(7.70674, \frac{1}{16})$, $(7.70674, \frac{5}{16})$, $(7.70674, \frac{1}{16})$, $(7.70674, \frac{5}{16})$\}

\item $c = 2$, $(d_i,\theta_i)$ = \{$(1., 0)$, $(9.89898, 0)$, $(8.89898, \frac{1}{6})$, $(5.44949, \frac{1}{4})$, $(5.44949, \frac{1}{4})$, $(1., \frac{1}{2})$, $(9.89898, \frac{1}{2})$, $(8.89898, -\frac{1}{3})$, $(5.44949, -\frac{1}{4})$, $(5.44949, -\frac{1}{4})$, $(10.899, \frac{1}{8})$, $(10.899, \frac{3}{8})$, $(8.89898, -\frac{1}{3})$, $(4.44949, 0)$, $(4.44949, 0)$, $(8.89898, -\frac{1}{3})$, $(4.44949, 0)$, $(4.44949, 0)$\}

\item $c = \frac{5}{2}$, $(d_i,\theta_i)$ = \{$(1., 0)$, $(9.89898, 0)$, $(8.89898, \frac{1}{6})$, $(5.44949, \frac{1}{4})$, $(5.44949, \frac{1}{4})$, $(1., \frac{1}{2})$, $(9.89898, \frac{1}{2})$, $(8.89898, -\frac{1}{3})$, $(5.44949, -\frac{1}{4})$, $(5.44949, -\frac{1}{4})$, $(12.5851, -\frac{13}{48})$, $(6.29253, \frac{1}{16})$, $(6.29253, \frac{1}{16})$, $(7.70674, \frac{7}{16})$, $(7.70674, \frac{3}{16})$, $(7.70674, \frac{7}{16})$, $(7.70674, \frac{3}{16})$\}

\item $c = 3$, $(d_i,\theta_i)$ = \{$(1., 0)$, $(9.89898, 0)$, $(8.89898, \frac{1}{6})$, $(5.44949, \frac{1}{4})$, $(5.44949, \frac{1}{4})$, $(1., \frac{1}{2})$, $(9.89898, \frac{1}{2})$, $(8.89898, -\frac{1}{3})$, $(5.44949, -\frac{1}{4})$, $(5.44949, -\frac{1}{4})$, $(10.899, \frac{1}{2})$, $(10.899, \frac{1}{4})$, $(8.89898, -\frac{5}{24})$, $(4.44949, \frac{1}{8})$, $(4.44949, \frac{1}{8})$, $(8.89898, -\frac{5}{24})$, $(4.44949, \frac{1}{8})$, $(4.44949, \frac{1}{8})$\}

\item $c = \frac{7}{2}$, $(d_i,\theta_i)$ = \{$(1., 0)$, $(9.89898, 0)$, $(8.89898, \frac{1}{6})$, $(5.44949, \frac{1}{4})$, $(5.44949, \frac{1}{4})$, $(1., \frac{1}{2})$, $(9.89898, \frac{1}{2})$, $(8.89898, -\frac{1}{3})$, $(5.44949, -\frac{1}{4})$, $(5.44949, -\frac{1}{4})$, $(12.5851, -\frac{7}{48})$, $(6.29253, \frac{3}{16})$, $(6.29253, \frac{3}{16})$, $(7.70674, \frac{5}{16})$, $(7.70674, -\frac{7}{16})$, $(7.70674, \frac{5}{16})$, $(7.70674, -\frac{7}{16})$\}

\item $c = 4$, $(d_i,\theta_i)$ = \{$(1., 0)$, $(9.89898, 0)$, $(8.89898, \frac{1}{6})$, $(5.44949, \frac{1}{4})$, $(5.44949, \frac{1}{4})$, $(1., \frac{1}{2})$, $(9.89898, \frac{1}{2})$, $(8.89898, -\frac{1}{3})$, $(5.44949, -\frac{1}{4})$, $(5.44949, -\frac{1}{4})$, $(10.899, -\frac{3}{8})$, $(10.899, \frac{3}{8})$, $(8.89898, -\frac{1}{12})$, $(4.44949, \frac{1}{4})$, $(4.44949, \frac{1}{4})$, $(8.89898, -\frac{1}{12})$, $(4.44949, \frac{1}{4})$, $(4.44949, \frac{1}{4})$\}

\item $c = \frac{9}{2}$, $(d_i,\theta_i)$ = \{$(1., 0)$, $(9.89898, 0)$, $(8.89898, \frac{1}{6})$, $(5.44949, \frac{1}{4})$, $(5.44949, \frac{1}{4})$, $(1., \frac{1}{2})$, $(9.89898, \frac{1}{2})$, $(8.89898, -\frac{1}{3})$, $(5.44949, -\frac{1}{4})$, $(5.44949, -\frac{1}{4})$, $(12.5851, -\frac{1}{48})$, $(6.29253, \frac{5}{16})$, $(6.29253, \frac{5}{16})$, $(7.70674, -\frac{5}{16})$, $(7.70674, \frac{7}{16})$, $(7.70674, -\frac{5}{16})$, $(7.70674, \frac{7}{16})$\}

\item $c = 5$, $(d_i,\theta_i)$ = \{$(1., 0)$, $(9.89898, 0)$, $(8.89898, \frac{1}{6})$, $(5.44949, \frac{1}{4})$, $(5.44949, \frac{1}{4})$, $(1., \frac{1}{2})$, $(9.89898, \frac{1}{2})$, $(8.89898, -\frac{1}{3})$, $(5.44949, -\frac{1}{4})$, $(5.44949, -\frac{1}{4})$, $(10.899, \frac{1}{2})$, $(10.899, -\frac{1}{4})$, $(8.89898, \frac{1}{24})$, $(4.44949, \frac{3}{8})$, $(4.44949, \frac{3}{8})$, $(8.89898, \frac{1}{24})$, $(4.44949, \frac{3}{8})$, $(4.44949, \frac{3}{8})$\}

\item $c = \frac{11}{2}$, $(d_i,\theta_i)$ = \{$(1., 0)$, $(9.89898, 0)$, $(8.89898, \frac{1}{6})$, $(5.44949, \frac{1}{4})$, $(5.44949, \frac{1}{4})$, $(1., \frac{1}{2})$, $(9.89898, \frac{1}{2})$, $(8.89898, -\frac{1}{3})$, $(5.44949, -\frac{1}{4})$, $(5.44949, -\frac{1}{4})$, $(12.5851, \frac{5}{48})$, $(6.29253, \frac{7}{16})$, $(6.29253, \frac{7}{16})$, $(7.70674, -\frac{7}{16})$, $(7.70674, -\frac{3}{16})$, $(7.70674, -\frac{7}{16})$, $(7.70674, -\frac{3}{16})$\}

\item $c = 6$, $(d_i,\theta_i)$ = \{$(1., 0)$, $(9.89898, \frac{1}{2})$, $(8.89898, \frac{1}{6})$, $(5.44949, \frac{1}{4})$, $(5.44949, \frac{1}{4})$, $(1., \frac{1}{2})$, $(9.89898, 0)$, $(8.89898, -\frac{1}{3})$, $(5.44949, -\frac{1}{4})$, $(5.44949, -\frac{1}{4})$, $(10.899, -\frac{3}{8})$, $(10.899, -\frac{1}{8})$, $(8.89898, \frac{1}{6})$, $(4.44949, \frac{1}{2})$, $(4.44949, \frac{1}{2})$, $(8.89898, \frac{1}{6})$, $(4.44949, \frac{1}{2})$, $(4.44949, \frac{1}{2})$\}

\item $c = \frac{13}{2}$, $(d_i,\theta_i)$ = \{$(1., 0)$, $(9.89898, \frac{1}{2})$, $(8.89898, \frac{1}{6})$, $(5.44949, \frac{1}{4})$, $(5.44949, \frac{1}{4})$, $(1., \frac{1}{2})$, $(9.89898, 0)$, $(8.89898, -\frac{1}{3})$, $(5.44949, -\frac{1}{4})$, $(5.44949, -\frac{1}{4})$, $(12.5851, \frac{11}{48})$, $(6.29253, -\frac{7}{16})$, $(6.29253, -\frac{7}{16})$, $(7.70674, -\frac{1}{16})$, $(7.70674, -\frac{5}{16})$, $(7.70674, -\frac{1}{16})$, $(7.70674, -\frac{5}{16})$\}

\item $c = 7$, $(d_i,\theta_i)$ = \{$(1., 0)$, $(9.89898, \frac{1}{2})$, $(8.89898, \frac{1}{6})$, $(5.44949, \frac{1}{4})$, $(5.44949, \frac{1}{4})$, $(1., \frac{1}{2})$, $(9.89898, 0)$, $(8.89898, -\frac{1}{3})$, $(5.44949, -\frac{1}{4})$, $(5.44949, -\frac{1}{4})$, $(10.899, 0)$, $(10.899, -\frac{1}{4})$, $(8.89898, \frac{7}{24})$, $(4.44949, -\frac{3}{8})$, $(4.44949, -\frac{3}{8})$, $(8.89898, \frac{7}{24})$, $(4.44949, -\frac{3}{8})$, $(4.44949, -\frac{3}{8})$\}

\item $c = \frac{15}{2}$, $(d_i,\theta_i)$ = \{$(1., 0)$, $(9.89898, \frac{1}{2})$, $(8.89898, \frac{1}{6})$, $(5.44949, \frac{1}{4})$, $(5.44949, \frac{1}{4})$, $(1., \frac{1}{2})$, $(9.89898, 0)$, $(8.89898, -\frac{1}{3})$, $(5.44949, -\frac{1}{4})$, $(5.44949, -\frac{1}{4})$, $(12.5851, \frac{17}{48})$, $(6.29253, -\frac{5}{16})$, $(6.29253, -\frac{5}{16})$, $(7.70674, -\frac{3}{16})$, $(7.70674, \frac{1}{16})$, $(7.70674, -\frac{3}{16})$, $(7.70674, \frac{1}{16})$\}

\end{enumerate}

\paragraph*{Rank 10; \#68}

\begin{enumerate}

\item $c = 0$, $(d_i,\theta_i)$ = \{$(1., 0)$, $(9.89898, 0)$, $(5.44949, \frac{1}{4})$, $(5.44949, \frac{1}{4})$, $(8.89898, -\frac{1}{6})$, $(1., \frac{1}{2})$, $(9.89898, \frac{1}{2})$, $(5.44949, -\frac{1}{4})$, $(5.44949, -\frac{1}{4})$, $(8.89898, \frac{1}{3})$, $(10.899, -\frac{1}{8})$, $(10.899, \frac{1}{8})$, $(8.89898, -\frac{5}{12})$, $(4.44949, \frac{1}{4})$, $(4.44949, \frac{1}{4})$, $(8.89898, -\frac{5}{12})$, $(4.44949, \frac{1}{4})$, $(4.44949, \frac{1}{4})$\}

\item $c = \frac{1}{2}$, $(d_i,\theta_i)$ = \{$(1., 0)$, $(9.89898, 0)$, $(5.44949, \frac{1}{4})$, $(5.44949, \frac{1}{4})$, $(8.89898, -\frac{1}{6})$, $(1., \frac{1}{2})$, $(9.89898, \frac{1}{2})$, $(5.44949, -\frac{1}{4})$, $(5.44949, -\frac{1}{4})$, $(8.89898, \frac{1}{3})$, $(12.5851, -\frac{17}{48})$, $(6.29253, \frac{5}{16})$, $(6.29253, \frac{5}{16})$, $(7.70674, \frac{3}{16})$, $(7.70674, -\frac{1}{16})$, $(7.70674, \frac{3}{16})$, $(7.70674, -\frac{1}{16})$\}

\item $c = 1$, $(d_i,\theta_i)$ = \{$(1., 0)$, $(9.89898, 0)$, $(5.44949, \frac{1}{4})$, $(5.44949, \frac{1}{4})$, $(8.89898, -\frac{1}{6})$, $(1., \frac{1}{2})$, $(9.89898, \frac{1}{2})$, $(5.44949, -\frac{1}{4})$, $(5.44949, -\frac{1}{4})$, $(8.89898, \frac{1}{3})$, $(10.899, 0)$, $(10.899, \frac{1}{4})$, $(8.89898, -\frac{7}{24})$, $(4.44949, \frac{3}{8})$, $(4.44949, \frac{3}{8})$, $(8.89898, -\frac{7}{24})$, $(4.44949, \frac{3}{8})$, $(4.44949, \frac{3}{8})$\}

\item $c = \frac{3}{2}$, $(d_i,\theta_i)$ = \{$(1., 0)$, $(9.89898, 0)$, $(5.44949, \frac{1}{4})$, $(5.44949, \frac{1}{4})$, $(8.89898, -\frac{1}{6})$, $(1., \frac{1}{2})$, $(9.89898, \frac{1}{2})$, $(5.44949, -\frac{1}{4})$, $(5.44949, -\frac{1}{4})$, $(8.89898, \frac{1}{3})$, $(12.5851, -\frac{11}{48})$, $(6.29253, \frac{7}{16})$, $(6.29253, \frac{7}{16})$, $(7.70674, \frac{1}{16})$, $(7.70674, \frac{5}{16})$, $(7.70674, \frac{1}{16})$, $(7.70674, \frac{5}{16})$\}

\item $c = 2$, $(d_i,\theta_i)$ = \{$(1., 0)$, $(9.89898, 0)$, $(5.44949, \frac{1}{4})$, $(5.44949, \frac{1}{4})$, $(8.89898, \frac{1}{3})$, $(1., \frac{1}{2})$, $(9.89898, \frac{1}{2})$, $(5.44949, -\frac{1}{4})$, $(5.44949, -\frac{1}{4})$, $(8.89898, -\frac{1}{6})$, $(10.899, \frac{3}{8})$, $(10.899, \frac{1}{8})$, $(8.89898, -\frac{1}{6})$, $(4.44949, \frac{1}{2})$, $(4.44949, \frac{1}{2})$, $(8.89898, -\frac{1}{6})$, $(4.44949, \frac{1}{2})$, $(4.44949, \frac{1}{2})$\}

\item $c = \frac{5}{2}$, $(d_i,\theta_i)$ = \{$(1., 0)$, $(9.89898, 0)$, $(5.44949, \frac{1}{4})$, $(5.44949, \frac{1}{4})$, $(8.89898, \frac{1}{3})$, $(1., \frac{1}{2})$, $(9.89898, \frac{1}{2})$, $(5.44949, -\frac{1}{4})$, $(5.44949, -\frac{1}{4})$, $(8.89898, -\frac{1}{6})$, $(12.5851, -\frac{5}{48})$, $(6.29253, -\frac{7}{16})$, $(6.29253, -\frac{7}{16})$, $(7.70674, \frac{7}{16})$, $(7.70674, \frac{3}{16})$, $(7.70674, \frac{7}{16})$, $(7.70674, \frac{3}{16})$\}

\item $c = 3$, $(d_i,\theta_i)$ = \{$(1., 0)$, $(9.89898, 0)$, $(5.44949, \frac{1}{4})$, $(5.44949, \frac{1}{4})$, $(8.89898, \frac{1}{3})$, $(1., \frac{1}{2})$, $(9.89898, \frac{1}{2})$, $(5.44949, -\frac{1}{4})$, $(5.44949, -\frac{1}{4})$, $(8.89898, -\frac{1}{6})$, $(10.899, \frac{1}{4})$, $(10.899, \frac{1}{2})$, $(8.89898, -\frac{1}{24})$, $(4.44949, -\frac{3}{8})$, $(4.44949, -\frac{3}{8})$, $(8.89898, -\frac{1}{24})$, $(4.44949, -\frac{3}{8})$, $(4.44949, -\frac{3}{8})$\}

\item $c = \frac{7}{2}$, $(d_i,\theta_i)$ = \{$(1., 0)$, $(9.89898, 0)$, $(5.44949, \frac{1}{4})$, $(5.44949, \frac{1}{4})$, $(8.89898, \frac{1}{3})$, $(1., \frac{1}{2})$, $(9.89898, \frac{1}{2})$, $(5.44949, -\frac{1}{4})$, $(5.44949, -\frac{1}{4})$, $(8.89898, -\frac{1}{6})$, $(12.5851, \frac{1}{48})$, $(6.29253, -\frac{5}{16})$, $(6.29253, -\frac{5}{16})$, $(7.70674, \frac{5}{16})$, $(7.70674, -\frac{7}{16})$, $(7.70674, \frac{5}{16})$, $(7.70674, -\frac{7}{16})$\}

\item $c = 4$, $(d_i,\theta_i)$ = \{$(1., 0)$, $(9.89898, 0)$, $(5.44949, \frac{1}{4})$, $(5.44949, \frac{1}{4})$, $(8.89898, \frac{1}{3})$, $(1., \frac{1}{2})$, $(9.89898, \frac{1}{2})$, $(5.44949, -\frac{1}{4})$, $(5.44949, -\frac{1}{4})$, $(8.89898, -\frac{1}{6})$, $(10.899, -\frac{3}{8})$, $(10.899, \frac{3}{8})$, $(8.89898, \frac{1}{12})$, $(4.44949, -\frac{1}{4})$, $(4.44949, -\frac{1}{4})$, $(8.89898, \frac{1}{12})$, $(4.44949, -\frac{1}{4})$, $(4.44949, -\frac{1}{4})$\}

\item $c = \frac{9}{2}$, $(d_i,\theta_i)$ = \{$(1., 0)$, $(9.89898, 0)$, $(5.44949, \frac{1}{4})$, $(5.44949, \frac{1}{4})$, $(8.89898, \frac{1}{3})$, $(1., \frac{1}{2})$, $(9.89898, \frac{1}{2})$, $(5.44949, -\frac{1}{4})$, $(5.44949, -\frac{1}{4})$, $(8.89898, -\frac{1}{6})$, $(12.5851, \frac{7}{48})$, $(6.29253, -\frac{3}{16})$, $(6.29253, -\frac{3}{16})$, $(7.70674, -\frac{5}{16})$, $(7.70674, \frac{7}{16})$, $(7.70674, -\frac{5}{16})$, $(7.70674, \frac{7}{16})$\}

\item $c = 5$, $(d_i,\theta_i)$ = \{$(1., 0)$, $(9.89898, 0)$, $(5.44949, \frac{1}{4})$, $(5.44949, \frac{1}{4})$, $(8.89898, \frac{1}{3})$, $(1., \frac{1}{2})$, $(9.89898, \frac{1}{2})$, $(5.44949, -\frac{1}{4})$, $(5.44949, -\frac{1}{4})$, $(8.89898, -\frac{1}{6})$, $(10.899, -\frac{1}{4})$, $(10.899, \frac{1}{2})$, $(8.89898, \frac{5}{24})$, $(4.44949, -\frac{1}{8})$, $(4.44949, -\frac{1}{8})$, $(8.89898, \frac{5}{24})$, $(4.44949, -\frac{1}{8})$, $(4.44949, -\frac{1}{8})$\}

\item $c = \frac{11}{2}$, $(d_i,\theta_i)$ = \{$(1., 0)$, $(9.89898, 0)$, $(5.44949, \frac{1}{4})$, $(5.44949, \frac{1}{4})$, $(8.89898, \frac{1}{3})$, $(1., \frac{1}{2})$, $(9.89898, \frac{1}{2})$, $(5.44949, -\frac{1}{4})$, $(5.44949, -\frac{1}{4})$, $(8.89898, -\frac{1}{6})$, $(12.5851, \frac{13}{48})$, $(6.29253, -\frac{1}{16})$, $(6.29253, -\frac{1}{16})$, $(7.70674, -\frac{7}{16})$, $(7.70674, -\frac{3}{16})$, $(7.70674, -\frac{7}{16})$, $(7.70674, -\frac{3}{16})$\}

\item $c = 6$, $(d_i,\theta_i)$ = \{$(1., 0)$, $(9.89898, \frac{1}{2})$, $(5.44949, \frac{1}{4})$, $(5.44949, \frac{1}{4})$, $(8.89898, \frac{1}{3})$, $(1., \frac{1}{2})$, $(9.89898, 0)$, $(5.44949, -\frac{1}{4})$, $(5.44949, -\frac{1}{4})$, $(8.89898, -\frac{1}{6})$, $(10.899, -\frac{3}{8})$, $(10.899, -\frac{1}{8})$, $(8.89898, \frac{1}{3})$, $(4.44949, 0)$, $(4.44949, 0)$, $(8.89898, \frac{1}{3})$, $(4.44949, 0)$, $(4.44949, 0)$\}

\item $c = \frac{13}{2}$, $(d_i,\theta_i)$ = \{$(1., 0)$, $(9.89898, \frac{1}{2})$, $(5.44949, \frac{1}{4})$, $(5.44949, \frac{1}{4})$, $(8.89898, \frac{1}{3})$, $(1., \frac{1}{2})$, $(9.89898, 0)$, $(5.44949, -\frac{1}{4})$, $(5.44949, -\frac{1}{4})$, $(8.89898, -\frac{1}{6})$, $(12.5851, \frac{19}{48})$, $(6.29253, \frac{1}{16})$, $(6.29253, \frac{1}{16})$, $(7.70674, -\frac{1}{16})$, $(7.70674, -\frac{5}{16})$, $(7.70674, -\frac{1}{16})$, $(7.70674, -\frac{5}{16})$\}

\item $c = 7$, $(d_i,\theta_i)$ = \{$(1., 0)$, $(9.89898, \frac{1}{2})$, $(5.44949, \frac{1}{4})$, $(5.44949, \frac{1}{4})$, $(8.89898, \frac{1}{3})$, $(1., \frac{1}{2})$, $(9.89898, 0)$, $(5.44949, -\frac{1}{4})$, $(5.44949, -\frac{1}{4})$, $(8.89898, -\frac{1}{6})$, $(10.899, -\frac{1}{4})$, $(10.899, 0)$, $(8.89898, \frac{11}{24})$, $(4.44949, \frac{1}{8})$, $(4.44949, \frac{1}{8})$, $(8.89898, \frac{11}{24})$, $(4.44949, \frac{1}{8})$, $(4.44949, \frac{1}{8})$\}

\item $c = \frac{15}{2}$, $(d_i,\theta_i)$ = \{$(1., 0)$, $(9.89898, \frac{1}{2})$, $(5.44949, \frac{1}{4})$, $(5.44949, \frac{1}{4})$, $(8.89898, \frac{1}{3})$, $(1., \frac{1}{2})$, $(9.89898, 0)$, $(5.44949, -\frac{1}{4})$, $(5.44949, -\frac{1}{4})$, $(8.89898, -\frac{1}{6})$, $(12.5851, -\frac{23}{48})$, $(6.29253, \frac{3}{16})$, $(6.29253, \frac{3}{16})$, $(7.70674, -\frac{3}{16})$, $(7.70674, \frac{1}{16})$, $(7.70674, -\frac{3}{16})$, $(7.70674, \frac{1}{16})$\}

\end{enumerate}

\paragraph*{Rank 10; \#69}

\begin{enumerate}

\item $c = 0$, $(d_i,\theta_i)$ = \{$(1., 0)$, $(0.101021, 0)$, $(-0.898979, \frac{1}{6})$, $(0.55051, \frac{1}{4})$, $(0.55051, \frac{1}{4})$, $(1., \frac{1}{2})$, $(0.101021, \frac{1}{2})$, $(-0.898979, -\frac{1}{3})$, $(0.55051, -\frac{1}{4})$, $(0.55051, -\frac{1}{4})$, $(1.10102, -\frac{1}{8})$, $(1.10102, \frac{1}{8})$, $(-0.898979, -\frac{1}{12})$, $(-0.44949, \frac{1}{4})$, $(-0.44949, \frac{1}{4})$, $(-0.898979, -\frac{1}{12})$, $(-0.44949, \frac{1}{4})$, $(-0.44949, \frac{1}{4})$\}

\item $c = \frac{1}{2}$, $(d_i,\theta_i)$ = \{$(1., 0)$, $(0.101021, 0)$, $(-0.898979, \frac{1}{6})$, $(0.55051, \frac{1}{4})$, $(0.55051, \frac{1}{4})$, $(1., \frac{1}{2})$, $(0.101021, \frac{1}{2})$, $(-0.898979, -\frac{1}{3})$, $(0.55051, -\frac{1}{4})$, $(0.55051, -\frac{1}{4})$, $(-1.27135, -\frac{1}{48})$, $(-0.635674, \frac{5}{16})$, $(-0.635674, \frac{5}{16})$, $(0.778539, \frac{3}{16})$, $(0.778539, -\frac{1}{16})$, $(0.778539, \frac{3}{16})$, $(0.778539, -\frac{1}{16})$\}

\item $c = 1$, $(d_i,\theta_i)$ = \{$(1., 0)$, $(0.101021, 0)$, $(-0.898979, \frac{1}{6})$, $(0.55051, \frac{1}{4})$, $(0.55051, \frac{1}{4})$, $(1., \frac{1}{2})$, $(0.101021, \frac{1}{2})$, $(-0.898979, -\frac{1}{3})$, $(0.55051, -\frac{1}{4})$, $(0.55051, -\frac{1}{4})$, $(1.10102, 0)$, $(1.10102, \frac{1}{4})$, $(-0.898979, \frac{1}{24})$, $(-0.44949, \frac{3}{8})$, $(-0.44949, \frac{3}{8})$, $(-0.898979, \frac{1}{24})$, $(-0.44949, \frac{3}{8})$, $(-0.44949, \frac{3}{8})$\}

\item $c = \frac{3}{2}$, $(d_i,\theta_i)$ = \{$(1., 0)$, $(0.101021, 0)$, $(-0.898979, \frac{1}{6})$, $(0.55051, \frac{1}{4})$, $(0.55051, \frac{1}{4})$, $(1., \frac{1}{2})$, $(0.101021, \frac{1}{2})$, $(-0.898979, -\frac{1}{3})$, $(0.55051, -\frac{1}{4})$, $(0.55051, -\frac{1}{4})$, $(-1.27135, \frac{5}{48})$, $(-0.635674, \frac{7}{16})$, $(-0.635674, \frac{7}{16})$, $(0.778539, \frac{1}{16})$, $(0.778539, \frac{5}{16})$, $(0.778539, \frac{1}{16})$, $(0.778539, \frac{5}{16})$\}

\item $c = 2$, $(d_i,\theta_i)$ = \{$(1., 0)$, $(0.101021, 0)$, $(-0.898979, \frac{1}{6})$, $(0.55051, \frac{1}{4})$, $(0.55051, \frac{1}{4})$, $(1., \frac{1}{2})$, $(0.101021, \frac{1}{2})$, $(-0.898979, -\frac{1}{3})$, $(0.55051, -\frac{1}{4})$, $(0.55051, -\frac{1}{4})$, $(1.10102, \frac{3}{8})$, $(1.10102, \frac{1}{8})$, $(-0.898979, \frac{1}{6})$, $(-0.44949, \frac{1}{2})$, $(-0.44949, \frac{1}{2})$, $(-0.898979, \frac{1}{6})$, $(-0.44949, \frac{1}{2})$, $(-0.44949, \frac{1}{2})$\}

\item $c = \frac{5}{2}$, $(d_i,\theta_i)$ = \{$(1., 0)$, $(0.101021, 0)$, $(-0.898979, \frac{1}{6})$, $(0.55051, \frac{1}{4})$, $(0.55051, \frac{1}{4})$, $(1., \frac{1}{2})$, $(0.101021, \frac{1}{2})$, $(-0.898979, -\frac{1}{3})$, $(0.55051, -\frac{1}{4})$, $(0.55051, -\frac{1}{4})$, $(-1.27135, \frac{11}{48})$, $(-0.635674, -\frac{7}{16})$, $(-0.635674, -\frac{7}{16})$, $(0.778539, \frac{7}{16})$, $(0.778539, \frac{3}{16})$, $(0.778539, \frac{7}{16})$, $(0.778539, \frac{3}{16})$\}

\item $c = 3$, $(d_i,\theta_i)$ = \{$(1., 0)$, $(0.101021, 0)$, $(-0.898979, \frac{1}{6})$, $(0.55051, \frac{1}{4})$, $(0.55051, \frac{1}{4})$, $(1., \frac{1}{2})$, $(0.101021, \frac{1}{2})$, $(-0.898979, -\frac{1}{3})$, $(0.55051, -\frac{1}{4})$, $(0.55051, -\frac{1}{4})$, $(1.10102, \frac{1}{2})$, $(1.10102, \frac{1}{4})$, $(-0.898979, \frac{7}{24})$, $(-0.44949, -\frac{3}{8})$, $(-0.44949, -\frac{3}{8})$, $(-0.898979, \frac{7}{24})$, $(-0.44949, -\frac{3}{8})$, $(-0.44949, -\frac{3}{8})$\}

\item $c = \frac{7}{2}$, $(d_i,\theta_i)$ = \{$(1., 0)$, $(0.101021, 0)$, $(-0.898979, \frac{1}{6})$, $(0.55051, \frac{1}{4})$, $(0.55051, \frac{1}{4})$, $(1., \frac{1}{2})$, $(0.101021, \frac{1}{2})$, $(-0.898979, -\frac{1}{3})$, $(0.55051, -\frac{1}{4})$, $(0.55051, -\frac{1}{4})$, $(-1.27135, \frac{17}{48})$, $(-0.635674, -\frac{5}{16})$, $(-0.635674, -\frac{5}{16})$, $(0.778539, \frac{5}{16})$, $(0.778539, -\frac{7}{16})$, $(0.778539, \frac{5}{16})$, $(0.778539, -\frac{7}{16})$\}

\item $c = 4$, $(d_i,\theta_i)$ = \{$(1., 0)$, $(0.101021, 0)$, $(-0.898979, \frac{1}{6})$, $(0.55051, \frac{1}{4})$, $(0.55051, \frac{1}{4})$, $(1., \frac{1}{2})$, $(0.101021, \frac{1}{2})$, $(-0.898979, -\frac{1}{3})$, $(0.55051, -\frac{1}{4})$, $(0.55051, -\frac{1}{4})$, $(1.10102, \frac{3}{8})$, $(1.10102, -\frac{3}{8})$, $(-0.898979, \frac{5}{12})$, $(-0.44949, -\frac{1}{4})$, $(-0.44949, -\frac{1}{4})$, $(-0.898979, \frac{5}{12})$, $(-0.44949, -\frac{1}{4})$, $(-0.44949, -\frac{1}{4})$\}

\item $c = \frac{9}{2}$, $(d_i,\theta_i)$ = \{$(1., 0)$, $(0.101021, 0)$, $(-0.898979, \frac{1}{6})$, $(0.55051, \frac{1}{4})$, $(0.55051, \frac{1}{4})$, $(1., \frac{1}{2})$, $(0.101021, \frac{1}{2})$, $(-0.898979, -\frac{1}{3})$, $(0.55051, -\frac{1}{4})$, $(0.55051, -\frac{1}{4})$, $(-1.27135, \frac{23}{48})$, $(-0.635674, -\frac{3}{16})$, $(-0.635674, -\frac{3}{16})$, $(0.778539, -\frac{5}{16})$, $(0.778539, \frac{7}{16})$, $(0.778539, -\frac{5}{16})$, $(0.778539, \frac{7}{16})$\}

\item $c = 5$, $(d_i,\theta_i)$ = \{$(1., 0)$, $(0.101021, 0)$, $(-0.898979, \frac{1}{6})$, $(0.55051, \frac{1}{4})$, $(0.55051, \frac{1}{4})$, $(1., \frac{1}{2})$, $(0.101021, \frac{1}{2})$, $(-0.898979, -\frac{1}{3})$, $(0.55051, -\frac{1}{4})$, $(0.55051, -\frac{1}{4})$, $(1.10102, -\frac{1}{4})$, $(1.10102, \frac{1}{2})$, $(-0.898979, -\frac{11}{24})$, $(-0.44949, -\frac{1}{8})$, $(-0.44949, -\frac{1}{8})$, $(-0.898979, -\frac{11}{24})$, $(-0.44949, -\frac{1}{8})$, $(-0.44949, -\frac{1}{8})$\}

\item $c = \frac{11}{2}$, $(d_i,\theta_i)$ = \{$(1., 0)$, $(0.101021, 0)$, $(-0.898979, \frac{1}{6})$, $(0.55051, \frac{1}{4})$, $(0.55051, \frac{1}{4})$, $(1., \frac{1}{2})$, $(0.101021, \frac{1}{2})$, $(-0.898979, -\frac{1}{3})$, $(0.55051, -\frac{1}{4})$, $(0.55051, -\frac{1}{4})$, $(-1.27135, -\frac{19}{48})$, $(-0.635674, -\frac{1}{16})$, $(-0.635674, -\frac{1}{16})$, $(0.778539, -\frac{7}{16})$, $(0.778539, -\frac{3}{16})$, $(0.778539, -\frac{7}{16})$, $(0.778539, -\frac{3}{16})$\}

\item $c = 6$, $(d_i,\theta_i)$ = \{$(1., 0)$, $(0.101021, \frac{1}{2})$, $(-0.898979, \frac{1}{6})$, $(0.55051, \frac{1}{4})$, $(0.55051, \frac{1}{4})$, $(1., \frac{1}{2})$, $(0.101021, 0)$, $(-0.898979, -\frac{1}{3})$, $(0.55051, -\frac{1}{4})$, $(0.55051, -\frac{1}{4})$, $(1.10102, -\frac{3}{8})$, $(1.10102, -\frac{1}{8})$, $(-0.898979, -\frac{1}{3})$, $(-0.44949, 0)$, $(-0.44949, 0)$, $(-0.898979, -\frac{1}{3})$, $(-0.44949, 0)$, $(-0.44949, 0)$\}

\item $c = \frac{13}{2}$, $(d_i,\theta_i)$ = \{$(1., 0)$, $(0.101021, \frac{1}{2})$, $(-0.898979, \frac{1}{6})$, $(0.55051, \frac{1}{4})$, $(0.55051, \frac{1}{4})$, $(1., \frac{1}{2})$, $(0.101021, 0)$, $(-0.898979, -\frac{1}{3})$, $(0.55051, -\frac{1}{4})$, $(0.55051, -\frac{1}{4})$, $(-1.27135, -\frac{13}{48})$, $(-0.635674, \frac{1}{16})$, $(-0.635674, \frac{1}{16})$, $(0.778539, -\frac{1}{16})$, $(0.778539, -\frac{5}{16})$, $(0.778539, -\frac{1}{16})$, $(0.778539, -\frac{5}{16})$\}

\item $c = 7$, $(d_i,\theta_i)$ = \{$(1., 0)$, $(0.101021, \frac{1}{2})$, $(-0.898979, \frac{1}{6})$, $(0.55051, \frac{1}{4})$, $(0.55051, \frac{1}{4})$, $(1., \frac{1}{2})$, $(0.101021, 0)$, $(-0.898979, -\frac{1}{3})$, $(0.55051, -\frac{1}{4})$, $(0.55051, -\frac{1}{4})$, $(1.10102, 0)$, $(1.10102, -\frac{1}{4})$, $(-0.898979, -\frac{5}{24})$, $(-0.44949, \frac{1}{8})$, $(-0.44949, \frac{1}{8})$, $(-0.898979, -\frac{5}{24})$, $(-0.44949, \frac{1}{8})$, $(-0.44949, \frac{1}{8})$\}

\item $c = \frac{15}{2}$, $(d_i,\theta_i)$ = \{$(1., 0)$, $(0.101021, \frac{1}{2})$, $(-0.898979, \frac{1}{6})$, $(0.55051, \frac{1}{4})$, $(0.55051, \frac{1}{4})$, $(1., \frac{1}{2})$, $(0.101021, 0)$, $(-0.898979, -\frac{1}{3})$, $(0.55051, -\frac{1}{4})$, $(0.55051, -\frac{1}{4})$, $(-1.27135, -\frac{7}{48})$, $(-0.635674, \frac{3}{16})$, $(-0.635674, \frac{3}{16})$, $(0.778539, -\frac{3}{16})$, $(0.778539, \frac{1}{16})$, $(0.778539, -\frac{3}{16})$, $(0.778539, \frac{1}{16})$\}

\end{enumerate}

\paragraph*{Rank 10; \#70}

\begin{enumerate}

\item $c = 0$, $(d_i,\theta_i)$ = \{$(1., 0)$, $(0.101021, 0)$, $(0.55051, \frac{1}{4})$, $(0.55051, \frac{1}{4})$, $(-0.898979, -\frac{1}{6})$, $(1., \frac{1}{2})$, $(0.101021, \frac{1}{2})$, $(0.55051, -\frac{1}{4})$, $(0.55051, -\frac{1}{4})$, $(-0.898979, \frac{1}{3})$, $(1.10102, -\frac{1}{8})$, $(1.10102, \frac{1}{8})$, $(-0.898979, \frac{1}{12})$, $(-0.44949, -\frac{1}{4})$, $(-0.44949, -\frac{1}{4})$, $(-0.898979, \frac{1}{12})$, $(-0.44949, -\frac{1}{4})$, $(-0.44949, -\frac{1}{4})$\}

\item $c = \frac{1}{2}$, $(d_i,\theta_i)$ = \{$(1., 0)$, $(0.101021, 0)$, $(0.55051, \frac{1}{4})$, $(0.55051, \frac{1}{4})$, $(-0.898979, -\frac{1}{6})$, $(1., \frac{1}{2})$, $(0.101021, \frac{1}{2})$, $(0.55051, -\frac{1}{4})$, $(0.55051, -\frac{1}{4})$, $(-0.898979, \frac{1}{3})$, $(-1.27135, \frac{7}{48})$, $(-0.635674, -\frac{3}{16})$, $(-0.635674, -\frac{3}{16})$, $(0.778539, \frac{3}{16})$, $(0.778539, -\frac{1}{16})$, $(0.778539, \frac{3}{16})$, $(0.778539, -\frac{1}{16})$\}

\item $c = 1$, $(d_i,\theta_i)$ = \{$(1., 0)$, $(0.101021, 0)$, $(0.55051, \frac{1}{4})$, $(0.55051, \frac{1}{4})$, $(-0.898979, -\frac{1}{6})$, $(1., \frac{1}{2})$, $(0.101021, \frac{1}{2})$, $(0.55051, -\frac{1}{4})$, $(0.55051, -\frac{1}{4})$, $(-0.898979, \frac{1}{3})$, $(1.10102, \frac{1}{4})$, $(1.10102, 0)$, $(-0.898979, \frac{5}{24})$, $(-0.44949, -\frac{1}{8})$, $(-0.44949, -\frac{1}{8})$, $(-0.898979, \frac{5}{24})$, $(-0.44949, -\frac{1}{8})$, $(-0.44949, -\frac{1}{8})$\}

\item $c = \frac{3}{2}$, $(d_i,\theta_i)$ = \{$(1., 0)$, $(0.101021, 0)$, $(0.55051, \frac{1}{4})$, $(0.55051, \frac{1}{4})$, $(-0.898979, -\frac{1}{6})$, $(1., \frac{1}{2})$, $(0.101021, \frac{1}{2})$, $(0.55051, -\frac{1}{4})$, $(0.55051, -\frac{1}{4})$, $(-0.898979, \frac{1}{3})$, $(-1.27135, \frac{13}{48})$, $(-0.635674, -\frac{1}{16})$, $(-0.635674, -\frac{1}{16})$, $(0.778539, \frac{1}{16})$, $(0.778539, \frac{5}{16})$, $(0.778539, \frac{1}{16})$, $(0.778539, \frac{5}{16})$\}

\item $c = 2$, $(d_i,\theta_i)$ = \{$(1., 0)$, $(0.101021, 0)$, $(0.55051, \frac{1}{4})$, $(0.55051, \frac{1}{4})$, $(-0.898979, \frac{1}{3})$, $(1., \frac{1}{2})$, $(0.101021, \frac{1}{2})$, $(0.55051, -\frac{1}{4})$, $(0.55051, -\frac{1}{4})$, $(-0.898979, -\frac{1}{6})$, $(1.10102, \frac{3}{8})$, $(1.10102, \frac{1}{8})$, $(-0.898979, \frac{1}{3})$, $(-0.44949, 0)$, $(-0.44949, 0)$, $(-0.898979, \frac{1}{3})$, $(-0.44949, 0)$, $(-0.44949, 0)$\}

\item $c = \frac{5}{2}$, $(d_i,\theta_i)$ = \{$(1., 0)$, $(0.101021, 0)$, $(0.55051, \frac{1}{4})$, $(0.55051, \frac{1}{4})$, $(-0.898979, \frac{1}{3})$, $(1., \frac{1}{2})$, $(0.101021, \frac{1}{2})$, $(0.55051, -\frac{1}{4})$, $(0.55051, -\frac{1}{4})$, $(-0.898979, -\frac{1}{6})$, $(-1.27135, \frac{19}{48})$, $(-0.635674, \frac{1}{16})$, $(-0.635674, \frac{1}{16})$, $(0.778539, \frac{7}{16})$, $(0.778539, \frac{3}{16})$, $(0.778539, \frac{7}{16})$, $(0.778539, \frac{3}{16})$\}

\item $c = 3$, $(d_i,\theta_i)$ = \{$(1., 0)$, $(0.101021, 0)$, $(0.55051, \frac{1}{4})$, $(0.55051, \frac{1}{4})$, $(-0.898979, \frac{1}{3})$, $(1., \frac{1}{2})$, $(0.101021, \frac{1}{2})$, $(0.55051, -\frac{1}{4})$, $(0.55051, -\frac{1}{4})$, $(-0.898979, -\frac{1}{6})$, $(1.10102, \frac{1}{4})$, $(1.10102, \frac{1}{2})$, $(-0.898979, \frac{11}{24})$, $(-0.44949, \frac{1}{8})$, $(-0.44949, \frac{1}{8})$, $(-0.898979, \frac{11}{24})$, $(-0.44949, \frac{1}{8})$, $(-0.44949, \frac{1}{8})$\}

\item $c = \frac{7}{2}$, $(d_i,\theta_i)$ = \{$(1., 0)$, $(0.101021, 0)$, $(0.55051, \frac{1}{4})$, $(0.55051, \frac{1}{4})$, $(-0.898979, \frac{1}{3})$, $(1., \frac{1}{2})$, $(0.101021, \frac{1}{2})$, $(0.55051, -\frac{1}{4})$, $(0.55051, -\frac{1}{4})$, $(-0.898979, -\frac{1}{6})$, $(-1.27135, -\frac{23}{48})$, $(-0.635674, \frac{3}{16})$, $(-0.635674, \frac{3}{16})$, $(0.778539, \frac{5}{16})$, $(0.778539, -\frac{7}{16})$, $(0.778539, \frac{5}{16})$, $(0.778539, -\frac{7}{16})$\}

\item $c = 4$, $(d_i,\theta_i)$ = \{$(1., 0)$, $(0.101021, 0)$, $(0.55051, \frac{1}{4})$, $(0.55051, \frac{1}{4})$, $(-0.898979, \frac{1}{3})$, $(1., \frac{1}{2})$, $(0.101021, \frac{1}{2})$, $(0.55051, -\frac{1}{4})$, $(0.55051, -\frac{1}{4})$, $(-0.898979, -\frac{1}{6})$, $(1.10102, -\frac{3}{8})$, $(1.10102, \frac{3}{8})$, $(-0.898979, -\frac{5}{12})$, $(-0.44949, \frac{1}{4})$, $(-0.44949, \frac{1}{4})$, $(-0.898979, -\frac{5}{12})$, $(-0.44949, \frac{1}{4})$, $(-0.44949, \frac{1}{4})$\}

\item $c = \frac{9}{2}$, $(d_i,\theta_i)$ = \{$(1., 0)$, $(0.101021, 0)$, $(0.55051, \frac{1}{4})$, $(0.55051, \frac{1}{4})$, $(-0.898979, \frac{1}{3})$, $(1., \frac{1}{2})$, $(0.101021, \frac{1}{2})$, $(0.55051, -\frac{1}{4})$, $(0.55051, -\frac{1}{4})$, $(-0.898979, -\frac{1}{6})$, $(-1.27135, -\frac{17}{48})$, $(-0.635674, \frac{5}{16})$, $(-0.635674, \frac{5}{16})$, $(0.778539, -\frac{5}{16})$, $(0.778539, \frac{7}{16})$, $(0.778539, -\frac{5}{16})$, $(0.778539, \frac{7}{16})$\}

\item $c = 5$, $(d_i,\theta_i)$ = \{$(1., 0)$, $(0.101021, 0)$, $(0.55051, \frac{1}{4})$, $(0.55051, \frac{1}{4})$, $(-0.898979, \frac{1}{3})$, $(1., \frac{1}{2})$, $(0.101021, \frac{1}{2})$, $(0.55051, -\frac{1}{4})$, $(0.55051, -\frac{1}{4})$, $(-0.898979, -\frac{1}{6})$, $(1.10102, -\frac{1}{4})$, $(1.10102, \frac{1}{2})$, $(-0.898979, -\frac{7}{24})$, $(-0.44949, \frac{3}{8})$, $(-0.44949, \frac{3}{8})$, $(-0.898979, -\frac{7}{24})$, $(-0.44949, \frac{3}{8})$, $(-0.44949, \frac{3}{8})$\}

\item $c = \frac{11}{2}$, $(d_i,\theta_i)$ = \{$(1., 0)$, $(0.101021, 0)$, $(0.55051, \frac{1}{4})$, $(0.55051, \frac{1}{4})$, $(-0.898979, \frac{1}{3})$, $(1., \frac{1}{2})$, $(0.101021, \frac{1}{2})$, $(0.55051, -\frac{1}{4})$, $(0.55051, -\frac{1}{4})$, $(-0.898979, -\frac{1}{6})$, $(-1.27135, -\frac{11}{48})$, $(-0.635674, \frac{7}{16})$, $(-0.635674, \frac{7}{16})$, $(0.778539, -\frac{7}{16})$, $(0.778539, -\frac{3}{16})$, $(0.778539, -\frac{7}{16})$, $(0.778539, -\frac{3}{16})$\}

\item $c = 6$, $(d_i,\theta_i)$ = \{$(1., 0)$, $(0.101021, \frac{1}{2})$, $(0.55051, \frac{1}{4})$, $(0.55051, \frac{1}{4})$, $(-0.898979, \frac{1}{3})$, $(1., \frac{1}{2})$, $(0.101021, 0)$, $(0.55051, -\frac{1}{4})$, $(0.55051, -\frac{1}{4})$, $(-0.898979, -\frac{1}{6})$, $(1.10102, -\frac{1}{8})$, $(1.10102, -\frac{3}{8})$, $(-0.898979, -\frac{1}{6})$, $(-0.44949, \frac{1}{2})$, $(-0.44949, \frac{1}{2})$, $(-0.898979, -\frac{1}{6})$, $(-0.44949, \frac{1}{2})$, $(-0.44949, \frac{1}{2})$\}

\item $c = \frac{13}{2}$, $(d_i,\theta_i)$ = \{$(1., 0)$, $(0.101021, \frac{1}{2})$, $(0.55051, \frac{1}{4})$, $(0.55051, \frac{1}{4})$, $(-0.898979, \frac{1}{3})$, $(1., \frac{1}{2})$, $(0.101021, 0)$, $(0.55051, -\frac{1}{4})$, $(0.55051, -\frac{1}{4})$, $(-0.898979, -\frac{1}{6})$, $(-1.27135, -\frac{5}{48})$, $(-0.635674, -\frac{7}{16})$, $(-0.635674, -\frac{7}{16})$, $(0.778539, -\frac{1}{16})$, $(0.778539, -\frac{5}{16})$, $(0.778539, -\frac{1}{16})$, $(0.778539, -\frac{5}{16})$\}

\item $c = 7$, $(d_i,\theta_i)$ = \{$(1., 0)$, $(0.101021, \frac{1}{2})$, $(0.55051, \frac{1}{4})$, $(0.55051, \frac{1}{4})$, $(-0.898979, \frac{1}{3})$, $(1., \frac{1}{2})$, $(0.101021, 0)$, $(0.55051, -\frac{1}{4})$, $(0.55051, -\frac{1}{4})$, $(-0.898979, -\frac{1}{6})$, $(1.10102, -\frac{1}{4})$, $(1.10102, 0)$, $(-0.898979, -\frac{1}{24})$, $(-0.44949, -\frac{3}{8})$, $(-0.44949, -\frac{3}{8})$, $(-0.898979, -\frac{1}{24})$, $(-0.44949, -\frac{3}{8})$, $(-0.44949, -\frac{3}{8})$\}

\item $c = \frac{15}{2}$, $(d_i,\theta_i)$ = \{$(1., 0)$, $(0.101021, \frac{1}{2})$, $(0.55051, \frac{1}{4})$, $(0.55051, \frac{1}{4})$, $(-0.898979, \frac{1}{3})$, $(1., \frac{1}{2})$, $(0.101021, 0)$, $(0.55051, -\frac{1}{4})$, $(0.55051, -\frac{1}{4})$, $(-0.898979, -\frac{1}{6})$, $(-1.27135, \frac{1}{48})$, $(-0.635674, -\frac{5}{16})$, $(-0.635674, -\frac{5}{16})$, $(0.778539, -\frac{3}{16})$, $(0.778539, \frac{1}{16})$, $(0.778539, -\frac{3}{16})$, $(0.778539, \frac{1}{16})$\}

\end{enumerate}

\section{\label{app:Z2modular extension} List of \texorpdfstring{$\mathbb{Z}_2$}{Z\_2} modular extensions}

Below, we present lists of modular extensions of primitive $\mathbb{Z}_2$-BFCs. Some of the names of unitary $\mathbb{Z}_2$-BFCs are retrieved from Ref.~\cite{LanKonWen2017:symmetryenriched}. Non-unitary modular extensions are marked with asterisk. Similarly to the fermionic modular extensions, each $\mathbb{Z}_2$ modular extension has its partner with the opposite signs on the quantum dimensions of the added objects. For brevity, we only present one of each pair.

\paragraph*{\texorpdfstring{$3_2^{\zeta^1_2}$}{3\_2\^{zeta\^1\_2}}}

$c = 2$, $(d_i,\theta_i)$ = \{$(1., 0)$, $(1., 0)$, $(2., \frac{1}{3})$\}

\begin{enumerate}

\item $c = 2$, $(d_i,\theta_i)$ = \{$(1., 0)$, $(1., 0)$, $(2., \frac{1}{3})$, $(1.73205, \frac{1}{8})$, $(1.73205, -\frac{3}{8})$\}

\item $c = 2$, $(d_i,\theta_i)$ = \{$(1., 0)$, $(1., 0)$, $(2., \frac{1}{3})$, $(1.73205, -\frac{1}{8})$, $(1.73205, \frac{3}{8})$\}

\end{enumerate}

\paragraph*{\texorpdfstring{$3_{-2}^{\zeta^1_2}$}{3\_{-2}\^{zeta\^1\_2}}}

$c = 6$, $(d_i,\theta_i)$ = \{$(1., 0)$, $(1., 0)$, $(2., -\frac{1}{3})$\}

\begin{enumerate}

\item $c = 6$, $(d_i,\theta_i)$ = \{$(1., 0)$, $(1., 0)$, $(2., -\frac{1}{3})$, $(1.73205, \frac{3}{8})$, $(1.73205, -\frac{1}{8})$\}

\item $c = 6$, $(d_i,\theta_i)$ = \{$(1., 0)$, $(1., 0)$, $(2., -\frac{1}{3})$, $(1.73205, -\frac{3}{8})$, $(1.73205, \frac{1}{8})$\}

\end{enumerate}

\paragraph*{\texorpdfstring{$4_1^{\zeta^1_2}$}{4\_1\^{zeta\^1\_2}}}

$c = 1$, $(d_i,\theta_i)$ = \{$(1., 0)$, $(1., \frac{1}{4})$, $(1., 0)$, $(1., \frac{1}{4})$\}

\begin{enumerate}

\item $c = 1$, $(d_i,\theta_i)$ = \{$(1., 0)$, $(1., \frac{1}{4})$, $(1., 0)$, $(1., \frac{1}{4})$, $(1., \frac{5}{16})$, $(1., -\frac{3}{16})$, $(1., \frac{5}{16})$, $(1., -\frac{3}{16})$\}

\item $c = 1$, $(d_i,\theta_i)$ = \{$(1., 0)$, $(1., \frac{1}{4})$, $(1., 0)$, $(1., \frac{1}{4})$, $(1., \frac{1}{16})$, $(1., -\frac{7}{16})$, $(1., \frac{1}{16})$, $(1., -\frac{7}{16})$\}

\end{enumerate}

\paragraph*{\texorpdfstring{$4_{-1}^{\zeta^1_2}$}{4\_{-1}\^{zeta\^1\_2}}}

$c = 7$, $(d_i,\theta_i)$ = \{$(1., 0)$, $(1., -\frac{1}{4})$, $(1., 0)$, $(1., -\frac{1}{4})$\}

\begin{enumerate}

\item $c = 7$, $(d_i,\theta_i)$ = \{$(1., 0)$, $(1., -\frac{1}{4})$, $(1., 0)$, $(1., -\frac{1}{4})$, $(1., \frac{7}{16})$, $(1., -\frac{1}{16})$, $(1., -\frac{1}{16})$, $(1., \frac{7}{16})$\}

\item $c = 7$, $(d_i,\theta_i)$ = \{$(1., 0)$, $(1., -\frac{1}{4})$, $(1., 0)$, $(1., -\frac{1}{4})$, $(1., -\frac{5}{16})$, $(1., \frac{3}{16})$, $(1., -\frac{5}{16})$, $(1., \frac{3}{16})$\}

\end{enumerate}

\paragraph*{\texorpdfstring{$4_0^{\zeta^1_2}$}{4\_0\^{zeta\^1\_2}}}

$c = 0$, $(d_i,\theta_i)$ = \{$(1., 0)$, $(1., 0)$, $(2., -\frac{1}{5})$, $(2., \frac{1}{5})$\}

\begin{enumerate}

\item $c = 0$, $(d_i,\theta_i)$ = \{$(1., 0)$, $(1., 0)$, $(2., -\frac{1}{5})$, $(2., \frac{1}{5})$, $(2.23607, -\frac{1}{4})$, $(2.23607, \frac{1}{4})$\}

\item $c = 0$, $(d_i,\theta_i)$ = \{$(1., 0)$, $(1., 0)$, $(2., -\frac{1}{5})$, $(2., \frac{1}{5})$, $(2.23607, 0)$, $(2.23607, \frac{1}{2})$\}

\end{enumerate}

\paragraph*{\texorpdfstring{$4_4^{\zeta^1_2}$}{4\_4\^{zeta\^1\_2}}}

$c = 4$, $(d_i,\theta_i)$ = \{$(1., 0)$, $(1., 0)$, $(2., -\frac{2}{5})$, $(2., \frac{2}{5})$\}

\begin{enumerate}

\item $c = 4$, $(d_i,\theta_i)$ = \{$(1., 0)$, $(1., 0)$, $(2., -\frac{2}{5})$, $(2., \frac{2}{5})$, $(2.23607, 0)$, $(2.23607, \frac{1}{2})$\}

\item $c = 4$, $(d_i,\theta_i)$ = \{$(1., 0)$, $(1., 0)$, $(2., -\frac{2}{5})$, $(2., \frac{2}{5})$, $(2.23607, \frac{1}{4})$, $(2.23607, -\frac{1}{4})$\}

\end{enumerate}

\paragraph*{\texorpdfstring{$5_1^{\zeta^1_2}$}{5\_1\^{zeta\^1\_2}}}

$c = 0$, $(d_i,\theta_i)$ = \{$(1., 0)$, $(1., \frac{1}{2})$, $(1., 0)$, $(1., \frac{1}{2})$, $(2., 0)$\}

\begin{enumerate}

\item $c = 0$, $(d_i,\theta_i)$ = \{$(1., 0)$, $(1., \frac{1}{2})$, $(1., 0)$, $(1., \frac{1}{2})$, $(2., 0)$, $(1.41421, \frac{1}{16})$, $(1.41421, \frac{7}{16})$, $(1.41421, -\frac{7}{16})$, $(1.41421, -\frac{1}{16})$\}

\item $c = 0$, $(d_i,\theta_i)$ = \{$(1., 0)$, $(1., \frac{1}{2})$, $(1., 0)$, $(1., \frac{1}{2})$, $(2., 0)$, $(1.41421, \frac{3}{16})$, $(1.41421, \frac{5}{16})$, $(1.41421, -\frac{5}{16})$, $(1.41421, -\frac{3}{16})$\}

\end{enumerate}

\paragraph*{\texorpdfstring{$5_1^{\zeta^1_2*}$}{5\_1\^{zeta\^1\_2*}}}

$c = 0$, $(d_i,\theta_i)$ = \{$(1., 0)$, $(1., \frac{1}{2})$, $(1., 0)$, $(1., \frac{1}{2})$, $(-2., 0)$\}

\begin{enumerate}

\item $c = 0$, $(d_i,\theta_i)$ = \{$(1., 0)$, $(1., \frac{1}{2})$, $(1., 0)$, $(1., \frac{1}{2})$, $(-2., 0)$, $(1.41421, \frac{1}{16})$, $(-1.41421, \frac{7}{16})$, $(1.41421, -\frac{7}{16})$, $(-1.41421, -\frac{1}{16})$\}

\item $c = 0$, $(d_i,\theta_i)$ = \{$(1., 0)$, $(1., \frac{1}{2})$, $(1., 0)$, $(1., \frac{1}{2})$, $(-2., 0)$, $(1.41421, \frac{3}{16})$, $(-1.41421, \frac{5}{16})$, $(1.41421, -\frac{5}{16})$, $(-1.41421, -\frac{3}{16})$\}

\end{enumerate}

\paragraph*{\texorpdfstring{$5_1^{\zeta^1_2}$}{5\_1\^{zeta\^1\_2}}}

$c = 1$, $(d_i,\theta_i)$ = \{$(1., 0)$, $(1., \frac{1}{2})$, $(1., 0)$, $(1., \frac{1}{2})$, $(2., \frac{1}{8})$\}

\begin{enumerate}

\item $c = 1$, $(d_i,\theta_i)$ = \{$(1., 0)$, $(1., \frac{1}{2})$, $(1., 0)$, $(1., \frac{1}{2})$, $(2., \frac{1}{8})$, $(1.41421, \frac{1}{16})$, $(1.41421, -\frac{7}{16})$, $(1.41421, -\frac{7}{16})$, $(1.41421, \frac{1}{16})$\}

\item $c = 1$, $(d_i,\theta_i)$ = \{$(1., 0)$, $(1., \frac{1}{2})$, $(1., 0)$, $(1., \frac{1}{2})$, $(2., \frac{1}{8})$, $(1.41421, \frac{5}{16})$, $(1.41421, -\frac{3}{16})$, $(1.41421, \frac{5}{16})$, $(1.41421, -\frac{3}{16})$\}

\item $c = 1$, $(d_i,\theta_i)$ = \{$(1., 0)$, $(1., \frac{1}{2})$, $(1., 0)$, $(1., \frac{1}{2})$, $(2., \frac{1}{8})$, $(1.41421, \frac{7}{16})$, $(1.41421, -\frac{1}{16})$, $(1.41421, -\frac{5}{16})$, $(1.41421, \frac{3}{16})$\}

\end{enumerate}

\paragraph*{\texorpdfstring{$5_1^{\zeta^1_2*}$}{5\_1\^{zeta\^1\_2*}}}

$c = 1$, $(d_i,\theta_i)$ = \{$(1., 0)$, $(1., \frac{1}{2})$, $(1., 0)$, $(1., \frac{1}{2})$, $(-2., \frac{1}{8})$\}

\begin{enumerate}

\item $c = 1$, $(d_i,\theta_i)$ = \{$(1., 0)$, $(1., \frac{1}{2})$, $(1., 0)$, $(1., \frac{1}{2})$, $(-2., \frac{1}{8})$, $(-1.41421, \frac{7}{16})$, $(-1.41421, -\frac{1}{16})$, $(1.41421, -\frac{5}{16})$, $(1.41421, \frac{3}{16})$\}

\item $c = 1$, $(d_i,\theta_i)$ = \{$(1., 0)$, $(1., \frac{1}{2})$, $(1., 0)$, $(1., \frac{1}{2})$, $(-2., \frac{1}{8})$, $(-1.41421, \frac{1}{16})$, $(-1.41421, -\frac{7}{16})$, $(1.41421, -\frac{7}{16})$, $(1.41421, \frac{1}{16})$\}

\item $c = 1$, $(d_i,\theta_i)$ = \{$(1., 0)$, $(1., \frac{1}{2})$, $(1., 0)$, $(1., \frac{1}{2})$, $(-2., \frac{1}{8})$, $(-1.41421, \frac{5}{16})$, $(-1.41421, -\frac{3}{16})$, $(1.41421, \frac{5}{16})$, $(1.41421, -\frac{3}{16})$\}

\end{enumerate}

\paragraph*{\texorpdfstring{$5_2^{\zeta^1_2}$}{5\_2\^{zeta\^1\_2}}}

$c = 2$, $(d_i,\theta_i)$ = \{$(1., 0)$, $(1., \frac{1}{2})$, $(1., 0)$, $(1., \frac{1}{2})$, $(2., \frac{1}{4})$\}

\begin{enumerate}

\item $c = 2$, $(d_i,\theta_i)$ = \{$(1., 0)$, $(1., \frac{1}{2})$, $(1., 0)$, $(1., \frac{1}{2})$, $(2., \frac{1}{4})$, $(1.41421, \frac{1}{16})$, $(1.41421, -\frac{7}{16})$, $(1.41421, \frac{3}{16})$, $(1.41421, -\frac{5}{16})$\}

\item $c = 2$, $(d_i,\theta_i)$ = \{$(1., 0)$, $(1., \frac{1}{2})$, $(1., 0)$, $(1., \frac{1}{2})$, $(2., \frac{1}{4})$, $(1.41421, -\frac{3}{16})$, $(1.41421, \frac{5}{16})$, $(1.41421, -\frac{1}{16})$, $(1.41421, \frac{7}{16})$\}

\end{enumerate}

\paragraph*{\texorpdfstring{$5_2^{\zeta^1_2*}$}{5\_2\^{zeta\^1\_2}*}}

$c = 2$, $(d_i,\theta_i)$ = \{$(1., 0)$, $(1., \frac{1}{2})$, $(1., 0)$, $(1., \frac{1}{2})$, $(-2., \frac{1}{4})$\}

\begin{enumerate}

\item $c = 2$, $(d_i,\theta_i)$ = \{$(1., 0)$, $(1., \frac{1}{2})$, $(1., 0)$, $(1., \frac{1}{2})$, $(-2., \frac{1}{4})$, $(-1.41421, -\frac{3}{16})$, $(-1.41421, \frac{5}{16})$, $(1.41421, -\frac{1}{16})$, $(1.41421, \frac{7}{16})$\}

\item $c = 2$, $(d_i,\theta_i)$ = \{$(1., 0)$, $(1., \frac{1}{2})$, $(1., 0)$, $(1., \frac{1}{2})$, $(-2., \frac{1}{4})$, $(-1.41421, \frac{1}{16})$, $(-1.41421, -\frac{7}{16})$, $(1.41421, \frac{3}{16})$, $(1.41421, -\frac{5}{16})$\}

\end{enumerate}

\paragraph*{\texorpdfstring{$5_3^{\zeta^1_2}$}{5\_3\^{zeta\^1\_2}}}

$c = 3$, $(d_i,\theta_i)$ = \{$(1., 0)$, $(1., \frac{1}{2})$, $(1., 0)$, $(1., \frac{1}{2})$, $(2., \frac{3}{8})$\}

\begin{enumerate}

\item $c = 3$, $(d_i,\theta_i)$ = \{$(1., 0)$, $(1., \frac{1}{2})$, $(1., 0)$, $(1., \frac{1}{2})$, $(2., \frac{3}{8})$, $(1.41421, \frac{7}{16})$, $(1.41421, -\frac{1}{16})$, $(1.41421, \frac{7}{16})$, $(1.41421, -\frac{1}{16})$\}

\item $c = 3$, $(d_i,\theta_i)$ = \{$(1., 0)$, $(1., \frac{1}{2})$, $(1., 0)$, $(1., \frac{1}{2})$, $(2., \frac{3}{8})$, $(1.41421, \frac{3}{16})$, $(1.41421, -\frac{5}{16})$, $(1.41421, \frac{3}{16})$, $(1.41421, -\frac{5}{16})$\}

\item $c = 3$, $(d_i,\theta_i)$ = \{$(1., 0)$, $(1., \frac{1}{2})$, $(1., 0)$, $(1., \frac{1}{2})$, $(2., \frac{3}{8})$, $(1.41421, \frac{5}{16})$, $(1.41421, -\frac{3}{16})$, $(1.41421, \frac{1}{16})$, $(1.41421, -\frac{7}{16})$\}

\end{enumerate}

\paragraph*{\texorpdfstring{$5_3^{\zeta^1_2*}$}{5\_3\^{zeta\^1\_2*}}}

$c = 3$, $(d_i,\theta_i)$ = \{$(1., 0)$, $(1., \frac{1}{2})$, $(1., 0)$, $(1., \frac{1}{2})$, $(-2., \frac{3}{8})$\}

\begin{enumerate}

\item $c = 3$, $(d_i,\theta_i)$ = \{$(1., 0)$, $(1., \frac{1}{2})$, $(1., 0)$, $(1., \frac{1}{2})$, $(-2., \frac{3}{8})$, $(-1.41421, \frac{3}{16})$, $(-1.41421, -\frac{5}{16})$, $(1.41421, \frac{3}{16})$, $(1.41421, -\frac{5}{16})$\}

\item $c = 3$, $(d_i,\theta_i)$ = \{$(1., 0)$, $(1., \frac{1}{2})$, $(1., 0)$, $(1., \frac{1}{2})$, $(-2., \frac{3}{8})$, $(-1.41421, \frac{5}{16})$, $(-1.41421, -\frac{3}{16})$, $(1.41421, \frac{1}{16})$, $(1.41421, -\frac{7}{16})$\}

\item $c = 3$, $(d_i,\theta_i)$ = \{$(1., 0)$, $(1., \frac{1}{2})$, $(1., 0)$, $(1., \frac{1}{2})$, $(-2., \frac{3}{8})$, $(-1.41421, \frac{7}{16})$, $(-1.41421, -\frac{1}{16})$, $(1.41421, \frac{7}{16})$, $(1.41421, -\frac{1}{16})$\}

\end{enumerate}

\paragraph*{\texorpdfstring{$5_{4}^{\zeta^1_2}$}{5\_{4}\^{zeta\^1\_2}}}

$c = 4$, $(d_i,\theta_i)$ = \{$(1., 0)$, $(1., \frac{1}{2})$, $(1., 0)$, $(1., \frac{1}{2})$, $(2., \frac{1}{2})$\}

\begin{enumerate}

\item $c = 4$, $(d_i,\theta_i)$ = \{$(1., 0)$, $(1., \frac{1}{2})$, $(1., 0)$, $(1., \frac{1}{2})$, $(2., \frac{1}{2})$, $(1.41421, \frac{1}{16})$, $(1.41421, \frac{7}{16})$, $(1.41421, -\frac{7}{16})$, $(1.41421, -\frac{1}{16})$\}

\item $c = 4$, $(d_i,\theta_i)$ = \{$(1., 0)$, $(1., \frac{1}{2})$, $(1., 0)$, $(1., \frac{1}{2})$, $(2., \frac{1}{2})$, $(1.41421, \frac{3}{16})$, $(1.41421, \frac{5}{16})$, $(1.41421, -\frac{5}{16})$, $(1.41421, -\frac{3}{16})$\}

\end{enumerate}

\paragraph*{\texorpdfstring{$5_{4}^{\zeta^1_2*}$}{5\_{4}\^{zeta\^1\_2*}}}

$c = 4$, $(d_i,\theta_i)$ = \{$(1., 0)$, $(1., \frac{1}{2})$, $(1., 0)$, $(1., \frac{1}{2})$, $(-2., \frac{1}{2})$\}

\begin{enumerate}

\item $c = 4$, $(d_i,\theta_i)$ = \{$(1., 0)$, $(1., \frac{1}{2})$, $(1., 0)$, $(1., \frac{1}{2})$, $(-2., \frac{1}{2})$, $(1.41421, \frac{1}{16})$, $(-1.41421, \frac{7}{16})$, $(1.41421, -\frac{7}{16})$, $(-1.41421, -\frac{1}{16})$\}

\item $c = 4$, $(d_i,\theta_i)$ = \{$(1., 0)$, $(1., \frac{1}{2})$, $(1., 0)$, $(1., \frac{1}{2})$, $(-2., \frac{1}{2})$, $(1.41421, \frac{3}{16})$, $(-1.41421, \frac{5}{16})$, $(1.41421, -\frac{5}{16})$, $(-1.41421, -\frac{3}{16})$\}

\end{enumerate}

\paragraph*{\texorpdfstring{$5_{-3}^{\zeta^1_2}$}{5\_{-3}\^{zeta\^1\_2}}}

$c = 5$, $(d_i,\theta_i)$ = \{$(1., 0)$, $(1., \frac{1}{2})$, $(1., 0)$, $(1., \frac{1}{2})$, $(2., -\frac{3}{8})$\}

\begin{enumerate}

\item $c = 5$, $(d_i,\theta_i)$ = \{$(1., 0)$, $(1., \frac{1}{2})$, $(1., 0)$, $(1., \frac{1}{2})$, $(2., -\frac{3}{8})$, $(1.41421, -\frac{1}{16})$, $(1.41421, \frac{7}{16})$, $(1.41421, \frac{3}{16})$, $(1.41421, -\frac{5}{16})$\}

\item $c = 5$, $(d_i,\theta_i)$ = \{$(1., 0)$, $(1., \frac{1}{2})$, $(1., 0)$, $(1., \frac{1}{2})$, $(2., -\frac{3}{8})$, $(1.41421, \frac{5}{16})$, $(1.41421, -\frac{3}{16})$, $(1.41421, \frac{5}{16})$, $(1.41421, -\frac{3}{16})$\}

\item $c = 5$, $(d_i,\theta_i)$ = \{$(1., 0)$, $(1., \frac{1}{2})$, $(1., 0)$, $(1., \frac{1}{2})$, $(2., -\frac{3}{8})$, $(1.41421, \frac{1}{16})$, $(1.41421, -\frac{7}{16})$, $(1.41421, -\frac{7}{16})$, $(1.41421, \frac{1}{16})$\}

\end{enumerate}

\paragraph*{\texorpdfstring{$5_5^{\zeta^1_2*}$}{5\_{-3}\^{zeta\^1\_2*}}}

$c = 5$, $(d_i,\theta_i)$ = \{$(1., 0)$, $(1., \frac{1}{2})$, $(1., 0)$, $(1., \frac{1}{2})$, $(-2., -\frac{3}{8})$\}

\begin{enumerate}

\item $c = 5$, $(d_i,\theta_i)$ = \{$(1., 0)$, $(1., \frac{1}{2})$, $(1., 0)$, $(1., \frac{1}{2})$, $(-2., -\frac{3}{8})$, $(-1.41421, \frac{5}{16})$, $(-1.41421, -\frac{3}{16})$, $(1.41421, \frac{5}{16})$, $(1.41421, -\frac{3}{16})$\}

\item $c = 5$, $(d_i,\theta_i)$ = \{$(1., 0)$, $(1., \frac{1}{2})$, $(1., 0)$, $(1., \frac{1}{2})$, $(-2., -\frac{3}{8})$, $(-1.41421, \frac{1}{16})$, $(-1.41421, -\frac{7}{16})$, $(1.41421, -\frac{7}{16})$, $(1.41421, \frac{1}{16})$\}

\item $c = 5$, $(d_i,\theta_i)$ = \{$(1., 0)$, $(1., \frac{1}{2})$, $(1., 0)$, $(1., \frac{1}{2})$, $(-2., -\frac{3}{8})$, $(-1.41421, -\frac{1}{16})$, $(-1.41421, \frac{7}{16})$, $(1.41421, \frac{3}{16})$, $(1.41421, -\frac{5}{16})$\}

\end{enumerate}

\paragraph*{\texorpdfstring{$5_{-2}^{\zeta^1_2}$}{5\_{-2}\^{zeta\^1\_2}}}

$c = 6$, $(d_i,\theta_i)$ = \{$(1., 0)$, $(1., \frac{1}{2})$, $(1., 0)$, $(1., \frac{1}{2})$, $(2., -\frac{1}{4})$\}

\begin{enumerate}

\item $c = 6$, $(d_i,\theta_i)$ = \{$(1., 0)$, $(1., \frac{1}{2})$, $(1., 0)$, $(1., \frac{1}{2})$, $(2., -\frac{1}{4})$, $(1.41421, -\frac{7}{16})$, $(1.41421, \frac{1}{16})$, $(1.41421, -\frac{5}{16})$, $(1.41421, \frac{3}{16})$\}

\item $c = 6$, $(d_i,\theta_i)$ = \{$(1., 0)$, $(1., \frac{1}{2})$, $(1., 0)$, $(1., \frac{1}{2})$, $(2., -\frac{1}{4})$, $(1.41421, \frac{5}{16})$, $(1.41421, -\frac{3}{16})$, $(1.41421, \frac{7}{16})$, $(1.41421, -\frac{1}{16})$\}

\end{enumerate}

\paragraph*{\texorpdfstring{$5_6^{\zeta^1_2*}$}{5\_{-2}\^{zeta\^1\_2*}}}

$c = 6$, $(d_i,\theta_i)$ = \{$(1., 0)$, $(1., \frac{1}{2})$, $(1., 0)$, $(1., \frac{1}{2})$, $(-2., -\frac{1}{4})$\}

\begin{enumerate}

\item $c = 6$, $(d_i,\theta_i)$ = \{$(1., 0)$, $(1., \frac{1}{2})$, $(1., 0)$, $(1., \frac{1}{2})$, $(-2., -\frac{1}{4})$, $(-1.41421, \frac{5}{16})$, $(-1.41421, -\frac{3}{16})$, $(1.41421, \frac{7}{16})$, $(1.41421, -\frac{1}{16})$\}

\item $c = 6$, $(d_i,\theta_i)$ = \{$(1., 0)$, $(1., \frac{1}{2})$, $(1., 0)$, $(1., \frac{1}{2})$, $(-2., -\frac{1}{4})$, $(-1.41421, -\frac{7}{16})$, $(-1.41421, \frac{1}{16})$, $(1.41421, -\frac{5}{16})$, $(1.41421, \frac{3}{16})$\}

\end{enumerate}

\paragraph*{\texorpdfstring{$5_{-1}^{\zeta^1_2}$}{5\_{-1}\^{zeta\^1\_2}}}

$c = 7$, $(d_i,\theta_i)$ = \{$(1., 0)$, $(1., \frac{1}{2})$, $(1., 0)$, $(1., \frac{1}{2})$, $(2., -\frac{1}{8})$\}

\begin{enumerate}

\item $c = 7$, $(d_i,\theta_i)$ = \{$(1., 0)$, $(1., \frac{1}{2})$, $(1., 0)$, $(1., \frac{1}{2})$, $(2., -\frac{1}{8})$, $(1.41421, -\frac{3}{16})$, $(1.41421, \frac{5}{16})$, $(1.41421, -\frac{7}{16})$, $(1.41421, \frac{1}{16})$\}

\item $c = 7$, $(d_i,\theta_i)$ = \{$(1., 0)$, $(1., \frac{1}{2})$, $(1., 0)$, $(1., \frac{1}{2})$, $(2., -\frac{1}{8})$, $(1.41421, -\frac{5}{16})$, $(1.41421, \frac{3}{16})$, $(1.41421, -\frac{5}{16})$, $(1.41421, \frac{3}{16})$\}

\item $c = 7$, $(d_i,\theta_i)$ = \{$(1., 0)$, $(1., \frac{1}{2})$, $(1., 0)$, $(1., \frac{1}{2})$, $(2., -\frac{1}{8})$, $(1.41421, \frac{7}{16})$, $(1.41421, -\frac{1}{16})$, $(1.41421, \frac{7}{16})$, $(1.41421, -\frac{1}{16})$\}

\end{enumerate}

\paragraph*{\texorpdfstring{$5_7^{\zeta^1_2*}$}{5\_{-1}\^{zeta\^1\_2*}}}

$c = 7$, $(d_i,\theta_i)$ = \{$(1., 0)$, $(1., \frac{1}{2})$, $(1., 0)$, $(1., \frac{1}{2})$, $(-2., -\frac{1}{8})$\}

\begin{enumerate}

\item $c = 7$, $(d_i,\theta_i)$ = \{$(1., 0)$, $(1., \frac{1}{2})$, $(1., 0)$, $(1., \frac{1}{2})$, $(-2., -\frac{1}{8})$, $(-1.41421, \frac{7}{16})$, $(-1.41421, -\frac{1}{16})$, $(1.41421, \frac{7}{16})$, $(1.41421, -\frac{1}{16})$\}

\item $c = 7$, $(d_i,\theta_i)$ = \{$(1., 0)$, $(1., \frac{1}{2})$, $(1., 0)$, $(1., \frac{1}{2})$, $(-2., -\frac{1}{8})$, $(-1.41421, -\frac{3}{16})$, $(-1.41421, \frac{5}{16})$, $(1.41421, -\frac{7}{16})$, $(1.41421, \frac{1}{16})$\}

\item $c = 7$, $(d_i,\theta_i)$ = \{$(1., 0)$, $(1., \frac{1}{2})$, $(1., 0)$, $(1., \frac{1}{2})$, $(-2., -\frac{1}{8})$, $(-1.41421, -\frac{5}{16})$, $(-1.41421, \frac{3}{16})$, $(1.41421, -\frac{5}{16})$, $(1.41421, \frac{3}{16})$\}

\end{enumerate}

\paragraph*{\texorpdfstring{$5_{\frac{12}{5}}^{\zeta^1_2}$}{5\_{frac{12}{5}}\^{zeta\^1\_2}}}

$c = \frac{12}{5}$, $(d_i,\theta_i)$ = \{$(1., 0)$, $(2.61803, \frac{1}{5})$, $(1., 0)$, $(2.61803, \frac{1}{5})$, $(3.23607, -\frac{2}{5})$\}

\begin{enumerate}

\item $c = \frac{12}{5}$, $(d_i,\theta_i)$ = \{$(1., 0)$, $(2.61803, \frac{1}{5})$, $(1., 0)$, $(2.61803, \frac{1}{5})$, $(3.23607, -\frac{2}{5})$, $(3.07768, -\frac{3}{8})$, $(1.90211, -\frac{7}{40})$, $(3.07768, \frac{1}{8})$, $(1.90211, \frac{13}{40})$\}

\item $c = \frac{12}{5}$, $(d_i,\theta_i)$ = \{$(1., 0)$, $(2.61803, \frac{1}{5})$, $(1., 0)$, $(2.61803, \frac{1}{5})$, $(3.23607, -\frac{2}{5})$, $(3.07768, -\frac{1}{8})$, $(1.90211, \frac{3}{40})$, $(3.07768, \frac{3}{8})$, $(1.90211, -\frac{17}{40})$\}

\end{enumerate}

\paragraph*{\texorpdfstring{$5_{-\frac{12}{5}}^{\zeta^1_2}$}{5\_{-frac{12}{5}}\^{zeta\^1\_2}}}

$c = \frac{28}{5}$, $(d_i,\theta_i)$ = \{$(1., 0)$, $(2.61803, -\frac{1}{5})$, $(1., 0)$, $(2.61803, -\frac{1}{5})$, $(3.23607, \frac{2}{5})$\}

\begin{enumerate}

\item $c = \frac{28}{5}$, $(d_i,\theta_i)$ = \{$(1., 0)$, $(2.61803, -\frac{1}{5})$, $(1., 0)$, $(2.61803, -\frac{1}{5})$, $(3.23607, \frac{2}{5})$, $(1.90211, -\frac{13}{40})$, $(3.07768, -\frac{1}{8})$, $(1.90211, \frac{7}{40})$, $(3.07768, \frac{3}{8})$\}

\item $c = \frac{28}{5}$, $(d_i,\theta_i)$ = \{$(1., 0)$, $(2.61803, -\frac{1}{5})$, $(1., 0)$, $(2.61803, -\frac{1}{5})$, $(3.23607, \frac{2}{5})$, $(1.90211, \frac{17}{40})$, $(3.07768, -\frac{3}{8})$, $(1.90211, -\frac{3}{40})$, $(3.07768, \frac{1}{8})$\}

\end{enumerate}

\paragraph*{\texorpdfstring{$5_{\frac{36}{5}}^{\zeta^1_2*}$}{5\_{frac{36}{5}}\^{zeta\^1\_2*}}}

$c = \frac{36}{5}$, $(d_i,\theta_i)$ = \{$(1., 0)$, $(0.381966, -\frac{2}{5})$, $(1., 0)$, $(0.381966, -\frac{2}{5})$, $(-1.23607, -\frac{1}{5})$\}

\begin{enumerate}

\item $c = \frac{36}{5}$, $(d_i,\theta_i)$ = \{$(1., 0)$, $(0.381966, -\frac{2}{5})$, $(1., 0)$, $(0.381966, -\frac{2}{5})$, $(-1.23607, -\frac{1}{5})$, $(-0.726543, -\frac{1}{8})$, $(1.17557, \frac{19}{40})$, $(-0.726543, \frac{3}{8})$, $(1.17557, -\frac{1}{40})$\}

\item $c = \frac{36}{5}$, $(d_i,\theta_i)$ = \{$(1., 0)$, $(0.381966, -\frac{2}{5})$, $(1., 0)$, $(0.381966, -\frac{2}{5})$, $(-1.23607, -\frac{1}{5})$, $(-0.726543, -\frac{3}{8})$, $(1.17557, \frac{9}{40})$, $(-0.726543, \frac{1}{8})$, $(1.17557, -\frac{11}{40})$\}

\end{enumerate}

\paragraph*{\texorpdfstring{$5_{\frac{4}{5}}^{\zeta^1_2*}$}{5\_{frac{4}{5}}\^{zeta\^1\_2*}}}

$c = \frac{4}{5}$, $(d_i,\theta_i)$ = \{$(1., 0)$, $(0.381966, \frac{2}{5})$, $(1., 0)$, $(0.381966, \frac{2}{5})$, $(-1.23607, \frac{1}{5})$\}

\begin{enumerate}

\item $c = \frac{4}{5}$, $(d_i,\theta_i)$ = \{$(1., 0)$, $(0.381966, \frac{2}{5})$, $(1., 0)$, $(0.381966, \frac{2}{5})$, $(-1.23607, \frac{1}{5})$, $(1.17557, \frac{11}{40})$, $(-0.726543, -\frac{1}{8})$, $(1.17557, -\frac{9}{40})$, $(-0.726543, \frac{3}{8})$\}

\item $c = \frac{4}{5}$, $(d_i,\theta_i)$ = \{$(1., 0)$, $(0.381966, \frac{2}{5})$, $(1., 0)$, $(0.381966, \frac{2}{5})$, $(-1.23607, \frac{1}{5})$, $(1.17557, \frac{1}{40})$, $(-0.726543, -\frac{3}{8})$, $(1.17557, -\frac{19}{40})$, $(-0.726543, \frac{1}{8})$\}

\end{enumerate}

\paragraph*{\texorpdfstring{$6_{\frac{1}{2}}^{\zeta^1_2}$}{6\_{frac{1}{2}}\^{zeta\^1\_2}}}

$c = \frac{1}{2}$, $(d_i,\theta_i)$ = \{$(1., 0)$, $(1.41421, \frac{1}{16})$, $(1., \frac{1}{2})$, $(1., 0)$, $(1.41421, \frac{1}{16})$, $(1., \frac{1}{2})$\}

\begin{enumerate}

\item $c = \frac{1}{2}$, $(d_i,\theta_i)$ = \{$(1., 0)$, $(1.41421, \frac{1}{16})$, $(1., \frac{1}{2})$, $(1., 0)$, $(1.41421, \frac{1}{16})$, $(1., \frac{1}{2})$, $(1., \frac{1}{8})$, $(1., -\frac{3}{8})$, $(1.41421, -\frac{1}{16})$, $(1.41421, \frac{7}{16})$, $(1., \frac{1}{8})$, $(1., -\frac{3}{8})$\}

\item $c = \frac{1}{2}$, $(d_i,\theta_i)$ = \{$(1., 0)$, $(1.41421, \frac{1}{16})$, $(1., \frac{1}{2})$, $(1., 0)$, $(1.41421, \frac{1}{16})$, $(1., \frac{1}{2})$, $(1., \frac{3}{8})$, $(1., -\frac{1}{8})$, $(1.41421, \frac{3}{16})$, $(1.41421, -\frac{5}{16})$, $(1., \frac{3}{8})$, $(1., -\frac{1}{8})$\}

\end{enumerate}

\paragraph*{\texorpdfstring{$6_{\frac{1}{2}}^{\zeta^1_2*}$}{6\_{frac{1}{2}}\^{zeta\^1\_2*}}}

$c = \frac{1}{2}$, $(d_i,\theta_i)$ = \{$(1., 0)$, $(-1.41421, \frac{1}{16})$, $(1., \frac{1}{2})$, $(1., 0)$, $(-1.41421, \frac{1}{16})$, $(1., \frac{1}{2})$\}

\begin{enumerate}

\item $c = \frac{1}{2}$, $(d_i,\theta_i)$ = \{$(1., 0)$, $(-1.41421, \frac{1}{16})$, $(1., \frac{1}{2})$, $(1., 0)$, $(-1.41421, \frac{1}{16})$, $(1., \frac{1}{2})$, $(-1., \frac{1}{8})$, $(-1., -\frac{3}{8})$, $(1.41421, -\frac{1}{16})$, $(1.41421, \frac{7}{16})$, $(-1., \frac{1}{8})$, $(-1., -\frac{3}{8})$\}

\item $c = \frac{1}{2}$, $(d_i,\theta_i)$ = \{$(1., 0)$, $(-1.41421, \frac{1}{16})$, $(1., \frac{1}{2})$, $(1., 0)$, $(-1.41421, \frac{1}{16})$, $(1., \frac{1}{2})$, $(-1., \frac{3}{8})$, $(-1., -\frac{1}{8})$, $(1.41421, \frac{3}{16})$, $(1.41421, -\frac{5}{16})$, $(-1., \frac{3}{8})$, $(-1., -\frac{1}{8})$\}

\end{enumerate}

\paragraph*{\texorpdfstring{$6_{\frac{3}{2}}^{\zeta^1_2}$}{6\_{frac{3}{2}}\^{zeta\^1\_2}}}

$c = \frac{3}{2}$, $(d_i,\theta_i)$ = \{$(1., 0)$, $(1.41421, \frac{3}{16})$, $(1., \frac{1}{2})$, $(1., 0)$, $(1.41421, \frac{3}{16})$, $(1., \frac{1}{2})$\}

\begin{enumerate}

\item $c = \frac{3}{2}$, $(d_i,\theta_i)$ = \{$(1., 0)$, $(1.41421, \frac{3}{16})$, $(1., \frac{1}{2})$, $(1., 0)$, $(1.41421, \frac{3}{16})$, $(1., \frac{1}{2})$, $(1., \frac{3}{8})$, $(1., -\frac{1}{8})$, $(1.41421, \frac{5}{16})$, $(1.41421, -\frac{3}{16})$, $(1., -\frac{1}{8})$, $(1., \frac{3}{8})$\}

\item $c = \frac{3}{2}$, $(d_i,\theta_i)$ = \{$(1., 0)$, $(1.41421, \frac{3}{16})$, $(1., \frac{1}{2})$, $(1., 0)$, $(1.41421, \frac{3}{16})$, $(1., \frac{1}{2})$, $(1., \frac{1}{8})$, $(1., -\frac{3}{8})$, $(1.41421, -\frac{7}{16})$, $(1.41421, \frac{1}{16})$, $(1., \frac{1}{8})$, $(1., -\frac{3}{8})$\}

\end{enumerate}

\paragraph*{\texorpdfstring{$6_{\frac{3}{2}}^{\zeta^1_2*}$}{6\_{frac{3}{2}}\^{zeta\^1\_2*}}}

$c = \frac{3}{2}$, $(d_i,\theta_i)$ = \{$(1., 0)$, $(-1.41421, \frac{3}{16})$, $(1., \frac{1}{2})$, $(1., 0)$, $(-1.41421, \frac{3}{16})$, $(1., \frac{1}{2})$\}

\begin{enumerate}

\item $c = \frac{3}{2}$, $(d_i,\theta_i)$ = \{$(1., 0)$, $(-1.41421, \frac{3}{16})$, $(1., \frac{1}{2})$, $(1., 0)$, $(-1.41421, \frac{3}{16})$, $(1., \frac{1}{2})$, $(-1., \frac{1}{8})$, $(-1., -\frac{3}{8})$, $(1.41421, -\frac{7}{16})$, $(1.41421, \frac{1}{16})$, $(-1., \frac{1}{8})$, $(-1., -\frac{3}{8})$\}

\item $c = \frac{3}{2}$, $(d_i,\theta_i)$ = \{$(1., 0)$, $(-1.41421, \frac{3}{16})$, $(1., \frac{1}{2})$, $(1., 0)$, $(-1.41421, \frac{3}{16})$, $(1., \frac{1}{2})$, $(-1., \frac{3}{8})$, $(-1., -\frac{1}{8})$, $(1.41421, \frac{5}{16})$, $(1.41421, -\frac{3}{16})$, $(-1., -\frac{1}{8})$, $(-1., \frac{3}{8})$\}

\end{enumerate}

\paragraph*{\texorpdfstring{$6_{\frac{5}{2}}^{\zeta^1_2}$}{6\_{frac{5}{2}}\^{zeta\^1\_2}}}

$c = \frac{5}{2}$, $(d_i,\theta_i)$ = \{$(1., 0)$, $(1.41421, \frac{5}{16})$, $(1., \frac{1}{2})$, $(1., 0)$, $(1.41421, \frac{5}{16})$, $(1., \frac{1}{2})$\}

\begin{enumerate}

\item $c = \frac{5}{2}$, $(d_i,\theta_i)$ = \{$(1., 0)$, $(1.41421, \frac{5}{16})$, $(1., \frac{1}{2})$, $(1., 0)$, $(1.41421, \frac{5}{16})$, $(1., \frac{1}{2})$, $(1., -\frac{3}{8})$, $(1., \frac{1}{8})$, $(1.41421, -\frac{5}{16})$, $(1.41421, \frac{3}{16})$, $(1., \frac{1}{8})$, $(1., -\frac{3}{8})$\}

\item $c = \frac{5}{2}$, $(d_i,\theta_i)$ = \{$(1., 0)$, $(1.41421, \frac{5}{16})$, $(1., \frac{1}{2})$, $(1., 0)$, $(1.41421, \frac{5}{16})$, $(1., \frac{1}{2})$, $(1., -\frac{1}{8})$, $(1., \frac{3}{8})$, $(1.41421, -\frac{1}{16})$, $(1.41421, \frac{7}{16})$, $(1., \frac{3}{8})$, $(1., -\frac{1}{8})$\}

\end{enumerate}

\paragraph*{\texorpdfstring{$6_{\frac{5}{2}}^{\zeta^1_2*}$}{6\_{frac{5}{2}}\^{zeta\^1\_2*}}}

$c = \frac{5}{2}$, $(d_i,\theta_i)$ = \{$(1., 0)$, $(-1.41421, \frac{5}{16})$, $(1., \frac{1}{2})$, $(1., 0)$, $(-1.41421, \frac{5}{16})$, $(1., \frac{1}{2})$\}

\begin{enumerate}

\item $c = \frac{5}{2}$, $(d_i,\theta_i)$ = \{$(1., 0)$, $(-1.41421, \frac{5}{16})$, $(1., \frac{1}{2})$, $(1., 0)$, $(-1.41421, \frac{5}{16})$, $(1., \frac{1}{2})$, $(-1., -\frac{1}{8})$, $(-1., \frac{3}{8})$, $(1.41421, -\frac{1}{16})$, $(1.41421, \frac{7}{16})$, $(-1., \frac{3}{8})$, $(-1., -\frac{1}{8})$\}

\item $c = \frac{5}{2}$, $(d_i,\theta_i)$ = \{$(1., 0)$, $(-1.41421, \frac{5}{16})$, $(1., \frac{1}{2})$, $(1., 0)$, $(-1.41421, \frac{5}{16})$, $(1., \frac{1}{2})$, $(-1., -\frac{3}{8})$, $(-1., \frac{1}{8})$, $(1.41421, -\frac{5}{16})$, $(1.41421, \frac{3}{16})$, $(-1., \frac{1}{8})$, $(-1., -\frac{3}{8})$\}

\end{enumerate}

\paragraph*{\texorpdfstring{$6_{\frac{7}{2}}^{\zeta^1_2}$}{6\_{frac{7}{2}}\^{zeta\^1\_2}}}

$c = \frac{7}{2}$, $(d_i,\theta_i)$ = \{$(1., 0)$, $(1.41421, \frac{7}{16})$, $(1., \frac{1}{2})$, $(1., 0)$, $(1.41421, \frac{7}{16})$, $(1., \frac{1}{2})$\}

\begin{enumerate}

\item $c = \frac{7}{2}$, $(d_i,\theta_i)$ = \{$(1., 0)$, $(1.41421, \frac{7}{16})$, $(1., \frac{1}{2})$, $(1., 0)$, $(1.41421, \frac{7}{16})$, $(1., \frac{1}{2})$, $(1., \frac{1}{8})$, $(1., -\frac{3}{8})$, $(1.41421, \frac{5}{16})$, $(1.41421, -\frac{3}{16})$, $(1., -\frac{3}{8})$, $(1., \frac{1}{8})$\}

\item $c = \frac{7}{2}$, $(d_i,\theta_i)$ = \{$(1., 0)$, $(1.41421, \frac{7}{16})$, $(1., \frac{1}{2})$, $(1., 0)$, $(1.41421, \frac{7}{16})$, $(1., \frac{1}{2})$, $(1., -\frac{1}{8})$, $(1., \frac{3}{8})$, $(1.41421, \frac{1}{16})$, $(1.41421, -\frac{7}{16})$, $(1., \frac{3}{8})$, $(1., -\frac{1}{8})$\}

\end{enumerate}

\paragraph*{\texorpdfstring{$6_{\frac{7}{2}}^{\zeta^1_2*}$}{6\_{frac{7}{2}}\^{zeta\^1\_2*}}}

$c = \frac{7}{2}$, $(d_i,\theta_i)$ = \{$(1., 0)$, $(-1.41421, \frac{7}{16})$, $(1., \frac{1}{2})$, $(1., 0)$, $(-1.41421, \frac{7}{16})$, $(1., \frac{1}{2})$\}

\begin{enumerate}

\item $c = \frac{7}{2}$, $(d_i,\theta_i)$ = \{$(1., 0)$, $(-1.41421, \frac{7}{16})$, $(1., \frac{1}{2})$, $(1., 0)$, $(-1.41421, \frac{7}{16})$, $(1., \frac{1}{2})$, $(-1., \frac{1}{8})$, $(-1., -\frac{3}{8})$, $(1.41421, \frac{5}{16})$, $(1.41421, -\frac{3}{16})$, $(-1., -\frac{3}{8})$, $(-1., \frac{1}{8})$\}

\item $c = \frac{7}{2}$, $(d_i,\theta_i)$ = \{$(1., 0)$, $(-1.41421, \frac{7}{16})$, $(1., \frac{1}{2})$, $(1., 0)$, $(-1.41421, \frac{7}{16})$, $(1., \frac{1}{2})$, $(-1., -\frac{1}{8})$, $(-1., \frac{3}{8})$, $(1.41421, \frac{1}{16})$, $(1.41421, -\frac{7}{16})$, $(-1., \frac{3}{8})$, $(-1., -\frac{1}{8})$\}

\end{enumerate}

\paragraph*{\texorpdfstring{$6_{-\frac{7}{2}}^{\zeta^1_2}$}{6\_{-frac{7}{2}}\^{zeta\^1\_2}}}

$c = \frac{9}{2}$, $(d_i,\theta_i)$ = \{$(1., 0)$, $(1.41421, -\frac{7}{16})$, $(1., \frac{1}{2})$, $(1., 0)$, $(1.41421, -\frac{7}{16})$, $(1., \frac{1}{2})$\}

\begin{enumerate}

\item $c = \frac{9}{2}$, $(d_i,\theta_i)$ = \{$(1., 0)$, $(1.41421, -\frac{7}{16})$, $(1., \frac{1}{2})$, $(1., 0)$, $(1.41421, -\frac{7}{16})$, $(1., \frac{1}{2})$, $(1., -\frac{3}{8})$, $(1., \frac{1}{8})$, $(1.41421, \frac{7}{16})$, $(1.41421, -\frac{1}{16})$, $(1., -\frac{3}{8})$, $(1., \frac{1}{8})$\}

\item $c = \frac{9}{2}$, $(d_i,\theta_i)$ = \{$(1., 0)$, $(1.41421, -\frac{7}{16})$, $(1., \frac{1}{2})$, $(1., 0)$, $(1.41421, -\frac{7}{16})$, $(1., \frac{1}{2})$, $(1., -\frac{1}{8})$, $(1., \frac{3}{8})$, $(1.41421, -\frac{5}{16})$, $(1.41421, \frac{3}{16})$, $(1., \frac{3}{8})$, $(1., -\frac{1}{8})$\}

\end{enumerate}

\paragraph*{\texorpdfstring{$6_{\frac{9}{2}}^{\zeta^1_2*}$}{6\_{frac{9}{2}}\^{zeta\^1\_2*}}}

$c = \frac{9}{2}$, $(d_i,\theta_i)$ = \{$(1., 0)$, $(-1.41421, -\frac{7}{16})$, $(1., \frac{1}{2})$, $(1., 0)$, $(-1.41421, -\frac{7}{16})$, $(1., \frac{1}{2})$\}

\begin{enumerate}

\item $c = \frac{9}{2}$, $(d_i,\theta_i)$ = \{$(1., 0)$, $(-1.41421, -\frac{7}{16})$, $(1., \frac{1}{2})$, $(1., 0)$, $(-1.41421, -\frac{7}{16})$, $(1., \frac{1}{2})$, $(-1., -\frac{1}{8})$, $(-1., \frac{3}{8})$, $(1.41421, -\frac{5}{16})$, $(1.41421, \frac{3}{16})$, $(-1., \frac{3}{8})$, $(-1., -\frac{1}{8})$\}

\item $c = \frac{9}{2}$, $(d_i,\theta_i)$ = \{$(1., 0)$, $(-1.41421, -\frac{7}{16})$, $(1., \frac{1}{2})$, $(1., 0)$, $(-1.41421, -\frac{7}{16})$, $(1., \frac{1}{2})$, $(-1., -\frac{3}{8})$, $(-1., \frac{1}{8})$, $(1.41421, \frac{7}{16})$, $(1.41421, -\frac{1}{16})$, $(-1., -\frac{3}{8})$, $(-1., \frac{1}{8})$\}

\end{enumerate}

\paragraph*{\texorpdfstring{$6_{-\frac{5}{2}}^{\zeta^1_2}$}{6\_{-frac{5}{2}}\^{zeta\^1\_2}}}

$c = \frac{11}{2}$, $(d_i,\theta_i)$ = \{$(1., 0)$, $(1.41421, -\frac{5}{16})$, $(1., \frac{1}{2})$, $(1., 0)$, $(1.41421, -\frac{5}{16})$, $(1., \frac{1}{2})$\}

\begin{enumerate}

\item $c = \frac{11}{2}$, $(d_i,\theta_i)$ = \{$(1., 0)$, $(1.41421, -\frac{5}{16})$, $(1., \frac{1}{2})$, $(1., 0)$, $(1.41421, -\frac{5}{16})$, $(1., \frac{1}{2})$, $(1., -\frac{3}{8})$, $(1., \frac{1}{8})$, $(1.41421, \frac{1}{16})$, $(1.41421, -\frac{7}{16})$, $(1., -\frac{3}{8})$, $(1., \frac{1}{8})$\}

\item $c = \frac{11}{2}$, $(d_i,\theta_i)$ = \{$(1., 0)$, $(1.41421, -\frac{5}{16})$, $(1., \frac{1}{2})$, $(1., 0)$, $(1.41421, -\frac{5}{16})$, $(1., \frac{1}{2})$, $(1., -\frac{1}{8})$, $(1., \frac{3}{8})$, $(1.41421, -\frac{3}{16})$, $(1.41421, \frac{5}{16})$, $(1., -\frac{1}{8})$, $(1., \frac{3}{8})$\}

\end{enumerate}

\paragraph*{\texorpdfstring{$6_{\frac{11}{2}}^{\zeta^1_2*}$}{6\_{frac{11}{2}}\^{zeta\^1\_2*}}}

$c = \frac{11}{2}$, $(d_i,\theta_i)$ = \{$(1., 0)$, $(-1.41421, -\frac{5}{16})$, $(1., \frac{1}{2})$, $(1., 0)$, $(-1.41421, -\frac{5}{16})$, $(1., \frac{1}{2})$\}

\begin{enumerate}

\item $c = \frac{11}{2}$, $(d_i,\theta_i)$ = \{$(1., 0)$, $(-1.41421, -\frac{5}{16})$, $(1., \frac{1}{2})$, $(1., 0)$, $(-1.41421, -\frac{5}{16})$, $(1., \frac{1}{2})$, $(-1., -\frac{3}{8})$, $(-1., \frac{1}{8})$, $(1.41421, \frac{1}{16})$, $(1.41421, -\frac{7}{16})$, $(-1., -\frac{3}{8})$, $(-1., \frac{1}{8})$\}

\item $c = \frac{11}{2}$, $(d_i,\theta_i)$ = \{$(1., 0)$, $(-1.41421, -\frac{5}{16})$, $(1., \frac{1}{2})$, $(1., 0)$, $(-1.41421, -\frac{5}{16})$, $(1., \frac{1}{2})$, $(-1., -\frac{1}{8})$, $(-1., \frac{3}{8})$, $(1.41421, -\frac{3}{16})$, $(1.41421, \frac{5}{16})$, $(-1., -\frac{1}{8})$, $(-1., \frac{3}{8})$\}

\end{enumerate}

\paragraph*{\texorpdfstring{$6_{-\frac{3}{2}}^{\zeta^1_2}$}{6\_{-frac{3}{2}}\^{zeta\^1\_2}}}

$c = \frac{13}{2}$, $(d_i,\theta_i)$ = \{$(1., 0)$, $(1.41421, -\frac{3}{16})$, $(1., \frac{1}{2})$, $(1., 0)$, $(1.41421, -\frac{3}{16})$, $(1., \frac{1}{2})$\}

\begin{enumerate}

\item $c = \frac{13}{2}$, $(d_i,\theta_i)$ = \{$(1., 0)$, $(1.41421, -\frac{3}{16})$, $(1., \frac{1}{2})$, $(1., 0)$, $(1.41421, -\frac{3}{16})$, $(1., \frac{1}{2})$, $(1., \frac{1}{8})$, $(1., -\frac{3}{8})$, $(1.41421, \frac{3}{16})$, $(1.41421, -\frac{5}{16})$, $(1., -\frac{3}{8})$, $(1., \frac{1}{8})$\}

\item $c = \frac{13}{2}$, $(d_i,\theta_i)$ = \{$(1., 0)$, $(1.41421, -\frac{3}{16})$, $(1., \frac{1}{2})$, $(1., 0)$, $(1.41421, -\frac{3}{16})$, $(1., \frac{1}{2})$, $(1., \frac{3}{8})$, $(1., -\frac{1}{8})$, $(1.41421, \frac{7}{16})$, $(1.41421, -\frac{1}{16})$, $(1., \frac{3}{8})$, $(1., -\frac{1}{8})$\}

\end{enumerate}

\paragraph*{\texorpdfstring{$6_{\frac{13}{2}}^{\zeta^1_2*}$}{6\_{frac{13}{2}}\^{zeta\^1\_2*}}}

$c = \frac{13}{2}$, $(d_i,\theta_i)$ = \{$(1., 0)$, $(-1.41421, -\frac{3}{16})$, $(1., \frac{1}{2})$, $(1., 0)$, $(-1.41421, -\frac{3}{16})$, $(1., \frac{1}{2})$\}

\begin{enumerate}

\item $c = \frac{13}{2}$, $(d_i,\theta_i)$ = \{$(1., 0)$, $(-1.41421, -\frac{3}{16})$, $(1., \frac{1}{2})$, $(1., 0)$, $(-1.41421, -\frac{3}{16})$, $(1., \frac{1}{2})$, $(-1., \frac{3}{8})$, $(-1., -\frac{1}{8})$, $(1.41421, \frac{7}{16})$, $(1.41421, -\frac{1}{16})$, $(-1., \frac{3}{8})$, $(-1., -\frac{1}{8})$\}

\item $c = \frac{13}{2}$, $(d_i,\theta_i)$ = \{$(1., 0)$, $(-1.41421, -\frac{3}{16})$, $(1., \frac{1}{2})$, $(1., 0)$, $(-1.41421, -\frac{3}{16})$, $(1., \frac{1}{2})$, $(-1., \frac{1}{8})$, $(-1., -\frac{3}{8})$, $(1.41421, \frac{3}{16})$, $(1.41421, -\frac{5}{16})$, $(-1., -\frac{3}{8})$, $(-1., \frac{1}{8})$\}

\end{enumerate}

\paragraph*{\texorpdfstring{$6_{-\frac{1}{2}}^{\zeta^1_2}$}{6\_{-frac{1}{2}}\^{zeta\^1\_2}}}

$c = \frac{15}{2}$, $(d_i,\theta_i)$ = \{$(1., 0)$, $(1.41421, -\frac{1}{16})$, $(1., \frac{1}{2})$, $(1., 0)$, $(1.41421, -\frac{1}{16})$, $(1., \frac{1}{2})$\}

\begin{enumerate}

\item $c = \frac{15}{2}$, $(d_i,\theta_i)$ = \{$(1., 0)$, $(1.41421, -\frac{1}{16})$, $(1., \frac{1}{2})$, $(1., 0)$, $(1.41421, -\frac{1}{16})$, $(1., \frac{1}{2})$, $(1., -\frac{3}{8})$, $(1., \frac{1}{8})$, $(1.41421, -\frac{3}{16})$, $(1.41421, \frac{5}{16})$, $(1., -\frac{3}{8})$, $(1., \frac{1}{8})$\}

\item $c = \frac{15}{2}$, $(d_i,\theta_i)$ = \{$(1., 0)$, $(1.41421, -\frac{1}{16})$, $(1., \frac{1}{2})$, $(1., 0)$, $(1.41421, -\frac{1}{16})$, $(1., \frac{1}{2})$, $(1., \frac{3}{8})$, $(1., -\frac{1}{8})$, $(1.41421, -\frac{7}{16})$, $(1.41421, \frac{1}{16})$, $(1., \frac{3}{8})$, $(1., -\frac{1}{8})$\}

\end{enumerate}

\paragraph*{\texorpdfstring{$6_{\frac{15}{2}}^{\zeta^1_2*}$}{6\_{frac{15}{2}}\^{zeta\^1\_2*}}}

$c = \frac{15}{2}$, $(d_i,\theta_i)$ = \{$(1., 0)$, $(-1.41421, -\frac{1}{16})$, $(1., \frac{1}{2})$, $(1., 0)$, $(-1.41421, -\frac{1}{16})$, $(1., \frac{1}{2})$\}

\begin{enumerate}

\item $c = \frac{15}{2}$, $(d_i,\theta_i)$ = \{$(1., 0)$, $(-1.41421, -\frac{1}{16})$, $(1., \frac{1}{2})$, $(1., 0)$, $(-1.41421, -\frac{1}{16})$, $(1., \frac{1}{2})$, $(-1., -\frac{3}{8})$, $(-1., \frac{1}{8})$, $(1.41421, -\frac{3}{16})$, $(1.41421, \frac{5}{16})$, $(-1., -\frac{3}{8})$, $(-1., \frac{1}{8})$\}

\item $c = \frac{15}{2}$, $(d_i,\theta_i)$ = \{$(1., 0)$, $(-1.41421, -\frac{1}{16})$, $(1., \frac{1}{2})$, $(1., 0)$, $(-1.41421, -\frac{1}{16})$, $(1., \frac{1}{2})$, $(-1., \frac{3}{8})$, $(-1., -\frac{1}{8})$, $(1.41421, -\frac{7}{16})$, $(1.41421, \frac{1}{16})$, $(-1., \frac{3}{8})$, $(-1., -\frac{1}{8})$\}

\end{enumerate}

\paragraph*{\texorpdfstring{$6_1^{\zeta^1_2}$}{6\_1\^{zeta\^1\_2}}; self-dual}

$c = 1$, $(d_i,\theta_i)$ = \{$(1., 0)$, $(1., -\frac{1}{4})$, $(1., 0)$, $(1., -\frac{1}{4})$, $(2., \frac{1}{12})$, $(2., \frac{1}{3})$\}

\begin{enumerate}

\item $c = 1$, $(d_i,\theta_i)$ = \{$(1., 0)$, $(1., -\frac{1}{4})$, $(1., 0)$, $(1., -\frac{1}{4})$, $(2., \frac{1}{12})$, $(2., \frac{1}{3})$, $(1.73205, \frac{3}{8})$, $(1.73205, -\frac{3}{8})$, $(1.73205, -\frac{1}{8})$, $(1.73205, \frac{1}{8})$\}

\end{enumerate}

\paragraph*{\texorpdfstring{$6_1^{\zeta^1_2}$}{6\_1\^{zeta\^1\_2}}}

$c = 1$, $(d_i,\theta_i)$ = \{$(1., 0)$, $(1., -\frac{1}{4})$, $(1., 0)$, $(1., -\frac{1}{4})$, $(2., \frac{1}{12})$, $(2., \frac{1}{3})$\}

\begin{enumerate}

\item $c = 1$, $(d_i,\theta_i)$ = \{$(1., 0)$, $(1., -\frac{1}{4})$, $(1., 0)$, $(1., -\frac{1}{4})$, $(2., \frac{1}{12})$, $(2., \frac{1}{3})$, $(1.73205, \frac{1}{16})$, $(1.73205, \frac{1}{16})$, $(1.73205, -\frac{7}{16})$, $(1.73205, -\frac{7}{16})$\}

\item $c = 1$, $(d_i,\theta_i)$ = \{$(1., 0)$, $(1., -\frac{1}{4})$, $(1., 0)$, $(1., -\frac{1}{4})$, $(2., \frac{1}{12})$, $(2., \frac{1}{3})$, $(1.73205, \frac{5}{16})$, $(1.73205, \frac{5}{16})$, $(1.73205, -\frac{3}{16})$, $(1.73205, -\frac{3}{16})$\}

\end{enumerate}

\paragraph*{\texorpdfstring{$6_1^{\zeta^1_2*}$}{6\_1\^{zeta\^1\_2*}}}

$c = 1$, $(d_i,\theta_i)$ = \{$(1., 0)$, $(-1., -\frac{1}{4})$, $(1., 0)$, $(-1., -\frac{1}{4})$, $(-2., \frac{1}{12})$, $(2., \frac{1}{3})$\}

\begin{enumerate}

\item $c = 1$, $(d_i,\theta_i)$ = \{$(1., 0)$, $(-1., -\frac{1}{4})$, $(1., 0)$, $(-1., -\frac{1}{4})$, $(-2., \frac{1}{12})$, $(2., \frac{1}{3})$, $(-1.73205, \frac{3}{8})$, $(1.73205, -\frac{3}{8})$, $(-1.73205, -\frac{1}{8})$, $(1.73205, \frac{1}{8})$\}

\end{enumerate}

\paragraph*{\texorpdfstring{$6_3^{\zeta^1_2}$}{6\_3\^{zeta\^1\_2}}; self-dual}

$c = 3$, $(d_i,\theta_i)$ = \{$(1., 0)$, $(1., \frac{1}{4})$, $(1., 0)$, $(1., \frac{1}{4})$, $(2., -\frac{5}{12})$, $(2., \frac{1}{3})$\}

\begin{enumerate}

\item $c = 3$, $(d_i,\theta_i)$ = \{$(1., 0)$, $(1., \frac{1}{4})$, $(1., 0)$, $(1., \frac{1}{4})$, $(2., -\frac{5}{12})$, $(2., \frac{1}{3})$, $(1.73205, \frac{1}{8})$, $(1.73205, \frac{3}{8})$, $(1.73205, -\frac{3}{8})$, $(1.73205, -\frac{1}{8})$\}

\end{enumerate}

\paragraph*{\texorpdfstring{$6_3^{\zeta^1_2}$}{6\_3\^{zeta\^1\_2}}}

$c = 3$, $(d_i,\theta_i)$ = \{$(1., 0)$, $(1., \frac{1}{4})$, $(1., 0)$, $(1., \frac{1}{4})$, $(2., -\frac{5}{12})$, $(2., \frac{1}{3})$\}

\begin{enumerate}

\item $c = 3$, $(d_i,\theta_i)$ = \{$(1., 0)$, $(1., \frac{1}{4})$, $(1., 0)$, $(1., \frac{1}{4})$, $(2., -\frac{5}{12})$, $(2., \frac{1}{3})$, $(1.73205, \frac{3}{16})$, $(1.73205, \frac{3}{16})$, $(1.73205, -\frac{5}{16})$, $(1.73205, -\frac{5}{16})$\}

\item $c = 3$, $(d_i,\theta_i)$ = \{$(1., 0)$, $(1., \frac{1}{4})$, $(1., 0)$, $(1., \frac{1}{4})$, $(2., -\frac{5}{12})$, $(2., \frac{1}{3})$, $(1.73205, \frac{7}{16})$, $(1.73205, \frac{7}{16})$, $(1.73205, -\frac{1}{16})$, $(1.73205, -\frac{1}{16})$\}

\end{enumerate}

\paragraph*{\texorpdfstring{$6_3^{\zeta^1_2*}$}{6\_3\^{zeta\^1\_2*}}}

$c = 3$, $(d_i,\theta_i)$ = \{$(1., 0)$, $(-1., \frac{1}{4})$, $(1., 0)$, $(-1., \frac{1}{4})$, $(-2., -\frac{5}{12})$, $(2., \frac{1}{3})$\}

\begin{enumerate}

\item $c = 3$, $(d_i,\theta_i)$ = \{$(1., 0)$, $(-1., \frac{1}{4})$, $(1., 0)$, $(-1., \frac{1}{4})$, $(-2., -\frac{5}{12})$, $(2., \frac{1}{3})$, $(-1.73205, \frac{1}{8})$, $(1.73205, \frac{3}{8})$, $(-1.73205, -\frac{3}{8})$, $(1.73205, -\frac{1}{8})$\}

\end{enumerate}

\paragraph*{\texorpdfstring{$6_{-3}^{\zeta^1_2}$}{6\_{-3}\^{zeta\^1\_2}}; self-dual}

$c = 5$, $(d_i,\theta_i)$ = \{$(1., 0)$, $(1., -\frac{1}{4})$, $(1., 0)$, $(1., -\frac{1}{4})$, $(2., \frac{5}{12})$, $(2., -\frac{1}{3})$\}

\begin{enumerate}

\item $c = 5$, $(d_i,\theta_i)$ = \{$(1., 0)$, $(1., -\frac{1}{4})$, $(1., 0)$, $(1., -\frac{1}{4})$, $(2., \frac{5}{12})$, $(2., -\frac{1}{3})$, $(1.73205, -\frac{1}{8})$, $(1.73205, \frac{1}{8})$, $(1.73205, \frac{3}{8})$, $(1.73205, -\frac{3}{8})$\}

\end{enumerate}

\paragraph*{\texorpdfstring{$6_{-3}^{\zeta^1_2}$}{6\_{-3}\^{zeta\^1\_2}}}

$c = 5$, $(d_i,\theta_i)$ = \{$(1., 0)$, $(1., -\frac{1}{4})$, $(1., 0)$, $(1., -\frac{1}{4})$, $(2., \frac{5}{12})$, $(2., -\frac{1}{3})$\}

\begin{enumerate}

\item $c = 5$, $(d_i,\theta_i)$ = \{$(1., 0)$, $(1., -\frac{1}{4})$, $(1., 0)$, $(1., -\frac{1}{4})$, $(2., \frac{5}{12})$, $(2., -\frac{1}{3})$, $(1.73025, \frac{1}{16})$, $(1.73025, \frac{1}{16})$, $(1.73025, -\frac{7}{16})$, $(1.73025, -\frac{7}{16})$\}

\item $c = 5$, $(d_i,\theta_i)$ = \{$(1., 0)$, $(1., -\frac{1}{4})$, $(1., 0)$, $(1., -\frac{1}{4})$, $(2., \frac{5}{12})$, $(2., -\frac{1}{3})$, $(1.73025, \frac{5}{16})$, $(1.73025, \frac{5}{16})$, $(1.73025, -\frac{3}{16})$, $(1.73025, -\frac{3}{16})$\}

\end{enumerate}

\paragraph*{\texorpdfstring{$6_5^{\zeta^1_2*}$}{6\_5\^{zeta\^1\_2*}}}

$c = 5$, $(d_i,\theta_i)$ = \{$(1., 0)$, $(-1., -\frac{1}{4})$, $(1., 0)$, $(-1., -\frac{1}{4})$, $(-2., \frac{5}{12})$, $(2., -\frac{1}{3})$\}

\begin{enumerate}

\item $c = 5$, $(d_i,\theta_i)$ = \{$(1., 0)$, $(-1., -\frac{1}{4})$, $(1., 0)$, $(-1., -\frac{1}{4})$, $(-2., \frac{5}{12})$, $(2., -\frac{1}{3})$, $(-1.73205, -\frac{1}{8})$, $(1.73205, \frac{1}{8})$, $(-1.73205, \frac{3}{8})$, $(1.73205, -\frac{3}{8})$\}

\end{enumerate}

\paragraph*{\texorpdfstring{$6_{-1}^{\zeta^1_2}$}{6\_{-1}\^{zeta\^1\_2}}; self-dual}

$c = 7$, $(d_i,\theta_i)$ = \{$(1., 0)$, $(1., \frac{1}{4})$, $(1., 0)$, $(1., \frac{1}{4})$, $(2., -\frac{1}{12})$, $(2., -\frac{1}{3})$\}

\begin{enumerate}

\item $c = 7$, $(d_i,\theta_i)$ = \{$(1., 0)$, $(1., \frac{1}{4})$, $(1., 0)$, $(1., \frac{1}{4})$, $(2., -\frac{1}{12})$, $(2., -\frac{1}{3})$, $(1.73205, -\frac{3}{8})$, $(1.73205, -\frac{1}{8})$, $(1.73205, \frac{1}{8})$, $(1.73205, \frac{3}{8})$\}

\end{enumerate}

\paragraph*{\texorpdfstring{$6_{-1}^{\zeta^1_2}$}{6\_{-1}\^{zeta\^1\_2}}}

$c = 7$, $(d_i,\theta_i)$ = \{$(1., 0)$, $(1., \frac{1}{4})$, $(1., 0)$, $(1., \frac{1}{4})$, $(2., -\frac{1}{12})$, $(2., -\frac{1}{3})$\}

\begin{enumerate}

\item $c = 7$, $(d_i,\theta_i)$ = \{$(1., 0)$, $(1., \frac{1}{4})$, $(1., 0)$, $(1., \frac{1}{4})$, $(2., -\frac{1}{12})$, $(2., -\frac{1}{3})$, $(1.73205, \frac{3}{16})$, $(1.73205, \frac{3}{16})$, $(1.73205, -\frac{5}{16})$, $(1.73205, -\frac{5}{16})$\}

\item $c = 7$, $(d_i,\theta_i)$ = \{$(1., 0)$, $(1., \frac{1}{4})$, $(1., 0)$, $(1., \frac{1}{4})$, $(2., -\frac{1}{12})$, $(2., -\frac{1}{3})$, $(1.73205, \frac{7}{16})$, $(1.73205, \frac{7}{16})$, $(1.73205, -\frac{1}{16})$, $(1.73205, -\frac{1}{16})$\}

\end{enumerate}

\paragraph*{\texorpdfstring{$6_7^{\zeta^1_2*}$}{6\_7\^{zeta\^1\_2*}}}

$c = 7$, $(d_i,\theta_i)$ = \{$(1., 0)$, $(-1., \frac{1}{4})$, $(1., 0)$, $(-1., \frac{1}{4})$, $(-2., -\frac{1}{12})$, $(2., -\frac{1}{3})$\}

\begin{enumerate}

\item $c = 7$, $(d_i,\theta_i)$ = \{$(1., 0)$, $(-1., \frac{1}{4})$, $(1., 0)$, $(-1., \frac{1}{4})$, $(-2., -\frac{1}{12})$, $(2., -\frac{1}{3})$, $(-1.73205, -\frac{3}{8})$, $(1.73205, -\frac{1}{8})$, $(-1.73205, \frac{1}{8})$, $(1.73205, \frac{3}{8})$\}

\end{enumerate}

\paragraph*{\texorpdfstring{$6_0^{\zeta^1_2}$}{6\_0\^{zeta\^1\_2}}}

$c = 0$, $(d_i,\theta_i)$ = \{$(1., 0)$, $(1., 0)$, $(2., 0)$, $(2., 0)$, $(2., \frac{1}{3})$, $(2., -\frac{1}{3})$\}

\begin{enumerate}

\item $c = 0$, $(d_i,\theta_i)$ = \{$(1., 0)$, $(1., 0)$, $(2., 0)$, $(2., 0)$, $(2., \frac{1}{3})$, $(2., -\frac{1}{3})$, $(3., -\frac{1}{4})$, $(3., \frac{1}{4})$\}

\item $c = 0$, $(d_i,\theta_i)$ = \{$(1., 0)$, $(1., 0)$, $(2., 0)$, $(2., 0)$, $(2., \frac{1}{3})$, $(2., -\frac{1}{3})$, $(3., 0)$, $(3., \frac{1}{2})$\}

\end{enumerate}

\paragraph*{\texorpdfstring{$6_0^{\zeta^1_2}$}{6\_0\^{zeta\^1\_2}}}

$(d_i,\theta_i)$ = \{$(1., 0)$, $(1., 0)$, $(2., 0)$, $(2., \frac{4}{9})$, $(2., -\frac{2}{9})$, $(2., \frac{1}{9})$\}

\begin{enumerate}

\item $c = 0$, $(d_i,\theta_i)$ = \{$(1., 0)$, $(1., 0)$, $(2., 0)$, $(2., \frac{4}{9})$, $(2., -\frac{2}{9})$, $(2., \frac{1}{9})$, $(3., -\frac{1}{4})$, $(3., \frac{1}{4})$\}

\item $c = 0$, $(d_i,\theta_i)$ = \{$(1., 0)$, $(1., 0)$, $(2., 0)$, $(2., \frac{4}{9})$, $(2., -\frac{2}{9})$, $(2., \frac{1}{9})$, $(3., 0)$, $(3., \frac{1}{2})$\}

\end{enumerate}

\paragraph*{\texorpdfstring{$6_0^{\zeta^1_2}$}{6\_0\^{zeta\^1\_2}}}

$c = 0$, $(d_i,\theta_i)$ = \{$(1., 0)$, $(1., 0)$, $(2., 0)$, $(2., -\frac{4}{9})$, $(2., \frac{2}{9})$, $(2., -\frac{1}{9})$\}

\begin{enumerate}

\item $c = 0$, $(d_i,\theta_i)$ = \{$(1., 0)$, $(1., 0)$, $(2., 0)$, $(2., -\frac{4}{9})$, $(2., \frac{2}{9})$, $(2., -\frac{1}{9})$, $(3., -\frac{1}{4})$, $(3., \frac{1}{4})$\}

\item $c = 0$, $(d_i,\theta_i)$ = \{$(1., 0)$, $(1., 0)$, $(2., 0)$, $(2., -\frac{4}{9})$, $(2., \frac{2}{9})$, $(2., -\frac{1}{9})$, $(3., 0)$, $(3., \frac{1}{2})$\}

\end{enumerate}

\paragraph*{\texorpdfstring{$6_4^{\zeta^1_2}$}{6\_4\^{zeta\^1\_2}}}

$c = 4$, $(d_i,\theta_i)$ = \{$(1., 0)$, $(1., 0)$, $(2., \frac{1}{3})$, $(2., \frac{1}{3})$, $(2., -\frac{1}{3})$, $(2., -\frac{1}{3})$\}

\begin{enumerate}

\item $c = 4$, $(d_i,\theta_i)$ = \{$(1., 0)$, $(1., 0)$, $(2., \frac{1}{3})$, $(2., \frac{1}{3})$, $(2., -\frac{1}{3})$, $(2., -\frac{1}{3})$, $(3., \frac{1}{4})$, $(3., -\frac{1}{4})$\}

\item $c = 4$, $(d_i,\theta_i)$ = \{$(1., 0)$, $(1., 0)$, $(2., \frac{1}{3})$, $(2., \frac{1}{3})$, $(2., -\frac{1}{3})$, $(2., -\frac{1}{3})$, $(3., 0)$, $(3., \frac{1}{2})$\}

\end{enumerate}

\end{widetext}

\bibliographystyle{apsrev4-2}
\bibliography{ref}

\begin{thebibliography}{55}%
\makeatletter
\providecommand \@ifxundefined [1]{%
 \@ifx{#1\undefined}
}%
\providecommand \@ifnum [1]{%
 \ifnum #1\expandafter \@firstoftwo
 \else \expandafter \@secondoftwo
 \fi
}%
\providecommand \@ifx [1]{%
 \ifx #1\expandafter \@firstoftwo
 \else \expandafter \@secondoftwo
 \fi
}%
\providecommand \natexlab [1]{#1}%
\providecommand \enquote  [1]{``#1''}%
\providecommand \bibnamefont  [1]{#1}%
\providecommand \bibfnamefont [1]{#1}%
\providecommand \citenamefont [1]{#1}%
\providecommand \href@noop [0]{\@secondoftwo}%
\providecommand \href [0]{\begingroup \@sanitize@url \@href}%
\providecommand \@href[1]{\@@startlink{#1}\@@href}%
\providecommand \@@href[1]{\endgroup#1\@@endlink}%
\providecommand \@sanitize@url [0]{\catcode `\\12\catcode `\$12\catcode
  `\&12\catcode `\#12\catcode `\^12\catcode `\_12\catcode `\%12\relax}%
\providecommand \@@startlink[1]{}%
\providecommand \@@endlink[0]{}%
\providecommand \url  [0]{\begingroup\@sanitize@url \@url }%
\providecommand \@url [1]{\endgroup\@href {#1}{\urlprefix }}%
\providecommand \urlprefix  [0]{URL }%
\providecommand \Eprint [0]{\href }%
\providecommand \doibase [0]{https://doi.org/}%
\providecommand \selectlanguage [0]{\@gobble}%
\providecommand \bibinfo  [0]{\@secondoftwo}%
\providecommand \bibfield  [0]{\@secondoftwo}%
\providecommand \translation [1]{[#1]}%
\providecommand \BibitemOpen [0]{}%
\providecommand \bibitemStop [0]{}%
\providecommand \bibitemNoStop [0]{.\EOS\space}%
\providecommand \EOS [0]{\spacefactor3000\relax}%
\providecommand \BibitemShut  [1]{\csname bibitem#1\endcsname}%
\let\auto@bib@innerbib\@empty
\bibitem [{\citenamefont {Wen}(1989)}]{Wen1989:degeneracy}%
  \BibitemOpen
  \bibfield  {author} {\bibinfo {author} {\bibfnamefont {X.~G.}\ \bibnamefont
  {Wen}},\ }\href {https://doi.org/10.1103/PhysRevB.40.7387} {\bibfield
  {journal} {\bibinfo  {journal} {Phys. Rev. B}\ }\textbf {\bibinfo {volume}
  {40}},\ \bibinfo {pages} {7387} (\bibinfo {year} {1989})}\BibitemShut
  {NoStop}%
\bibitem [{\citenamefont {Wen}(1990)}]{Wen1990:rigidstates}%
  \BibitemOpen
  \bibfield  {author} {\bibinfo {author} {\bibfnamefont {X.-G.}\ \bibnamefont
  {Wen}},\ }\href {https://doi.org/https://doi.org/10.1142/S0217979290000139}
  {\bibfield  {journal} {\bibinfo  {journal} {Int. J. Mod. Phys. B}\ }\textbf
  {\bibinfo {volume} {4}},\ \bibinfo {pages} {239} (\bibinfo {year}
  {1990})}\BibitemShut {NoStop}%
\bibitem [{\citenamefont {Keski-Vakkuri}\ and\ \citenamefont
  {Wen}(1993)}]{KesWen1993:fqh}%
  \BibitemOpen
  \bibfield  {author} {\bibinfo {author} {\bibfnamefont {E.}~\bibnamefont
  {Keski-Vakkuri}}\ and\ \bibinfo {author} {\bibfnamefont {X.-G.}\ \bibnamefont
  {Wen}},\ }\href {https://doi.org/10.1142/s0217979293003644} {\bibfield
  {journal} {\bibinfo  {journal} {Int. J. Mod. Phys. B}\ }\textbf {\bibinfo
  {volume} {07}},\ \bibinfo {pages} {4227} (\bibinfo {year}
  {1993})}\BibitemShut {NoStop}%
\bibitem [{\citenamefont {Wen}(2002)}]{Wen2002:quantumorder}%
  \BibitemOpen
  \bibfield  {author} {\bibinfo {author} {\bibfnamefont {X.-G.}\ \bibnamefont
  {Wen}},\ }\href
  {https://doi.org/https://doi.org/10.1016/S0375-9601(02)00808-3} {\bibfield
  {journal} {\bibinfo  {journal} {Phys. Lett. A}\ }\textbf {\bibinfo {volume}
  {300}},\ \bibinfo {pages} {175} (\bibinfo {year} {2002})}\BibitemShut
  {NoStop}%
\bibitem [{\citenamefont {Wen}(2017)}]{Wen2017:zoo}%
  \BibitemOpen
  \bibfield  {author} {\bibinfo {author} {\bibfnamefont {X.-G.}\ \bibnamefont
  {Wen}},\ }\href {https://doi.org/10.1103/RevModPhys.89.041004} {\bibfield
  {journal} {\bibinfo  {journal} {Rev. Mod. Phys.}\ }\textbf {\bibinfo {volume}
  {89}},\ \bibinfo {pages} {041004} (\bibinfo {year} {2017})}\BibitemShut
  {NoStop}%
\bibitem [{\citenamefont {Tsui}\ \emph {et~al.}(1982)\citenamefont {Tsui},
  \citenamefont {Stormer},\ and\ \citenamefont {Gossard}}]{TsuStoGos1982:fqhe}%
  \BibitemOpen
  \bibfield  {author} {\bibinfo {author} {\bibfnamefont {D.~C.}\ \bibnamefont
  {Tsui}}, \bibinfo {author} {\bibfnamefont {H.~L.}\ \bibnamefont {Stormer}},\
  and\ \bibinfo {author} {\bibfnamefont {A.~C.}\ \bibnamefont {Gossard}},\
  }\href {https://doi.org/10.1103/PhysRevLett.48.1559} {\bibfield  {journal}
  {\bibinfo  {journal} {Phys. Rev. Lett.}\ }\textbf {\bibinfo {volume} {48}},\
  \bibinfo {pages} {1559} (\bibinfo {year} {1982})}\BibitemShut {NoStop}%
\bibitem [{\citenamefont {Stormer}\ \emph {et~al.}(1999)\citenamefont
  {Stormer}, \citenamefont {Tsui},\ and\ \citenamefont
  {Gossard}}]{StoTsuGos1999:fqhe}%
  \BibitemOpen
  \bibfield  {author} {\bibinfo {author} {\bibfnamefont {H.~L.}\ \bibnamefont
  {Stormer}}, \bibinfo {author} {\bibfnamefont {D.~C.}\ \bibnamefont {Tsui}},\
  and\ \bibinfo {author} {\bibfnamefont {A.~C.}\ \bibnamefont {Gossard}},\
  }\href {https://doi.org/10.1103/RevModPhys.71.S298} {\bibfield  {journal}
  {\bibinfo  {journal} {Rev. Mod. Phys.}\ }\textbf {\bibinfo {volume} {71}},\
  \bibinfo {pages} {S298} (\bibinfo {year} {1999})}\BibitemShut {NoStop}%
\bibitem [{\citenamefont {Kitaev}(2003)}]{Kit2003:faulttolerant}%
  \BibitemOpen
  \bibfield  {author} {\bibinfo {author} {\bibfnamefont {A.~Y.}\ \bibnamefont
  {Kitaev}},\ }\href
  {https://doi.org/https://doi.org/10.1016/S0003-4916(02)00018-0} {\bibfield
  {journal} {\bibinfo  {journal} {Annals of Physics}\ }\textbf {\bibinfo
  {volume} {303}},\ \bibinfo {pages} {2} (\bibinfo {year} {2003})}\BibitemShut
  {NoStop}%
\bibitem [{\citenamefont {Savary}\ and\ \citenamefont
  {Balents}(2017)}]{SavBal2017:qsl}%
  \BibitemOpen
  \bibfield  {author} {\bibinfo {author} {\bibfnamefont {L.}~\bibnamefont
  {Savary}}\ and\ \bibinfo {author} {\bibfnamefont {L.}~\bibnamefont
  {Balents}},\ }\href {https://doi.org/10.1088/0034-4885/80/1/016502}
  {\bibfield  {journal} {\bibinfo  {journal} {Reports on Progress in Physics}\
  }\textbf {\bibinfo {volume} {80}},\ \bibinfo {pages} {016502} (\bibinfo
  {year} {2017})}\BibitemShut {NoStop}%
\bibitem [{\citenamefont {Witten}(1988)}]{Wit1988:tqft}%
  \BibitemOpen
  \bibfield  {author} {\bibinfo {author} {\bibfnamefont {E.}~\bibnamefont
  {Witten}},\ }\href {https://doi.org/10.1007/BF01223371} {\bibfield  {journal}
  {\bibinfo  {journal} {Commun. Math. Phys.}\ }\textbf {\bibinfo {volume}
  {117}},\ \bibinfo {pages} {353} (\bibinfo {year} {1988})}\BibitemShut
  {NoStop}%
\bibitem [{\citenamefont {Moore}\ and\ \citenamefont
  {Seiberg}(1989)}]{MooSei1989:cft}%
  \BibitemOpen
  \bibfield  {author} {\bibinfo {author} {\bibfnamefont {G.}~\bibnamefont
  {Moore}}\ and\ \bibinfo {author} {\bibfnamefont {N.}~\bibnamefont
  {Seiberg}},\ }\href {https://doi.org/10.1007/BF01238857} {\bibfield
  {journal} {\bibinfo  {journal} {Commun. Math. Phys.}\ }\textbf {\bibinfo
  {volume} {123}},\ \bibinfo {pages} {177} (\bibinfo {year}
  {1989})}\BibitemShut {NoStop}%
\bibitem [{\citenamefont {Nayak}\ \emph {et~al.}(2008)\citenamefont {Nayak},
  \citenamefont {Simon}, \citenamefont {Stern}, \citenamefont {Freedman},\ and\
  \citenamefont {Das~Sarma}}]{NaySimSte-etal2008:tqc}%
  \BibitemOpen
  \bibfield  {author} {\bibinfo {author} {\bibfnamefont {C.}~\bibnamefont
  {Nayak}}, \bibinfo {author} {\bibfnamefont {S.~H.}\ \bibnamefont {Simon}},
  \bibinfo {author} {\bibfnamefont {A.}~\bibnamefont {Stern}}, \bibinfo
  {author} {\bibfnamefont {M.}~\bibnamefont {Freedman}},\ and\ \bibinfo
  {author} {\bibfnamefont {S.}~\bibnamefont {Das~Sarma}},\ }\href
  {https://doi.org/10.1103/RevModPhys.80.1083} {\bibfield  {journal} {\bibinfo
  {journal} {Rev. Mod. Phys.}\ }\textbf {\bibinfo {volume} {80}},\ \bibinfo
  {pages} {1083} (\bibinfo {year} {2008})}\BibitemShut {NoStop}%
\bibitem [{\citenamefont {Kitaev}\ and\ \citenamefont
  {Laumann}(2009)}]{KitLau2009:tqc}%
  \BibitemOpen
  \bibfield  {author} {\bibinfo {author} {\bibfnamefont {A.}~\bibnamefont
  {Kitaev}}\ and\ \bibinfo {author} {\bibfnamefont {C.}~\bibnamefont
  {Laumann}},\ }\href@noop {} {\bibinfo {title} {Topological phases and quantum
  computation}} (\bibinfo {year} {2009}),\ \Eprint
  {https://arxiv.org/abs/0904.2771} {arXiv:0904.2771 [cond-mat.mes-hall]}
  \BibitemShut {NoStop}%
\bibitem [{\citenamefont {Wang}(2010)}]{Wan2010:tqc}%
  \BibitemOpen
  \bibfield  {author} {\bibinfo {author} {\bibfnamefont {Z.}~\bibnamefont
  {Wang}},\ }\href@noop {} {\emph {\bibinfo {title} {Topological quantum
  computation}}},\ \bibinfo {number} {112}\ (\bibinfo  {publisher} {American
  Mathematical Society},\ \bibinfo {year} {2010})\BibitemShut {NoStop}%
\bibitem [{\citenamefont {Stern}\ and\ \citenamefont
  {Lindner}(2013)}]{SteLin2013:tqc}%
  \BibitemOpen
  \bibfield  {author} {\bibinfo {author} {\bibfnamefont {A.}~\bibnamefont
  {Stern}}\ and\ \bibinfo {author} {\bibfnamefont {N.~H.}\ \bibnamefont
  {Lindner}},\ }\href {https://doi.org/10.1126/science.1231473} {\bibfield
  {journal} {\bibinfo  {journal} {Science}\ }\textbf {\bibinfo {volume}
  {339}},\ \bibinfo {pages} {1179} (\bibinfo {year} {2013})}\BibitemShut
  {NoStop}%
\bibitem [{\citenamefont {Bernevig}\ and\ \citenamefont
  {Neupert}(2015)}]{BerNeu2015:category}%
  \BibitemOpen
  \bibfield  {author} {\bibinfo {author} {\bibfnamefont {A.}~\bibnamefont
  {Bernevig}}\ and\ \bibinfo {author} {\bibfnamefont {T.}~\bibnamefont
  {Neupert}},\ }\href@noop {} {\bibinfo {title} {Topological superconductors
  and category theory}} (\bibinfo {year} {2015}),\ \Eprint
  {https://arxiv.org/abs/1506.05805} {arXiv:1506.05805 [cond-mat.str-el]}
  \BibitemShut {NoStop}%
\bibitem [{\citenamefont {Wen}(2015)}]{Wen2015:bosonic}%
  \BibitemOpen
  \bibfield  {author} {\bibinfo {author} {\bibfnamefont {X.-G.}\ \bibnamefont
  {Wen}},\ }\href {https://doi.org/10.1093/nsr/nwv077} {\bibfield  {journal}
  {\bibinfo  {journal} {Natl. Sci. Rev.}\ }\textbf {\bibinfo {volume} {3}},\
  \bibinfo {pages} {68} (\bibinfo {year} {2015})}\BibitemShut {NoStop}%
\bibitem [{\citenamefont {Lan}\ \emph {et~al.}(2016{\natexlab{a}})\citenamefont
  {Lan}, \citenamefont {Kong},\ and\ \citenamefont
  {Wen}}]{LanKonWen2016:fermionic}%
  \BibitemOpen
  \bibfield  {author} {\bibinfo {author} {\bibfnamefont {T.}~\bibnamefont
  {Lan}}, \bibinfo {author} {\bibfnamefont {L.}~\bibnamefont {Kong}},\ and\
  \bibinfo {author} {\bibfnamefont {X.-G.}\ \bibnamefont {Wen}},\ }\href
  {https://doi.org/10.1103/PhysRevB.94.155113} {\bibfield  {journal} {\bibinfo
  {journal} {Phys. Rev. B}\ }\textbf {\bibinfo {volume} {94}},\ \bibinfo
  {pages} {155113} (\bibinfo {year} {2016}{\natexlab{a}})}\BibitemShut
  {NoStop}%
\bibitem [{\citenamefont {Kong}\ and\ \citenamefont
  {Zhang}(2022)}]{KonZha2022:invitation}%
  \BibitemOpen
  \bibfield  {author} {\bibinfo {author} {\bibfnamefont {L.}~\bibnamefont
  {Kong}}\ and\ \bibinfo {author} {\bibfnamefont {Z.-H.}\ \bibnamefont
  {Zhang}},\ }\href@noop {} {\bibinfo {title} {An invitation to topological
  orders and category theory}} (\bibinfo {year} {2022}),\ \Eprint
  {https://arxiv.org/abs/2205.05565} {arXiv:2205.05565 [cond-mat.str-el]}
  \BibitemShut {NoStop}%
\bibitem [{\citenamefont {Ng}\ \emph {et~al.}(2023{\natexlab{a}})\citenamefont
  {Ng}, \citenamefont {Rowell}, \citenamefont {Wang},\ and\ \citenamefont
  {Wen}}]{NgRowWanWen2023:reconstruction}%
  \BibitemOpen
  \bibfield  {author} {\bibinfo {author} {\bibfnamefont {S.-H.}\ \bibnamefont
  {Ng}}, \bibinfo {author} {\bibfnamefont {E.~C.}\ \bibnamefont {Rowell}},
  \bibinfo {author} {\bibfnamefont {Z.}~\bibnamefont {Wang}},\ and\ \bibinfo
  {author} {\bibfnamefont {X.-G.}\ \bibnamefont {Wen}},\ }\bibfield  {journal}
  {\bibinfo  {journal} {Commun. Math. Phys.}\ }\href
  {https://doi.org/10.1007/s00220-023-04775-w} {10.1007/s00220-023-04775-w}
  (\bibinfo {year} {2023}{\natexlab{a}})\BibitemShut {NoStop}%
\bibitem [{\citenamefont {Bruillard}\ \emph {et~al.}(2017)\citenamefont
  {Bruillard}, \citenamefont {Galindo}, \citenamefont {Hagge}, \citenamefont
  {Ng}, \citenamefont {Plavnik}, \citenamefont {Rowell},\ and\ \citenamefont
  {Wang}}]{BruGalHag-etal2017:16fold}%
  \BibitemOpen
  \bibfield  {author} {\bibinfo {author} {\bibfnamefont {P.}~\bibnamefont
  {Bruillard}}, \bibinfo {author} {\bibfnamefont {C.}~\bibnamefont {Galindo}},
  \bibinfo {author} {\bibfnamefont {T.}~\bibnamefont {Hagge}}, \bibinfo
  {author} {\bibfnamefont {S.-H.}\ \bibnamefont {Ng}}, \bibinfo {author}
  {\bibfnamefont {J.~Y.}\ \bibnamefont {Plavnik}}, \bibinfo {author}
  {\bibfnamefont {E.~C.}\ \bibnamefont {Rowell}},\ and\ \bibinfo {author}
  {\bibfnamefont {Z.}~\bibnamefont {Wang}},\ }\bibfield  {journal} {\bibinfo
  {journal} {J. Math. Phys.}\ }\textbf {\bibinfo {volume} {58}},\ \href
  {https://doi.org/10.1063/1.4982048} {10.1063/1.4982048} (\bibinfo {year}
  {2017})\BibitemShut {NoStop}%
\bibitem [{\citenamefont {Cho}\ \emph {et~al.}(2023)\citenamefont {Cho},
  \citenamefont {Kim}, \citenamefont {Seo},\ and\ \citenamefont
  {You}}]{ChoKimSeoYou2022:classification}%
  \BibitemOpen
  \bibfield  {author} {\bibinfo {author} {\bibfnamefont {G.~Y.}\ \bibnamefont
  {Cho}}, \bibinfo {author} {\bibfnamefont {H.-C.}\ \bibnamefont {Kim}},
  \bibinfo {author} {\bibfnamefont {D.}~\bibnamefont {Seo}},\ and\ \bibinfo
  {author} {\bibfnamefont {M.}~\bibnamefont {You}},\ }\href
  {https://doi.org/10.1103/PhysRevB.108.115103} {\bibfield  {journal} {\bibinfo
   {journal} {Phys. Rev. B}\ }\textbf {\bibinfo {volume} {108}},\ \bibinfo
  {pages} {115103} (\bibinfo {year} {2023})}\BibitemShut {NoStop}%
\bibitem [{\citenamefont {Lan}\ \emph {et~al.}(2017)\citenamefont {Lan},
  \citenamefont {Kong},\ and\ \citenamefont
  {Wen}}]{LanKonWen2017:symmetryenriched}%
  \BibitemOpen
  \bibfield  {author} {\bibinfo {author} {\bibfnamefont {T.}~\bibnamefont
  {Lan}}, \bibinfo {author} {\bibfnamefont {L.}~\bibnamefont {Kong}},\ and\
  \bibinfo {author} {\bibfnamefont {X.-G.}\ \bibnamefont {Wen}},\ }\href
  {https://doi.org/10.1103/PhysRevB.95.235140} {\bibfield  {journal} {\bibinfo
  {journal} {Phys. Rev. B}\ }\textbf {\bibinfo {volume} {95}},\ \bibinfo
  {pages} {235140} (\bibinfo {year} {2017})}\BibitemShut {NoStop}%
\bibitem [{\citenamefont {Cho}\ \emph {et~al.}(2020)\citenamefont {Cho},
  \citenamefont {Gang},\ and\ \citenamefont {Kim}}]{ChoGanKim2020:mtheory}%
  \BibitemOpen
  \bibfield  {author} {\bibinfo {author} {\bibfnamefont {G.~Y.}\ \bibnamefont
  {Cho}}, \bibinfo {author} {\bibfnamefont {D.}~\bibnamefont {Gang}},\ and\
  \bibinfo {author} {\bibfnamefont {H.-C.}\ \bibnamefont {Kim}},\ }\href
  {https://doi.org/10.1007/jhep11(2020)115} {\bibfield  {journal} {\bibinfo
  {journal} {J. High Energy Phys.}\ }\textbf {\bibinfo {volume} {2020}}\bibinfo
   {number} { (11)}}\BibitemShut {NoStop}%
\bibitem [{\citenamefont {Alekseyev}\ \emph {et~al.}(2023)\citenamefont
  {Alekseyev}, \citenamefont {Bruns}, \citenamefont {Palcoux},\ and\
  \citenamefont {Petrov}}]{integralrank12}%
  \BibitemOpen
\bibfield  {number} {  }\bibfield  {author} {\bibinfo {author} {\bibfnamefont
  {M.~A.}\ \bibnamefont {Alekseyev}}, \bibinfo {author} {\bibfnamefont
  {W.}~\bibnamefont {Bruns}}, \bibinfo {author} {\bibfnamefont
  {S.}~\bibnamefont {Palcoux}},\ and\ \bibinfo {author} {\bibfnamefont {F.~V.}\
  \bibnamefont {Petrov}},\ }\href@noop {} {\bibinfo {title} {Classification of
  modular data of integral modular fusion categories up to rank 12}} (\bibinfo
  {year} {2023}),\ \Eprint {https://arxiv.org/abs/2302.01613} {arXiv:2302.01613
  [math.QA]} \BibitemShut {NoStop}%
\bibitem [{\citenamefont {Ng}\ \emph {et~al.}(2023{\natexlab{b}})\citenamefont
  {Ng}, \citenamefont {Rowell},\ and\ \citenamefont
  {Wen}}]{NgRowWen2023:rank11}%
  \BibitemOpen
  \bibfield  {author} {\bibinfo {author} {\bibfnamefont {S.-H.}\ \bibnamefont
  {Ng}}, \bibinfo {author} {\bibfnamefont {E.~C.}\ \bibnamefont {Rowell}},\
  and\ \bibinfo {author} {\bibfnamefont {X.-G.}\ \bibnamefont {Wen}},\
  }\href@noop {} {\bibinfo {title} {Classification of modular data up to rank
  11}} (\bibinfo {year} {2023}{\natexlab{b}}),\ \Eprint
  {https://arxiv.org/abs/2308.09670} {arXiv:2308.09670 [math.QA]} \BibitemShut
  {NoStop}%
\bibitem [{\citenamefont {Mignard}\ and\ \citenamefont
  {Schauenburg}(2021)}]{MigSch2021:notdeterminedby}%
  \BibitemOpen
  \bibfield  {author} {\bibinfo {author} {\bibfnamefont {M.}~\bibnamefont
  {Mignard}}\ and\ \bibinfo {author} {\bibfnamefont {P.}~\bibnamefont
  {Schauenburg}},\ }\href {https://doi.org/10.1007/s11005-021-01395-0}
  {\bibfield  {journal} {\bibinfo  {journal} {Lett. Math. Phys.}\ }\textbf
  {\bibinfo {volume} {111}},\ \bibinfo {pages} {60} (\bibinfo {year}
  {2021})}\BibitemShut {NoStop}%
\bibitem [{\citenamefont {Wen}\ and\ \citenamefont
  {Wen}(2019)}]{WenWen2019:highergenus}%
  \BibitemOpen
  \bibfield  {author} {\bibinfo {author} {\bibfnamefont {X.}~\bibnamefont
  {Wen}}\ and\ \bibinfo {author} {\bibfnamefont {X.-G.}\ \bibnamefont {Wen}},\
  }\href@noop {} {\bibinfo {title} {Distinguish modular categories and 2+1d
  topological orders beyond modular data: Mapping class group of higher genus
  manifold}} (\bibinfo {year} {2019}),\ \Eprint
  {https://arxiv.org/abs/1908.10381} {arXiv:1908.10381 [cond-mat.str-el]}
  \BibitemShut {NoStop}%
\bibitem [{\citenamefont {Bonderson}\ \emph {et~al.}(2018)\citenamefont
  {Bonderson}, \citenamefont {Rowell}, \citenamefont {Wang},\ and\
  \citenamefont {Zhang}}]{BonRowWanZha2018:congruence}%
  \BibitemOpen
  \bibfield  {author} {\bibinfo {author} {\bibfnamefont {P.}~\bibnamefont
  {Bonderson}}, \bibinfo {author} {\bibfnamefont {E.}~\bibnamefont {Rowell}},
  \bibinfo {author} {\bibfnamefont {Z.}~\bibnamefont {Wang}},\ and\ \bibinfo
  {author} {\bibfnamefont {Q.}~\bibnamefont {Zhang}},\ }\href
  {https://doi.org/10.2140/pjm.2018.296.257} {\bibfield  {journal} {\bibinfo
  {journal} {Pac. J. Math.}\ }\textbf {\bibinfo {volume} {296}},\ \bibinfo
  {pages} {257} (\bibinfo {year} {2018})}\BibitemShut {NoStop}%
\bibitem [{\citenamefont {Ng}\ \emph {et~al.}(2023{\natexlab{c}})\citenamefont
  {Ng}, \citenamefont {Wang},\ and\ \citenamefont
  {Wilson}}]{NgWanWil2023;symmetric}%
  \BibitemOpen
  \bibfield  {author} {\bibinfo {author} {\bibfnamefont {S.-H.}\ \bibnamefont
  {Ng}}, \bibinfo {author} {\bibfnamefont {Y.}~\bibnamefont {Wang}},\ and\
  \bibinfo {author} {\bibfnamefont {S.}~\bibnamefont {Wilson}},\ }\href
  {https://doi.org/https://doi.org/10.1090/proc/16205} {\bibfield  {journal}
  {\bibinfo  {journal} {Proc. Am. Math. Soc.}\ }\textbf {\bibinfo {volume}
  {151}},\ \bibinfo {pages} {1415} (\bibinfo {year}
  {2023}{\natexlab{c}})}\BibitemShut {NoStop}%
\bibitem [{\citenamefont {Ng}\ \emph {et~al.}(2023{\natexlab{d}})\citenamefont
  {Ng}, \citenamefont {Rowell},\ and\ \citenamefont
  {Wen}}]{ng2023classification}%
  \BibitemOpen
  \bibfield  {author} {\bibinfo {author} {\bibfnamefont {S.-H.}\ \bibnamefont
  {Ng}}, \bibinfo {author} {\bibfnamefont {E.~C.}\ \bibnamefont {Rowell}},\
  and\ \bibinfo {author} {\bibfnamefont {X.-G.}\ \bibnamefont {Wen}},\
  }\href@noop {} {\bibinfo {title} {Classification of modular data up to rank
  11}} (\bibinfo {year} {2023}{\natexlab{d}}),\ \Eprint
  {https://arxiv.org/abs/2308.09670} {arXiv:2308.09670 [math.QA]} \BibitemShut
  {NoStop}%
\bibitem [{\citenamefont {Lan}\ \emph {et~al.}(2016{\natexlab{b}})\citenamefont
  {Lan}, \citenamefont {Kong},\ and\ \citenamefont {Wen}}]{LanKonWen2016:mex}%
  \BibitemOpen
  \bibfield  {author} {\bibinfo {author} {\bibfnamefont {T.}~\bibnamefont
  {Lan}}, \bibinfo {author} {\bibfnamefont {L.}~\bibnamefont {Kong}},\ and\
  \bibinfo {author} {\bibfnamefont {X.-G.}\ \bibnamefont {Wen}},\ }\href
  {https://doi.org/10.1007/s00220-016-2748-y} {\bibfield  {journal} {\bibinfo
  {journal} {Commun. Math. Phys.}\ }\textbf {\bibinfo {volume} {351}},\
  \bibinfo {pages} {709} (\bibinfo {year} {2016}{\natexlab{b}})}\BibitemShut
  {NoStop}%
\bibitem [{\citenamefont {Rowell}\ \emph {et~al.}(2023)\citenamefont {Rowell},
  \citenamefont {Solomon},\ and\ \citenamefont
  {Zhang}}]{RowSolZha2023:neargroup}%
  \BibitemOpen
  \bibfield  {author} {\bibinfo {author} {\bibfnamefont {E.~C.}\ \bibnamefont
  {Rowell}}, \bibinfo {author} {\bibfnamefont {H.}~\bibnamefont {Solomon}},\
  and\ \bibinfo {author} {\bibfnamefont {Q.}~\bibnamefont {Zhang}},\
  }\href@noop {} {\bibinfo {title} {On near-group centers and super-modular
  categories}} (\bibinfo {year} {2023}),\ \Eprint
  {https://arxiv.org/abs/2305.09108} {arXiv:2305.09108 [math.QA]} \BibitemShut
  {NoStop}%
\bibitem [{\citenamefont {Bakalov}\ and\ \citenamefont
  {Kirillov}(2001)}]{BakKir2001:lectures}%
  \BibitemOpen
  \bibfield  {author} {\bibinfo {author} {\bibfnamefont {B.}~\bibnamefont
  {Bakalov}}\ and\ \bibinfo {author} {\bibfnamefont {A.}~\bibnamefont
  {Kirillov}},\ }\href@noop {} {\emph {\bibinfo {title} {Lectures on tensor
  categories and modular functors}}},\ Vol.~\bibinfo {volume} {21}\ (\bibinfo
  {publisher} {American Mathematical Society},\ \bibinfo {year}
  {2001})\BibitemShut {NoStop}%
\bibitem [{\citenamefont {Mac~Lane}(2013)}]{Mac2013:workingmathematician}%
  \BibitemOpen
  \bibfield  {author} {\bibinfo {author} {\bibfnamefont {S.}~\bibnamefont
  {Mac~Lane}},\ }\href@noop {} {\emph {\bibinfo {title} {Categories for the
  working mathematician}}},\ Vol.~\bibinfo {volume} {5}\ (\bibinfo  {publisher}
  {Springer Science \& Business Media},\ \bibinfo {year} {2013})\BibitemShut
  {NoStop}%
\bibitem [{\citenamefont {Kitaev}\ and\ \citenamefont
  {Preskill}(2006)}]{KitPre2006:entropy}%
  \BibitemOpen
  \bibfield  {author} {\bibinfo {author} {\bibfnamefont {A.}~\bibnamefont
  {Kitaev}}\ and\ \bibinfo {author} {\bibfnamefont {J.}~\bibnamefont
  {Preskill}},\ }\href {https://doi.org/10.1103/PhysRevLett.96.110404}
  {\bibfield  {journal} {\bibinfo  {journal} {Phys. Rev. Lett.}\ }\textbf
  {\bibinfo {volume} {96}},\ \bibinfo {pages} {110404} (\bibinfo {year}
  {2006})}\BibitemShut {NoStop}%
\bibitem [{\citenamefont {Levin}\ and\ \citenamefont
  {Wen}(2006)}]{LevWen2006:entropy}%
  \BibitemOpen
  \bibfield  {author} {\bibinfo {author} {\bibfnamefont {M.}~\bibnamefont
  {Levin}}\ and\ \bibinfo {author} {\bibfnamefont {X.-G.}\ \bibnamefont
  {Wen}},\ }\href {https://doi.org/10.1103/PhysRevLett.96.110405} {\bibfield
  {journal} {\bibinfo  {journal} {Phys. Rev. Lett.}\ }\textbf {\bibinfo
  {volume} {96}},\ \bibinfo {pages} {110405} (\bibinfo {year}
  {2006})}\BibitemShut {NoStop}%
\bibitem [{\citenamefont {Johnson-Freyd}\ and\ \citenamefont
  {Reutter}(2023)}]{JohReu2023:mex}%
  \BibitemOpen
  \bibfield  {author} {\bibinfo {author} {\bibfnamefont {T.}~\bibnamefont
  {Johnson-Freyd}}\ and\ \bibinfo {author} {\bibfnamefont {D.}~\bibnamefont
  {Reutter}},\ }\bibfield  {journal} {\bibinfo  {journal} {J. Am. Math. Soc.}\
  }\href {https://doi.org/10.1090/jams/1023} {10.1090/jams/1023} (\bibinfo
  {year} {2023})\BibitemShut {NoStop}%
\bibitem [{\citenamefont {Barkeshli}\ \emph {et~al.}(2019)\citenamefont
  {Barkeshli}, \citenamefont {Bonderson}, \citenamefont {Cheng},\ and\
  \citenamefont {Wang}}]{BarBonCheWan2019:prb}%
  \BibitemOpen
  \bibfield  {author} {\bibinfo {author} {\bibfnamefont {M.}~\bibnamefont
  {Barkeshli}}, \bibinfo {author} {\bibfnamefont {P.}~\bibnamefont
  {Bonderson}}, \bibinfo {author} {\bibfnamefont {M.}~\bibnamefont {Cheng}},\
  and\ \bibinfo {author} {\bibfnamefont {Z.}~\bibnamefont {Wang}},\ }\href
  {https://doi.org/10.1103/PhysRevB.100.115147} {\bibfield  {journal} {\bibinfo
   {journal} {Phys. Rev. B}\ }\textbf {\bibinfo {volume} {100}},\ \bibinfo
  {pages} {115147} (\bibinfo {year} {2019})}\BibitemShut {NoStop}%
\bibitem [{\citenamefont {Lan}(2018)}]{lan2018classification}%
  \BibitemOpen
  \bibfield  {author} {\bibinfo {author} {\bibfnamefont {T.}~\bibnamefont
  {Lan}},\ }\href@noop {} {\bibinfo {title} {A classification of (2+1)d
  topological phases with symmetries}} (\bibinfo {year} {2018}),\ \Eprint
  {https://arxiv.org/abs/1801.01210} {arXiv:1801.01210 [cond-mat.str-el]}
  \BibitemShut {NoStop}%
\bibitem [{\citenamefont {Delmastro}\ \emph {et~al.}(2021)\citenamefont
  {Delmastro}, \citenamefont {Gaiotto},\ and\ \citenamefont
  {Gomis}}]{DelGaiGom2021:anomaly}%
  \BibitemOpen
  \bibfield  {author} {\bibinfo {author} {\bibfnamefont {D.}~\bibnamefont
  {Delmastro}}, \bibinfo {author} {\bibfnamefont {D.}~\bibnamefont {Gaiotto}},\
  and\ \bibinfo {author} {\bibfnamefont {J.}~\bibnamefont {Gomis}},\ }\href
  {https://doi.org/10.1007/jhep11(2021)142} {\bibfield  {journal} {\bibinfo
  {journal} {J. High Energy Phys.}\ }\textbf {\bibinfo {volume} {2021}}\bibinfo
   {number} { (11)}}\BibitemShut {NoStop}%
\bibitem [{Note1()}]{Note1}%
  \BibitemOpen
\bibfield  {number} {  }\bibinfo {note} {In Ref.~\cite {DelGaiGom2021:anomaly},
  there are also \protect \emph {puncture states} in the R-R sector, which are
  odd under fermion parity. These states are irrelevant for our purposes ---
  they do not contribute to the modular extension, or give rise to fermionic
  RCFT characters --- so we ignore them.}\BibitemShut {Stop}%
\bibitem [{\citenamefont {Hsieh}\ \emph {et~al.}(2021)\citenamefont {Hsieh},
  \citenamefont {Nakayama},\ and\ \citenamefont
  {Tachikawa}}]{HsiNakTac2021:fermionicminimal}%
  \BibitemOpen
  \bibfield  {author} {\bibinfo {author} {\bibfnamefont {C.-T.}\ \bibnamefont
  {Hsieh}}, \bibinfo {author} {\bibfnamefont {Y.}~\bibnamefont {Nakayama}},\
  and\ \bibinfo {author} {\bibfnamefont {Y.}~\bibnamefont {Tachikawa}},\ }\href
  {https://doi.org/10.1103/PhysRevLett.126.195701} {\bibfield  {journal}
  {\bibinfo  {journal} {Phys. Rev. Lett.}\ }\textbf {\bibinfo {volume} {126}},\
  \bibinfo {pages} {195701} (\bibinfo {year} {2021})}\BibitemShut {NoStop}%
\bibitem [{\citenamefont {Kulp}(2021)}]{Kul2021:fermionicminimal}%
  \BibitemOpen
  \bibfield  {author} {\bibinfo {author} {\bibfnamefont {J.}~\bibnamefont
  {Kulp}},\ }\href {https://doi.org/10.1007/jhep03(2021)124} {\bibfield
  {journal} {\bibinfo  {journal} {J. High Energy Phys.}\ }\textbf {\bibinfo
  {volume} {2021}}\bibinfo  {number} { (3)}}\BibitemShut {NoStop}%
\bibitem [{\citenamefont {Duan}\ \emph {et~al.}(2023)\citenamefont {Duan},
  \citenamefont {Lee}, \citenamefont {Lee},\ and\ \citenamefont
  {Li}}]{DuaLeeLeeLi2023:classification}%
  \BibitemOpen
\bibfield  {number} {  }\bibfield  {author} {\bibinfo {author} {\bibfnamefont
  {Z.}~\bibnamefont {Duan}}, \bibinfo {author} {\bibfnamefont {K.}~\bibnamefont
  {Lee}}, \bibinfo {author} {\bibfnamefont {S.}~\bibnamefont {Lee}},\ and\
  \bibinfo {author} {\bibfnamefont {L.}~\bibnamefont {Li}},\ }\href
  {https://doi.org/10.1007/jhep02(2023)079} {\bibfield  {journal} {\bibinfo
  {journal} {J. High Energy Phys.}\ }\textbf {\bibinfo {volume} {2023}}\bibinfo
   {number} { (2)}}\BibitemShut {NoStop}%
\bibitem [{\citenamefont {Bae}\ and\ \citenamefont
  {Lee}(2021)}]{BaeLee2021:supersymmetry}%
  \BibitemOpen
\bibfield  {number} {  }\bibfield  {author} {\bibinfo {author} {\bibfnamefont
  {J.}~\bibnamefont {Bae}}\ and\ \bibinfo {author} {\bibfnamefont
  {S.}~\bibnamefont {Lee}},\ }\bibfield  {journal} {\bibinfo  {journal}
  {{SciPost} Phys.}\ }\textbf {\bibinfo {volume} {11}},\ \href
  {https://doi.org/10.21468/scipostphys.11.5.091}
  {10.21468/scipostphys.11.5.091} (\bibinfo {year} {2021})\BibitemShut
  {NoStop}%
\bibitem [{\citenamefont {Bae}\ \emph {et~al.}(2021{\natexlab{a}})\citenamefont
  {Bae}, \citenamefont {Duan}, \citenamefont {Lee}, \citenamefont {Lee},\ and\
  \citenamefont {Sarkis}}]{BaeDuaLee-etal2021:frcft}%
  \BibitemOpen
  \bibfield  {author} {\bibinfo {author} {\bibfnamefont {J.-B.}\ \bibnamefont
  {Bae}}, \bibinfo {author} {\bibfnamefont {Z.}~\bibnamefont {Duan}}, \bibinfo
  {author} {\bibfnamefont {K.}~\bibnamefont {Lee}}, \bibinfo {author}
  {\bibfnamefont {S.}~\bibnamefont {Lee}},\ and\ \bibinfo {author}
  {\bibfnamefont {M.}~\bibnamefont {Sarkis}},\ }\bibfield  {journal} {\bibinfo
  {journal} {Prog. Theor. Exp. Phys.}\ }\textbf {\bibinfo {volume} {2021}},\
  \href {https://doi.org/10.1093/ptep/ptab033} {10.1093/ptep/ptab033} (\bibinfo
  {year} {2021}{\natexlab{a}})\BibitemShut {NoStop}%
\bibitem [{\citenamefont {Bae}\ \emph {et~al.}(2022)\citenamefont {Bae},
  \citenamefont {Duan}, \citenamefont {Lee}, \citenamefont {Lee},\ and\
  \citenamefont {Sarkis}}]{BaeDuaLee-etal2022:bootstrap}%
  \BibitemOpen
  \bibfield  {author} {\bibinfo {author} {\bibfnamefont {J.-B.}\ \bibnamefont
  {Bae}}, \bibinfo {author} {\bibfnamefont {Z.}~\bibnamefont {Duan}}, \bibinfo
  {author} {\bibfnamefont {K.}~\bibnamefont {Lee}}, \bibinfo {author}
  {\bibfnamefont {S.}~\bibnamefont {Lee}},\ and\ \bibinfo {author}
  {\bibfnamefont {M.}~\bibnamefont {Sarkis}},\ }\href
  {https://doi.org/10.1007/jhep01(2022)089} {\bibfield  {journal} {\bibinfo
  {journal} {J. High Energy Phys.}\ }\textbf {\bibinfo {volume} {2022}}\bibinfo
   {number} { (1)}}\BibitemShut {NoStop}%
\bibitem [{\citenamefont {Benjamin}\ and\ \citenamefont
  {Lin}(2020)}]{BenLin2020:lessons}%
  \BibitemOpen
\bibfield  {number} {  }\bibfield  {author} {\bibinfo {author} {\bibfnamefont
  {N.}~\bibnamefont {Benjamin}}\ and\ \bibinfo {author} {\bibfnamefont {Y.-H.}\
  \bibnamefont {Lin}},\ }\bibfield  {journal} {\bibinfo  {journal} {{SciPost}
  Phys.}\ }\textbf {\bibinfo {volume} {9}},\ \href
  {https://doi.org/10.21468/scipostphys.9.5.065} {10.21468/scipostphys.9.5.065}
  (\bibinfo {year} {2020})\BibitemShut {NoStop}%
\bibitem [{\citenamefont {Bae}\ \emph {et~al.}(2021{\natexlab{b}})\citenamefont
  {Bae}, \citenamefont {Duan},\ and\ \citenamefont {Lee}}]{bae2021energy}%
  \BibitemOpen
  \bibfield  {author} {\bibinfo {author} {\bibfnamefont {J.-B.}\ \bibnamefont
  {Bae}}, \bibinfo {author} {\bibfnamefont {Z.}~\bibnamefont {Duan}},\ and\
  \bibinfo {author} {\bibfnamefont {S.}~\bibnamefont {Lee}},\ }\href@noop {}
  {\bibinfo {title} {Can the energy bound $e \geq 0$ imply supersymmetry?}}
  (\bibinfo {year} {2021}{\natexlab{b}}),\ \Eprint
  {https://arxiv.org/abs/2112.14130} {arXiv:2112.14130 [hep-th]} \BibitemShut
  {NoStop}%
\bibitem [{\citenamefont {Kawabata}\ and\ \citenamefont
  {Yahagi}(2023)}]{KawYah2023:codecft}%
  \BibitemOpen
  \bibfield  {author} {\bibinfo {author} {\bibfnamefont {K.}~\bibnamefont
  {Kawabata}}\ and\ \bibinfo {author} {\bibfnamefont {S.}~\bibnamefont
  {Yahagi}},\ }\href {https://doi.org/10.1007/jhep05(2023)096} {\bibfield
  {journal} {\bibinfo  {journal} {J. High Energy Phys.}\ }\textbf {\bibinfo
  {volume} {2023}}\bibinfo  {number} { (5)}}\BibitemShut {NoStop}%
\bibitem [{\citenamefont {Chen}\ \emph {et~al.}(2013)\citenamefont {Chen},
  \citenamefont {Gu}, \citenamefont {Liu},\ and\ \citenamefont
  {Wen}}]{Chen_2013}%
  \BibitemOpen
\bibfield  {number} {  }\bibfield  {author} {\bibinfo {author} {\bibfnamefont
  {X.}~\bibnamefont {Chen}}, \bibinfo {author} {\bibfnamefont {Z.-C.}\
  \bibnamefont {Gu}}, \bibinfo {author} {\bibfnamefont {Z.-X.}\ \bibnamefont
  {Liu}},\ and\ \bibinfo {author} {\bibfnamefont {X.-G.}\ \bibnamefont {Wen}},\
  }\bibfield  {journal} {\bibinfo  {journal} {Phys. Rev. B}\ }\textbf {\bibinfo
  {volume} {87}},\ \href {https://doi.org/10.1103/physrevb.87.155114}
  {10.1103/physrevb.87.155114} (\bibinfo {year} {2013})\BibitemShut {NoStop}%
\bibitem [{\citenamefont {Cheng}\ and\ \citenamefont
  {Williamson}(2020)}]{Cheng_2020}%
  \BibitemOpen
  \bibfield  {author} {\bibinfo {author} {\bibfnamefont {M.}~\bibnamefont
  {Cheng}}\ and\ \bibinfo {author} {\bibfnamefont {D.~J.}\ \bibnamefont
  {Williamson}},\ }\bibfield  {journal} {\bibinfo  {journal} {Phys. Rev.
  Research}\ }\textbf {\bibinfo {volume} {2}},\ \href
  {https://doi.org/10.1103/physrevresearch.2.043044}
  {10.1103/physrevresearch.2.043044} (\bibinfo {year} {2020})\BibitemShut
  {NoStop}%
\bibitem [{\citenamefont {Wang}\ and\ \citenamefont {Chen}(2017)}]{Wang_2017}%
  \BibitemOpen
  \bibfield  {author} {\bibinfo {author} {\bibfnamefont {Z.}~\bibnamefont
  {Wang}}\ and\ \bibinfo {author} {\bibfnamefont {X.}~\bibnamefont {Chen}},\
  }\bibfield  {journal} {\bibinfo  {journal} {Physical Review B}\ }\textbf
  {\bibinfo {volume} {95}},\ \href {https://doi.org/10.1103/physrevb.95.115142}
  {10.1103/physrevb.95.115142} (\bibinfo {year} {2017})\BibitemShut {NoStop}%
\bibitem [{\citenamefont {You}(2023)}]{you2023gapped}%
  \BibitemOpen
  \bibfield  {author} {\bibinfo {author} {\bibfnamefont {M.}~\bibnamefont
  {You}},\ }\href@noop {} {\bibinfo {title} {Gapped boundaries of fermionic
  topological orders and higher central charges}} (\bibinfo {year} {2023}),\
  \Eprint {https://arxiv.org/abs/2311.01096} {arXiv:2311.01096
  [cond-mat.str-el]} \BibitemShut {NoStop}%
\end{thebibliography}%

\end{document}